\documentclass[11pt,a4paper]{article}
\pdfoutput=1

\usepackage{jheppub-nosort,amsthm}
\PassOptionsToPackage{svgnames}{xcolor}
\usepackage{xcolor}
\usepackage{amsmath,amssymb,amsthm,amscd}
\usepackage{physics}
\usepackage{tikz}
\tikzstyle{arrow} = [->]
\usetikzlibrary{decorations.pathmorphing}
\tikzset{node distance=2cm, auto}
\tikzset{snake it/.style={decorate, decoration=snake}}
\usepackage{epsfig}
\usepackage{psfrag,comment}
\usepackage{graphicx}
\usepackage{caption}
\usepackage{subcaption}
\usepackage{mathrsfs}
\usepackage[normalem]{ulem}
\usepackage{bm}
\usepackage{lscape}
\usepackage{diagbox}
\usepackage{upgreek}
\usepackage{float}
\usepackage{url}
\newcommand\doilink[1]{\href{http://dx.doi.org/#1}{#1}}
\newcommand\arxivlink[1]{\href{http://arxiv.org/abs/#1}{#1}}

\newcommand{\CB}{{\mathcal B}}
\newcommand{\CC}{{\mathcal C}}

\newcommand{\CE}{{\mathcal E}}
\newcommand{\CF}{{\mathcal F}}
\newcommand{\CG}{{\mathcal G}}

\newcommand{\CK}{{\mathcal K}}

\newcommand{\CN}{{\mathcal N}}
\newcommand{\CO}{{\mathcal O}}
\newcommand{\CP}{{\mathcal P}}
\newcommand{\CQ}{{\mathcal Q}}

\newcommand{\CW}{{\mathcal W}}

\newcommand{\CZ}{{\mathcal Z}}

\newcommand{\NCC}{{\mathscr C}}

\newcommand{\NCO}{{\mathscr O}}
\newcommand{\NCP}{{\mathscr P}}

\newcommand{\NCZ}{{\mathscr Z}}

\def\BN{{\mathbb N}}
\def\BZ{{\mathbb Z}}
\def\BR{{\mathbb R}}
\def\BC{{\mathbb C}}
\def\BP{{\mathbb P}}
\def\BT{{\mathbb T}}
\def\BS{{\mathbb S}}
\def\BH{{\mathbb H}}
\def\BL{{\mathbb L}}
\def\BK{{\mathbb K}}

\newcommand{\be}{\begin{equation}}
\newcommand{\ee}{\end{equation}}
\newcommand{\ba}{\begin{aligned}}
\newcommand{\ea}{\end{aligned}}
\newcommand{\bea}{\begin{eqnarray}}
\newcommand{\eea}{\end{eqnarray}}
\newcommand{\bean}{\begin{eqnarray*}}
\newcommand{\eean}{\end{eqnarray*}}

\newdimen\tableauside\tableauside=1.0ex
\newdimen\tableaurule\tableaurule=0.4pt
\newdimen\tableaustep
\def\phantomhrule#1{\hbox{\vbox to0pt{\hrule height\tableaurule width#1\vss}}}
\def\phantomvrule#1{\vbox{\hbox to0pt{\vrule width\tableaurule height#1\hss}}}
\def\sqr{\vbox{%
  \phantomhrule\tableaustep
  \hbox{\phantomvrule\tableaustep\kern\tableaustep\phantomvrule\tableaustep}%
  \hbox{\vbox{\phantomhrule\tableauside}\kern-\tableaurule}}}
\def\squares#1{\hbox{\count0=#1\noindent\loop\sqr
  \advance\count0 by-1 \ifnum\count0>0\repeat}}
\def\tableau#1{\vcenter{\offinterlineskip
  \tableaustep=\tableauside\advance\tableaustep by-\tableaurule
  \kern\normallineskip\hbox
    {\kern\normallineskip\vbox
      {\gettableau#1 0 }%
     \kern\normallineskip\kern\tableaurule}%
  \kern\normallineskip\kern\tableaurule}}
\def\gettableau#1{\ifnum#1=0\let\next=\null\else
\squares{#1}\let\next=\gettableau\fi\next}
\def\r{\right\rangle}

\def\1{\mathbf{1}}
\def\0{|\1\r}

\newcommand{\rme}{{\mathrm{e}}}
\newcommand{\rmi}{{\mathrm{i}}}
\newcommand{\rmd}{{\mathrm{d}}}

\renewcommand{\mod}{\mbox{~mod~}}

\def\XXint#1#2#3{{\setbox0=\hbox{$#1{#2#3}{\int}$}
     \vcenter{\hbox{$#2#3$}}\kern-.5\wd0}}

\newenvironment{lcases}
  {\left\lbrace\begin{aligned}}
  {\end{aligned}\right.}

\newcommand{\PI}{{\textsf{P}$_{\textsf{I}}$}}

\newcommand{\YL}{{\textsf{YL}}}

\makeatletter
\newsavebox\myboxA
\newsavebox\myboxB
\newlength\mylenA

\newcommand*\widebar[2][0.75]{%
    \sbox{\myboxA}{$\m@th#2$}%
    \setbox\myboxB\null
    \ht\myboxB=\ht\myboxA%
    \dp\myboxB=\dp\myboxA%
    \wd\myboxB=#1\wd\myboxA
    \sbox\myboxB{$\m@th\overline{\copy\myboxB}$}
    \setlength\mylenA{\the\wd\myboxA}
    \addtolength\mylenA{-\the\wd\myboxB}%
    \ifdim\wd\myboxB<\wd\myboxA%
       \rlap{\hskip 0.8\mylenA\usebox\myboxB}{\usebox\myboxA}%
    \else
        \hskip -0.5\mylenA\rlap{\usebox\myboxA}{\hskip 0.5\mylenA\usebox\myboxB}%
    \fi}
\makeatother

\definecolor{cottoncandy}{rgb}{1.0, 0.74, 0.85}
\definecolor{cornellred}{rgb}{0.7, 0.11, 0.11}
\definecolor{darktangerine}{rgb}{1.0, 0.66, 0.07}
\definecolor{deepsaffron}{rgb}{1.0, 0.8, 0.6}

\title{Exact Solutions to Matrix Models and String Theories: The Local Construction}

\author[a]{Jasper~Kager,}
\affiliation[a]{CAMGSD, Departamento de Matem\'atica, Instituto Superior T\'ecnico,\\ Universidade de Lisboa, 1049-001 Lisboa, Portugal}
\emailAdd{jasper.kager@}

\author[a]{Jo\~ao~Rodrigues,}
\emailAdd{joao.carlos.rodrigues@}

\author[a]{Ricardo~Schiappa,}
\emailAdd{ricardo.schiappa@}

\author[b]{Maximilian~Schwick,}
\affiliation[b]{Albert Einstein Center for Fundamental Physics, Institute for Theoretical Physics,\\ University of Bern, CH-3012 Bern, Switzerland}
\emailAdd{maximilian.schwick@unibe.ch}

\author[a]{Noam~Tamarin\,}
\emailAdd{noam.tamarin@tecnico.ulisboa.pt}

\abstract{Exact nonperturbative solutions to hermitian one-matrix models, their topological string duals, as well as their double-scaling limits to multicritical and minimal string theories, may be obtained via the use of resurgent transseries. These solutions are generically resonant, entailing both eigenvalues and anti-eigenvalues, or, equivalently, both D-branes and negative-tension D-branes---but are otherwise intricate to write down, having been previously addressed on a case-by-case approach. This work shows how there is a general and rather compact way to write down all these exact and fully nonperturbative transseries solutions in closed-form, immediately starting from the spectral geometry of the matrix model or string theory at hand, in the form of a discrete Fourier or Zak transform for their partition functions. This structure is inherently associated to the existence of anti-eigenvalues or negative-tension D-branes. The validity of these solutions is testable across all values of the parameters---from weak to strong 't~Hooft coupling, from small to large $N$; equivalently, from semi-classical to deeply quantum regimes---and many such nonperturbative tests are performed against diverse examples ranging from matrix models to non-critical strings, fully validating our analytical and exact expressions. In particular, anti-eigenvalues or negative-tension D-branes are absolutely required to find sharp numerical matches. In order to study these solutions globally across their phase diagrams, however, complete non-linear Stokes data is still needed---which will be addressed in a complementary follow-up paper.}


\begin{document}

\maketitle

\vfill

\eject

\allowdisplaybreaks

\section{Introduction and Summary}
\label{sec:intro-summary}

Exact yet non-trivial solutions to modern physical theories have played prominent roles in the past. Textbook famous are Schwarzschild's solution for spherically symmetric black holes in general relativity \cite{s16} or Schr\"odinger's solution for the spectrum of the hydrogen atom in quantum mechanics \cite{s26}. Quantum field theory already poses a larger obstacle, as exact nonperturbative solutions are harder to come by: whilst interesting examples exist in two dimensions, results are much sparser in four dimensions. On this note, unsurprisingly, the string theoretic scenario comes across a bit grimmer. Whereas motivated by \cite{m97} one might wonder of exact nonperturbative solutions to string theory in anti-de~Sitter spacetimes---in both string length and coupling---, our work only takes the first steps in such direction by tackling much simpler settings: constructing \textit{exact}, \textit{fully nonperturbative} solutions to (large classes of) \textit{minimal} and \textit{topological} string theories.

There are two key aspects behind our constructions. One is that all the models we solve are based upon an underlying random \textit{matrix model}, and the second is that these matrix models precisely turn out to be fully and exactly solvable via the use of \textit{resurgence} and \textit{transseries}.

Since its rise to fame in the pioneering work by Wigner---on the statistical study of heavy-nuclei spectra \cite{w55}---random matrix theory has been at the forefront in the study of complex quantum systems. This prominent role gained further strength from the Bohigas--Giannoni--Schmit conjecture \cite{bgs84}, which turned random matrices ubiquitous across the study of classically-chaotic quantum-systems. As zero-dimensional gauge-theories time-and-again addressed from the standpoint of the large $N$ expansion \cite{th74}, it was later only natural when matrix models made their way into (double-scaled) solutions of (non-critical) string theories \cite{gm90a, ds90, bk90, d90, gm90b}. This was in fact one of the earliest examples of a string-theoretic holographic duality \cite{m97}; the nature of which was extended off-criticality and made fully clear in \cite{dv02a, dv02b, emo07}---where the topological-string B-model on certain non-compact Calabi--Yau target-space geometries (built upon the matrix-model spectral-curve) was found to undergo a large $N$ duality into the corresponding (off-critical) hermitian one-matrix model. With holography now at center-stage, another large and very interesting class of double-scaled matrix-models quickly unfolded, the minimal string theories in \cite{ss03, kopss04, mmss04, ss04b}. Finally, saturating the Maldacena--Shenker--Stanford bound \cite{mss15} black holes likely are maximally quantum-chaotic systems, hence inherently also describable by random matrix theory. Perhaps one of the best illustrations of this aspect is their near-horizon near-extremal description via Jackiw--Teitelboim two-dimensional dilaton gravity \cite{t83, j85, ap14} that itself ends up localizing on yet another very interesting class of double-scaled matrix models, albeit which start departing from the above hermitian realm \cite{sss19, sw19, tw23}.

Nonperturbative corrections to the (double-scaled) perturbative-free-energy large-$N$ expansion were first found in \cite{d91, d92}, in the guise of eigenvalue tunneling. These nonperturbative effects turn the large-$N$ perturbative-expansion \textit{asymptotic}, with the closed string-theoretic counterpart \cite{gp88} realized by D-brane nonperturbative effects \cite{s90, p94, p95}. In the context of minimal string theory, these have been very thoroughly studied and divide into ZZ- \cite{zz01} and FZZT- \cite{fzz00, t00} branes. Large-order tests of such asymptotic behavior were first carried through in the non-critical or multicritical context \cite{gz90b, gz91, ez93}. But the \textit{resurgent} nature of the large $N$---correspondingly, the closed-string---expansion was only uncovered over a decade later.

Asymptotic series are resurgent if their Borel transforms admit endless analytic continuation \cite{e81, e84, e93}. This seemingly simple definition has far reaching ramifications. Singularities on the Borel plane of a given asymptotic series encode the aforementioned nonperturbative effects---eigenvalue tunneling, D-branes, whatever else---with complete information: their locations yield the weights of the transmonomials we need to add to the original asymptotic-series in order to construct a transseries-solution; and their ``generalized residues'' (singularities are usually branch-points; see, \textit{e.g.}, \cite{abs18}) yield the corresponding nonperturbative asymptotic-series associated to said transmonomials. All together---with all transmonomial contributions, from all Borel singularities---this construct gives rise to an exact nonperturbative transseries-solution of whatever problem at hand. Because Borel singularities have complete information, and given we can run this game for any chosen asymptotic series in the full transseries, then these nonperturbative contributions ``talk to each other'' via resurgence---they are \textit{resurgent transseries} \cite{e81, e84, e93}.

It is in the above context that the \textit{resurgent} nature of the large $N$, or closed string, asymptotic expansions has been unveiling over the years---albeit in a long series of case-by-case computations. Once the eigenvalue-tunneling description of nonperturbative matrix-model effects was understood off-criticality \cite{m06, msw07, msw08, ps09}, resurgence could finally be used to systematically assemble all this information coherently \cite{m08}. As first found for the Painlev\'e~I problem \cite{gikm10, kmr10}, and subsequently systematically investigated in many off-critical and double-scaled hermitian matrix model examples\footnote{The topological string theories we address in the present work are exclusively associated to the simpler target-space Dijkgraaf--Vafa classes of geometries \cite{dv02a}. Resurgent transseries have also been studied for more complicated local Calabi--Yau geometries such as toric manifolds, \textit{e.g.}, \cite{m06, msw07, ps09, dmp11, cesv13, gmz14, cesv14, c15, csv16, cms16, cms17, gm21, gm22a, gm22b, gkkm23, im23, ms24}, where the same conclusion holds.} \cite{asv11, sv13, as13, gs21, bssv22, mss22, sst23, eggls23}, all string-theoretic or large-$N$ transseries solutions at play are \textit{resonant}. Resonance has been shown to be a generic feature of multicritical and minimal string theories \cite{gs21} and of hermitian one-matrix models \cite{mss22}, and it is likely a property of any string-theoretic expansion \cite{sst23}. What this means, at the transseries level, is that all instanton actions must always appear in symmetric pairs. At the physical level, what this means is that there are always nonperturbative effects associated to anti-eigenvalues \cite{mss22} or negative-tension D-branes \cite{sst23}, and that this is a required feature of matrix-model/string-theoretic resurgence. But if resonance is generic, as are anti-eigenvalues/negative-branes, can we not also find how to write-down \textit{generic solutions} based on this simple yet universal structural input?

This is what we shall do in the present work, showing how all the above transseries examples (and, in fact, any other cases one might be interested-in; based upon spectral curves associated to hermitian one-matrix models, their topological string duals, or their limits towards multicritical and minimal string theories, or even arbitrary two-dimensional topological gravities), they are all writable in rather compact closed-forms by making use of a discrete Fourier or Zak transform representation for their partition functions. Our exact solutions will be presented in section~\ref{sec:resurgent-Z-transseries} (the avid reader may enjoy a sneak preview in formulae \eqref{eq:pinchedcubic}, \eqref{eq:pinchedcubic-dsl}, or \eqref{eq:pinchedmulti}). In particular, they yield solutions to the full matrix integral as computed from finite-difference (off-criticality) or differential (double-scaled) string equations, which will be set-up in section~\ref{sec:strong-coupling-phases}. Precisely because we can track back these constructions to solutions of a string equation, this means that, from the point-of-view of the matrix model, they are being constructed around the spectral-geometry \textit{one}-cut background, which is discussed in section~\ref{sec:strong-coupling-phases}. But this is really just a (very) convenient choice of reference background: as will be discussed in \cite{ss26}, the fully resurgent, resonant, nonperturbative large $N$ expansion is in fact background-independent and \textit{any} other choice of \textit{multi}-cut reference-solution would have worked as well.

It is important to stress that the discrete Fourier/Zak transform representations for the partition-function transseries we build in this paper are just that: rather generic and very compact closed-forms for the \textit{transseries solutions}. In other words, and as any transseries alone, they are just \textit{local} solutions. They depend on transseries parameters which encode initial/boundary data\footnote{In particular, we can fix these data and study particular solutions; for example the physical matrix-model solution where eigenvalues satisfy the Coulomb-gas equilibrium physical requirements. Of course we can also study more general solutions, which are not the equilibrium ones; say, in the spirit of (generalized) Boutroux classifications for string-equation solutions---see the discussion in subsection~\ref{subsec:DSL-phases}.}, but which also undergo Stokes transitions connecting all these local solutions, in distinct Stokes wedges, to each other. As such, and in order to become fully \textit{global} solutions, valid everywhere across their phase diagrams---from weak to strong 't~Hooft coupling, crossing phase boundaries; from the semi-classical large $N$ to the deeply-quantum small $N$ regimes---these resurgent transseries still require complete non-linear Stokes data. These resurgence data enable the necessary Stokes phenomena via jumps (or connection formulae) in the transseries parameters, at the Stokes lines (see, \textit{e.g.}, \cite{abs18}); but for our problems they are given by an infinite number of (transcendental) coefficients \cite{gikm10, asv11, sv13, as13, gs21}, whose generating functions have only been recently uncovered for the Painlev\'e~I/II cases \cite{bssv22} (and semi-classically explained via matrix-integrals in \cite{mss22} and boundary conformal field theory in \cite{sst23}, but therein without immediate generating functions). The computation of these Stokes data for the large classes of random matrix models we address will be done in a complementary paper \cite{krsst26b}. These generic resurgent Stokes data will include their generating functions, alongside their connection formulae implementing Stokes transitions---hence they will take our present \textit{local} constructions and glue them across the full moduli space of distinct Stokes regions, establishing \textit{global} solutions; valid for any value of $N$ or of the string coupling $g_{\text{s}}$ and any value of the 't~Hooft coupling $t \in \BC$.

This paper is the first in a series of tentatively seven papers fully constructing exact solutions for large classes of matrix models and string theories. As just mentioned, the complementary paper \cite{krsst26b} will move us from local to global solutions, at the level of the partition-function (hence, the free-energy), by computing complete non-linear resurgent Stokes data. These exact and global solutions for matrix-model/string-theory free-energies are shown to be background independent in \cite{ss26}, in the sense that they yield the same (exact) function regardless of the reference background we decide to start from. Further, their two-dimensional topological gravity versions interestingly connect with Argyres--Douglas theories \cite{ad95, apsw95}, as will be discussed in \cite{ss26, krst26a}. There are also many interplays between the Stokes data of these global solutions, their monodromy properties (in the sense of generalized Painlev\'e properties; see, \textit{e.g.}, \cite{bssv22}), and wall-crossing formulae and cluster algebras, to be further explored in \cite{krst26b}. Having set solid ground with the construction of exact partition functions, one may move towards other observables and their correlation functions. In \cite{krs26} we extend our analysis to exact, nonperturbative FZZT D-brane amplitudes and their correlation functions, which are essentially multi-determinant insertions in the matrix integral; and finally in \cite{ccfks26} we extend our analysis to exact, nonperturbative multi-resolvent correlation functions. This essentially solves any random hermitian one-matrix model, their topological-string duals \cite{dv02a, dv02b}, and their double-scaled limits---including any multicritical, minimal, or topological-gravity string-equation built upon arbitrary choices of KdV times and corresponding Gelf'fand--Dikii polynomials \cite{gd75, mss91, ss03, dw18, gs21}. It also includes the JT matrix model of \cite{sss19}, which will be addressed in \cite{krst26a, krs26, ccfks26} (in particular, this could provide the remainder resonant sectors in the KdV transseries analysis in \cite{hmo25}). One other interesting aspect is that, via the Miura map, one can in principle start from our classes of hermitian matrix models and obtain results for unitary matrix models along the modified KdV hierarchy; see, \textit{e.g.}, \cite{kms03b, v23}. It would be very interesting to make this map explicit in the future. In particular, this could also open a very interesting door possibly connecting to the recent work in, \textit{e.g.}, \cite{emms22a, emms22b, e23, emm23, cemm24, cmt24, emt24}. Finally, a reinterpretation of our partition-functions as Fredholm determinants associated to specific quantum-mechanical problems, and the study of the relation between their corresponding  spectra and our specific-heat poles in the spirit of, \textit{e.g.}, \cite{dt00, m09, m10} would also be an extremely interesting venue of future research.

The contents of this paper are summarized as follows. We begin in section~\ref{sec:strong-coupling-phases} by asking ourselves what must analytical nonperturbative solutions to matrix models and string theories exactly do for us. A lot of non-trivial numerical information concerning matrix models exists or can be directly produced, which we shall compute in this section---and that should now be fully reproduced from an analytical standpoint. Of particular interest will be the strong-coupling regimes (strong 't~Hooft coupling and/or small $N$; correspondingly large string-coupling $g_{\text{s}}$) of matrices and strings, where traditionally analytical methods struggle to produce sharp results. Having set-up the matrix-model stage, we first discuss their (off-critical) solutions via orthogonal polynomials in subsection~\ref{subsec:OP-phases}. For our canonical examples of cubic and quartic matrix models, solutions to the orthogonal-polynomial recursion-coefficients string-equations \eqref{eq:prestringequationcubticmatrixmodel} and \eqref{eq:prestringequationquarticmatrixmodel} unveil a rich set of diverse large $N$ phases, with rather distinct properties as a quick glance at figures~\ref{fig:CMMrDataChaosPlot},~\ref{fig:CMMrNDataAndOPRoots}, and~\ref{fig:QMMrNDataAndOPRoots} illustrates. The double-scaled version of these results is then addressed in subsection~\ref{subsec:DSL-phases}. We are now dealing with differential string-equations of which we focus upon the cases of the Painlev\'e~I and Yang--Lee non-linear ordinary differential equations in \eqref{eq:Painleve1Equation} and \eqref{eq:YangLeeEquation}. Their rich set of diverse phases is now best illustrated by the locations of their (movable) poles, in a generalized version of the famous Boutroux classification of Painlev\'e~I solutions \cite{b13, b14}. Another quick peek at figures~\ref{fig:PISolutionsNoZeros} and~\ref{fig:YLSolutionsNoZeros} clearly illustrates this point. It is important to stress how complete nonperturbative information is absolutely vital in order to analytically reach these diverse phases, either off-criticality (the oscillations in figures~\ref{fig:CMMrDataChaosPlot},~\ref{fig:CMMrNDataAndOPRoots}, and~\ref{fig:QMMrNDataAndOPRoots}) or in the double-scaling limit (the poles in figures~\ref{fig:PISolutionsNoZeros} and~\ref{fig:YLSolutionsNoZeros}). Having discussed string equations, we turn to spectral geometry in subsection~\ref{subsec:SG-phases}. The aforementioned distinct phases are now understood as different arrangements of eigenvalue configurations, as described by (hyperelliptic) multi-cut spectral curves as in figure~\ref{fig:multicutPotential} or equation \eqref{eq:spectral-curve-moment-function}, and as illustrated in figures~\ref{fig:SpecGeofig:eigcmm} and~\ref{fig:SpecGeofig:eigqmm} for the cubic and quartic matrix models, respectively. A global vision of all this rich  set of distinct large $N$ phases is finally made clear in subsection~\ref{subsec:stokes-vs-phases}; with phase diagrams for any value of the 't~Hooft coupling $t \in \BC$, for cubic and quartic matrix models, illustrated in figures~\ref{fig:CMM Phase Diagram} and~\ref{fig:QMM Phase Diagram}. Expectations concerning the partition-function asymptotics across all these distinct phases are at last discussed in subsection~\ref{subsec:Z-phases}. When discarding for resonance (to simplify, and to use previous results in the literature), there are basically two types of asymptotics to expect: ``regular'' large $N$ asymptotics \cite{msw08}, as in \eqref{eq:partition-function-Z-msw08}, and ``oscillatory'' large $N$ asymptotics \cite{bde00, e08, em08}, as in \eqref{eq:EM-2-cut-Z}; and all these different asymptotics should be completely encoded in our exact transseries solutions. Section~\ref{sec:strong-coupling-phases} is supported by two appendices. In appendix~\ref{app:elliptic-theta-modular} we briefly review the basics of hyperelliptic spectral curves, relevant from subsection~\ref{subsec:SG-phases} onwards; as well as the basics of their related theta-functions, relevant for the partition function from subsection~\ref{subsec:Z-phases} onwards. In appendix~\ref{appendix:SpectralGeometryMatrixModels} we construct multi-cut spectral-curves for the cubic (in sub-appendix~\ref{subappendix:cubicmatrixmodel}) and quartic (in sub-appendix~\ref{subappendix:quarticmatrixmodel}) matrix models, which are required from subsection~\ref{subsec:SG-phases} onwards.

Moving on towards section~\ref{sec:resurgent-Z-transseries}, we are ready to present our main proposal for resonant, resurgent-transseries solutions for large classes of matrix models and string theories. The first point to highlight upon is the genericness of resonance. Due to the existence of anti-eigenvalues \cite{mss22} or negative-tension D-branes \cite{sst23} the complete resonant transseries is much larger than naively expected at first, as illustrated in figure~\ref{fig:twopararectmtransseriesgrid} (where the naive expectation would correspond to just the first ``purely {\color{blue}blue} grid''). In equations, this amounts to \eqref{eq:genericresonanttransseries}. But precisely because instanton actions are herein organized in \textit{symmetric} pairs, there is another way to write this transseries. Instead of the ``rectangular framing'' \eqref{eq:genericresonanttransseries} one may use the resonant kernel direction \cite{asv11, abs18, gs21, bssv22} to rewrite the transseries in ``diagonal framing'' \eqref{eq:PreDiagonalFramingTransseries}, as illustrated in figure~\ref{fig:rectangulartodiagonal}. This might seem like a meaningless change of transmonomial organization, except for the fact that on top one may still exchange perturbative-genus and nonperturbative-instanton sums, in which case the latter sum can actually be evaluated in closed-form and partition-function transseries hence see their overall structure get dramatically simplified. In subsection~\ref{subsec:resurgent-Z-transasymptotics} we illustrate this procedure with many explicit such examples, as are the cases of the Painlev\'e~I equation \eqref{eq:PIDiscreteFourierNC}, the Yang--Lee equation \eqref{eq:YLFullDFT}, the cubic matrix model \eqref{eq:CMMDiscreteFourierNC}, and the quartic matrix model \eqref{eq:QMMDiscreteFourierNC}. Other examples will be featured in \cite{krst26a, krst26b}. Clearly, all final expressions for their partition functions are of the exact same type, which directly leads to subsection~\ref{subsec:resurgent-Z}. Herein, a very simple eigenvalue--anti-eigenvalue tunneling argument---generalizing the discussion in \cite{msw08}---immediately leads to the general discrete Fourier/Zak transform expression for the partition-function, which was repeatedly found throughout the examples of the preceding subsection. This can be summarized in formulae \eqref{eq:pinchedcubic}, \eqref{eq:pinchedcubic-dsl}, and \eqref{eq:pinchedmulti}. These expressions also ``explain'' the origin of anti-eigenvalues or negative-tension branes as arising from ``fractional'' eigenvalue tunneling. All that it takes in order to explicitly write-down our formulae in any different example is a matrix-model or string-theoretic spectral-curve, upon which---once having implemented the \textit{unpinching} procedure explained in the main text---one can run the topological recursion \cite{eo07a} to generate all required data; hence these expressions are final. We point out that partition functions in the guise of discrete Fourier transforms have previously been discussed in the literature, and we compare such discussions to our own proposal in subsection~\ref{subsec:from-theta-to-dual-to-transseries}. Finally, one neat punch-line from our formulae is to produce explicit solutions for all multicritical string equations along the KdV hierarchy, which we do in subsection~\ref{subsec:KdV-solutions}. Section~\ref{sec:resurgent-Z-transseries} is supported by two appendices: appendix~\ref{app:elliptic-theta-modular} with theta-function basics relevant for the partition-function transseries; and appendix~\ref{app:transasymptotic-transseries} where we have hidden most intricate transseries computations. In appendix~\ref{app:transasymptotic-transseries} we describe all rather technical calculations which support the examples in subsection~\ref{subsec:resurgent-Z-transasymptotics}: the Painlev\'e~I equation (in sub-appendix~\ref{subapp:transasymptotic-transseries-PI}), the Yang--Lee equation (in sub-appendix~\ref{subapp:transasymptotic-transseries-YL}), the cubic matrix model (in sub-appendix~\ref{subapp:transasymptotic-transseries-CMM}), and the quartic matrix model  (in sub-appendix~\ref{subapp:transasymptotic-transseries-QMM}).

As explained in section~\ref{sec:resurgent-Z-transseries}, our starting point is simple: a spectral-curve for a matrix model or string theory, where the data that goes into the discrete Fourier/Zak transform is essentially generated by running the topological recursion \cite{eo07a} upon said \textit{unpinched}, \textit{multi}-cut spectral-curve. But this results in the transseries obtainable from the \textit{one}-cut string equation. There are hence \textit{two} free energies at play, the \textit{multi}-cut and the \textit{one}-cut, and, as such, in section~\ref{sec:topological-recursion} we swiftly explore their interplays. We discuss these two viewpoints in subsection~\ref{subsec:two-viewpoints-NP-sectors} in the particular two-cut case, and obtain explicit formulae for how to move in-between them in subsection~\ref{subsec:secondformulaF}. The generalization to multi-cuts is then briefly addressed in subsection~\ref{subsec:multi-cuts-F}. Section~\ref{sec:topological-recursion} is supported by appendix~\ref{appendix:Allordersaddlepoint} where the required (and rather involuted) computations concerning the cubic matrix model---which is the sole example we discuss in this section---are stored.

One of our main points is that our proposal must be testable, from perturbative to nonperturbative regimes, from weak to strong 't~Hooft coupling, from classical to deeply quantum realms. We start such tests in section~\ref{sec:checks-tests-numerics}, only to succeed purely at the local level. This allows us to end this paper with a cliffhanger which hopefully will spark the interest of the reader into \cite{krsst26b}, where we move from local to global solutions and hence fully test our proposal everywhere on their corresponding phase diagrams. The first point to address is how to move from our formal, asymptotic series, with both exponential suppressed \textit{and}
exponentially enhanced multi-instanton contributions, into actual \textit{finite} numbers which may be compared to numerics. This is discussed in subsection~\ref{subsec:numerical-resummation-details}, and is a simple procedure: asymptotic series are dealt-with via standard Borel resummation (see, \textit{e.g.}, \cite{abs18}) whereas multi-instanton sums are actually \textit{convergent}---and, should one wish, may be evaluated in closed-form. We illustrate these regions of convergence in figures~\ref{fig:HMM Convergence Regions} and~\ref{fig:CMMPosNegMixTensionConvergenceRegions} for the cubic matrix model, and in figure~\ref{fig:PI Convergence Regions} for the case of the Painlev\'e~I equation. In comparing our analytical results to numerics, one further requires an adequate numerical method. For us, this will be the Fornberg--Weideman algorithm \cite{fw11, fw14, sv22} which we briefly describe in subsection~\ref{subsec:numerical-dsl-details}. Knowing how to evaluate our local solutions and knowing how to produce numerical results for the systems they are supposed to fully describe, then all that is left to do is to actually carry through the comparisons. This is first done in subsection~\ref{subsec:matrix-model-numerics} for the case of cubic and quartic hermitian matrix models, and then continued in subsection~\ref{subsec:dsl-string-eq-numerics} for the non-critical string-theoretic examples described by the Painlev\'e~I and the Yang--Lee equations. Figure~\ref{fig: CMM failure unfixed} shows a comparison between our transseries results for the cubic matrix model, and its finite-$N$ orthogonal-polynomial recursion coefficients. The match starts drifting away from being excellent as soon as the backward Stokes line is crossed (and herein ignored). A similar comparison is done for the quartic matrix model in figure~\ref{fig: QMM failure unfixed}, where now the mismatch after the backward Stokes line is even more dramatic. These two tests clearly illustrate how one \textit{cannot} ignore Stokes transitions, and, in particular, how one \textit{cannot} ignore Stokes transitions which turn-on \textit{negative-tension} transmonomial contributions to the transseries. Finally, closing the circle back to our opening plots in figures~\ref{fig:CMMrNDataAndOPRoots} and~\ref{fig:QMMrNDataAndOPRoots}, we test the capability of our transseries in reproducing all oscillations of orthogonal-polynomial recursion-coefficients in figure~\ref{fig:HMMFiniteNVsQuadraticTransasymptotics} (for both cubic and quartic models). The results are in line with all aforementioned checks, again highlighting how Stokes transitions must be a key component of our complete story. The double-scaled tests of subsection~\ref{subsec:dsl-string-eq-numerics} compare predictions for the movable double-pole singularities of multicritical string equations. For the case of the Painlev\'e~I equation this is illustrated in figure~\ref{fig:PIFailureSolutions}, where the red-dot analytical-predictions are spot-on into the numerical contour-plot circles, as long as one does not cross Stokes lines. The difference between considering and not-considering negative-tension contributions is also rather evident, with the former producing much better results. This is even more evident for the Yang--Lee problem illustrated in figures~\ref{fig:YLFailureSolutionsNoStokesYesNeg} (with negative-tension transmonomial contributions) and~\ref{fig:YLFailureSolutionsNoStokesNoNeg} (without negative-tension contributions). Entire wedges of correctly-predicted zero-pole-pairs in figure~\ref{fig:YLFailureSolutionsNoStokesYesNeg} are completely absent in figure~\ref{fig:YLFailureSolutionsNoStokesNoNeg}. All effects of both forward and backward Stokes transitions will be addressed and included in \cite{krsst26b} producing globally exact solutions to matrix models and string theories.

\section{On the Strong-Coupling Phases of Matrices and Strings}
\label{sec:strong-coupling-phases}

All models we solve are, either directly or even slightly indirectly via double-scaling, based upon an underlying random matrix set-up (\textit{i.e.}, upon a spectral curve). It is hence natural to begin with a few matrix model conventions, and to both first review and then further explore their different phases---which must obviously be all fully reproducible from any proposed exact solution.

Given a potential $V(M)$, the partition function $\CZ$ for the corresponding $N \times N$ hermitian one-matrix model is given by
\bea
\label{eq:partitionfunctionhermitianmatrix}
\mathcal{Z} \left(N, g_{\text{s}}\right) &=& \frac{1}{\text{vol}\left(\text{U}(N)\right)} \int \rmd M\, \rme^{-\frac{1}{g_{\text{s}}} \tr V(M)} = \\
\label{eq:partitionfunctioneigenvalues}
&=& \frac{1}{N!} \int \prod_{i=1}^{N} \frac{\rmd \lambda_i}{2\pi}\, \Delta^2 (\lambda)\, \rme^{-\frac{1}{g_{\text{s}}} \sum\limits_{i=1}^{N} V(\lambda_i)},
\eea
\noindent
where $g_{\text{s}}$ is the string coupling, we have normalized by the volume of the gauge group, and in the second line we have fixed diagonal gauge with eigenvalues $\left\{ \lambda_i \right\}$ and Vandermonde determinant $\Delta (\lambda) = \prod_{i<j} \left( \lambda_i - \lambda_j \right)$. All multi-trace correlation functions further follow from their generating functions: the $h$-point multi-resolvent correlators (herein $(\text{c})=$ connected)
\be
\label{eq:multiresolventcorrelationfunctionshermitianmatrix}
\CW_h \left( p_1, \ldots p_h \right) = \ev{\tr \frac{1}{p_1-M}\, \cdots\, \tr \frac{1}{p_h-M}}_{(\text{c})}.
\ee
\noindent
In evaluating all these quantities, one is usually interested in the 't~Hooft large $N$ limit \cite{th74}, taken at some fixed 't~Hooft coupling $t = g_{\text{s}} N$. In this limit, both free energy $\CF = \log \CZ$ and multi-resolvent correlation functions \eqref{eq:multiresolventcorrelationfunctionshermitianmatrix} commonly have asymptotic, perturbative genus expansions\footnote{Using the notational conventions in \cite{mss22}, we denote matrix-integral results with the use of curly fonts, as in $\mathcal{Z}$, and string-equation transseries results regularly, as in $Z$.}
\bea
\label{eq:genusgfreeenergies}
\CF &\simeq& \sum_{g=0}^{+\infty} \CF_g (t)\, g_{\text{s}}^{2g-2}, \\
\label{eq:genusgmultiresolvents}
\CW_h \left( p_1, \ldots p_h \right) &\simeq& \sum_{g=0}^{+\infty} \CW_{g;h} \left( p_1, \ldots p_h; t \right) g_{\text{s}}^{2g-2+h},
\eea
\noindent
albeit this is \textit{not} always the case \cite{d91, bde00, mpp09}, as we discuss in this section. Across the years there have been many excellent reviews on random matrices, of which we highlight, \textit{e.g.}, \cite{dgz93, m04, ekr15}.

There are two complementary ways to solve matrix models. One is to make use of the measure in the matrix integral \eqref{eq:partitionfunctionhermitianmatrix} in order to define orthogonal polynomials whose recursion coefficients satisfy finite-difference string equations \cite{biz80, iz92}. This approach will be discussed in subsection~\ref{subsec:OP-phases}, alongside its (multicritical/minimal) double-scaled version in subsection~\ref{subsec:DSL-phases} (now dealing with differential string equations). The other is to make use of the resolvent $\CW_1 (p)$ in order to define a spectral curve \cite{bipz78} which recursively generates genus-$g$ coefficients in \eqref{eq:genusgfreeenergies}-\eqref{eq:genusgmultiresolvents} via the loop equations \cite{ackm93, eo07a}. This approach will be discussed in subsection~\ref{subsec:SG-phases}. Both these approaches allow us to understand the rich phase diagrams that hermitian matrix models and their double-scaling limits display, and which subsequently yield varied asymptotic structures which we wish to exactly describe in this work. This set-up will be discussed in subsection~\ref{subsec:stokes-vs-phases}. What one might expect concerning the partition-function asymptotics is left to subsection~\ref{subsec:Z-phases}.

\subsection{Large $N$ Phases from Orthogonal Polynomials}
\label{subsec:OP-phases}

Let us first address orthogonal polynomials \cite{biz80, iz92}. So far we have said nothing concerning the integration contours of the $N$-dimensional eigenvalue integral \eqref{eq:partitionfunctioneigenvalues}, but these are simply taken as steepest-descent contours associated to the saddle-points of the potential. If we were to choose the same integration contour $\CC$ for all eigenvalues, then matrix models are solvable by introducing polynomials $\left\{ p_n (z) \right\}$ orthogonal with respect to the positive-definite measure that follows naturally from \eqref{eq:partitionfunctionhermitianmatrix},
\begin{equation}
\rmd \mu (z) = \rme^{-\frac{1}{g_{\text{s}}} V(z)}\, \frac{\rmd z}{2\pi}.
\end{equation}
\noindent
Orthogonality translates to
\be
\label{eq:OPhnrelation}
\int_{\CC} \rmd\mu (z)\, p_n (z)\, p_m (z) = h_n\, \delta_{nm}, \qquad n,m \geq 0,
\ee
\noindent
where one further considers a monic normalization as $p_n (z) = z^n + \cdots$. In this set-up, these polynomials satisfy simple recursion relations of the form
\be
\label{eq:OPrecursionrelationforpns}
p_{n+1} (z) = \left( z+s_n \right) p_n (z) - r_n\, p_{n-1} (z),
\ee
\noindent
where the recursion coefficients $s_n$ vanish for even potentials whilst the $r_n$ are dictated by the matrix-integral potential via a finite-difference string equation. They also relate to the above normalization coefficients $h_n$ via
\be
r_n = \frac{h_n}{h_{n-1}}, \qquad n \geq 1.
\ee
\noindent
The reason why matrix models are now solvable is that if we know either the $h_n$ or $r_n$ coefficients, then \cite{biz80} we know the partition function \eqref{eq:partitionfunctionhermitianmatrix} as
\be
\label{eq:partitionfunctionintermsofrs}
\CZ_{N} = \prod_{n=0}^{N-1} h_n = h_0^N \prod_{n=1}^{N} r_n^{N-n}.
\ee
\noindent
Conversely, if we know the partition function we also know these coefficients via
\be
\label{eq:rsintermsofpartitionfunction}
r_N = \frac{\CZ_{N+1}\, \CZ_{N-1}}{\CZ_{N}^2}.
\ee

There are two prototypical matrix models we shall consider in great detail in this work. These are the cubic matrix model, \textit{e.g.}, \cite{biz80, msw08, kmr10, mss22}, defined via the potential
\be
\label{eq:CubicMatrixModelPotential}
V_{\text{cubic}} (x) = \frac{1}{2}\, x^2 - \frac{\lambda_{\text{c}}}{6}\, x^3,
\ee
\noindent
and the quartic matrix model, \textit{e.g.},  \cite{biz80, msw07, m08, asv11, sv13, csv15, mss22}, defined via the potential
\be
\label{eq:QuarticMatrixModelPotential}
V_{\text{quartic}} (x) = \frac{1}{2}\, x^2 - \frac{\lambda_{\text{q}}}{24}\, x^4.
\ee
\noindent
Because we always address these models separately, we will henceforth just write $\lambda_{\text{c}}, \lambda_{\text{q}} \equiv \lambda$ for the coupling in the matrix potentials. The recursion coefficients of the cubic matrix model satisfy \cite{biz80, mss22}
\begin{equation}
\label{eq:prestringequationcubticmatrixmodel}
\frac{1}{2} r_n \left( \sqrt{1 - \lambda^2\, r_n - \lambda^2 r_{n-1}} + \sqrt{1 - \lambda^2\, r_n - \lambda^2\,r_{n+1}} \right) = n g_{\text{s}},
\end{equation}
\noindent
while those of the quartic matrix model satisfy \cite{biz80, mss22}
\begin{equation}
\label{eq:prestringequationquarticmatrixmodel}
r_n \left( 1 - \frac{\lambda}{6} \left( r_{n-1} + r_n + r_{n+1} \right) \right) = n g_{\text{s}}.
\end{equation}
\noindent
In the 't~Hooft large $N$ limit the recursion coefficients become functions $r_n \mapsto R (t;g_{\text{s}})$ which immediately satisfy finite-difference string equations (see, \textit{e.g.}, \cite{m08, asv11, mss22} for finer details). For the cubic matrix model the ``pre-string equation'' \eqref{eq:prestringequationcubticmatrixmodel} hence becomes
\be
\label{eq:cubicstringequationthooftlimit}
\frac{1}{2} R (t) \left( \sqrt{1 - \lambda^2\, R (t) - \lambda^2\, R (t-g_{\text{s}})} + \sqrt{1 - \lambda^2\, R (t) - \lambda^2\, R (t+g_{\text{s}})} \right) = t,
\ee
\noindent
whereas for the quartic matrix model the ``pre-string equation'' \eqref{eq:prestringequationquarticmatrixmodel} becomes
\be
\label{eq:quarticstringequationthooftlimit}
R (t) \left( 1 - \frac{\lambda}{6}\, \Big( R (t-g_{\text{s}}) + R (t) + R (t+g_{\text{s}}) \Big) \right) = t.
\ee
\noindent
Given the $r_n$ coefficients determine the partition function via \eqref{eq:partitionfunctionintermsofrs}, they also immediately yield the matrix model free energy. In the 't~Hooft large $N$ limit this follows by simple use of the Euler--MacLaurin formula (see, \textit{e.g.}, \cite{m04}) resulting in\footnote{Recall the common practice where one always normalizes the free energy by the Gaussian, $\CF - \CF_{\text{G}} \equiv \log \frac{\CZ}{\CZ_{\text{G}}}$.}
\begin{equation}
\label{eq:freeenergyfromstringequation}
\mathcal{F} (t+g_{\text{s}}) - 2 \mathcal{F} (t) + \mathcal{F} (t-g_{\text{s}}) = \log \frac{R (t)}{t}.
\end{equation}
\noindent
Both string equations \eqref{eq:cubicstringequationthooftlimit}-\eqref{eq:quarticstringequationthooftlimit} may be solved via resurgent transseries on which we shall give more details in section~\ref{sec:resurgent-Z-transseries}. For the moment let us just say that for both examples their corresponding solutions are two-parameter $\sigma_1,\sigma_2$, resonant, resurgent transseries, writeable as
\be
\label{eq:twoparameterresurgenttransseriesforR}
R \left( t,g_{\text{s}}; \sigma_1,\sigma_2 \right) = \sum_{n=0}^{+\infty} \sum_{m=0}^{+\infty} \sigma_1^n \sigma_2^m\, \rme^{- \left( n-m \right) \frac{A (t)}{g_{\text{s}}}}\, R^{(n|m)} (t,g_{\text{s}}).
\ee
\noindent
Herein $A(t)$ is the instanton action and the full transseries is assembled via an infinite number of asymptotic series in $g_{\text{s}}$, $R^{(n|m)}$, labeled by $n$ and $m$, each consisting of an infinite set of ($t$-dependent) coefficients (see section~\ref{sec:resurgent-Z-transseries}, \textit{e.g.}, \eqref{eq:CubicPartitionFunctionRectangular} and \eqref{eq:QuarticPartitionFunctionRectangular} for cubic and quartic matrix model partition-function transseries, respectively). This is a large amount of \textit{model-dependent} data which needs to be computationally generated by plugging the above \textit{ansatz} into the corresponding string equation, say either \eqref{eq:cubicstringequationthooftlimit} or \eqref{eq:quarticstringequationthooftlimit}, and then iteratively solving the resulting intricate infinite set of coupled equations. This has been done\footnote{Transseries data in these examples can and have also been obtained via (matching) matrix integral calculations in \cite{msw07, msw08, mss22}. Albeit this is \textit{no longer} a model-dependent calculation, with final results given in terms of a spectral curve, it is of course a much slower procedure as illustrated by the intricate calculations in \cite{mss22, krsst26b}.} for cubic and quartic matrix models in \cite{m08, msw08, kmr10, asv11, sv13, csv15, mss22}. In this way, whereas these generic transseries are fully exact, nonperturbative solutions\footnote{To be completely precise, they still require Stokes data---more on this later, albeit mostly in \cite{krsst26b}.} to the string equations, they are still only obtainable on a case-by-case basis. One of our main results in section~\ref{sec:resurgent-Z-transseries} is precisely to find a \textit{model-independent} closed-form expression for these transseries solutions. Its sole input is the spectral curve, hence all required data is automatically dictated by the topological recursion \cite{eo07a}. This applies to both hermitian one-matrix models (their Dijkgraaf--Vafa topological string duals \cite{dv02a} included) and their double-scaling limits (multicritical models, minimal strings \cite{ss03} and topological gravity included). But at this early stage the question we first need to address is: what qualifies as a truly (even at finite $N$) nonperturbative behavior of the matrix model, which we would like to see the transseries \textit{exactly} and \textit{globally} reproduce?

\begin{figure}
	\centering
	\begin{subfigure}{0.47\textwidth}
		\includegraphics[width=\textwidth]{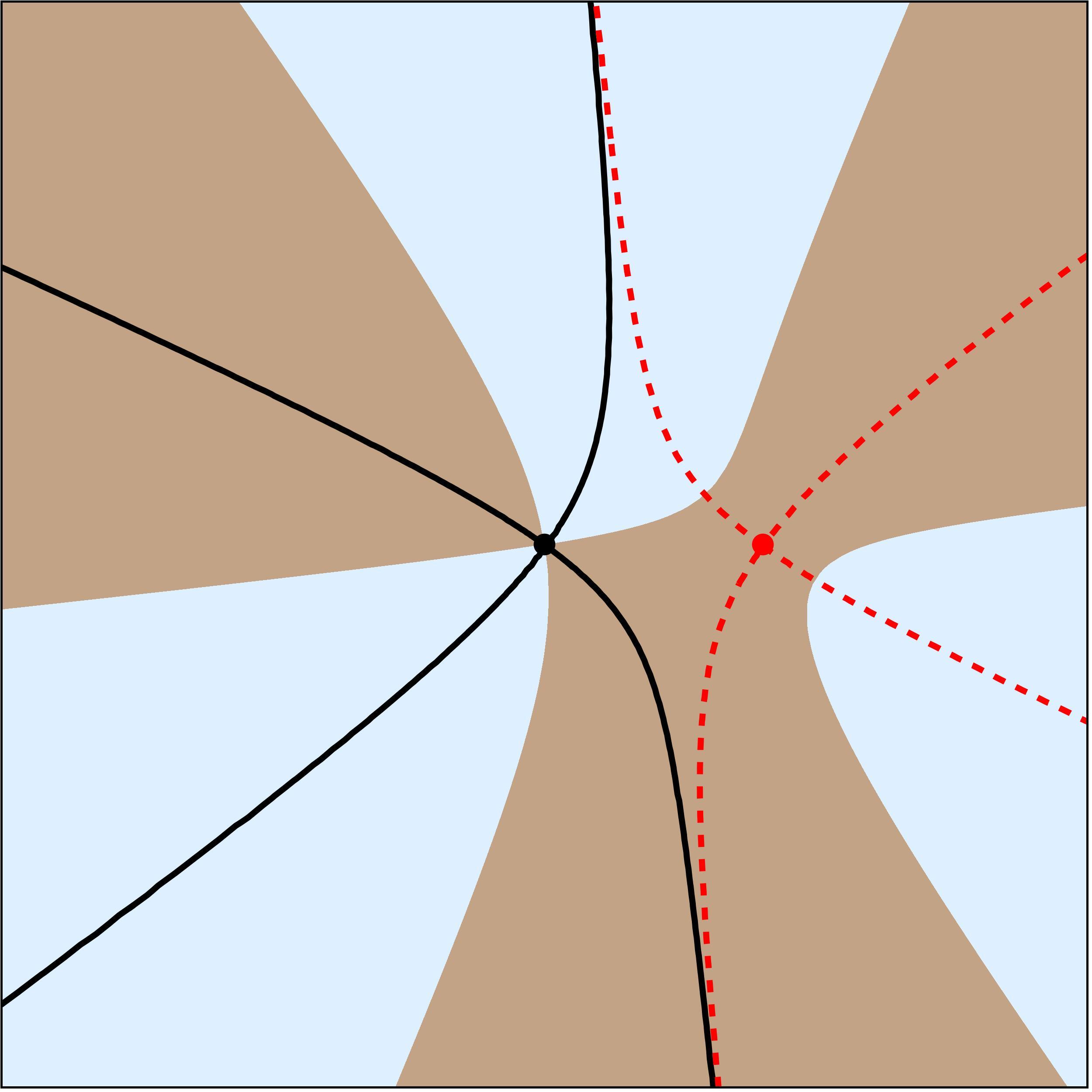}
		\caption{The cubic matrix model potential \eqref{eq:CubicMatrixModelPotential}.}
		\label{fig:CubicMatrixModelSteepestDescentContours}
	\end{subfigure}
	\hfill
	\begin{subfigure}{0.47\textwidth}
		\includegraphics[width=\textwidth]{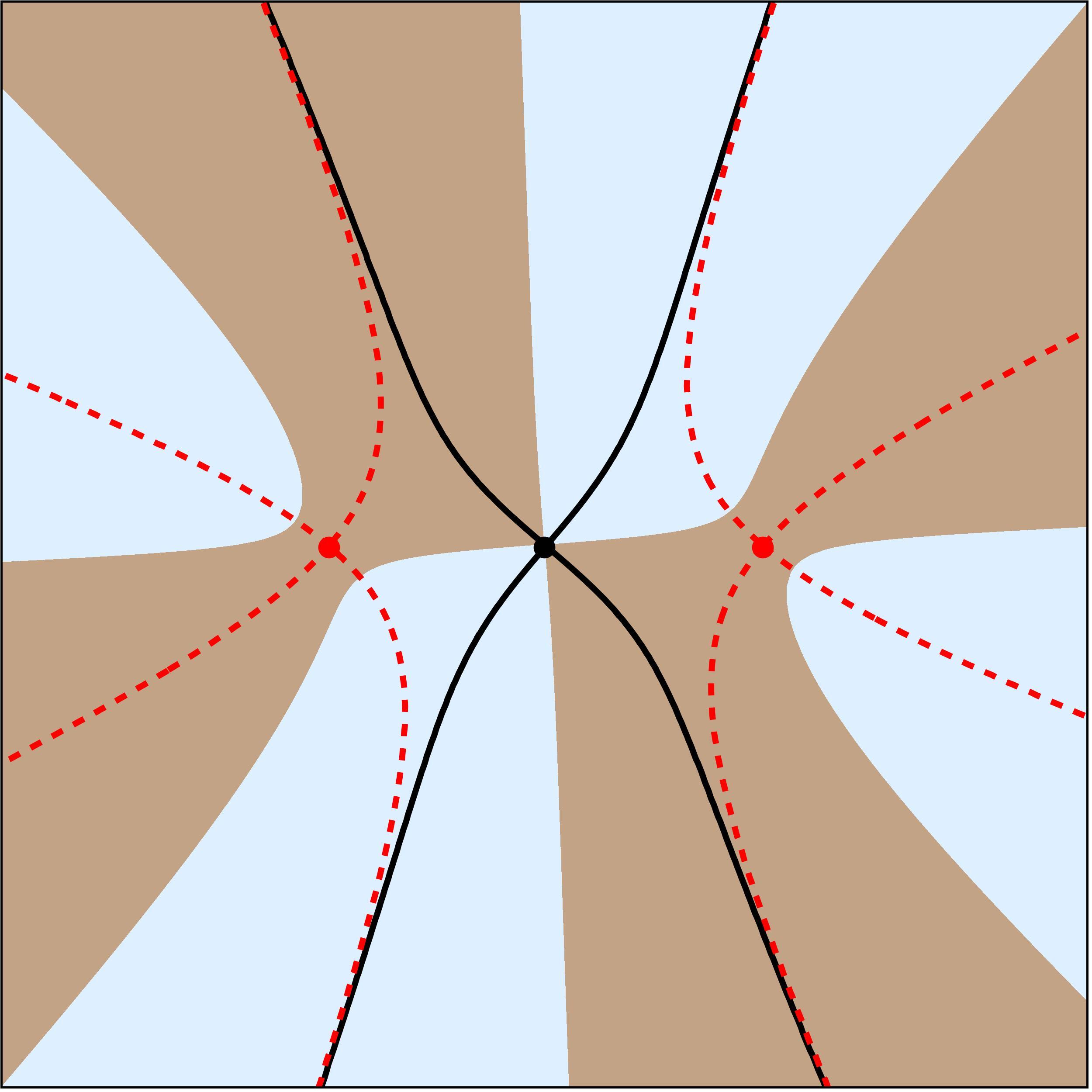}
		\caption{The quartic matrix model potential \eqref{eq:QuarticMatrixModelPotential}.}
		\label{fig:QuarticMatrixModelSteepestDescentContours}		
	\end{subfigure}	
	\caption{Steepest-descent and steepest-ascent contours for cubic and quartic potentials \eqref{eq:CubicMatrixModelPotential} and \eqref{eq:QuarticMatrixModelPotential}, plotted in the complex $x$-plane (with $\lambda=1$). The \textcolor{brown}{brown}/\textcolor{blue}{blue} (or ``land/sea'') background indicates regions where $\Re V(x)$ is positive/negative. On top, the perturbative contours are plotted in solid-black, whilst the nonperturbative contours are in dashed-red (steepest-descent contours going-off to infinity along the ``land'', steepest-ascent along the ``sea'').}
	\label{fig:Cubic-AND-QuarticMatrixModelSteepestDescentContours}
\end{figure}

For this, let us see how far can one go in solving both the cubic and quartic matrix models via orthogonal polynomials, \textit{i.e.}, solving the pre-string equations \eqref{eq:prestringequationcubticmatrixmodel} and \eqref{eq:prestringequationquarticmatrixmodel}. What these equations do is to iteratively generate higher recursion-coefficients $r_n$ data, given the first few. Equivalently, one may introduce the moments of the matrix model by
\begin{equation}
\mathfrak{m}_k = \int_{\boldsymbol{w} \cdot \boldsymbol{\mathcal{C}}} \rmd\mu(z)\, z^k,
\end{equation}
\noindent
where integration is carried out over steepest-descent contours $\boldsymbol{\CC}$, weighed by contour weights $\boldsymbol{w}$, which will be made precise in the following. Then, just use \eqref{eq:OPrecursionrelationforpns} and \eqref{eq:OPhnrelation} to obtain, for both our matrix models,
\begin{align}
r_1 &= \frac{\mathfrak{m}_2}{\mathfrak{m}_0} - \left(\frac{\mathfrak{m}_1}{\mathfrak{m}_0}\right)^2, & r_2 &= \frac{\mathfrak{m}_0 \left( \mathfrak{m}_4 \left( \mathfrak{m}_2\, \mathfrak{m}_0 - \mathfrak{m}_1^2 \right) - \mathfrak{m}_3 \left( \mathfrak{m}_3\, \mathfrak{m}_0 - 2 \mathfrak{m}_2\, \mathfrak{m}_1 \right) - \mathfrak{m}_2^3 \right)}{\left( \mathfrak{m}_2\, \mathfrak{m}_0 - \mathfrak{m}_1^2 \right)^2}, & r_3 &= \cdots, \\
s_0 &= -\frac{\mathfrak{m}_1}{\mathfrak{m}_0}, & s_1 &= - \frac{\mathfrak{m}_3\, \mathfrak{m}_0^2 - 2 \mathfrak{m}_2\, \mathfrak{m}_1\, \mathfrak{m}_0 + \mathfrak{m}_1^3 }{\mathfrak{m}_0 \left( \mathfrak{m}_2\, \mathfrak{m}_0 - \mathfrak{m}_1^2\right)}, & s_2 &= \cdots,
\end{align}
\noindent
and so on (see, \textit{e.g.}, \cite{csv15} for further details and further explicit $r_n$-coefficients for the quartic model). The reason why this is useful is that one can be rather precise on the nature of these moments---in fact, explicitly compute them for our two examples. For the cubic potential \eqref{eq:CubicMatrixModelPotential}, saddle-points are located at $z=0$ and $z=2/\lambda$, correspondingly giving rise to two\footnote{These are the ``perturbative'' contour, $w_{\text{left}} \equiv w_{\text{p}}$, through the ``perturbative'' saddle $z=0$; alongside the ``nonperturbative'' contour, $w_{\text{right}} \equiv w_{\text{np}}$, through the ``nonperturbative'' saddle $z=2/\lambda$ (see, \textit{e.g.}, \cite{abs18} for more on this sort of analysis).} steepest-descent contours $\mathcal{C}_{1,2}$, which we illustrate in figure~\ref{fig:CubicMatrixModelSteepestDescentContours}. With their associated weights denoted by $w_{\text{left},\text{right}}$, so that $\boldsymbol{w} \cdot \boldsymbol{\mathcal{C}} = w_{\text{left}}\, \CC_1 + w_{\text{right}}\, \CC_2$, the cubic moments may finally be explicitly written as
\begin{equation}
\label{eq:mkw1w2cubic}
\mathfrak{m}_k = w_{\text{left}} \left( \mathfrak{m}_{k;1} - \mathfrak{m}_{k;0} \right) + w_{\text{right}} \left( \mathfrak{m}_{k;2} - \mathfrak{m}_{k;1} \right),
\end{equation}
\noindent
where we have split the contour integrations into their ``radial'' components $\mathfrak{m}_{k;\ell}$ along the three $\ell = 0,1,2$ basic rays (again, see figure~\ref{fig:CubicMatrixModelSteepestDescentContours}), and which are given by
\bea
\mathfrak{m}_{k;\ell} &=& \int_{0}^{\rme^{\rmi\theta_{\ell}} \cdot \infty} \rmd \mu(z)\, z^k = \\
&=& \frac{1}{3} P_{k;\ell} \left\{ \Gamma \left(\frac{k+1}{3}\right) {}_2F_2 \left( \frac{k+1}{6}, \frac{k+4}{6}; \frac{1}{3}, \frac{2}{3}; \left( \frac{2^{2/3}\, b_{\ell}}{3} \right)^3 \right) + \right. \nonumber \\
&&
+ b_{\ell}\, \Gamma \left(\frac{k+3}{3}\right) {}_2F_2 \left( \frac{k+3}{6}, \frac{k+6}{6}; \frac{4}{3}, \frac{2}{3}; \left(\frac{2^{2/3}\, b_{\ell}}{3} \right)^3 \right) + \nonumber \\
&&
\left. + \frac{1}{2}\, b_{\ell}^2\, \Gamma \left(\frac{k+5}{3}\right) {}_2F_2 \left( \frac{k+5}{6}, \frac{k+8}{6}; \frac{4}{3}, \frac{5}{3}; \left(\frac{2^{2/3}\, b_{\ell}}{3} \right)^3 \right) \right\}.
\eea
\noindent
Notice how if $\lambda/g_{\text{s}}$ is real, then $\theta_{0,1,2} = \frac{\pi}{3}, \pi, \frac{5\pi}{3}$ in rough accordance with the main directions along the ``land'' in figure~\ref{fig:CubicMatrixModelSteepestDescentContours}. Otherwise, the three basic rays are:
\be
\theta_{\ell} = \frac{1}{3} \left( \left( 2\ell+1 \right) \pi - \arg \left( \frac{\lambda}{g_{\text{s}}} \right) \right), \qquad \ell = 0,1,2.
\ee
\noindent
Furthermore, ${}_2 F_2 \left( a_1,a_2;,b_1,b_2;z \right)$ is a generalized hypergeometric function (see \cite{olbc10}) and, for compactness of notation, we have further introduced the quantities:
\bea
P_{k;\ell} &=& 6^{\frac{k+1}{3}} \left( \frac{g_{\text{s}}}{\lambda} \right)^{\frac{k+1}{3}} \rme^{\frac{\rmi\pi}{3} \left( k+1 \right) \left( 2\ell+1 \right)}, \\
b_{\ell} &=& -\frac{6^{2/3}}{2}\, \frac{\rme^{\frac{2\pi\rmi}{3} \left( 2\ell+1 \right)}}{\left( \lambda^2 g_{\text{s}} \right)^{\frac{1}{3}}}.
\eea
\noindent
These formulae now allow us to use the pre-string equation \eqref{eq:prestringequationcubticmatrixmodel} to compute as much $r_n$ data as we want, and, in this process, to further compute the finite $N$ partition function via \eqref{eq:partitionfunctionintermsofrs}. 

The quartic matrix model case is completely analogous. The quartic potential \eqref{eq:QuarticMatrixModelPotential} now has three saddle-points, located at $z=0$ and $z=\pm\sqrt{6/\lambda}$, which give rise to three\footnote{These are the ``perturbative'' contour, $w_{\text{middle}} \equiv w_{\text{p}}$, through the ``perturbative'' saddle $z=0$; now alongside two (symmetric) ``nonperturbative'' contours, $w_{\text{side}} \equiv w_{\text{np}}$, through the ``nonperturbative'' saddles $z=\pm\sqrt{6/\lambda}$ (again, see, \textit{e.g.}, \cite{abs18}).} steepest-descent contours $\mathcal{C}_{1,2,3}$, which we illustrate in figure~\ref{fig:QuarticMatrixModelSteepestDescentContours}. The corresponding weights should now be denoted by $w_{1,2,3}$, howbeit, due to the $\BZ_2$-symmetry of the quartic potential, we will always take the contour weights to be symmetric; \textit{i.e.}, denote by $\mathcal{C}_{\text{middle}}$ and $\mathcal{C}_{\text{side}}$ the contours going through the middle-saddle and both side-saddles, respectively, with associated weights $w_{\text{middle}}$ and $w_{\text{side}}$. The quartic moments are then explicitly given by
\begin{equation}
\label{eq:mkwmiddlewsidequartic}
\mathfrak{m}_k = w_{\text{middle}} \left( \mathfrak{m}_{k,1} - \mathfrak{m}_{k,3} \right) +  w_{\text{side}} \left( \mathfrak{m}_{k,0} + \mathfrak{m}_{k,1} - \mathfrak{m}_{k,2} - \mathfrak{m}_{k,3} \right).
\end{equation}
\noindent
As for the cubic, we have split the contour integrations into their ``radial'' components $\mathfrak{m}_{k;\ell}$ along the now four basic rays $\ell=0,1,2,3$ (see figure~\ref{fig:QuarticMatrixModelSteepestDescentContours}). These are now simpler:
\bea
\mathfrak{m}_{k;\ell} &=& \int_{0}^{\rme^{\rmi\theta_{\ell}} \cdot \infty} \rmd \mu(z)\, z^k = \nonumber \\
&=& \frac{1}{4} P_{k;\ell} \left\{ \Gamma \left( \frac{k+1}{4} \right) {}_1F_1 \left( \frac{k+1}{4}; \frac{1}{2}; \frac{b_{\ell}^2}{4} \right) + b_{\ell}\, \Gamma \left( \frac{k+3}{4} \right) {}_1F_1 \left( \frac{k+3}{4}; \frac{3}{2}; \frac{b_{\ell}^2}{4} \right) \right\}.
\label{eq:QMMMomentsExpression}
\eea
\noindent
Again, when $\lambda/g_{\text{s}}$ is real we have $\theta_{0,1,2,3} = \frac{\pi}{4}, \frac{3\pi}{4}, \frac{5\pi}{4}, \frac{7\pi}{4}$ in rough accordance with the main ``land'' directions in figure~\ref{fig:QuarticMatrixModelSteepestDescentContours}. If not, the four basic rays are:
\be
\theta_{\ell} = \frac{1}{4} \left( \left( 2\ell+1 \right) \pi - \arg \left( \frac{\lambda}{g_{\text{s}}} \right) \right), \qquad \ell = 0,1,2,3.
\ee
\noindent
Finally, ${}_1 F_1 \left( a; b; z \right)$ is the Kummer confluent hypergeometric function (see \cite{olbc10}) and, for compactness of notation (and without confusing with the corresponding cubic quantities), we have further introduced:
\bea
P_{k;\ell} &=& \abs{\frac{24 g_{\text{s}}}{\lambda}}^{\frac{1}{4} \left(k+1\right)} \rme^{\rmi \left( k+1 \right) \theta_{\ell}}, \\
b_{\ell} &=& - \frac{1}{2g_{\text{s}}}\, \abs{\frac{24 g_{\text{s}}}{\lambda}}^{\frac{1}{2}} \rme^{2\rmi\theta_{\ell}}.
\eea
\noindent
Again, these formulae allow us to use the pre-string equation \eqref{eq:prestringequationquarticmatrixmodel} to compute as much $r_n$ data as computationally feasible in reasonable time, and consequently, the finite $N$ partition function.

We have hence obtained an iterative way to explicitly construct the exact finite $N$ partition function \eqref{eq:partitionfunctionhermitianmatrix} in our two canonical examples \eqref{eq:CubicMatrixModelPotential} and \eqref{eq:QuarticMatrixModelPotential}, by making use of orthogonal polynomials. Of course that any alternative exact solution via large $N$ resurgent transseries must, necessarily, reproduce these very same results. As a technical remark, we also point out that although we only need $r_n$ coefficients for $n\leq N$ in order to construct the partition function \eqref{eq:partitionfunctionintermsofrs}, it is of course completely fine from a purely algorithmic point-of-view to compute $r_n$ data for $n>N$ as well (which from the resurgent transseries large-$N$ point-of-view instead would correspond to computing $R ( x,g_{\text{s}} )$ in \eqref{eq:twoparameterresurgenttransseriesforR} for\footnote{We are being slightly superficial with some details. In taking the 't~Hooft limit we should have first considered $r_n \mapsto R (x;g_{\text{s}})$ with $x = n g_{\text{s}} \in [0,t]$ and only then have taken the large $N$ limit. See \cite{m08, asv11} for precise details.} $x>t$). 

\begin{figure}
	\centering
	\begin{subfigure}[t]{\textwidth}
		\includegraphics[width=0.495\textwidth]{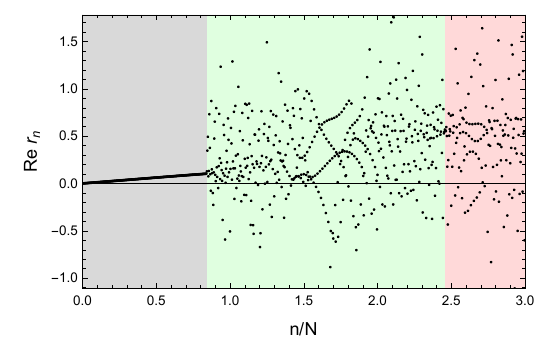}\hfill
		\includegraphics[width=0.495\textwidth]{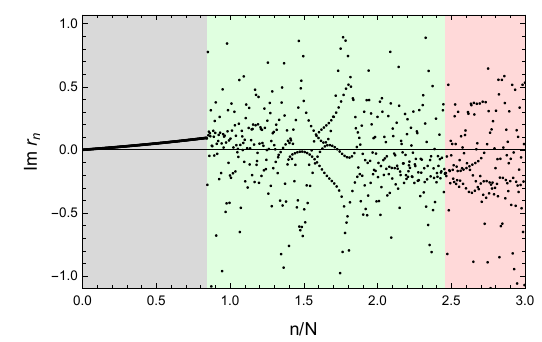}
	\end{subfigure}
	\bigskip
	\begin{subfigure}[b]{\textwidth}
		\includegraphics[width=0.495\textwidth]{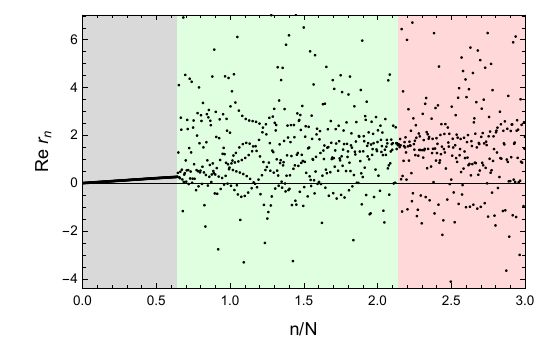}\hfill
		\includegraphics[width=0.495\textwidth]{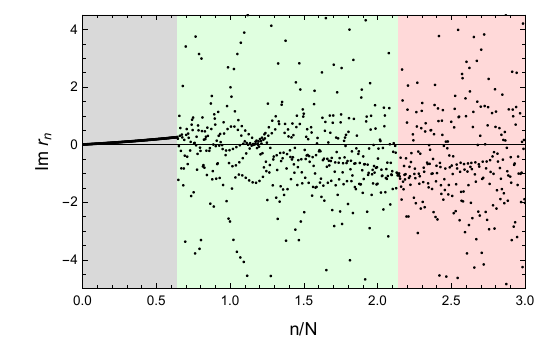}
	\end{subfigure}	
	\caption{Plots of the real (left) and imaginary (right) parts of $r_n$ data of the cubic (top) and quartic (bottom) matrix models, for varying $n$, as obtained from the pre-string equations \eqref{eq:prestringequationcubticmatrixmodel} and \eqref{eq:prestringequationquarticmatrixmodel}. We chose $N=300$, $\lambda=1$ for both plots, and $t = 0.15\, \rme^{\rmi\pi/5}$ and $t = 0.5\, \rme^{\rmi\pi/5}$ for the cubic and quartic models, respectively. The data very clearly change behaviors at certain points, hinting at the existence of different phases in the complex $t$-plane for both models (also highlighted by the gray, green, pink background colors which will be explained in the main text).}
	\label{fig:CMMrDataChaosPlot}
\end{figure}

We illustrate the behavior of the resulting $r_n$ ``raw data'' for both cubic and quartic matrix models in figure~\ref{fig:CMMrDataChaosPlot}. In order to understand the complexity at play, a crucial observation immediately arises from studying the behavior of the $r_n$ coefficients---in either of our examples---across the complex 't~Hooft plane, \textit{i.e.}, for any $t \in \BC$: we qualitatively observe that the data drastically \textit{change behavior} at certain points (illustrated by the distinct background colors to be explained in the following), which is a first hint of the non-trivial phase transitions that we will have to sharply address in the remainder of this work. This sort of behaviors and their associated large $N$ phases were already studied since the early days of studying orthogonal-polynomial recursion-coefficients, see, \textit{e.g.}, \cite{ddjt90, l92, j91, s92, bdjt93}, albeit a chaotic structure was misidentified\footnote{The behavior is in fact always either periodic or quasi-periodic as later shown in \cite{bde00}.} and the precise form of their (full) asymptotics was lacking. Herein we will exactly describe the full asymptotics of the partition function, for large classes of models and in all their possible phases.

\begin{figure}
	\centering
	\begin{subfigure}[t]{0.495\textwidth}
         \includegraphics[width=\textwidth]{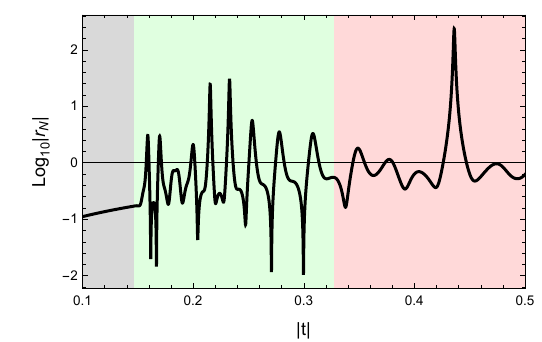}
         \label{fig:CMMrNAbsPlot}
	\end{subfigure}	  
	\hfill
	\begin{subfigure}[t]{0.495\textwidth}
         \includegraphics[width=\textwidth]{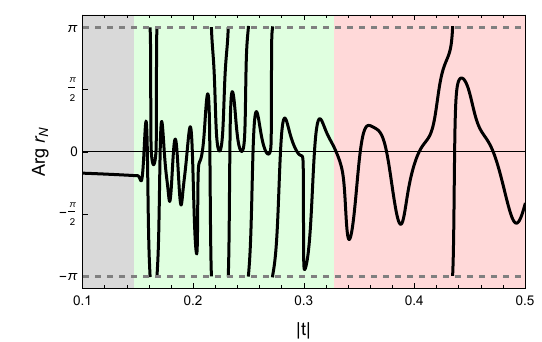}
         \label{fig:CMMrNArgPlot}
	\end{subfigure}  
	\bigskip     
	\begin{subfigure}[b]{0.325\textwidth}
         \includegraphics[width=\textwidth]{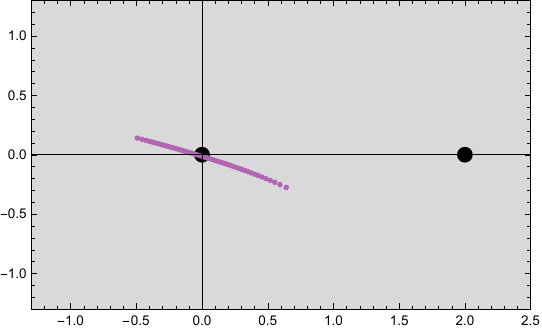}
         \label{fig:CMM1CutOPRoots}
	\end{subfigure}
	\hfill
	\begin{subfigure}[b]{0.325\textwidth}
         \includegraphics[width=\textwidth]{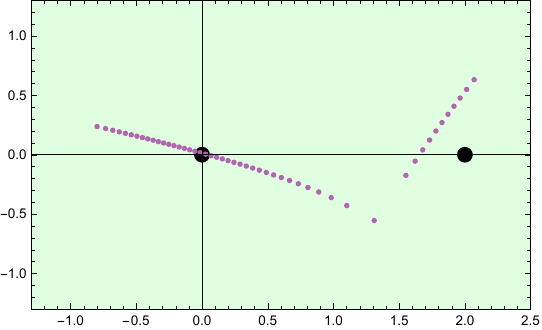}
         \label{fig:CMM2CutOPRoots}
	\end{subfigure}
	\hfill
	\begin{subfigure}[b]{0.325\textwidth}
         \includegraphics[width=\textwidth]{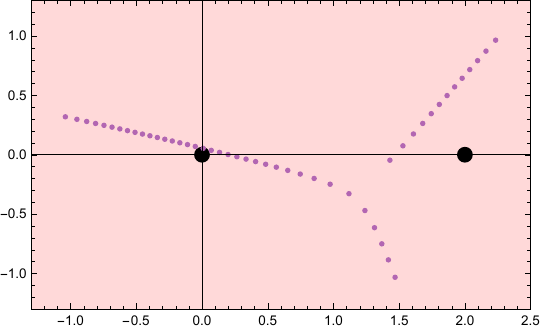}
         \label{fig:CMMTrivalentOPRoots}
	\end{subfigure}
	\caption{Another illustration of the existence of different phases for the cubic matrix model. The top row shows the logarithmic norm and argument of $r_{50}(t)$, for $\arg(t)=-\pi/5$ and varying $\abs{t}$. It distinctively shows the phase transition from a phase of stable behavior (indicated in gray) to one characterized by rapid oscillations (indicated in green), ensued by a change in frequency of oscillations (the transition from green to pink). The bottom row shows what is actually driving the different phases. It displays the roots of the orthogonal polynomial $p_{50}(z)$ for $\abs{t}=0.1$ (bottom left), $\abs{t}=0.36$ (middle) and $\abs{t}=0.7$ (bottom right), clearly clustering into different-looking configurations. As we shall discuss in subsection~\ref{subsec:SG-phases} these are providing strong support for the gray-to-green phase transition, as well as for the green-to-pink transition which is less obvious from the $r_{50}(t)$ data alone.}
        \label{fig:CMMrNDataAndOPRoots}
\end{figure}

\begin{figure}
	\centering
	\begin{subfigure}[t]{0.495\textwidth}
         \includegraphics[width=\textwidth]{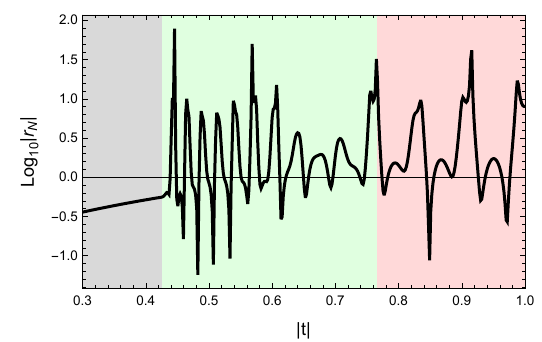}
         \label{fig:QMMrNAbsPlot}
	\end{subfigure}	  
	\hfill
	\begin{subfigure}[t]{0.495\textwidth}
         \includegraphics[width=\textwidth]{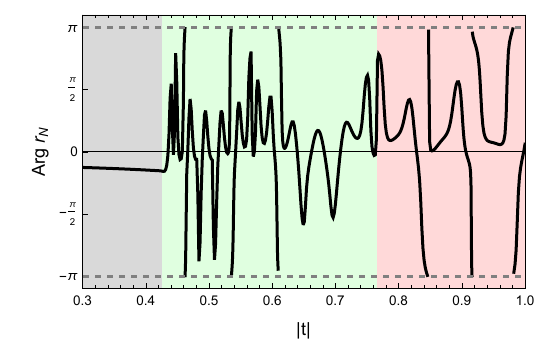}
         \label{fig:QMMrNArgPlot}
	\end{subfigure}    
	\bigskip     
	\begin{subfigure}[b]{0.325\textwidth}
         \includegraphics[width=\textwidth]{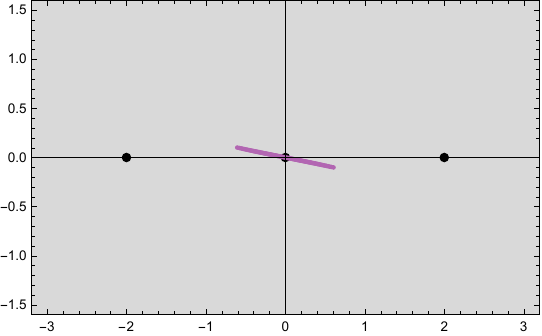}
         \label{fig:QMM1CutOPRoots}
	\end{subfigure}
	\hfill
	\begin{subfigure}[b]{0.325\textwidth}
         \includegraphics[width=\textwidth]{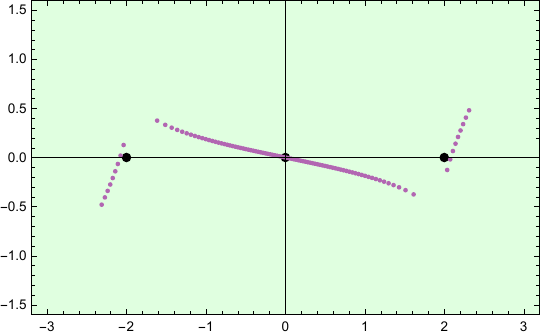}
         \label{fig:QMM3CutOPRoots}
	\end{subfigure}
	\hfill
	\begin{subfigure}[b]{0.325\textwidth}
         \includegraphics[width=\textwidth]{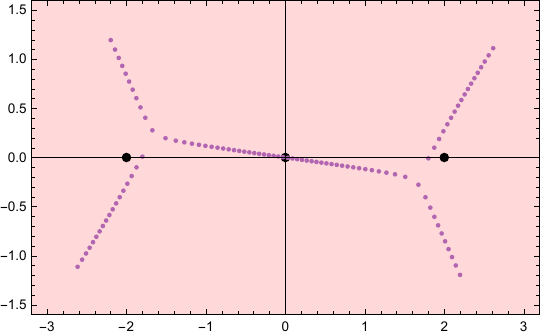}
         \label{fig:QMMTrivalentOPRoots}
	\end{subfigure}
	\caption{Another illustration of the existence of different phases for the quartic matrix model. The top row shows the logarithmic norm and argument of $r_{100}(t)$, for $\arg(t)=-\pi/10$ and varying $\abs{t}$. It distinctively shows the phase transition from a phase of stable behavior (indicated in gray) to one characterized by rapid oscillations (indicated in green), ensued by a change in frequency of oscillations (the transition from green to pink). The bottom row shows what is actually driving the different phases. It displays the roots of the orthogonal polynomial $p_{100}(z)$ for $\abs{t}=0.1$ (bottom left), $\abs{t}=0.7$ (middle) and $\abs{t}=2.1$ (bottom right), clearly clustering into different-looking configurations. As we shall discuss in subsection~\ref{subsec:SG-phases} these are providing strong support for the gray-to-green phase transition, as well as for the green-to-pink transition which is less obvious from the $r_{100}(t)$ data alone.}
        \label{fig:QMMrNDataAndOPRoots}
\end{figure}

Whereas figure~\ref{fig:CMMrDataChaosPlot} illustrates a sharp change in behavior from the ``gray region'' to the ``green region'', the change to the ``pink region'' is not noticeable to the naked eye (and at first the reader might wonder if there really is any difference). In this case, a second, perhaps more direct, piece of evidence illustrating the complexity of phase transitions at play can be obtained by instead varying the 't~Hooft coupling $t$ whilst keeping $N$ fixed; and in this setting studying the behavior of $r_N (t)$ (in particular, the logarithm of its absolute value, alongside its phase). This is illustrated for the cubic matrix model in figure~\ref{fig:CMMrNDataAndOPRoots} and for the quartic matrix model in figure~\ref{fig:QMMrNDataAndOPRoots}. The top two plots of figure~\ref{fig:CMMrNDataAndOPRoots} for the cubic matrix model hence display a cleaner visualization of the associated phase transitions as compared to figure~\ref{fig:CMMrDataChaosPlot}. It is clear how $r_N (t)$ is well-behaved for sufficiently small $\abs{t}$, but also how at some point a phase transition occurs that leads to rapid oscillations that persist as the norm of $\abs{t}$ increases further. These oscillations then change frequency with $\abs{t}$ increasing the furthest. To get a better grasp of what is driving these phase transitions, we also address the behavior of the roots\footnote{As we shall see in subsection~\ref{subsec:SG-phases}, at large $N$ these roots match the locations of the random-matrix eigenvalues.} of the associated orthogonal polynomials. For the cubic model, this is illustrated in the bottom row of figure~\ref{fig:CMMrNDataAndOPRoots}. In the ``gray region'', where $r_N (t)$ behaves simply, the orthogonal polynomial roots line up in a single cut. A second cut opens up at the phase transition point, as one transitions to the ``green region''. These two cuts continue to grow in size as $\abs{t}$ increases, until they eventually merge to create a trivalent-tree configuration upon reaching the ``pink region''. We will dedicate subsection~\ref{subsec:SG-phases} to quantitatively show how to explain these qualitative observations concerning the one-cut gray-phase, the two-cuts green-phase, and the trivalent-tree pink-phase, from the point-of-view of spectral geometry. Further note that any exact solutions we shall construct obviously need to be exact \textit{everywhere}, hence obviously need to describe the behavior of the partition function (of  $r_N (t)$ via \eqref{eq:rsintermsofpartitionfunction}) at all values of the parameters---which our resurgent transseries in fact sharply do. As to the discussion concerning the quartic model in figure~\ref{fig:QMMrNDataAndOPRoots}, it is completely analogous and we shall not repeat it.

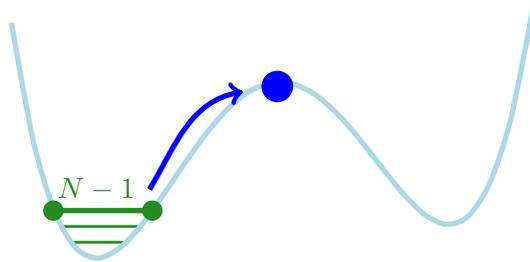
\begin{figure}
\centering
\begin{tikzpicture}
	\begin{scope}[scale=0.7, shift={({3},{-6})}]
	\draw[color=ForestGreen, line width=2pt] (-0.88,0.9) -- (1.15,0.9);
	\draw[color=ForestGreen, line width=1pt] (-0.7,0.6) -- (0.88,0.6);
	\draw[color=ForestGreen, line width=1pt] (-0.51,0.3) -- (0.6,0.3);
	 \draw[scale=2, domain=-0.8:4.1, smooth, variable=\x, LightBlue, line width=2pt] plot ({\x}, {2*1.14*\x*\x - 2*0.671*\x*\x*\x + 0.2*\x*\x*\x*\x});
	\draw[ForestGreen, fill=ForestGreen] (-0.81,0.9) circle (1.1ex);
	 \draw[ForestGreen, fill=ForestGreen] (1.05,0.9) circle (1.1ex);
	 \node[ForestGreen] at (0, 1.3) {$N-1$};
	 \fill[blue, line width=2pt] (3.4,3.25) circle (1.8ex);
	 \draw[blue, line width=2pt, ->] (1, 1.3) to[out=60, in=190] (2.75, 3.15);
	\end{scope}
\end{tikzpicture}
\caption{Visualization of eigenvalue tunneling from the main cut to the first extremal point of a given potential (plotted in {\color{blue}blue}) with $3$ saddles. This is how matrix integrals realize Stokes transitions and this mechanism underlies the appearance of different phases in the matrix integral.}
\label{fig:FiniteNeigenvaluetunneling}
\end{figure}

One final comment as a prelude to what is to come next concerns the underlying reason for why changes in the distributions of roots\footnote{Corresponding to changes in the distributions of random-matrix eigenvalues; see subsection~\ref{subsec:SG-phases}.} of orthogonal polynomials end-up producing changes in the actual (asymptotic) behavior of the matrix model partition function---which we understand as changes of the dominant large $N$ phase. This is illustrated in figure~\ref{fig:FiniteNeigenvaluetunneling} and basically simply describes what was explained in \cite{d91, d92, msw07, msw08, mss22}: under Stokes phenomena, eigenvalues tunnel from the perturbative saddle to a nonperturbative saddle of the matrix integral \eqref{eq:partitionfunctioneigenvalues}, hence changing the asymptotics of the partition function at the same time that they reorganize themselves upon the complex plane (\textit{i.e.}, at the same time that they cluster into distinct configurations, as illustrated in the bottom rows of either figures~\ref{fig:CMMrNDataAndOPRoots} and~\ref{fig:QMMrNDataAndOPRoots}).

\subsection{Double-Scaled Phases from String Equations}
\label{subsec:DSL-phases}

Still within the realm of orthogonal polynomials and their associated string-equations, one may consider double-scaling limits towards multicritical models \cite{gm90a, ds90, bk90, d90, gm90b}, minimal strings \cite{m03, kms03b, ss03, kopss04, mmss04, ss04b}, or two-dimensional topological gravities \cite{dw90, w91, k92, iz92, dw18}, all now describable by \textit{differential} string equations (see, \textit{e.g.}, \cite{gs21} for a recent discussion in the present resurgence context). As a very quick recap, recall that if for a given matrix model potential \eqref{eq:partitionfunctionhermitianmatrix} the moment function of its spectral curve (see subsection~\ref{subsec:SG-phases}, \textit{e.g.}, \eqref{eq:moment-function-one-cut} or \eqref{eq:moment-function-multi-cut}) has a zero of order $k-1$ at the end-point of the cut, this model is said to define a multicritical point of order $k$ \cite{dgz93} (or a $(2,2k-1)$ multicritical model if emphasizing the CFT minimal-model matter \cite{bpz84} of the corresponding string theory). Near such a multicritical point, characterized by coupling $t_{\text{critical}}$, the genus-$g$ free energies \eqref{eq:genusgfreeenergies} all diverge\footnote{The $\CF_g (t)$ are typically convergent on a disk, up to singular values of the K\"ahler moduli where they diverge; see, \textit{e.g.}, \cite{dgz93, cgmps06}. These critical points are common to all genera, and include, \textit{e.g.}, conifold points alongside our multicritical points (for generic topological strings, there are many special points, \textit{e.g.}, \cite{cogp91, bcov93b, gv95}, and mirror symmetry allows for analytic continuation past these points, \textit{e.g.}, \cite{w93, agm93a, agm93b}). As we shall discuss in subsection~\ref{subsec:stokes-vs-phases} our nonperturbative phase transitions are associated to Stokes phenomena and driven by eigenvalue-tunneling/D-brane contributions; and neatly these matrix-model (multi-)critical points precisely fall on top of anti-Stokes lines.} with the same $g$-dependent scaling as $t \to t_{\text{critical}}$. The asymptotic expansion \eqref{eq:genusgfreeenergies} only remains regular at criticality, \textit{i.e.}, only keeps its full perturbative components in the form
\be
F_{\text{ds}} \simeq \sum_{g=0}^{+\infty} F_g\, \kappa_{\text{s}}^{2g-2},
\ee
\noindent
if at the same time one takes the double-scaling limit where $g_{\text{s}} \to 0$ as $t \to t_{\text{critical}}$, whilst at fixed double-scaled string-coupling $\kappa_{\text{s}}$,
\be
\label{eq:multicriticalpoint}
\kappa_{\text{s}} = g_{\text{s}} \left( t - t_{\text{critical}} \right)^{-\frac{2k+1}{2k}}.
\ee
\noindent
In this limit, finite-difference string equations such as \eqref{eq:cubicstringequationthooftlimit} or \eqref{eq:quarticstringequationthooftlimit} become non-linear ordinary differential equations (ODE) for the specific heat $u (z)$ (the double-scaled version of the recursion coefficients \eqref{eq:twoparameterresurgenttransseriesforR}, as $R ( t,g_{\text{s}} ) \underset{\tiny{\text{DSL}}}{\longmapsto} u (z)$), in the variable $z$ given by
\be
z = \kappa^{-\frac{2k}{2k+1}}_{\text{s}}.
\ee
\noindent
As to the free energy, the double-scaled version of \eqref{eq:freeenergyfromstringequation} is now the simpler\footnote{This is the normalization where string equations follow straight from the Gel'fand--Dikii KdV potentials as in \eqref{eq:orderkstringequationfromgelfanddikii}, double-scaled from matrix-model even-potentials; see, \textit{e.g.}, \cite{gs21}---but pay attention to the caveat below.}
\be
\label{eq:FdsEVENnormalization}
F_{\text{ds}}'' (z) = -\frac{1}{2} u (z).
\ee

Double-scaled string equations for the generic order-$k$ or $(2,2k-1)$ multicritical model are very simple\footnote{String equations are simpler to write for multicritical models than for minimal strings, whereas spectral curves are simpler to write for minimal strings than for multicritical models; see, \textit{e.g.}, \cite{gs21}.} to write down \cite{gm90b}; they are non-linear ODEs of order $(2k-2)$ determined by the Gel'fand--Dikii $R_{\ell} \left[ u(z) \right]$ KdV potentials\footnote{Not to be confused with the matrix-model large-$N$ recursion coefficients, in spite of similar-looking notation.} \cite{gd75} as
\be
\label{eq:orderkstringequationfromgelfanddikii}
\left(-1\right)^{k} \frac{2^{k+1}\, k!}{\left( 2k-1 \right)!!}\, R_{k} \left[ u (z) \right] = z.
\ee
\noindent
Several examples of Gel'fand--Dikii KdV potentials alongside the multicritical and minimal string equations they build appeared in \cite{gs21} (see as well \cite{t07} for a complementary point-of-view). Herein let us focus on two cases (which however already exemplify the whole complexity present in the full KdV hierarchy, to be further explored in \cite{krst26a, krst26b}). These are the $k=2$ or $(2,3)$ multicritical theory, where \eqref{eq:orderkstringequationfromgelfanddikii} reduces to the Painlev\'e~I (\PI) equation, see, \textit{e.g.}, from the early days \cite{p02, p06, b13, b14} to the resurgence days \cite{gikm10, asv11, d16, bssv22}; and the $k=3$ or $(2,5)$ multicritical theory, where now \eqref{eq:orderkstringequationfromgelfanddikii} reduces to the Yang--Lee (\YL) equation, see, \textit{e.g.}, \cite{bmp90, gz90b, c17, gs21}.

The \PI~equation is given by\footnote{Here is the caveat on normalization, as this is \textit{not} the $k=2$ element of the multicritical hierarchy from \eqref{eq:orderkstringequationfromgelfanddikii}. This normalization is associated to double-scaling from a matrix-model odd-potential and we are using it  mostly for historical reasons: it is the one in \cite{msw07, msw08, asv11, bssv22, mss22} where large amounts of data were generated which are recycled herein. It implies that, for \PI, \eqref{eq:FdsEVENnormalization} should be written as $F_{\text{\PI}}'' (z) = - u (z)$. Details on various conventions at play may be found in \cite{bssv22}, which have no real effect on the analytics of the solution discussed in the following.}
\begin{equation}
\label{eq:Painleve1Equation}
u^2 (z) - \frac{1}{6} u^{\prime\prime} (z) = z.
\end{equation}
\noindent
The complete resurgent transseries solution to this equation, \textit{i.e.}, as adequately supplemented by its full non-linear Stokes data, was recently solved in \cite{bssv22}. Setting $x=z^{-\frac{5}{4}}$ the (double-scaled) string coupling, $A=\frac{8\sqrt{3}}{5}$ the instanton action and $\alpha_{\text{\PI}}=\frac{4}{\sqrt{3}}$ the logarithmic resummation constant (to be discussed in section~\ref{sec:resurgent-Z-transseries} and appendix~\ref{app:transasymptotic-transseries}), such transseries is given by
\begin{equation}
\label{eq:Painleve1SOlution}
u \left( x; \sigma_1, \sigma_2 \right) = x^{-\frac{2}{5}} \sum_{n=0}^{+\infty} \sum_{m=0}^{+\infty} \sigma_1^n \sigma_2^m\, \rme^{- \left( n-m \right) \frac{A}{x}}\, x^{-\frac{1}{2} \alpha_{\text{\PI}} \left( n-m \right) \sigma_1 \sigma_2}\, u^{(n|m)} (x),
\end{equation}
\noindent
where we are being a bit more detailed than in \eqref{eq:twoparameterresurgenttransseriesforR}, including the form of the transseries asymptotic sectors $u^{(n|m)}$ and their starting powers $\beta_{nm}$ as
\be
\label{eq:P1(n|m)sectors}
u^{(n|m)} (x) \simeq \sum_{g=0}^{+\infty} u_{g}^{(n|m)}\, x^{g+\beta_{nm}}
\ee
\noindent
and $\beta_{nm} = \frac{1}{2} \left(n+m\right) - \left[\frac{\min[n,m]-n\,\delta_{nm}}{2}\right]_{\text{I}}$. Akin to \eqref{eq:twoparameterresurgenttransseriesforR}, the $u_{g}^{(n|m)}$ are computed iteratively.

Next, the \YL~equation is given by
\begin{equation}
\label{eq:YangLeeEquation}
u^3 (z) - u (z)\, u^{\prime\prime} (z) - \frac{1}{2} \left( u^{\prime} (z) \right)^2 + \frac{1}{10} u^{(4)} (z) = z.
\end{equation}
\noindent
It turns out that we can solve this equation in much the same way as we did for \PI, also producing a resurgent transseries solution for $u (z)$. This time setting $x=z^{-\frac{7}{6}}$ as the (double-scaled) string coupling, as well as the appropriate logarithmic resummation constant (see section~\ref{sec:resurgent-Z-transseries} and appendix~\ref{app:transasymptotic-transseries}) $\alpha = 2 \rmi \sqrt{1+\frac{\rmi}{\sqrt{5}}}$, and instanton actions $A_1=\frac{6}{7} \sqrt{5+\rmi \sqrt{5}}$ and $A_2=\frac{6}{7} \sqrt{5-\rmi \sqrt{5}}$ (see \cite{gs21}), we find that this transseries solution is given by\footnote{Throughout we use standard vector notation often; \textit{e.g.}, $\boldsymbol{\sigma} = \left( \sigma_1, \sigma_2, \ldots \right)$, $\boldsymbol{n} = \left( n_1, n_2, \ldots \right)$, and so on.}
\begin{equation}
\label{eq:YangLeeSolution}
u \left( x; \boldsymbol{\sigma} \right) = x^{-\frac{2}{7}} \sum_{\boldsymbol{n}=0}^{+\infty} \boldsymbol{\sigma}^{\boldsymbol{n}}\, \rme^{- \left( n_1-n_2 \right) \frac{A_1}{x}}\, \rme^{- \left( n_3-n_4 \right) \frac{A_2}{x}}\, x^{\frac{\alpha}{2} \left( n_1-n_2 \right) \sigma_1 \sigma_2}\, x^{\frac{\bar{\alpha}}{2} \left( n_3-n_4 \right) \sigma_3 \sigma_4}\, u^{(\boldsymbol{n})} (x),
\end{equation}
\noindent
with transseries sectors
\be
\label{eq:YL(n|m)sectors}
u^{(\boldsymbol{n})} (x) \simeq \sum_{g=0}^{+\infty} u_{g}^{(\boldsymbol{n})}\, x^{g+\beta_{\boldsymbol{n}}}
\ee
\noindent
and starting powers
\begin{equation}
\beta_{\boldsymbol{n}} = \frac{1}{2} \sum_{i=1}^{4} n_i - \left[ \frac{1}{2} \Big( \min[n_1,n_2] + \min[n_3,n_4] - \left( n_1\, \delta_{n_1 n_2} + n_3\, \delta_{n_3 n_4} \right) \delta_{n_1 n_2}\, \delta _{n_3 n_4} \Big) \right]_{\text{I}}. 
\end{equation}

It is never enough stressing how these transseries are fully exact, nonperturbative solutions to their corresponding string equations---at least once non-linear Stokes data is added to the package \cite{krsst26b}. In this way, they must capture the full structure of general solutions. It is therefore only natural if we now ask the exact same question we had in subsection~\ref{subsec:OP-phases}: what qualifies as truly nonperturbative behavior, which we would like to see the transseries exactly and globally reproduce? One key property these general solutions have is the Painlev\'e property \cite{p02, p06}: they have \textit{no} movable\footnote{Movable singularities are singularities which depend upon initial/boundary-data; see, \textit{e.g.}, \cite{bssv22}.} \textit{multi}-valued singularities. In other words, their movable singularities are hence restricted to being \textit{poles}, and for the full KdV hierarchy they are\footnote{Start with the string equation \eqref{eq:orderkstringequationfromgelfanddikii} for the $k$th member in the hierarchy, differentiate in $z$ and rewrite making use of the Gel'fand--Dikii recursion relation \cite{gd75}
\be
\label{eq:GDrecursionforpoles}
R_{k}' = \frac{1}{4} R_{k-1}''' - u\, R_{k-1}' - \frac{1}{2} u^{\prime} R_{k-1}.
\ee
\noindent
Next assume this $k$th KdV solution may be expanded around some movable singularity $z_0$, as an arbitrary-degree Laurent-series. Regularity of the resulting relation then strictly implies degree $2$ (\textit{i.e.}, solely double poles), and further fixes the corresponding residue as belonging to the set $2, \ldots, j \left(j-1\right), \ldots, k \left(k-1\right)$ \cite{gz90b}.} just \textit{double} poles \cite{gz90b, s04}. The fascinating result in \cite{b13, b14} is that these double-poles always accumulate in one or more radial wedges on the complex $z$-plane---eventually leaving some other wedges empty---, and that these ``pizza-slice'' patterns allow for a qualitative \textit{classification} of solutions into distinct families: the Boutroux classification. This is a very well-known story for \PI~(details in the following). Herein we shall begin the equivalent story for \YL, as well as for generic solutions along the KdV hierarchy (see as well \cite{krsst26b, krst26a, krst26b}). The immediate punch-line is that these features must be \textit{fully} captured in a rather concrete and visual manner by the above transseries solutions. At the heart of how this is accomplished is the transmonomial structure \cite{e81, e93} of the transseries solutions, \eqref{eq:Painleve1SOlution} or \eqref{eq:YangLeeSolution} or else, which is always of the type
\be
\prod_{i=1}^{2k-2} \sigma_i^{n_i}\, \rme^{- n_i \frac{A_i}{x}},
\ee
\noindent
regardless of dealing with instantons or negative instantons \cite{mss22, sst23}. The two key players in these transmonomials are the instanton actions and the transseries parameters.

\begin{figure}
    \centering
    \begin{tikzpicture}[scale=0.6]    
        \draw[gray, line width=0.9pt, ->] (-7,0) -- (7,0);
        \draw[gray, line width=0.9pt, ->] (0,-7) -- (0,7);
        
        \draw[gray, line width=2pt] (-6,6) -- (-5.6,6);
        \draw[gray, line width=2pt] (-5.6,6) -- (-5.6,6.4);
        \draw[gray, fill=gray] (-5.6,6) circle (0.8pt);
        \node at (-5.9,6.3) {$z$};
        
        \draw[blue, line width=2.8pt] (0,0) -- (7,0);
        \node[blue] at (7, 0.75) {$\underline{\mathfrak{S}}_{0}$};
        
        \draw[orange, line width=2.8pt] (0,0) -- ({7*cos(144)},{7*sin(144)});
        \node[orange] at ({7.5*cos(151.2)}, {7.5*sin(151.2)}) {$\underline{\mathfrak{S}}_{\pi}$};
        
        \draw[orange, line width=2.8pt] (0,0) -- ({7*cos(216)},{7*sin(216)});
        \node[orange] at ({7.5*cos(208.8)}, {7.5*sin(208.8)}) {$\underline{\mathfrak{S}}_{-\pi}$};        
    \end{tikzpicture}
    \caption{Stokes automorphisms for \PI. In \textcolor{blue}{blue} we depict the forward Stokes automorphism, and in \textcolor{orange}{orange} the backward ones. Subscripts indicate the corresponding locations on the $x = z^{-\frac{5}{4}}$ plane, whereas our picture represents the complex $z$-plane, where $\arg (z) = -\frac{4}{5} \arg (x)$.}
    \label{fig:P1stokesAutomorphismsZplane}
\end{figure}
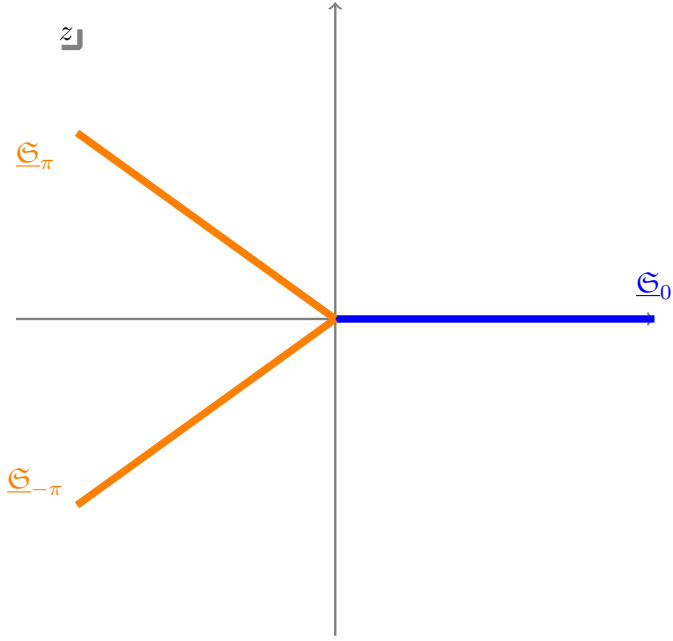

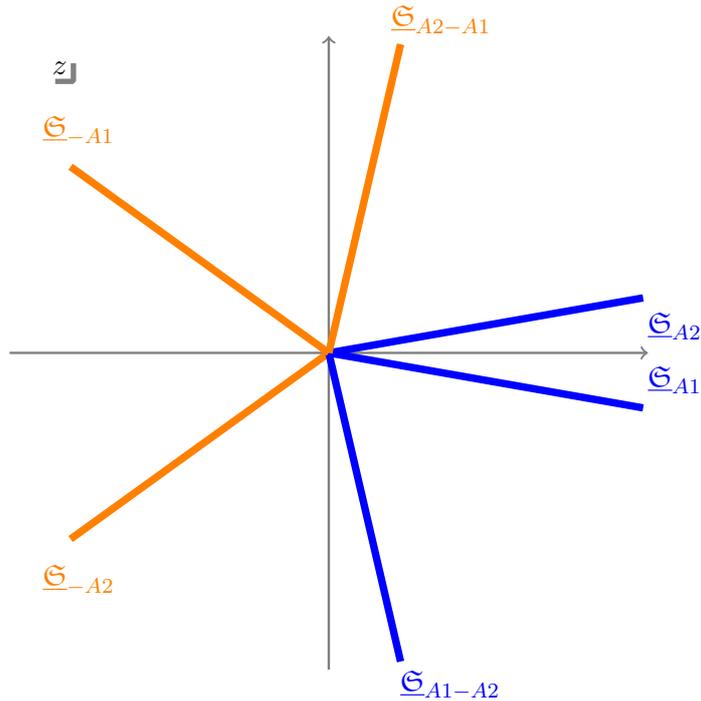
\begin{figure}
    \centering
    \begin{tikzpicture}[scale=0.6]
        \draw[gray, line width=0.9pt, ->] (-7,0) -- (7,0);
        \draw[gray, line width=0.9pt, ->] (0,-7) -- (0,7);
        
        \draw[gray, line width=2pt] (-6,6) -- (-5.6,6);
        \draw[gray, line width=2pt] (-5.6,6) -- (-5.6,6.4);
        \draw[gray, fill=gray] (-5.6,6) circle (0.8pt);
        \node at (-5.9,6.3) {$z$};
        
        \draw[orange, line width=2.8pt] (0,0) -- ({7*cos(-144)},{7*sin(-144)});
        \node[orange] at ({7.5*cos(-144+7)},{7.5*sin(-144+7)+0.1}) {$\underline{\mathfrak{S}}_{-A2}$};
        
        \draw[blue, line width=2.8pt] (0,0) -- ({7*cos(-77)},{7*sin(-77)});
        \node[blue] at ({7.5*cos(-77+7)+0.1},{7.5*sin(-77+7)-0.3}) {$\underline{\mathfrak{S}}_{A1-A2}$};
        
        \draw[blue, line width=2.8pt] (0,0) -- ({7*cos(-10)},{7*sin(-10)});
        \node[blue] at ({7.5*cos(-10+7)+0.1},{7.5*sin(-10+7)-0.2}) {$\underline{\mathfrak{S}}_{A1}$};
        
        \draw[blue, line width=2.8pt] (0,0) -- ({7*cos(10)},{7*sin(10)});
        \node[blue] at ({7.5*cos(10-7)+0.1},{7.5*sin(10-7)+0.2}) {$\underline{\mathfrak{S}}_{A2}$};
        
        \draw[orange, line width=2.8pt] (0,0) -- ({7*cos(77)},{7*sin(77)});
        \node[orange] at ({7.5*cos(77-7)-0.1},{7.5*sin(77-7)+0.3}) {$\underline{\mathfrak{S}}_{A2-A1}$};
        
        \draw[orange, line width=2.8pt] (0,0) -- ({7*cos(144)},{7*sin(144)});
        \node[orange] at ({7.5*cos(144-7)},{7.5*sin(144-7)-0.2}) {$\underline{\mathfrak{S}}_{-A1}$};
    \end{tikzpicture}
    \caption{Stokes automorphisms for \YL. In \textcolor{blue}{blue} we depict the forward Stokes automorphisms, and in \textcolor{orange}{orange} the backward ones. Subscripts indicate locations on the $x=z^{-\frac{7}{6}}$ plane such that the Stokes lines lie at $\arg(x)$. Our picture representes the complex $z$-plane.}
    \label{fig:YLstokesAutomorphismsZplane}
\end{figure}

Instanton actions (or, more generally, linear combinations thereof) encode the location of both Stokes and anti-Stokes lines. Roughly speaking, Stokes lines signal rays along which specific instanton sectors are turned on and start contributing (even if exponentially weakly) to the transseries; and anti-Stokes lines signal rays where these new contributions become of the same order as the original expansion and hence (eventually exponentially) change its behavior \cite{s64}. This of course amounts to a \textit{nonperturbative} change in behavior, and in fact to the would-be appearance or disappearance of the aforementioned poles at the crossing of anti-Stokes lines (albeit this story will only become fully explicit in \cite{krsst26b}, where we will use fully global transseries solutions to sharply analytically-reproduce numerical-algorithmical results for the locations of these poles). For \PI, Stokes lines are located at $\arg(x)=0$ and $\arg(x)=\pi$ (alongside the $n\pi$, $n \in \BZ$ multiples thereof as we rotate around either $x$ or $z$ complex planes), as shown in figure~\ref{fig:P1stokesAutomorphismsZplane} where we display the Stokes automorphisms $\underline{\mathfrak{S}}_{\arg (x)}$ which implement the corresponding Stokes jumps upon the transseries; see as well, \textit{e.g.}, \cite{k94, c97, jk01, fw11, n13, cch13, bssv22}. For \YL, due to the complicated nature of the actions (see \cite{gs21} and sections~\ref{sec:resurgent-Z-transseries} and~\ref{sec:checks-tests-numerics}), it is more convenient to display the Stokes automorphisms as labeled by $\underline{\mathfrak{S}}_{x}$. We visualize these Stokes lines on the complex $z$-plane in figure~\ref{fig:YLstokesAutomorphismsZplane} (see \cite{krsst26b} for complete details).

Transseries parameters enable Stokes transitions on transseries via the bridge equations \cite{e81, e84, e93}, where Stokes automorphisms get implemented by keeping transseries structures fixed but having transseries parameters \textit{jump} according to Stokes data, as in $\boldsymbol{\sigma} \mapsto \underline{\pmb{\BS}}_{\theta} (\boldsymbol{\sigma}) \sim \boldsymbol{\sigma} + \boldsymbol{S}$ \cite{e81, e84}. These jumps, as dictated by (complete) non-linear Stokes data, were recently fully described for \PI~in \cite{bssv22} to where we refer the reader for further details, and will be much deeply delved into in \cite{krsst26b, krst26a, krst26b}. One common concern arises when transseries parameters associated to negative instantons get turned on, where one worries about the role of enhanced exponentials in obtaining actual numerical values from a transseries solution. This should not be a concern as what one is really dealing with for either specific-heat or free-energy is a power-\textit{series} in these transmonomials. Even in practice, where truncations might be needed, there are different tools one can use which we elaborate upon in the following (and in \cite{krsst26b}). For the purpose of this discussion it suffices to say that by choosing sufficiently small values of $\sigma$ (for an appropriately enhanced direction), one can easily and immediately obtain reasonable, rather finite results for the partition-function even by as simple a method as optimal truncation.

Having introduced the idea of the Boutroux classification, one may ask how (or if) does it connect to the discussion on matrix-model phases in the previous subsection. After all, the present discussion should just be a ``double-scaled version'' of that earlier discussion. Well, as mentioned, and for the full KdV hierarchy, movable singularities of the string-equation solutions---of the specific heat---are double-poles with residue $2$. But then, via \eqref{eq:FdsEVENnormalization}, these translate to simple zeroes of the partition function (see \cite{bssv22} as well), which is to say, they translate to Yang--Lee \cite{ly52a, ly52b} or Fisher \cite{f65} zeroes describing different phases of these double-scaled multicritical and minimal-string models. Different ``pizza-slice'' Boutroux distributions are hence in direct correspondence with the different physics of the string equations along the KdV hierarchy.

\begin{figure}
\centering
\begin{subfigure}[b]{0.325\textwidth}
    \centering
    \includegraphics[width=\textwidth]{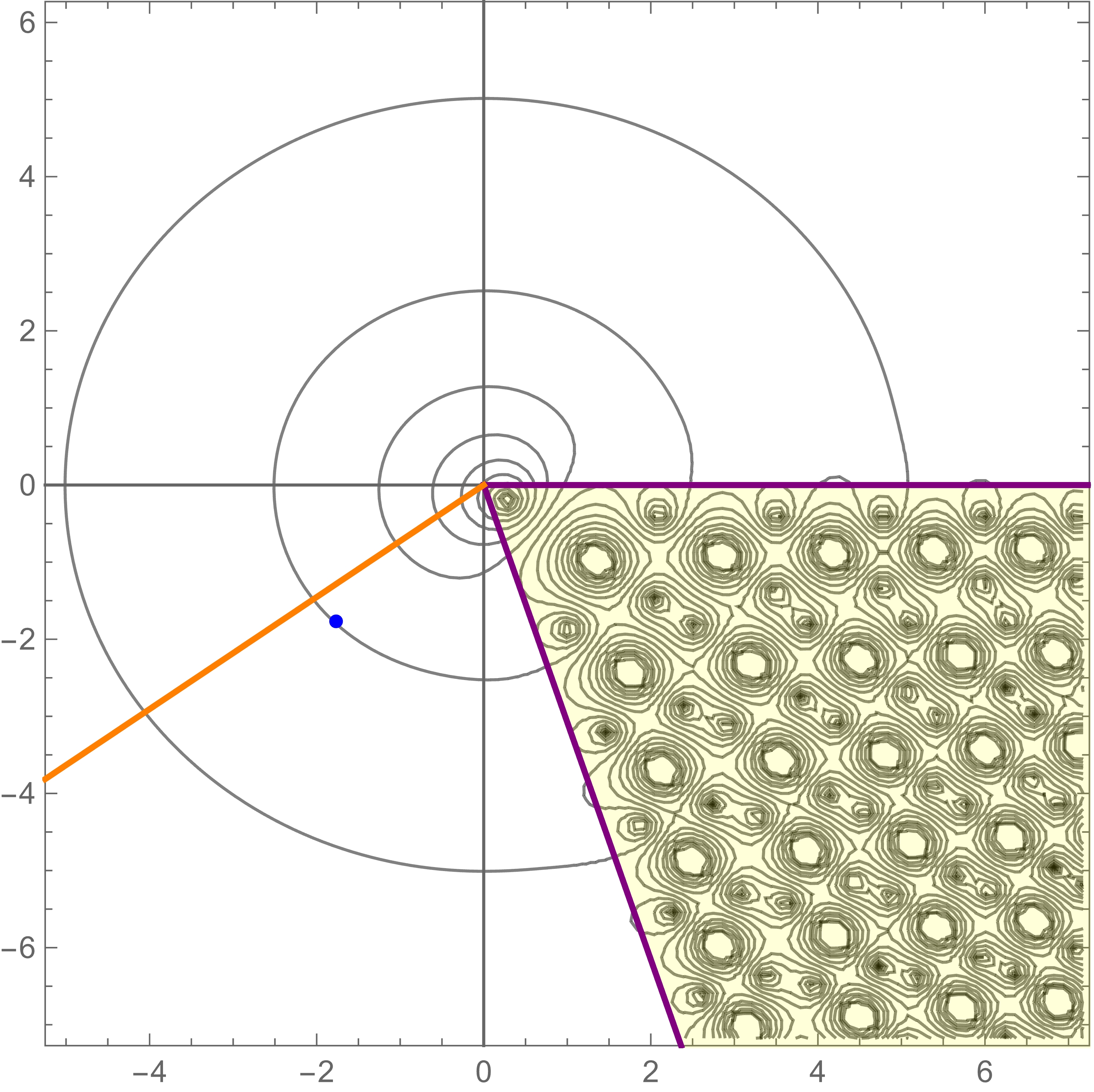}
    \caption{Tritronqu\'ee: $z_0 = \frac{5}{2}\, \rme^{-\frac{3\pi\rmi}{4}}$, \\ $(\sigma_1,\sigma_2) = (0,-0.371)$.}
\end{subfigure}
\hfill
\begin{subfigure}[b]{0.325\textwidth}
    \centering
    \includegraphics[width=\textwidth]{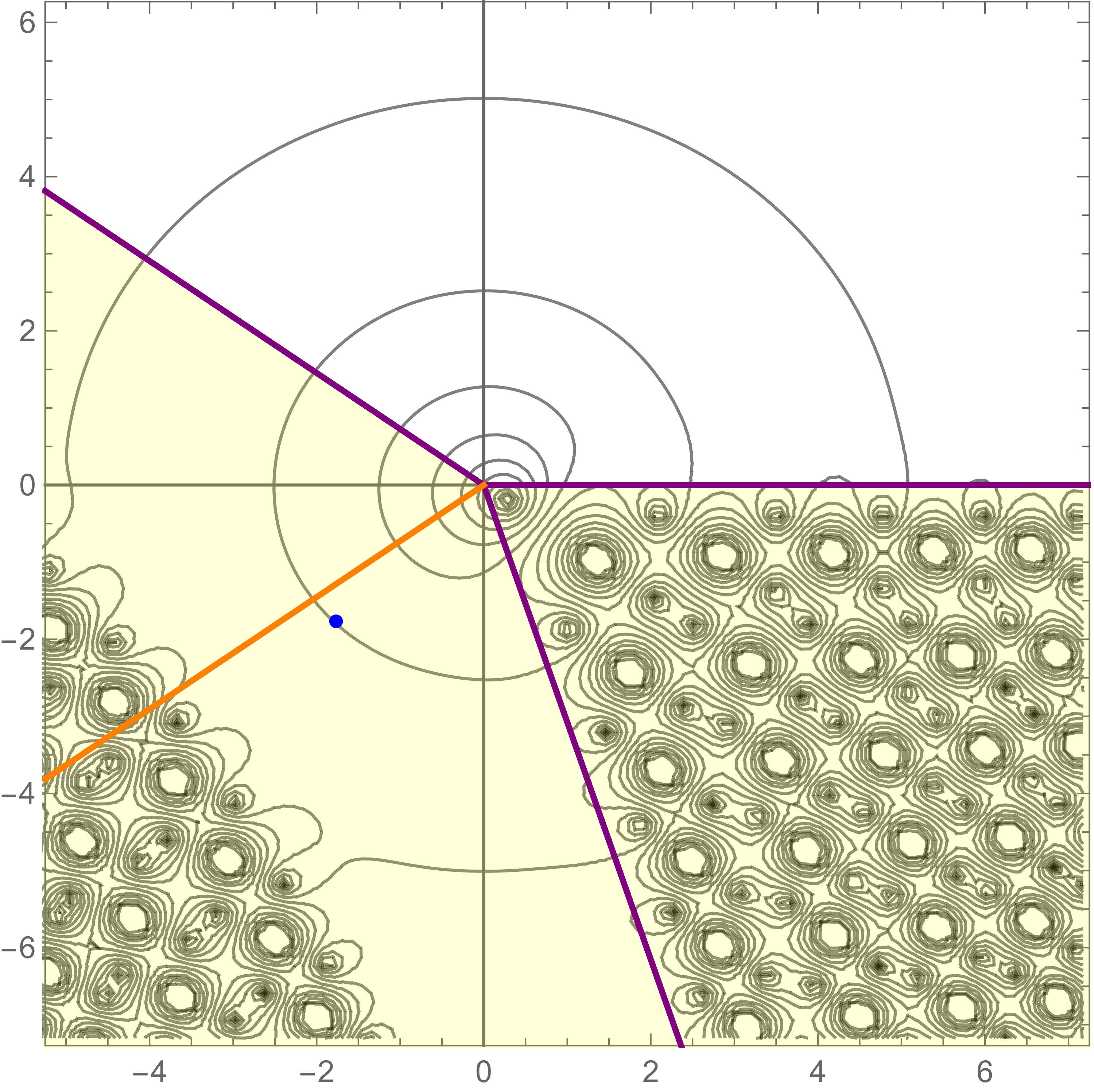}
    \caption{Tronqu\'ee: $z_0 = \frac{5}{2}\, \rme^{-\frac{3\pi\rmi}{4}}$, \\ $(\sigma_1,\sigma_2) \sim (10^{-8},-0.371)$.}
\end{subfigure}
\hfill
\begin{subfigure}[b]{0.325\textwidth}
    \centering
    \includegraphics[width=\textwidth]{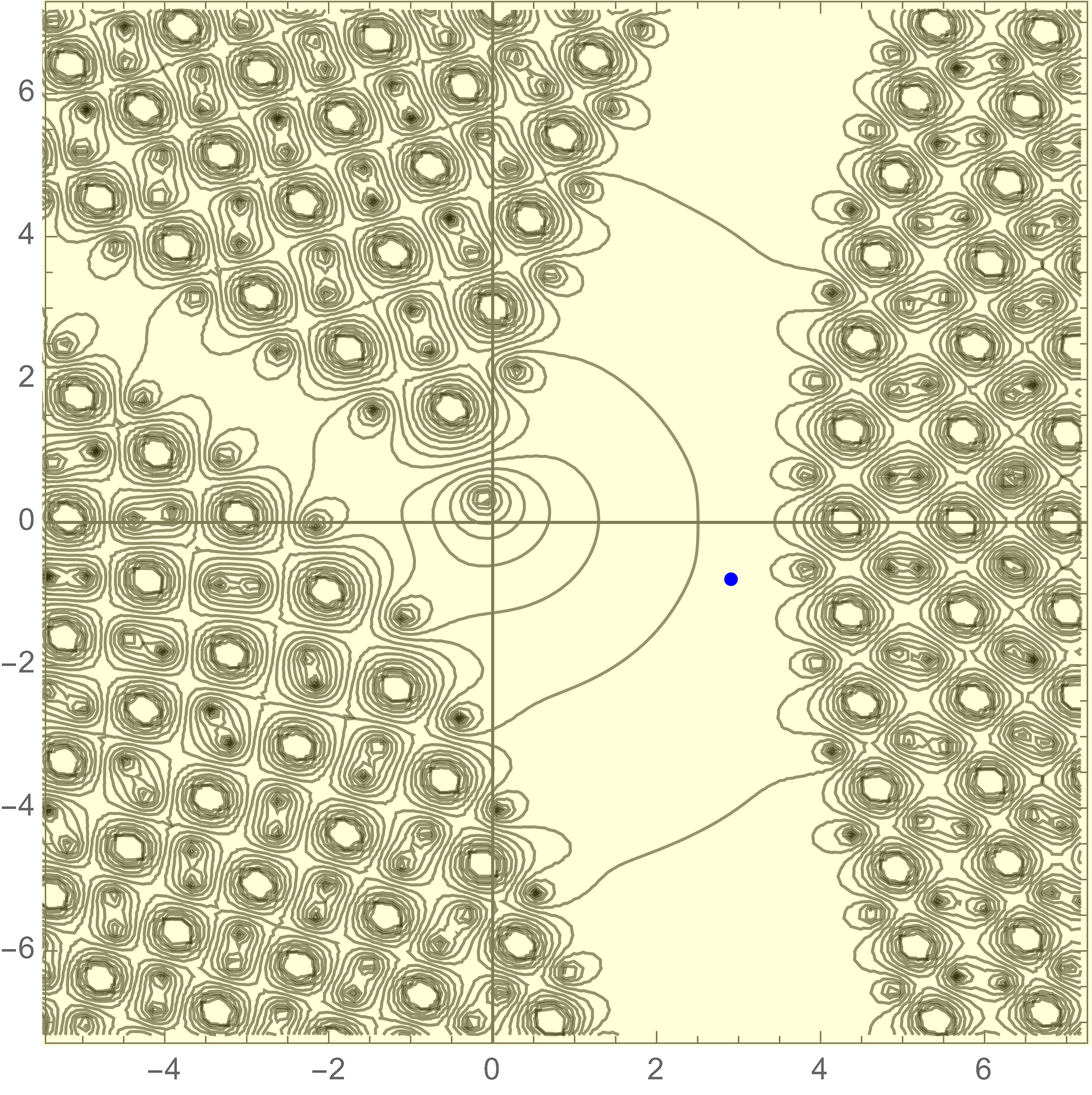}
    \caption{Generic: $z_0 = 3\, \rme^{-\frac{\rmi \pi}{12}}$, \\ $(\sigma_1,\sigma_2) = (10^{-2},10^{-6})$.}
\end{subfigure}
    \caption{Numerical plots of \PI~solutions, illustrating the different Boutroux families. Poles of the specific heat appear as circles around which the contour-plot lines cluster. The initial point for the numerical algorithm is $z_0$ and is indicated in the figures by the \textcolor{blue}{blue} disk. \textcolor{orange}{Orange} lines indicate the backward Stokes automorphisms, relevant for these solutions (see the discussion in \cite{krsst26b}). \textcolor{purple}{Purple} lines indicate where arrays of poles asymptote towards (obtained by numerically probing various solutions with similar initial conditions). As we will later see, these lines are in fact anti-Stokes lines. The tritronqu\'ee and tronqu\'ee solutions are actually $0$ and $1$ parameter transseries respectively, although they are \textit{locally} described by $1$ and $2$ parameters (see the text).}
    \label{fig:PISolutionsNoZeros}
\end{figure}

As mentioned and referenced above (see \cite{bssv22} for further references), the Boutroux classification is a very well-known story for \PI; whereas for the KdV hierarchy understanding and classifying the types of solutions is, in general, an open problem (with recent progress for \YL~in \cite{gkk13}). What Boutroux showed for \PI~\cite{b13, b14} is that there exist three types of solutions, based on the locations of their poles on the complex $z$-plane:
\begin{enumerate}
\item \textit{General} solutions, which have poles on all five wedges of the complex plane;
\item \textit{Tronqu\'ee} solutions, which have poles in three out of five wedges of the complex $z$-plane, and are regular in the remaining two;
\item \textit{Tritronqu\'ee} solutions, which have poles in a single wedge and are regular elsewhere.
\end{enumerate}
\noindent
Using numerical methods similar to those in \cite{fw11, fw14, sv22} (which will receive much more attention later-on in section~\ref{sec:checks-tests-numerics} and in \cite{krsst26b}), one can plot different \PI~solutions and see these various classes of solutions arise, as illustrated\footnote{Let us clarify the data in these figures. They are generated numerically, based on the algorithms in \cite{fw11, sv22}, for which they need an initial point $z_0$; and in this sense they amount to double-scaled specific-heat ``raw data''. In section~\ref{sec:checks-tests-numerics} and in \cite{krsst26b} we will reproduce these results using transseries, for which one first needs to translate from initial/boundary data $z_0$ into transseries parameters. Albeit this will be done later, we already present such dictionary herein. Note that when transseries parameters have a simple exact form, we write an equality to describe them; but where they are complicated, we use the $\sim$ symbol to indicate their ``effective'' value.} in figure~\ref{fig:PISolutionsNoZeros}. What we shall show in section~\ref{sec:checks-tests-numerics} is that all these generic numerical results may be compared to (analytical) transseries solutions with remarkable precision---at least up to an important caveat. What we construct in the present paper are exact albeit \textit{local} solutions. In other words, the transseries solutions only match the numerical results in \textit{certain regions} of the complex plane. In order to fully reproduce the numerical solutions \textit{all over} the complex plane one needs to augment our local transseries solutions with \textit{global} Stokes data. Then, upon Stokes transitions, we indeed find precise matches between our resurgent transseries and the numerical solutions \textit{all over} the complex plane---which will be demonstrated in the follow-up paper \cite{krsst26b}. Another crucial factor which will be required in order to globally reproduce all solutions correctly is the existence and incorporation of \textit{negative instantons} in the transseries solutions \cite{mss22, sst23} (which we already do in the present paper). Interestingly enough and somewhat unexpectedly, this is required not only for the most generic solutions, but even for the more mundane tritronqu\'ee and tronqu\'ee solutions of \PI.

Another comment regarding figure~\ref{fig:PISolutionsNoZeros} is that although the tritronqu\'ee and tronqu\'ee solutions will later appear as $1$ and $2$ parameter transseries, they are in fact still just $0$ and $1$ parameter \textit{transcendents}, respectively, as defined in the following. This is an observation already made in \cite{sv22}, and is a consequence of the fact that transseries parameters are only \textit{local} descriptions of the solution. In fact, in \cite{krsst26b} we demonstrate that, upon crossing the backward Stokes line (plotted in orange in the figures), these two solutions indeed are characterized by boundary data $(\sigma_1,\sigma_2) = (0,0)$ and $(\sigma_1,\sigma_2) = (10^{-8},0)$, respectively. In other words, one should regard the \textit{transcendents} as the transseries representation characterized by the \textit{minimal} amount of transseries parameters, for the given solution.

\begin{figure}
\centering
\begin{subfigure}[b]{0.48\textwidth}
    \centering
    \includegraphics[width=\textwidth]{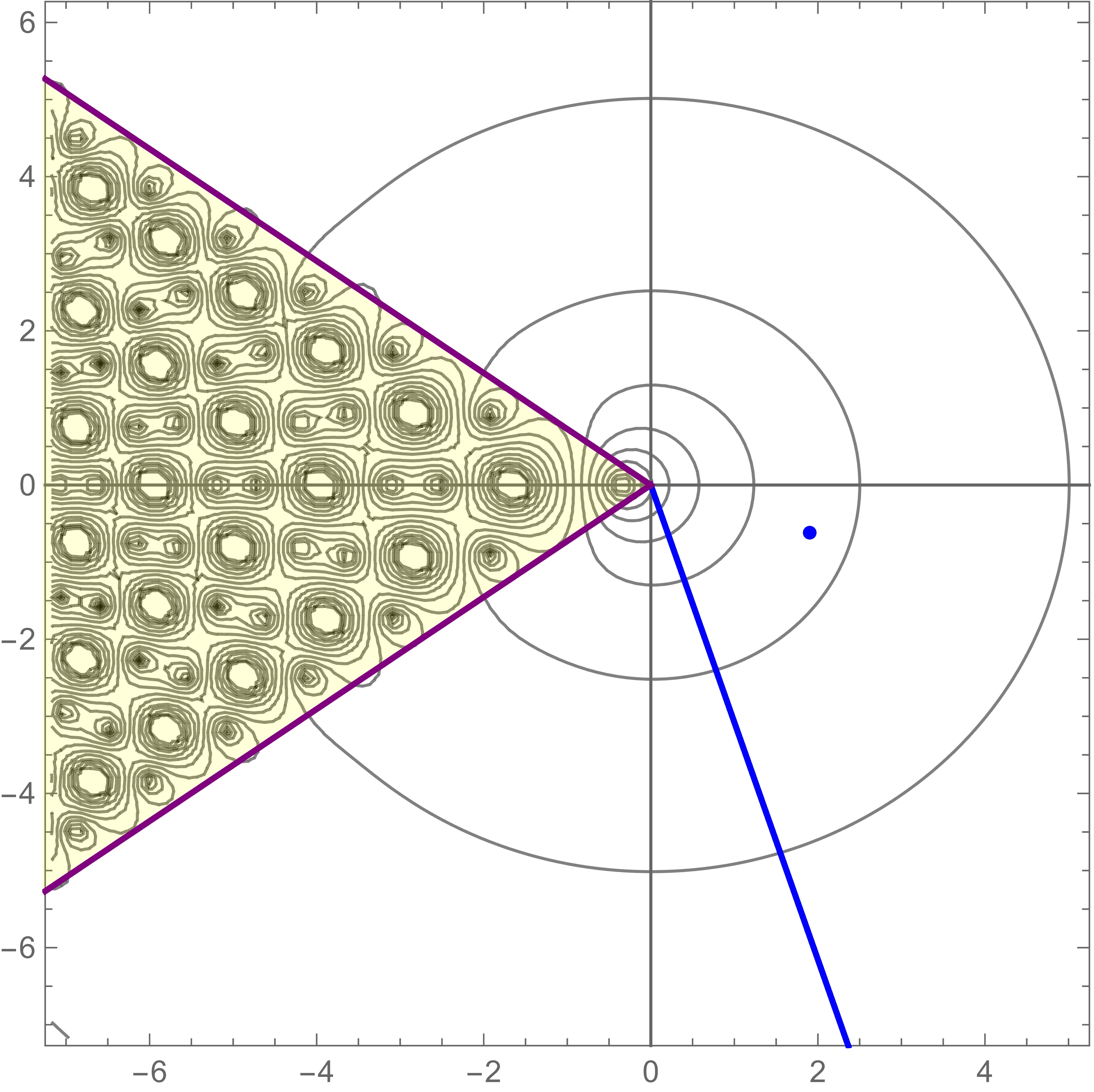}
    \caption{Tritronqu\'ee: $z_0 = 2\, \rme^{-\frac{\rmi \pi}{10}}$, $(\sigma_1,\sigma_2)=(0,0)$.}
\end{subfigure}
\hfill
\begin{subfigure}[b]{0.48\textwidth}
    \centering
    \includegraphics[width=\textwidth]{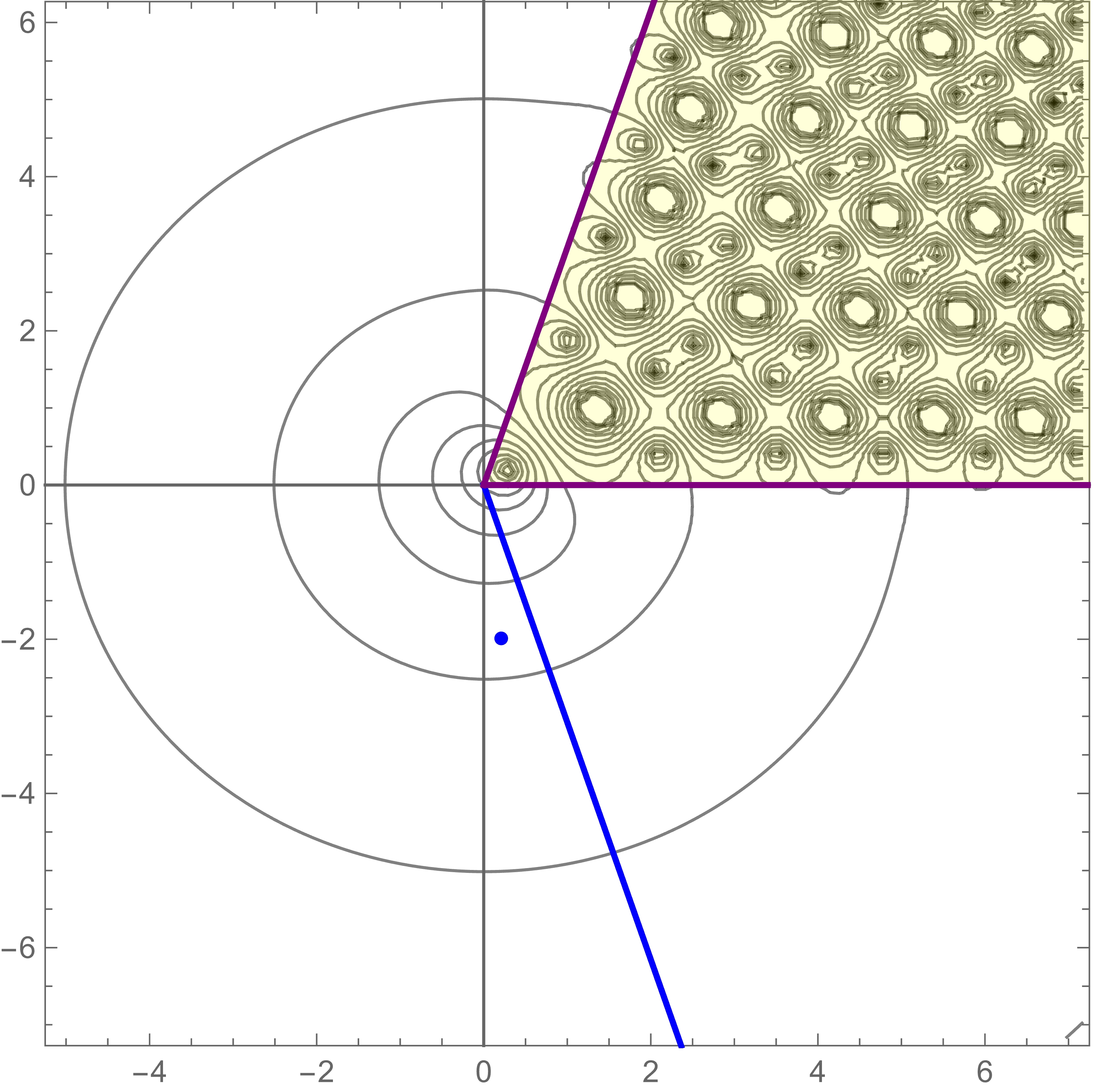}
    \caption{Tritronqu\'ee: $z_0 = 2\, \rme^{-\frac{7\pi \rmi}{15}}$, $(\sigma_1,\sigma_2)=(0,0)$.}
\end{subfigure}
    \caption{Numerical plots of \PI~solutions, focusing upon the tritronqu\'ee case. Features in the plots are as in figure~\ref{fig:PISolutionsNoZeros}, with the addition of the \textcolor{blue}{blue} lines which now indicate the forward Stokes automorphisms relevant for these solutions (see the discussion in \cite{krsst26b}). The main point is that these solutions are reproducible by the \textit{second} Riemann sheet of our transseries solution, as is explained in the main text (and as will be made much clearer in \cite{krsst26b}).}
    \label{fig:PINewBranchTritronqueeSolutionsNoZeros}
\end{figure}

There is one final very interesting subtlety we still want to mention, to which we shall return in  detail later-on and in \cite{krsst26b}. It stems from a few observations concerning transseries solutions: ones in \cite{bssv22, sv22} within the context of \PI, and one other in \cite{v23} within the context of the Hastings--McLeod solution \cite{hm80} of the homogeneous Painlev\'e~II equation; which go as follows. There are regions of the complex plane that, for certain \textit{special} transseries solutions, have at least one of the transseries parameters $(\sigma_1,\sigma_2)$ diverge, and hence end-up producing a \textit{singular transseries}; \textit{i.e.}, the transseries is seemingly \textit{no longer} an appropriate representation of the function in these regions. For example, trying to reproduce plots as in figure~\ref{fig:PISolutionsNoZeros} for the tritronqu\'ee solutions with poles in either the first or third ``pizza slice'' wedges at first naively seems to be out of reach. But if our claim of having obtained exact solutions is correct, then this cannot be the case. The apparent problem is one of (in)adequate analytic continuation, and was solved in \cite{v23}: the exact description of \textit{global} solutions not only requires Stokes data \cite{krsst26b} but in the process the transseries solution also needs to undergo changes of branches within the Riemann surface which describes the solution to the classical string equation. In the \PI~example, the \textit{classical} string equation from \eqref{eq:Painleve1Equation} is simply (see \cite{gs21} for more examples)
\be
\label{eq:DSL-P1-classical-string-eq}
u^2 (z)= z,
\ee
\noindent
which involves a square-root double-sheeted surface, and which translates into the transseries solution \eqref{eq:Painleve1SOlution} as the leading $x^{-\frac{2}{5}} = \sqrt{z}$ term. Because the transseries description is, in this way, multi-sheeted, when the transseries parameters are singular on one sheet, we simply rotate upon the complex $z$-plane to describe the solution on a different\footnote{Note that this part needs to be partially exploratory, as one can rotate in either direction, \textit{i.e.}, rotating in either way gives the \textit{same} pole distribution. Choosing between which direction to rotate can be convenient for implementation purposes only (as we might end up with transseries parameters associated with better convergence properties of instanton summation in one of the directions). Knowledge of the exact Stokes automorphisms of the problem are of great help for these explorations; see \cite{krsst26b}.} sheet. In practice, one may either change the genus-zero perturbative-sector coefficient $u_0^{(0|0)}$ in \eqref{eq:Painleve1SOlution} as $u_0 = +1 \mapsto u_0 = -1$ (which of course has appreciable consequences on the full structure of the transseries), or as perhaps a simpler approach, one can take our original transseries solution \eqref{eq:Painleve1SOlution} and implement the analytic continuation $z \mapsto z \cdot \rme^{2\pi\rmi}$. In any case, doing either of this, and as discussed in \cite{krsst26b}, we are immediately able to obtain the two ``missing'' tritronqu\'ee solutions (shown in figure~\ref{fig:PINewBranchTritronqueeSolutionsNoZeros}). This is a feature which is easy to spot in this case of \PI, but which is of course also of relevance for all double-scaled and off-critical matrix models.

\begin{figure}
\centering
\begin{subfigure}[b]{0.325\textwidth}
    \centering
    \includegraphics[width=\textwidth]{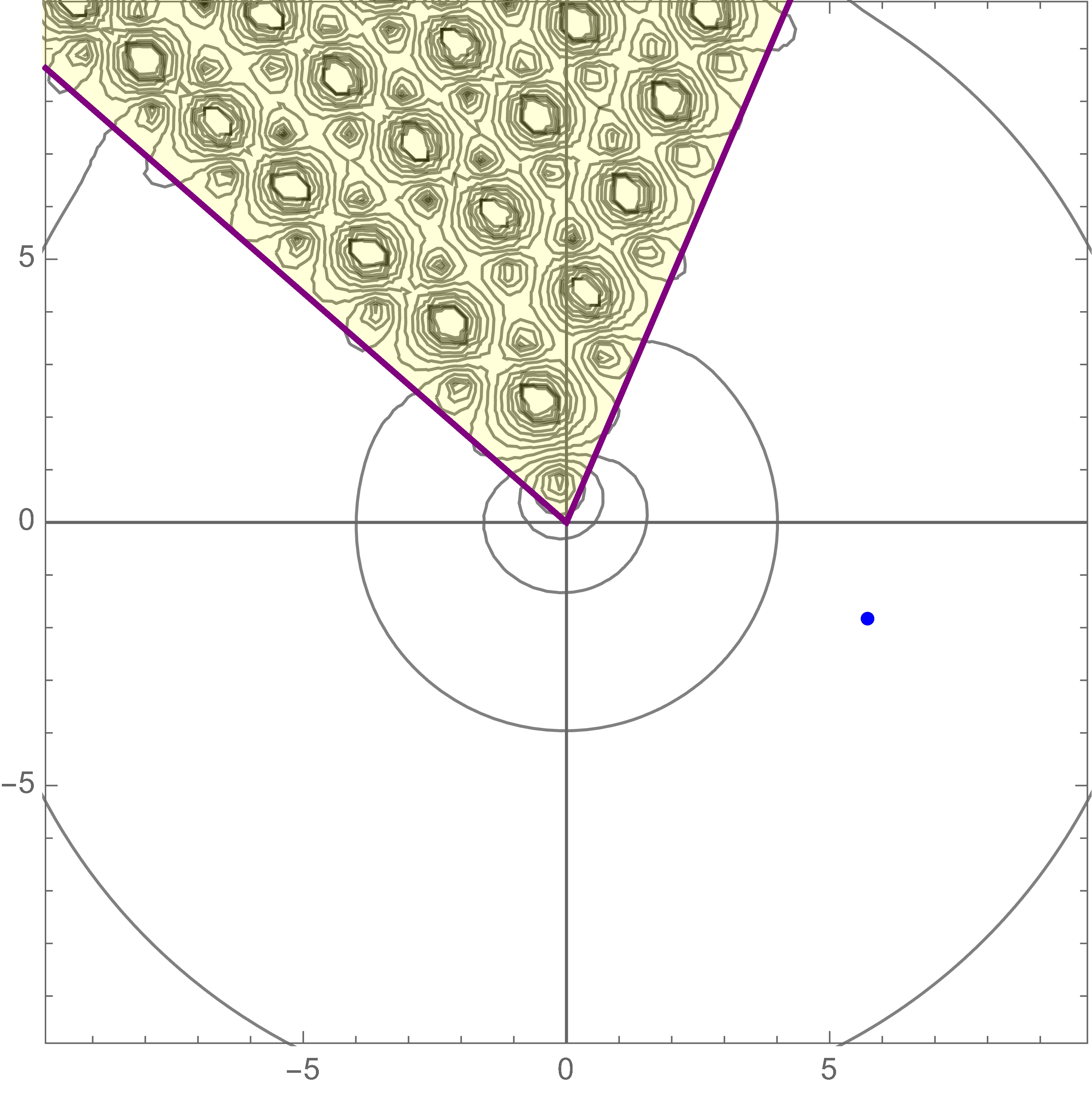}
    \caption{$z_0 = 6\, \rme^{-\frac{\rmi\pi}{10}}$, \\ $\boldsymbol{\sigma} = \left( 0,0,0,0 \right)$.}
	\label{fig:YLSolutionsNoZeros0000}
\end{subfigure}
\hfill
\begin{subfigure}[b]{0.325\textwidth}
    \centering
    \includegraphics[width=\textwidth]{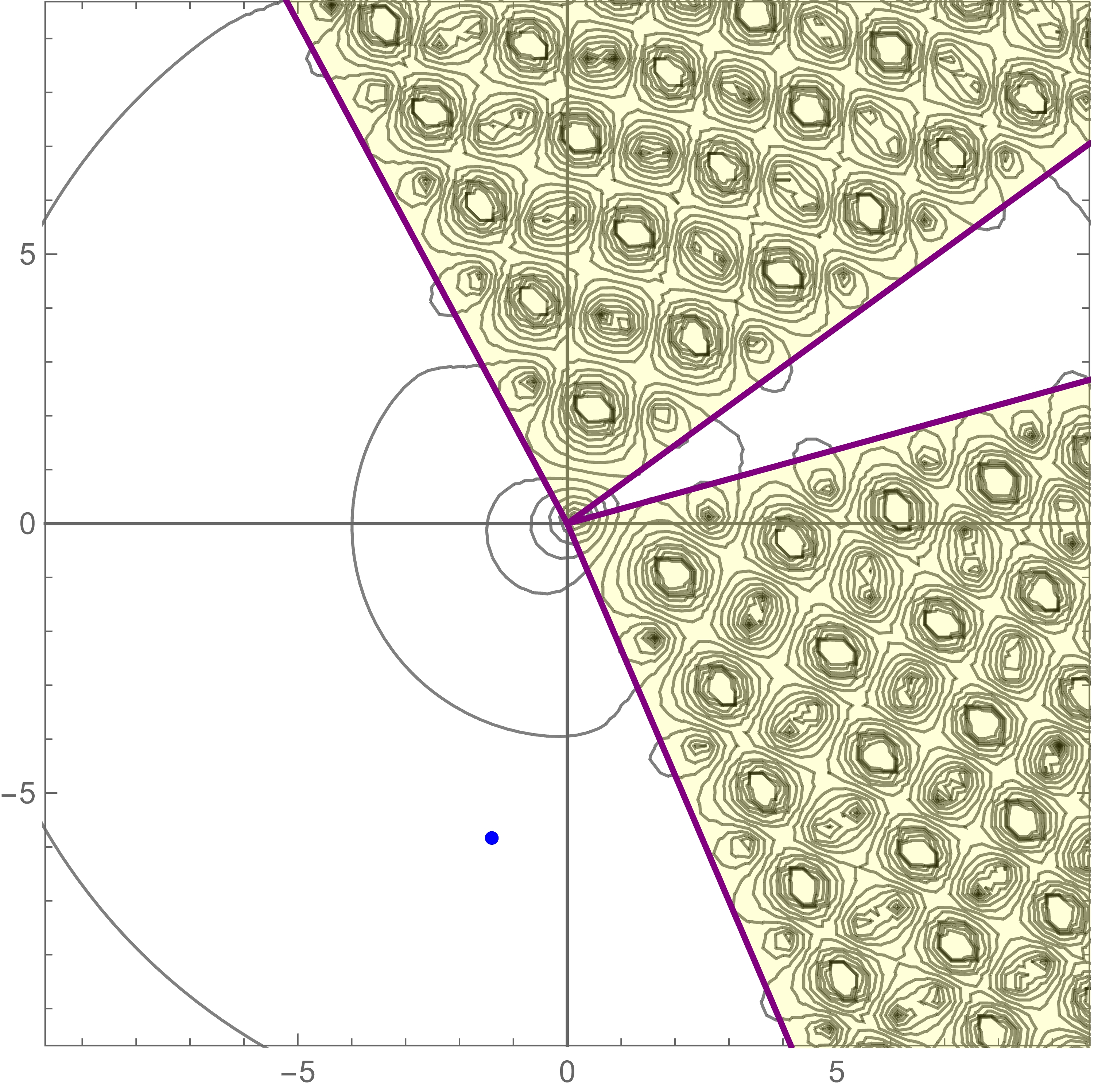}
    \caption{$z_0 = 6\, \rme^{- \frac{23\rmi\pi}{40}}$, \\ $\boldsymbol{\sigma} = \left( 0,0,0,\mathbf{S} \right)$.}
	\label{fig:YLSolutionsNoZeros000S}
\end{subfigure}
\hfill
\begin{subfigure}[b]{0.325\textwidth}
    \centering
    \includegraphics[width=\textwidth]{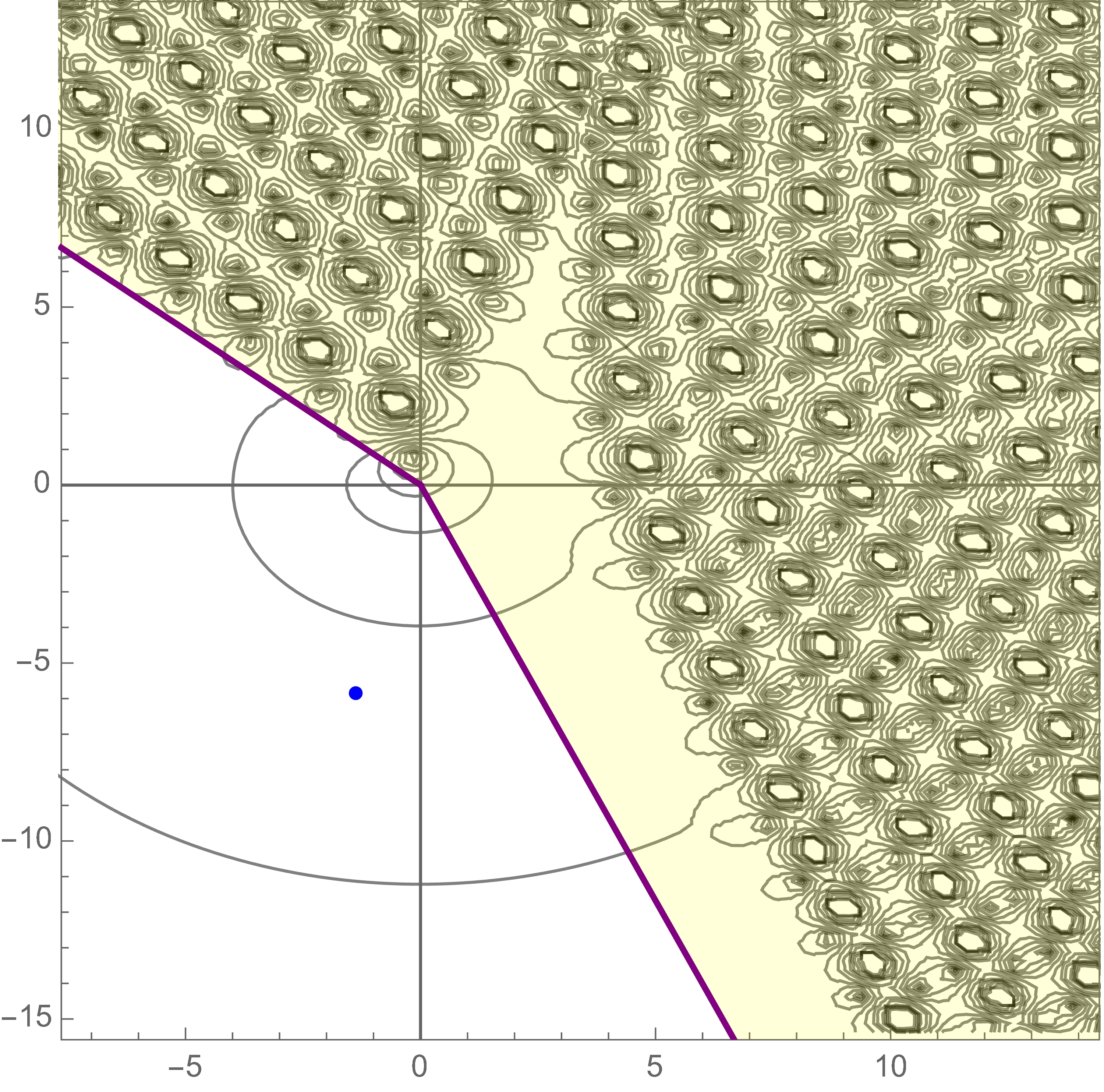}
    \caption{$z_0 = 6\, \rme^{- \frac{23\rmi\pi}{40}}$, \\ $\boldsymbol{\sigma} = \left( 0,0,0,\frac{1}{10^4} \right)$.}
	\label{fig:YLSolutionsNoZeros0004}
\end{subfigure}

\bigskip

\begin{subfigure}[b]{0.325\textwidth}
    \centering
    \includegraphics[width=\textwidth]{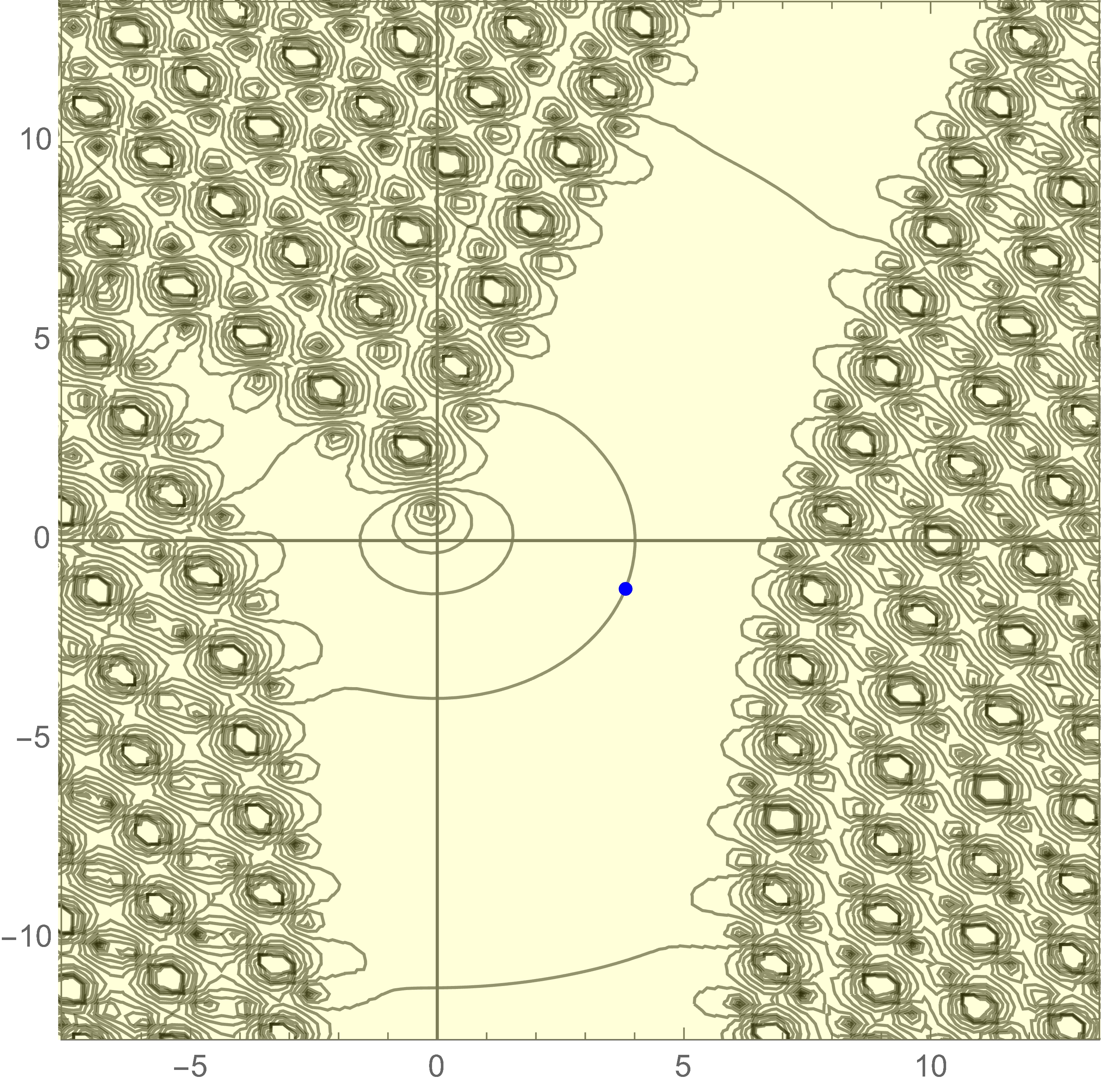}
    \caption{$z_0 = 4\, \rme^{- \frac{\rmi\pi}{10}}$, \\ $\boldsymbol{\sigma} = \left( \frac{1}{10^4},\frac{1}{10^8},0,0 \right)$.}
	\label{fig:YLSolutionsNoZeros4800}
\end{subfigure}
\hfill
\begin{subfigure}[b]{0.325\textwidth}
    \centering
    \includegraphics[width=\textwidth]{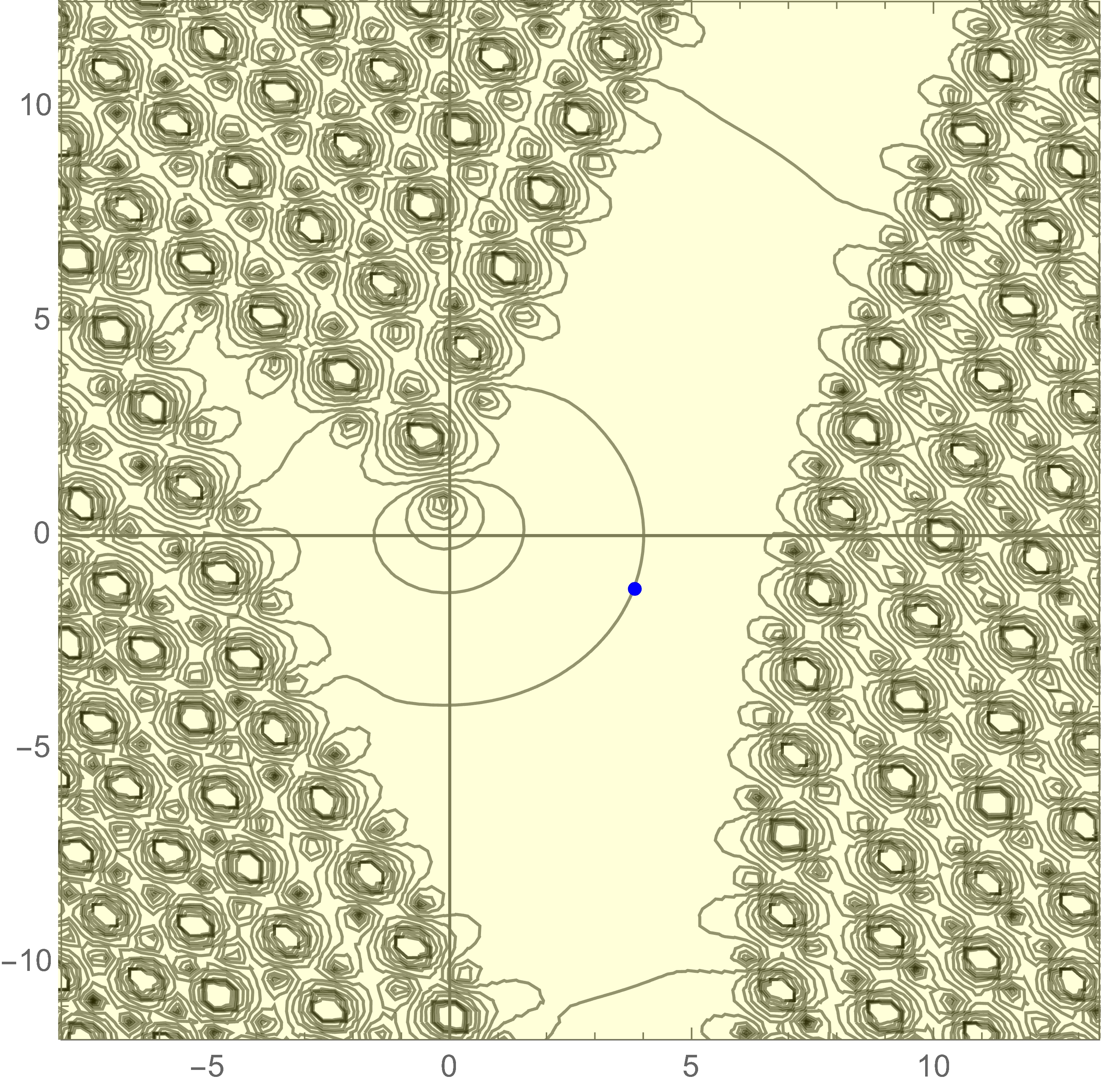}
    \caption{$z_0 = 4\, \rme^{- \frac{\rmi\pi}{10}}$, \\ $\boldsymbol{\sigma} = \left( \frac{1}{10^4},\frac{1}{10^8},\frac{1}{10^5},0 \right)$.}
	\label{fig:YLSolutionsNoZeros4850}
\end{subfigure}
\hfill
\begin{subfigure}[b]{0.325\textwidth}
    \centering
    \includegraphics[width=\textwidth]{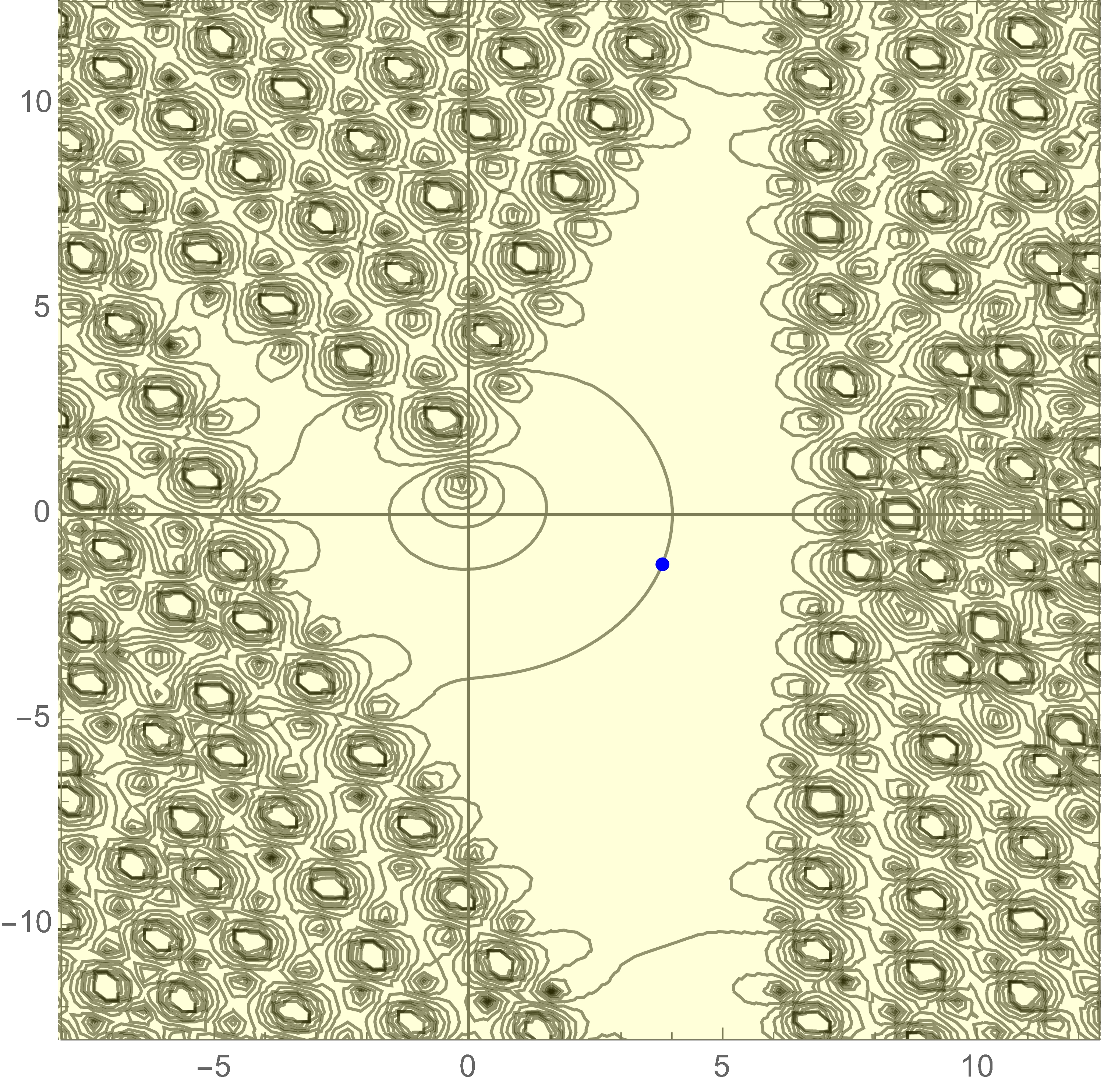}
    \caption{$z_0 = 4\, \rme^{- \frac{\rmi\pi}{10}}$, \\ $\boldsymbol{\sigma} = \left( \frac{1}{10^4},\frac{1}{10^8},\frac{1}{10^4},\frac{1}{10^8} \right)$.}
	\label{fig:YLSolutionsNoZeros4848}
\end{subfigure}
    \caption{Numerical plots of \YL~solutions, illustrating new solution classes, or different \textit{generalized} Boutroux families. As in previous plots, poles of the specific heat are circles around which contour-plot lines cluster; and the initial point for the numerical algorithm is the \textcolor{blue}{blue} disk $z_0$. No forward or backward Stokes lines are depicted as they are a bit convoluted (but see figure~\ref{fig:YLstokesAutomorphismsZplane}); they will be fully presented in \cite{krsst26b} where the global reconstructions are carried through in detail. \textcolor{purple}{Purple} lines indicate where arrays of poles asymptote towards; which are in fact anti-Stokes lines. We further made use of the canonical Stokes coefficient $\boldsymbol{S} = - \frac{1}{\sqrt{\pi}} \sqrt[4]{- \frac{5}{24} - \rmi \frac{\sqrt{5}}{24}}$ (this is the second canonical Stokes coefficient complementing the first one already computed in \cite{gs21}).}
    \label{fig:YLSolutionsNoZeros}
\end{figure}

Even though the Boutroux classification of \PI~solutions is very well known and understood, the same is very far from being the case for the entire KdV hierarchy; in fact let alone already starting at its second element. For this case, the \YL~case\footnote{The \YL~equation is sometimes referred to in the mathematical literature as the \PI$^2$ equation.}, only a few classes of solutions have been identified in the literature \cite{gkk13}, in particular the ones denoted by tritronqu\'ee of type 1 and tritronqu\'ee of type 2, again based on the locations of their poles on the complex $z$-plane:
\begin{enumerate}
\item \textit{Tritronqu\'ee} solutions of \textit{Type 1}, as displayed in subfigure~\ref{fig:YLSolutionsNoZeros0000} of figure~\ref{fig:YLSolutionsNoZeros}, are the \YL~analogues of the \PI~tritronqu\'ee solutions, which have poles in a single wedge of the complex $z$-plane, and are regular everywhere else;
\item \textit{Tritronqu\'ee} solutions of \textit{Type 2}, as displayed in subfigure~\ref{fig:YLSolutionsNoZeros000S} of figure~\ref{fig:YLSolutionsNoZeros}, have poles in \textit{two} wedges of the complex $z$-plane and are regular everywhere else, in spite of being zero-parameter transcendents.
\end{enumerate}
\noindent
Now, in spite of not much being known about the classification of possible solutions, nothing stops us from just probing them: first numerically in precisely the same way as we did for \PI; but, most importantly, subsequently with our exact transseries solutions. By doing this we are in fact opening the door to the \textit{classification} of generalized Boutroux families of solutions for the \textit{entire} KdV hierarchy. For example: one can use the transseries parameters of our exact solutions being ``on'' or ``off'' as a first (albeit incomplete!) non-trivial systematization of a would-be generalized Boutroux classification for all KdV solutions. This is further discussed in \cite{krsst26b}. For the moment, focusing on the numerical approach, figure~\ref{fig:YLSolutionsNoZeros} illustrates several new types of solutions, organized as in \textit{generalized} Boutroux families, which display a remarkably varied behavior of pole locations given distinct boundary conditions\footnote{Note how in the figures it is already becoming somewhat clear how boundary conditions arising from several non-zero transseries parameters generically correspond to solutions with more ``pizza-slice'' pole regions.}. Later on in section~\ref{sec:checks-tests-numerics} we will investigate how we can understand (and verify the existence of) these numerical results directly from transseries solutions to the \YL~equation. Again, as was the case for \PI, the match will only work locally at first. To globally reconstruct these solutions, as discussed in \cite{krsst26b}, we will further require once again both the use of \textit{all} relevant Stokes automorphisms, as well as both positive and negative instanton sectors. On top, one will encounter for \YL~the exact same issue of singular transseries as just mentioned in the case for \PI. In particular, this will occur for the tritronqu\'ee type 2 solution shown in subfigure~\ref{fig:YLSolutionsNoZeros000S} of figure~\ref{fig:YLSolutionsNoZeros}, an issue which can once again be overcome in precisely the same way as for \PI, via adequate analytic continuation. In fact, these features of Stokes transitions and analytic continuation encode all the non-trivial steps required to properly establish our exact transseries solutions as being \textit{globally} and \textit{everywhere} valid for the whole KdV hierarchy.

\subsection{Large $N$ Phases from Spectral Geometry}
\label{subsec:SG-phases}

Let us now turn to address the spectral geometry approach \cite{bipz78, ackm93, a96, e04, eo07a} and how it immediately leads to a very natural (and visual) understanding \cite{d91} of all the large $N$ phases we have encountered in the two previous subsections, \ref{subsec:OP-phases} and~\ref{subsec:DSL-phases}. Going back to the steepest-descent integration contours of the $N$-dimensional eigenvalue integral \eqref{eq:partitionfunctioneigenvalues}, in subsection~\ref{subsec:OP-phases} we considered the case where all eigenvalues followed the same integration contour $\CC$. In such scenario, the eigenvalue spectral density $\varrho (\lambda)$ has support upon a \textit{single}-cut $\NCC = (x_1,x_2) \subset \BC$. The planar one-point resolvent $\CW_{0;1} (p)$ \eqref{eq:genusgmultiresolvents}---which is essentially the Hilbert transform of $\varrho (\lambda)$---then simply defines the genus-zero or one-cut matrix-model spectral-curve for this configuration as
\be
\label{eq:one-cut-spectral-curve=background-around-which-string-equations-are-constructed}
y (z) = V'(z) - 2 t\, \CW_{0;1} (z) \equiv M(z) \sqrt{ \left( z-x_1 \right) \left( z-x_2 \right) }.
\ee
\noindent
Herein, and for polynomial potentials, the moment function $M (z)$ is
\be
\label{eq:moment-function-one-cut}
M (z) = \oint_{(0)} \frac{\rmd w}{2\pi\rmi}\, \frac{V' (1/w)}{1 - w z}\, \frac{1}{\sqrt{ \left( 1 - x_1 w \right) \left( 1 - x_2 w \right) }},
\ee
\noindent
where the two end-points $x_1$, $x_2$ of the single-cut $\NCC$ satisfy a system of two equations
\be
\label{eq:one-cut-end-points}
\oint_{\NCC} \frac{\rmd w}{2\pi\rmi}\, \frac{w^n\, V' (w)}{\sqrt{ \left( w-x_1 \right) \left( w-x_2 \right) }} = 2t\, \delta_{n,1}, \qquad n=0,1.
\ee
\noindent
For the cubic matrix model \eqref{eq:CubicMatrixModelPotential} the single cut is given by\footnote{The sum of the end-points of the cut satisfies a cubic equation, $\left(x_1+x_2\right) \left( 2 - \lambda \left(x_1+x_2\right) \right) \left( 4 - \lambda \left(x_1+x_2\right) \right) = 16 \lambda t$, with their difference then given by $\left(x_1-x_2\right)^2 = \frac{2}{\lambda} \left(x_1+x_2\right) \left( 4 - \lambda \left(x_1+x_2\right) \right)$.} $\NCC = (x_1,x_2)$ with
\be
\label{eq:cubic-spectral-curve-one-cut}
y(z) = \left( 1 - \frac{\lambda}{4} \left( 2 z + x_1 + x_2 \right) \right) \sqrt{ \left( z-x_1 \right) \left( z-x_2 \right) }.
\ee
\noindent
For the quartic matrix model \eqref{eq:QuarticMatrixModelPotential} the single cut is given by $\NCC = (-2\alpha,2\alpha)$ with\footnote{This is the branch which reduces to the Gaussian matrix model in the limit where the quartic coupling $\lambda \to 0$.}
\be
\label{eq:quartic-cut-endpoint-one-cut}
\alpha^2 = \frac{1}{\lambda} \left( 1 - \sqrt{ 1-2\lambda t } \right)
\ee
\noindent
and\footnote{In this simple example, one may also explicitly write the eigenvalue density straight from the spectral curve as
\be
\varrho (z) = \frac{1}{2\pi t} \Im y(z) = \frac{1}{2\pi t} \left( 1 - \frac{\lambda}{6} \left( z^2 + 2 \alpha^2 \right) \right) \sqrt{4 \alpha^2 - z^2}, \qquad z \in \NCC.
\ee
\noindent
Further, the one-cut planar free energy for this quartic matrix model, normalized by the Gaussian as usual, is readily obtained as $\CF_{0} (t) = \frac{1}{24} \left( t - \alpha^2 \right) \left( 9 t - \alpha^2 \right) + \frac{1}{2} t^2 \log \frac{\alpha^2}{t}$.}
\be
\label{eq:quartic-spectral-curve-one-cut}
y(z) = \left( 1 - \frac{\lambda}{6} \left( z^2 + 2 \alpha^2 \right) \right) \sqrt{ z^2-4\alpha^2 }.
\ee
\noindent
The force that eigenvalues feel is dictated by the holomorphic effective potential $V_{\text{h;eff}} (z)$, which itself follows from the spectral curve as $V_{\text{h;eff}}^{\prime} (z) = y (z)$. It also appears at leading order in the large $N$ expansion of the matrix integral \eqref{eq:partitionfunctioneigenvalues}, as
\be
\mathcal{Z} \sim \int \prod_{i=1}^{N} \rmd \lambda_i\, \exp \left( - \frac{1}{g_{\text{s}}} \sum_{i=1}^{N} V_{\text{h;eff}} (\lambda_i) + \cdots \right),
\ee
\noindent
hence eigenvalue integration contours are essentially dictated by the spectral network of $V_{\text{h;eff}} (z)$. For the one-cut cubic matrix model \eqref{eq:CubicMatrixModelPotential},
\bea
\label{eq:cubic-V-holo-eff-one-cut}
V_{\text{h;eff}} (z) &=& \frac{1}{3} \left( z + \frac{1}{4} \left( 2 z + x_1 + x_2 \right) \left( 1 - \lambda z \right) \right) \sqrt{\left(z-x_1\right) \left(z-x_2\right)} - \\
&& - 4 t \log \left( \sqrt{z-x_1} + \sqrt{z-x_2} \right) + 2t \log \left( x_2-x_1 \right), \nonumber
\eea
\noindent
whereas for the one-cut quartic matrix model \eqref{eq:QuarticMatrixModelPotential},
\be
\label{eq:quartic-V-holo-eff-one-cut}
V_{\text{h;eff}} (z) = \frac{1}{2} z \left( 1 - \frac{\lambda}{12} \left( z^2+2\alpha^2 \right) \right) \sqrt{z^2-4\alpha^2} - 4t \log \left( \sqrt{z+2\alpha} + \sqrt{z-2\alpha} \right) + 2t \log 4 \alpha.
\ee

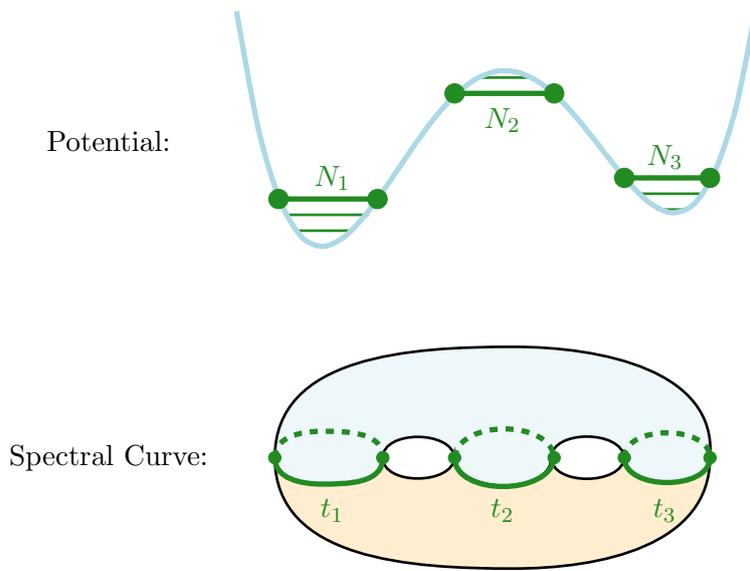
\begin{figure}
\centering
\begin{tikzpicture}
	\begin{scope}[scale=0.7, shift={({0},{2})}]
	\node at (-4,2) {Potential:};
	\draw[color=ForestGreen, line width=2pt] (-0.88,0.9) -- (1.15,0.9);
	\draw[color=ForestGreen, line width=1pt] (-0.7,0.6) -- (0.88,0.6);
	\draw[color=ForestGreen, line width=1pt] (-0.51,0.3) -- (0.6,0.3);
	\draw[color=ForestGreen, line width=2pt] (2.5,2.9) -- (4.32,2.9);
	\draw[color=ForestGreen, line width=1pt] (3,3.2) -- (4,3.2);
	\draw[color=ForestGreen, line width=2pt] (5.69,1.3) -- (7.3,1.3);
	\draw[color=ForestGreen, line width=1pt] (5.95,1) -- (7.2,1);
	\draw[color=ForestGreen, line width=1pt] (6.3,0.7) -- (6.85,0.7);
	 \draw[scale=2, domain=-0.8:4.1, smooth, variable=\x, LightBlue, line width=2pt] plot ({\x}, {2*1.14*\x*\x - 2*0.671*\x*\x*\x + 0.2*\x*\x*\x*\x});
	\draw[ForestGreen, fill=ForestGreen] (-0.81,0.9) circle (1.1ex);
	 \draw[ForestGreen, fill=ForestGreen] (1.05,0.9) circle (1.1ex);
	 \draw[ForestGreen, fill=ForestGreen] (2.5,2.9) circle (1.1ex);
	 \draw[ForestGreen, fill=ForestGreen] (4.37,2.9) circle (1.1ex);
	 \draw[ForestGreen, fill=ForestGreen] (5.69,1.3) circle (1.1ex);
	 \draw[ForestGreen, fill=ForestGreen] (7.3,1.3) circle (1.1ex);
	 \node[ForestGreen] at (0.2,1.3) {$N_1$};
	\node[ForestGreen] at (3.4,2.4) {$N_2$};
	\node[ForestGreen] at (6.45,1.7) {$N_3$};
	\end{scope}
	\begin{scope}[scale=0.7,  shift={({0},{-2})}]
	\node at (-4,0) {Spectral Curve:};
	\draw[fill=LightBlue,fill opacity=0.2, line width=1pt] (1.15,0)   to [out=90,in=95] (2.5,0)
	to [out=270,in=270] (4.32,0)
	to [out=90,in=95] (5.69,0)
	to [out=270,in=270] (7.3,0)
	to [out=85,in=0] (3.6,2.1)
    to [out=180,in=90] (-0.88,0)
    to [out=270, in=180] (0, -0.5)
    to [out=0, in=270] cycle;
    \draw[fill=darktangerine,fill opacity=0.2, line width=1pt] (-0.88,0)
    to [out=270,in=180] (3.6,-2.1)
    to [out=0,in=270] (7.3,0)
    to [out=270,in=270] (5.69,0)
    to [out=265,in=270] (4.32,0)
    to [out=270,in=275] (2.5,0)
    to [out=265,in=270] (1.15,0)
    to [out=270, in=0] (0, -0.5)
    to [out=180, in=270] cycle;
    \draw[color=ForestGreen, line width=2pt] (-0.88,0) to [out=270, in=180] (0, -0.5)
    to [out=0, in=270] (1.15,0);
    \draw[dashed, color=ForestGreen, line width=2pt] (-0.88,0) to [out=90, in=180] (0, 0.5)
    to [out=0, in=90] (1.15,0);
\draw[ForestGreen, fill=ForestGreen] (-0.88,0) circle (.7ex);
\draw[ForestGreen, fill=ForestGreen] (1.15,0) circle (.7ex);
\draw[color=ForestGreen, line width=2pt] (2.5,0) to [out=270, in=265] (4.37,0);
    \draw[dashed, color=ForestGreen, line width=2pt] (2.5,0) to [out=85, in=95] (4.37,0);
\draw[ForestGreen, fill=ForestGreen] (2.5,0) circle (.7ex);
\draw[ForestGreen, fill=ForestGreen] (4.37,0) circle (.7ex);
\draw[color=ForestGreen, line width=2pt] (5.69,0) to [out=270, in=270] (7.3,0);
    \draw[dashed, color=ForestGreen, line width=2pt] (5.69,0) to [out=90, in=90] (7.3,0);
\draw[ForestGreen, fill=ForestGreen] (5.69,0) circle (.7ex);
\draw[ForestGreen, fill=ForestGreen] (7.3,0) circle (.7ex);
\node[ForestGreen] at (0.2,-1) {$t_1$};
\node[ForestGreen] at (3.4,-1) {$t_2$};
\node[ForestGreen] at (6.45,-1) {$t_3$};
\end{scope}
\end{tikzpicture}
	\caption{Visualization of the distribution of the $N$ eigenvalues across the $s$ extremal points of a given potential (plotted in {\color{blue}blue} with $s=3$ saddles). The upper plot displays the potential $V(z)$ with $N_i$ eigenvalues clustering around saddle $i$; whereas the lower plot depicts the corresponding spectral curve, with $s=3$ cuts each of size $t_i=g_{\text{s}} N_i$. The latter picture also illustrates (part of) the dual topological-string background geometry within the formulation of \cite{dv02a}.}
	\label{fig:multicutPotential}
\end{figure}

Let us next consider the case where the potential in the $N$-dimensional eigenvalue integral \eqref{eq:partitionfunctioneigenvalues} has $s$ extrema, so that distinct sets of eigenvalues may  now follow distinct choices of integration contours \cite{a96, bde00, msw08}. Distributing the $N$ eigenvalues across the $s$ extrema by assigning $N_i$ eigenvalues to extrema $i$, such that $N_1 + N_2 + \cdots + N_s = N$, our matrix integral is rewritten as
\bea
\label{eq:partitionfunctioneigenvaluesmulticut}
\mathcal{Z} \left( N_1, \ldots, N_s; g_{\text{s}} \right) &=& \frac{1}{N_1! N_2! \cdots N_s!} \int_{\CC_1}\, \prod_{i_1=1}^{N_1} \frac{\rmd \lambda_{i_1}}{2\pi} \int_{\CC_2}\, \prod_{i_2=N_1+1}^{N_1+N_2} \frac{\rmd \lambda_{i_2}}{2\pi} \cdots \\
&&
\hspace{120pt}
\cdots \int_{\CC_s}\, \prod_{i_s=N-N_s+1}^{N} \frac{\rmd \lambda_{i_s}}{2\pi}\, \Delta^2 (\lambda)\, \rme^{-\frac{1}{g_{\text{s}}} \sum \limits_{i=1}^{N} V(\lambda_i)}, \nonumber
\eea
\noindent
where we have further introduced a combinatorial pre-factor $\frac{N!}{N_1! N_2! \cdots N_s!}$ counting the number of equivalent ways to distribute the $N$ eigenvalues across the $s$ saddles; see, \textit{e.g.}, \cite{msw08}. This is visualized in figure~\ref{fig:multicutPotential}. In this current scenario, and within the 't~Hooft large $N$ expansion, the eigenvalue spectral density $\varrho (\lambda)$ now has support upon a \textit{multi}-cut configuration\footnote{We are implicitly assuming that the number of cuts equals the number of critical points. It is also perfectly fine to consider multi-cut configurations where there are less cuts than critical points (albeit more than one).}
\be
\label{eq:the-multi-cut-set}
\NCC = \bigcup_{i=1}^{s}\, \left( x_{2i-1}, x_{2i} \right)
\ee
\noindent
whose spectral curve generically is a double-sheeted hyperelliptic curve (of genus $s-1$) \cite{a96}
\be
\label{eq:spectral-curve-moment-function}
y (z) = M(z) \sqrt{ \upsigma (z) }, \qquad \upsigma (z) \equiv \prod_{k=1}^{2s} \left( z-x_k \right).
\ee
\noindent
Chosen canonical $A$ and $B$ cycles (see appendix~\ref{app:elliptic-theta-modular} for all hyperelliptic conventions), the special geometry relations hold
\be
\label{eq:specialgeometryrelations}
t_{i} = \frac{1}{4\pi\rmi} \oint_{A_{i}} \text{d}z\, y(z), \qquad \frac{\partial \CF_{0}}{\partial t_{k}} = \oint_{B_{k}} \text{d}z\, y(z),
\ee
\noindent
where $i = 1, \ldots, s$. This computes the partial 't~Hooft couplings $t_i=g_{\text{s}} N_i$ (with $t_1 + t_2 + \cdots + t_s = t$) alongside the planar or classical free energy, $\CF_0 (t)$, sometimes denoted as the pre-potential in topological string language. Higher $g_{\text{s}}$ corrections in \eqref{eq:genusgfreeenergies}, $\CF_g (t)$, follow via, \textit{e.g.}, the topological recursion \cite{e04, eo07a}. This hyperelliptic spectral curve is also visualized in figure~\ref{fig:multicutPotential}.

Herein for \eqref{eq:spectral-curve-moment-function} the moment function $M (z)$ is
\be
\label{eq:moment-function-multi-cut}
M (z) = \oint_{(\infty)} \frac{\rmd w}{2\pi\rmi}\, \frac{V' (w)}{\left(z-w\right) \sqrt{\upsigma (w)}},
\ee
\noindent
but where the standard asymptotics of the planar resolvent now only yield $s+1$ independent equations for the $2s$ end-points $\left\{ x_1, \ldots, x_{2s} \right\}$ of the multi-cut $\NCC$. These are \cite{a96}
\be
\label{eq:SpecGeo2a}
\oint_{\NCC} \frac{\rmd w}{2\pi\rmi}\, \frac{w^n\, V' (w)}{\sqrt{\upsigma (w)}} = 2t\, \delta_{n,s}, \qquad n=0,1,\ldots,s.
\ee
\noindent
For the random-matrix model, the remaining $s-1$ conditions still required to fix all $\left\{ x_1, \ldots, x_{2s} \right\}$ are \textit{equilibrium} conditions, \textit{i.e.}, all cuts need to be energetically equipotential \cite{d92}. In the simpler scenario where all cuts are upon the real axis---with eigenvalue fractions in each cut real, and the respective eigenvalue densities well-defined---, this equipotential requirement that the real part of the holomorphic effective potential is the same along the several cuts translates to the precise number of ``missing'' equations:
\be
V_{\text{h;eff}} (x_{2i}) = V_{\text{h;eff}} (x_{2i+1}), \qquad i=1,\ldots,s-1.
\ee
\noindent
But as soon as eigenvalues venture into the complex plane, as they often do, it is best to rewrite the previous equilibrium-conditions as the equivalent requirement of vanishing $B$-cycles,
\be
\label{eq:equilibrium-condition-via-B-cycles}
\int_{x_{2i}}^{x_{2i+1}} \rmd z\, y(z) = 0, \qquad i=1,\ldots,s-1.
\ee
\noindent
These equations are now immediately generalizable to the \textit{Boutroux conditions} \cite{d92, bm06, bt16}
\be
\label{eq:SpecGeo2b}
\frac{1}{t} \oint_{\gamma} \rmd z\, y (z) \in \rmi\BR,
\ee
\noindent
which define an admissible\footnote{In equilibrium, if more than one such admissible solution exists, then one must still enforce a minimal action configuration, \textit{i.e.}, pick the solution whose action has smallest real part \cite{d91, d92}.} spectral curve; and where $\gamma$ is any closed cycle (encircling any pair of branch points) on the curve. This now works for any\footnote{In particular it works for our initial simpler scenario: when all cuts are upon the real axis, where $A$-periods are purely imaginary and $B$-periods purely real, then \eqref{eq:SpecGeo2b} correspondingly tells us that indeed eigenvalue fractions are real and that \eqref{eq:equilibrium-condition-via-B-cycles} must hold.} distribution of eigenvalues upon the complex plane, and we are done: our hyperelliptic curve of genus $s-1$ has a $(2s-2)$-dimensional basis\footnote{For a trivalent-tree configuration, the usual split into $A$ and $B$ cycles is somewhat misleading, and one should just think of homologically independent contours; see, \textit{e.g.}, \cite{em08}.} of cycles (see appendix~\ref{app:elliptic-theta-modular}), hence the Boutroux condition spits out $2s-2$ \textit{real} conditions which are the $s-1$ \textit{complex} conditions that were still required on top of \eqref{eq:SpecGeo2a}.

In this way, the saddle-point equations-of-motion have been uniquely solved, as \eqref{eq:SpecGeo2a} and \eqref{eq:SpecGeo2b} provide a complete system of equations for all end-points. Do note that by fixing eigenvalue equilibrium conditions we are taking the matrix-model ``grand-canonical'' point-of-view, where the natural physical equilibrium configuration is the dominant eigenvalue configuration within the partition function \cite{bde00}. An alternative approach is to fix the partial 't~Hooft couplings $t_i$ in \eqref{eq:specialgeometryrelations} as \textit{input} (these are indeed $s-1$ independent data), which leads to the standard ``canonical'' point-of-view typical of topological string settings \cite{dv02a}. This will be further discussed in \cite{ss26}.

As alluded in subsection~\ref{subsec:OP-phases}, in this paper and its close companion \cite{krsst26b} the cubic and quartic matrix models \eqref{eq:CubicMatrixModelPotential} and \eqref{eq:QuarticMatrixModelPotential} serve as prototypical examples for illustrating our results---and we now delve into the spectral geometry of these models (albeit most technical details will be deferred to appendix~\ref{appendix:SpectralGeometryMatrixModels}, supported by appendix~\ref{app:elliptic-theta-modular}). In particular, we need to be explicit on their \eqref{eq:SpecGeo2a} and \eqref{eq:SpecGeo2b}; and then solving these equations analytically, whenever possible, and numerically otherwise. This will allow us to probe the phase diagrams of these models, computing their spectral curves and plotting their respective spectral networks in all phases. 

\paragraph{The Cubic Matrix Model:} 

\begin{figure}
    \centering
    \begin{subfigure}[b]{0.48\textwidth}
        \centering
        \includegraphics[width=\linewidth]{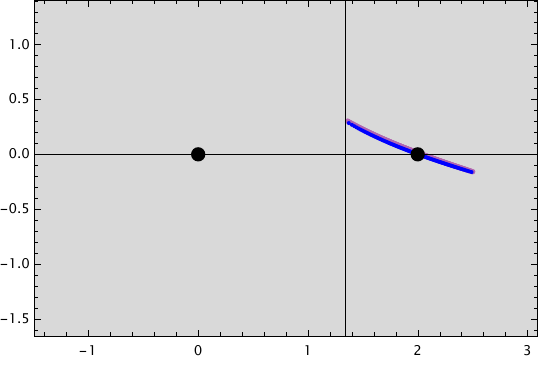}
        \caption{The one-cut region (left-most).}
		\label{fig:SpecGeofig:eigcmmSUBA}
    \end{subfigure}
    \hspace{5pt}
    \begin{subfigure}[b]{0.48\textwidth}
        \centering
        \includegraphics[width=\linewidth]{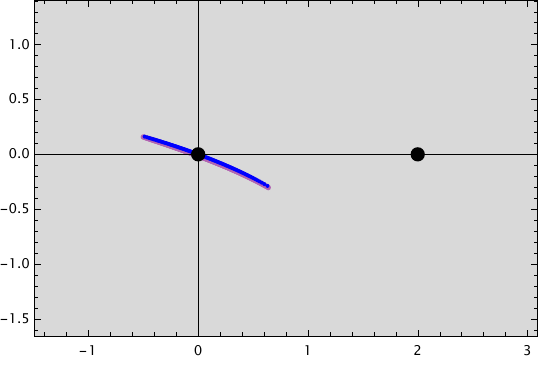}
        \caption{The one-cut region (right-most).}
		\label{fig:SpecGeofig:eigcmmSUBB}
    \end{subfigure}
    
    \vspace{0.7cm}
    
    \begin{subfigure}[b]{0.48\textwidth}
        \centering
        \includegraphics[width=\linewidth]{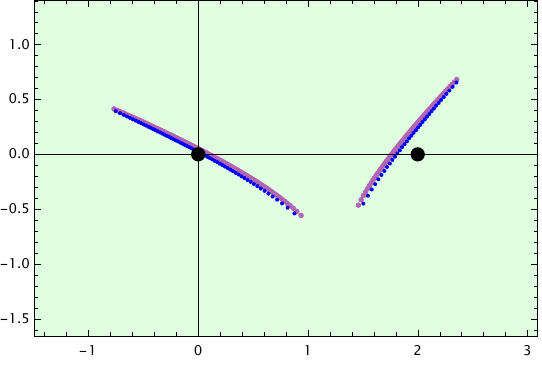}
        \caption{The two-cut region.}
		\label{fig:SpecGeofig:eigcmmSUBC}
    \end{subfigure}
\hspace{5pt}
    \begin{subfigure}[b]{0.48\textwidth}
        \centering
        \includegraphics[width=\linewidth]{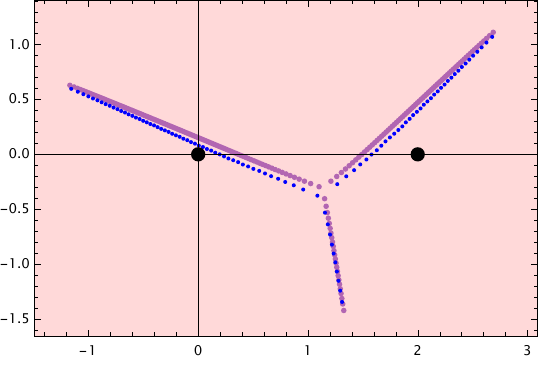}
        \caption{The trivalent-tree region.}
		\label{fig:SpecGeofig:eigcmmSUBD}
	\end{subfigure}
	\caption{Figures displaying the rather precise match between $N = 200$ eigenvalue location plots (obtained from \eqref{eq:ZEoM}), and the roots of the orthogonal polynomial $p_{200}(z)$ as in \eqref{eq:OPasDETcorrelator} (and constructed via the recursion relations \eqref{eq:OPrecursionrelationforpns}). These matches are performed at sample values of the 't~Hooft parameter $t$ in different phases of the cubic matrix-model phase-diagram (see subsection~\ref{subsec:stokes-vs-phases}, in particular for the ``left-most'' and ``right-most'' one-cut regions). The nomenclature for the phases is immediate from the plots. Eigenvalue-locations are depicted by the \textcolor{blue}{blue} dots, whereas orthogonal-polynomial roots are depicted as the \textcolor{purple}{purple} dots. The saddles of the ``classical'' matrix model potential \eqref{eq:CubicMatrixModelPotential} are illustrated via larger black dots.}
	\label{fig:SpecGeofig:eigcmm}
\end{figure}

Consider the cubic matrix model \eqref{eq:CubicMatrixModelPotential}, for which the orthogonal-polynomial results in subsection~\ref{subsec:OP-phases} indicated that there are three distinct phases; dubbed the gray, green, pink regions in figure~\ref{fig:CMMrNDataAndOPRoots}. These were found by looking at the behavior of the recursion coefficients $r_n$ with varying 't~Hooft coupling, alongside the distributions of the roots of their corresponding orthogonal polynomials. It is straightforward to compare those results against the present spectral-curve formulation, where eigenvalue densities generically have support upon multi-cut configurations. Starting at finite $N$, the computation of saddle-points of the $N$-dimensional matrix integral \eqref{eq:partitionfunctioneigenvalues} reduces to solving the ``equations of motion'' for a Coulomb gas of $N$ eigenvalues (see, \textit{e.g.}, \cite{m04}),
\be
\label{eq:ZEoM}
\frac{1}{2t}\, V'(\lambda_i) = \frac{1}{N} \underbrace{\sum_{k=1}^N}_{k \neq i} \frac{1}{\lambda_i-\lambda_k}, \qquad i=1,\ldots,N.
\ee
\noindent
An alternative (and computationally efficient) approach to finding the locations $\left\{ \lambda_i \right\}$ of the random-matrix eigenvalues (\textit{i.e.}, of solving the equations of motion \eqref{eq:ZEoM}) is via the roots of the orthogonal polynomials $p_{n} (z)$ in \eqref{eq:OPhnrelation}. In fact, orthogonal polynomials may be computed directly in the matrix model via $n$-dimensional determinant-correlators (see, \textit{e.g.}, \cite{ekr15})
\be
\label{eq:OPasDETcorrelator}
p_n (z) = \ev{ \det \left( z-M \right) }_{n \times n} \approx \prod_{i=1}^{n} \left( z-\lambda_i \right) + \cdots.
\ee
\noindent
As such, at large $n$ and $N$, solutions to the eigenvalue equations of motion \eqref{eq:ZEoM} should precisely match the roots of the orthogonal polynomials \eqref{eq:OPasDETcorrelator} (more and less rigorous proofs may be found in, \textit{e.g.}, \cite{bm06, b07, b.s07, bt11, aam13b, hkl13, bt16}). As a solid check, in figure~\ref{fig:SpecGeofig:eigcmm} we display a match between $N=200$ eigenvalue locations from \eqref{eq:ZEoM} and the corresponding orthogonal-polynomial roots as in \eqref{eq:OPasDETcorrelator}, for all phases. The plots immediately justify the nomenclature for the three phases: the one-cut, the two-cut, and the trivalent-tree (further indicating the link to spectral geometry).

These eigenvalue or root distributions are directly obtainable from spectral geometry. Since each phase is inherently defined by the number of cuts or, more precisely, the number of branch-points in the spectral curve, one simply has to determine the corresponding cut end-points and moment function; fixed by equations \eqref{eq:SpecGeo2a} and \eqref{eq:SpecGeo2b}. These amount to several integrals---generically intricate and not very illuminating for all except the one-cut case. Accordingly, below we only discuss the one-cut computations and refer the interested reader to appendix~\ref{appendix:SpectralGeometryMatrixModels} for details on explicit computations associated to the other phases.  

\begin{figure}
    \centering
    \includegraphics[width=0.7\linewidth]{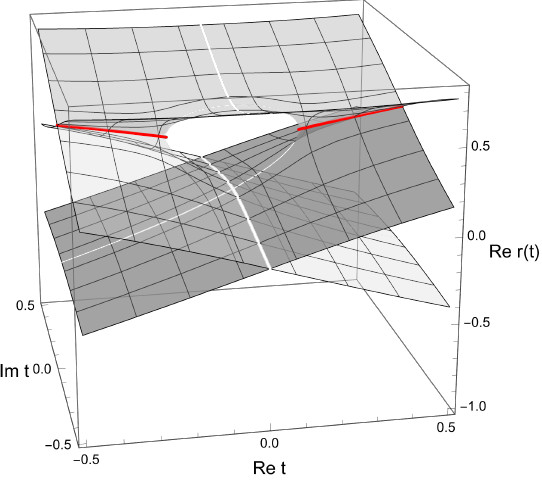}
    \caption{Solution of the planar string equation \eqref{eq:cubic-MM-classical-string-eq}, explicitly given by \eqref{eq:CMMPlanarSolution1}-\eqref{eq:CMMPlanarSolution2}-\eqref{eq:CMMPlanarSolution3}. The three solution-branches analytically continue into each other through the \textcolor{red}{red} branch cuts. Our choice of principal branch \eqref{eq:CMMPlanarSolution3} is displayed in the darkest shade of \textcolor{gray}{gray}. It analytically continues into \eqref{eq:CMMPlanarSolution1}, illustrated in a slightly lighter shade of \textcolor{gray}{gray}. The branch \eqref{eq:CMMPlanarSolution1} then has a second branch cut that analytically continues into the branch \eqref{eq:CMMPlanarSolution2} shown in translucent \textcolor{gray}{gray}.}
    \label{fig:CMMPlanarBranching}
\end{figure}

\begin{figure}
	\centering
	\begin{subfigure}[b]{0.48\textwidth}
        \centering
        \includegraphics[width=\linewidth,height=5cm]{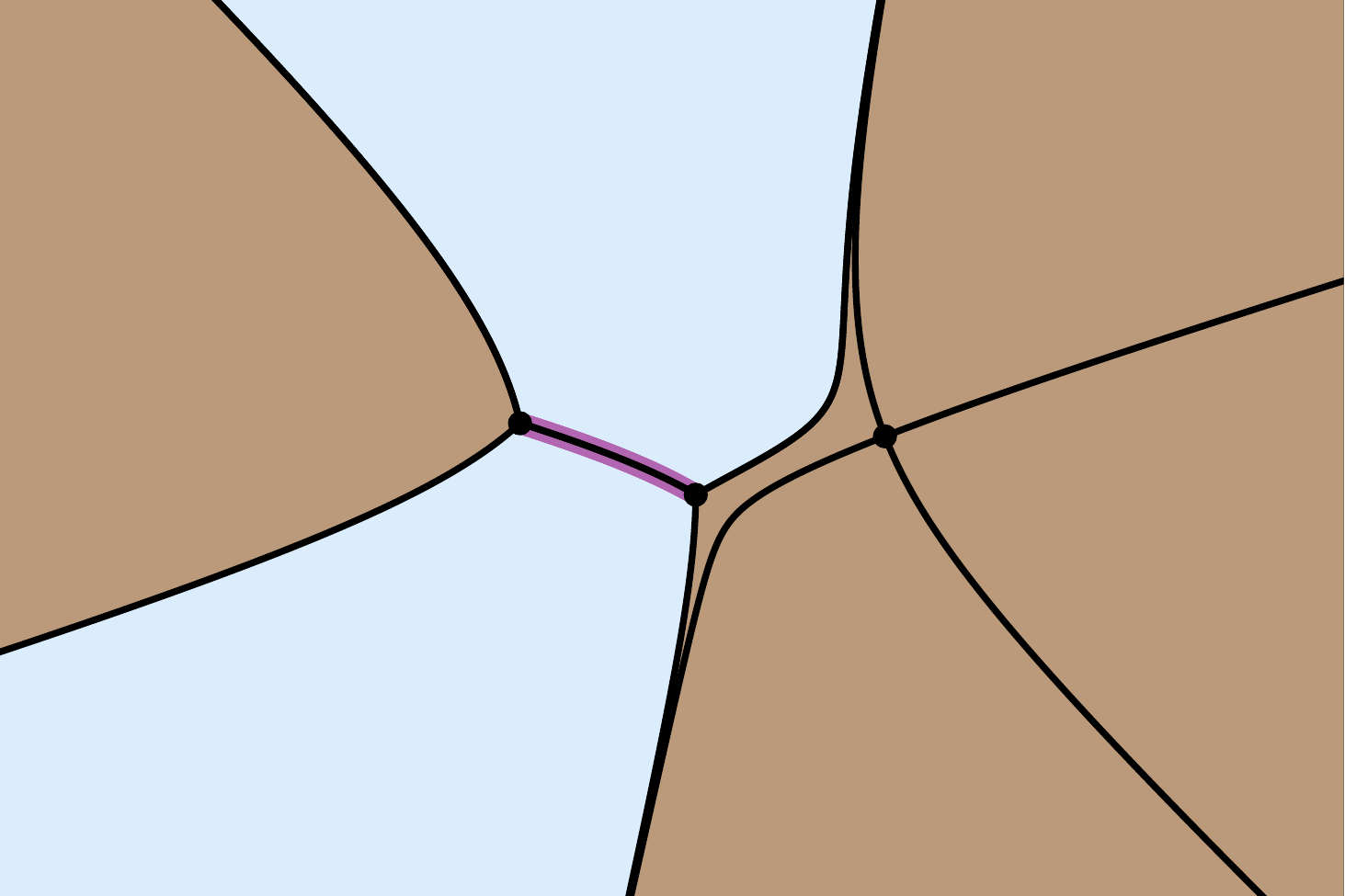}
        \caption{The one-cut region (left-most).}
        \label{fig:SpecGeofig:one-cut-right-spectral-networks}
    \end{subfigure}
    \hspace{5pt}
    \begin{subfigure}[b]{0.48\textwidth}
        \centering
        \includegraphics[width=\linewidth,height=5cm]{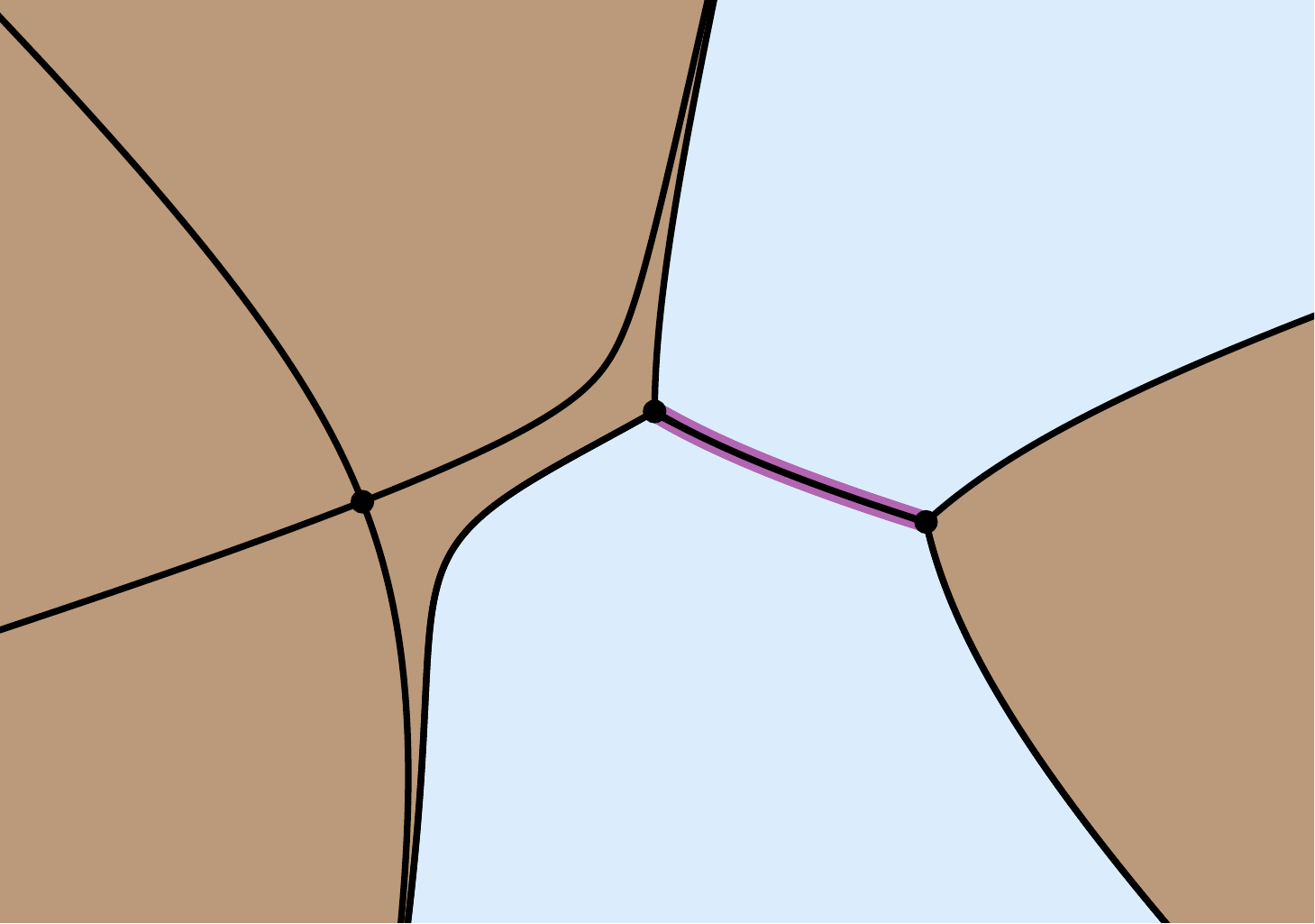}
        \caption{The one-cut region (right-most).}
        \label{fig:SpecGeofig:one-cut-left-spectral-networks}
    \end{subfigure}

    \vspace{0.7cm}
    
    \begin{subfigure}[b]{0.48\textwidth}
        \centering
        \includegraphics[width=\linewidth]{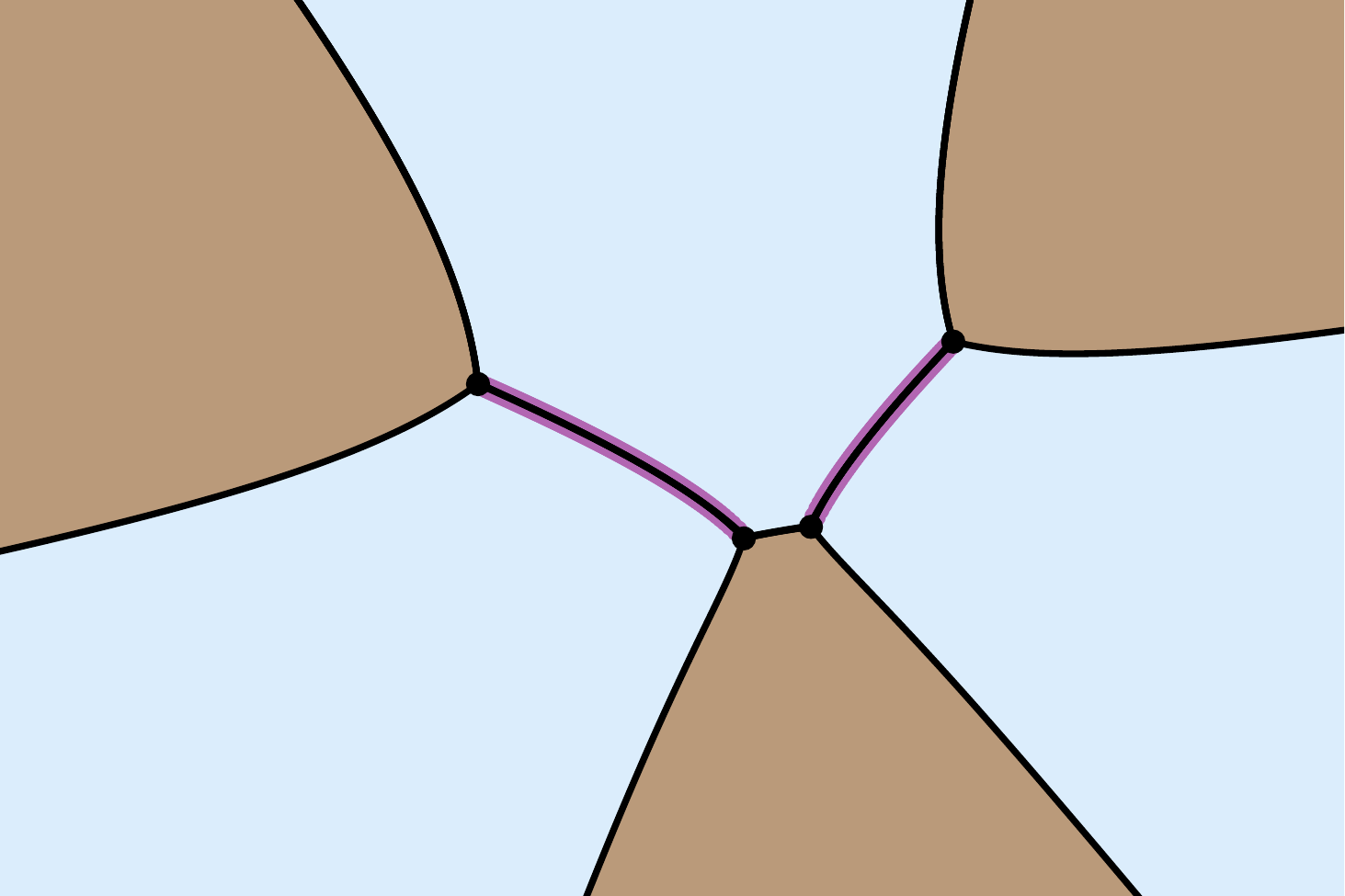}
        \caption{The two-cut region.}
        \label{fig:SpecGeofig:two-cut-spectral-networks}
    \end{subfigure}
	\hspace{5pt}
    \begin{subfigure}[b]{0.48\textwidth}
        \centering
        \includegraphics[width=\linewidth]{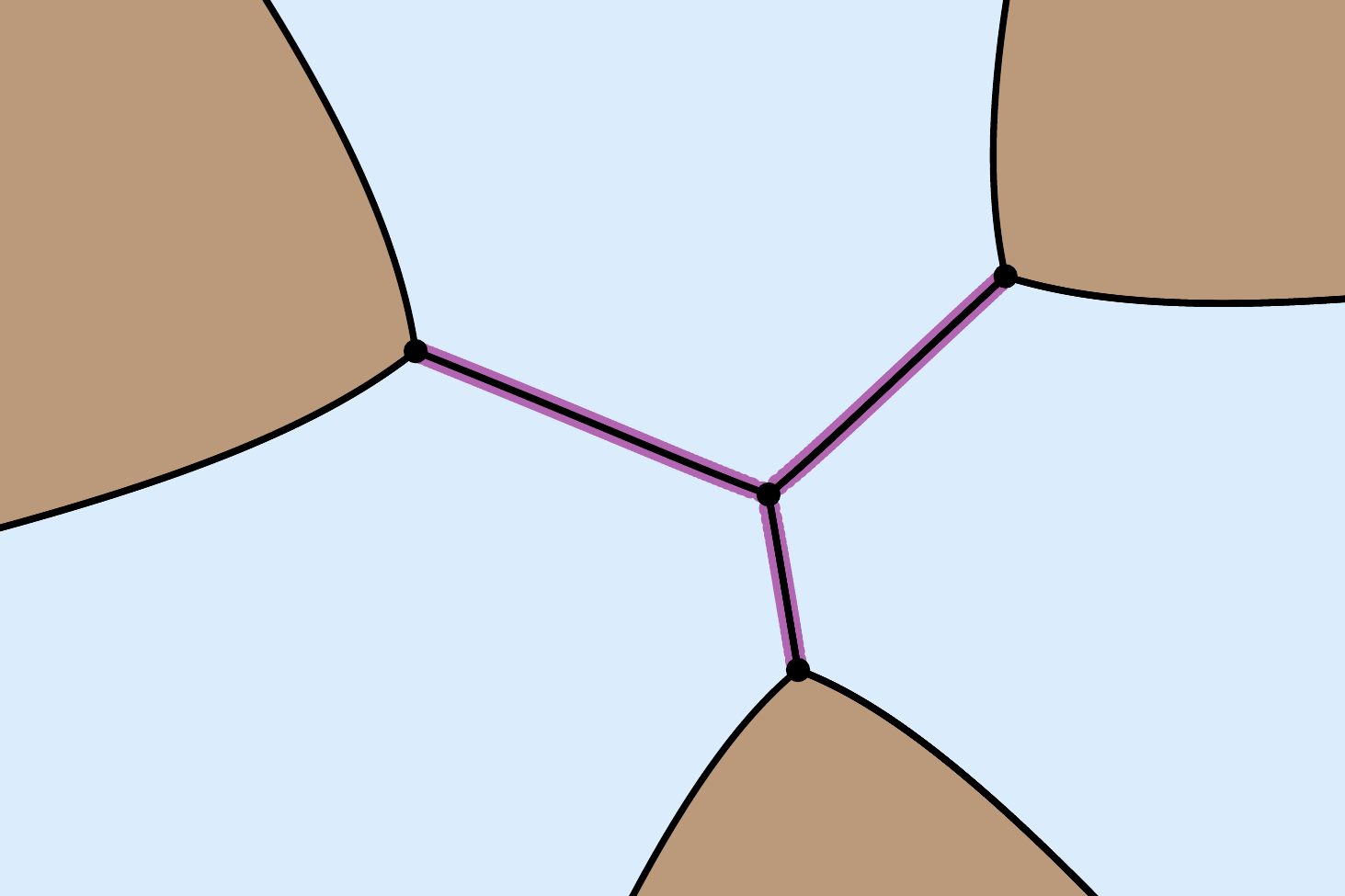}
        \caption{The trivalent-tree region.}
        \label{fig:SpecGeofig:trivalent-spectral-networks}
	\end{subfigure}
	\caption{The real part of the holomorphic effective potential, $\Re V_{\text{h;eff}} (z)$, in the complex $z$-plane. The \textcolor{brown}{brown}/\textcolor{blue}{blue} (or ``land/sea'') regions are where $\Re V_{\text{h;eff}} (z)$ is positive/negative; the solid-black lines depict the spectral network $\Re V_{\text{h;eff}} (z) = 0$ (including ``shorelines'' and ``bridges'', the latter corresponding to the spectral-curve cuts). The black dots on ``land'' are the pinched cycles or nonperturbative saddles \cite{msw07, msw08}. Roots of $p_{200} (z)$ are plotted as \textcolor{purple}{purple} dots on top of the spectral network showing a very sharp match with the ``bridges''. The figures are at $\lambda=1$ and sample values of the 't~Hooft parameter $t$ illustrating the different phases of the cubic matrix-model phase-diagram (see subsection~\ref{subsec:stokes-vs-phases}).}
	    \label{fig:cubics-spectral-networks}
\end{figure}

In the one-cut case $s =1$ and \eqref{eq:one-cut-end-points} is enough to fully specify the spectral geometry \eqref{eq:one-cut-spectral-curve=background-around-which-string-equations-are-constructed}. The end-points of the single cut $\NCC = (x_1,x_2)$ for the cubic spectral curve \eqref{eq:cubic-spectral-curve-one-cut} explicitly satisfy
\begin{eqnarray}
\lambda \left( 3 x_1^2 + 2 x_1 x_2 + 3 x_2^2 \right) - 8 \left( x_1 + x_2 \right) &=& 0, \\ 
\lambda \left( 5 x_1^3 + 3 x_1 x_2 \left( x_1 + x_2 \right) + 5 x_2^3 \right) - 4 \left( 3 x_1^2 + 2 x_1 x_2 + 3 x_2^2 \right) &=& - 64 t.
\end{eqnarray}
\noindent
This cubic system has three solutions, which may be written as
\be
\label{eq:SpecGeo12}
\begin{lcases}
x_1 (t) &= \frac{1}{\lambda} \left( 1 - 2 \lambda \sqrt{r_i(t)} - \sqrt{1-2\lambda^2 r_i(t)}\right) \\ 
x_2(t) &= \frac{1}{\lambda} \left( 1 + 2 \lambda \sqrt{r_i(t)} - \sqrt{1-2\lambda^2 r_i(t)} \right)
\end{lcases},
\ee
\noindent
where, for $i=1,2,3$, the functions $\left\{ r_i(t) \right\}$ are the three solutions of the algebraic equation
\be
\label{eq:cubic-MM-classical-string-eq}
r_i^2 \left( 1 - 2 \lambda^2\, r_i \right) = t^2.
\ee
\noindent
This is the planar, classical solution of the cubic-model string-equation \eqref{eq:cubicstringequationthooftlimit}. It should now not be too hard to guess that the same issues present in the double-scaling limit, and discussed around \eqref{eq:DSL-P1-classical-string-eq}, are back to haunt us off-criticality as well. In other words, also for the cubic matrix model there will be regions of the complex plane that end up producing a singular transseries in \eqref{eq:twoparameterresurgenttransseriesforR}. The solution to this conundrum is now known: this occurs because the transseries solution described by \eqref{eq:twoparameterresurgenttransseriesforR} undergoes changes of branches within the Riemann surface which describes the classical solution \eqref{eq:cubic-MM-classical-string-eq}, and these need to be properly taken into account via adequate analytic continuation. This is illustrated by the three-sheeted plot in figure~\ref{fig:CMMPlanarBranching}, displaying the following explicit solution to the classical string equation \eqref{eq:cubic-MM-classical-string-eq}. Indeed, in order to be completely explicit, the functions $\left\{ r_i(t) \right\}$ for $i=1,2,3$ are:
\bea
\label{eq:CMMPlanarSolution1}
r_1 (t) &=& \frac{1}{6\lambda^2} \left( 1 - \rho - \frac{1}{\rho} \right), \\
\label{eq:CMMPlanarSolution2}
r_2 (t) &=& \frac{1}{12\lambda^2} \left( 2 + \left( 1-\rmi \sqrt{3}\right) \rho + \frac{1+\rmi \sqrt{3}}{\rho} \right), \\
\label{eq:CMMPlanarSolution3}
r_3 (t) &=& \frac{1}{12\lambda^2} \left(2 + \left( 1+\rmi \sqrt{3}\right) \rho + \frac{1-\rmi \sqrt{3}}{\rho} \right),
\eea
\noindent
where
\be
\rho = \sqrt[3]{6 \sqrt{3 \lambda^4 t^2 \left( 27 \lambda^4 t^2 - 1 \right)} + 54 \lambda^4 t^2 - 1}.
\ee
\noindent
As we shall see in figure~\ref{fig:cubics-spectral-networks}, the solutions $r_2(t)$ and $r_3(t)$ describe the end-points of two distinct one-cut configurations. Specifically, $r_2(t)$ corresponds to the end-points of the one-cut configuration shown in plot~\ref{fig:SpecGeofig:eigcmmSUBB} of figure~\ref{fig:SpecGeofig:eigcmm}, whilst $r_3(t)$ corresponds to the end-points of the one-cut configuration shown in plot~\ref{fig:SpecGeofig:eigcmmSUBA} of the same figure. The solution $r_1(t)$ is a bit more subtle, as it describes the end-points of an energetically unstable configuration, engineered by fixing the contour weights of the cubic matrix model (see equation \eqref{eq:mkw1w2cubic}) to be $w_1 = 1 = w_2$. This configuration should be understood as the cubic matrix model analogue of the energetically unstable configurations identified in \cite{bt11} for the quartic matrix model.

Finally, the moment function follows from \eqref{eq:moment-function-one-cut} as in \eqref{eq:cubic-spectral-curve-one-cut},
\be
\label{eq:SpecGeo8}
M (z) = 1 - \frac{\lambda}{4} \left( 2 z + x_1 + x_2 \right) .
\ee
\noindent
The root of this moment function, signaling the location of the nonperturbative saddle \cite{msw07}, is simply
\be
\label{eq:CMMxstarsaddle}
x^\star = \frac{2}{\lambda} - \frac{1}{2} \left( x_1 + x_2 \right).
\ee

All ingredients are now on the table. Having complete information for the cubic spectral curve \eqref{eq:cubic-spectral-curve-one-cut} immediately and explicitly spits out its corresponding holomorphic effective potential \eqref{eq:cubic-V-holo-eff-one-cut}, and we may proceed to plot its spectral network (sometimes denoted by zero-level set) $\Re V_{\text{h;eff}} (z) = 0$ (generalizing to the matrix model the corresponding cubic-potential plot in figure~\ref{fig:CubicMatrixModelSteepestDescentContours}). This generically trivalent network is plotted in black, whereas regions where $\Re V_{\text{h;eff}} (z)$ is positive are colored in brown (the ``land'') and regions where $\Re V_{\text{h;eff}} (z)$ is negative are colored in blue (the ``sea''). In this\footnote{This very appealing nomenclature was introduced in \cite{bm06, b07, bt11} and we follow it throughout. Using such terminology, admissible eigenvalue integration-contours move from land-at-infinity to land-at-infinity ``without getting wet''---which in fact already happened with the steepest-descent contours back in figure~\ref{fig:Cubic-AND-QuarticMatrixModelSteepestDescentContours}.} way, the ``bridges'' across the ``sea'', in-between different ``continents'', are precisely the cuts of the spectral curve. In particular, that the spectral network is generically composed of \textit{trivalent} nodes, located at the several branch points, is a direct consequence\footnote{That the most general distributions of eigenvalues upon the complex plane correspond to trivalent tree-like structures was first pointed out in \cite{d92}. This trivalent property of the full spectral network was later placed on firm mathematical ground with the work in \cite{bm06, b07, bt11, bt16}. After all these works, the trivalent property was mathematically readdressed in \cite{bs16} for the cubic and in \cite{bgm21} for the quartic matrix models. The example of the cubic matrix model to illustrate the generic trivalent phase was also further readdressed in \cite{gjk21}.} of the fact that all our spectral curves are \textit{hyperelliptic} \eqref{eq:spectral-curve-moment-function} (see as well appendix~\ref{app:elliptic-theta-modular}). Indeed, this implies that near each (non-degenerate) branch-point $x_k$ in \eqref{eq:the-multi-cut-set}, the curve \eqref{eq:spectral-curve-moment-function} behaves as
\be
y (z) \sim \CO (z-x_k)^{1/2} \quad \text{ as } \quad z \to x_k,
\ee
\noindent
in which case the (real part of the) holomorphic effective potential must behave as
\be
\Re V_{\text{h;eff}} (z) \sim \CO (z-x_k)^{3/2} \quad \text{ as } \quad z \to x_k.
\ee
\noindent
Hence \textit{three} zero-level curves emanate from each branch-point giving rise to the \textit{trivalent} structure of the spectral network. This is illustrated in several plots. In figures~\ref{fig:SpecGeofig:one-cut-right-spectral-networks} and~\ref{fig:SpecGeofig:one-cut-left-spectral-networks}, we show the spectral network of the cubic matrix model for two sample values of the 't~Hooft parameter $t$ in one-cut regions (see the next subsection as well). We also show a very precise match between the roots of the corresponding orthogonal polynomial $p_{200}(z)$ and the ``bridges'' of the spectral network---where they end up accumulating at large degree (see proofs in \cite{b07}). In the one-cut case this corresponds to one each for $r_2$ and $r_3$. Supported by the results in appendix~\ref{appendix:SpectralGeometryMatrixModels} we may proceed to address the multi-cut holomorphic effective potential obtained from the multi-cut spectral curve \eqref{eq:spectral-curve-moment-function}. Checks for the two-cut and trivalent phases are illustrated in figures~\ref{fig:SpecGeofig:two-cut-spectral-networks} and~\ref{fig:SpecGeofig:trivalent-spectral-networks}, again with excellent agreement. This fully validates our spectral networks, the match of orthogonal-polynomial root-distributions and eigenvalue-distributions, as well our generic ability to predict and describe all of them.

\paragraph{The Quartic Matrix Model:} 

\begin{figure}
	\centering
    \begin{subfigure}[b]{0.48\textwidth}
        \centering
        \includegraphics[width=\linewidth]{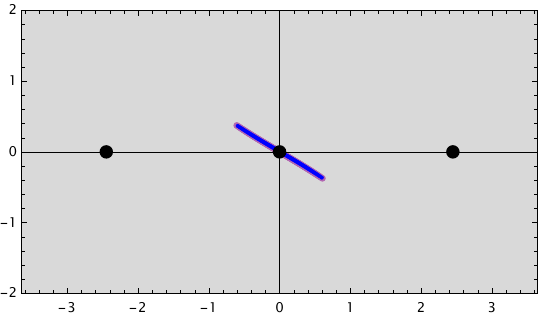}
        \caption{The one-cut region.}
    \end{subfigure}
    \hspace{5pt}
    \begin{subfigure}[b]{0.48\textwidth}
        \centering
        \includegraphics[width=\linewidth]{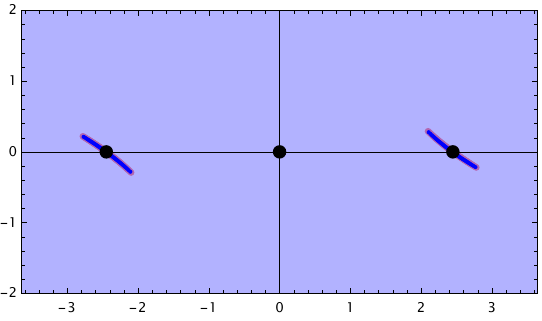}
        \caption{The (symmetric) two-cut region.}
    \end{subfigure}

    \vspace{0.7cm}
    
    \begin{subfigure}[b]{0.48\textwidth}
        \centering
        \includegraphics[width=\linewidth]{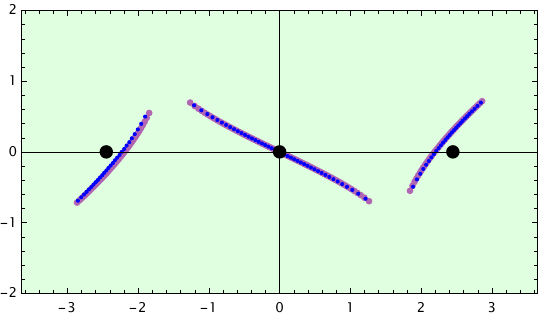}
        \caption{The (symmetric) three-cut region.}
    \end{subfigure}
	\hspace{5pt}
    \begin{subfigure}[b]{0.48\textwidth}
        \centering
        \includegraphics[width=\linewidth]{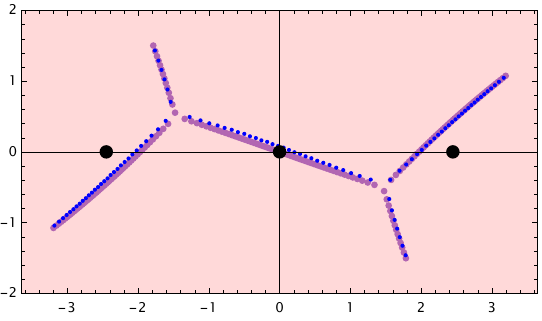}
        \caption{The trivalent-tree region.}
    \end{subfigure}
    \caption{Figures displaying the rather precise match between $N = 200$ eigenvalue location plots (obtained from \eqref{eq:ZEoM}), and the roots of the orthogonal polynomial $p_{200}(z)$ as in \eqref{eq:OPasDETcorrelator} (and constructed via the recursion relations \eqref{eq:OPrecursionrelationforpns}). These matches are performed at sample values of the 't~Hooft parameter $t$ in different phases of the quartic matrix-model phase-diagram (see subsection~\ref{subsec:stokes-vs-phases}). The nomenclature for the phases is immediate from the plots. Eigenvalue-locations are depicted by the \textcolor{blue}{blue} dots, orthogonal-polynomial roots as the \textcolor{purple}{purple} dots. The saddles of the ``classical'' matrix model potential \eqref{eq:QuarticMatrixModelPotential} are illustrated via larger black dots.}
    \label{fig:SpecGeofig:eigqmm}
\end{figure}

Next consider the quartic matrix model \eqref{eq:QuarticMatrixModelPotential}, for which the orthogonal-polynomial results in subsection~\ref{subsec:OP-phases} indicated that there are at least three distinct phases; dubbed the gray, green, pink regions in figure~\ref{fig:QMMrNDataAndOPRoots} (as we shall soon see, there are actually \textit{four} phases). These were found by looking at the behavior of the recursion coefficients $r_n$ with varying 't~Hooft coupling, alongside the distributions of the roots of their corresponding orthogonal polynomials. It is again straightforward to compare those results with the present spectral-curve formulation, with eigenvalues generically supported upon multi-cut configurations. As a solid check, in figure~\ref{fig:SpecGeofig:eigqmm} we display a match between $N=200$ eigenvalue locations from \eqref{eq:ZEoM} and the corresponding orthogonal-polynomial roots as in \eqref{eq:OPasDETcorrelator}, for all phases. The plots immediately justify the nomenclature for the four phases: the one-cut, the symmetric two-cut, the symmetric three-cut, and the trivalent-tree (further indicating the link to spectral geometry).

\begin{figure}
    \centering
    \includegraphics[width=0.7\linewidth]{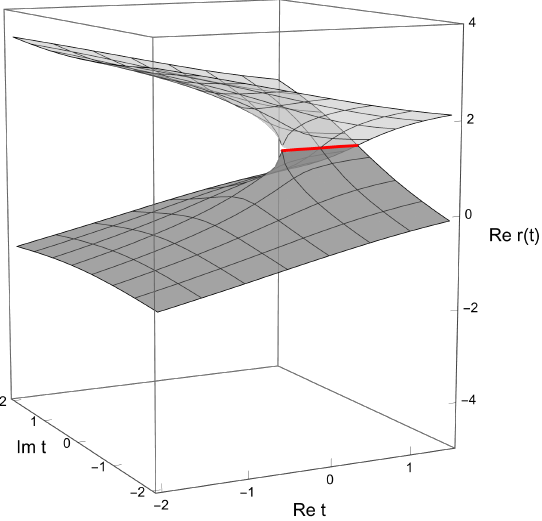}
    \caption{Solution of the planar string equation \eqref{eq:quartic-MM-classical-string-eq}, explicitly given by \eqref{eq:QMMPlanarSolution1}-\eqref{eq:QMMPlanarSolution2}. The two solution-branches analytically continue into each other through the \textcolor{red}{red} branch cut. Our choice of principal branch \eqref{eq:QMMPlanarSolution1} is displayed in the darkest shade of \textcolor{gray}{gray}. It analytically continues into \eqref{eq:QMMPlanarSolution2}, illustrated in the slightly lighter shade of \textcolor{gray}{gray}.}
    \label{fig:QMMPlanarBranching}
\end{figure}

\begin{figure}
    \centering
    \begin{subfigure}[b]{0.48\textwidth}
        \centering
        \includegraphics[width=\linewidth]{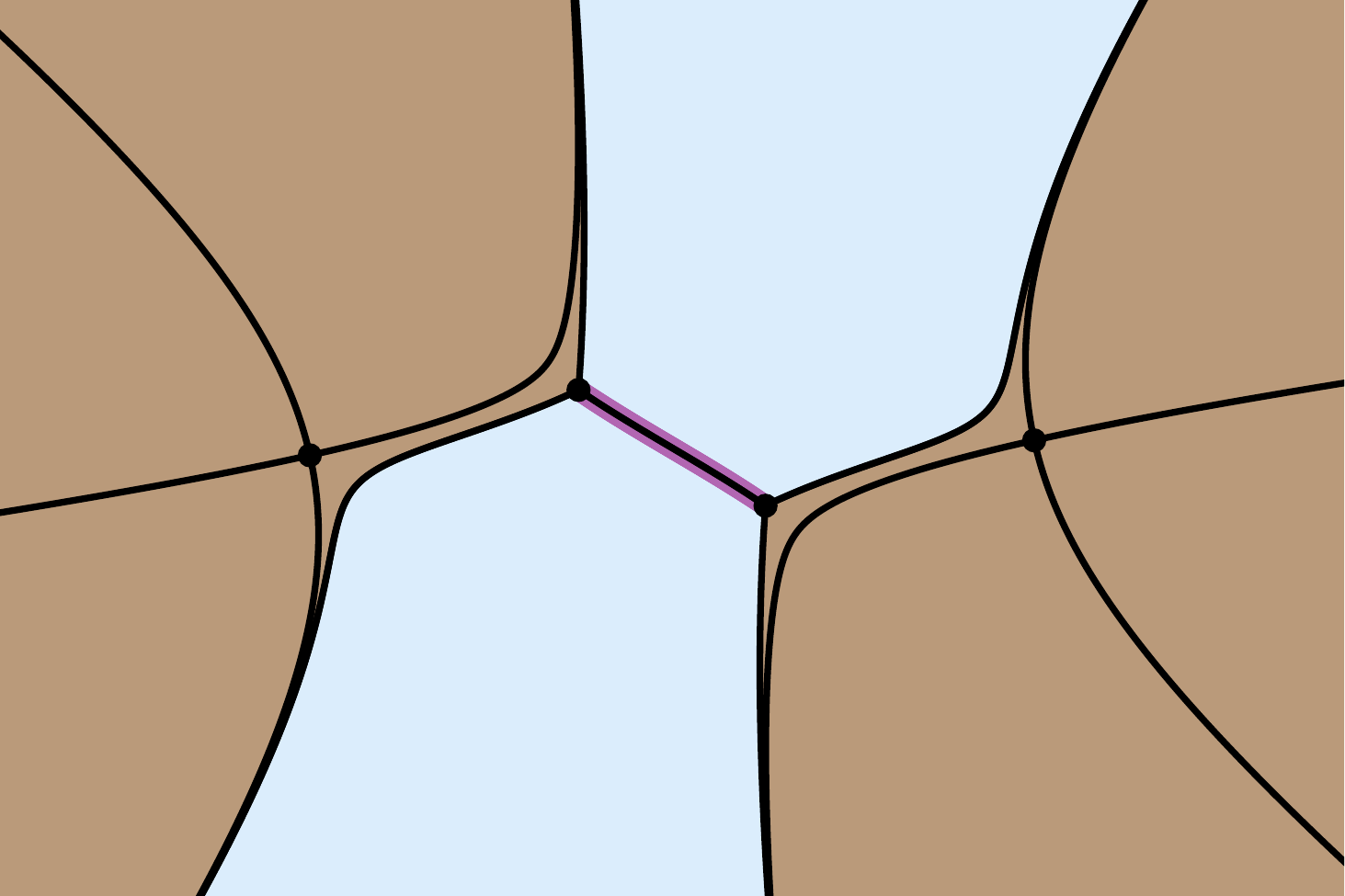}
        \caption{The one-cut region.}
        \label{fig:SpecGeofig:one-cut-right-spectral-networks-QMM}
    \end{subfigure}
    \hspace{5pt}
    \begin{subfigure}[b]{0.48\textwidth}
        \centering
        \includegraphics[width=\linewidth]{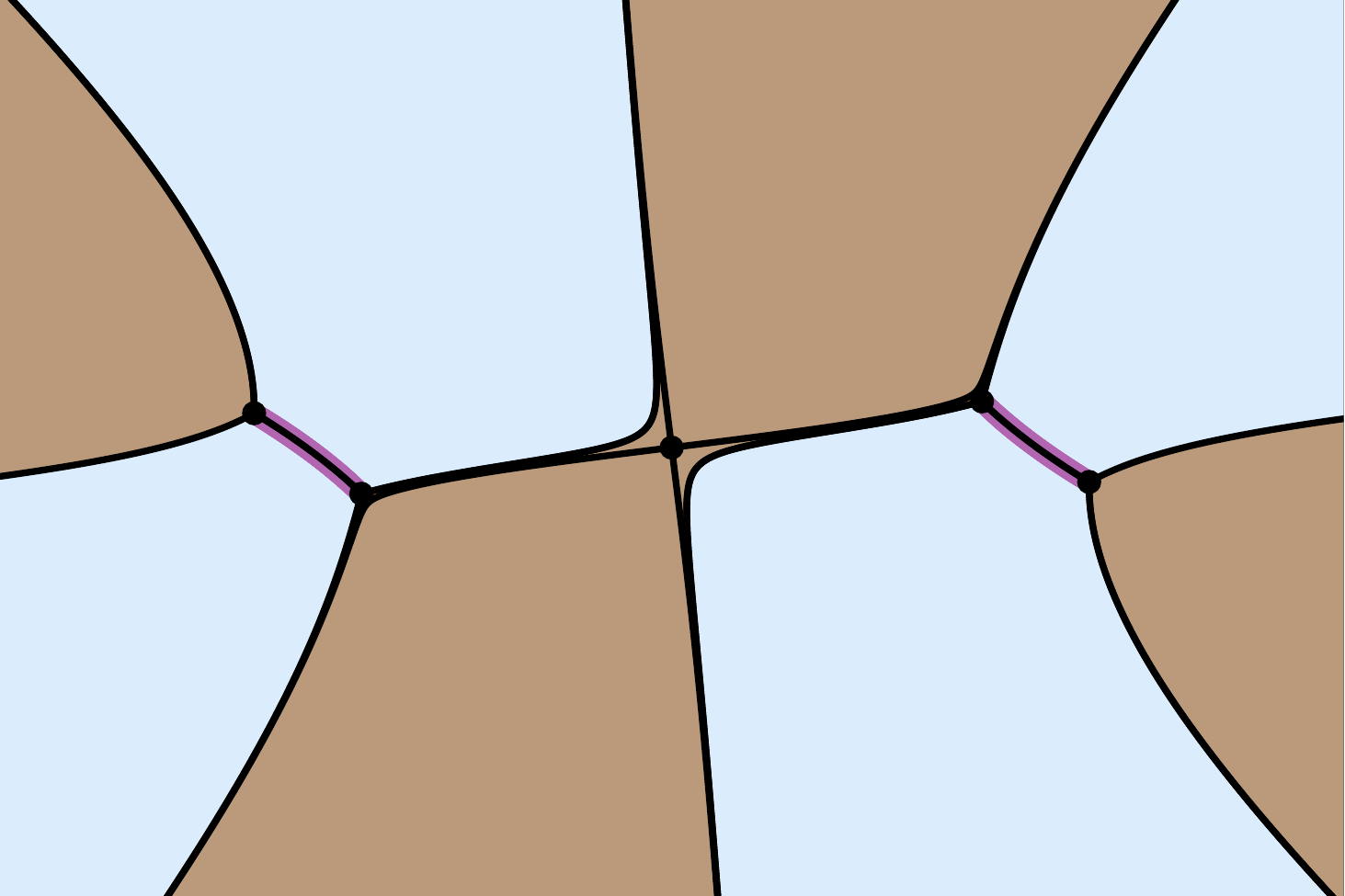}
        \caption{The (symmetric) two-cut region.}
        \label{fig:SpecGeofig:Two-cut-spectral-networks-QMM}
    \end{subfigure}

    \vspace{0.7cm}

    \begin{subfigure}[b]{0.48\textwidth}
        \centering
        \includegraphics[width=\linewidth]{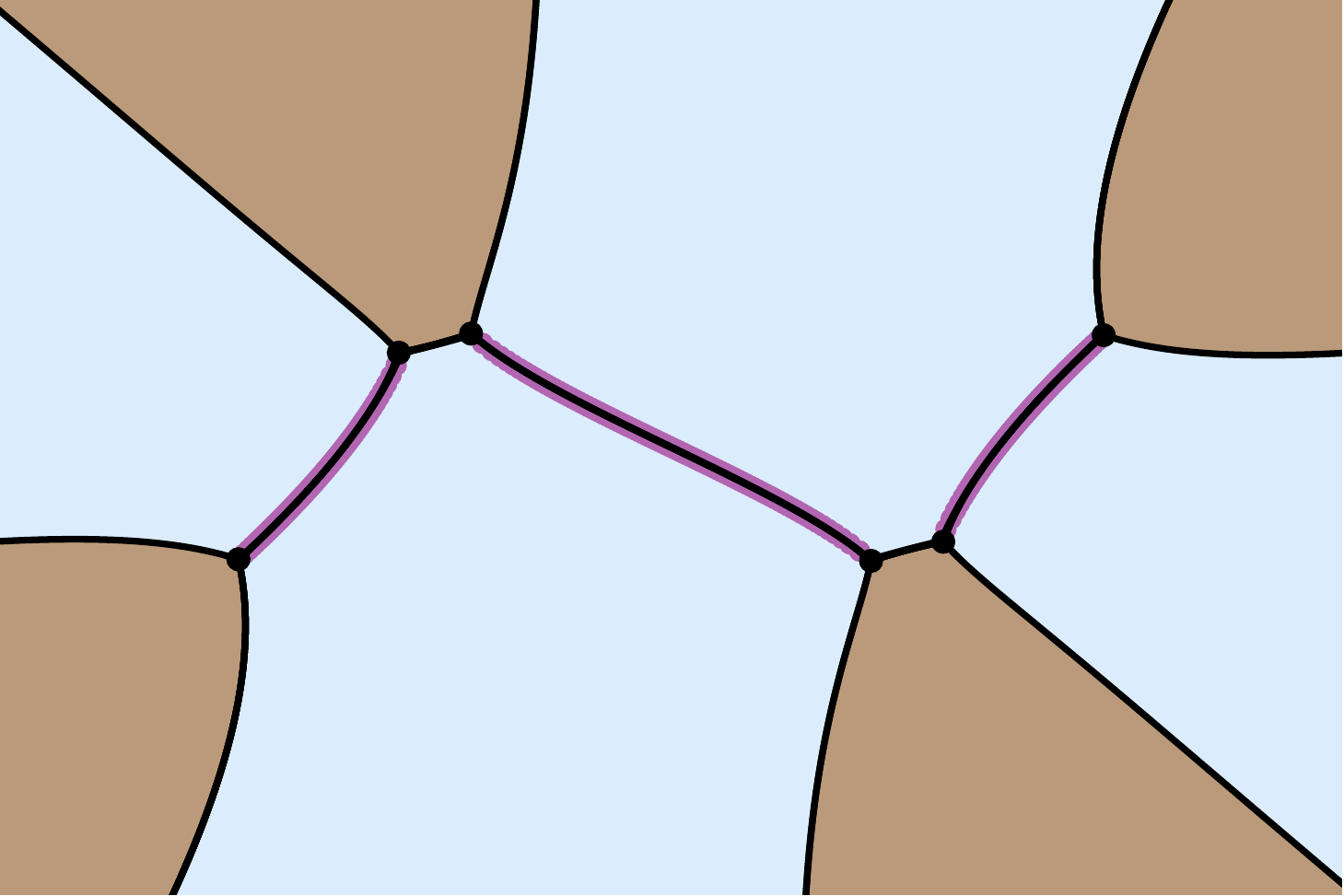}
        \caption{The (symmetric) three-cut region.}
         \label{fig:SpecGeofig:three-cut-right-spectral-networks-QMM}
    \end{subfigure}
    \hspace{5pt}    
    \begin{subfigure}[b]{0.48\textwidth}
        \centering
        \includegraphics[width=\linewidth]{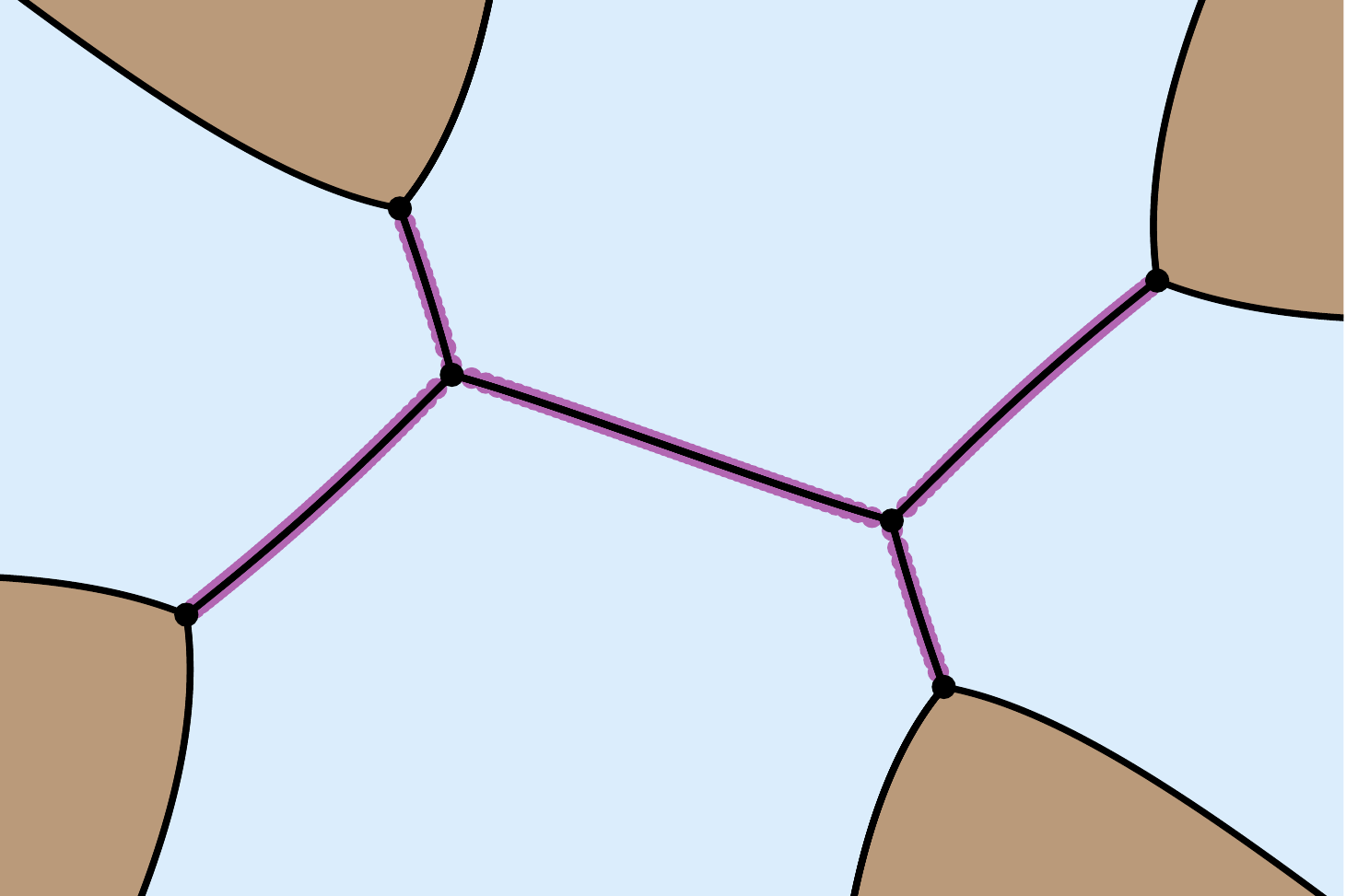}
        \caption{The trivalent-tree region.}
        \label{fig:SpecGeofig:trivalent-spectral-networks-QMM}
    \end{subfigure}
    \caption{The real part of the holomorphic effective potential, $\Re V_{\text{h;eff}} (z)$, in the complex $z$-plane. The \textcolor{brown}{brown}/\textcolor{blue}{blue} (or ``land/sea'') regions are where $\Re V_{\text{h;eff}} (z)$ is positive/negative; the solid-black lines depict the spectral network $\Re V_{\text{h;eff}} (z) = 0$ (including ``shorelines'' and ``bridges'', the latter corresponding to the spectral-curve cuts). The black dots on ``land'' are the pinched cycles or nonperturbative saddles \cite{msw07, sv13}. Roots of $p_{200} (z)$ are plotted as \textcolor{purple}{purple} dots on top of the spectral network showing a very sharp match with the ``bridges''. The figures are at $\lambda=1$ and sample values of the 't~Hooft parameter $t$ illustrating the different phases of the quartic matrix-model phase-diagram (see subsection~\ref{subsec:stokes-vs-phases}).}
	    \label{fig:quartic-spectral-networks}
\end{figure}

As for the cubic model, these eigenvalue or root distributions are directly obtainable from spectral geometry. This requires going through the computation of the cut end-points and moment function, fixed by equations \eqref{eq:SpecGeo2a} and \eqref{eq:SpecGeo2b}, for the quartic model. Again, we shall only discuss the one-cut computations below, referring the interested reader to appendix~\ref{appendix:SpectralGeometryMatrixModels} for details on explicit computations associated to the other phases. In the one-cut case $s=1$ and \eqref{eq:one-cut-end-points} is enough to fully specify the spectral geometry \eqref{eq:one-cut-spectral-curve=background-around-which-string-equations-are-constructed}. The end-points of the single cut $\NCC = (x_1,x_2)$ for the quartic spectral curve \eqref{eq:quartic-spectral-curve-one-cut} explicitly satisfy
\begin{eqnarray}
5 \lambda \left( x_1^3 + x_2^3 \right) + 3 \lambda x_1 x_2 \left( x_1 + x_2 \right) - 48 \left( x_1 + x_2 \right) &=& 0, \\ 
35 \lambda \left( x_1^4 + x_2^4 \right) + 20 \lambda x_1 x_2 \left( x_1^2 + x_2^2 \right) + 18 \lambda x_1^2 x_2^2 - 96 \left( 3 x_1^2 + 2 x_1 x_2 + 3 x_2^2 \right) &=& - 1536 t.
\end{eqnarray}
\noindent
This quartic system has two (symmetric) solutions, which may be written as
\be
\begin{lcases}
x_1(t) &= -2 \sqrt{r_i(t)} \\
x_2(t) &= 2 \sqrt{r_i(t)}
\end{lcases},
\ee
\noindent
where, for $i=1,2$, the functions $\left\{ r_i(t) \right\}$ are the two solutions of the algebraic equation
\be
\label{eq:quartic-MM-classical-string-eq}
r_i \left( 1 - \frac{\lambda}{2}\, r_i \right) = t.
\ee
\noindent
This is the planar, classical solution of the quartic-model string-equation \eqref{eq:quarticstringequationthooftlimit}, and we are back to the by-now familiar issues of handling potentially singular transseries arising from the multi-sheetness structure of their required planar input. For the quartic matrix model this is illustrated in the two-sheeted plot in figure~\ref{fig:QMMPlanarBranching}, displaying the following explicit solution to the classical string equation \eqref{eq:quartic-MM-classical-string-eq}. Being completely explicit, the functions $\left\{ r_i(t) \right\}$ for $i=1,2$ are:
\bea
\label{eq:QMMPlanarSolution1}
r_1 (t) &=& \frac{1}{\lambda} \left( 1 - \sqrt{ 1-2\lambda t } \right), \\
\label{eq:QMMPlanarSolution2}
r_2 (t) &=& \frac{1}{\lambda} \left( 1 + \sqrt{ 1-2\lambda t } \right).
\eea
\noindent
As we shall see in plot~\ref{fig:SpecGeofig:one-cut-right-spectral-networks-QMM} of figure~\ref{fig:quartic-spectral-networks}, the solution $r_1(t)$ describes the end-points of the one-cut configuration within the one-cut phase. As for the cubic  model, the other solution $r_2(t)$ is slightly subtle; describing the end-points of an energetically unstable configuration---engineered by fixing the contour weights of the quartic matrix model (see equation \eqref{eq:mkwmiddlewsidequartic}) to be $w_{\text{middle}} = - w_{\text{side}} = 1$. Within the trivalent region of the complex $t$-plane we then find that this configuration describes a single-cut \textit{non}-trivalent singular spectral curve corresponding to one of the energetically unstable configurations identified in \cite{bt11}.

Finally, the moment function follows from \eqref{eq:moment-function-one-cut} as in \eqref{eq:quartic-spectral-curve-one-cut},
\be
\label{eq:SpecGeo14}
M (z) = 1 - \frac{\lambda}{6} \left( z^2 + \frac{1}{2}\, x_1^2 \right).
\ee
\noindent
The roots of this moment function, signaling the location of the nonperturbative saddles \cite{msw07}, are simply
\be
\label{eq:2-saddles-quartic}
x^\star_1 = \sqrt{\frac{2}{\lambda} \left( 3 - \frac{\lambda}{4}\, x_1^2 \right)}, \qquad x^\star_2 = -x_1^\star.
\ee

Having all the information for the quartic spectral curve \eqref{eq:quartic-spectral-curve-one-cut} immediately spits out its corresponding holomorphic effective potential \eqref{eq:quartic-V-holo-eff-one-cut} in fully explicit fashion, and we may proceed to plot its spectral network $\Re V_{\text{h;eff}} (z) = 0$ (generalizing to the matrix model the corresponding quartic-potential plot in figure~\ref{fig:QuarticMatrixModelSteepestDescentContours}). This is illustrated in several plots. In figure \ref{fig:SpecGeofig:one-cut-right-spectral-networks-QMM} we display the one-cut spectral network of the quartic matrix model at an adequate sample value of the 't~Hooft parameter. This also shows a very precise match between the roots of the corresponding orthogonal polynomial $p_{200}(z)$ and the ``bridges'' of the spectral network---where they accumulate at large degree \cite{b07}. Making use of the results in appendix~\ref{appendix:SpectralGeometryMatrixModels} we may proceed to address the multi-cut holomorphic effective potential obtained from the multi-cut spectral curve \eqref{eq:spectral-curve-moment-function}. Checks for the symmetric two- and three-cut phases are illustrated in figures~\ref{fig:SpecGeofig:Two-cut-spectral-networks-QMM} and~\ref{fig:SpecGeofig:three-cut-right-spectral-networks-QMM}, and for the trivalent phase in figure~\ref{fig:SpecGeofig:trivalent-spectral-networks-QMM}, all with excellent agreement. This fully validates our spectral networks, the match of orthogonal-polynomial root-distributions and eigenvalue-distributions, as well our generic ability to predict and describe all of them.

\subsection{Large $N$ Phase Transitions and Stokes Phenomena}
\label{subsec:stokes-vs-phases}

Having addressed what we dubbed \textit{phases} of matrix models from several different---and slightly empirical---angles, we may now both be more precise on what we actually \textit{mean} by phases and how transseries already naturally incorporate them all in the guise of Stokes phenomena. We also need to address the global problem of determining which values of $t \in \BC$ fall into which phases, \textit{i.e.}, plot the large-$N$ random-matrix phase-diagram upon the complex 't~Hooft $t$-plane.

In the framework of Yang--Lee \cite{ly52a, ly52b} and Fisher \cite{f65} zeroes, grand-canonical partition functions are entire functions (on the complex plane of either fugacity or temperature, respectively) essentially characterized by their zeroes---which translate to singularities of the free energy and then poles of the specific heat; as already briefly discussed in the double-scaling context of the Boutroux classification in subsection~\ref{subsec:DSL-phases}. As such, these singularities of the free energy signal the occurrence of phase transitions (see, \textit{e.g.}, \cite{bdl05} for a review). Initially having focused on partition-function roots coalescing into lines that, in the thermodynamic limit, approach the positive real axis (of the fugacity/temperature complex plane), this research-program later uncovered that the distributions of these zeroes may be far more complicated, occupying whole regions within these planes. A very limited list of such examples includes the Ising model \cite{sk84, sc84, ajjmm17}, scalar fields near criticality \cite{admt17}, polymer adsorption \cite{bj17}, and wetting transitions \cite{ps93}. As we shall see (albeit mostly in \cite{krsst26b}), it turns out that our matrix models display rather similar-looking patterns of Yang--Lee--Fisher zeroes as in these references (which is somewhat natural due to random matrix universality, alongside the thermodynamic nature of the large $N$ limit \cite{h81}).

The example in \cite{ps93} is of particular relevance. Therein, phase transitions arise as a consequence of transseries Stokes jumps, and Yang--Lee--Fisher zeroes accumulate at phase boundaries---which coincide with partition-function anti-Stokes lines. In this picture, zeroes of the partition function are precisely ``created'' by Stokes phenomenon, where the emerging previously beyond-all-orders exponentials have conspired to change the asymptotics\footnote{In the ``classical'' Airy case, the asymptotics on positive versus negative real axes changes from monotonous to oscillatory \cite{s64}. To some extent, the difference between the Airy case and the matrix model case is that instead of Stokes phenomenon giving rise to trigonometric behavior it now gives rise to theta- or elliptic-function behavior.}. The simplest realization of these phenomena in our context was already illustrated in the Boutroux pictures of figures~\ref{fig:PISolutionsNoZeros}, \ref{fig:PINewBranchTritronqueeSolutionsNoZeros}, and~\ref{fig:YLSolutionsNoZeros}, where the Painlev\'e-type poles yield partition-function zeroes, and where these zeroes occupied ``pizza slices'' on the complex $z$-plane and ended up asymptoting towards anti-Stokes lines at their boundaries. An analogous albeit more intricate distribution of partition-function zeroes will occur off-criticality \cite{krsst26b}. In this way, random-matrix transseries provide a generalization of the mechanism already at work in the simpler example of \cite{ps93}: the usual Stokes phenomenon.

In order to be more quantitative, let us consider our matrix-model partition-function \eqref{eq:partitionfunctionhermitianmatrix} but in the limit where the 't~Hooft parameter vanishes. In this case, the Vandermonde repulsion in \eqref{eq:partitionfunctioneigenvalues} is equally suppressed and the multi-cut canonical partition function \eqref{eq:partitionfunctioneigenvaluesmulticut} factorizes as
\be
\lim_{t \to 0} \mathcal{Z} \left( N_1, \ldots, N_s; g_{\text{s}} \right) \sim \prod_{i=1}^{s} \left( \int_{\CC_i} \frac{\rmd \lambda}{2\pi}\, \rme^{-\frac{1}{g_{\text{s}}} V(\lambda)} \right)^{N_i} \sim \exp \left\{ - \frac{1}{g_{\text{s}}} \sum_{i=1}^{s} N_i\, V(\lambda_i^{\star}) \right\},
\ee
\noindent
where in the last step we used the steepest-descent approximation around the $\lambda_i^{\star}$ saddle (these are now just like the cubic or quartic saddles in figure~\ref{fig:Cubic-AND-QuarticMatrixModelSteepestDescentContours}). Exchanging one eigenvalue in-between critical points $\lambda_i^{\star}$ and $\lambda_j^{\star}$ produces a transmonomial factor as
\be
\lim_{t \to 0} \mathcal{Z} \left( \ldots, N_i+1, \ldots, N_j-1, \ldots \right) \sim \rme^{-\frac{1}{g_{\text{s}}} \left( V(\lambda_i^{\star}) - V(\lambda_j^{\star}) \right)}\, \lim_{t \to 0} \mathcal{Z} \left( \ldots, N_i, \ldots, N_j, \ldots \right),
\ee
\noindent
from where the ``classical'' tunneling-action follows as $A (t=0) = V(\lambda_i^{\star}) - V(\lambda_j^{\star})$. At finite $t \in \BC$ 't~Hooft coupling this expression gets naturally generalized as \cite{d91, d92, msw07, msw08}
\be
\label{eq:rmt-inst-action}
A (t) = V_{\text{h;eff}} (\lambda_i^{\star}) - V_{\text{h;eff}} (\lambda_j^{\star}) = \int_{\lambda_j^{\star}}^{\lambda_i^{\star}} \rmd z\, y(z).
\ee
\noindent
Of course now $\lambda_i^{\star}, \lambda_j^{\star}$ are in different \textit{cuts} (either open or pinched \cite{msw07, msw08}). This eigenvalue-tunneling instanton-action was earlier illustrated in figure~\ref{fig:FiniteNeigenvaluetunneling}, when conveying the idea that as eigenvalues jump around, eigenvalue-configurations with distinct number of cuts and configurations may also switch dominance---what precisely occurs at phase boundaries. This \eqref{eq:rmt-inst-action} is also the instanton action which appears in the basic transmonomial building-blocks of the matrix-model transseries \eqref{eq:twoparameterresurgenttransseriesforR} \cite{m08, msw08, asv11, sv13}, that we wish to consider from the point-of-view of the large $N$ limit (with $N \in \BR$) as
\be
\label{eq:gstoNtransmonomial}
\rme^{- \frac{A (t)}{g_{\text{s}}}} \equiv \rme^{- N \frac{A (t)}{t}}.
\ee
\noindent
The natural definitions of Stokes and anti-Stokes lines which follow from these transmonomials (see, \textit{e.g.}, \cite{abs18}) are hence immediately obtained from eigenvalue tunneling. Indeed, a pinched-cycle starts opening up into a full-fledged cut once it reaches equilibrium with the other cuts as in the conditions \eqref{eq:equilibrium-condition-via-B-cycles}, which via \eqref{eq:SpecGeo2b} translates into
\begin{equation}
\label{eq:anti-stokes-in-the-matrix-model}
\Re \left( \frac{1}{t} \oint_{B} \rmd z\, y (z) \right) = \Re \frac{A(t)}{t} = 0.
\end{equation}
\noindent
This last condition is nothing but the definition of an anti-Stokes line\footnote{Stokes lines of \eqref{eq:twoparameterresurgenttransseriesforR} correspond to the ``orthogonal'' condition
\be
\Im \frac{A(t)}{t} = 0.
\ee
} of the matrix-model transseries \eqref{eq:twoparameterresurgenttransseriesforR}, and hence completely fixes all large $N$ phase boundaries in the 't~Hooft complex plane $t \in \BC$. Let us make these arguments explicit in our two main examples. 

\paragraph{The Cubic Matrix Model:} 

\begin{figure}
    \centering
    \includegraphics[width=0.7\linewidth]{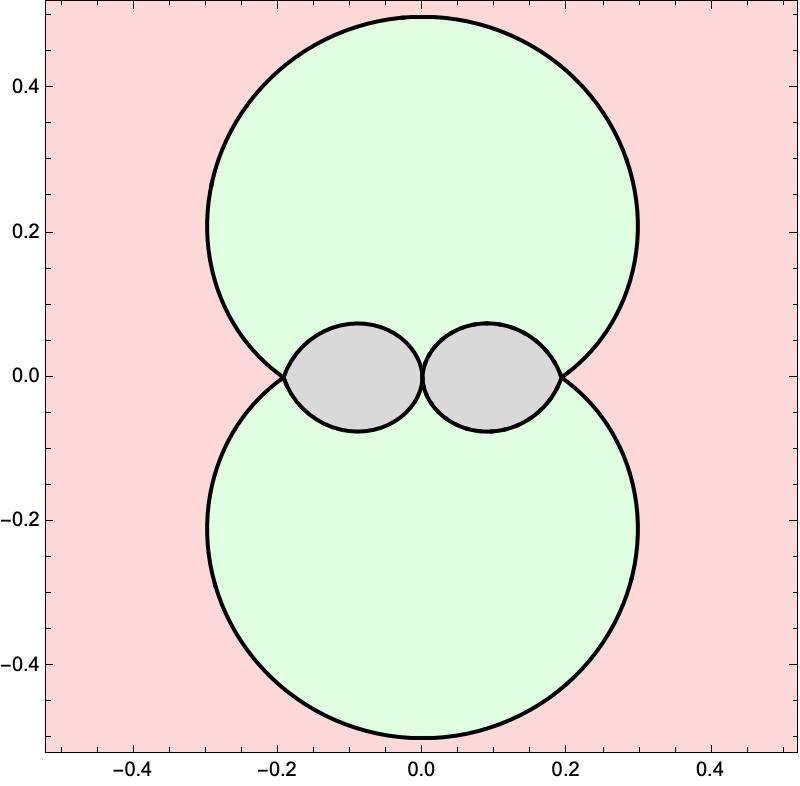}
    \caption{The large $N$ phase diagram of the cubic matrix model \eqref{eq:CubicMatrixModelPotential} for all 't~Hooft coupling $t \in \BC$. The colors indicating the several dominant eigenvalue-configurations are in correspondence with what was discussed in subsections~\ref{subsec:OP-phases} and~\ref{subsec:SG-phases}, in particular with figure~\ref{fig:SpecGeofig:eigcmm} (one-cut, two-cut, and trivalent-tree phases, in gray, green, and pink, respectively). The phase boundaries or anti-Stokes lines (as predicted by the instanton actions) are given by the black lines. Note how the \PI~critical point of \eqref{eq:multicriticalpoint} is herein clearly located at $t=\frac{1}{3\sqrt{3}\lambda^2}$ (with $\lambda=1$).}
    \label{fig:CMM Phase Diagram}
\end{figure}

\begin{figure}
    \centering
    \includegraphics[width=0.6\linewidth]{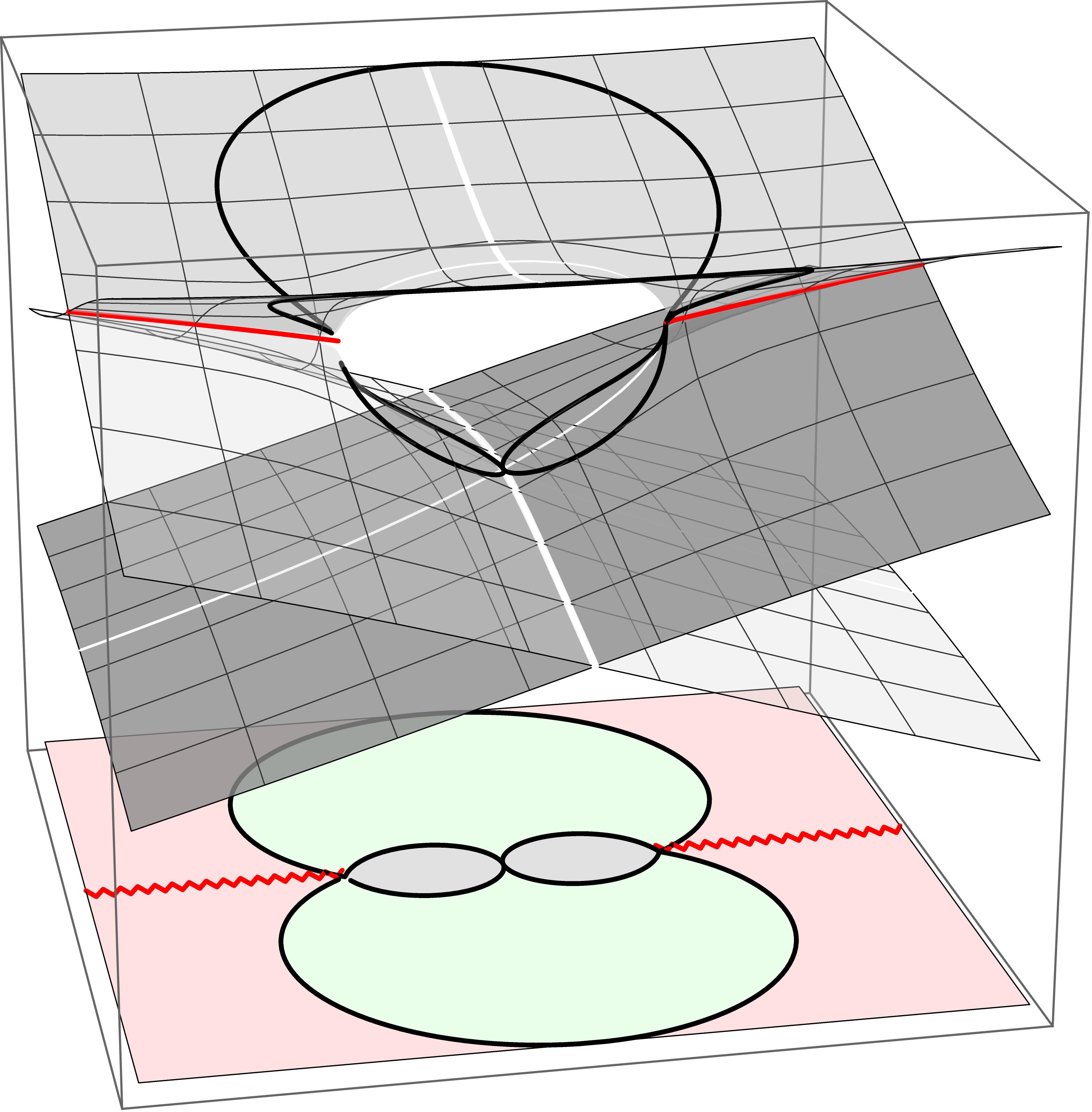}
    \caption{Planar solution to the cubic matrix model, as displayed in figure~\ref{fig:CMMPlanarBranching}, alongside the cubic-model phase-diagram just introduced in figure~\ref{fig:CMM Phase Diagram}. In order to show which branch gives rise to which phase boundary, we plot the black phase-boundaries on their appropriate branches.}
    \label{fig:CMMPlanarBranchingPhaseDiagram}
\end{figure}

\begin{figure}
     \centering
     \begin{subfigure}[b]{0.32\textwidth}
         \centering
         \includegraphics[width=\textwidth]{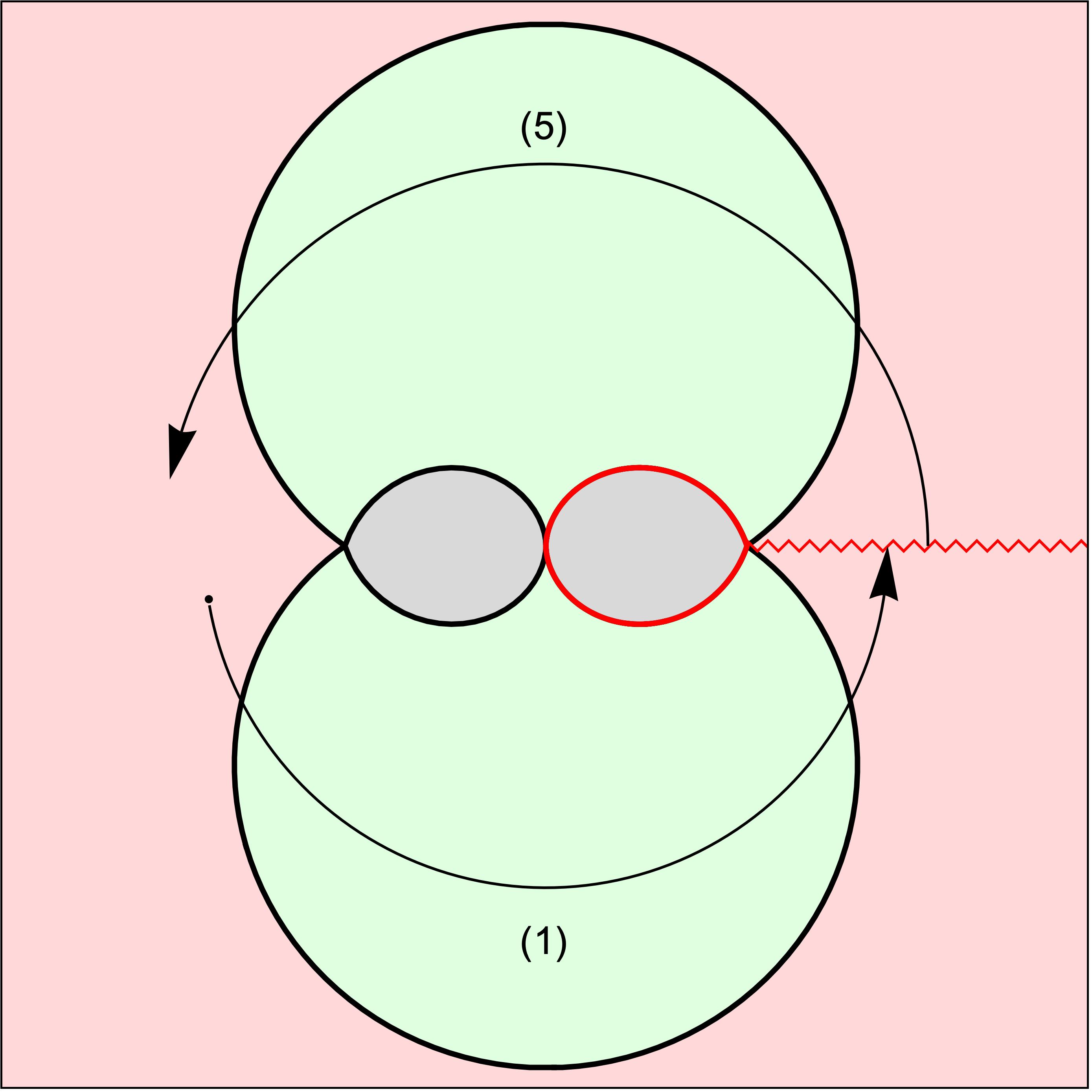}
         \caption{Right-most one-cut branch.}
         \label{fig: CMM r Right}
     \end{subfigure}
     \hfill
     \begin{subfigure}[b]{0.32\textwidth}
         \centering
         \includegraphics[width=\textwidth]{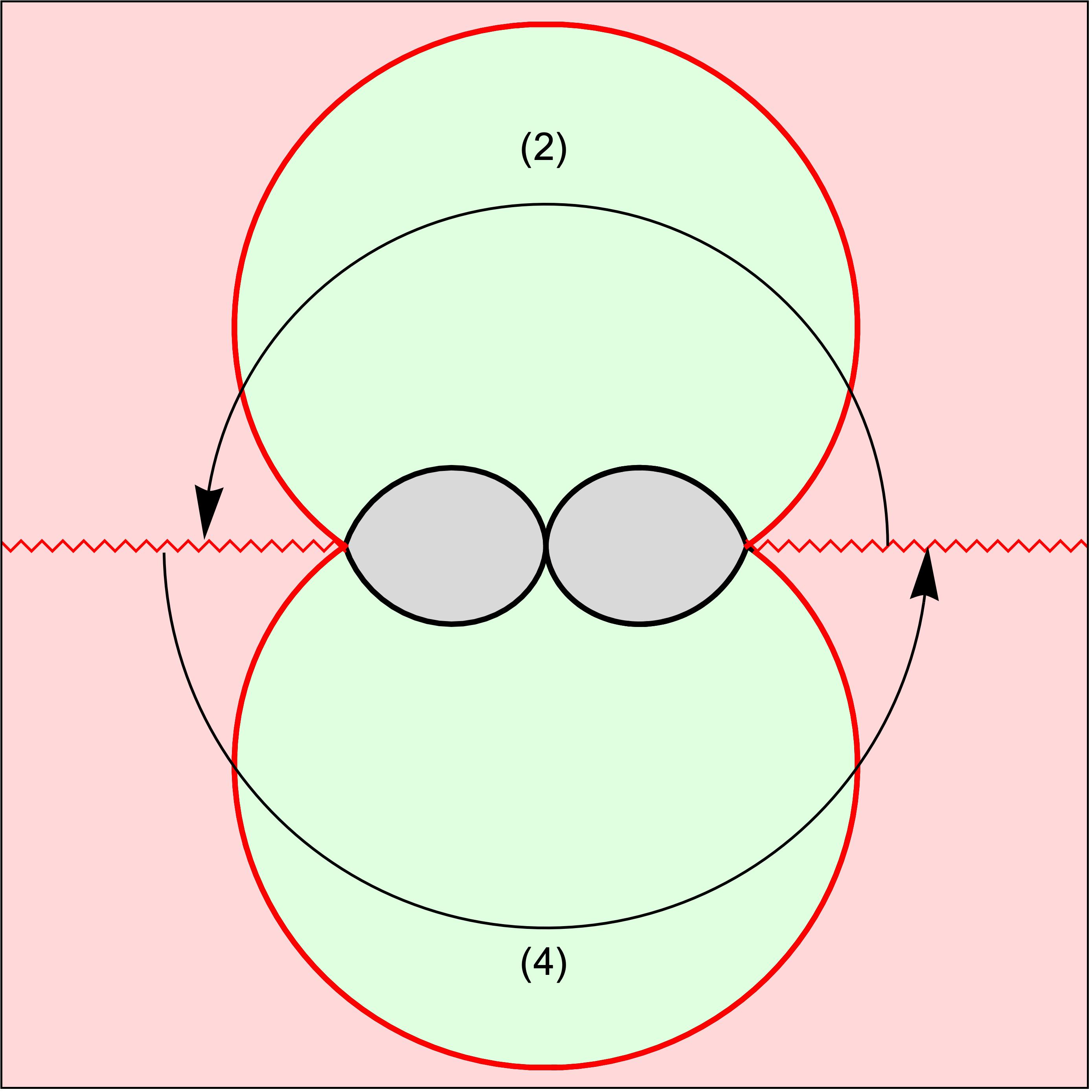}
         \caption{Trivalent branch.}
         \label{fig: CMM r Triv}
     \end{subfigure}
     \hfill
     \begin{subfigure}[b]{0.32\textwidth}
         \centering
         \includegraphics[width=\textwidth]{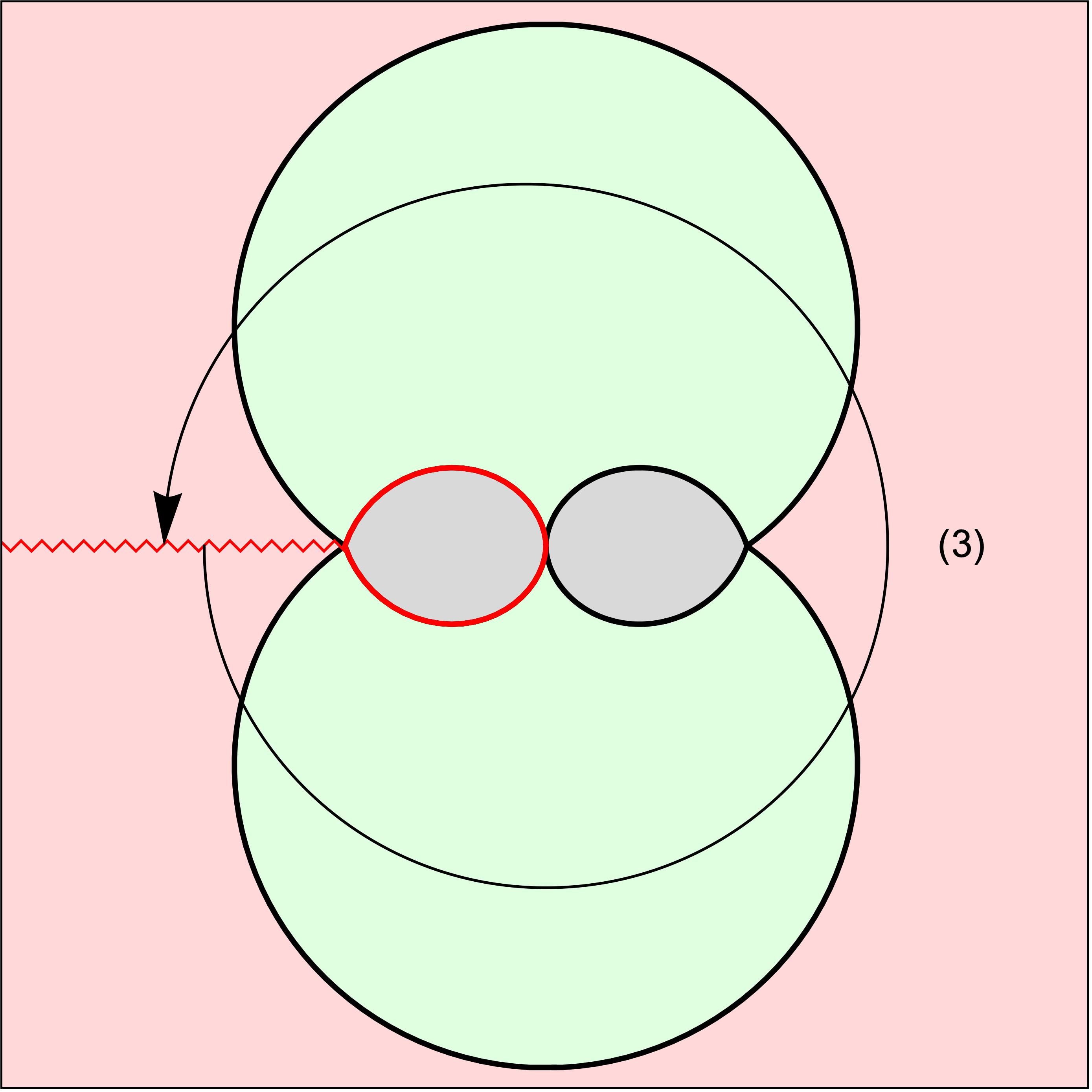}
         \caption{Left-most one-cut branch.}
         \label{fig: CMM r Left}
     \end{subfigure}
        \caption{Multi-branched structure of $r(t)$ for the cubic matrix model, following the trajectory: from $(1)$ in~\ref{fig: CMM r Right} through $(2)$ in~\ref{fig: CMM r Triv} through $(3)$ in~\ref{fig: CMM r Left} through $(4)$ back in~\ref{fig: CMM r Triv} and through $(5)$ back in~\ref{fig: CMM r Right}, completing the multi-sheeted path. On each branch, we indicate in bright-\textcolor{red}{red} both the branch-cuts (wavy) of $r(t)$ and the anti-Stokes line (solid) of the corresponding instanton action. The branches are all connected via the trajectory that started on the left-hand-side figure (and which we further choose to be the \textit{principal-sheet} in this work) and loops through all the remaining branch-cuts so as to fill-out all sheets of the complete Riemann surface.}
        \label{fig: CMM r branching}
\end{figure}

The eigenvalue-tunneling instanton-action \eqref{eq:rmt-inst-action} is easily obtained, say, from the one-cut holomorphic effective potential we computed in \eqref{eq:cubic-V-holo-eff-one-cut} (but see as well sub-appendix~\ref{subappendix:cubicmatrixmodel} for multi-cut solutions of the cubic model). To be more precise, recall that this action is explicitly written in terms of the solutions to the classical string equation \eqref{eq:cubic-MM-classical-string-eq}, \textit{i.e.}, in terms of the functions $\left\{ r_i(t) \right\}$ for $i=1,2,3$. In this way, the matrix-model anti-Stokes requirements \eqref{eq:anti-stokes-in-the-matrix-model} amount to \textit{three} conditions
\begin{equation}
\Re \frac{A (t)}{t} \equiv \Re \frac{A \left( r_i (t) \right)}{t} = 0,
\end{equation}
\noindent
where each of these conditions yields a \textit{single} anti-Stokes line or phase boundary; namely: $i=1$ gives the boundary from two-cut to trivalent phases; $i=2$ gives the left-most boundary from one-cut to two-cut phases; $i=3$ gives the right-most boundary from one-cut to two-cut phases. Overall, this yields the phase boundaries\footnote{The phase diagram of the cubic matrix model was previously---albeit only partially, as still missing a complete description of its trivalent phase---addressed in, \textit{e.g.}, \cite{d91, d92, mpp09, aam13b}.} which are displayed in figure~\ref{fig:CMM Phase Diagram}. It is also important to recall the multi-sheeted structure of the solutions to the planar string equation \eqref{eq:cubic-MM-classical-string-eq}, illustrated in figure~\ref{fig:CMMPlanarBranching}---which now gets supplemented by the phase diagram of figure~\ref{fig:CMM Phase Diagram} into figure~\ref{fig:CMMPlanarBranchingPhaseDiagram}. This structure carries through to the present discussion, as the \textit{different} $\left\{ r_i(t) \right\}$ variables may be analytically continued \textit{into each other} on top of the phase diagram. This is shown in figure~\ref{fig: CMM r branching}. In this way, by adequately addressing the multi-branched structure of our problem via appropriate analytic continuation, the complete cubic matrix-model phase-diagram follows from using a \textit{single} instanton action. Having its precise analytic structure in mind, this is then the instanton action which makes its way through to the matrix-model transseries \eqref{eq:twoparameterresurgenttransseriesforR}.

\begin{figure}
\centering
	\begin{tikzpicture}
    \def\hspacing{0.3};
	\begin{scope}[scale=0.7, shift={({-7},{-6})}]
    \draw[color=ForestGreen, line width=1pt] (-2+0.1,2.68-\hspacing) -- (4-0.8,2.68-\hspacing);
    \draw[color=ForestGreen, line width=1pt] (-2+0.2,2.68-2*\hspacing) -- (4-1.2,2.68-2*\hspacing);
    \draw[color=ForestGreen, line width=1pt] (-2+0.3,2.68-3*\hspacing) -- (4-1.6,2.68-3*\hspacing);
    \draw[color=ForestGreen, line width=1pt] (-2+0.5,2.68-4*\hspacing) -- (4-1.9,2.68-4*\hspacing);
    \draw[color=ForestGreen, line width=1pt] (-2+0.6,2.68-5*\hspacing) -- (4-2.2,2.68-5*\hspacing);
    \draw[color=ForestGreen, line width=1pt] (-2+0.8,2.68-6*\hspacing) -- (4-2.5,2.68-6*\hspacing);
    \draw[color=ForestGreen, line width=1pt] (-2+1,2.68-7*\hspacing) -- (4-2.8,2.68-7*\hspacing);
    \draw[color=ForestGreen, line width=1pt] (-2+1.3,2.68-8*\hspacing) -- (4-3.2,2.68-8*\hspacing);
	 \draw[scale=2, domain=-1.3:3, smooth, variable=\x, LightBlue, line width=2pt] plot ({\x}, {\x*\x-2/6*\x*\x*\x});
	\draw[ForestGreen, fill=ForestGreen] (-2,2.68) circle (1.1ex);
	 \draw[ForestGreen, fill=ForestGreen] (4,2.68) circle (1.1ex);
	 \node[ForestGreen] at (0,3.4) {$N-2$};
\draw[color=ForestGreen, line width=2pt] (-2,2.68) -- (4,2.68);
\draw[line width = 2pt, color = blue, ->] (4,2.68+0.5) to[out = 70, in = 180+50] (4+0.7,2.68+0.5+1);
\draw[line width = 2pt, color = blue, ->] (4,2.68-0.5) to[out = -70, in = 180-50] (4+0.7,2.68-0.5-1);
 \draw[blue, fill=blue] (4+0.7+0.2,2.68-0.5-1-0.3) circle (1.1ex);
 \draw[blue, fill=blue] (4+0.7+0.2,2.68+0.5+1+0.3) circle (1.1ex);
 \draw[line width = 2pt, color = red, ->] (4.5,2.68+0.3) to[out = -10, in = 90] (6.1,0.5);
 \draw[blue, fill=blue] (5.95,0.1) circle (1.1ex);
\def\hspacing{1.2};
\def\vspacing{-0.7};
 \draw[line width = 2pt, color = red] (5+\hspacing,3+\vspacing) -- (5+0.5+\hspacing,3+0.5+\vspacing);
  \draw[line width = 2pt, color = red] (5+0.5+\hspacing,3+\vspacing) -- (5+\hspacing,3+0.5+\vspacing);
	\end{scope}
	\end{tikzpicture}
\caption{Visualization of eigenvalue leakage at the boundary between the one-cut and the trivalent phases of the cubic matrix model (upon reaching the \PI~critical point). One of the cut end-points now coincides with the unstable saddle, where conventional tunneling of eigenvalues along the real line would induce an instability (marked by the \textcolor{red}{red} cross, which would represent eigenvalues tunneling into oblivion). To avoid this scenario, the dominant configuration instead involves eigenvalues leaking into more stable regions of the complex plane, depicted by the pair of \textcolor{blue}{blue} arrows illustrating the guiding of eigenvalues away from the real potential curve.}
\label{fig:Eigenvaluetunnelingleakage}
\end{figure}
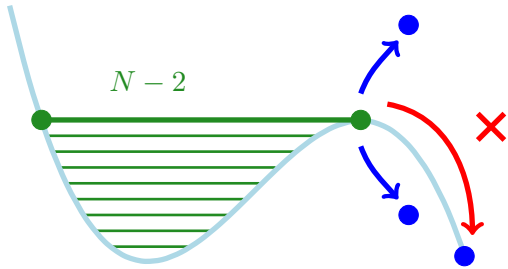

Let us add a small comment on the physical picture concerning the behavior of eigenvalues at these diverse phase boundaries. The usual one- to two-cuts phase-transition is clearly understood in terms of eigenvalue ``tunneling'' \cite{d91, d92, msw07, msw08, mss22}, as partially illustrated in the transition from figure~\ref{fig:FiniteNeigenvaluetunneling} to~\ref{fig:multicutPotential}. Eigenvalues from one cut (say, in figure~\ref{fig:FiniteNeigenvaluetunneling}) can tunnel across the potential barrier into a pinched cycle, and start filling it in so as to create another cut (say, as the resulting figure~\ref{fig:multicutPotential}). As for the one- or two-cuts to trivalent phase-transition, this is probably best understood in terms of eigenvalue ``leakage'', and as illustrated in figure~\ref{fig:Eigenvaluetunnelingleakage}. Past the point where the single-cut has been completely filled with eigenvalues, up to the \PI~critical point, eigenvalues can no longer cross upon the real axis---as that would lead onto an instability---but instead start leaking into the complex plane, in complex conjugate pairs so as to form a trivalent spectral\footnote{In the matrix model of \cite{gjk21}, the states corresponding to this (trivalent) complex part of the spectrum were interpreted as metastable black hole states.} arrangement \cite{gjk21}.

\paragraph{The Quartic Matrix Model:} 

\begin{figure}
    \centering
    \includegraphics[width=0.7\linewidth]{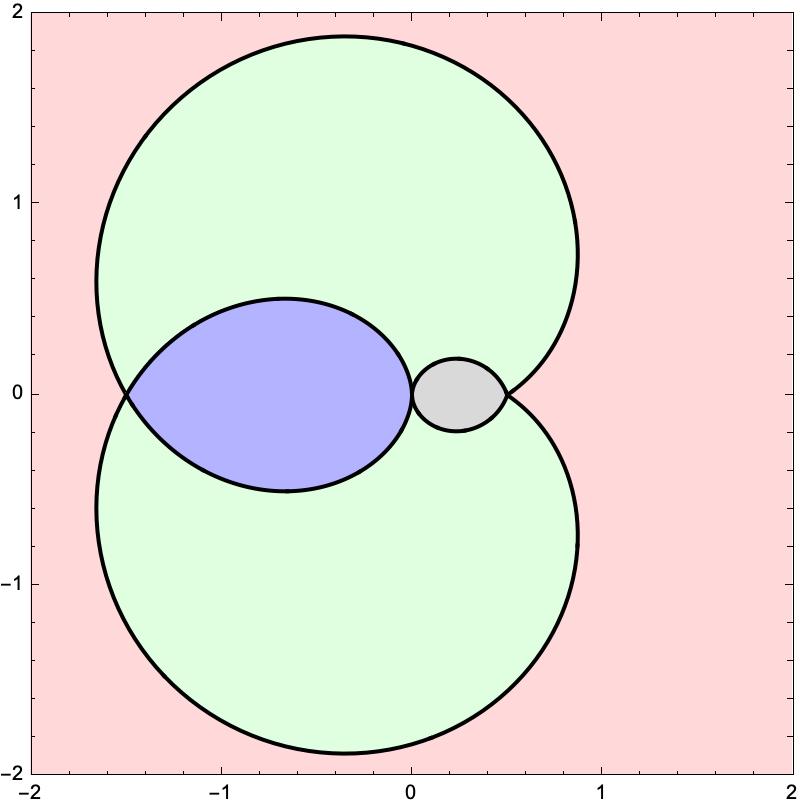}
    \caption{The large $N$ phase diagram of the quartic matrix model \eqref{eq:QuarticMatrixModelPotential} for all 't~Hooft coupling $t \in \BC$. Colors indicating the several dominant eigenvalue-configurations are in correspondence with what was discussed in subsections~\ref{subsec:OP-phases} and~\ref{subsec:SG-phases}, in particular with figure~\ref{fig:SpecGeofig:eigqmm} (one-, two-, three-cut, and trivalent phases, in gray, blue, green, and pink, respectively). The phase boundaries or anti-Stokes lines (as predicted by the instanton actions) are given by the black lines. Note how the \PI~critical point of \eqref{eq:multicriticalpoint} is herein clearly located at $t=\frac{1}{2\lambda}$ (with $\lambda=1$).}
    \label{fig:QMM Phase Diagram}
\end{figure}

\begin{figure}
    \centering
    \includegraphics[width=0.6\linewidth]{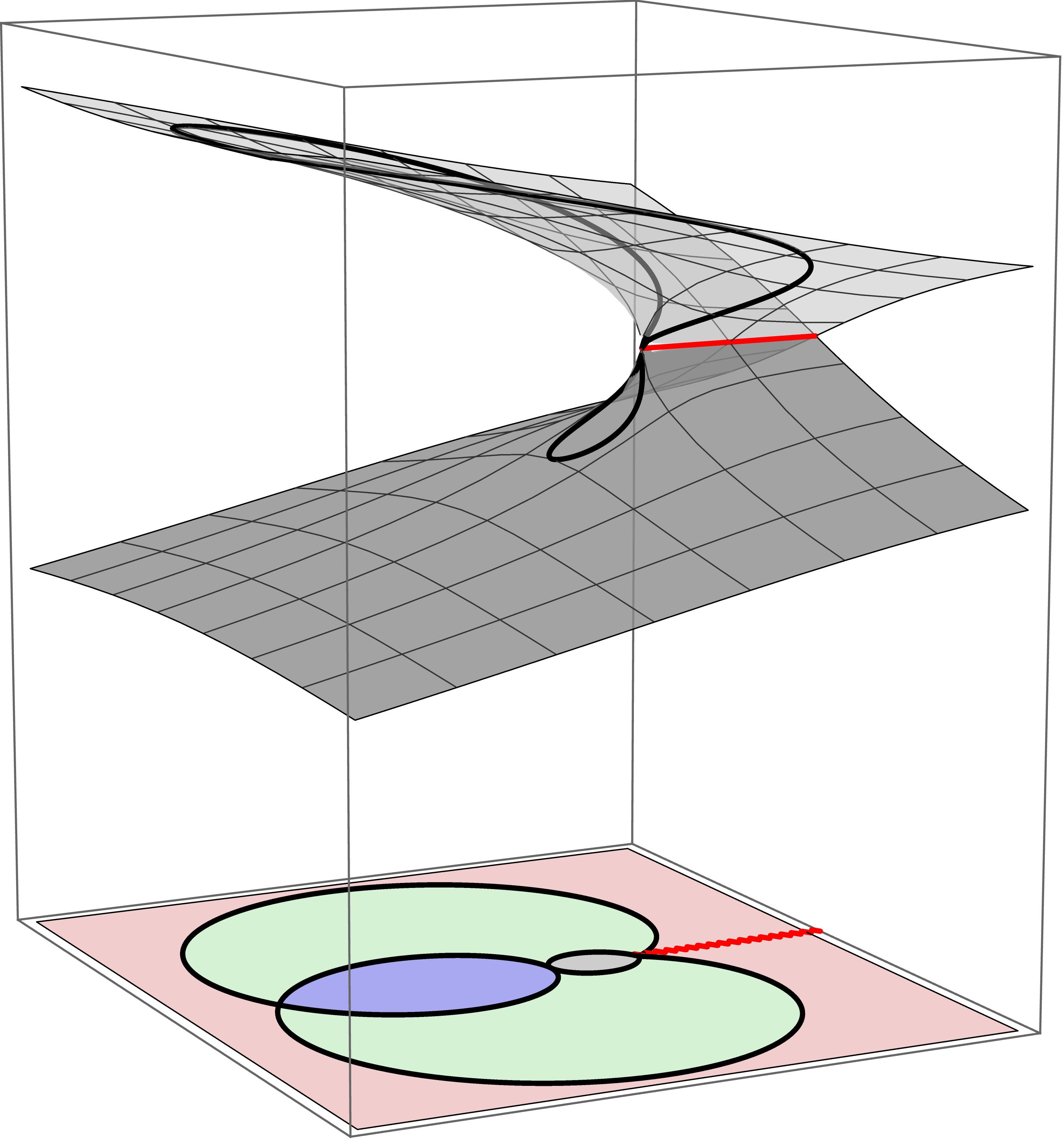}
    \caption{Planar solution to the quartic matrix model, as displayed in figure~\ref{fig:QMMPlanarBranching}, alongside the quartic-model phase-diagram just introduced in figure~\ref{fig:QMM Phase Diagram}. In order to show which branch gives rise to which phase boundary, we plot the black phase-boundaries on their appropriate branches.}
    \label{fig:QMMPlanarBranchingPhaseDiagram}
\end{figure}

\begin{figure}
     \centering
     \begin{subfigure}[b]{0.48\textwidth}
         \centering
         \includegraphics[width=\textwidth]{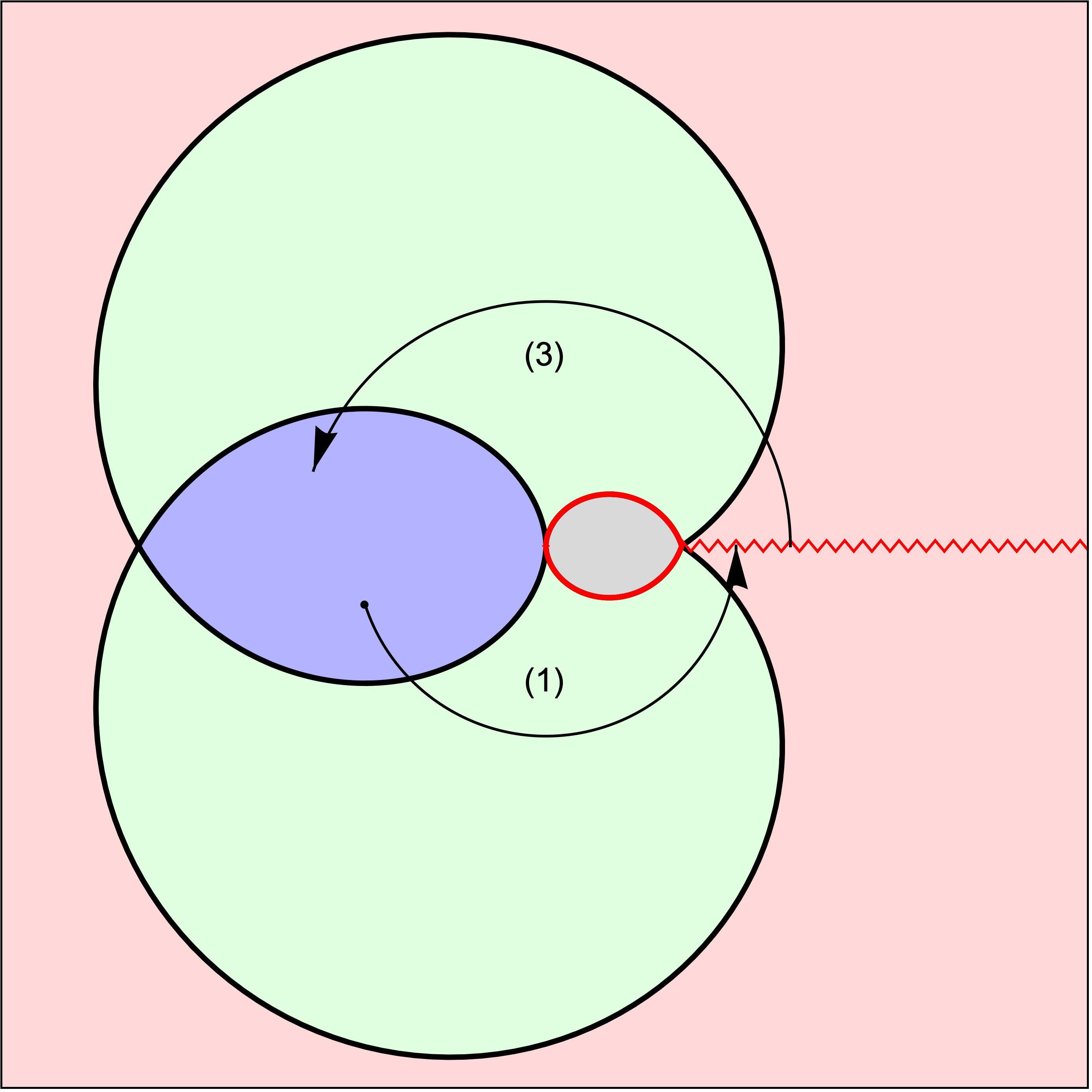}
         \caption{One-cut branch.}
         \label{fig: QMM r One cut}
     \end{subfigure}
     \hfill
     \begin{subfigure}[b]{0.48\textwidth}
         \centering
         \includegraphics[width=\textwidth]{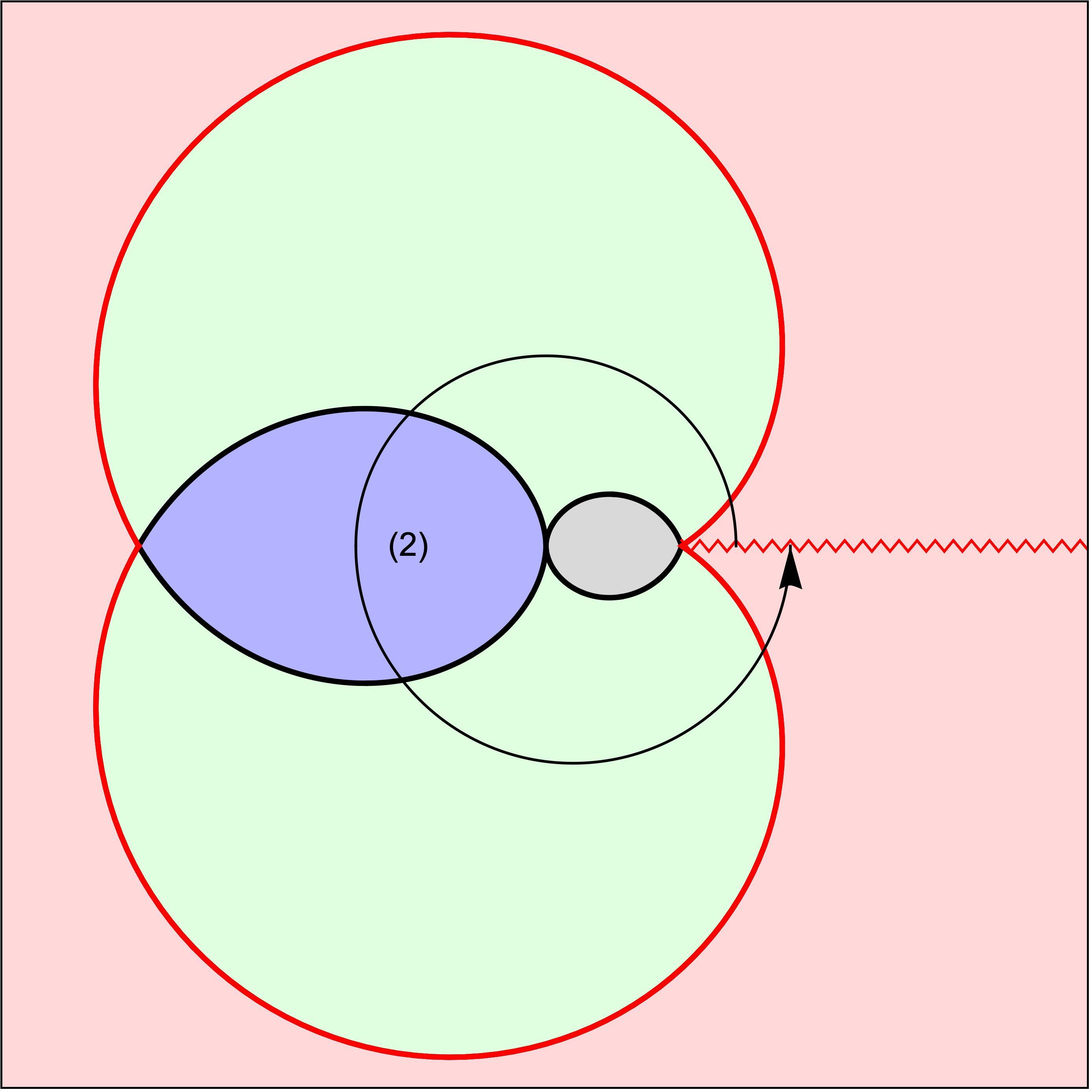}
         \caption{Trivalent branch.}
         \label{fig: QMM r Triv}
     \end{subfigure}
        \caption{Multi-branched structure of $r(t)$ for the quartic matrix model, following the trajectory: from $(1)$ in~\ref{fig: QMM r One cut} through $(2)$ in~\ref{fig: QMM r Triv} and through $(3)$ back in~\ref{fig: QMM r One cut}, completing the multi-sheeted path (this complete structure first appeared in \cite{csv15}). On each branch, we indicate in bright-\textcolor{red}{red} both the branch-cuts (wavy) of $r(t)$ and the anti-Stokes line (solid) of the corresponding instanton action. The branches are all connected via the trajectory that started on the left-figure (and which we further choose to be the \textit{principal-sheet} in this work) and loops through all the remaining branch-cuts so as to fill-out all sheets of the complete Riemann surface. Note that we do not encounter the two-cut region in this way (see \cite{ss26} for the two-cut region).}
        \label{fig:QMM r branching}
\end{figure}

Just like for the cubic model, the eigenvalue-tunneling instanton-action \eqref{eq:rmt-inst-action} is easily obtained, say, from the one-cut holomorphic effective potential we computed in \eqref{eq:quartic-V-holo-eff-one-cut} (but see as well sub-appendix~\ref{subappendix:quarticmatrixmodel} for multi-cut solutions of the quartic model). Because quartic-model formulae are rather compact (as compared to cubic-model formulae) we can also be herein a bit more explicit than in the previous cubic-paragraph. Having identified the nonperturbative saddle in \eqref{eq:quartic-spectral-curve-one-cut} as (note that there are two non-trivial saddles)
\be
x_{\star}^2 = \frac{6}{\lambda} - 2\alpha^2,
\ee
\noindent
and using the branch which reduces to the Gaussian model as $\lambda \to 0$, \eqref{eq:quartic-cut-endpoint-one-cut}, the instanton action\footnote{With vanishing 't~Hooft coupling, the cut collapses as $2\alpha(t) \sim 2\sqrt{t} + \cdots$, nonperturbative saddles are at $x_{\star} (t) \sim \pm \sqrt{6/\lambda} + \cdots$, and instanton action $A(t) \sim 3/2\lambda + \cdots$, all matching the quartic integral example in \cite{abs18}.} is then, explicitly \cite{msw07},
\bea
A_{\text{1-cut}} (t) &=& V_{\text{h;eff}} (x_{\star}) - V_{\text{h;eff}} (2\alpha) = \nonumber \\
&=& \frac{1}{4} x_{\star} \sqrt{x_{\star}^2-4\alpha^2} - 4t \log \left( \sqrt{x_{\star}-2\alpha} + \sqrt{x_{\star}+2\alpha} \right) + 2t \log 4\alpha.
\eea
\noindent
Of course this is not the whole story. In the quartic matrix model, the instanton action is explicitly written in terms of the solutions to the classical string equation \eqref{eq:quartic-MM-classical-string-eq}, \textit{i.e.}, in terms of the functions $\left\{ r_i(t) \right\}$ for $i=1,2$. In this way, the matrix-model anti-Stokes requirements \eqref{eq:anti-stokes-in-the-matrix-model} now amount to \textit{two} conditions
\begin{equation}
\Re \frac{A (t)}{t} \equiv \Re \frac{A \left( r_i (t) \right)}{t} = 0,
\end{equation}
\noindent
where each of them yields a \textit{single} anti-Stokes line or phase boundary; namely: $i=1$ gives the boundary from three-cut to trivalent phases; $i=2$ gives the boundary from one-cut to three-cut phases. On top of these, one still has to consider the $\BZ_2$-symmetric two-cut \textit{Stokes phase} discussed in \cite{sv13} and herein described in sub-appendix~\ref{subappendix:quarticmatrixmodel}. One finds \cite{sv13}
\be
A_{\text{2-cut}} (t) = V_{\text{h;eff}} (x_{\star}) - V_{\text{h;eff}} (x_1) = \frac{1}{4} x_1 x_2 - t \log \frac{x_2-x_1}{x_2+x_1}.
\ee
\noindent
The anti-Stokes requirement on this action spits out the final phase boundary, from two-cut to three-cut phases. Overall, this yields the phase boundaries\footnote{This quartic phase diagram was very thoroughly addressed, both at and off eigenvalue equilibrium in \cite{bt11, bt16}.} which are displayed in figure~\ref{fig:QMM Phase Diagram}. As for the cubic, also in the quartic model it is important to recall the multi-sheeted structure of the solutions to the planar string equation \eqref{eq:quartic-MM-classical-string-eq}, illustrated in figure~\ref{fig:QMMPlanarBranching}---which now gets supplemented by the phase diagram of figure~\ref{fig:QMM Phase Diagram} into figure~\ref{fig:QMMPlanarBranchingPhaseDiagram}. Again, this structure carries through to the present discussion, as the \textit{different} $\left\{ r_i(t) \right\}$ variables may be analytically continued \textit{into each other} on top of the phase diagram (see as well \cite{bt16}). This is shown in figure~\ref{fig:QMM r branching}, starting from the one-cut Stokes phase. In this way, by adequately addressing the multi-branched structure of our problem via appropriate analytic continuation, the quartic matrix-model phase-diagram follows from using a \textit{single} instanton action, with the exception of the two-cut region\footnote{How to reach the two-cut region starting from the one-cut transseries \eqref{eq:twoparameterresurgenttransseriesforR} will be discussed in detail in \cite{ss26}.}. Having its precise analytic structure in mind, this is then the instanton action which makes its way through to the matrix-model transseries \eqref{eq:twoparameterresurgenttransseriesforR}.

\medskip

The above picture unifies phase transitions and Stokes phenomena within random matrix models in the exact same spirit as the analysis in \cite{ps93}; and this will be our context going forward. But for the statistical-mechanical reader, the derivation of the phase diagrams in figures~\ref{fig:CMM Phase Diagram} and~\ref{fig:QMM Phase Diagram} may carry a bit too much of a mathematical-analysis flavor, and instead ask for a suitable \textit{order parameter}; physically\footnote{Albeit phases were physically established based on features of the underlying spectral curve, encoding the dominant large $N$ eigenvalue configuration of the matrix model.} establishing the phase transitions at play by tracking its behavior across different values of the 't~Hooft coupling. Breaks in regularity of such order parameter will naturally correspond to values of $t$ lying on phase transition lines. Now, any function derived from the large $N$ expansion of the matrix model can serve as an order parameter, as such functions can be, in principle, fully determined by the spectral curve and are thus sensitive to its changes. In what follows, we define our order parameter as
\be
\label{eq:SpecGeo9}
\NCO(t) = \frac{1}{t} \left (\frac{\partial \CF_0}{\partial t_1} (t) - \frac{\partial \CF_0}{\partial t_2}(t) \right) = \frac{1}{t} \oint_{B} \rmd z\, y(z),
\ee
\noindent
where $t_1$ and $t_2$ are the partial 't~Hooft couplings associated with two cuts in the matrix model and $B$ is the $B$-cycle of the spectral curve that connects them. This is of course a completely natural definition given \eqref{eq:SpecGeo2b}, \eqref{eq:anti-stokes-in-the-matrix-model}, and our preceding discussions (in particular, the feature that phase boundaries were defined as the set of values of $t \in \BC$ for which the spectral curve features vanishing periods along one or more cycles). As we did for spectral geometry and instanton actions, we will compute the order parameter \eqref{eq:SpecGeo9} analytically, whenever possible, and numerically otherwise; following its behavior across various values of $t$ (in every phase, as were previously defined) and highlighting visible breaks in regularity signaling phase transitions.

\paragraph{The Cubic Order-Parameter:}

\begin{figure}
    \centering
    \includegraphics[width=0.65\linewidth]{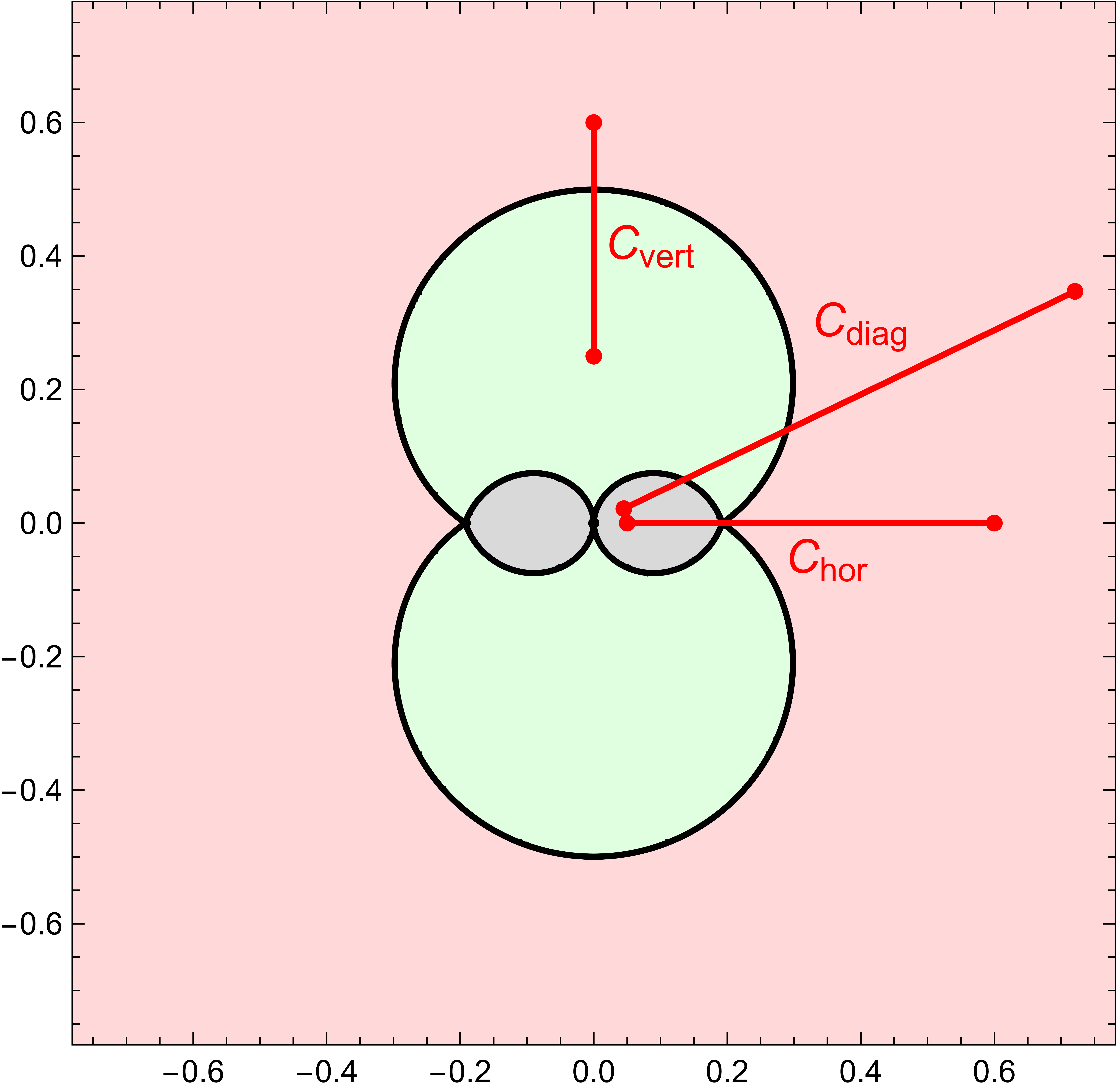}
    \caption{Depiction of the paths we chose to explore the behavior of the order parameter $\NCO(t)$ in \eqref{eq:cubicaction} across the phase diagram of the cubic model from figure~\ref{fig:CMM Phase Diagram}. The path $C_{\text{hor}}$ is upon the real axis and precisely crosses the critical point at which the matrix-model double-scales to \PI. The path $C_{\text{diag}}$ crosses both phase transition lines. The path $C_{\text{vert}}$ is upon the imaginary axis and only crosses the phase transition line separating the two-cut and trivalent phases.}
    \label{fig:SpecGeofig6}
\end{figure}

Start with the order parameter \eqref{eq:SpecGeo9} in the one-cut case. The resulting expression is precisely the instanton action for the cubic matrix model transseries, and is explicitly given by
\bea
\NCO(t) &=& \frac{1}{48 t \lambda}\, \sqrt{ \left( \lambda \left( 3 x_1 + x_2 \right) - 4 \right) \left( \lambda \left( x_1 + 3 x_2 \right) - 4 \right)} \left( \frac{8}{\lambda} + \lambda \left( 3 x_1^2 + 2 x_1 x_2 + 3 x_2^2 \right) - 8 \left( x_1 + x_2 \right) \right) + \nonumber \\
&&
+ \frac{1}{48 t \lambda}\, \left( 6 \lambda \left( x_1 - x_2 \right)^2 \left( \lambda \left( x_1 + x_2 \right) - 2 \right)\, \text{arctanh}\, \sqrt{\frac{\lambda \left( x_1 +3 x_2 \right) - 4}{\lambda \left( 3 x_1 + x_2 \right) - 4}} \right).
\label{eq:cubicaction}
\eea
\noindent
The analogous expression for the other phases is equally lengthy and not very illuminating, and we refer the reader to sub-appendix~\ref{subappendix:cubicmatrixmodel} for this expression and computation.

\begin{figure}
    \centering
   \begin{subfigure}{0.48\textwidth}
        \centering
    \includegraphics[width=\textwidth]{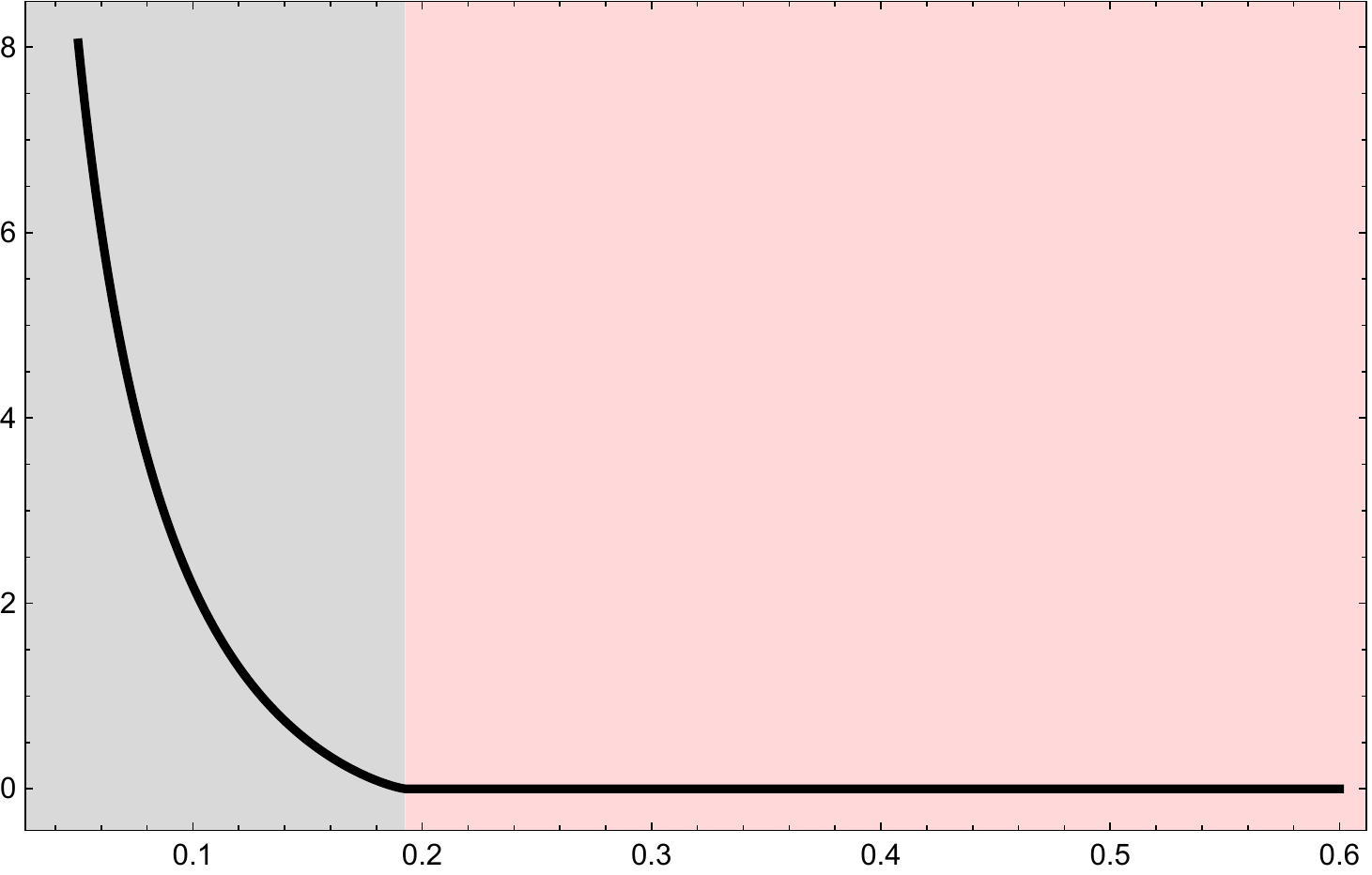}
    \caption{Real part of $\NCO(t)$.}
    \end{subfigure}
    \hspace{5pt}
    \begin{subfigure}{0.48\textwidth}
        \centering
    \includegraphics[width=\textwidth]{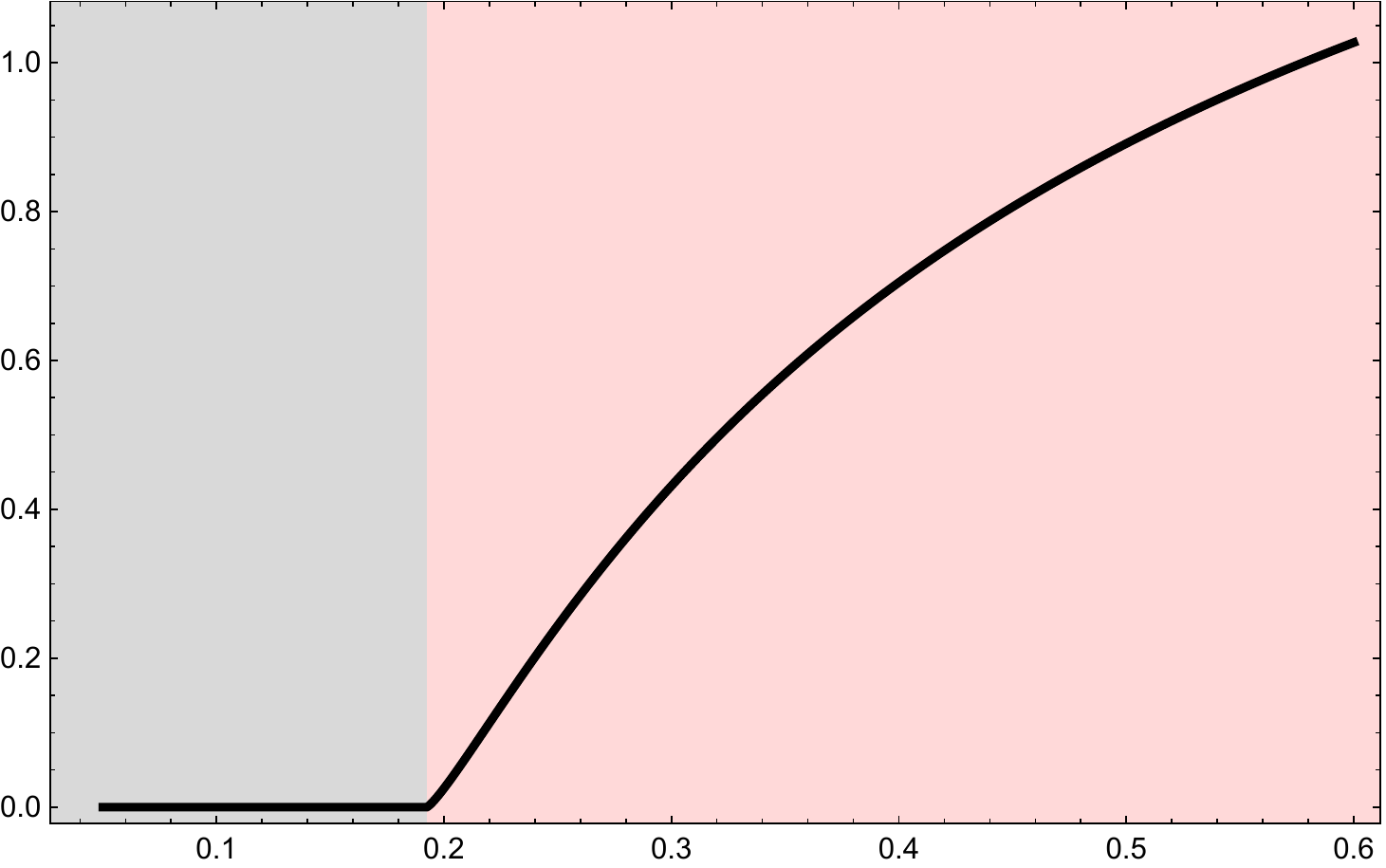}
    \caption{Imaginary part of $\NCO(t)$.}
    \end{subfigure}
    \caption{Evaluation of the order parameter $\NCO(t)$ along the path $C_{\text{hor}}$, displayed in figure~\ref{fig:SpecGeofig6}.}
    \label{fig:SpecGeofig5}
\end{figure}
%
\begin{figure}
    \centering
   \begin{subfigure}{0.48\textwidth}
        \centering
    \includegraphics[width=\textwidth]{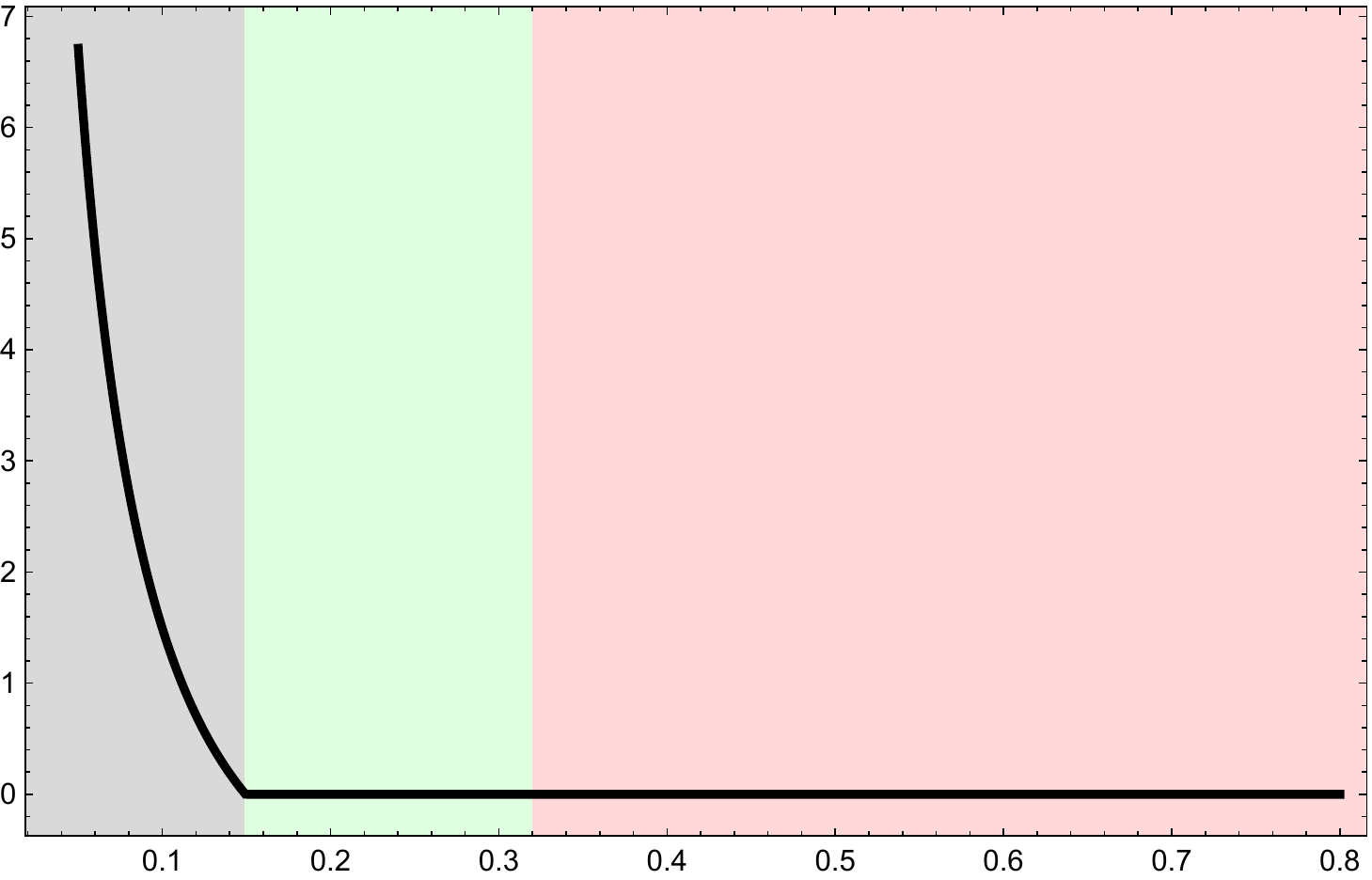}
    \caption{Real part of $\NCO(t)$.}
    \end{subfigure}
    \hspace{5pt}
   \begin{subfigure}{0.48\textwidth}
        \centering
    \includegraphics[width=\textwidth]{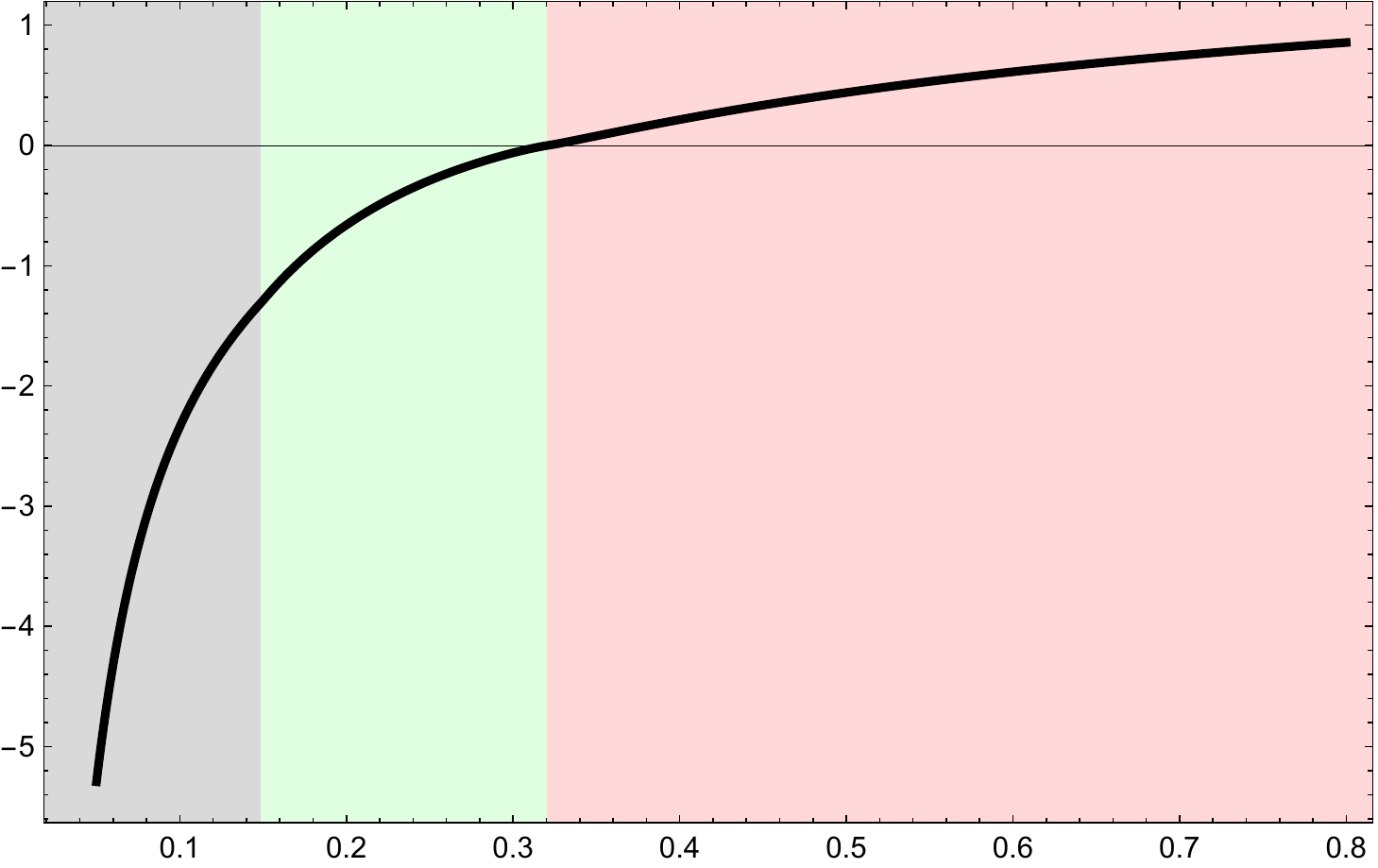}
    \caption{Imaginary part of $\NCO(t)$.}
    \end{subfigure}
    \caption{Evaluation of the order parameter $\NCO(t)$ along the path $C_{\text{diag}}$, displayed in figure~\ref{fig:SpecGeofig6}.}
    \label{fig:SpecGeofig3}
\end{figure}
%
\begin{figure}
    \centering
   \begin{subfigure}{0.48\textwidth}
        \centering
    \includegraphics[width=\textwidth]{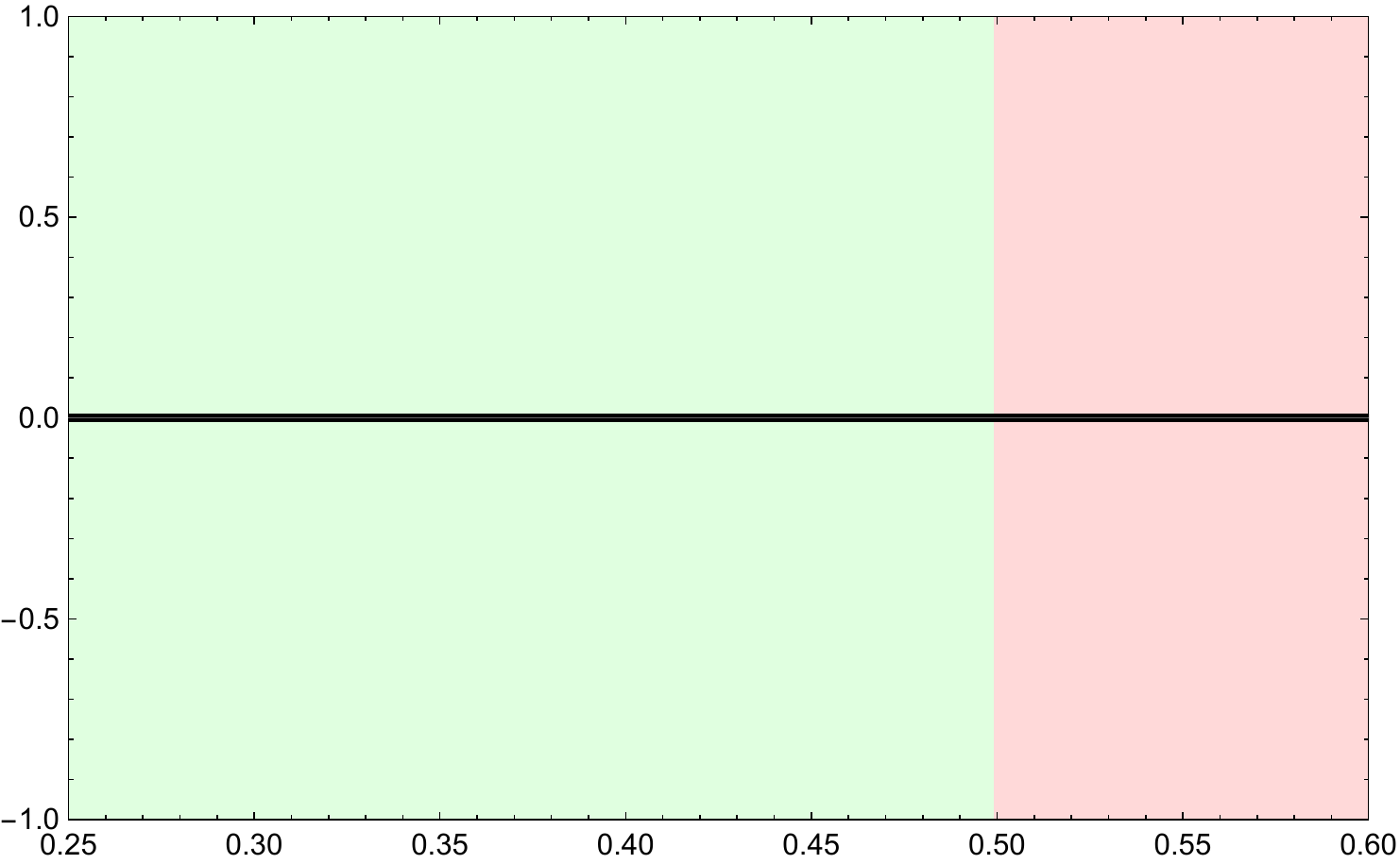}
    \caption{Real part of $\NCO(t)$.}
    \end{subfigure}
    \hspace{5pt}
   \begin{subfigure}{0.48\textwidth}
        \centering
    \includegraphics[width=\textwidth]{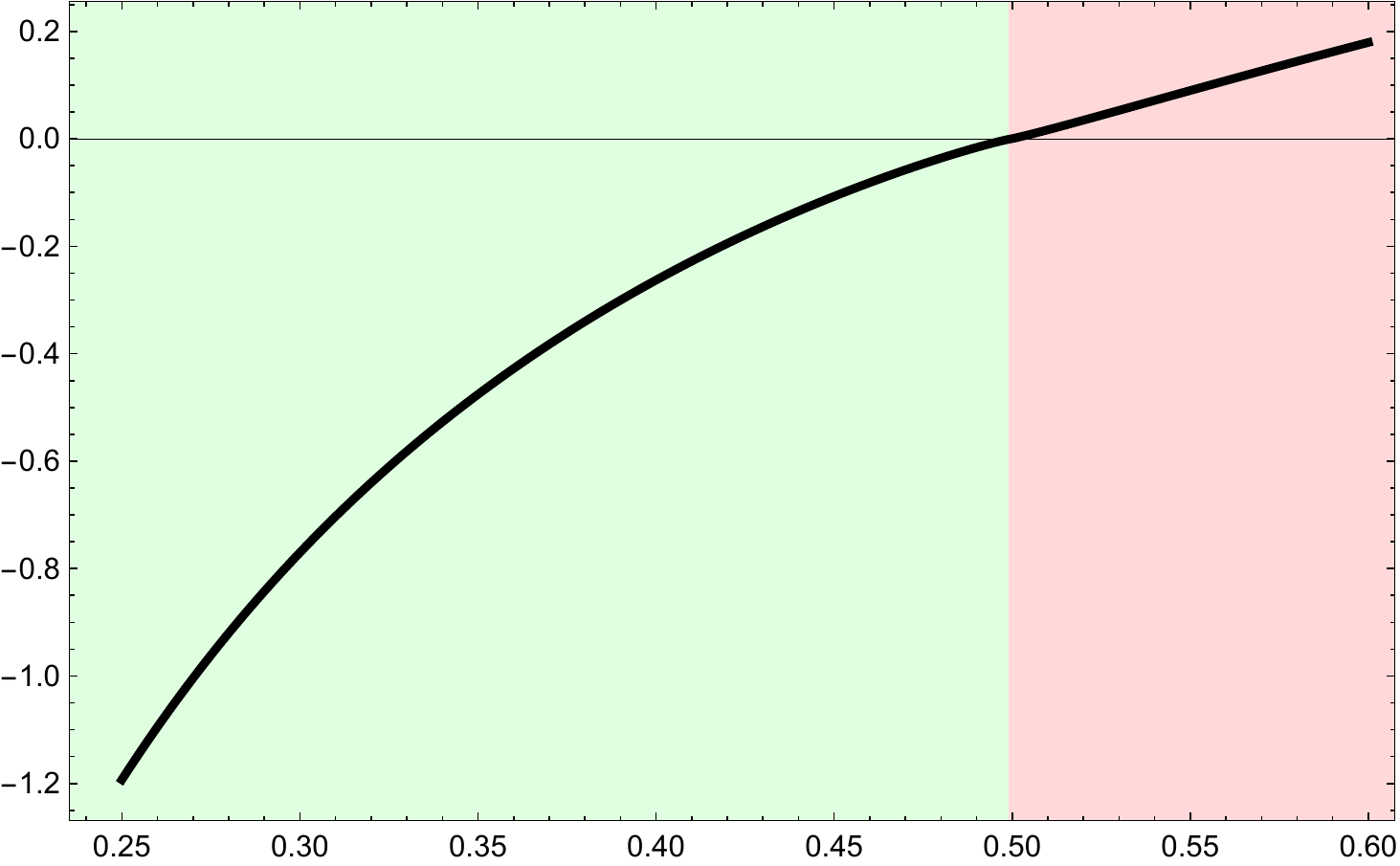}
    \caption{Imaginary part of $\NCO(t)$.}
    \end{subfigure}
    \caption{Evaluation of the order parameter $\NCO(t)$ along the path $C_{\text{vert}}$, displayed in figure~\ref{fig:SpecGeofig6}.}
    \label{fig:SpecGeofig4}
\end{figure}

\begin{figure}
    \centering
   \begin{subfigure}{0.48\textwidth}
        \centering
    \includegraphics[width=\textwidth]{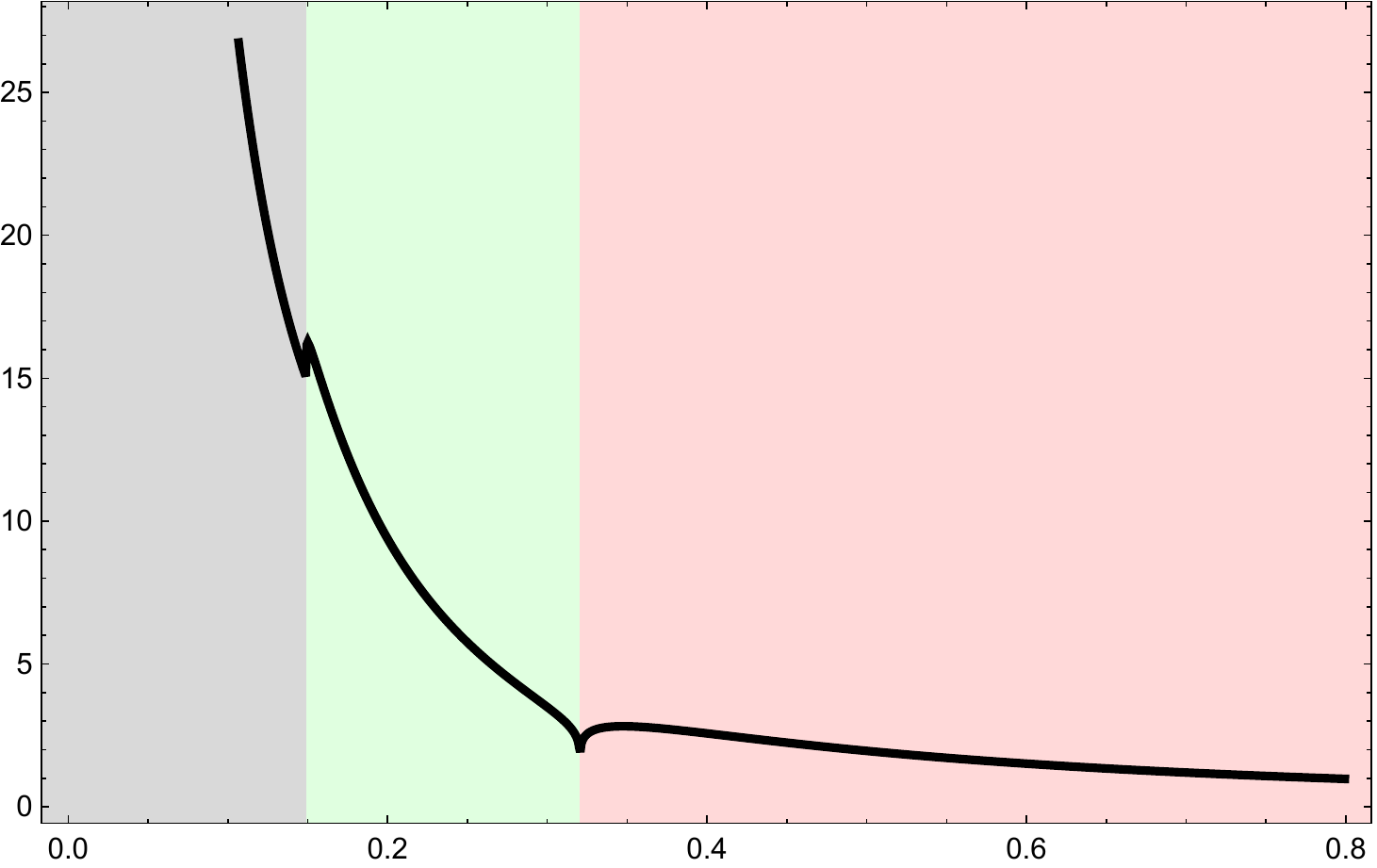}
    \caption{Along the path $C_{\text{diag}}$.}
    \end{subfigure}
    \hspace{5pt}
   \begin{subfigure}{0.48\textwidth}
        \centering
    \includegraphics[width=\textwidth]{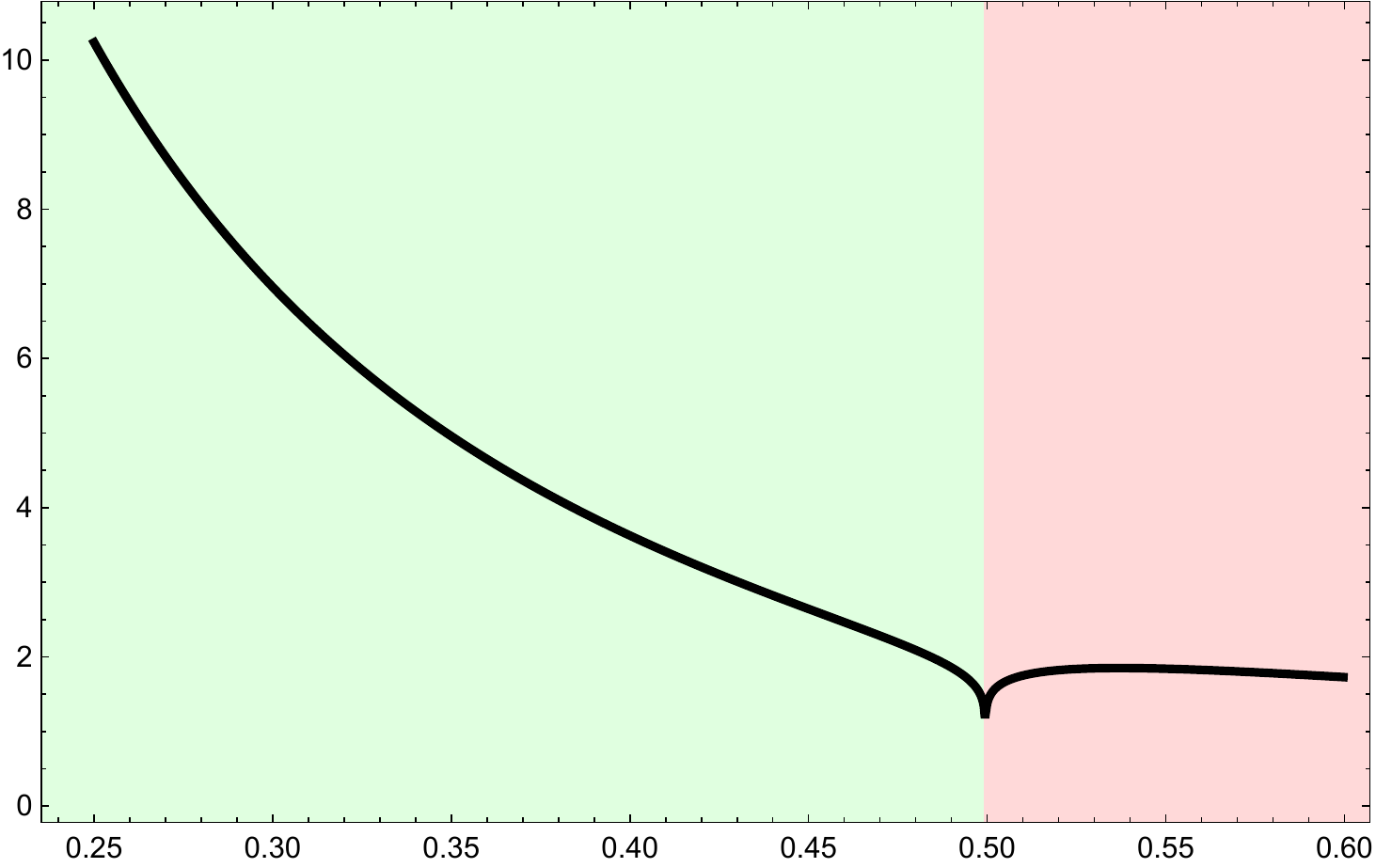}
    \caption{Along the path $C_{\text{vert}}$.}
    \end{subfigure}
    \caption{Evaluation of the imaginary part of the first-order derivative of the order parameter $\NCO(t)$ in \eqref{eq:cubicaction} along the paths $C_{\text{diag}}$ and $C_{\text{vert}}$, displayed in figure~\ref{fig:SpecGeofig6}.}
    \label{fig:SpecGeofig7}
\end{figure}

Let us pick a few paths upon the cubic phase diagram of figure~\ref{fig:CMM Phase Diagram}, as illustrated in figure~\ref{fig:SpecGeofig6}. Then, in figures~\ref{fig:SpecGeofig5}, \ref{fig:SpecGeofig3}, and~\ref{fig:SpecGeofig4}, we plot the cubic order-parameter as we probe the complex 't~Hooft plane along these aforementioned paths. As shown in figures~\ref{fig:SpecGeofig5} and~\ref{fig:SpecGeofig3}, $\NCO (t)$ reveals a clear break in regularity whenever leaving the one-cut region. At this transition, its real part becomes locally constant and equal to zero (reflecting the Boutroux property of the underlying spectral curves) whilst its first-order derivative discontinuously jumps. This is particularly noticeable when crossing the \PI~double-scaling critical point as seen in figure~\ref{fig:SpecGeofig5}, as this discontinuous jump then occurs for both real and imaginary parts of the order parameter. Crossing from the two-cut to the trivalent phase displays milder behavior, as shown in figures~\ref{fig:SpecGeofig3} and~\ref{fig:SpecGeofig4}. Both real and imaginary parts of the order parameter clearly feature continuous first-order derivatives at the phase transition line, indicating that we are in the presence of a higher-order phase transition. In order to test this, we plot the (imaginary part of the) \textit{first-order derivative} of $\NCO(t)$ along the paths $C_{\text{diag}}$ and $C_{\text{vert}}$ in figure~\ref{fig:SpecGeofig7}. These plots make clear how the order parameter indeed features a discontinuous \textit{second-order} derivative along the green-to-pink phase transition.

\paragraph{The Quartic Order-Parameter:}

\begin{figure}
    \centering
    \includegraphics[width=0.65\linewidth]{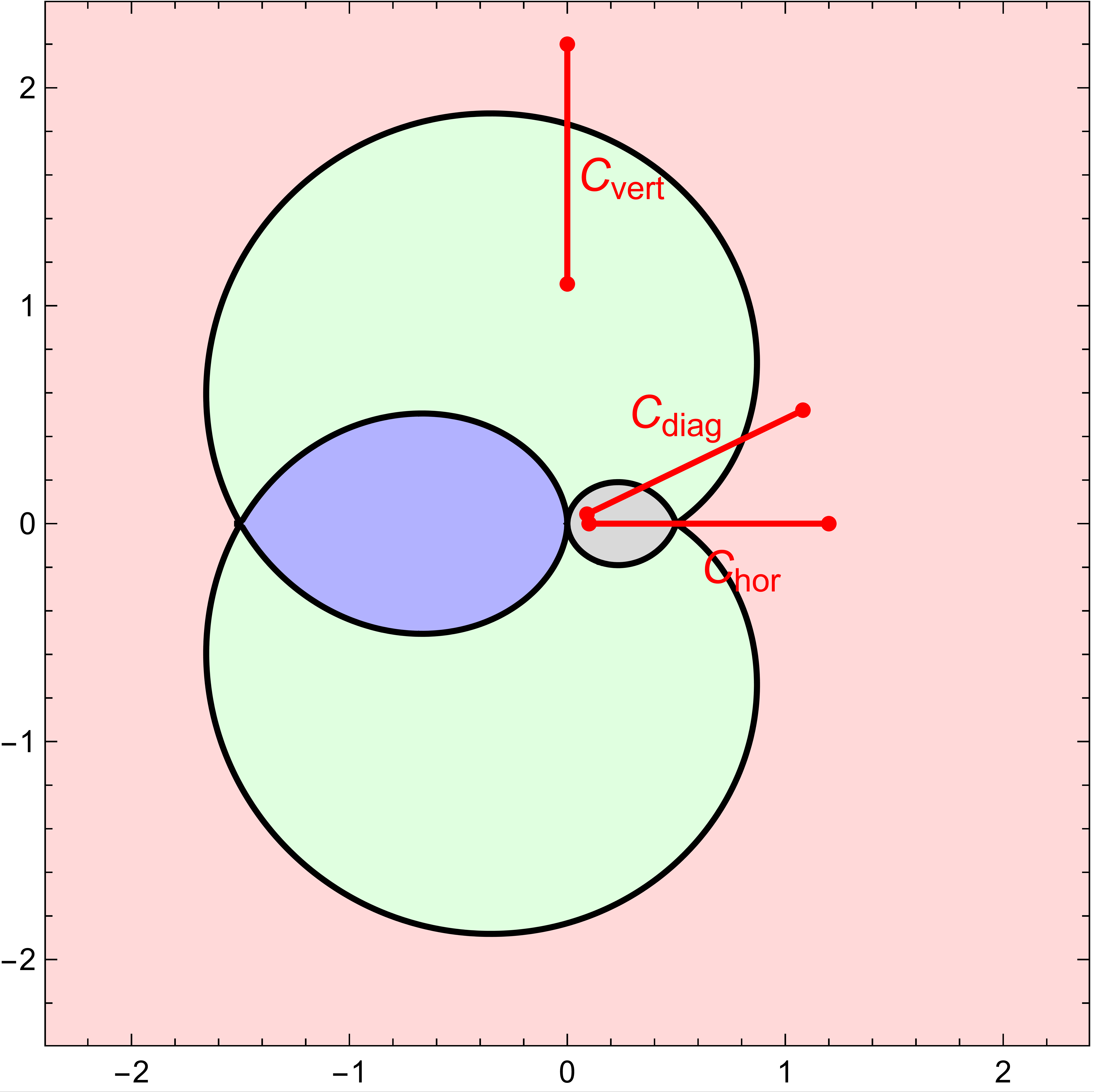}
    \caption{Depiction of paths we chose to explore the behavior of the order parameter $\NCO(t)$ in \eqref{eq:quarticaction} across the phase diagram of the quartic model from figure~\ref{fig:QMM Phase Diagram}. The path $C_{\text{hor}}$ is upon the real axis and precisely crosses the critical point at which the matrix-model double-scales to \PI. The path $C_{\text{diag}}$ crosses both phase transition lines. The path $C_{\text{vert}}$ is upon the imaginary axis and only crosses the phase transition line separating the two-cut and trivalent phases.}
    \label{fig:SpecGeofig13}
\end{figure}

As in the previous case, start with the order parameter \eqref{eq:SpecGeo9} in the one-cut case. The resulting expression is the instanton action for the quartic matrix model transseries, given by
\be
\label{eq:quarticaction}
\NCO(t) = \frac{\sqrt{3}}{2} \sqrt{\sqrt{1-2t}+2}\, \sqrt[4]{1-2t} - 2t \log \left( \sqrt{3}\,  \sqrt[4]{1-2t} + \sqrt{\sqrt{1-2t}+2} \right) + t \log \left( 2 - 2 \sqrt{1-2t} \right).
\ee
\noindent
For analogous three-cut and trivalent expressions, we refer the reader to sub-appendix~\ref{subappendix:quarticmatrixmodel}.

\begin{figure}
    \centering
   \begin{subfigure}{0.48\textwidth}
        \centering
    \includegraphics[width=\textwidth]{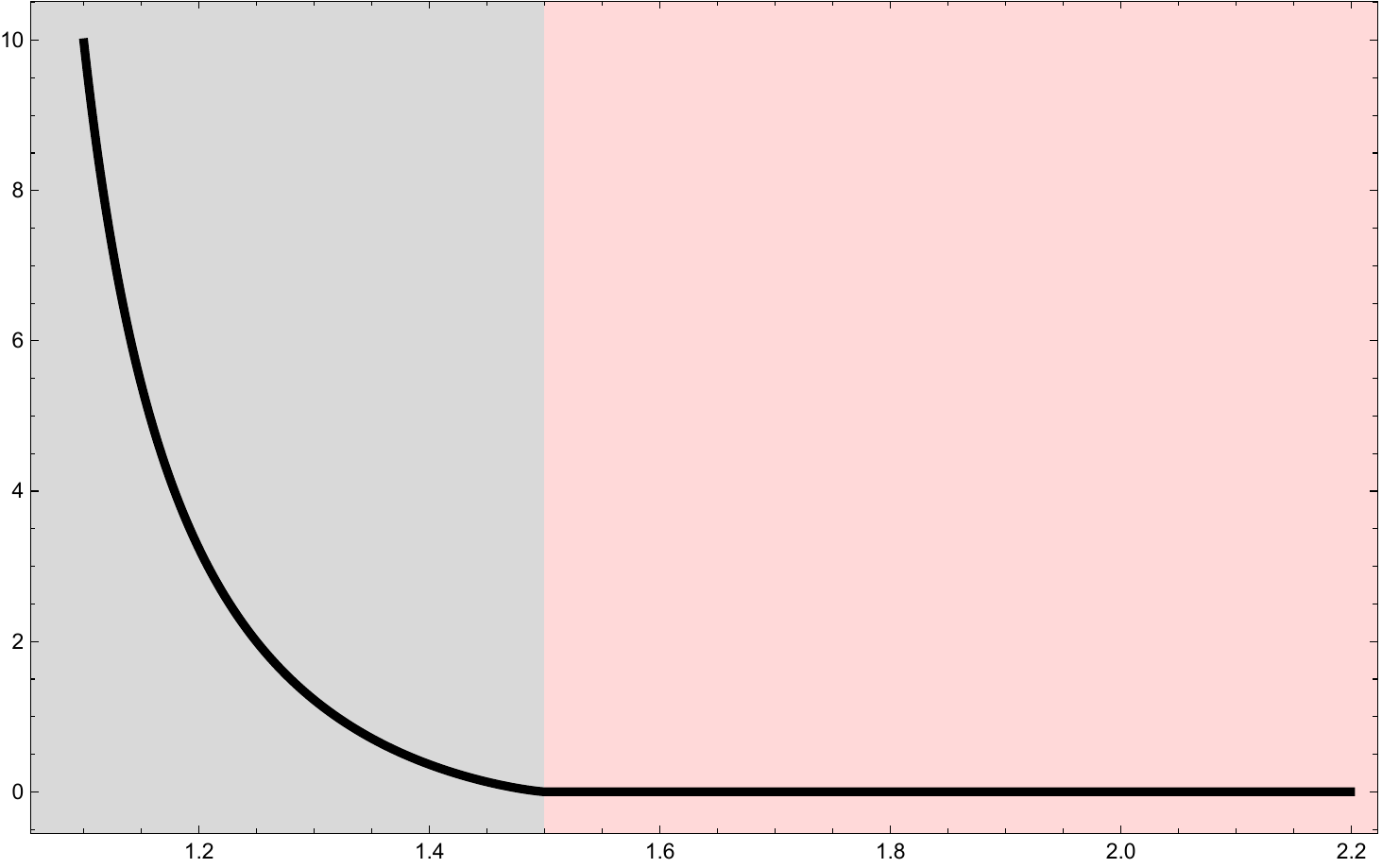}
    \caption{Real part of $\NCO(t)$.}
    \end{subfigure}
    \hspace{5pt}
    \begin{subfigure}{0.48\textwidth}
        \centering
    \includegraphics[width=\textwidth]{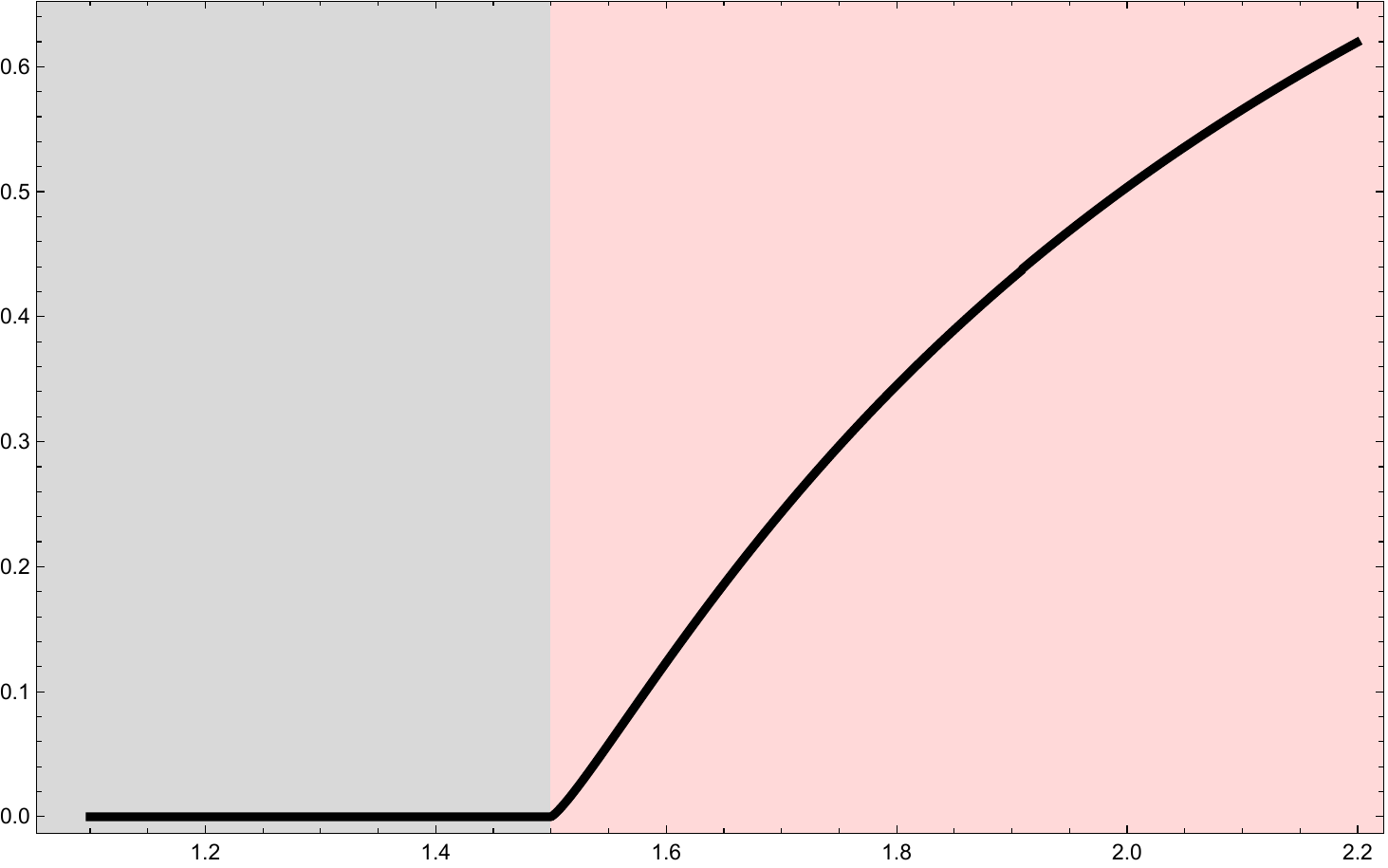}
    \caption{Imaginary part of $\NCO(t)$.}
    \end{subfigure}
    \caption{Evaluation of the order parameter $\NCO(t)$ along the path $C_{\text{hor}}$, displayed in figure~\ref{fig:SpecGeofig13}.}
    \label{fig:SpecGeofig12}
\end{figure}
%
\begin{figure}
    \centering
   \begin{subfigure}{0.48\textwidth}
        \centering
    \includegraphics[width=\textwidth]{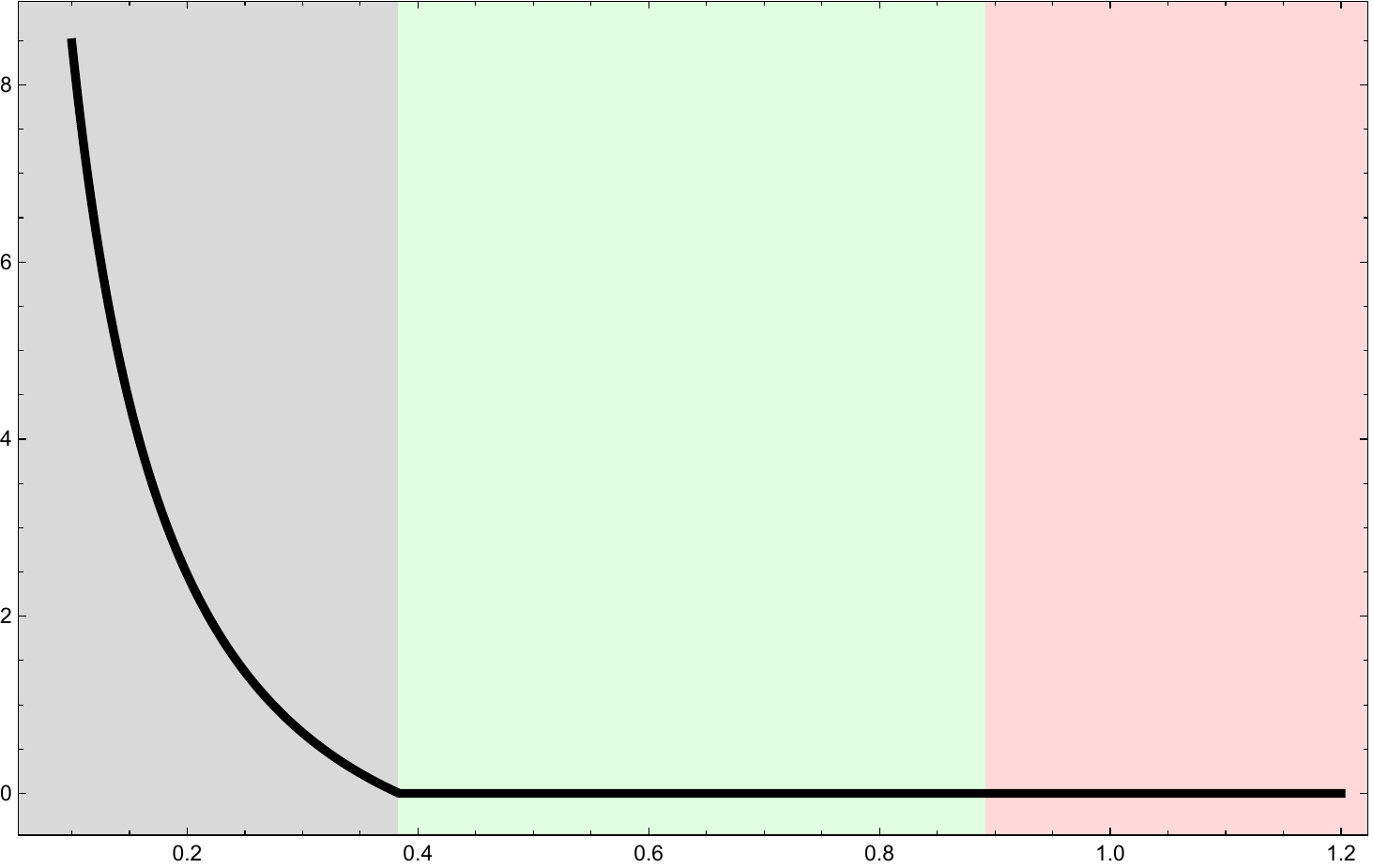}
    \caption{Real part of $\NCO(t)$.}
    \end{subfigure}
    \hspace{5pt}
    \begin{subfigure}{0.48\textwidth}
        \centering
    \includegraphics[width=\textwidth]{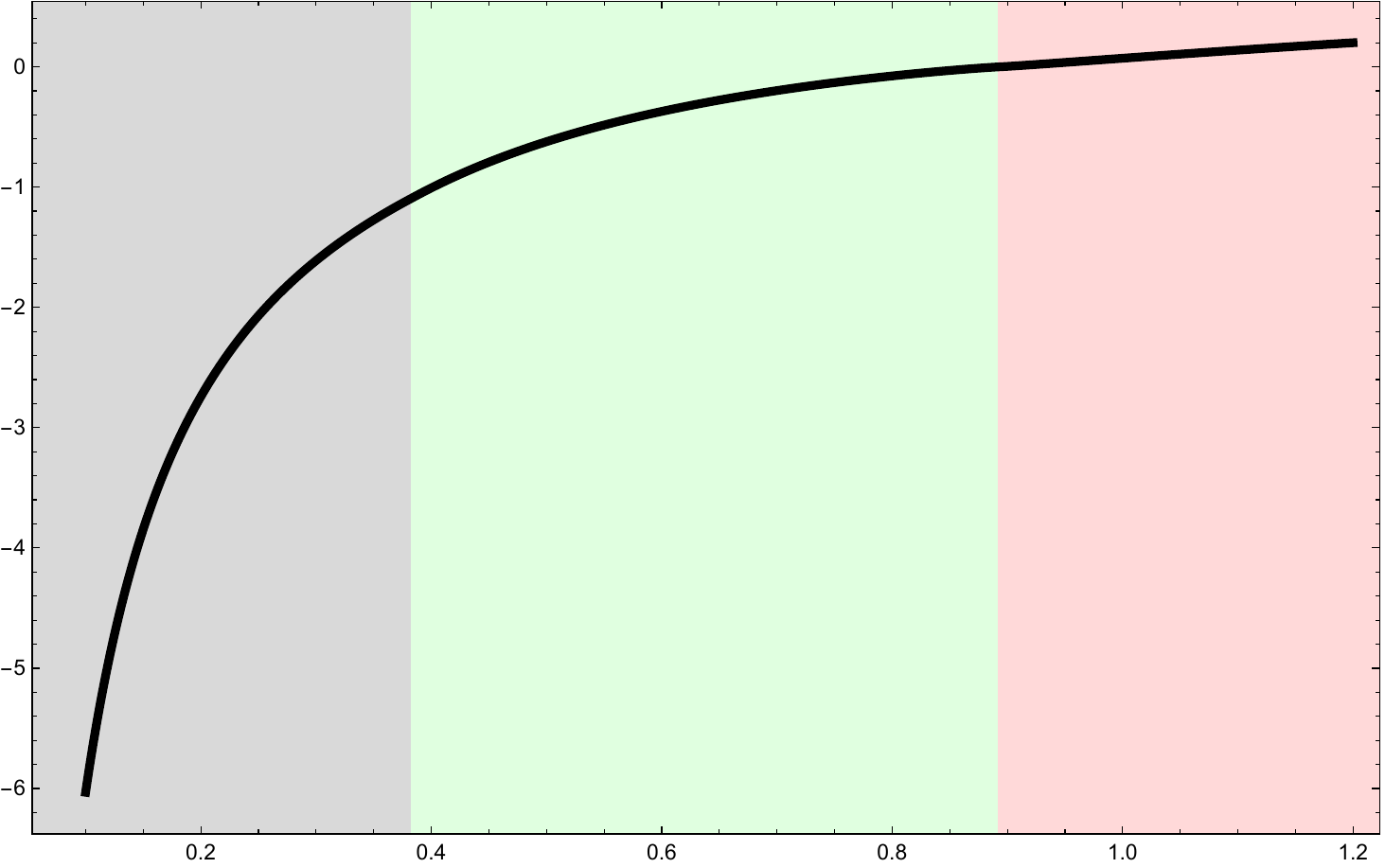}
    \caption{Imaginary part of $\NCO(t)$.}
    \end{subfigure}
    \caption{Evaluation of the order parameter $\NCO(t)$ along the path $C_{\text{diag}}$, displayed in figure~\ref{fig:SpecGeofig13}.}
    \label{fig:SpecGeofig10}
\end{figure}
%
\begin{figure}
    \centering
   \begin{subfigure}{0.48\textwidth}
        \centering
    \includegraphics[width=\textwidth]{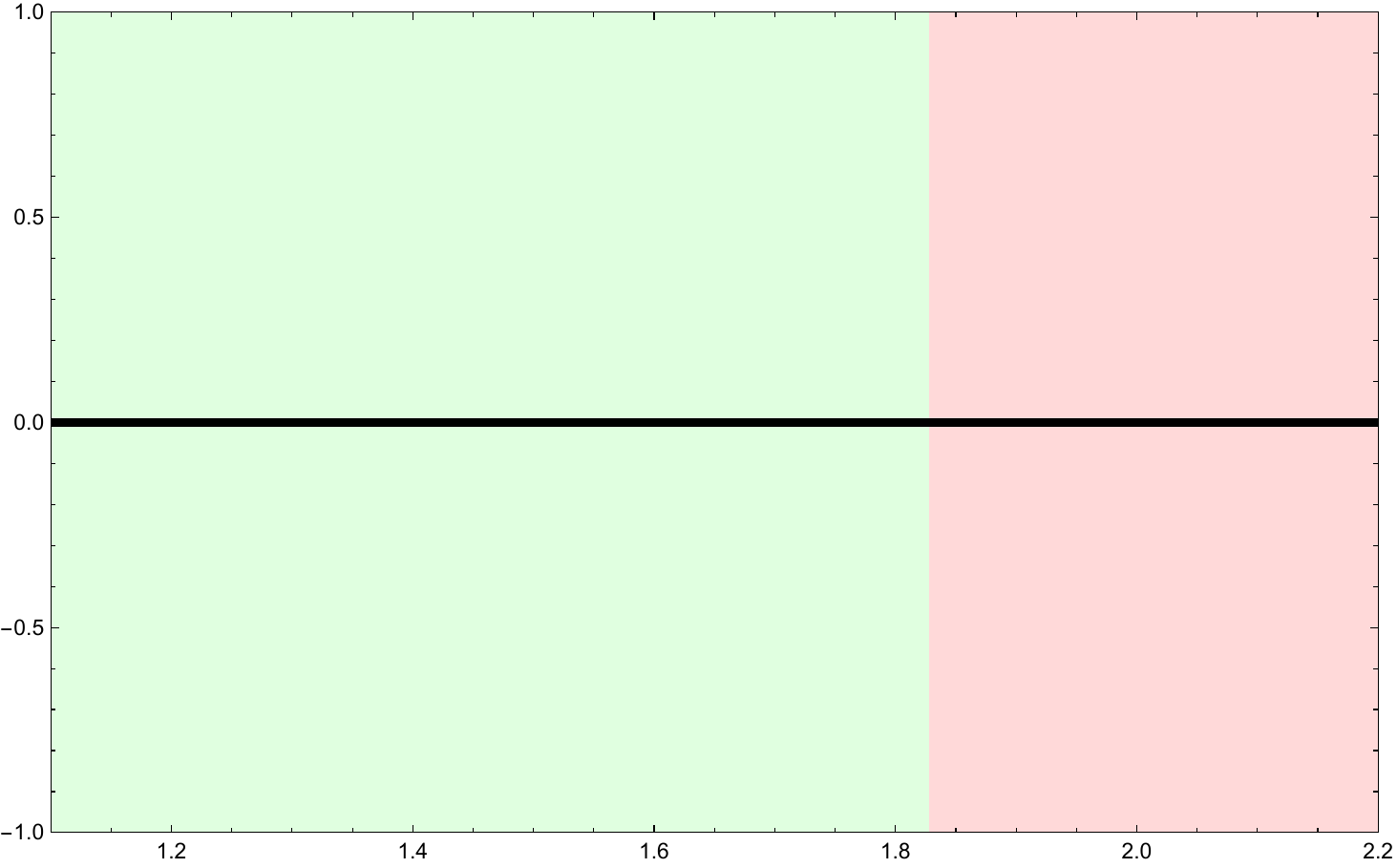}
    \caption{Real part of $\NCO(t)$.}
    \end{subfigure}
    \hspace{5pt}
    \begin{subfigure}{0.48\textwidth}
        \centering
    \includegraphics[width=\textwidth]{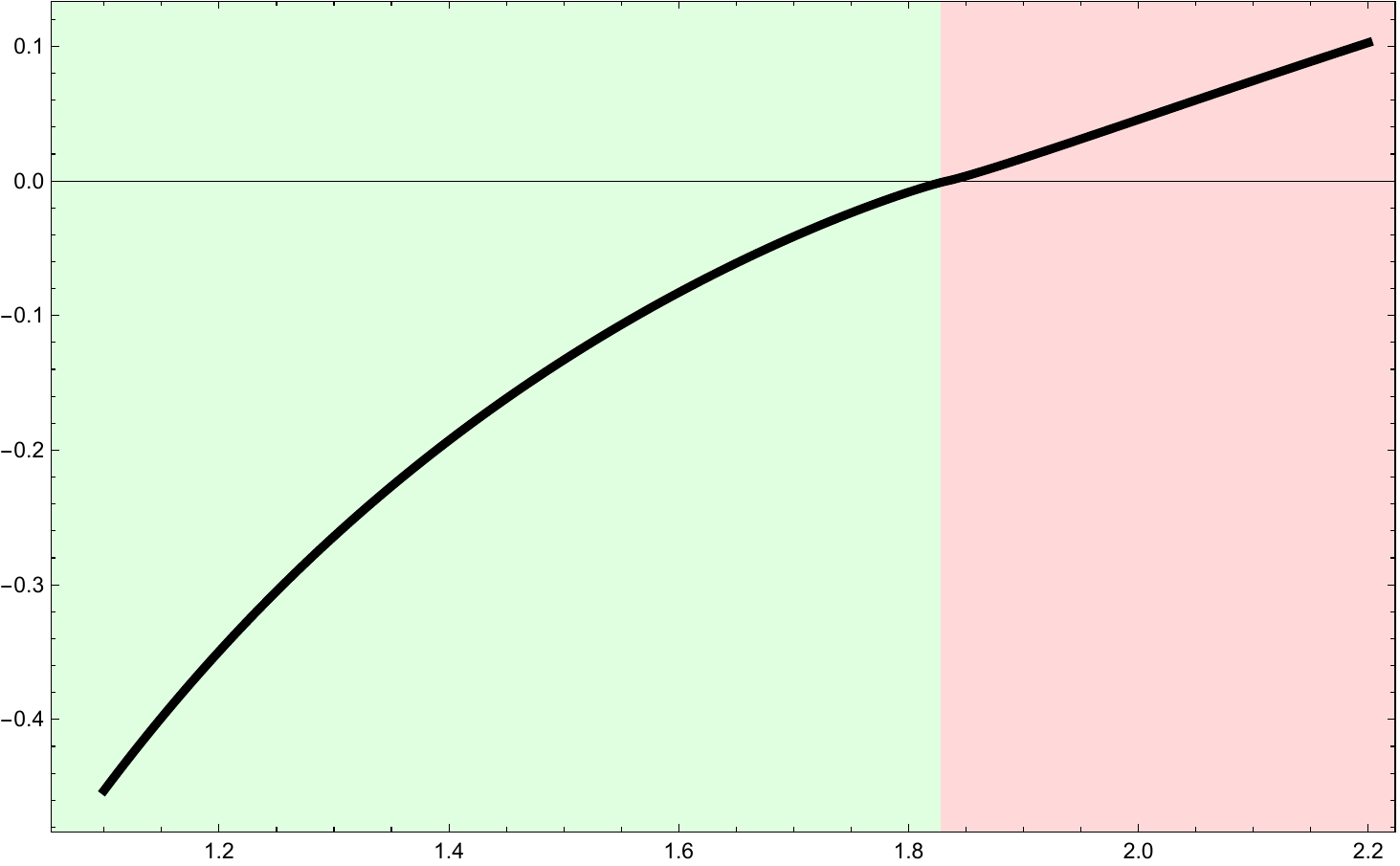}
    \caption{Imaginary part of $\NCO(t)$.}
    \end{subfigure}
    \caption{Evaluation of the order parameter $\NCO(t)$ along the path $C_{\text{vert}}$, displayed in figure~\ref{fig:SpecGeofig13}.}
    \label{fig:SpecGeofig11}
\end{figure}

\begin{figure}
    \centering
   \begin{subfigure}{0.48\textwidth}
        \centering
    \includegraphics[width=\textwidth]{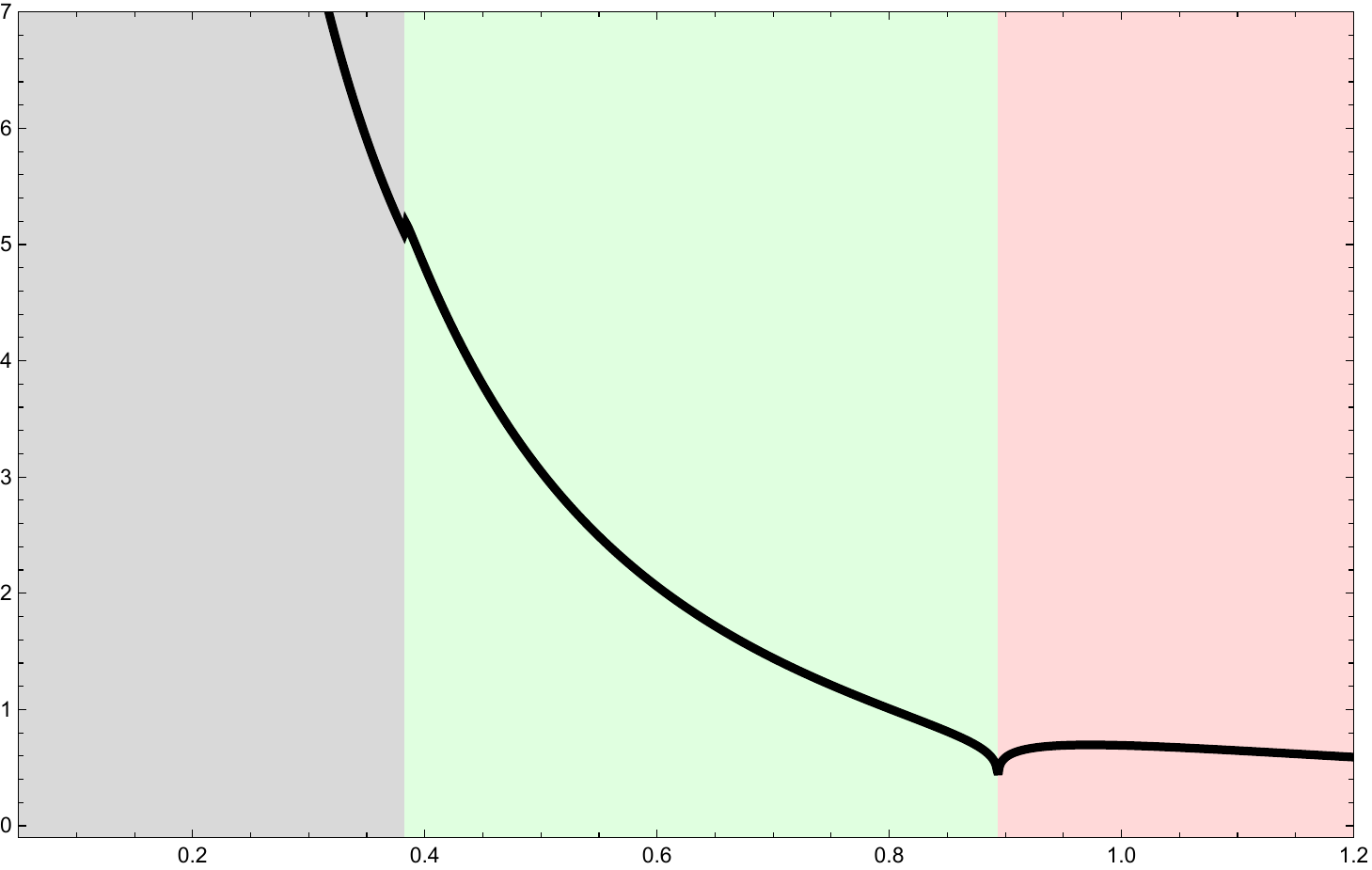}
    \caption{Along the path $C_{\text{diag}}$.}
    \end{subfigure}
    \hspace{5pt}
    \begin{subfigure}{0.48\textwidth}
        \centering
    \includegraphics[width=\textwidth]{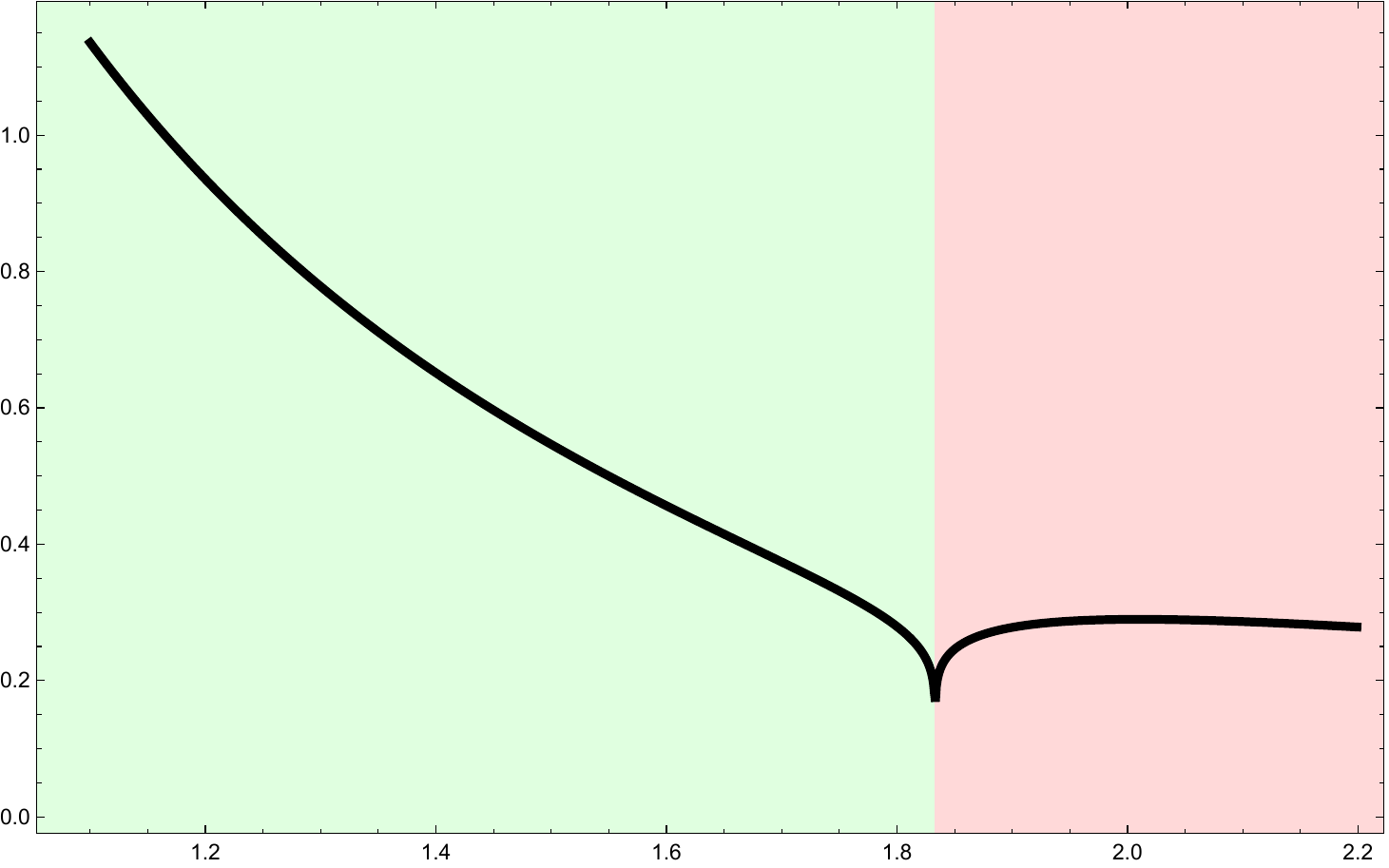}
    \caption{Along the path $C_{\text{vert}}$.}
    \end{subfigure}
    \caption{Evaluation of the imaginary part of the first-order derivative of the order parameter $\NCO(t)$ in \eqref{eq:quarticaction} along the paths $C_{\text{diag}}$ and $C_{\text{vert}}$, displayed in figure~\ref{fig:SpecGeofig13}.}
    \label{fig:SpecGeofig14}
\end{figure}

Pick a few paths upon the quartic phase diagram of figure~\ref{fig:QMM Phase Diagram}, as illustrated in figure~\ref{fig:SpecGeofig13} and in the same guise as for the cubic. In figures~\ref{fig:SpecGeofig12}, \ref{fig:SpecGeofig10}, and~\ref{fig:SpecGeofig11}, we plot the quartic order-parameter along these paths. The discussion is very similar to the cubic case. As shown in figures~\ref{fig:SpecGeofig12} and~\ref{fig:SpecGeofig10}, $\NCO (t)$ reveals a break in regularity whenever leaving the one-cut region: its real part becomes zero whilst its first-order derivative discontinuously jumps. Again, this is particularly noticeable when crossing the \PI~double-scaling critical point as seen in figure~\ref{fig:SpecGeofig12}. Crossing from the three-cut to the trivalent phase displays milder behavior, as shown in figures~\ref{fig:SpecGeofig10} and~\ref{fig:SpecGeofig11}, indicating that we are in the presence of a higher-order phase transition. To test this, we plot the first-order derivative of $\NCO(t)$ along the paths $C_{\text{diag}}$ and $C_{\text{vert}}$ in figure~\ref{fig:SpecGeofig14}. These plots illustrate how the order parameter  features a discontinuous second-order derivative along the green-to-pink phase transition.

\subsection{On the Large $N$ Asymptotics of the Partition Function}
\label{subsec:Z-phases}

At this stage, the large $N$ \textit{phases} of our random matrix models are well explored and understood, with both numerical and analytical input, at- or off-criticality, and all the way from weak to strong 't~Hooft coupling---in fact for any $t \in \BC$. It should also be clear how most of their different features will feed into the resurgent transseries we wish to explicit and fully construct, in particular the warning on the subtleties associated to analytic continuation. One can hence turn to our main goal in this work, addressing the complete, exact, resurgent, large $N$ \textit{asymptotics} of the partition function \eqref{eq:partitionfunctionhermitianmatrix}. But in order to do so, we need to start somewhere. In this subsection we will briefly review the ``pre-resurgence setting'' (closely following the very clean discussion in \cite{mpp09}), in order to set the stage for our main construction to be presented in the next section~\ref{sec:resurgent-Z-transseries}.

It has been known since at least \cite{d91} that the usual 't~Hooft limit does not always exist: the various ``regular'' (Stokes) large $N$ phases are separated by ``non-regular'' (anti-Stokes and trivalent) domains---which is very clear in figures~\ref{fig:CMM Phase Diagram} and~\ref{fig:QMM Phase Diagram}. As discussed earlier in subsection~\ref{subsec:OP-phases}, these ``non-regular'' domains are regions of oscillations of the orthogonal-polynomial recursion-coefficients; in subsection~\ref{subsec:DSL-phases}, these ``non-regular'' domains are regions of poles of the solutions to the (double-scaled) string equations; and in the previous subsection~\ref{subsec:stokes-vs-phases}, these ``non-regular'' domains are regions of accumulation of Yang--Lee--Fisher zeroes---which means that more than just denoting phase boundaries as anti-Stokes lines, we can dub the whole green\footnote{Because Yang--Lee--Fisher zeroes also accumulate in the pink regions, we could also dub them anti-Stokes phases. It should be clear by now that the better name is just to keep calling them trivalent phases.} regions in figures~\ref{fig:CMM Phase Diagram} and~\ref{fig:QMM Phase Diagram} as \textit{anti-Stokes phases}. Similarly, as a regular large $N$ limit exists in the gray (and blue \cite{sv13}) regions of these figures, we can dub them as \textit{Stokes-phase} regions. But how exactly does the behavior of the large $N$ partition function change as we change phases?

Now, the setting of \cite{mpp09} addresses \textit{some} of the different large $N$ phases albeit\footnote{This is in contrast to what we wish to do, which is to address \textit{all} phases and in a \textit{unified} manner.} \textit{separately}. In particular, it addresses what they call a \textit{boundary point}, describing the asymptotics in the \textit{one}-cut ``gray'' region, depicted in figures~\ref{fig:SpecGeofig:eigcmm} and~\ref{fig:CMM Phase Diagram} for the cubic matrix model, and in figures~\ref{fig:SpecGeofig:eigqmm} and~\ref{fig:QMM Phase Diagram} for the quartic matrix model. Herein, the large $N$ asymptotics is just the conventional genus expansion \cite{th74} corrected by exponentially-suppressed instantons \cite{msw07, m08, msw08, ps09}. It also addresses what they call an \textit{interior point}, describing the asymptotics in the \textit{multi}-cut ``green'' regions of the same figures. Herein, the asymptotics no longer is a genus expansion around some fixed background, it is oscillatory \cite{bde00, e08, em08}. Let us briefly recall why. 

Go back to the multi-cut matrix integral \eqref{eq:partitionfunctioneigenvaluesmulticut} which we now wish to regard as a \textit{canonical} configuration, \textit{i.e.}, with a fixed $\left\{ N_i \right\}$ distribution, sitting inside the \textit{grand-canonical} partition function where we sum over all possible such distributions across the $s$ extrema \cite{bde00, msw08}
\be
\label{eq:BDE-sum}
\NCZ \left( \upzeta_1, \ldots, \upzeta_s; g_{\text{s}} \right) = \sum_{N_1 + \cdots + N_s = N} \upzeta_1^{N_1} \cdots \upzeta_s^{N_s}\, \mathcal{Z} \left( N_1, \ldots, N_s; g_{\text{s}} \right),
\ee
\noindent
and where the $\left\{ \upzeta_i \right\}$ are the associated fugacities. Due to the summing over all possible configurations this is sometimes denoted a ``background independent'' partition function \cite{e08, em08}, although in our large $N$ resurgent context background independence will be discussed in \cite{ss26}. In order to write the above explicitly, one first needs to select a reference background. Focusing on the simplest $s=2$ case there are essentially two types of reference backgrounds \cite{mpp09}: the \textit{boundary} backgrounds $\left\{ N_i^\star \right\} = (N,0)$ or $(0,N)$; otherwise, the \textit{interior} backgrounds.

Discarding resonance \cite{gikm10, asv11, sv13, as13, gs21} (equivalently, discarding anti-eigenvalues \cite{mss22} or discarding negative-tension branes \cite{sst23}), the asymptotics around the $(N,0)$ \textit{boundary} background is essentially the one in \cite{msw08}, which reads
\bea
\label{eq:partition-function-Z-msw08}
\NCZ ( \upzeta ) &=& \CZ_{\text{pert}} (N,0)\, \Bigg( 1 + \sum_{\ell \ge 1} \CZ_{\text{G}} (\ell)\, \upzeta^\ell\, \widehat{q}^{\frac{\ell^2}{2}}\, \exp \left( - \frac{\ell A (t)}{g_{\text{s}}} \right) \times \\
&&
\hspace{150pt}
\times \left\{ 1 - g_{\text{s}} \left( \ell\, \partial_s \widehat{\CF}_1 (t) + \frac{\ell^3}{6}\, \partial_s^3 \widehat{\CF}_0 (t) \right) + \CO (g_{\text{s}}^2) \right\} \Bigg). \nonumber
\eea
\noindent
We have normalized the fugacities $\upzeta_1=1$, $\upzeta_2=\upzeta$; introduced the shorthand $\widehat{q} = \exp \left( \partial_s^{2}\widehat{\CF}_0 \right)$ where $s = \frac{1}{2} \left( t_1 - t_2 \right)$ and derivatives are evaluated at $t_1=t$, $t_2=0$; and all genus-$g$ free energies as in \eqref{eq:genusgfreeenergies} have required regularization so as to remove their conifold singularity upon full pinching of the second cut (\textit{i.e.}, upon reaching the one-cut ``boundary'' background starting from two-cuts). This regularization is denoted by the hat-symbol and amounts to \cite{msw08}
\be
\label{eq:gaussianregularizationF}
\widehat{\CF}_{g} ( t_1,t_2 ) = \CF_{g} ( t_1,t_2 ) - \CF_{g}^{\text{G}} ( t_2 ).
\ee
\noindent
As explained in \cite{msw08}, this singular component, removed in the large $N$ expansion, must still be counterbalanced by the addition of its exact contribution to the partition function, in the form of the exact Gaussian\footnote{We recall that the exact Gaussian partition function is given by
\be
\label{eq:exactGaussianZell}
\CZ_{\text{G}} (\ell) = \frac{g_{\text{s}}^{\ell^2/2}}{( 2\pi )^{\ell/2}}\, G_2 (\ell+1),
\ee
\noindent
where $G_2$ is the Barnes function, \textit{e.g.}, \cite{olbc10}.} partition function $\CZ_{\text{G}} (\ell)$. The above ``boundary'' asymptotic expansion, corresponding to a string-theoretic genus expansion augmented by exponentially-suppressed corrections, is valid as long as
\be
\Re \frac{A(t)}{g_{\text{s}}} > 0,
\ee
\noindent
which basically entails we have not yet reached the anti-Stokes phase boundary as in \eqref{eq:anti-stokes-in-the-matrix-model}. As for the asymptotics around the $(N_1^\star, N_2^\star)$ \textit{interior} background, it is instead the one in \cite{bde00, e08, em08}, which now reads
\be
\label{eq:EM-2-cut-Z}
\NCZ ( \upzeta ) = \CZ_{\text{pert}} (N_1^\star,N_2^\star) \left\{ \Uptheta_{\mathsf{0}\upzeta} + g_{\text{s}} \left( \partial_{\boldsymbol{z}} \Uptheta_{\mathsf{0}\upzeta}\, \partial_s \CF_1 + \frac{1}{6}\, \partial_{\boldsymbol{z}}^3 \Uptheta_{\mathsf{0}\upzeta}\, \partial_s^3 \CF_0 \right) + \CO (g_{\text{s}}^2) \right\}.
\ee
\noindent
We have now normalized the fugacities $\upzeta_1=1$, $\upzeta_2=\rme^{2\pi\rmi \upzeta}$ and introduced the theta-function from \cite{e08} as a Riemann theta-function with characteristics (see \eqref{eq:riemanntheta}-\eqref{eq:riemannthetacharacteristics} in appendix~\ref{app:elliptic-theta-modular}) given by
\be
\label{eq:theta-for-EM-2-cut-Z}
\Uptheta_{\mathsf{0}\upzeta} \left( \boldsymbol{z} | \boldsymbol{\tau} \right) = \vartheta { \begin{bmatrix} 0 \\ \upzeta \end{bmatrix} } \left( \left. - \frac{A}{2\pi\rmi g_{\text{s}}} \right| \frac{\partial_s^{2} \CF_0}{2\pi\rmi} \right).
\ee
\noindent
Note that this is a different expansion from the earlier one; in particular there is no regularization procedure and, more importantly, nor is there\footnote{The free energy $\CF \simeq \frac{1}{g_{\text{s}}^2}\, \CF_0 + \CF_1 + \log \Uptheta_{\mathsf{0}\upzeta} + \cdots$ is now oscillatory due to the theta-function \cite{bde00}.} an 't~Hooft expansion. This ``interior'' asymptotic expansion is valid as long as
\be
\Re \frac{1}{t} \oint_{\gamma} \rmd z\, y (z) = 0,
\ee
\noindent
which is the Boutroux condition in \eqref{eq:equilibrium-condition-via-B-cycles}-\eqref{eq:SpecGeo2b}, \textit{i.e.}, which basically entails we are within the anti-Stokes region (and as computed within the two-cut scenario). This resulting random-matrix oscillatory phase has been addressed in \cite{bde00, e08, em08, bt11, bt16}.

The overall picture is now complete, with different phases leading to different asymptotics of the random-matrix partition-function. In the above discussion one generically starts in the two-cut configuration and then writes the partition function according to the required boundary/interior specialization of interest. This leads to somewhat independent results, and still misses the trivalent phase. In this work we are aiming at a unified picture of all phase asymptotics, starting off from the one-cut string-equation construct and crossing phases solely and fully as dictated by standard Stokes phenomena, without any need for performing separate analyses. On top, we will also be able to distinguish ``equilibrium'' and ``off-equilibrium'' solutions. For example, in the matrix model analyses of subsections~\ref{subsec:SG-phases} and~\ref{subsec:stokes-vs-phases} we were careful to discuss the Boutroux \textit{equilibrium} condition (and much less detailed on the canonical-partition-function topological-string standpoint \cite{dv02a}) with the resulting cubic and quartic matrix-model phase-diagrams describing \textit{equilibrium} configurations. In contrast, in the double-scaled analyses of subsection~\ref{subsec:DSL-phases}, we allowed for arbitrary boundary- or initial-value conditions for the string equations, in order to discuss the Boutroux classification of \textit{general} solutions---hence solutions at eigenvalue equilibrium \textit{or not}; depending on the initial choices of transseries parameters. This classification should in principle of course also be extended to the off-critical case, whereupon we set ourselves free to choose any transseries parameters, and the corresponding solutions will no longer describe eigenvalue equilibrium configurations but rather be subject to a matrix-model Boutroux classification as well (how to set up these generalized classifications, for both matrix models and string equations along the KdV hierarchy, will also be discussed in \cite{krsst26b, krst26a}). It goes without saying that our transseries---alongside its Stokes and anti-Stokes networks---are still fully and completely valid regardless of what solution we decide to work with (physical, a particular solution; or non-physical, arbitrary general solutions). Finally, we can still consider classes of solutions where all transseries parameters are ``free'' and we will also see how in these cases transseries data amounts to spectral geometry moduli. Let us make all this precise next.

\section{(Resonant) Resurgent-Transseries Description of Matrices and Strings}
\label{sec:resurgent-Z-transseries}

Let us now turn our attention to the resurgent transseries point-of-view, in order to present our main conjecture. On the literature up to today, resurgent transseries for diverse hermitian matrix models and minimal and topological strings have been built on a case-by-case very-long list of examples, \textit{e.g.}, \cite{m08, msw08, ps09, gikm10, kmr10, asv11, sv13, as13, csv15, d16, i19, gs21, bssv22, mss22, jr26}. On trying to abstract general structures from these many examples, perhaps the most significant has been that of resonance \cite{gikm10, asv11, sv13, as13, abs18, gs21, bssv22, mss22, sst23}, which seems to be ubiquitous across string theoretic examples \cite{sst23}. A separate clue arises from (spectral geometry) multi-cut computations pinpointing the presence of theta-function behavior within anti-Stokes regions of the phase diagrams, as just discussed in subsection~\ref{subsec:Z-phases} \cite{bde00, e08, em08, bt16}. Are we doomed to having to deal with this plethora of examples, or might it be instead possible to write down some compact-looking, closed-form, general solution? As we shall see in this section, and somewhat surprisingly, this seemingly intricate problem turns out to have a general solution which may be written in a remarkably simple form, and encompassing (local\footnote{As we have already repeatedly stressed, what we achieve in this work is a \textit{local} construction, \textit{i.e.}, the transseries solution. This turns out to be the easier part of the whole problem. As will be seen in the follow-up paper \cite{krsst26b}, going \textit{global} still requires computing full Stokes data, which is the harder part of the problem.}) solutions to all hermitian one-matrix models alongside their topological-string duals, and including their double-scaling limits towards multicritical models, minimal strings, and generic topological gravities anywhere along the full KdV hierarchy. Further aspects of the construction will appear in \cite{krsst26b, ss26, krst26a, krst26b}.

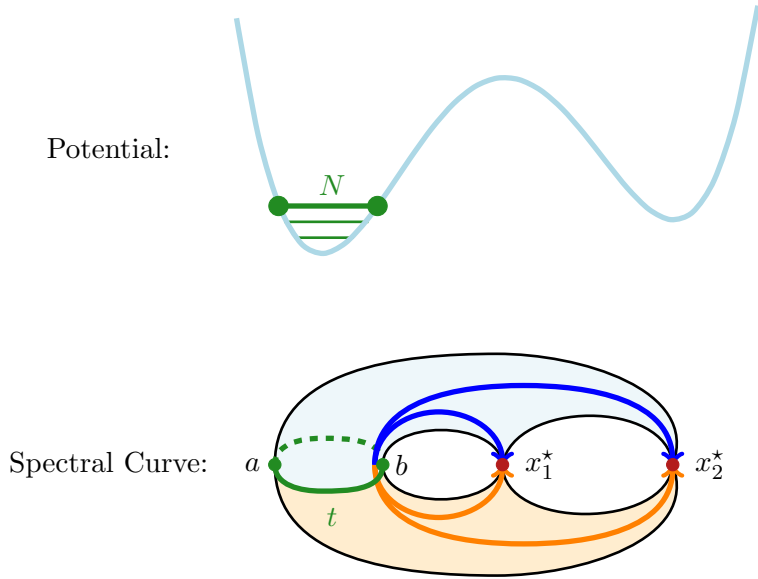
\begin{figure}
\centering
\begin{tikzpicture}
	\begin{scope}[scale=0.7, shift={({0},{2})}]
	\node at (-4,2) {Potential:};
	\draw[color=ForestGreen, line width=2pt] (-0.88,0.9) -- (1.15,0.9);
	\draw[color=ForestGreen, line width=1pt] (-0.7,0.6) -- (0.88,0.6);
	\draw[color=ForestGreen, line width=1pt] (-0.51,0.3) -- (0.6,0.3);
	 \draw[scale=2, domain=-0.8:4.1, smooth, variable=\x, LightBlue, line width=2pt] plot ({\x}, {2*1.14*\x*\x - 2*0.671*\x*\x*\x + 0.2*\x*\x*\x*\x});
	\draw[ForestGreen, fill=ForestGreen] (-0.81,0.9) circle (1.1ex);
	 \draw[ForestGreen, fill=ForestGreen] (1.05,0.9) circle (1.1ex);
	 \node[ForestGreen] at (0.2,1.3) {$N$};
	\end{scope}
	\begin{scope}[scale=0.7,  shift={({0},{-2})}]
	\node at (-4,0) {Spectral Curve:};
	\draw[fill=LightBlue,fill opacity=0.2, line width=1pt] (1.15,0)   to [out=90,in=95] (3.4,0)
	to [out=80,in=95] (6.6,0)
	to [out=70,in=0] (3.3,2.1)
    to [out=180,in=90] (-0.88,0)
    to [out=270, in=180] (0, -0.5)
    to [out=0, in=270] cycle;
    \draw[fill=darktangerine,fill opacity=0.2, line width=1pt] (-0.88,0)
    to [out=270,in=180] (3.3,-2.1)
    to [out=0,in=290] (6.6,0)
    to [out=265,in=270] (3.4,0)
    to [out=265,in=270] (1.15,0)
    to [out=270, in=0] (0, -0.5)
    to [out=180, in=270] cycle;
    \draw[color=orange, line width=2pt, ->] (1,0) to [out=270, in=180] (2.2, -1) to[out=0, in=270] (3.4, 0);
    \draw[color=orange, line width=2pt,->] (1,0) to [out=270, in=180] (3.8, -1.5) to[out=0, in=270] (6.6, 0);
    \draw[color=ForestGreen, line width=2pt] (-0.88,0) to [out=270, in=180] (0, -0.5)
    to [out=0, in=270] (1.15,0);
    \draw[dashed, color=ForestGreen, line width=2pt] (-0.88,0) to [out=90, in=180] (0, 0.5)
    to [out=0, in=90] (1.15,0);
    \draw[color=blue, line width=2pt, ->] (1,0) to [out=90, in=180] (2.2, 1) to[out=0, in=90] (3.4, 0);
    \draw[color=blue, line width=2pt,->] (1,0) to [out=90, in=180] (3.8, 1.5) to[out=0, in=90] (6.6, 0);
\draw[ForestGreen, fill=ForestGreen] (-0.88,0) circle (.7ex);
\draw[ForestGreen, fill=ForestGreen] (1.15,0) circle (.7ex);
\node[ForestGreen] at (0.2,-1) {$t$};
\draw[cornellred, fill=cornellred] (3.4,0) circle (.7ex);
\draw[cornellred, fill=cornellred] (6.6,0) circle (.7ex); 
\node at (-1.3, 0) {$a$};
\node at (4.1, 0) {$x_{1}^{\star}$}; 
\node at (7.3, 0) {$x_{2}^{\star}$};
\node at (1.5, 0) {$b$}; 
\end{scope}
\end{tikzpicture}
	\caption{Visualization of the one-cut eigenvalue configuration for a given potential (plotted in {\color{blue}blue} with $s=3$ saddles; compare with figure~\ref{fig:multicutPotential}). The top plot displays the potential with the accumulation of all $N$ eigenvalues in the energetically lowest saddle. The bottom plot depicts the corresponding spectral curve, with one cut of size $t$ and two pinched-cycles. Herein, the $A$ half-cycle is visualized in solid {\color{ForestGreen}green} whereas the $B$ half-cycles are shown in {\color{blue}blue} and {\color{orange}orange}.}
	\label{fig:OneCutPotential}
\end{figure}

Assume we are given some string equation as in section~\ref{sec:strong-coupling-phases}. This could literally be \textit{any} problem, off-criticality or double-scaled. One expects to find an explicit resurgent-transseries solution for the free energy of such problem. Because we start with a string equation, this implies we start life in the one-cut region; a setting visualized in figure~\ref{fig:OneCutPotential}. Our solution is an asymptotic series in small string coupling $g_{\text{s}}$ (with 't~Hooft parameter $t$). Crucially, these transseries are \textit{resonant} which implies there are \textit{two} natural ways to write them \cite{asv11, gs21, bssv22}---which we shall use heavily in the following---, namely what we call ``rectangular framing'' and ``diagonal framing''.

\begin{figure}
\centering
\def\xstep{1.9}
\def\ystep{2}
\definecolor{dgray}{rgb}{0.43, 0.5, 0.5}
\begin{tikzpicture}[grayframe/.style={
		rectangle,
		draw=gray,
		fill=white,
		text width=1.7em,
		align=center,
		rounded corners,
		minimum height=1em
	},
	box/.style={rectangle,draw=black,thick, minimum height=0.8cm, minimum width =1.2cm},
	]
	\foreach \x in {0,1,...,5}{
    \foreach \y in {0,1,...,5}
        \node[box, fill=white] at (2*1.4+1.2*\x,2*1+0.8*\y){$F^{({\color{blue}\x} | {\color{orange}2})}_{{\color{dgray}\y}}$};
}
	\foreach \x in {0,1,...,5}{
    \foreach \y in {0,1,...,5}
        \node[box, fill=white] at (1.4+1.2*\x,1+0.8*\y){$F^{({\color{blue}\x} | {\color{orange}1})}_{{\color{dgray}\y}}$};
}
	\foreach \x in {0,1,...,5}{
    \foreach \y in {0,1,...,5}
        \node[box, fill=white] at (1.2*\x,0.8*\y){$F^{({\color{blue}\x} | {\color{orange}0})}_{{\color{dgray}\y}}$};
}
\draw[blue, line width=2pt, ->] (-1,-1) -- (7, -1);
\draw[dgray, line width=2pt, ->] (-1,-1) -- (-1, 5);
\draw[orange, line width=2pt, ->] (7,-1) -- (7+2.5*1.4, -1+2.5*1);
\node[dgray] at (-1.5, 2) {$g$};
\node[blue] at (3, -1.5) {$n_1$};
\node[orange] at (9.5, 0) {$\check{n}_1$};
	\end{tikzpicture}
	\caption{The semi-positive grid representing a two-parameter resonant transseries in \textit{rectangular} framing. The entries in each box are the different coefficients in the asymptotic series \eqref{eq:GenericFreeEnergyTransseriesSector} (with no logarithms). The {\color{blue}blue} entry in each coefficient represents the instanton number whilst the {\color{orange}orange} entry is tied to the negative-instanton number \cite{mss22, sst23}. The {\color{dgray}gray} subscript relates to the perturbative order. Compare this figure to its diagonal-framing version in figures~\ref{fig:rectangulartodiagonal} and~\ref{fig:twoparamdiagtransseriesgrid}.}
	\label{fig:twopararectmtransseriesgrid}
\end{figure}
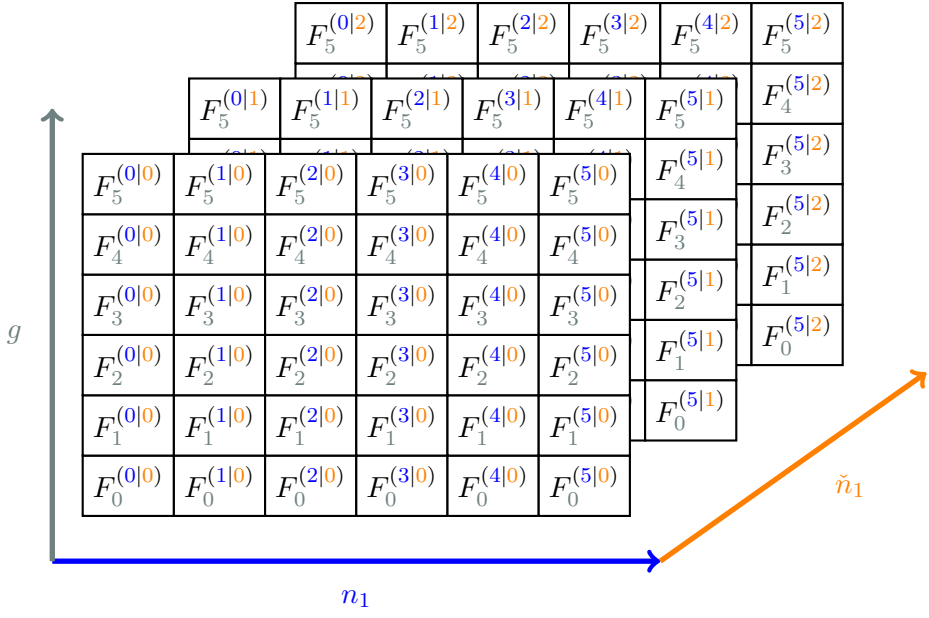

\begin{itemize}
\item \textbf{Rectangular Framing:} This is the standard formulation of resurgent transseries which has appeared at great length in the literature. In vector notation\footnote{Already briefly explained next to \eqref{eq:YangLeeSolution}; see \cite{abs18, gs21, bssv22} for further details within transseries and Stokes data.}, the free-energy resurgent-transeries, as a formal expansion in small string coupling $g_{\text{s}}$, has the form\footnote{Recall we are using the notational conventions in \cite{mss22}, where matrix-integral results use curly fonts, as in $\mathcal{F}$, and string-equation transseries results are regular, as in $F$.}
\be
\label{eq:generictransseriesmatrixmodels}
F \left( g_{\text{s}}; \boldsymbol{\sigma} \right) = \sum_{\boldsymbol{n}\in\mathbb{N}_{0}^{2\kappa}} \boldsymbol{\sigma}^{\boldsymbol{n}}\, \rme^{-\frac{\boldsymbol{n}\cdot\boldsymbol{A}}{g_{\text{s}}}}\, F^{(\boldsymbol{n})} (g_{\text{s}}),
\ee
\noindent
with $2\kappa$ transseries parameters $\boldsymbol{\sigma}$. The $2\kappa$ components of the instanton action $\boldsymbol{A}$ are given by linear combinations of (half-) cycles\footnote{See, \textit{e.g.}, \cite{emms22a, sst23} for discussions of half-cycles in matrix models versus full-cycles in minimal string theories.} on the one-cut spectral-curve and the number $\kappa$ of such combinations is example dependent: those can be $A$- and $B$-cycles; and in the example of a potential with three saddles the corresponding half-cycles are illustrated in figure~\ref{fig:OneCutPotential}. The aforementioned critical feature of resonance implies that such cycles always appears in \textit{symmetric} pairs \cite{gs21, mss22, sst23}. For the components of the instanton action this explicitly means we have $\boldsymbol{A} = \left( A_1,  -A_1, \cdots, A_{\kappa}, -A_{\kappa} \right)$. In this way, \eqref{eq:generictransseriesmatrixmodels} should be better rewritten as\footnote{To make contact with slightly distinct notations, \textit{e.g.}, \cite{asv11, sv13, gs21, bssv22, mss22, sst23}, transseries asymptotic sectors are denoted by $F^{(n_1|\widetilde{n}_1) \cdots (n_{\kappa}|\widetilde{n}_{\kappa})} \equiv F^{(n_1|\widetilde{n}_1| \cdots |n_{\kappa}|\widetilde{n}_{\kappa})} \equiv F^{(n_1,\widetilde{n}_1, \ldots, n_{\kappa}, \widetilde{n}_{\kappa})}$, with translations absolutely evident.}
\begin{equation}
\label{eq:genericresonanttransseries}
F \left( g_{\text{s}}; \boldsymbol{\sigma} \right) = \sum_{\boldsymbol{n}, \widetilde{\boldsymbol{n}}}\, \prod_{i=1}^{\kappa} \sigma_i^{n_i} \widetilde{\sigma}_i^{\widetilde{n}_i}\, \rme^{- \left( n_i - \widetilde{n}_i \right) \frac{A_i}{g_{\text{s}}}}\, F^{(n_1 | \widetilde{n}_1 | \cdots | n_{\kappa} | \widetilde{n}_{\kappa})} (g_{\text{s}}).
\end{equation}
\noindent
The transseries sectors are asymptotic series of the form (compare\footnote{In these equations the logarithmic sectors have already been resummed within the transseries, \eqref{eq:Painleve1SOlution} or \eqref{eq:YangLeeSolution}.} with \eqref{eq:P1(n|m)sectors} or \eqref{eq:YL(n|m)sectors})
\begin{equation}
\label{eq:GenericFreeEnergyTransseriesSector}
F^{(n_1 | \widetilde{n}_1 | \cdots | n_{\kappa} | \widetilde{n}_{\kappa})} (g_{\text{s}}) \simeq \sum_{g=0}^{+\infty} g_{\text{s}}^{g+\beta_{(n_1 | \widetilde{n}_1 | \cdots | n_{\kappa} | \widetilde{n}_{\kappa})}} \underbrace{\sum_{k=0}^{k_{\text{max}}} \log^k \left( g_{\text{s}} \right)\, F^{(n_1 | \widetilde{n}_1 | \cdots | n_{\kappa} | \widetilde{n}_{\kappa})[k]}_g}_{\equiv\, F^{(n_1 | \widetilde{n}_1 | \cdots | n_{\kappa} | \widetilde{n}_{\kappa})}_g},
\end{equation}
\noindent
where the starting-genus $\beta_{\boldsymbol{n}}$ and range of logarithmic sectors $k_{\text{max}}$ are\footnote{There are however some generic, problem-independent features to the logarithmic sectors of (double scaled) matrix-model transseries. For example, resonant sectors that have for each $i$ either $n_i=0$ or $m_i=0$ or $n_i=m_i$ have no logarithmic dependence. For details see \cite{gikm10, asv11, sv13, gs21, bssv22, mss22, sst23}.} problem-specific. The structure of a two-parameter resonant transseries is visualized in figure~\ref{fig:twopararectmtransseriesgrid}, and several very explicit examples of this construction will be presented in the following subsection~\ref{subsec:resurgent-Z-transasymptotics}.

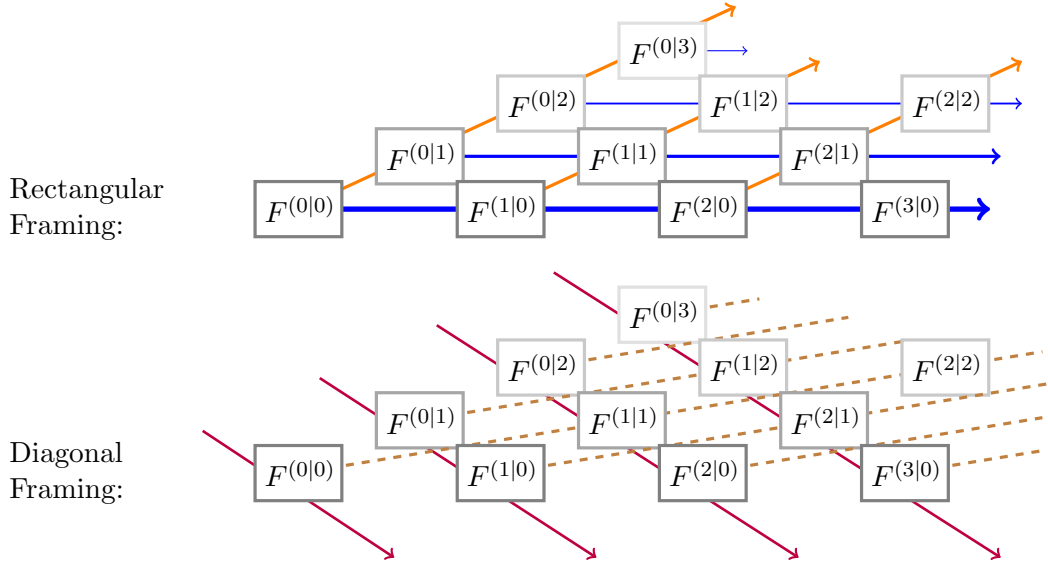
\begin{figure}
\centering
\def\xstep{1.9}
\def\ystep{2}
\definecolor{lightgray}{gray}{0.65}
\definecolor{llightgray}{gray}{0.79}
\definecolor{lllightgray}{gray}{0.88}
\begin{tikzpicture}[grayframe/.style={
		rectangle,
		draw=gray,
		fill=white,
		text width=2.3em,
		align=center,
		minimum height=1.9em
	},
	lgrayframe/.style={
		rectangle,
		draw=lightgray,
		fill=white,
		text width=2.3em,
		align=center,
		minimum height=1.9em
	},
	llgrayframe/.style={
		rectangle,
		draw=llightgray,
		fill=white,
		text width=2.3em,
		align=center,
		minimum height=1.9em
	},
	lllgrayframe/.style={
		rectangle,
		draw=lllightgray,
		fill=white,
		text width=2.3em,
		align=center,
		minimum height=1.9em
	},
	line width=1.2
	]
	\begin{scope}[scale=1.4,  shift={({0},{0})}]
	\node[text width=2cm] at (-2, 0) {Rectangular\\ Framing:};
	\draw[orange, ->] (0,0) -- (3*1.37,1.4*1.37);
	\draw[orange, ->] (2*\xstep,0) -- (2*\xstep+3,1.4);
	\draw[orange, ->] (\xstep,0) -- (\xstep+3,1.4);
	\draw[line width=0.7, blue, ->] (2*0.6*\xstep,2*0.25*\ystep) -- (6.8,2*0.25*\ystep);
	\draw[line width=0.4, blue, ->] (3*0.6*\xstep,3*0.25*\ystep) -- (3*0.6*\xstep+0.8,3*0.25*\ystep);
	\node[llgrayframe] at (2*\xstep*0.6,2*\ystep*0.25) {$F^{(0|2)}$};
	\node[lllgrayframe] at (3*\xstep*0.6,3*\ystep*0.25) {$F^{(0|3)}$};
	\node[llgrayframe] at (\xstep*2.2,\ystep*0.5) {$F^{(1|2)}$};
	\node[llgrayframe] at (\xstep*3.2,\ystep*0.5) {$F^{(2|2)}$};
	\draw[line width=1.2, blue, ->] (0.6*\xstep,0.25*\ystep) -- (6.6,0.25*\ystep);
	\node[lgrayframe] at (\xstep*0.6,\ystep*0.25) {$F^{(0|1)}$};
	\node[lgrayframe] at (\xstep*1.6,\ystep*0.25) {$F^{(1|1)}$};
	\node[lgrayframe] at (\xstep*2.6,\ystep*0.25) {$F^{(2|1)}$};
	\draw[line width=2, blue, ->] (0,0) -- (6.5,0);
	\node[grayframe] at (0,0) {$F^{(0|0)}$};
	\node[grayframe] at (\xstep,0) {$F^{(1|0)}$};
	\node[grayframe] at (2*\xstep,0) {$F^{(2|0)}$};
	\node[grayframe] at (3*\xstep,0) {$F^{(3|0)}$};
	\end{scope}
	\begin{scope}[scale=1.4,  shift={({0},{-2.5})}]
	\node[text width=2cm] at (-2, 0) {Diagonal \\ Framing:};
	\draw[purple, ->, line width=1] (0.2, 0.9) -- (2.8, -0.8);
	\draw[purple, ->, line width=1] (1.3, 1.4) -- (4.7, -0.8);
	\draw[purple, ->, line width=1] (0.2+2*1.1, 0.9+1) -- (2.8+2*1.9, -0.8);
	\draw[purple, ->, line width=1] (0.2-1.1, 0.9-0.5) -- (2.8-1.9, -0.8);
	\draw[brown, dashed] (3*\xstep*0.6,3*\ystep*0.25) -- (\xstep*1.6*0.3+3*\xstep*0.6,\ystep*0.25*0.3+3*\ystep*0.25);
	\node[lllgrayframe] at (3*\xstep*0.6,3*\ystep*0.25) {$F^{(0|3)}$};
	\draw[brown, dashed] (2*\xstep*0.6,2*\ystep*0.25) -- (\xstep*1.6*0.95+2*\xstep*0.6,\ystep*0.25*0.95+2*\ystep*0.25);
	\node[llgrayframe] at (2*\xstep*0.6,2*\ystep*0.25) {$F^{(0|2)}$};
	\draw[brown, dashed] (\xstep*0.6,\ystep*0.25) -- (\xstep*1.6*1.499+\xstep*0.6,\ystep*0.25*1.499+\ystep*0.25);
	\node[llgrayframe] at (\xstep*2.2,\ystep*0.5) {$F^{(1|2)}$};
	\node[lgrayframe] at (\xstep*0.6,\ystep*0.25) {$F^{(0|1)}$};
	\draw[brown, dashed] (0,0) -- (\xstep*1.6*2.3,\ystep*0.25*2.3);
	\node[llgrayframe] at (\xstep*3.2,\ystep*0.5) {$F^{(2|2)}$};
	\node[lgrayframe] at (\xstep*1.6,\ystep*0.25) {$F^{(1|1)}$};
	\draw[brown, dashed] (\xstep,0) -- (\xstep*1.6*1.7+\xstep,\ystep*0.25*1.7);
	\node[lgrayframe] at (\xstep*2.6,\ystep*0.25) {$F^{(2|1)}$};
	\node[grayframe] at (0,0) {$F^{(0|0)}$};
	\node[grayframe] at (\xstep,0) {$F^{(1|0)}$};
	\draw[brown, dashed] (2*\xstep,0) -- (\xstep*1.6*1.05+2*\xstep,\ystep*0.25*1.05);
	\node[grayframe] at (2*\xstep,0) {$F^{(2|0)}$};
	\draw[brown, dashed] (3*\xstep,0) -- (\xstep*1.6*0.433+3*\xstep,\ystep*0.25*0.433);
	\node[grayframe] at (3*\xstep,0) {$F^{(3|0)}$};
	\end{scope}
	\end{tikzpicture}
	\caption{Diagonal versus rectangular transseries framings. We illustrate the transseries sectors in \eqref{eq:genericresonanttransseries} for a two-parameter resonant transseries. The top plot is (a compact illustration of) the rectangular framing in figure~\ref{fig:twopararectmtransseriesgrid}. The bottom plot shows the reorganization into diagonal framing, where we now group transseries blocks according to their exponential weight rather than according to instanton numbers. The ``diagonal directions'' are in {\color{brown}dashed brown}, and the transseries lattice is now rearranged along the {\color{purple}purple} lines, as visualized in figure~\ref{fig:twoparamdiagtransseriesgrid}.}
	\label{fig:rectangulartodiagonal}
\end{figure}

\item \textbf{Diagonal Framing:} The existence of resonance allows for an alternative, rather convenient reorganization of the transseries, precisely along the direction of the resonant kernel \cite{abs18, gs21, bssv22}. This essentially amounts to regrouping transmonomial contributions in \eqref{eq:genericresonanttransseries} solely according to their exponential weights (this is a very simple reorganization, see \cite{gs21} for some examples). Introducing the ``moduli'' $\upmu_i \equiv \sigma_i \widetilde{\sigma}_i$, one immediately finds \cite{gs21, bssv22}
\bea
\label{eq:PreDiagonalFramingTransseries}
F \left( g_{\text{s}}; \boldsymbol{\sigma} \right) &=& \sum_{\ell_1, \cdots, \ell_{\kappa} = 0}^{+\infty}\, \prod\limits_{i=1}^{\kappa} \upmu_i^{\ell_i}\, F^{(\ell_1 | \ell_1 | \cdots | \ell_{\kappa} | \ell_{\kappa})} (g_{\text{s}}) + \\
&&+ \sum_{{n_1, \cdots, n_{\kappa} = 0}\atop{(n_1, \cdots, n_{\kappa}) \neq \boldsymbol{0}}}^{+\infty}\, \prod_{i=1}^{\kappa} \sigma_i^{n_i}\, \rme^{- n_i \frac{A_i}{g_{\text{s}}}}\, \sum_{\ell_1, \cdots, \ell_{\kappa} = 0}^{+\infty}\, \prod_{i=1}^{\kappa} \upmu_i^{\ell_i}\, F^{(\ell_1 + n_1 | \ell_1 | \cdots | \ell_{\kappa} + n_{\kappa} | \ell_{\kappa})} (g_{\text{s}}) + \nonumber \\
&& + \cdots + \nonumber \\
&& + \sum_{{\widetilde{n}_1, \cdots, \widetilde{n}_{\kappa} = 0}\atop{(\widetilde{n}_1, \cdots, \widetilde{n}_{\kappa}) \neq \boldsymbol{0}}}^{+\infty}\, \prod_{i=1}^{\kappa} \widetilde{\sigma}_i^{\widetilde{n}_i}\, \rme^{+ \widetilde{n}_i \frac{A_i}{g_{\text{s}}}}\, \sum_{\ell_1, \cdots, \ell_{\kappa} = 0}^{+\infty}\, \prod_{i=1}^{\kappa} \upmu_i^{\ell_i}\, F^{(\ell_1 | \ell_1 + \widetilde{n}_1 | \cdots | \ell_{\kappa} | \ell_{\kappa} + \widetilde{n}_{\kappa})} (g_{\text{s}}), \nonumber
\eea
\noindent
where in the first line we have included all terms without any transmonomial contributions, and in the second and last lines the ones with all falling and all growing transmonomials, respectively (and all ``mixed'' ones hiding in the dots). The change of framing from rectangular to diagonal is illustrated in figure~\ref{fig:rectangulartodiagonal}. The key point is that this seemingly more intricate reorganization in \eqref{eq:PreDiagonalFramingTransseries} can be again rewritten as a transseries with solely $2\kappa$ parameters whilst keeping the aforementioned ``moduli''---albeit we need to make $\kappa$ choices of whether to keep (original) transseries parameters $\left\{ \sigma_i \right\}$ or $\left\{ \widetilde{\sigma}_i \right\}$. To illustrate this point let us make the choice\footnote{Bare in mind that we could have made any other possible combination of choices---we will in fact see how these possible choices of ``coordinates'' are crucial for the discussions in subsections~\ref{subsec:resurgent-Z-transasymptotics} and~\ref{subsec:resurgent-Z}, and in \cite{ss26}.} of just keeping $\left\{ \sigma_i \right\}$. One must then replace all $\widetilde{\sigma}_i \mapsto \upmu/\sigma_i$, yielding
\begin{align}
\label{eq:diagonalframingF}
F \left( g_{\text{s}}; \boldsymbol{\sigma} \right) = \sum_{\boldsymbol{\ell} \in \mathbb{Z}^{\kappa}}\, \prod\limits_{i=1}^{\kappa} \left( \sigma_i\, \rme^{ -\frac{A_i}{g_{\text{s}}}}  \left( \frac{f_i(t)}{g_{\text{s}}} \right)^{\alpha_i \upmu_i} \right)^{\ell_i}  F^{\langle \boldsymbol{\ell} \rangle} ( g_{\text{s}}; \boldsymbol{\upmu} ).
\end{align}
\noindent
This is again a rather compact-looking expression, but we have still left some further simplifying steps under the rug. First, the extra factor $f_i(t)/g_{\text{s}}$ (where $f_i(t)$ is an yet unspecified function of the 't~Hooft coupling). This appears via resummation of all logarithmic sectors which appeared in \eqref{eq:GenericFreeEnergyTransseriesSector} (and some such resummations were already featured in \eqref{eq:Painleve1SOlution} and \eqref{eq:YangLeeSolution}). The $\alpha_i$ are their associated, problem-dependent logarithmic-resummation constants (\textit{e.g.}, see equation (5.40) of \cite{asv11} for further details; or else, see equation \eqref{eq:ASV-log-resonance} in the upcoming case of \PI). Note that this resulting resummation structure is a non-trivial result which albeit turns out to be a standard feature of transseries for generic hermitian matrix models as we shall see in the examples in the following subsection~\ref{subsec:resurgent-Z-transasymptotics}; see as well \cite{mss22}. Very likely, this property also holds in some form or another for broader classes of \textit{resonant} transseries. Second, the asymptotic sectors $F^{\langle \boldsymbol{\ell} \rangle} ( g_{\text{s}}; \boldsymbol{\upmu} )$ are actually \textit{polynomial} in the $\left\{ \upmu_i \right\}$ variables---again which we shall uncover in the upcoming examples. Finally, the overall summation in this formula is now over \textit{all} integers, $\mathbb{Z}$. The transseries sectors are asymptotic series of the form
\begin{equation}
\label{eq:genericdiagonalsectors}
F^{\langle \boldsymbol{\ell} \rangle} ( g_{\text{s}}; \boldsymbol{\upmu} ) \simeq \sum_{g=0}^{+\infty} g_{\text{s}}^{g+\beta_{\boldsymbol{\ell}}} F^{\langle \boldsymbol{\ell} \rangle}_{g} ( \boldsymbol{\upmu} ),
\end{equation}
\noindent
where one can compare with \eqref{eq:GenericFreeEnergyTransseriesSector}---there are now \textit{no} logarithmic sectors---and where we already stated that the $\left\{ \upmu_i \right\}$ dependence is polynomial. Further notice the trade-off: our transseries sectors are now labeled by only $\kappa$ instanton numbers (rather than $2\kappa$ as in the rectangular case) although on the other hand each index is now summed over the full integers (rather than just positive). This is illustrated for a two parameter transseries in figure \ref{fig:twoparamdiagtransseriesgrid}, and very  explicit examples of these constructions will be presented next.
\end{itemize}

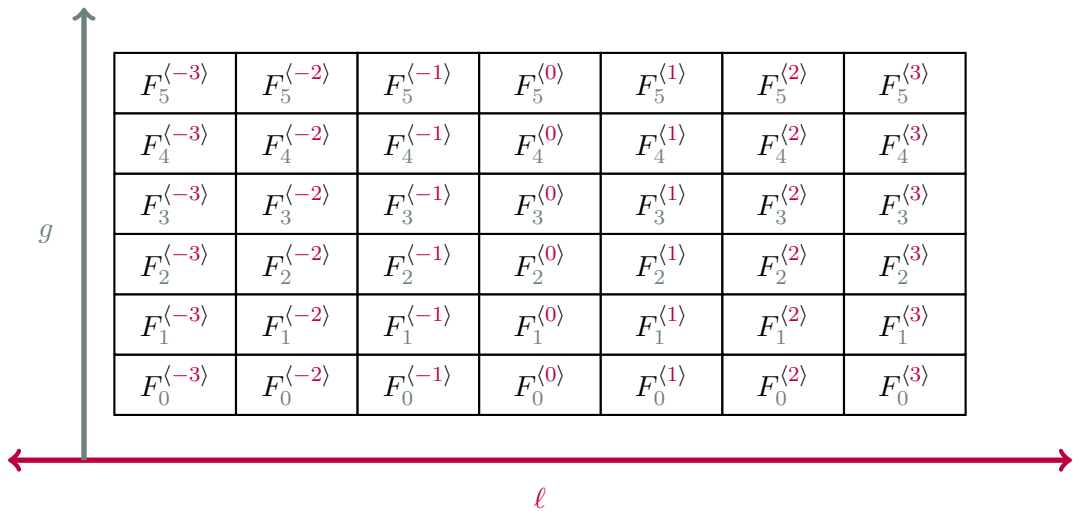
\begin{figure}
\centering
\def\xstep{1.9}
\def\ystep{2}
\definecolor{dgray}{rgb}{0.43, 0.5, 0.5}
\begin{tikzpicture}[grayframe/.style={
		rectangle,
		draw=gray,
		fill=white,
		text width=1.7em,
		align=center,
		rounded corners,
		minimum height=1em
	},
	box/.style={rectangle,draw=black,thick, minimum height=0.8cm, minimum width =1.6cm},
	]
	\foreach \x in {-3,-2,...,3}{
    \foreach \y in {0,1,...,5}
        \node[box, fill=white] at (1.6*\x,0.8*\y){$F^{\langle{\color{purple}\x}\rangle}_{{\color{dgray}\y}}$};
}
\draw[purple, line width=2pt, <->] (-7,-1) -- (7, -1);
\draw[dgray, line width=2pt, ->] (-6,-1) -- (-6, 5);
\node[dgray] at (-6.5, 2) {$g$};
\node[purple] at (0, -1.5) {$\ell$};
	\end{tikzpicture}\\
	\caption{The infinite grid representing a two-parameter resonant transseries in \textit{diagonal} framing. Note how we now extend to both plus and minus infinity along the ${\color{purple}\ell}$ direction (unlike the semi-positive figure~\ref{fig:twopararectmtransseriesgrid}). The entries in each box are the different coefficients in the asymptotic series \eqref{eq:genericdiagonalsectors}. When comparing to figure~\ref{fig:twopararectmtransseriesgrid} the grid is now just two-dimensional albeit all sectors have acquired $\left\{ \upmu_i \right\}$-dependence. The {\color{dgray}gray} subscript still relates to the perturbative order. Compare this figure to its rectangular-framing version in figures~\ref{fig:twopararectmtransseriesgrid} and~\ref{fig:rectangulartodiagonal}.}
	\label{fig:twoparamdiagtransseriesgrid}
\end{figure}

The above discussion was rather generic albeit we did have in mind the free-energy asymptotic expansion \eqref{eq:genusgfreeenergies}. In this way, if \eqref{eq:generictransseriesmatrixmodels} describes the full nonperturbative free energy, then the full nonperturbative partition function is just the transseries obtained via exponentiation,
\begin{equation}
\label{eq:partitionfunctionexponentiation}
Z \left( g_{\text{s}}; \boldsymbol{\sigma} \right) = \exp F ( g_{\text{s}};  \boldsymbol{\sigma} ).
\end{equation}
\noindent
Our main result, which we start unveiling in the following subsection~\ref{subsec:resurgent-Z-transasymptotics} by addressing a few exploratory examples, is that, once in diagonal framing, the partition-function transseries of any hermitian matrix-model or their double-scaling limits can be rewritten in the form of a so-called discrete Fourier transform or Zak transform. This further intertwines with the mathematical concept of (quadratic) transasymptotic resummation \cite{c97, cc01, cch13, asv17a, asv17b}. Building upon these examples, the ensuing subsection~\ref{subsec:resurgent-Z} presents a general construction of the partition-function for generic hermitian matrix models, which stands as our main conjecture. This construction will be thoroughly tested \cite{krsst26b, krst26a, krst26b}, and further compared with results in the literature in subsection~\ref{subsec:from-theta-to-dual-to-transseries}. A very interesting application which immediately follows from these results is to write down solutions for all string-equations along the KdV hierarchy, which we do in subsection~\ref{subsec:KdV-solutions}.

Our transseries, such as \eqref{eq:generictransseriesmatrixmodels} but hence necessarily also \eqref{eq:diagonalframingF}, are always initially constructed out of string-equations, which is to say they always start life in the one-cut setting. In order to reach all other phases discussed throughout section~\ref{sec:strong-coupling-phases}, and as already explained, the transseries will have to undergo Stokes phenomena. This is simply implemented by the resurgent properties \cite{e81, e84, e93} of \eqref{eq:generictransseriesmatrixmodels}, where its transseries parameters jump at the crossing of Stokes lines and according to their Stokes data, in the schematic form $\boldsymbol{\sigma} \mapsto \underline{\pmb{\BS}}_{\theta} (\boldsymbol{\sigma}) \sim \boldsymbol{\sigma} + \boldsymbol{S}$. In diagonal framing for \eqref{eq:diagonalframingF} this was recently fully worked out for the \PI~example in \cite{bssv22}. And as applied to our generic matrix-model transseries, it will be fully discussed in \cite{krsst26b}.

\subsection{Transseries Framings and Multi-Parameter Resonant Transasymptotics}
\label{subsec:resurgent-Z-transasymptotics}

Having discussed the generics of transseries framings, let us now illustrate how to write both rectangular and diagonal solutions in several examples. These are not the only natural ways to write our solutions. In the road towards our main conjecture, we will also write these partition functions in the guise of a (quadratic) transseries transasymptotic resummation, as a discrete Fourier transform, and via the use of theta-functions. We will focus on four canonical, prototypical examples in this work (and present several other examples in \cite{krst26a}). Our examples are:
\begin{center}
\begin{tabular}{ll}
\textbullet\,\, The Painlev\'e~I equation \eqref{eq:Painleve1Equation}, \qquad \qquad & \textbullet\,\, The cubic matrix model \eqref{eq:CubicMatrixModelPotential}, \\ [5pt]
\textbullet\,\, The Yang--Lee equation \eqref{eq:YangLeeEquation}, & \textbullet\,\, The quartic matrix model \eqref{eq:QuarticMatrixModelPotential}.\\
\end{tabular}
\end{center}

Let us hence begin this subsection by briefly reviewing the transasymptotics construction \cite{c97, cc01, cch13, asv17a, asv17b}, with emphasis in its two natural implementations: \textit{linear} and \textit{quadratic}. The basic question is extremely simple: while the perturbative string-coupling expansions, say in \eqref{eq:GenericFreeEnergyTransseriesSector} or \eqref{eq:genericdiagonalsectors}, are typically asymptotic, one may ask if perhaps the instanton series appearing in \eqref{eq:generictransseriesmatrixmodels}, \eqref{eq:genericresonanttransseries}, \eqref{eq:PreDiagonalFramingTransseries} or \eqref{eq:diagonalframingF} might be better behaved? If this were to be the case one could try exchanging\footnote{We are always working with formal power-series anyways, in which case exchanging sums is always allowed.} summations and then summing the latter first. Generically, this need not be true: these series could in principle also be asymptotic, but in which case via resurgence they would need to give rise to double-exponential\footnote{These would also be denoted as exponential-height-two transmonomials; see, \textit{e.g.}, \cite{abs18}.} transmonomial contributions to the transseries---which we know are not there as per the construction of our transseries as complete solutions\footnote{Recall how these solutions are built with a transseries \textit{ans\"atze}, \textit{e.g.}, \cite{m08, gikm10, asv11, sv13}. The original string equation is whence turned into an infinite set of coupled equations, but where in particular instanton sectors satisfy (inhomogeneous) \textit{linear} equations. As such, each isolated instanton sector is ``relatively mild''.} to some string equation. As we shall see, in our classes of problems the instanton series are indeed convergent and can be explicitly evaluated in closed-form.

\begin{figure}
	\begin{subfigure}[t]{0.47\textwidth}
\centering
\definecolor{ao}{rgb}{0.0, 0.5, 0.0}
\definecolor{dgray}{rgb}{0.43, 0.5, 0.5}
	\begin{tikzpicture}[grayframe/.style={
		rectangle,
		draw=gray,
		fill=white,
		text width=1.7em,
		align=center,
		rounded corners,
		minimum height=1em
	},
	box/.style={rectangle,draw=black,thick, minimum height=0.8cm, minimum width =1cm},
	]
	\foreach \x in {0,1,...,5}{
    \foreach \y in {0,1,...,5}
        \node[box] at (\x,0.8*\y){$0$};
}
\node[box, fill=ao] at (0,0.8*0){};
\node[box, fill=ao] at (0,0.8*1){};
\node[box, fill=ao] at (0,0.8*2){};
\node[box, fill=ao] at (0,0.8*3){};
\node[box, fill=ao] at (0,0.8*4){};
\node[box, fill=ao] at (0,0.8*5){};
\node[box, fill=ao] at (1,0.8*1){};
\node[box, fill=ao] at (1,0.8*2){};
\node[box, fill=ao] at (1,0.8*3){};
\node[box, fill=ao] at (1,0.8*4){};
\node[box, fill=ao] at (1,0.8*5){};
\node[box, fill=ao] at (2,0.8*2){};
\node[box, fill=ao] at (2,0.8*3){};
\node[box, fill=ao] at (2,0.8*4){};
\node[box, fill=ao] at (2,0.8*5){};
\node[box, fill=ao] at (3,0.8*3){};
\node[box, fill=ao] at (3,0.8*4){};
\node[box, fill=ao] at (3,0.8*5){};
\node[box, fill=ao] at (4,0.8*4){};
\node[box, fill=ao] at (4,0.8*5){};
\node[box, fill=ao] at (5,0.8*5){};
\draw[blue, line width=2pt, ->] (-1,-1) -- (6, -1);
\draw[dgray, line width=2pt, ->] (-1,-1) -- (-1, 5);
\node[blue] at (6.1, -1.2) {$n$};
\foreach \y in {0,1,...,5}
\node[dgray] at (-1.3,0.8*\y){$g_{\text{s}}^{\y}$};
\foreach \x in {0,1,...,5}
\node[blue] at (\x,-1.5){$\x$};
	\end{tikzpicture}
	\caption{Linear $\beta$-structure: $\beta_n=n$.}
	\label{fig:gridbetalin}
	\end{subfigure}
\hspace{0.2cm}
	\begin{subfigure}[t]{0.47\textwidth}
\centering
\definecolor{ao}{rgb}{0.0, 0.5, 0.0}
\definecolor{dgray}{rgb}{0.43, 0.5, 0.5}
	\begin{tikzpicture}[grayframe/.style={
		rectangle,
		draw=gray,
		fill=white,
		text width=1.7em,
		align=center,
		rounded corners,
		minimum height=1em
	},
	box/.style={rectangle,draw=black,thick, minimum height=0.8cm, minimum width =1cm},
	]
	\foreach \x in {0,1,...,5}{
    \foreach \y in {0,1,...,5}
        \node[box] at (\x,0.8*\y){$0$};
}
\node[box, fill=ao] at (0,0.8*0){};
\node[box, fill=ao] at (0,0.8*1){};
\node[box, fill=ao] at (0,0.8*2){};
\node[box, fill=ao] at (0,0.8*3){};
\node[box, fill=ao] at (0,0.8*4){};
\node[box, fill=ao] at (0,0.8*5){};
\node[box, fill=ao] at (1,0.8*0){};
\node[box, fill=ao] at (1,0.8*1){};
\node[box, fill=ao] at (1,0.8*2){};
\node[box, fill=ao] at (1,0.8*3){};
\node[box, fill=ao] at (1,0.8*4){};
\node[box, fill=ao] at (1,0.8*5){};
\node[box, fill=ao] at (2,0.8*1){};
\node[box, fill=ao] at (2,0.8*2){};
\node[box, fill=ao] at (2,0.8*3){};
\node[box, fill=ao] at (2,0.8*4){};
\node[box, fill=ao] at (2,0.8*5){};
\node[box, fill=ao] at (3,0.8*3){};
\node[box, fill=ao] at (3,0.8*4){};
\node[box, fill=ao] at (3,0.8*5){};
\draw[blue, line width=2pt, ->] (-1,-1) -- (6, -1);
\draw[dgray, line width=2pt, ->] (-1,-1) -- (-1, 5);
\node[blue] at (6.1, -1.2) {$n$};
\foreach \y in {0,1,...,5}
\node[dgray] at (-1.3,0.8*\y){$g_{\text{s}}^{\y}$};
\foreach \x in {0,1,...,5}
\node[blue] at (\x,-1.5){$\x$};
	\end{tikzpicture}\\
	\caption{Quadratic $\beta$-structure: $\beta_n=\frac{1}{2} n \left(n-1\right)$.}
	\label{fig:gridbetaquadratic}
	\end{subfigure}
	\caption{Visualizing the ``$\beta$-structure'' of a transseries of the form \eqref{eq:genericresonanttransseries}. Our grids display transseries sectors along one instanton direction. However, in comparison to the grids in figures~\ref{fig:twopararectmtransseriesgrid} and~\ref{fig:twoparamdiagtransseriesgrid}, the vertical axis lists the full exponent of the string-coupling including the $\beta$-shift. This implies that some lattice nodes are not populated, which allows for the visualization the $\beta$-structure. The non-zero transseries sectors are represented by the colored \textcolor{ForestGreen}{green} squares \cite{asv17a, asv17b}.}
	\label{fig:gridbetastructure}
\end{figure}
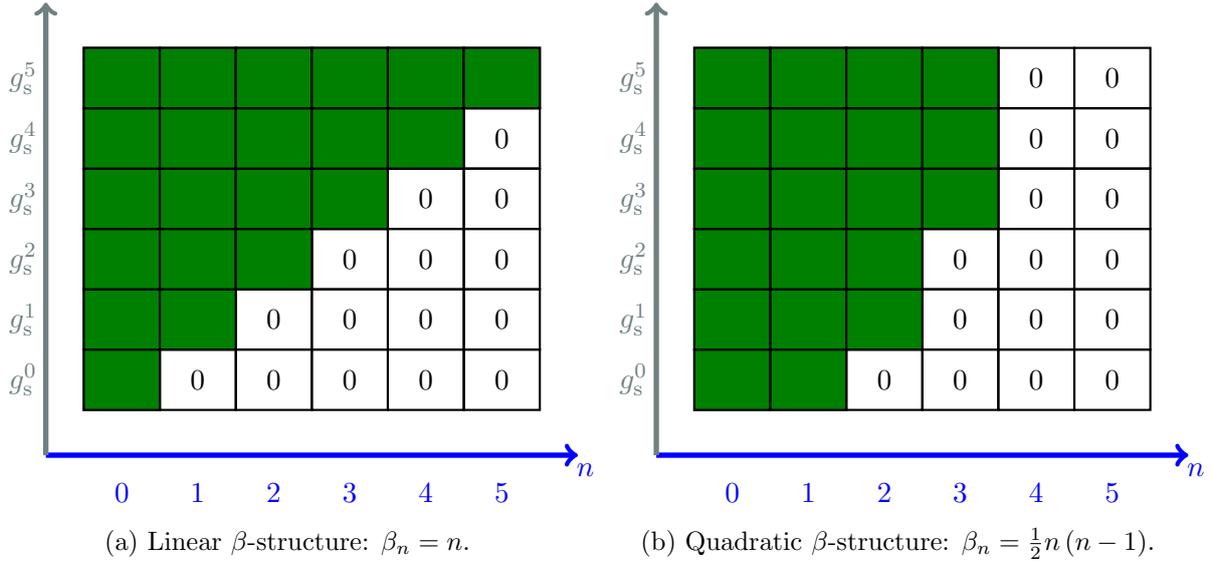

To make this concrete one must first be precise about what it means to evaluate an instanton series. As explained in \cite{asv17a, asv17b}, this crucially depends upon the ``$\beta$-structure'' of the transseries in question (the starting genus in \eqref{eq:GenericFreeEnergyTransseriesSector} or \eqref{eq:genericdiagonalsectors}), as these values dictate the lowest non-vanishing instanton contributions. This is very simple to visualize following \cite{asv17a, asv17b} which redraws figures~\ref{fig:twopararectmtransseriesgrid} and~\ref{fig:twoparamdiagtransseriesgrid} by, instead of listing the perturbative-order $g$ along the vertical axis, actually listing the full exponent of the string-coupling including the $\beta$-shift. In this way, some nodes on these grids may be populated by zeroes alone---which makes the full $\beta$-structure manifest. This is illustrated in figure~\ref{fig:gridbetastructure}. Focus on a two-parameter resonant transseries and consider the two basic cases:
%
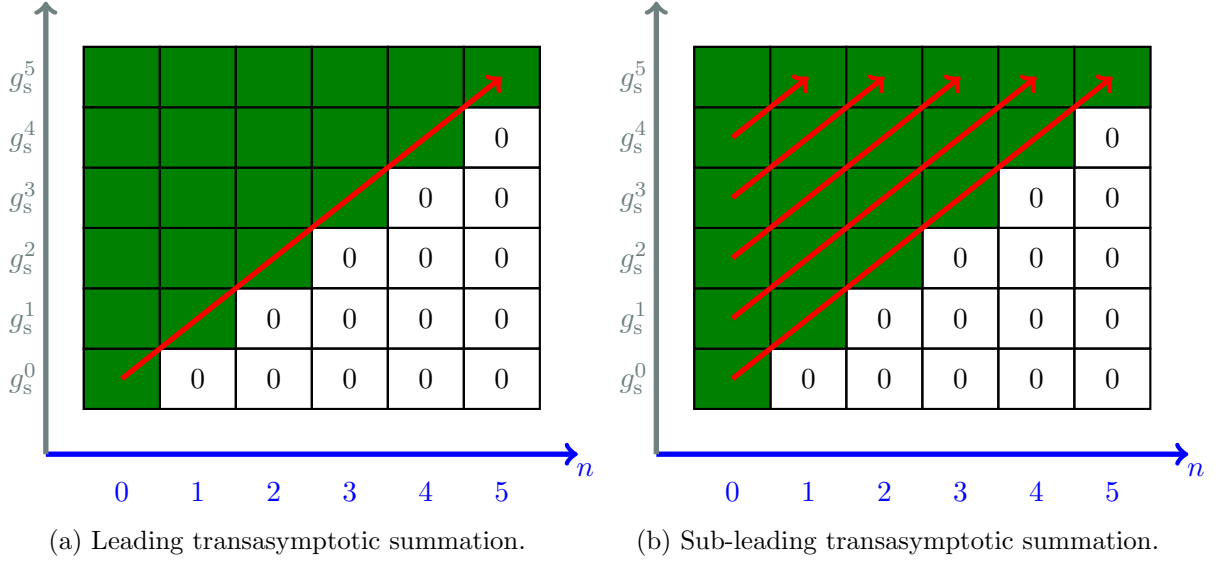
\begin{figure}
	\begin{subfigure}[t]{0.47\textwidth}
\centering
\definecolor{ao}{rgb}{0.0, 0.5, 0.0}
\definecolor{dgray}{rgb}{0.43, 0.5, 0.5}
\definecolor{dred}{rgb}{0.81, 0.09, 0.13}
\begin{tikzpicture}[grayframe/.style={
		rectangle,
		draw=gray,
		fill=white,
		text width=1.7em,
		align=center,
		rounded corners,
		minimum height=1em
	},
	box/.style={rectangle,draw=black,thick, minimum height=0.8cm, minimum width =1cm},
	]
	\foreach \x in {0,1,...,5}{
    \foreach \y in {0,1,...,5}
        \node[box] at (\x,0.8*\y){$0$};
}
\node[box, fill=ao] at (0,0.8*0){};
\node[box, fill=ao] at (0,0.8*1){};
\node[box, fill=ao] at (0,0.8*2){};
\node[box, fill=ao] at (0,0.8*3){};
\node[box, fill=ao] at (0,0.8*4){};
\node[box, fill=ao] at (0,0.8*5){};
\node[box, fill=ao] at (1,0.8*1){};
\node[box, fill=ao] at (1,0.8*2){};
\node[box, fill=ao] at (1,0.8*3){};
\node[box, fill=ao] at (1,0.8*4){};
\node[box, fill=ao] at (1,0.8*5){};
\node[box, fill=ao] at (2,0.8*2){};
\node[box, fill=ao] at (2,0.8*3){};
\node[box, fill=ao] at (2,0.8*4){};
\node[box, fill=ao] at (2,0.8*5){};
\node[box, fill=ao] at (3,0.8*3){};
\node[box, fill=ao] at (3,0.8*4){};
\node[box, fill=ao] at (3,0.8*5){};
\node[box, fill=ao] at (4,0.8*4){};
\node[box, fill=ao] at (4,0.8*5){};
\node[box, fill=ao] at (5,0.8*5){};
\draw[blue, line width=2pt, ->] (-1,-1) -- (6, -1);
\draw[dgray, line width=2pt, ->] (-1,-1) -- (-1, 5);
\node[blue] at (6.1, -1.2) {$n$};
\foreach \y in {0,1,...,5}
\node[dgray] at (-1.3,0.8*\y){$g_{\text{s}}^{\y}$};
\foreach \x in {0,1,...,5}
\node[blue] at (\x,-1.5){$\x$};
\draw[red, line width=2pt,->] (0,0) -- (5, 0.8*5);
	\end{tikzpicture}
	\caption{Leading transasymptotic summation.}
	\label{fig:lineartrans1}
\end{subfigure}
\hspace{0.2cm}
	\begin{subfigure}[t]{0.47\textwidth}
\centering
\definecolor{ao}{rgb}{0.0, 0.5, 0.0}
\definecolor{dgray}{rgb}{0.43, 0.5, 0.5}
\begin{tikzpicture}[grayframe/.style={
		rectangle,
		draw=gray,
		fill=white,
		text width=1.7em,
		align=center,
		rounded corners,
		minimum height=1em
	},
	box/.style={rectangle,draw=black,thick, minimum height=0.8cm, minimum width =1cm},
	]
	\foreach \x in {0,1,...,5}{
    \foreach \y in {0,1,...,5}
        \node[box] at (\x,0.8*\y){$0$};
}
\node[box, fill=ao] at (0,0.8*0){};
\node[box, fill=ao] at (0,0.8*1){};
\node[box, fill=ao] at (0,0.8*2){};
\node[box, fill=ao] at (0,0.8*3){};
\node[box, fill=ao] at (0,0.8*4){};
\node[box, fill=ao] at (0,0.8*5){};
\node[box, fill=ao] at (1,0.8*1){};
\node[box, fill=ao] at (1,0.8*2){};
\node[box, fill=ao] at (1,0.8*3){};
\node[box, fill=ao] at (1,0.8*4){};
\node[box, fill=ao] at (1,0.8*5){};
\node[box, fill=ao] at (2,0.8*2){};
\node[box, fill=ao] at (2,0.8*3){};
\node[box, fill=ao] at (2,0.8*4){};
\node[box, fill=ao] at (2,0.8*5){};
\node[box, fill=ao] at (3,0.8*3){};
\node[box, fill=ao] at (3,0.8*4){};
\node[box, fill=ao] at (3,0.8*5){};
\node[box, fill=ao] at (4,0.8*4){};
\node[box, fill=ao] at (4,0.8*5){};
\node[box, fill=ao] at (5,0.8*5){};
\draw[blue, line width=2pt, ->] (-1,-1) -- (6, -1);
\draw[dgray, line width=2pt, ->] (-1,-1) -- (-1, 5);
\node[blue] at (6.1, -1.2) {$n$};
\foreach \y in {0,1,...,5}
\node[dgray] at (-1.3,0.8*\y){$g_{\text{s}}^{\y}$};
\foreach \x in {0,1,...,5}
\node[blue] at (\x,-1.5){$\x$};
\draw[red, line width=2pt,->] (0,0) -- (5, 0.8*5);
\draw[red, line width=2pt,->] (0,0.8) -- (4, 0.8*5);
\draw[red, line width=2pt,->] (0,2*0.8) -- (3, 0.8*5);
\draw[red, line width=2pt,->] (0,3*0.8) -- (2, 0.8*5);
\draw[red, line width=2pt,->] (0,4*0.8) -- (1, 0.8*5);
	\end{tikzpicture}
	\caption{Sub-leading transasymptotic summation.}
	\label{fig:lineartrans2}
\end{subfigure}
	\caption{Visualization of the \textit{linear} transasymptotic summation, both at leading and sub-leading orders in the string coupling. We illustrate with the grid of transseries sectors in the one-instanton direction as in the previous figure~\ref{fig:gridbetalin}.}
	\label{fig:lineartransasymptotics}
\end{figure}
%
\begin{enumerate}
\item \textbf{Linear Transasymptotics:} \cite{c97, cc01, cch13} For simple illustration, consider the forward instanton direction of the transseries \eqref{eq:genericresonanttransseries} with two parameters. By inspection on several\footnote{Several examples are discussed at greater length in appendix~\ref{app:transasymptotic-transseries}.} specific-heat examples, the $\beta$-structure is usually linear \cite{msw08, gikm10, kmr10, asv11, sv13, mss22} as depicted in figure~\ref{fig:gridbetalin}; which is to say
\begin{equation}
\label{eq:betanlineartrans}
\beta(n) \sim n.
\end{equation}
\noindent
Let us next ask how to sum the lowest non-zero instanton contribution. For the generic transseries \eqref{eq:generictransseriesmatrixmodels} (albeit now focused on a string-equation solution as a practical example) this entails finding a closed-form for the sum
\be
\label{eq:lineartransasymptoticssum}
\sum_{n_1=1}^{+\infty}\, \underbrace{ \left( g_{\text{s}}^{b}\, \sigma\, \rme^{-\frac{1}{g_{\text{s}}}A_1} \right)^{n_1}}_{\equiv \upxi^{n_1}}\, R^{(n_1,0)}_{0}.
\ee
\noindent
This way of summing is illustrated in figure~\ref{fig:lineartransasymptotics}; at both leading and then sub-leading orders. To see how this calculation generically goes, let us perform the sum over instantons for the cubic matrix model\footnote{Of course in order to do so we have explicitly constructed the solutions to the string equations \eqref{eq:quarticstringequationthooftlimit} and \eqref{eq:freeenergyfromstringequation}. They are not important at this stage, the interested reader may see further details in sub-appendix~\ref{subapp:transasymptotic-transseries-CMM}.} \eqref{eq:CubicMatrixModelPotential}. At lowest genus and in the forward instanton direction we have $\beta_n = \frac{1}{2}\, n$, in which case from the transseries solution \eqref{eq:twoparameterresurgenttransseriesforR} for the recursion-coefficients string-equation \eqref{eq:cubicstringequationthooftlimit} one finds for the $R_0^{(n_1,0)}$ sum
\bea
\label{eq:linear-transasymptotics-example}
\frac{\sqrt{r}}{\sqrt[4]{1-3\lambda^2\, r}}\, \upxi - \frac{\lambda^2\, r}{2 \left(1-3\lambda^2\, r\right)^{3/2}}\, \upxi^2 + \frac{3\lambda^4\, r^{3/2}}{16 \left(1-3\lambda^2\, r\right)^{11/4}}\, \upxi^3 - \cdots &=& \\
&&
\hspace{-100pt}
= \frac{16 \sqrt{r}}{\sqrt[4]{1-3\lambda^2\, r}}\, \frac{1}{\left(4+\frac{\lambda^2 \sqrt{r}}{\left(1-3\lambda^2\, r\right)^{5/4}}\, \upxi \right)^2}, \nonumber
\eea
\noindent
where we have used as an adequate variable the solution $r$ to the classical string equation \eqref{eq:cubic-MM-classical-string-eq}. Note that for brevity we are only writing the first few terms in the instanton sum---albeit in order to find the correct analytic continuation we have computed the first 30 terms. In a similar manner we find for the free-energy $F^{(n_1,0)}_{0}$ sum via \eqref{eq:freeenergyfromstringequation},
\bea
\frac{\lambda^2 \sqrt{r}}{4 \left(1-3\lambda^2\, r\right)^{5/4}}\, \upxi - \frac{\lambda^4\, r}{32 \left(1-3\lambda^2\, r\right)^{5/2}}\, \upxi^2 + \frac{\lambda^6\, r^{3/2}}{192 \left(1-3\lambda^2\, r\right)^{15/4}}\, \upxi^3 + \cdots &=& \\
&&
\hspace{-100pt}
= \log \left( 1 + \frac{\lambda^2 \sqrt{r}}{4 \left(1-3\lambda^2\, r\right)^{5/4}}\, \upxi \right). \nonumber
\eea
\noindent
As a consistency check, these results exactly double-scale to the summation for \PI~performed in \cite{c97, cc01}. Of course in order to use linear transasymptotics to its full advantage we would further like to find closed-form expressions for \textit{all} the instanton sums
\be
\label{eq:lineartransasymptoticssumNM}
\sum_{n_1=1}^{+\infty} \left( g_{\text{s}}^{b}\, \sigma\, \rme^{-\frac{1}{g_{\text{s}}}A_1} \right)^{n_1} R^{(n_1,\widetilde{n}_1)}_{g},
\ee
\noindent
but which is not very illuminating generically. In addition, in resonant settings, this way of summing has the disadvantage that we would need to sum both instanton and negative-instanton directions one after each other---making resonance no longer manifest.
%
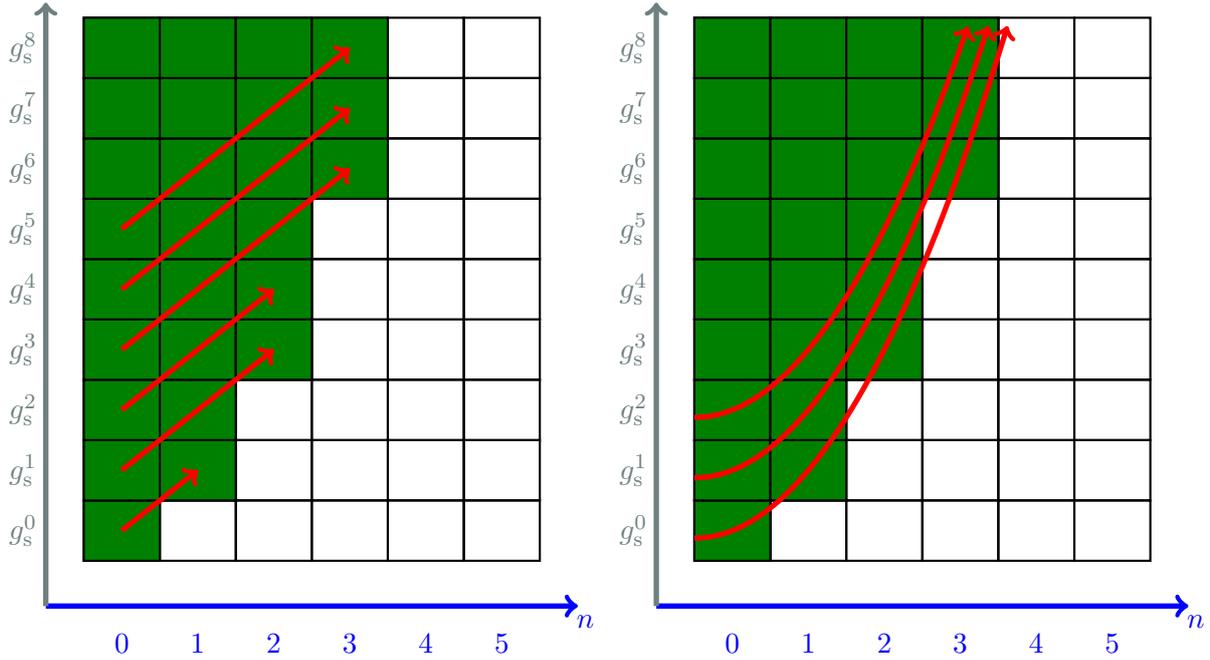
\begin{figure}
	\begin{subfigure}[t]{0.47\textwidth}
\centering
\definecolor{ao}{rgb}{0.0, 0.5, 0.0}
\definecolor{dgray}{rgb}{0.43, 0.5, 0.5}
\definecolor{dred}{rgb}{0.81, 0.09, 0.13}
\begin{tikzpicture}[grayframe/.style={
		rectangle,
		draw=gray,
		fill=white,
		text width=1.7em,
		align=center,
		rounded corners,
		minimum height=1em
	},
	box/.style={rectangle,draw=black,thick, minimum height=0.8cm, minimum width =1cm},
	]
	\foreach \x in {0,1,...,5}{
    \foreach \y in {0,1,...,8}
        \node[box] at (\x,0.8*\y){};
}
\foreach \y in {0,1,...,8}
        \node[box, fill=ao] at (0,0.8*\y){};
\foreach \y in {1,2,...,8}
        \node[box, fill=ao] at (1,0.8*\y){};
        \foreach \y in {3,4,...,8}
        \node[box, fill=ao] at (2,0.8*\y){};
        \foreach \y in {6,7,...,8}
        \node[box, fill=ao] at (3,0.8*\y){};
\draw[blue, line width=2pt, ->] (-1,-1) -- (6, -1);
\draw[dgray, line width=2pt, ->] (-1,-1) -- (-1, 7);
\node[blue] at (6.1, -1.2) {$n$};
\foreach \y in {0,1,...,8}
\node[dgray] at (-1.3,0.8*\y){$g_{\text{s}}^{\y}$};
\foreach \x in {0,1,...,5}
\node[blue] at (\x,-1.5){$\x$};
\draw[red, line width=2pt,->] (0,0) -- (1, 0.8*1);
\draw[red, line width=2pt,->] (0,0.8*1) -- (2, 0.8*3);
\draw[red, line width=2pt,->] (0,0.8*2) -- (2, 0.8*4);
\draw[red, line width=2pt,->] (0,0.8*3) -- (3, 0.8*6);
\draw[red, line width=2pt,->] (0,0.8*4) -- (3, 0.8*7);
\draw[red, line width=2pt,->] (0,0.8*5) -- (3, 0.8*8);
	\end{tikzpicture}
	\caption{Quadratic $\beta$-structure with linear summation.}
	\label{fig:quadratictrans1}
\end{subfigure}
\hspace{0.2cm}
	\begin{subfigure}[t]{0.47\textwidth}
\centering
\definecolor{ao}{rgb}{0.0, 0.5, 0.0}
\definecolor{dgray}{rgb}{0.43, 0.5, 0.5}
\begin{tikzpicture}[grayframe/.style={
		rectangle,
		draw=gray,
		fill=white,
		text width=1.7em,
		align=center,
		rounded corners,
		minimum height=1em
	},
	box/.style={rectangle,draw=black,thick, minimum height=0.8cm, minimum width =1cm},
	]
	\foreach \x in {0,1,...,5}{
    \foreach \y in {0,1,...,8}
        \node[box] at (\x,0.8*\y){};
}
\foreach \y in {0,1,...,8}
        \node[box, fill=ao] at (0,0.8*\y){};
\foreach \y in {1,2,...,8}
        \node[box, fill=ao] at (1,0.8*\y){};
        \foreach \y in {3,4,...,8}
        \node[box, fill=ao] at (2,0.8*\y){};
        \foreach \y in {6,7,...,8}
        \node[box, fill=ao] at (3,0.8*\y){};
\draw[blue, line width=2pt, ->] (-1,-1) -- (6, -1);
\draw[dgray, line width=2pt, ->] (-1,-1) -- (-1, 7);
\node[blue] at (6.1, -1.2) {$n$};
\foreach \y in {0,1,...,8}
\node[dgray] at (-1.3,0.8*\y){$g_{\text{s}}^{\y}$};
\foreach \x in {0,1,...,5}
\node[blue] at (\x,-1.5){$\x$};
	\draw[scale=1, domain=-0.5:3.62, smooth, variable=\x, red, line width=2pt, ->] plot ({\x}, {0.4*\x*\x+0.4*\x});
	\draw[scale=1, domain=-0.5:3.37, smooth, variable=\x, red, line width=2pt, ->] plot ({\x}, {0.4*\x*\x+0.4*\x+0.8});
	\draw[scale=1, domain=-0.5:3.1, smooth, variable=\x, red, line width=2pt, ->] plot ({\x}, {0.4*\x*\x+0.4*\x+1.6});
	\end{tikzpicture}
	\caption{Same structure with quadratic summation.}
	\label{fig:quadratictrans2}
\end{subfigure}
	\caption{Visualization of the \textit{quadratic} transasymptotic summation. We illustrate with  the grid of transseries sectors in the one-instanton direction as in the previous figure~\ref{fig:gridbetaquadratic}. In subfigure~\ref{fig:quadratictrans1} we show a quadratic $\beta$-structure with linear summation---herein the transasymptotics sums are finite. In subfigure~\ref{fig:quadratictrans2} we show the full quadratic transasymptotic summation---which turns out to be the most efficient way of resumming \textit{resonant} resurgent transseries \cite{asv17a, asv17b}.}
	\label{fig:quadratictransasymptotics}
\end{figure}
%
\item \textbf{Quadratic Transasymptotics:} \cite{asv17a, asv17b} In the previous case we were addressing either the specific-heat or the free-energy, which have linear $\beta$-structures. This ceases to be the case as one moves to the partition function. In fact, upon performing the exponentiation \eqref{eq:partitionfunctionexponentiation}, many cancellations take place and the resulting $\beta$-structure becomes much more restricted---it becomes \textit{quadratic} in the instanton numbers\footnote{This was already implicit in \cite{msw08} within the one-parameter case, where the partition-function $\beta$-structure quadratic-behavior arises from the need to regularize a conifold point (as already alluded to in \eqref{eq:gaussianregularizationF}).} \cite{msw08, mss22, sst23}. This is depicted in figure \ref{fig:quadratictransasymptotics}; and amounts to
\begin{equation}
\beta(n) \sim n^2.
\end{equation}
\noindent
If we keep forcing a linear summation, the sums become finite, and not much progress arises (see figure~\ref{fig:quadratictrans1}). The actual step forward occurs when also summing quadratically \cite{asv17a, asv17b}, as each sum is now effectively capturing an infinite amount of information (see figure~\ref{fig:quadratictrans2}). On top of this, all these sums may be directly performed in diagonal framing, which will be key for our formulae. To see how this calculation generically goes, let us again consider the example of the cubic matrix model. At lowest genus and in the forward instanton direction we have $\beta_n = \frac{1}{2}\, n^2$, in which case the partition-function sum becomes
\bea
\frac{\lambda^2 \sqrt{r}}{4 \left(1-3\lambda^2\, r \right)^{5/4}}\, \sqrt{g_{\text{s}}}\, \upxi - \frac{\lambda^8\, r^2}{256 \left(1-3 \lambda^2\, r\right)^5}\, g_{\text{s}}^2\, \upxi^2 - \frac{\lambda^{18}\, r^{9/2}}{131072 \left(1-3 \lambda^2\, r\right)^{45/4}}\, g_{\text{s}}^{9/2}\, \upxi^3 + \cdots &=& \nonumber \\
&&
\hspace{-280pt}
= \sum_{n=0}^{+\infty} \left(-\rmi\right)^{n}\, G_2 \left(n+1\right) \left( \frac{\rmi\lambda^2 \sqrt{r}}{4\left(1-3\lambda^2\, r\right)^{5/4}} \right)^{n^2} g_{\text{s}}^{\frac{n^2}{2}}\, \upxi^{n}.
\label{eq:Z-trans-G2-asympt}
\eea
\noindent
In contrast to linear transasymptotics, we have found a closed-form expression for each term in the sum but not yet for the full sum---which is now in fact asymptotic\footnote{The Barnes superfactorial growth is mitigated by the $n^2$ exponent in the coupling constant, so as to end up in precisely the standard factorial asymptotic growth. Indeed, one can write
\begin{equation}
\sum_{n=0}^{+\infty} G_2(n+1)\, g_{\text{s}}^{\frac{n^2}{2}} \left(\cdots\right) =\sum_{m=0}^{+\infty} G_2\left(\sqrt{2m}+1\right) g_{\text{s}}^{m} \left(\cdots\right),
\end{equation}
\noindent
where at large-$m$
\begin{equation}
G_2 \left(\sqrt{2m}+1\right) \simeq \exp \left( \frac{m}{2}\, \log m \right) \left(1+\cdots\right).
\end{equation}
\noindent
Using the Stirling approximation, we identify the leading large-$m$ behavior as $\sqrt{m!}$. Further splitting into even and odd powers of $g_{\text{s}}$, finally yields two series featuring standard factorial growth.}, albeit with no new novelties. In fact, it is important to stress that this asymptotic growth in the instanton direction does not give rise to any new nonperturbative effects as compared to the free energy; it is simply a rewrite of the standard growth due to \eqref{eq:partitionfunctionexponentiation}. Indeed, this exponentiation conspires to promote the (instanton direction) growth of the starting-genus power to $\sim n^2$. What this means is that the sum of the leading partition-function genus-coefficients along the instanton-direction actually probes ever deeper genera coefficients of the free energy, and thus is bound to inherit its typical genus asymptotic-growth. As such, the corresponding ``would be'' nonperturbative effect is not novel in any way and is in fact the very same as in the free energy. This will become clearer in the examples below and in appendix~\ref{app:transasymptotic-transseries}. Further, notice how the Barnes-function insertion should be rather reminiscent of the Gaussian partition-function insertion in \eqref{eq:partition-function-Z-msw08}.
\end{enumerate}
\noindent
Consequently, let us apply the above procedure to our many specific examples by spelling out the various transseries and explicitly evaluating all quadratic transasymptotic summations.

\paragraph{Painlev\'e~I:}

\begin{table}
\centering
\begin{tabular}{c|c}
$k$ & $\widetilde{D}_{k}(\nu)$\\\hline
$0$ & $1$ \\
$1$ & $-\frac{1}{96} \rmi \nu \left(94\nu^2+17\right)$ \\
$2$ & $-\frac{\left(8836\nu^4+34064\nu^2+14997\right) \nu^2}{18432}-\frac{7}{480}$ \\
$3$ & $ \frac{\rmi \nu \left(156847302\nu^4+124622833\nu^2+940 \left(4418\nu^2+48699\right) \nu^6+13059000\right)}{26542080}$ \\
$4$ & {\footnotesize{$\frac{5 \left(305174960717\nu^2+8 \left(8661460269\nu^2+22090 \left(2209\nu^2+47900\right) \nu^4+31269469550\right) \nu^4+84170674656\right) \nu^2+5424427008}{50960793600}$}} \\
\end{tabular}
\caption{Some examples of the $\widetilde{D}_k(\nu)$ polynomials for \PI. For our calculations in the main text and in sub-appendix~\ref{subapp:transasymptotic-transseries-PI}, we have explicitly computed them up until $k=45$.}
\label{tab:PIDkPolynomials}
\end{table}

Start with the \PI~equation in \eqref{eq:Painleve1Equation} alongside its transseries solution in \eqref{eq:Painleve1SOlution}. The explicit computations associated to the formulae we present in the following, including rectangular and diagonal formulations, are detailed in sub-appendix~\ref{subapp:transasymptotic-transseries-PI}, for specific heat, free energy, and partition function (for previous related discussions, see, \textit{e.g.}, \cite{gikm10, asv11, as13, bssv22}). 
\begin{itemize}
\item \textbf{Rectangular Framing:} Start with the partition-function in rectangular framing, 
\begin{equation}
\label{eq:PainleveIPartitionFunctionRectangular}
Z \left( x; \sigma_1,\sigma_2 \right) = \rme^{F_{\text{A}} \left(x;\upmu\right)} \sum_{n,m=0}^{+\infty} \sigma_{1}^{n} \sigma_{2}^{m}\, \rme^{- \left(n-m\right) \frac{A}{x}} \sum_{k=0}^{k_{nm}} \left( \frac{\log(x)}{2} \right)^{k} \sum_{g=0}^{+\infty} Z^{(n|m)[k]}_{2g}\, x^{g+\beta^{[k]}_{nm}},
\end{equation}
\noindent
where the pre-factor reads
\begin{equation}
\label{eq:FAPainleveI}
F_{\text{A}} \left( x; \sigma_{1}\sigma_{2} \right) = -\frac{4}{15 x^2} + \frac{16}{5x}\, \sigma_1\sigma_2 + \frac{4}{5} \left( \frac{1}{48} + \frac{5}{6} \left( \sigma_1 \sigma_2 \right)^2 \right) \log x,
\end{equation}
\noindent
and the $\beta$-structure is specified in sub-appendix~\ref{subapp:transasymptotic-transseries-PI}. \eqref{eq:PainleveIPartitionFunctionRectangular} is the partition-function analogue of \eqref{eq:generictransseriesmatrixmodels}, in which case all we said concerning rectangular versus diagonal framing, alongside linear and quadratic transasymptotic resummations, holds herein \textit{verbatim}.
\item \textbf{Diagonal Framing:} We can reformulate the above rectangular-framing transseries in diagonal form by following the procedure we just described. Use diagonal variable $\upmu \equiv \sigma_1 \sigma_2$ and introduce adequate resummation parameters (compare with \eqref{eq:defzeta} and \eqref{eq:XiMuRelation})
\begin{equation}
\label{eq:defxi}
\upxi_{1} \equiv \sqrt{x}\, \sigma_{1}\, \rme^{-\frac{A}{x}}\, x^{-\frac{2}{\sqrt{3}} \upmu}, \qquad \upxi_{2} \equiv \sqrt{x}\, \sigma_{2}\, \rme^{\frac{A}{x}}\, x^{\frac{2}{\sqrt{3}} \upmu}.
\end{equation}
\noindent
The term $x^{\frac{2}{\sqrt{3}} \upmu}$ absorbs all the logarithmic dependence in \eqref{eq:PainleveIPartitionFunctionRectangular} (see sub-appendix~\ref{subapp:transasymptotic-transseries-PI}). Then we find for the partition function in diagonal framing
\begin{equation}
\label{eq:PainleveIPartitionFunctionDiagonalFraming}
Z \left( x; \upxi_{1}, \upmu \right) = \rme^{F_{\text{A}} \left( x; \upmu \right)}\, \sum_{g=0}^{+\infty} \sum_{\alpha=-g}^{+\infty} \upxi_{1}^{\alpha}\, \mathsf{Z}^{(\alpha)}_{g} (\upmu)\, x^{g},
\end{equation}
\noindent
where all relevant quantities are introduced in sub-appendix~\ref{subapp:transasymptotic-transseries-PI}. Some comments are in order. As anticipated, \textit{e.g.}, the main-diagonal coefficients $\mathsf{Z}^{(0)}_{g} (\upmu)$ are now  polynomials in $\upmu$, of degree $3g$ (see some such examples in sub-appendix~\ref{subapp:transasymptotic-transseries-PI}, equations \eqref{eq:Zmu-poly-examples-1} through \eqref{eq:Zmu-poly-examples-9}). Notice also that negative values of $\alpha$ are included in the summation. This expression is the \PI~partition-function analogue of \eqref{eq:diagonalframingF}. Its coefficients satisfy a version of the backward-forward symmetry \cite{bssv22},
\begin{equation}
\mathsf{Z}^{(-\alpha)}_{g} (\upmu) = (-1)^{g+\alpha}\, \upmu^{\alpha}\, \mathsf{Z}^{(\alpha)}_{g-\alpha} (-\upmu), \qquad \alpha>0.
\end{equation}
\noindent
and they may be inverted in order to recover the original rectangular-framing coefficients\footnote{The notation $\left. \bullet \right|_{m}$ selects the degree-$m$ term in $\upmu$, out from the $\bullet$-expression.}
\bea
Z_{2g}^{(n|m)[0]} &=& \left. \mathsf{Z}^{(n-m)}_{g+\left\lfloor\frac{m+1}{2}\right\rfloor} (\upmu) \right|_{m}, \qquad n-m > 0, \\
Z_{2g}^{(n|n)[0]} &=& \left. \mathsf{Z}^{(0)}_{g+n} (\upmu) \right|_{n}.
\eea
\item \textbf{Quadratic Transasymptotics:} Having computed the partition function in diagonal framing we can turn to its quadratic transasymptotic formulation. For this endeavour we have computed $50$ instanton/negative-instanton contributions, up to genus $60$, for which we wanted to find closed-form expressions for $\mathsf{Z}^{(\alpha)}_{g} (\upmu)$ along the $\alpha$-direction. This is indeed possible and we conjecture (and check against our data---see the derivation leading up to \eqref{eq:diagonalframingresultapp} in sub-appendix~\ref{subapp:transasymptotic-transseries-PI}) the closed-form
\bea
\label{eq:diagonalframingresult}
Z \left( z; \upxi_{1}, \upmu \right) &=& \rme^{- \frac{4}{15} z^{\frac{5}{2}} + \frac{16}{5} \upmu  z^{\frac{5}{4}}}\, z^{- \frac{1}{48} - \frac{5}{6} \upmu^2}\, \sum_{g=0}^{+\infty}\, \sum_{n \in\mathbb{Z}} \upxi_{1}^{n}\, z^{- \frac{5}{4} \left( \frac{1}{2} n \left(n-1\right) + g \right)}\, \times \\
&&
\hspace{-65pt}
\times \underbrace{\widetilde{D}_{g} \left(n-\frac{2}{\sqrt{3}}\, \upmu\right) \left\{ \left(-1\right)^{g} \left(-\frac{1}{12}\right)^{n+\frac{g}{2}} \left(-96\sqrt{3}\right)^{-\frac{1}{2} n \left(n-1\right)} \frac{G_2 \left(1+n-\frac{2}{\sqrt{3}}\, \upmu\right)}{G_2 \left(1-\frac{2}{\sqrt{3}}\, \upmu\right) \Gamma \left(1-\frac{2}{\sqrt{3}}\, \upmu\right)^n} \right\}}_{\equiv \mathsf{Z}^{(n)}_{g+\frac{1}{2} n \left(n-1\right)} (\upmu)}, \nonumber
\eea
\noindent
We will start unveiling the role of the Barnes and Gamma functions in the next paragraph, as we rewrite this expression in the form of a discrete Fourier transform. The $\widetilde{D}_g (\nu)$ are degree $3g$ polynomials which follow from the partition-function in diagonal-framing, and the first few are displayed in table~\ref{tab:PIDkPolynomials} (to the extent that they have been computed, they exactly match the same-notation polynomials in \cite{blmst16}). It is clearly convenient to change their normalization, which we often do, as
\begin{equation}
\label{eq:Dktilde-versus-Dknontilde}
\widetilde{D}_{g} (\nu) \equiv \rmi^{g}\, 2^{g}\, 3^{\frac{g}{2}}\, D_{g} (\nu) \equiv \rmi^{g}\, 2^{g}\, 3^{\frac{g}{2}}\, \mathsf{Z}_{g}^{(0)} \left(-\sqrt{\frac{3}{4}}\, \nu\right).
\end{equation}
\item \textbf{Discrete Fourier Transform:} One of our key points is that the previous, quadratic-transasymptotically resummed partition-function \eqref{eq:diagonalframingresult}, can now be very conveniently rewritten as a discrete Fourier transform
\begin{equation}
\label{eq:PIDiscreteFourierNC}
Z \left( z; \uprho, \upmu \right) = \NCP (\upmu) \sum_{\ell\in\mathbb{Z}} \uprho^{\ell}\, \mathcal{Z} \left( z; \ell-\frac{2}{\sqrt{3}}\, \upmu \right),
\end{equation}
\noindent
where the above ``discrete Fourier modes'' are given by\footnote{Note that \eqref{eq:PIDiscreteFourierNC} exactly matches against equation (3.9) in \cite{blmst16} upon rotation of $x=z^{-5/4}$ by $-2\pi$.}
\be
\label{eq:DFTKernelPINC}
\mathcal{Z} \left( z; \nu \right) = z^{-\frac{1}{48}}\, \rme^{- \frac{4}{15} z^{\frac{5}{2}}} \left( -\frac{1}{96\sqrt{3}} \right)^{\frac{\nu^2}{2}} \exp \left( - \frac{8\sqrt{3}}{5}\, z^{\frac{5}{4}}\,  \nu \right) \frac{G_2 \left(1+\nu\right)}{\left(2\pi\right)^{\frac{\nu}{2}}}\, \sum_{k=0}^{+\infty} D_k (\nu)\, z^{-\frac{5}{8}\nu^2-\frac{5}{4}k},
\ee
\noindent
Let us point out some informative numerology in the above expressions. The $-\frac{4}{15}$ is the planar (perturbative) free energy whereas the $-\frac{1}{48}$ is its genus-one. The $\frac{8 \sqrt{3}}{5}$ is the instanton action, the $\frac{2}{\sqrt{3}}$ the logarithmic resummation constant, and the $96 \sqrt{3}$ the ``normalization'' of the Stokes-data generating-function \cite{bssv22}. Indeed, the ``discrete Fourier variable'' $\uprho$ is related to the original transseries-parameter $\sigma_1$ via
\begin{equation}
\label{eq:DFTrhoP1}
\uprho = \frac{\sigma_1}{S_1}\, \rme^{\frac{2\pi\rmi}{\sqrt{3}}\, \upmu}\, \frac{\left( 96\sqrt{3} \right)^{-\frac{2}{\sqrt{3}}\, \upmu}}{\Gamma \left( 1 - \frac{2}{\sqrt{3}}\, \upmu \right)}.
\end{equation}
\noindent
where we recall the \PI~canonical Stokes coefficient
\begin{equation}
S_{1} = -\rmi\, \frac{3^{1/4}}{2\sqrt{\pi}}.
\end{equation}
\noindent
Another of our key points is that this Fourier-variable is exactly given by the generating function\footnote{To make this precise, and successively using the notation in \cite{bssv22}, one has
\be
\left. \uprho_\pi \right|_{\text{there}} = \sigma_1\, N^{(-1)} \left( \upmu \right) = - \rmi\, \rme^{\frac{2\pi\rmi}{\sqrt{3}}\, \upmu}\, \sigma_1\, N^{(1)} (-\upmu) = - \rmi\, \rme^{\frac{2\pi\rmi}{\sqrt{3}}\, \upmu}\, \sigma_1 \underbrace{\left( - \rmi\, \frac{\sqrt[4]{3}}{2\sqrt{\pi}} \right)}_{\text{canonical Stokes}} \frac{\left( 96 \sqrt{3} \right)^{-\frac{2 \upmu}{\sqrt{3}}}}{\Gamma \left( 1 - \frac{2 \upmu}{\sqrt{3}} \right)}.
\ee} of non-linear resurgent Stokes data for \PI, as computed in \cite{bssv22}, upon the small change of normalization
\begin{equation}
\rmi\, S_1^2 \left. \uprho \right|_{\text{here}} = \left. \uprho_\pi \right|_{\text{there}}.
\end{equation}
\noindent
This illustrates how even if we are fully ready to give up on constructing transseries, due to the much neater compact expression in \eqref{eq:PIDiscreteFourierNC}, it is still the case that we need to address resurgent Stokes data in order to actually write down \eqref{eq:PIDiscreteFourierNC} in the first place (and, later, to achieve global rather than simply local solutions \cite{krsst26b}). These aspects will be further explored in the subsequent papers \cite{krsst26b, ss26, krst26a, krst26b}. Finally, there is a pre-factor in \eqref{eq:PIDiscreteFourierNC}, which is solely a function of $\upmu$ and reads
\begin{equation}
\label{eq:DFTnormalizationP1}
\NCP (\upmu) = \left( 3 \cdot 2^{\frac{10}{3}} \right)^{\upmu^2}\, \frac{\rme^{- \frac{2\pi\rmi}{3}\, \upmu^2}}{\left(2\pi\right)^{\frac{\upmu}{\sqrt{3}}}\, G_2 \left( 1 - \frac{2}{\sqrt{3}}\, \upmu \right)}.
\end{equation}
\noindent
Because this pre-factor is independent of the string-coupling---herein, independent of $z$---it just constitutes a choice of normalization of the partition function. Normalizations are, of course, just example-dependent convenient choices---and herein this choice arises from the way the \PI~specific-heat transseries was initially constructed \cite{gikm10, asv11, as13, bssv22}. For further details and notation we refer the reader to sub-appendix~\ref{subapp:transasymptotic-transseries-PI}.
%
\newcommand{\thetaf}[4]{
\vartheta{\begin{bmatrix} #1 \\ #2 \end{bmatrix} } \left( #3 \, | \, #4\right)}
%
\item \textbf{Theta Function:} Yet another way to repackage our partition function in \eqref{eq:PIDiscreteFourierNC} is via the use of Riemann theta-functions with characteristics (see \eqref{eq:riemanntheta}-\eqref{eq:riemannthetacharacteristics} in appendix~\ref{app:elliptic-theta-modular}). Expanding \eqref{eq:PIDiscreteFourierNC} and rewriting\footnote{Whether this is a well-defined power-series expansion in the one-cut case will be discussed in \cite{ss26}. At this stage this is purely a formal rewrite, with no good expansion parameter, and stands solely for comparison purposes.} one obtains\footnote{Notice how comparing with \eqref{eq:EM-2-cut-Z} and \eqref{eq:theta-for-EM-2-cut-Z} both theta-function characteristics are now turned on.}
\bea
\label{eq:PItheta}
Z \left( z; \uprho, \upmu \right) &=& z^{-\frac{1}{48}}\, \rme^{-\frac{4}{15} z^{5/2}}\, \uprho^{-\upmu}\, \NCP (\upmu) \sum\limits_{k=0}^{+\infty} f_k (z)\, \partial_{\chi}^k \thetaf{\upmu}{\frac{1}{2\pi\rmi} \log \uprho}{\chi}{\tau} = \\
&&
\hspace{-50pt}
= z^{-\frac{1}{48}}\, \rme^{-\frac{4}{15} z^{5/2}}\, \uprho^{-\upmu}\, \NCP (\upmu) \left\{ \thetaf{\upmu}{\frac{1}{2\pi\rmi} \log \uprho}{\chi}{\tau} + f_1 (z)\, \partial_{\chi} \thetaf{\upmu}{\frac{1}{2\pi\rmi} \log \uprho}{\chi}{\tau} + \cdots \right\}, \nonumber
\eea
\noindent
where the arguments in the theta-functions are
\bea
\chi &=& - \frac{1}{2\pi\rmi} \left( \frac{8\sqrt{3}}{5}\, z^{\frac{5}{4}} + \frac{1}{3}\, \log 2\pi \right), \\
\tau &=& - \frac{5}{8}\, \log z - \frac{1}{2}\, \log 96 \sqrt{3} + \frac{3\pi\rmi}{2},
\eea
\noindent
and the $f_k (z)$ are the coefficients of the expansion of the Fourier modes \eqref{eq:DFTKernelPINC} in powers of $\nu$ (and where $f_0 (z) = 1$). The emergence of a theta-function structure might make one tempted to directly compare this expression to the results in \cite{em08}; but whilst there are relations they are not exactly the same. As already discussed in subsection~\ref{subsec:Z-phases}, \cite{em08} starts life in the two-cut regime whereas our formulae start out in the one-cut. Details on the comparisons will appear in the following subsection~\ref{subsec:from-theta-to-dual-to-transseries}.
\end{itemize}

The bottom line is, as discussed above and in excruciating detail in sub-appendix~\ref{subapp:transasymptotic-transseries-PI}, that whatever form of writing the transseries one picks, they all contain the exact same information---be it rectangular framing, diagonal framing, linear or quadratic transasymptotics, the discrete Fourier transform or its theta-function version; and regardless if one is to address the specific heat, the free energy, or the partition function. What automatically pops-out from string equations are rectangular-framing transseries for the specific heat. And the most compact way to encode all that amount of information is via the discrete Fourier transform for the partition function. At the end of the day, showing that one matches the other is all one really cares about (for all other aforementioned incarnations immediately follow). Hence, going forward, we shall mostly focus on these two ``extreme points'' of this long chain of equivalences.

\paragraph{Yang--Lee:}

Start with the \YL~equation in \eqref{eq:YangLeeEquation} alongside its transseries solution in \eqref{eq:YangLeeSolution}. The explicit computations associated to the formulae we present in the following---mostly focusing on the rectangular-framing partition-function, which itself may be directly compared to the discrete Fourier transform we will present in the following---are detailed in sub-appendix~\ref{subapp:transasymptotic-transseries-YL}. Although this problem is of much greater complexity that the previous \PI~case, the set of calculations to be done is still essentially the same (albeit much larger). More details will appear in \cite{krsst26b, krst26a}.
\begin{itemize}
\item \textbf{Rectangular Framing:} Start with the partition-function in rectangular framing,
\be
\label{eq:YangLeePartitionFunctionRectangular}
Z \left( x; \boldsymbol{\sigma} \right) = \rme^{F_{\text{A}} \left(x;\upmu_1,\upmu_2\right)} \sum_{\boldsymbol{n}=0}^{+\infty} \boldsymbol{\sigma}^{\boldsymbol{n}}\, \rme^{- \left( n_1-n_2 \right) \frac{A_1}{x}}\, \rme^{- \left( n_3-n_4 \right) \frac{A_2}{x}} \sum_{k=0}^{k_{\boldsymbol{n}}} \left( \frac{\log x}{2} \right)^{k}\, \sum_{g=0}^{+\infty} Z^{(\boldsymbol{n})[k]}_{g}\, x^{g+\beta^{[k]}_{\boldsymbol{n}}},
\ee
\noindent
where $\upmu_1 \equiv \sigma_{1}\sigma_{2}$, $\upmu_2 \equiv \sigma_3 \sigma_4$, and the pre-factor reads 
\bea
\label{eq:FAYangLee}
F_{\text{A}} \left( x; \upmu_1,\upmu_2 \right) &=& -\frac{9}{56 x^2} + \frac{6 \left(1-\rmi \sqrt{5}\right)}{7 x}\, \upmu_1 + \frac{6 \left(1+\rmi \sqrt{5}\right)}{7 x}\, \upmu_2 + \\
&&
\hspace{50pt}
+ \left( \frac{1}{42} - \frac{1}{10} \rmi \left(\sqrt{5}-5\rmi\right) \upmu_1^2 + \frac{1}{10} \rmi \left(\sqrt{5}+5\rmi\right) \upmu_2^2 \right) \log x. \nonumber
\eea
\noindent
The $\beta^{[k]}_{\boldsymbol{n}}$-structure dictates the starting genus of the transseries, which is quadratic in the instanton numbers as discussed in table~\ref{tab:betaZYangLee} of sub-appendix~\ref{subapp:transasymptotic-transseries-YL}.
\item \textbf{Discrete Fourier Transform:} We jump over the two middle-steps, of diagonal framing and of quadratic transasymptotics, and move to the endgame. Of course one could have discussed these in detail, but due to the rise in algebraic complexity of the \YL~problem this would have been rather tedious and non-enlightening. Instead, one can take the more pragmatic approach of simply stating the final result for the discrete Fourier transform, from which these other two rewritings may be easily extracted. The partition function may hence now be very conveniently rewritten as (compare with \eqref{eq:PIDiscreteFourierNC})
\be
\label{eq:YLFullDFT}
Z \left( z; \boldsymbol{\uprho}, \boldsymbol{\upmu} \right) = \NCP (\boldsymbol{\upmu}) \sum_{\ell_1 \in \BZ} \sum_{\ell_2 \in \BZ}  \uprho^{\ell_1}_1 \uprho^{\ell_2}_2\, \mathcal{Z} \left( z; \ell_1-\frac{\alpha}{2}\, \upmu_1, \ell_2+\frac{\bar{\alpha}}{2}\, \upmu_2 \right),
\ee
\noindent
where $\alpha = 2 \rmi \sqrt{1+\frac{\rmi}{\sqrt{5}}}$ is the \YL~logarithmic resummation constant, which we already encountered below equation \eqref{eq:YangLeeEquation}. Written in this form, and just like what happened for \PI, the partition function seems deceivingly simple. To be clear on this repackaging, let us be very explicit concerning all the information encoded in the above formula. Akin to \PI, the pre-factor $\NCP (\boldsymbol{\upmu})$ is solely a function of the moduli---independent of $z$---and amounts to a choice of normalization\footnote{Again, normalizations are just example-dependent convenient choices; and the current choice is just the one corresponding to our specific transseries solution to the \YL~string equation \cite{krst26a}.} of the partition function. It is given by
\be
\NCP (\boldsymbol{\upmu}) = \text{B}^{\frac{1}{8} \alpha^2 \upmu_1^2}\, \text{C}^{\frac{1}{4} \alpha \bar{\alpha}\, \upmu_1 \upmu_2}\, \bar{\text{B}}^{\frac{1}{8} \bar{\alpha}^2 \upmu_2^2}\, \frac{\rme^{\frac{\rmi\pi}{8} \bar{\alpha}^2 \upmu_2^2}}{\left(2\pi\right)^{\frac{\alpha}{4} \upmu_1} G_2 \left( 1 - \frac{\alpha}{2}\, \upmu_1 \right) \left( 2\pi \right)^{-\frac{\bar{\alpha}}{4} \upmu_2} G_2 \left( 1 + \frac{\bar{\alpha}}{2}\, \upmu_2 \right)},
\ee
\noindent
where we denote the pure numbers $\text{B} = 8 \sqrt{125 - 95\rmi \sqrt{5}}$ and $\text{C} = - 11 - 2 \sqrt{30}$. The Fourier parameters $\uprho_1$ and $\uprho_2$ will relate to the generating functions of non-linear resurgent Stokes data of \YL, as will be discussed in \cite{krsst26b, krst26a}; and for the moment further relate to the original transseries parameters as
\be
\label{eq:DFTrhosYL}
\uprho_1 = - \sigma_2\, \sqrt{- \pi \alpha}\, \frac{\text{B}^{-\frac{\alpha}{2}\, \upmu_1}\, \text{C}^{-\frac{\bar{\alpha}}{2}\, \upmu_2}}{\Gamma \left( 1 - \frac{\alpha}{2}\, \upmu_1 \right)} \quad \text{ and } \quad 
\uprho_2 = -\rmi\, \rme^{\frac{\rmi \pi \bar{\alpha}\upmu_2}{2}}\, \sigma_3\, \sqrt{- \pi \bar{\alpha}}\, \frac{\bar{\text{B}}^{\frac{\bar{\alpha}}{2}\, \upmu_2}\, \text{C}^{\frac{\alpha}{2}\, \upmu_1}}{\Gamma \left( 1 + \frac{\bar{\alpha}}{2}\, \upmu_2 \right)}.
\ee
\noindent
Finally, we need to address the Fourier modes. For these, one finds
\bea
\mathcal{Z} \left( z; \nu_1, \nu_2 \right) &=& z^{-\frac{1}{36}}\, \rme^{-\frac{9}{56} z^{\frac{7}{3}}}\, \rme^{-\frac{\rmi\pi}{2} \nu_2^2}\, \text{B}^{-\frac{1}{2} \nu_1^2}\, \text{C} ^{\nu_1\nu_2}\, \bar{\text{B}}^{-\frac{1}{2} \nu_2^2}\, \exp \left( A_1\, z^{\frac{7}{6}}\, \nu_1 \right) \exp \left( - A_2\, z^{\frac{7}{6}}\, \nu_2 \right) \times \nonumber \\
&&
\times \frac{G_2 \left(1+\nu_1\right)}{\left(2\pi\right)^{\frac{\nu_1}{2}}}\, \frac{G_2 \left(1+\nu_2\right)}{\left(2\pi\right)^{\frac{\nu_2}{2}}}\, \sum_{k=0}^{+\infty} D_k (\nu_1,\nu_2)\, z^{- \frac{7}{12} \nu_1^2 - \frac{7}{12} \nu_2^2 - \frac{7}{6} k}.
\label{eq:curlyZYangLeeFullExpression}
\eea
\noindent
As for \PI, let us point out some informative numerology in the above expressions. The $-\frac{9}{56}$ is the planar (perturbative) free energy whereas the $-\frac{1}{36}$ is its genus-one. The $A_{1,2}$ are the instanton actions as in \eqref{eq:YangLeeSolution}, and the $\text{B}$, $\text{C}$ are the ``normalizations'' of the (canonical/non-canonical) Stokes-data generating functions \cite{krsst26b, krst26a}. Expressions are not symmetric in the $\upmu_{1,2}$ entries as our discrete Fourier transforms are (local) formulae written with respect to a specific (global) Stokes line, and different coordinates are relevant for different such Stokes lines \cite{bssv22, krsst26b, krst26a, krst26b}. Further, the functions $D_k (\nu_1,\nu_2)$ are again polynomials of degree $3k$, which now further enjoy the following two symmetries\footnote{These are precisely the expected backward-forward and complex-conjugacy symmetries of the coefficients in the original rectangular-framing transseries, which are now manifesting themselves as simple symmetries of the $D_k (\nu_1,\nu_2)$ polynomials. This is likely to be the structure for the entire multicritical hierarchy, and to some extent reveals the origin of these symmetries for the transseries coefficients; see as well the discussions in \cite{krst26a, krst26b}.}
\be
\label{eq:YLSymmetriesOfDkPolynomials}
D_k (\nu_1,\nu_2) = (-1)^{k}\, \overline{D_k (\bar{\nu}_2,\bar{\nu}_1)}, \qquad  D_k (\nu_1,\nu_2) = (-1)^{k}\, D_k (-\nu_1,-\nu_2).
\ee
\noindent
For completeness, and actually being able to compare the above form of the partition-function to its usual rectangular framing, let us give the first couple of $D_k (\nu_1,\nu_2)$ polynomials explicitly, as
\bea
\label{eq:DkPolynomialsYangLee-1}
D_1 (\nu_1,\nu_2) &=& \frac{\sqrt{650310 - 433326 \rmi \sqrt{5}}}{2160}\, \nu_1^3 - \frac{\sqrt{650310 + 433326 \rmi \sqrt{5}}}{2160}\, \nu_2^3 - \\
&&
\hspace{-60pt}
- \frac{1}{30} \sqrt{- 955 + 119 \rmi \sqrt{5}}\, \nu_1^2\, \nu_2 + \frac{1}{30} \sqrt{- 955 - 119 \rmi \sqrt{5}}\, \nu_1\, \nu_2^2 + \frac{\sqrt{75750 - 46254 \rmi \sqrt{5}}}{4320}\, \nu_1 - \nonumber \\
&&
\hspace{-60pt}
- \frac{\sqrt{75750 + 46254 \rmi \sqrt{5}}}{4320}\, \nu_2, \nonumber \\
\label{eq:DkPolynomialsYangLee-2}
D_2 (\nu_1,\nu_2) &=& \frac{\left(108385 - 72221 \rmi \sqrt{5}\right)}{1555200}\, \nu_1^6 + \frac{\left(108385 + 72221 \rmi \sqrt{5}\right)}{1555200}\, \nu_2^6 - \\
&&
\hspace{-60pt}
- \frac{\left(3637 \sqrt{5} + 11255 \rmi\right) }{10800 \sqrt{6}}\, \nu_1^5\, \nu_2 + \frac{\left( - 3637 \sqrt{5} + 11255 \rmi\right)}{10800 \sqrt{6}}\, \nu_1\, \nu_2^5 + \frac{\left(- 7695 - 1664 \rmi \sqrt{5}\right)}{10800}\, \nu_1^4\, \nu_2^2 + \nonumber \\
&&
\hspace{-60pt}
+ \frac{\left( - 7695 + 1664 \rmi \sqrt{5}\right)}{10800}\, \nu_1^2\, \nu_2^4 - \frac{191893}{25920 \sqrt{30}}\, \nu_1^3\, \nu_2^3 + \frac{\left( 13987 - 11933 \rmi \sqrt{5}\right)}{77760}\, \nu_1^4 + \nonumber \\
&&
\hspace{-60pt}
+ \frac{\left( 13987 + 11933 \rmi \sqrt{5}\right)}{77760}\, \nu_2^4 - \frac{\left(99407 \sqrt{5} + 142368 \rmi\right)}{51840 \sqrt{6}}\, \nu_1^3\, \nu_2 + \frac{\left( - 99407 \sqrt{5} + 142368 \rmi\right)}{51840 \sqrt{6}}\, \nu_1\, \nu_2^3 - \nonumber \\
&&
\hspace{-60pt}
- \frac{1879}{720}\, \nu_1^2\, \nu_2^2 + \frac{\left( 84375 - 78827 \rmi \sqrt{5}\right)}{2073600}\, \nu_1^2 + \frac{\left( 84375 + 78827 \rmi \sqrt{5}\right)}{2073600}\, \nu_2^2 - \frac{301789}{103680 \sqrt{30}}\, \nu_1\, \nu_2 + \frac{1}{240}. \nonumber
\eea
\noindent
Having outlined the various components making up the \YL~partition function, we have found perfect agreement between the discrete Fourier transform solution laid out above and the rectangular-framing partition-function transseries discussed in sub-appendix~\ref{subapp:transasymptotic-transseries-YL}.
\end{itemize}

\paragraph{Cubic Matrix Model:}

Moving to our off-criticality examples, start with the cubic matrix model \eqref{eq:CubicMatrixModelPotential} with string equation \eqref{eq:cubicstringequationthooftlimit} (alongside free energy counterpart \eqref{eq:freeenergyfromstringequation}) and with transseries solution of the form \eqref{eq:twoparameterresurgenttransseriesforR}. We have explicitly constructed solutions to all these equations (details and notations are outlined in sub-appendix~\ref{subapp:transasymptotic-transseries-CMM}; and for previous related discussions, see, \textit{e.g.}, \cite{msw08, kmr10, mss22}). Herein we mainly focus on the partition function, but, as for \PI, we still cover its different facets: starting from rectangular and diagonal formulations, towards quadratic transasymptotics including discrete-Fourier-transform and theta-function formulations.
\begin{itemize}
\item \textbf{Rectangular Framing:} Start with the partition-function in rectangular framing,
\bea
\label{eq:CubicPartitionFunctionRectangular}
Z \left( t, g_{\text{s}}; \sigma_1, \sigma_2 \right) &=& \rme^{F_{\text{A}} \left( g_{\text{s}}; \upmu \right)} \sum_{n,m=0}^{+\infty} \sigma_{1}^{n} \sigma_{2}^{m}\, \rme^{- \left(n-m\right) \frac{A (t)}{g_{\text{s}}}} \times \\
&&
\hspace{50pt}
\times \sum_{k=0}^{k_{n m}} \log \left( \frac{\left( 1 - 3 \lambda^2\, r \right)^5}{108 \lambda^8\, r^2\, g_{\text{s}}^2} \right)^{k} \sum_{g=0}^{+\infty} Z^{(n|m)[k]}_{2g} (t)\,  g_{\text{s}}^{g+\beta^{[k]}_{nm}}, \nonumber
\eea
\noindent
where the instanton action is encoded in \eqref{eq:cubicaction} (see as well sub-appendix~\ref{subappendix:cubicmatrixmodel}), we have used the classical string solution from \eqref{eq:cubic-MM-classical-string-eq} with $r \equiv r(t)$, and the global pre-factor reads
\be
F_{\text{A}} \left( g_{\text{s}}; \sigma_1 \sigma_2 \right) = \frac{1}{g_{\text{s}}^2}\, F^{(0|0)}_{0} (t) + F^{(0|0)}_1 (t) + \frac{1}{g_{\text{s}}}\, F^{(1|1)}_0 (t)\, \upmu + F^{(2|2)}_0 (t)\, \upmu^2,
\ee
\noindent
where these free-energy transseries-coefficients are presented in sub-appendix~\ref{subapp:transasymptotic-transseries-CMM}.
\item \textbf{Diagonal Framing:} We can reformulate the above rectangular-framing transseries in diagonal form by following the exact same procedure we described earlier for \PI. With the usual diagonal variable $\upmu \equiv \sigma_1 \sigma_2$, introduce adequate resummation parameters (compare with \eqref{eq:defxi}, or \eqref{eq:defzeta} and \eqref{eq:XiMuRelation})
\bea
\label{eq:zetaxiCMM-1}
\upzeta_{1} &=& \frac{\upxi_{1}}{\sqrt{g_s}} \equiv \sigma_{1}\, \rme^{-\frac{A(t)}{g_{\text{s}}}} \left(\frac{\left( 1 - 3 \lambda^2\, r \right)^5}{108 \lambda^8\, r^2\, g_{\text{s}}^2}\right)^{\frac{1}{2} \upmu}, \\
\label{eq:zetaxiCMM-2}
\upzeta_{2} &=& \frac{\upxi_{2}}{\sqrt{g_s}} \equiv \sigma_{2}\, \rme^{\frac{A(t)}{g_{\text{s}}}} \left(\frac{\left( 1 - 3 \lambda^2\, r \right)^5}{108 \lambda^8\, r^2\, g_{\text{s}}^2}\right)^{-\frac{1}{2} \upmu}.
\eea
\noindent
Very much in parallel with \PI, it follows that the partition function in diagonal framing is of the form (compare with \eqref{eq:PainleveIPartitionFunctionDiagonalFraming})
\begin{equation}
\label{eq:CubicPartitionFunctionDiagonalFraming}
Z \left( g_{\text{s}}; \upxi_{1}, \upmu \right) = \rme^{F_{\text{A}} \left( g_{\text{s}}; \upmu \right)}\, \sum_{g=0}^{+\infty} \sum_{\alpha=-g}^{+\infty} \upxi_{1}^{\alpha}\, \mathsf{Z}^{(\alpha)}_{g} (t,\upmu)\, g_{\text{s}}^{g}.
\end{equation}
\noindent
The main-diagonal coefficients $\mathsf{Z}^{(0)}_{g} (t,\upmu)$ are again polynomials in $\upmu$, of degree $3g$. This expression is the cubic-matrix-model partition-function analogue of \eqref{eq:diagonalframingF}.
\item \textbf{Quadratic Transasymptotics:} Having addressed diagonal framing, let us turn to the quadratic transasymptotic formulation of the cubic transseries. It is rather instructive to explicitly compare the present cubic matrix model with the earlier \PI~example (to which it naturally double scales; or, from a different angle, of which it is a natural deformation). Explicitly, denoting $C=-\frac{\sqrt{2}}{3^{1/4}}$, in the double-scaling limit one observes
\bea
\label{eq:cubic-to-pi-mu-DSL}
\sigma_{\text{cubic}} &\longrightarrow& C\, \sigma_{\text{\PI}}, \qquad \frac{\upmu_{\text{cubic}}}{C^2} \longrightarrow \upmu_{\text{\PI}}, \\
\frac{\upzeta_{\text{cubic}, 1,2}}{C} &\longrightarrow& \sigma_{\text{\PI}, 1,2}\, \rme^{\mp A_{\text{\PI}}\, z^{5/4}} \left( z^{5/4} \right)^{\pm \frac{2}{\sqrt{3}} \upmu_{\text{\PI}}} = \upzeta_{\text{\PI}, 1,2}.
\eea
\noindent
Further introducing the following (cubic/\PI) quantities, which will appear below,
\be
\label{eq:pcubicpP1}
p_{\text{cubic}} (t) = \frac{\rmi \lambda^2\, \sqrt{r}}{4 \left( 1 - 3 \lambda^2\, r \right)^{5/4}}, \qquad p_{\text{\PI}} = \frac{\rmi}{2^{5/2}\, 3^{3/4}} = \left( - \frac{1}{96\sqrt{3}} \right)^{\frac{1}{2}},
\ee
\noindent
they double-scale as
\be
g_{\text{s}}^{\frac{1}{2}}\, p_{\text{cubic}} (t) \longrightarrow z^{-\frac{5}{8}}\, p_{\text{\PI}}.
\ee
\noindent
Having computed five non-zero coefficients up to the $(3|3)$ sector, and two non-zero coefficients up to the $(6|6)$ sector, we conjecture (and check against our data) the closed-form (compare with \eqref{eq:PreparedFormForSummingQuadratically} and with \eqref{eq:diagonalframingresult})
\be
\label{eq:cubicPartitionFunction}
Z \left( t, g_{\text{s}}; \upxi_{1}, \upmu \right) = \rme^{F_{\text{A}} \left( g_{\text{s}}; \upmu \right)} \sum_{g=0}^{+\infty}\, \sum_{\alpha \in \BZ} \upxi_1^\alpha\, \mathsf{Z}_{g + \frac{1}{2} \alpha \left(\alpha-1\right)}^{(\alpha)} (t,\upmu)\, g_{\text{s}}^{g + \frac{1}{2} \alpha \left(\alpha-1\right)},
\ee
\noindent
where
\bea
\label{eq:cubicDFTKernel}
Z_{g + \frac{1}{2} \alpha \left(\alpha-1\right)}^{(\alpha)} (t,\upmu) &=& \left( \rmi\, \frac{2^{1/2}}{3^{1/4} C} \right)^\alpha \left( p_{\text{cubic}} (t)\, g_{\text{s}}^{\frac{1}{2}} \right)^{\alpha^2} \times \\
&&
\times D_{\text{cubic}, g} \left( \alpha-\frac{2}{\sqrt{3}}\, \frac{\upmu_{\text{cubic}}}{C^2} \right) \frac{G_2 \left( 1 + \alpha - \frac{2}{\sqrt{3}}\, \frac{\upmu_{\text{cubic}}}{C^2} \right)}{G_2 \left( 1 - \frac{2}{\sqrt{3}}\, \frac{\upmu_{\text{cubic}}}{C^2} \right) \Gamma \left( 1 - \frac{2}{\sqrt{3}}\, \frac{\upmu_{\text{cubic}}}{C^2} \right)^\alpha}. \nonumber
\eea
\noindent
The above cubic-model $D_{k} (\nu)$ polynomials double-scale to the corresponding ones in \PI, as
\begin{equation}
g_{\text{s}}^{g}\, D_{\text{cubic}, g}(\nu) \,\longrightarrow\, z^{-\frac{5}{4}g}\, D_{\text{\PI}, g} (\nu).
\end{equation}
\noindent
The first few cubic-model $D_{k} (\nu)$ polynomials read (compare with table~\ref{tab:PIDkPolynomials})
\bea
\label{eq:CMM-Dk0}
D_{0} (\nu) &=& 1, \\
\label{eq:CMM-Dk1}
D_{1} (\nu) &=& \frac{\nu}{96 r \left( 1 - 3 r \right)^{5/2}}\, \Big\{ - 16 \nu ^2 + \left( 366 \nu^2 + 57 \right) r^2 - 12 \left( 14 \nu^2 + 5 \right) r + 8 \Big\}, \\
\label{eq:CMM-Dk2}
D_{2} (\nu) &=& \frac{48 r^3 \left( 522 r^2 - 615 r + 175 \right)}{92160\, r^2 \left( 1 - 2 r \right) \left( 1 - 3 r \right)^5} - \frac{\nu^2}{4608\, r^2 \left( 1 - 3 r \right)^5}\, \bigg\{ \frac{1}{4} \left( 3 r \left( r \left( 3 r \left( 4803 r + \right. \right. \right. \right. \\
&&
\left. \left. \left. \left.
+ 8600 \right) - 23648 \right) + 3648 \right) - 832 \right) - \nu^2\, \Big\{ 12 r \left( r  \left( 5847 r^2 - 8412 r + 2480 \right) + 192 \right) - \nonumber \\
&&
- \nu^2 \left( 3 \left( 28 - 61 r \right) r + 8 \right)^2 - 256 \Big\} \bigg\}. \nonumber 
\eea
\noindent
We have set $\lambda=1$ for simplicity, and although we computed up to the next $D_{3} (\nu)$ polynomial, we are omitting its expression as it is too large to display herein.
\item \textbf{Discrete Fourier Transform:} One of our key points is that, just like in the two previous double-scaled examples, also off-criticality the quadratic-transasymptotically resummed partition-function \eqref{eq:cubicPartitionFunction} can now be very conveniently rewritten as a discrete Fourier transform (compare\footnote{Observe how now the logarithmic resummation factor is $2$ for the cubic model.} with \eqref{eq:PIDiscreteFourierNC})
\begin{equation}
\label{eq:CMMDiscreteFourierNC}
Z \left( g_{\text{s}}; \uprho, \upmu \right) = \NCP (\upmu) \sum_{\ell\in\mathbb{Z}} \uprho^{\ell}\, \mathcal{Z} \left( g_{\text{s}}; \ell-\upmu \right),
\end{equation}
\noindent
where the ``discrete Fourier modes'' are given by (compare with \eqref{eq:DFTKernelPINC})
\be
\label{eq:DFTKernelCNC}
\mathcal{Z} \left( g_{\text{s}}; \nu \right) = \rme^{\frac{1}{g_{\text{s}}^2}\, F_{-2}^{(0|0)}(t) + F_{0}^{(0|0)}(t)} \big( p_{\text{cubic}}(t) \big)^{\nu^2} \exp \left( - \frac{A(t)}{g_{\text{s}}}\, \nu \right) \frac{G_2 \left(1+\nu\right)}{\left(2\pi\right)^{\frac{\nu}{2}}}\, \sum_{k=0}^{+\infty} D_k (\nu)\, g_{\text{s}}^{\frac{1}{2}\nu^2+k},
\ee
\noindent
and where the above free-energy transseries-coefficients may be found in sub-appendix~\ref{subapp:transasymptotic-transseries-CMM}. Further, the function $p_{\text{cubic}}(t)$ is the exponential of the second derivative of the planar free energy. The ``discrete Fourier variable'' $\uprho$ is related to the original transseries-parameter $\sigma_1$ via (compare with \eqref{eq:DFTrhoP1})
\begin{equation}
\label{eq:cubicrho}
\uprho = \frac{\sigma_1}{S_1}\, \rme^{\rmi\pi\, \upmu}\, \frac{\left( 96\sqrt{3} \right)^{-\upmu}}{\Gamma \left( 1 - \upmu \right)},
\end{equation}
\noindent
where we recall the canonical Stokes coefficient for the cubic matrix model
\begin{equation}
S_{1} = -\frac{\rmi}{\sqrt{2\pi}}.
\end{equation}
\noindent
Another of our key points is that this Fourier-variable is exactly given by the generating function of non-linear resurgent Stokes data for the cubic matrix model, as computed in \cite{mss22}. Finally, the pre-factor in \eqref{eq:CMMDiscreteFourierNC}, which is solely a function of $\upmu$, reads (compare with \eqref{eq:DFTnormalizationP1})
\begin{equation}
\label{eq:DFTnormalizationCMM}
\NCP (\upmu) = \left( - 4 \rmi \right)^{\mu^2}\, \frac{1}{\left(2\pi\right)^{\frac{\upmu}{2}}\, G_2 \left( 1 - \upmu \right)}
\end{equation}
\noindent
As in the double-scaled examples, also herein the pre-factor is independent of the 't~Hooft and string couplings and amounts to a choice of normalization of the partition function, arising from the way one originally solves the cubic matrix-model string-equation.
%
\newcommand{\thetaf}[4]{
\vartheta{\begin{bmatrix} #1 \\ #2 \end{bmatrix} } \left( #3 \, | \, #4\right)}
%
\item \textbf{Theta Function:} In parallel with the \PI~discussion, also for the matrix model our partition function \eqref{eq:CMMDiscreteFourierNC} may be repackaged via the use of Riemann theta-functions with characteristics (see \eqref{eq:riemanntheta}-\eqref{eq:riemannthetacharacteristics} in appendix~\ref{app:elliptic-theta-modular}). Expanding \eqref{eq:CMMDiscreteFourierNC} and rewriting\footnote{Whether this is a well-defined power-series expansion in the one-cut case will be discussed in \cite{ss26}. At this stage this is purely a formal rewrite, with no good expansion parameter, and stands solely for comparison purposes.} one obtains\footnote{As in \eqref{eq:PItheta}, comparing with \eqref{eq:EM-2-cut-Z} and \eqref{eq:theta-for-EM-2-cut-Z} both theta-function characteristics are now turned on.} (compare with \eqref{eq:PItheta})
\bea
\label{eq:cubictheta}
Z \left( g_{\text{s}}; \uprho, \upmu \right) &=& \rme^{\frac{1}{g_{\text{s}}^2}\, F_{-2}^{(0)} + F_{0}^{(0)}} \uprho^{-\upmu}\, \NCP (\upmu) \sum_{k=0}^{+\infty} f_k (r, g_{\text{s}})\, \partial_{\chi}^k \thetaf{\upmu}{\frac{1}{2\pi\rmi} \log \uprho}{\chi}{\tau} = \\
&&
\hspace{-50pt}
= \rme^{\frac{1}{g_{\text{s}}^2}\, F_{-2}^{(0)} + F_{0}^{(0)}} \uprho^{-\upmu}\, \NCP (\upmu) \left\{ \thetaf{\upmu}{\frac{1}{2\pi\rmi} \log \uprho}{\chi}{\tau} + f_1 (r, g_{\text{s}})\, \partial_{\chi} \thetaf{\upmu}{\frac{1}{2\pi\rmi} \log \uprho}{\chi}{\tau} + \cdots \right\}, \nonumber
\eea
\noindent
where the arguments in the theta-functions are
\bea
\chi &=& - \frac{1}{2\pi\rmi} \left( \frac{A(t)}{g_{\text{s}}} + \frac{1}{3}\, \log 2\pi \right), \\
\tau &=& \log \Big( p_{\text{cubic}} (t)\, \sqrt{g_{\text{s}}} \Big),
\eea
\noindent
and, again, the $f_k (r, g_{\text{s}})$ are the coefficients of the expansion of the modes \eqref{eq:DFTKernelCNC} in powers of $\nu$; for instance,
\bea
f_0 (r, g_{\text{s}}) &=& 1, \\
f_1 (r, g_{\text{s}}) &=& -\frac{1}{2}+\frac{\rmi g_{\text{s}} \left(57 r^2-60 r+8\right)}{16 \sqrt{3} (1-3 r)^{5/2} r}.
\eea
\noindent
Note how in contrast with the theta-functions in \cite{em08} the $f_k (r, g_{\text{s}})$ do not scale with $g_{\text{s}}^k$. For further comparisons we again refer the reader to subsection~\ref{subsec:from-theta-to-dual-to-transseries}.
\end{itemize}

\paragraph{Quartic Matrix Model:}

Our final example is the quartic matrix model \eqref{eq:QuarticMatrixModelPotential} with string equation \eqref{eq:quarticstringequationthooftlimit} (and free energy counterpart \eqref{eq:freeenergyfromstringequation}) whose transseries solution is again of the form \eqref{eq:twoparameterresurgenttransseriesforR}. Explicit solutions to these equations are outlined in sub-appendix~\ref{subapp:transasymptotic-transseries-QMM}; and for previous related discussions, see, \textit{e.g.}, \cite{m08, asv11, sv13, csv15, mss22}. As usual, we first focus mainly upon rectangular and diagonal framings for the partition function, laying the ground towards its quadratic-transasymptotics and discrete-Fourier-transform upscaled versions.
\begin{itemize}
\item \textbf{Rectangular Framing:} Start with the partition-function in rectangular framing,
\bea
\label{eq:QuarticPartitionFunctionRectangular}
Z \left( t, g_{\text{s}}; \sigma_1, \sigma_2 \right) &=& \rme^{F_{\text{A}} \left( g_{\text{s}}; \upmu \right)} \sum_{n,m=0}^{+\infty} \sigma_{1}^{n} \sigma_{2}^{m}\, \rme^{- \left(n-m\right) \frac{A (t)}{g_{\text{s}}}} \times \\
&&
\hspace{20pt}
\times \sum_{k=0}^{k_{n m}} \log \left( \frac{\left( 3 - \lambda\, r \right)^3 \left( 3 - 3 \lambda\, r \right)^5}{15552 \lambda^6\, r^4\, g_{\text{s}}^2} \right)^{k}\, \sum_{g=0}^{+\infty} Z^{(n|m)[k]}_{2g} (t)\, g_{\text{s}}^{g+\beta^{[k]}_{nm}}, \nonumber
\eea
\noindent
where the instanton action is encoded in \eqref{eq:quarticaction} (see as well sub-appendix~\ref{subappendix:quarticmatrixmodel}), the starting genus $\beta_{nm}$ is discussed at length in appendix~\ref{subapp:transasymptotic-transseries-QMM}, and we have used the classical string solution from \eqref{eq:quartic-MM-classical-string-eq} with $r \equiv r(t)$. The global pre-factor reads the usual
\be
F_{\text{A}} \left( g_{\text{s}}; \sigma_1 \sigma_2 \right) = \frac{1}{g_{\text{s}}^2}\, F^{(0|0)}_{-2} (t) + F^{(0|0)}_{0} (t) + \frac{1}{g_{\text{s}}}\, F^{(1|1)}_{-1} (t)\, \upmu + F^{(2|2)}_{0} (t)\, \upmu^2.
\ee
\item \textbf{Diagonal Framing:} The diagonal framing reformulation follows in complete parallel with our earlier examples and is described at length in sub-appendix~\ref{subapp:transasymptotic-transseries-QMM}. Introduce adequate resummations parameters (compare with \eqref{eq:defxi}, or \eqref{eq:defzeta} and \eqref{eq:XiMuRelation}, and with\footnote{Cubic and quartic normalizations differ slightly: the double-scaling rescaling-constant $C$ herein has $\lambda$ dependence, whereas it did not for the cubic example. This materializes $\lambda$-factors next to $\upmu$ as compared to \eqref{eq:zetaxiCMM-1}-\eqref{eq:zetaxiCMM-2}.} \eqref{eq:zetaxiCMM-1}-\eqref{eq:zetaxiCMM-2})
\begin{equation}
\label{eq:upxi12forQuarticMAIN}
\upxi_{1} = \sqrt{g_{\text{s}}}\, \sigma_{1}\, \rme^{-\frac{A(t)}{g_{\text{s}}}} \left( \frac{f(t)}{5184 \lambda^2\, g_{\text{s}}^2} \right)^{\frac{\lambda}{12} \upmu}, \qquad \upxi_{2} = \sqrt{g_{\text{s}}}\, \sigma_{2}\, \rme^{\frac{A(t)}{g_{\text{s}}}} \left( \frac{f(t)}{5184 \lambda^2\, g_{\text{s}}^2}\right)^{-\frac{\lambda}{12} \upmu},
\end{equation}
\noindent
where
\begin{equation}
f(t) = \frac{\left( 3 - \lambda\, r \right)^3 \left( 3 - 3 \lambda\, r \right)^5}{3 \lambda^6\,  r^4},
\end{equation}
\noindent
and with which the diagonal formulation is simply (compare with \eqref{eq:PainleveIPartitionFunctionDiagonalFraming} and \eqref{eq:CubicPartitionFunctionDiagonalFraming})
\begin{equation}
Z \left( g_{\text{s}}; \upxi_{1}, \upmu \right) = \rme^{F_{\text{A}} \left( g_{\text{s}}; \upmu \right)}\, \sum_{g=0}^{+\infty} \sum_{\alpha=-g}^{+\infty} \upxi_{1}^{\alpha}\, \mathsf{Z}^{(\alpha)}_{g} (t,\upmu)\, g_{\text{s}}^{g}.
\end{equation}
\noindent
The main-diagonal coefficients $\mathsf{Z}^{(0)}_{g} (t,\upmu)$ are again polynomials in $\upmu$, of degree $3g$ (see some such examples in sub-appendix~\ref{subapp:transasymptotic-transseries-QMM}, equations \eqref{eq:Zmu-poly-examples-QMM-1} through \eqref{eq:Zmu-poly-examples-QMM-9}). This expression is the quartic-matrix-model partition-function analogue of \eqref{eq:diagonalframingF}.
\item \textbf{Quadratic Transasymptotics:} With the experience we have acquired so far it is now immediate to conjecture (and check against our data\footnote{For this purpose we have computed diagonal partition-function coefficients (albeit truncated at $\upmu$-order $5$) for $-6<\alpha<6$ up to the first $5$ non-vanishing $g_{\text{s}}$ orders, and for $-10<\alpha<10$ up to the first $2$ non-vanishing $g_{\text{s}}$ orders. In addition we have computed diagonal partition-function coefficients (now truncated at $\upmu$-order $9$) for $-3< \alpha < 3$ up to the first $3$ non-vanishing $g_{\text{s}}$ contributions. The truncations in $\upmu$ are for computational efficiency, and associated to the highest instanton numbers we computed. See sub-appendix~\ref{subapp:transasymptotic-transseries-QMM} for some selected examples.}) a closed-form expression for quadratic-transasymptotics of the quartic-model partition-function transseries. This expression is presented in sub-appendix~\ref{subapp:transasymptotic-transseries-QMM}; see equation \eqref{eq:QMM-quadratic-transasymptotics-APP} and its ensuing discussion. It is given by
\be
\label{eq:QMM-quadratic-transasymptotics}
Z \left( t, g_{\text{s}}; \upzeta_{1}, \upmu \right) = \rme^{F_{\text{A}} \left( g_{\text{s}}; \upmu \right)}\, \sum_{\alpha\in\mathbb{Z}} \upzeta_1^\alpha\, \mathsf{Z}^{(\alpha)} (t, g_{\text{s}}; \upmu)
\ee
\noindent
where, due to the $\BZ_2$-symmetry of the quartic matrix model, the coefficients have now a slightly more intricate structure, given by
\be
\label{eq:QMM-quadratic-transasymptotics-II}
\mathsf{Z}^{(\alpha)} (t,g_{\text{s}}; \upmu) \equiv 
\begin{cases}
\sum\limits_{s\in\mathbb{Z}} \sum\limits_{g=0}^{+\infty} \sum\limits_{\tilde{g}=0}^{g} \mathsf{Y}^{(\alpha-s)}_{\text{even},g-\tilde{g}+\frac{(\alpha-s)(\alpha-s-1)}{2}} (t,\upmu)\, \mathsf{Y}^{(s)}_{\text{even},\tilde{g}+\frac{s(s-1)}{2}} (t,\upmu), \quad & \alpha \text{ even}, \\
d \cdot \sum\limits_{s\in\mathbb{Z}} \sum\limits_{g=0}^{+\infty} \sum\limits_{\tilde{g}=0}^{g} \mathsf{Y}^{(\alpha-s)}_{\text{odd},g-\tilde{g}+\frac{(\alpha-s)(\alpha-s-1)}{2}} (t,\upmu)\, \mathsf{Y}^{(s)}_{\text{odd},\tilde{g}+\frac{s(s-1)}{2}} (t,\upmu), \quad & \alpha \text{ odd}.
\end{cases}
\ee
\noindent
Herein we have introduced $d^2 = \frac{3-\lambda r}{2 \lambda r}$, alongside (comparing with \eqref{eq:diagonalframingresult} and, specially, with \eqref{eq:cubicDFTKernel} we see an even/odd split is now required due to the $\BZ_2$-symmetry)
\bea
\mathsf{Y}^{(\alpha)}_{\text{even}/\text{odd},g+\frac{\alpha(\alpha-1)}{2}} (t,\upmu) &=& \left( \rmi\, \frac{2^{1/2}}{3^{1/4} C} \right)^{\alpha} \left( p_{\text{quartic}} (t)\, g_{\text{s}}^{\frac{1}{2}} \right)^{\frac{2}{\sqrt{3}}\, \frac{\upmu_{\text{quartic}}}{C^2} \left( 2\alpha - \frac{2}{\sqrt{3}}\, \frac{\upmu_{\text{quartic}}}{C^2} \right)} \times \nonumber \\
&\times& \left( 2\pi \right)^{\frac{\alpha+\upmu_{\text{quartic}}}{2}} \frac{\mathcal{Y}_{\text{even}/\text{odd},g} \left( \alpha-\frac{2}{\sqrt{3}}\, \frac{\upmu_{\text{quartic}}}{C^2} \right)}{G_2 \left( 1 - \frac{2}{\sqrt{3}}\, \frac{\upmu_{\text{quartic}}}{C^2} \right) \Gamma \left( 1 - \frac{2}{\sqrt{3}}\, \frac{\upmu_{\text{quartic}}}{C^2} \right)^\alpha}
\eea
\noindent
and
\be
\mathcal{Y}_{\text{even}/\text{odd},g} (\nu) = \frac{G_2 \left(1+\nu\right)}{\left(2\pi\right)^{\frac{\nu}{2}}} \left( p_{\text{quartic}} (t)\, g_{\text{s}}^{\frac{1}{2}} \right)^{\nu^2} \left( - \frac{\rmi}{2\sqrt{3}}\, g_{\text{s}} \right)^g D_{\text{even}/\text{odd}, g} (\nu).
\ee
\noindent
Further, where we are now denoting $C = -\frac{2 \cdot 3^{1/4}}{\sqrt{\lambda}}$ and
\begin{equation}
p_{\text{quartic}} (t) = \frac{\rmi \sqrt{3}\, \lambda^{3/2} r}{2 \left( 3 - 3 \lambda r \right)^{5/4} \left( 3 - \lambda r \right)^{3/4}},
\end{equation}
\noindent
and the $D_{\text{even}/\text{odd}, g}$ are polynomials in $\nu$. At lowest order these polynomials are trivial, $D_{\text{even}, 0} (\nu) = 1 = D_{\text{odd}, 0} (\nu)$, and at next-to-lowest order they are
\bea
\label{eq:quartic-dk-e}
D_{\text{even}, 1} (\nu) &=& - \frac{3 \rmi \sqrt{3} \left( 2 r^3 - 30 r^2 + 63 r - 18 \right)}{4 r \left( 3 - 3 r \right)^{5/2} \left( 3 - r \right)^{3/2}}\, \nu - \frac{3 \rmi \sqrt{3} \left( 29 r^3 - 90 r^2 + 36 r + 72 \right)}{2 r \left( 3 - 3 r \right)^{5/2} \left( 3 - r \right)^{3/2}}\, \nu^3, \\
\label{eq:quartic-dk-o}
D_{\text{odd}, 1} (\nu) &=& \frac{3 \rmi \sqrt{3} \left( 28 r^3 - 18 r^2 - 153 r + 126 \right)}{4 r \left( 3 - 3 r \right)^{5/2} \left( 3 - r \right)^{3/2}}\, \nu - \frac{3 \rmi \sqrt{3} \left( 29 r^3 - 90 r^2 + 36 r + 72 \right)}{2 r \left( 3 - 3 r \right)^{5/2} \left( 3 - r \right)^{3/2}}\, \nu^3,
\eea
\noindent
where we have set $\lambda=1$ for brevity. We have computed them up to $g=3$, but the terms quickly become too vast to display herein.
\item \textbf{Discrete Fourier Transform:} Our by-now-familiar key point is that, just like in the previous double-scaled and off-critical examples, the quadratic-transasymptotically resummed partition-function \eqref{eq:QMM-quadratic-transasymptotics}-\eqref{eq:QMM-quadratic-transasymptotics-II} may now be very conveniently rewritten as a discrete Fourier transform. Given the two nonperturbative saddles of the quartic matrix model \eqref{eq:2-saddles-quartic} one might at first expect this discrete Fourier transform to be a generalization of the \YL~case in \eqref{eq:YLFullDFT}. These saddles are, however, $\BZ_2$-symmetric and the quartic transseries indeed only has two transseries parameters \eqref{eq:QuarticPartitionFunctionRectangular}---which places us closer to comparison with \PI~and cubic examples, \eqref{eq:PIDiscreteFourierNC} and \eqref{eq:CMMDiscreteFourierNC}. One hence arrives at\footnote{Observe how now the logarithmic resummation factor is $\frac{\lambda}{3}$ for the quartic model.}
\begin{equation}
\label{eq:QMMDiscreteFourierNC}
Z \left( g_{\text{s}}; \uprho, \upmu \right) = \NCP (\upmu) \sum_{\boldsymbol{\ell}\in\mathbb{Z}^2}  \uprho^{\ell_1+\ell_2}\, \mathcal{Z} \left( g_{\text{s}}; \ell_1-\frac{2}{\sqrt{3}}\, \frac{\upmu}{C^2}, \ell_2-\frac{2}{\sqrt{3}}\, \frac{\upmu}{C^2} \right),
\end{equation}
\noindent
where the ``discrete Fourier modes'' read (compare with \eqref{eq:DFTKernelCNC})
\bea
\mathcal{Z} \left( g_{\text{s}}; \nu_1, \nu_2 \right) &=& \rme^{\frac{1}{g_{\text{s}}^2}\, F_{-2}^{(0|0)}(t) + F_{0}^{(0|0)}(t)} \big( p_{\text{quartic}}(t) \big)^{\nu_1^2+\nu_2^2}\, \rme^{- \frac{A(t)}{g_{\text{s}}}\, \left(\nu_1+\nu_2\right)}\, \mathcal{Z}_{\text{G}} \left(\nu_1\right) \mathcal{Z}_{\text{G}} \left(\nu_2\right) \times \nonumber \\
&\times&
\sum_{g=0}^{+\infty} \left( - \frac{\rmi}{2\sqrt{3}}\, g_{\text{s}}\right)^g \begin{cases}
\sum\limits_{\tilde{g}=0}^{g} D_{\text{even}, g-\tilde{g}} (\nu_1)\, D_{\text{even}, \tilde{g}} (\nu_2), \quad \nu_1-\nu_2 \text{ even}, \\
d \cdot \sum\limits_{\tilde{g}=0}^{g} D_{\text{odd}, g-\tilde{g}} (\nu_1)\, D_{\text{odd}, \tilde{g}} (\nu_2), \quad \nu_1-\nu_2 \text{ odd}.
\end{cases}
\label{eq:quartic-dft-kernel}
\eea
\noindent
The above free-energy transseries-coefficients are discussed in sub-appendix~\ref{subapp:transasymptotic-transseries-QMM}, the function $p_{\text{quartic}}(t)$ is the exponential of the second derivative of the planar free energy, and one may notice how in spite the condition in \eqref{eq:quartic-dft-kernel} depends on $\nu_1$ and $\nu_2$ it is still integer valued as $\nu_1-\nu_2=\ell_1-\ell_2\in\mathbb{Z}$ in \eqref{eq:QMMDiscreteFourierNC}. The ``discrete Fourier variable'' $\uprho$ is related to the original transseries-parameter $\sigma_1$ via (compare with \eqref{eq:DFTrhoP1} and \eqref{eq:cubicrho})
\be
\label{eq:quarticrho}
\uprho = \frac{\sigma_1}{S_1}\, \frac{\left( 96\sqrt{3} \right)^{-\frac{2}{\sqrt{3}}\, \frac{\upmu}{C^2}}}{\Gamma \left( 1 - \frac{2}{\sqrt{3}}\, \frac{\upmu}{C^2} \right)},
\ee
\noindent
where we recall the canonical Stokes coefficient for the quartic matrix model
\begin{equation}
S_{1} = \rmi\, \sqrt{\frac{3}{\pi \lambda}}.
\end{equation}
\noindent
Finally, there is a global, $t$-independent\footnote{To cancel the $t$ dependence notice that $F^{(2|2)}_0(t) = \frac{2}{3 C^4} \log \frac{p_{\text{quartic}}^4 (t)}{9}$ \cite{asv11}.} pre-factor which simply amounts to a choice of normalization of the partition function, arising from our original solution to the quartic matrix-model string-equation. It is herein given by (compare with \eqref{eq:DFTnormalizationP1} or \eqref{eq:DFTnormalizationCMM})
\be
\NCP (\upmu) = 3^{-\frac{\lambda^2}{36}}\, \frac{1}{\left(2\pi\right)^{\frac{\upmu}{2}}\, G_2 \left( 1 - \frac{\lambda}{6}\, \upmu \right)^2}.
\ee
\end{itemize}

\subsection{Resurgent-Transseries for All Matrix-Model Partition-Functions}
\label{subsec:resurgent-Z}

All these examples show a common structure emerge. String-theoretic resurgent transseries are \textit{resonant}, due to the presence of eigenvalue alongside anti-eigenvalue tunneling \cite{mss22}; equivalently, the presence of ZZ-branes alongside negative-tension ZZ-branes \cite{sst23}---at the end of the day this is what makes the discrete-Fourier-transform structure come alive when changing from rectangular to diagonal framings. Of course our results strictly only hold in the realm of hermitian matrix models (and their associated double-scaling and string theoretic facets) but one can wonder how far these formulae reign. Further note that the discrete-Fourier/Zak transform structure is a \textit{local} construction---as is the transseries itself---but as will be fully clear in \cite{krsst26b, krst26a, krst26b} our proposal includes complete Stokes data implementing \textit{all} Stokes transitions; \textit{i.e.}, our proposal is \textit{globally complete} in the sense that it can be extended to any (complex) values of the parameters.

\begin{figure}
	\centering
	\begin{tikzpicture}
\begin{scope}[scale=0.5, shift={({-10},{4})}]	
	\draw[fill=LightBlue,fill opacity=0.2, line width=1pt] (0,0) to [out=90,in=95] (4,0)
	to [out=85,in=0] (1,2.5)
    to [out=180,in=90] (-2,0)
    to [out=270, in=180] (-1, -0.5)
    to [out=0, in=270] cycle;
    \draw[fill=darktangerine,fill opacity=0.2, line width=1pt] (-2,0)
    to [out=270,in=180] (1,-2.5)
    to [out=0,in=275] (4,0)
    to [out=265,in=270] (0,0)
    to [out=270, in=0] (-1, -0.5)
    to [out=180, in=270] cycle;
    \draw[color=ForestGreen, line width=2pt] (-2,0) to [out=270, in=180] (-1, -0.5)
    to [out=0, in=270] (0,0);
    \draw[dashed, color=ForestGreen, line width=2pt] (-2,0) to [out=90, in=180] (-1, 0.5)
    to [out=0, in=90] (0,0);
\draw[ForestGreen, fill=ForestGreen] (-2,0) circle (.7ex);
\draw[ForestGreen, fill=ForestGreen] (0,0) circle (.7ex);
\draw[cornellred, fill=cornellred] (4,0) circle (.7ex);
\node at (-2.4, 0) {$a$}; 
\node at (0.4, 0) {$b$};
\node at (4.6, 0.1) {$x^{\star}$}; 
\end{scope} 
\begin{scope}[scale=0.5, shift={({0},{4})}]	
	\draw[fill=LightBlue,fill opacity=0.2, line width=1pt] (0,0) to [out=90,in=95] (4,0)
	to [out=85,in=0] (1,2.5)
    to [out=180,in=90] (-2,0)
    to [out=270, in=180] (-1, -0.5)
    to [out=0, in=270] cycle;
    \draw[fill=darktangerine,fill opacity=0.2, line width=1pt] (-2,0)
    to [out=270,in=180] (1,-2.5)
    to [out=0,in=275] (4,0)
    to [out=265,in=270] (0,0)
    to [out=270, in=0] (-1, -0.5)
    to [out=180, in=270] cycle;
    \draw[color=ForestGreen, line width=2pt] (-2,0) to [out=270, in=180] (-1, -0.5)
    to [out=0, in=270] (0,0);
    \draw[dashed, color=ForestGreen, line width=2pt] (-2,0) to [out=90, in=180] (-1, 0.5)
    to [out=0, in=90] (0,0);
    \draw[blue, line width=1.5pt, ->] (0,0) to [out=90, in=180] (1.8, 1.6);
    \draw[blue, line width=1.5pt] (1.8, 1.6) to [out=0, in=90] (4, 0);
\draw[ForestGreen, fill=ForestGreen] (-2,0) circle (.7ex);
\draw[ForestGreen, fill=ForestGreen] (0,0) circle (.7ex);
\draw[cornellred, fill=cornellred] (4,0) circle (.7ex);
\node at (-2.4, 0) {$a$}; 
\node at (0.4, 0) {$b$};
\node at (4.6, 0.1) {$x^{\star}$}; 
\end{scope} 
\begin{scope}[scale=0.5, shift={({10},{4})}]	
	\draw[fill=LightBlue,fill opacity=0.2, line width=1pt] (0,0) to [out=90,in=95] (4,0)
	to [out=85,in=0] (1,2.5)
    to [out=180,in=90] (-2,0)
    to [out=270, in=180] (-1, -0.5)
    to [out=0, in=270] cycle;
    \draw[fill=darktangerine,fill opacity=0.2, line width=1pt] (-2,0)
    to [out=270,in=180] (1,-2.5)
    to [out=0,in=275] (4,0)
    to [out=265,in=270] (0,0)
    to [out=270, in=0] (-1, -0.5)
    to [out=180, in=270] cycle;
    \draw[color=ForestGreen, line width=2pt] (-2,0) to [out=270, in=180] (-1, -0.5)
    to [out=0, in=270] (0,0);
    \draw[dashed, color=ForestGreen, line width=2pt] (-2,0) to [out=90, in=180] (-1, 0.5)
    to [out=0, in=90] (0,0);
    \draw[blue, line width=1.5pt, ->] (0,0) to [out=90, in=180] (1.8, 1.6);
    \draw[blue, line width=1.5pt] (1.8, 1.6) to [out=0, in=90] (4, 0);
        \draw[blue, line width=1.5pt, ->] (0,0) to [out=90, in=180] (1.8, 2);
    \draw[blue, line width=1.5pt] (1.8, 2) to [out=0, in=90] (4, 0);
\draw[ForestGreen, fill=ForestGreen] (-2,0) circle (.7ex);
\draw[ForestGreen, fill=ForestGreen] (0,0) circle (.7ex);
\draw[cornellred, fill=cornellred] (4,0) circle (.7ex);
\node at (-2.4, 0) {$a$}; 
\node at (0.4, 0) {$b$};
\node at (4.6, 0.1) {$x^{\star}$}; 
\end{scope} 
\begin{scope}[scale=0.5, shift={({-10},{-4})}]	
	\draw[fill=LightBlue,fill opacity=0.2, line width=1pt] (0,0) to [out=90,in=95] (4,0)
	to [out=85,in=0] (1,2.5)
    to [out=180,in=90] (-2,0)
    to [out=270, in=180] (-1, -0.5)
    to [out=0, in=270] cycle;
    \draw[fill=darktangerine,fill opacity=0.2, line width=1pt] (-2,0)
    to [out=270,in=180] (1,-2.5)
    to [out=0,in=275] (4,0)
    to [out=265,in=270] (0,0)
    to [out=270, in=0] (-1, -0.5)
    to [out=180, in=270] cycle;
    \draw[color=ForestGreen, line width=2pt] (-2,0) to [out=270, in=180] (-1, -0.5)
    to [out=0, in=270] (0,0);
    \draw[dashed, color=ForestGreen, line width=2pt] (-2,0) to [out=90, in=180] (-1, 0.5)
    to [out=0, in=90] (0,0);
\draw[ForestGreen, fill=ForestGreen] (-2,0) circle (.7ex);
\draw[ForestGreen, fill=ForestGreen] (0,0) circle (.7ex);
\draw[cornellred, fill=cornellred] (4,0) circle (.7ex);
\node at (-2.4, 0) {$a$}; 
\node at (0.4, 0) {$b$};
\node at (4.6, 0.1) {$x^{\star}$}; 
\end{scope} 
\begin{scope}[scale=0.5, shift={({0},{-4})}]	
	\draw[fill=LightBlue,fill opacity=0.2, line width=1pt] (0,0) to [out=90,in=95] (4,0)
	to [out=85,in=0] (1,2.5)
    to [out=180,in=90] (-2,0)
    to [out=270, in=180] (-1, -0.5)
    to [out=0, in=270] cycle;
    \draw[fill=darktangerine,fill opacity=0.2, line width=1pt] (-2,0)
    to [out=270,in=180] (1,-2.5)
    to [out=0,in=275] (4,0)
    to [out=265,in=270] (0,0)
    to [out=270, in=0] (-1, -0.5)
    to [out=180, in=270] cycle;
    \draw[color=ForestGreen, line width=2pt] (-2,0) to [out=270, in=180] (-1, -0.5)
    to [out=0, in=270] (0,0);
    \draw[dashed, color=ForestGreen, line width=2pt] (-2,0) to [out=90, in=180] (-1, 0.5)
    to [out=0, in=90] (0,0);
    \draw[blue, line width=1.5pt, ->] (0,0) to [out=90, in=180] (1.8, 1.6);
    \draw[blue, line width=1.5pt] (1.8, 1.6) to [out=0, in=90] (4, 0);
    \draw[orange, line width=1.5pt, ->] (0,0) to [out=270, in=180] (1.8, -1.6);
    \draw[orange, line width=1.5pt] (1.8, -1.6) to [out=0, in=270] (4, 0);
\draw[ForestGreen, fill=ForestGreen] (-2,0) circle (.7ex);
\draw[ForestGreen, fill=ForestGreen] (0,0) circle (.7ex);
\draw[cornellred, fill=cornellred] (4,0) circle (.7ex);
\node at (-2.4, 0) {$a$}; 
\node at (0.4, 0) {$b$};
\node at (4.6, 0.1) {$x^{\star}$}; 
\end{scope} 
\begin{scope}[scale=0.5, shift={({10},{-4})}]	
	\draw[fill=LightBlue,fill opacity=0.2, line width=1pt] (0,0) to [out=90,in=95] (4,0)
	to [out=85,in=0] (1,2.5)
    to [out=180,in=90] (-2,0)
    to [out=270, in=180] (-1, -0.5)
    to [out=0, in=270] cycle;
    \draw[fill=darktangerine,fill opacity=0.2, line width=1pt] (-2,0)
    to [out=270,in=180] (1,-2.5)
    to [out=0,in=275] (4,0)
    to [out=265,in=270] (0,0)
    to [out=270, in=0] (-1, -0.5)
    to [out=180, in=270] cycle;
    \draw[color=ForestGreen, line width=2pt] (-2,0) to [out=270, in=180] (-1, -0.5)
    to [out=0, in=270] (0,0);
    \draw[dashed, color=ForestGreen, line width=2pt] (-2,0) to [out=90, in=180] (-1, 0.5)
    to [out=0, in=90] (0,0);
    \draw[blue, line width=1.5pt, ->] (0,0) to [out=90, in=180] (1.8, 1.6);
    \draw[blue, line width=1.5pt] (1.8, 1.6) to [out=0, in=90] (4, 0);
    \draw[orange, line width=1.5pt, ->] (0,0) to [out=270, in=180] (1.8, -1.6);
    \draw[orange, line width=1.5pt] (1.8, -1.6) to [out=0, in=270] (4, 0);
        \draw[blue, line width=1.5pt, ->] (0,0) to [out=90, in=180] (1.8, 2);
    \draw[blue, line width=1.5pt] (1.8, 2) to [out=0, in=90] (4, 0);
    \draw[orange, line width=1.5pt, ->] (0,0) to [out=270, in=180] (1.8, -2);
    \draw[orange, line width=1.5pt] (1.8, -2) to [out=0, in=270] (4, 0);
\draw[ForestGreen, fill=ForestGreen] (-2,0) circle (.7ex);
\draw[ForestGreen, fill=ForestGreen] (0,0) circle (.7ex);
\draw[cornellred, fill=cornellred] (4,0) circle (.7ex);
\node at (-2.4, 0) {$a$}; 
\node at (0.4, 0) {$b$};
\node at (4.6, 0.1) {$x^{\star}$}; 
\end{scope}
\node at (-2, 2) {\Large $+\,\sigma_1$};
\node at (3, 2) {\Large $+\,\sigma_1^2$};
\node at (-2, -2) {\Large $+\,\upmu$};
\node at (3, -2) {\Large $+\,\upmu^2$};
\node at (8.1, 2) {\Large $+\,\cdots$};
\node at (8.1, -2) {\Large $+\,\cdots$};
\node at (-4.5, 3.8) {$(0|0)$ sector:};
\node at (-4.3, -0.2) {$(0|0)$ sector:};
\node at (0.5, 3.8) {$(1|0)$ sector:};
\node at (0.5, -0.2) {$(1|1)$ sector:};
\node at (5.5, 3.8) {$(2|0)$ sector:};
\node at (5.5, -0.2) {$(2|2)$ sector:};
	\end{tikzpicture}
\caption{Illustration of the $(\ell|0)$ (top) and $(\ell|\ell)$ (bottom) contributions to a resurgent \textit{resonant} transseries. These two directions (lowest horizontal rectangular-direction and main diagonal-direction, as in figure~\ref{fig:rectangulartodiagonal}), at lowest $g_{\text{s}}$ order, are remarkably similar and in fact only differ by half-cycles. Each picture shows the spectral-curve of the cubic matrix model, with its two sheets separated by the cut ({\color{ForestGreen}green}). The physical sheet ({\color{LightBlue}blue}) contains eigenvalue integration contours, whereas the involuted sheet ({\color{darktangerine}orange}) contains integration contours producing instanton actions with symmetric signs; hence the anti-eigenvalues \cite{mss22}.}
	\label{fig:comparisonforwardvsdiagonal}
\end{figure}
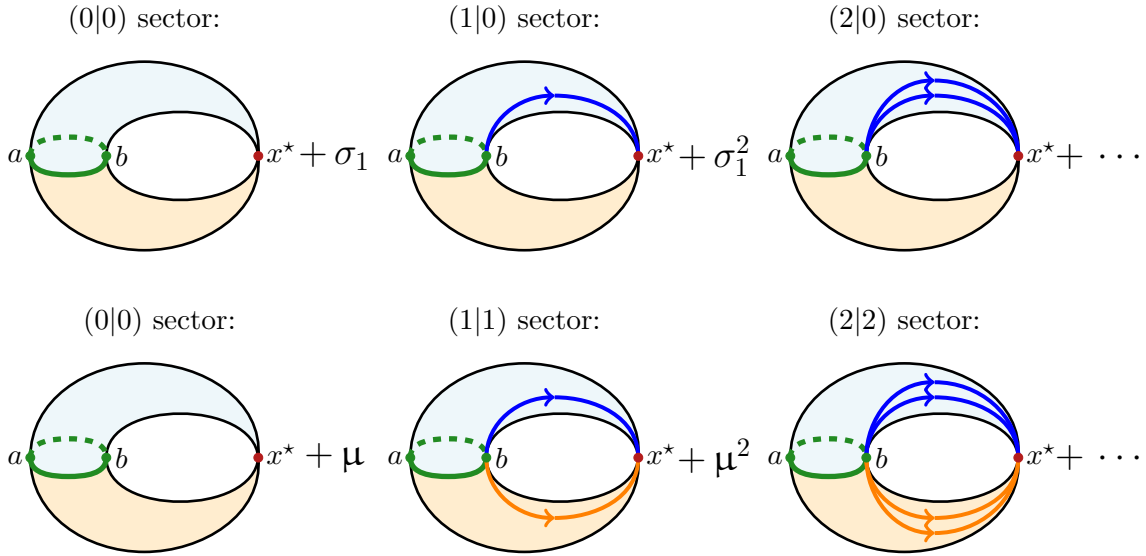

A general discrete-Fourier/Zak transform formulation of the matrix-model partition-function is almost trivial to obtain starting from spectral geometry and a straightforward extension of the results in \cite{msw08} in light of the existence of (resonant) anti-eigenvalues \cite{mss22}. But, to start, let us first drop anti-eigenvalues and briefly recall the \textit{non-resonant} formulation of \cite{msw08}. To simplify, consider a two-cut hermitian matrix model; following the spectral geometry analysis in subsection~\ref{subsec:SG-phases}, say, the cubic matrix model, but whilst in its \textit{one}-cut phase. Standard nonperturbative corrections to its perturbative asymptotic series are governed by eigenvalue tunneling to the sole nonperturbative saddle $x^{\star}$, as illustrated in the top-half of figure~\ref{fig:comparisonforwardvsdiagonal}. Tunneling $\ell$ eigenvalues to this saddle \cite{msw07} yields the (resurgent) $\ell$-instanton contribution to the partition function, $\CZ^{(\ell)}(t)$.

Now, this contribution may be equally described by \cite{msw08} considering the \textit{two}-cut free energy $\CF (t_1, t_2)$, with $t_1$ the modulus controlling the size of the perturbative cut and $t_2 \to 0$ the vanishing-size of the pinched-cut around the saddle-point $x^{\star}$. With respect to these data, the $\ell$-instanton contribution to the partition function may be written as \cite{msw08}
\be
\label{eq:InstantonsOneParameter}
\frac{\mathcal{Z}^{(\ell)}(t)}{\mathcal{Z}^{(0)}(t)} = \mathcal{Z}_{\text{G}}(\ell)\, \exp \left\{ \widehat{\mathcal{F}} (t-\ell g_{\text{s}}, \ell g_{\text{s}}) - \mathcal{F} (t) \right\} \quad \overset{\simeq}{\longleftrightarrow} \quad \frac{Z^{(\ell)}(t)}{Z^{(0)}(t)}.
\ee
\noindent
Due to the pinching cut, the free energy requires regularization (denoted by the hat-symbol, as we already alluded to in \eqref{eq:gaussianregularizationF}). Let us repeat it in here for convenience:
\be
\label{eq:regularizedtwocut}
\widehat{\CF}_{g} ( t_1,t_2 ) = \CF_{g} ( t_1,t_2 ) - \CF_{g}^{\text{G}} ( t_2 ),
\ee
\noindent
where the exact Gaussian partition function is given in \eqref{eq:exactGaussianZell}. Further, upon perturbative expansion of the free energies in the above expression, this leads to the asymptotics in \eqref{eq:partition-function-Z-msw08}. The asymptotic-equality arrow to the right-hand-side of \eqref{eq:InstantonsOneParameter} establishes the ``bridge'' between matrix-integral (curly notation) and string-equation (regular notation) quantities, in the line of \cite{mss22}. Pinched spectral curves were illustrated earlier in figure~\ref{fig:OneCutPotential} and herein in figure~\ref{fig:comparisonforwardvsdiagonal}. Equation \eqref{eq:InstantonsOneParameter} gives rise to instanton action \cite{msw08}
\be
\label{eq:InstantonActionmsw08}
\widehat{A} (t) = \partial_{s} \widehat{\CF}_0 \to \int_{b}^{x^{\star}} \rmd z\, y(z),
\ee
\noindent
where, as explained in subsection~\ref{subsec:Z-phases}, $s = \frac{1}{2} \left( t_1 - t_2 \right)$ and all quantities are evaluated at $\left( t_1,t_2 \right) \to \left( t,0 \right)$. Finally, and exclusively on what concerns \textit{non-resonant} (no sectors with anti-eigenvalues) multi-instanton transseries-sectors, these may all be directly reproduced from \eqref{eq:InstantonsOneParameter} by summing over all $\ell$ and all saddle-points $x^{\star}$ in question. Of course this is not nearly enough to construct the full resonant transseries; recall figure~\ref{fig:twopararectmtransseriesgrid}. Nonetheless, there is a simple and clean spectral-geometry interpretation of these non-resonant multi-instanton contributions:
\begin{itemize}
\item $B$-(half-)cycles are realized by tunneling eigenvalues out from the perturbative cut and into the nonperturbative (pinched) saddle at $x^{\star}$ \cite{msw07, msw08, mss22}.
\item $A$-(half-)cycles are realized by a choice of logarithmic sheet in the effective potential; which can also be interpreted as tunneling eigenvalues from one end of the cut to the other \cite{ps09, krsst26b}.
\end{itemize}

However, as we all well know by now, generic matrix-model\footnote{In fact, generic \textit{string}-theoretic transseries are resonant \cite{sst23}.} transseries are resonant and \eqref{eq:InstantonsOneParameter} does not give rise to the symmetric instanton action which is thereby required. In other words, to construct a resonant, resurgent partition-function for a generic hermitian matrix-model out from multi-cut spectral-geometry something else is needed. At large $N$, this translates to the inclusion of anti-eigenvalues or negative-tension branes, as we already know from string equations. One of our main points is that these effects have a remarkably simple explanation in the present multi-cut context, where they translate to the tunneling of ``fractional'' eigenvalues. To see this, and still within the two-cut setting with nonperturbative saddle $x^{\star}$, let us compare the integration cycles on the spectral curve with transseries sectors, in the following two instances:
\begin{itemize}
\item Contributions solely weighted by transseries parameter $\sigma_{1}^{\ell}$ are purely ``forward'' contributions, of the type $\mathcal{Z}^{(\ell|0)}(t)$ following from \eqref{eq:InstantonsOneParameter}, and interpretable in spectral geometry via $B$-cycles as illustrated in the upper part of figure~\ref{fig:comparisonforwardvsdiagonal}.
\item Contributions solely weighted by transseries parameter $\upmu^{\ell}$ are (necessarily resonant) purely ``diagonal'' contributions, of the type $\mathcal{Z}^{(\ell|\ell)}(t)$ (\textit{e.g.}, as in the first line of \eqref{eq:PreDiagonalFramingTransseries}). They are interpretable in terms of equal pairs of eigenvalue and anti-eigenvalue tunneling \cite{mss22}, with associated cycles illustrated in the lower part of figure~\ref{fig:comparisonforwardvsdiagonal}. 
\end{itemize}

Now the key point is that comparing the contributions solely weighted by $\sigma$ to those solely weighted by $\upmu$, as clearly shown in figure~\ref{fig:comparisonforwardvsdiagonal}, they only differ by half \textit{versus} full cycles. In other words, diagonal contributions to the resonant transseries must be produced very similarly to the pure instanton contributions in \eqref{eq:InstantonsOneParameter}. It is then a matter of a few simple checks for one to conjecture the \textit{fully resonant} transseries solution, extending the $\ell$-instanton sector contributions of \eqref{eq:InstantonsOneParameter} with the aforementioned ``diagonal'' sectors, as:
\begin{equation}
\label{eq:pinchedcubic}
\CZ \left( g_{\text{s}}; \uprho, \upmu \right) = \frac{1}{\mathcal{Z}_{\text{G}} \left( \frac{\upmu}{2\pi\rmi} \right)} \sum_{\ell\in\mathbb{Z}} \uprho^{\ell}\, \mathcal{Z}_{\text{G}} \left( \frac{\upmu}{2\pi\rmi}+\ell \right) \exp \left\{ \widehat{\mathcal{F}} \left( t - g_{\text{s}} \left( \frac{\upmu}{2\pi\rmi}+\ell \right), g_{\text{s}} \left( \frac{\upmu}{2\pi\rmi}+\ell \right) \right) \right\},
\end{equation}
\noindent
with
\be
\CZ \left( g_{\text{s}}; \uprho, \upmu \right) \,\,\, \overset{\simeq}{\longleftrightarrow} \,\,\, Z \left( g_{\text{s}}; \sigma_1, \sigma_2 \right).
\ee
\noindent
The inclusion of $\upmu \in \BC$ always coupled to $\ell \in \BZ$ can be interpreted as allowing for ``fractional'' eigenvalue tunneling (which is completely natural in the large $N$ limit), making this the ultimate source for the existence of anti-eigenvalues or negative-tension branes. In addition, for the convenience of a cleaner main formula, we have normalized $\upmu$ in such a way that the canonical Stokes coefficients in both forward and backward directions are $1$. The asymptotic-equality arrow establishes a ``bridge'' \cite{mss22} between matrix-integral quantities on the left-hand-side (curly notation)---\textit{which now include diagonal-framing transseries parameters $\uprho$ and $\upmu$ whose map to rectangular-framing will be specified below}---, and the string-equation transseries solution on the right-hand-side (regular notation). This allows for a very compact way to write \textit{complete} resurgent-transseries solutions to hermitian matrix models (all their stringy limits included), which only depends on (pinched\footnote{As explained, the pinching always gives rise to a conifold/Gaussian singularity which requires regularization as in \cite{msw08}. Now, strictly speaking, we should regularize Gaussian \textit{minima} with the usual $\mathcal{Z}_{\text{G}}(g_{\text{s}}, N)$, whereas Gaussian \textit{maxima} should be regularized with $\mathcal{Z}_{\text{G}}(-g_{\text{s}}, N)$ instead. As it turns out, both singular behaviors are in fact the very same, which means that a regularization as in \eqref{eq:regularizedtwocut} and hence into \eqref{eq:pinchedcubic} is \textit{always} justified.}) spectral geometry data as we shall specify in the following.

Let us understand the different ingredients in our proposal \eqref{eq:pinchedcubic}. The two-cut free-energy input, $\widehat{\CF} (t_1,t_2)$ (pinched and regularized as explained) is \textit{purely perturbative\footnote{It may at first seem surprising that the fully exact nonperturbative result \eqref{eq:pinchedcubic} is seemingly constructed out of perturbative data alone. This is not true. First, the perturbative data is associated to the \textit{unpinched} problem. Second, our proposal \eqref{eq:pinchedcubic} does include fully nonperturbative data, in the guise of \textit{Stokes data} which is already present via the Fourier parameter $\uprho$ (see the discussion below, as well as \cite{krsst26b}).} data}, hence recursively computed directly from the topological recursion \cite{eo07a, eo08, eo09} (also see the discussion in section~\ref{sec:topological-recursion}). The spectral curve required for this is the two-cut spectral curve together with a choice of cycles on it. Each distinct choice of cycles corresponds to a distinct way of writing the transseries in terms of corresponding $\left\{\uprho,\upmu\right\}$ variables---and understanding these various ``coordinate patches'' on the space of possible transseries parameters will be one of the main objectives of \cite{krsst26b}. For the examples presented herein we will simply make the canonical choice corresponding to \eqref{eq:InstantonActionmsw08}. In practice, however, the multi-cut topological recursion is hard to execute. In addition, the $g_{\text{s}}$-expansion of \eqref{eq:pinchedcubic} produces derivatives of the pinched multi-cut free-energies (akin to what happened back in \eqref{eq:partition-function-Z-msw08}). Explicitly,
\bea
\label{eq:zak-kernel}
\exp \left\{ \widehat{\mathcal{F}} \left( t - \nu g_{\text{s}}, \nu g_{\text{s}} \right) - \mathcal{F} (t) \right\}
&=& \rme^{- \nu\, \frac{\widehat{A}(t)}{g_{\text{s}}}}\, \exp \left( \frac{\nu^2}{2} \left.\partial^2_{s} \widehat{\mathcal{F}}_0 (t,s) \right|_{t_2=0} \right) \times \\
&&
\times \left\{ 1 - g_{\text{s}} \left( \nu \left. \partial_s \widehat{\mathcal{F}}_1 (t,s) \right|_{t_2=0} + \frac{\nu^3}{6} \left. \partial_s^{3} \widehat{\mathcal{F}}_0 (t,s) \right|_{t_2=0} \right) + \mathcal{O} (g_{\text{s}}^2) \right\}, \nonumber
\eea
\noindent
where we have immediately identified $\left. \partial_s \widehat{\mathcal{F}}_0 (t,s) \right|_{s=0}$ with the instanton action as in \eqref{eq:InstantonActionmsw08}. All these derivatives can be conveniently expressed in terms of the original one-cut spectral curve rather than the full multi-cut spectral curve \cite{msw08}; which results in an efficient albeit involved algorithmic-tool for computing the transseries up to any order in $g_{\text{s}}$. For example \cite{msw08},
\bea
\label{eq:msw08-ds2F0}
\left. \partial_s^2 \widehat{\mathcal{F}}_0 (t,s) \right|_{t_2=0} &=& \log \frac{\left(x_1-x_2\right)^2}{16\, M(x^{\star}) \left[ \left(x_1-x^{\star}\right) \left(x_2-x^{\star}\right) \right]^{5/2}}, \\
\label{eq:msw08-d3sF0}
\left. \partial_s^3 \widehat{\mathcal{F}}_0 (t,s) \right|_{t_2=0} &=& \frac{1}{\sqrt{\left(x^{\star}-x_1\right) \left(x^{\star}-x_2\right)}}\, \Bigg\{ \frac{8 \left(x^{\star}-x_1\right)^2}{\left(x_1-x_2\right)^2 \left(x^{\star}-x_2\right)^2\, M(x_2)} + \\
&&
\hspace{-40pt}
+ \frac{8 \left(x^{\star}-x_2\right)^2}{\left(x_1-x_2\right)^2 \left(x^{\star}-x_1\right)^2\, M(x_1)} + \frac{4\, M'(x^{\star})^2}{M(x^{\star})^3} - \frac{3\, M''(x^{\star})}{2\, M(x^{\star})^2} - \nonumber \\
&&
\hspace{-40pt}
- \frac{17 \left(x_1+x_2-2x^{\star}\right)\, M'(x^{\star})}{2 \left(x^{\star}-x_1\right) \left(x^{\star}-x_2\right)\, M(x^{\star})^2} + \frac{77x_1^2 + 118x_1 x_2 + 77x_2^2 - 272x^{\star} \left(x_1+x_2-x^{\star}\right)}{8 \left(x^{\star}-x_1\right)^2 \left(x^{\star}-x_2\right)^2\, M(x^{\star})} \Bigg\}, \nonumber\\
\label{eq:msw08-dsF1}
\left. \partial_s \widehat{\mathcal{F}}_1 (t,s) \right|_{t_2=0} &=& \frac{1}{\sqrt{\left(x^{\star}-x_1\right) \left(x^{\star}-x_2\right)}}\, \Bigg\{ \frac{M'(x^{\star})^2}{6\,M(x^{\star})^3} - \frac{M''(x^{\star})}{8\, M(x^{\star})^2} - \\
&&
\hspace{-40pt}
- \frac{\left(x_1+x_2-2x^{\star}\right)\, M'(x^{\star})}{24 \left(x^{\star}-x_1\right) \left(x^{\star}-x_2\right)\, M(x^{\star})^2} + \frac{19 \left(x_1^2+x_2^2\right) - 22x_1 x_2 - 16x^{\star} \left(x_1+x_2-x^{\star}\right)}{96 \left(x^{\star}-x_1\right)^2 \left(x^{\star}-x_2\right)^2\, M(x^{\star})} - \nonumber \\
&&
\hspace{-40pt}
- \frac{\left(x^{\star}-x_2\right) \left(x_2-3x_1+2x^{\star}\right)}{3 \left(x_1-x_2\right)^2 \left(x^{\star}-x_1\right)^2\, M(x_1)} - \frac{\left(x^{\star}-x_1\right) \left(x_1-3x_2+2x^{\star}\right)}{3 \left(x_1-x_2\right)^2 \left(x^{\star}-x_2\right)^2\, M(x_2)} - \nonumber \\
&&
\hspace{-40pt}
- \frac{\left(x^{\star}-x_2\right)\, M'(x_1)}{4 \left(x_1-x_2\right) \left(x^{\star}-x_1\right)\, M(x_1)^2} + \frac{\left(x^{\star}-x_1\right)\, M'(x_2)}{4 \left(x_1-x_2\right) \left(x^{\star}-x_2\right)\, M(x_2)^2} \Bigg\}. \nonumber
\eea
\noindent
We refer the reader to \cite{msw08} for details and further explicit expressions.

With these formulae in hand it is straightforward to make sense of our proposal \eqref{eq:pinchedcubic} also for the double-scaled case (previously addressed in subsection~\ref{subsec:DSL-phases}). The double-scaled spectral-curve now has the form
\begin{equation}
y_{\text{ds}} (z) = M_{\text{ds}} (z) \sqrt{z-a}.
\end{equation}
\noindent
Given this expression and using standard techniques in \cite{gs21, sst23}, one can double-scale\footnote{In the process one also has to take into account a factor of two that appears between double-scaled and matrix-model formulae, and which is due to half-cycles in the matrix model in comparison to the usual full-cycles for the double-scaled curve \cite{gs21, sst23}. See as well the comment after \eqref{eq:generictransseriesmatrixmodels} and \cite{emms22a}.} the above derivative formulae. With $x^{\star}$ the nonperturbative saddle, root of the above moment function, one finds\footnote{Recall that in the double-scaling limit $g_{\text{s}} \to 0$ as $t \to t_{\text{critical}}$ at fixed \eqref{eq:multicriticalpoint}.}
\bea
\label{eq:msw08-dsF0-DSL}
\left. \partial_{s} \widehat{\mathcal{F}}^{\text{ds}}_0 (s) \right|_{s=0} &=& - \oint_{B} \rmd z\, y_{\text{ds}} (z), \\
\label{eq:msw08-d2sF0-DSL}
\left. \partial_{s}^2 \widehat{\mathcal{F}}^{\text{ds}}_0 (s) \right|_{s=0} &=& - \log \left( - 32 \left(x^{\star}-a\right)^{5/2} M_{\text{ds}}^{\prime} (x^{\star}) \right), \\
\label{eq:msw08-d3sF0-DSL}
\left. \partial_{s}^3 \widehat{\mathcal{F}}^{\text{ds}}_0 (s) \right|_{s=0} &=& - \frac{8 \left(x^{\star}-a\right)^2 M_{\text{ds}}^{\prime\prime} (x^{\star})^2 + 77 M_{\text{ds}}^{\prime} (x^{\star})^2}{16 \left(x^{\star}-a\right)^{5/2} M_{\text{ds}}^{\prime} (x^{\star})^3} + \frac{2\left(x^{\star}-a\right) M_{\text{ds}}^{\prime\prime\prime} (x^{\star}) - 17 M_{\text{ds}}^{\prime\prime} (x^{\star})}{8 \left(x^{\star}-a\right)^{3/2} M_{\text{ds}}^{\prime} (x^{\star})^2} + \nonumber \\
&&
+ \frac{4}{\left(x^{\star}-a\right)^{3/2}M_{\text{ds}} (a)}, \\
\label{eq:msw08-dsF1-DSL}
\left. \partial_{s} \widehat{\mathcal{F}}^{\text{ds}}_1 (s) \right|_{s=0} &=& - \frac{4 \left(x^{\star}-a\right)^2 M_{\text{ds}}^{\prime\prime} (x^{\star})^2 + 19 M_{\text{ds}}^{\prime} (x^{\star})^2}{192 \left(x^{\star}-a\right)^{5/2} M_{\text{ds}}^{\prime} (x^{\star})^3} + \frac{2 \left(x^{\star}-a\right) M_{\text{ds}}^{\prime\prime\prime} (x^{\star}) - M_{\text{ds}}^{\prime\prime} (x^{\star})}{96 \left(x^{\star}-a\right)^{3/2} M_{\text{ds}}^{\prime} (x^{\star})^2} + \nonumber \\
&&
+\frac{M_{\text{ds}} (a) - 3 \left(x^{\star}-a\right) M_{\text{ds}}^{\prime} (a)}{24 \left(x^{\star}-a\right)^{3/2} M_{\text{ds}} (a)^2}.
\eea
\noindent
Herein $s$ arises from the $A$-cycle associated to the unpinched, nonperturbative cut\footnote{The definition of $s$ in the double-scaling limit has a $(-)$ sign in its definition, in comparison to off-criticality.}. Rewriting $\widehat{\mathcal{F}}_{\text{ds}} (s)$ as a function of $\nu$ hence simply amounts\footnote{This procedure is really in complete analogy to the off-critical one in \cite{msw08}, where the free energy is sometimes written in the variables $t_1,t_2$ and other times in the variables $t,s$.} to $\widehat{\mathcal{F}}_{\text{ds}} (\nu g_{\text{s}})$, which becomes the ``double-scaled multi-cut'' (or unpinched) free energy. As such, given the above formulae, also in the double-scaling limit the multi-cut free-energy is evaluated via derivatives of the one-cut (standard double-scaled) free energy; equivalently, the double-scaled analogue of \eqref{eq:zak-kernel} is
\bea
\label{eq:dsl-fourier-kernel}
\exp \left\{ \widehat{\mathcal{F}}_{\text{ds}} \left( \nu g_{\text{s}} \right) - \mathcal{F}_{\text{ds}} \right\} &=& \rme^{- \nu\, \frac{\widehat{A}_{\text{ds}}}{g_{\text{s}}}}\, \exp \left( \frac{\nu^2}{2} \left. \partial^2_{s} \widehat{\mathcal{F}}^{\text{ds}}_0 (s) \right|_{s=0} \right) \times \\
&&
\times \left\{ 1 + g_{\text{s}} \left( \nu \left. \partial_s \widehat{\mathcal{F}}_1^{\text{ds}} (s) \right|_{s=0} + \frac{\nu^3}{6} \left. \partial_s^{3} \widehat{\mathcal{F}}_0^{\text{ds}} (s) \right|_{s=0} \right) + \mathcal{O} (g_{\text{s}}^2) \right\}. \nonumber
\eea
\noindent
which of course only depends on $a$ and $x^{\star}$ from the spectral curve, via the earlier derivative formulae. This expansion can now be used to formulate the double-scaled analogue of the nonperturbative partition function \eqref{eq:pinchedcubic}. It simply reads
\begin{equation}
\label{eq:pinchedcubic-dsl}
\CZ_{\text{ds}} \left( g_{\text{s}}; \uprho, \upmu \right) = \frac{1}{\mathcal{Z}_{\text{G}} \left( \frac{\upmu}{2\pi\rmi} \right)} \sum_{\ell\in\mathbb{Z}} \uprho^{\ell}\, \mathcal{Z}_{\text{G}} \left( \frac{\upmu}{2\pi\rmi}+\ell \right) \exp \left\{ \widehat{\mathcal{F}}_{\text{ds}} \left( g_{\text{s}} \left( \frac{\upmu}{2\pi\rmi}+\ell \right) \right) \right\} \,\,\, \overset{\simeq}{\longleftrightarrow} \,\,\, Z_{\text{ds}} \left( g_{\text{s}}; \sigma_1, \sigma_2 \right).
\end{equation}

In both instances we have normalized the resulting partition function \eqref{eq:pinchedcubic} or \eqref{eq:pinchedcubic-dsl} with a Gaussian pre-factor\footnote{The choice of this normalizing pre-factor will also be discussed in greater detail in \cite{ss26}.} $\mathcal{Z}_{\text{G}} (\upmu)$, which is $t$ independent (and $z$ independent in the double-scaling limit). Such a normalization is always allowed by the difference-equation for the free-energy \eqref{eq:freeenergyfromstringequation}, which precisely only fixes the $t$ dependence allowing the normalization to depend on all else. As it turns out, this normalization is also a natural one from the standpoint of matrix integral\footnote{As we will see in the following, it is needed in order to exactly reproduce the diagonal matrix integrals without having to worry about any remaining pre-factors.} computations \cite{mss22}.

Finally, we need to specify the diagonal transseries coefficients $(\uprho,\upmu)$, how they relate to rectangular coefficients $(\sigma_1,\sigma_2)$, and how they undergo Stokes phenomena. The simple one is $\upmu$: it is defined as usual $\upmu \equiv \sigma_1 \sigma_2$ and as made explicit in \eqref{eq:pinchedcubic} it controls the size of the pinched cycle, from the set of integers to the realm of complex numbers. The hard one is $\uprho$: the Fourier parameter, which is initially motivated by the \PI~results in \cite{bssv22} and generically found to be of the form\footnote{The reader may wonder if there is an asymmetry in the map towards rectangular transseries parameters, as there is a specific selection of $\sigma_1$ in this expression. This is solely due to the local nature of the transseries, and as one moves global another $\uprho$ coordinate relating to $\sigma_2$ will also be needed, in order to cover the full moduli space. This already occurred in \cite{bssv22} for \PI, and will be explained in the present context in \cite{krsst26b}.}
\be
\label{eq:rectangular-map}
\uprho \sim \frac{\sigma_1}{\Gamma \left( 1+\alpha \upmu \right)}\, \rme^{\alpha \upmu}.
\ee
\noindent
As in \cite{bssv22}, its form is dictated by the generating function of non-linear Stokes data for the problem at hand \cite{krsst26b}, and herein $\alpha$ is a problem-dependent coefficient. Compare with \eqref{eq:DFTrhoP1} for \PI, \eqref{eq:DFTrhosYL} for \YL, \eqref{eq:cubicrho} for the cubic model, and \eqref{eq:quarticrho} for the quartic model. The spectral-geometry motivation for its functional form will appear in \cite{krsst26b}. As for $\alpha$, it may be computed from spectral geometry following the prescription to be discussed in subsection~\ref{subsec:KdV-solutions}.

It is immediate to check our proposal \eqref{eq:pinchedcubic}-\eqref{eq:pinchedcubic-dsl} by simply reversing the analyses in the previous subsection~\ref{subsec:resurgent-Z-transasymptotics}, on what concerns our several examples. Further, model-independent checks may also be accomplished by direct comparison against matrix integrals as computed in \cite{mss22} (additional direct matrix-integral checks will be presented in \cite{krsst26b}). Begin with the latter:
\begin{itemize}
\item \textbf{Matrix Integrals:} The first obvious and trivial check is when $\upmu=0$, in which case \eqref{eq:pinchedcubic} exactly agrees with \eqref{eq:InstantonsOneParameter} which has been extensively checked in \cite{msw08}. This takes care of all the purely-forward instantons, albeit little they may be of the full transseries. Next, let us address the contribution when $\ell=0$ in \eqref{eq:pinchedcubic}, \textit{i.e.}, the case when the transseries has no dependence on $\uprho$ (hence, no dependence on $\sigma_1$ alone) but only\footnote{This is almost like trading $\ell$ with $\upmu$ in \eqref{eq:InstantonsOneParameter} which would then lead to \eqref{eq:partition-function-Z-msw08} upon series-expansion, except for the important caveat that we are now interested in the power-series expansion in $\upmu$ rather than in $g_{\text{s}}$.} on $\upmu$---this is the main diagonal. In particular, comparison with matrix integrals for these diagonal sectors requires fixed Stokes data (which now is solely canonical Stokes coefficients; see \cite{mss22, sst23}), and in particular herein this requires fixing $\upmu = 1$ \cite{mss22} (recall that for the convenience of a clean main formula we have normalized our proposal \eqref{eq:pinchedcubic} such that these canonical Stokes data equal unity). Expanding \eqref{eq:pinchedcubic} \textit{to first order} in $\upmu$ allows for a direct comparison with the $(1|1)$ sector, which, following \cite{mss22}, we express as a ratio of partition functions. Explicitly,
\begin{equation}
\frac{\CZ^{(1|1)} (t, g_{\text{s}})}{\CZ^{(0|0)} (t, g_{\text{s}})} = - \underbrace{\left( \frac{\upmu}{2\pi\rmi} \right)}_{\upmu \to 1} \left\{ \frac{1}{g_{\text{s}}}\, \partial_s \widehat{\mathcal{F}}_0 (s) + g_{\text{s}}\, \partial_s \widehat{\mathcal{F}}_1 (s) + \cdots \right\}.
\end{equation}
\noindent
Comparison with the respective matrix integrals evaluated in \cite{mss22, sst23}, at lowest order in $g_{\text{s}}$, yields perfect agreement upon the use of \eqref{eq:InstantonActionmsw08}---indeed we just find $\frac{\rmi}{2\pi}\, \widehat{A}$, which is in perfect agreement with formula (3.70) in \cite{mss22}. At next-to-leading order in $g_{\text{s}}$ one can use the computations of $\partial_s \widehat{\mathcal{F}}_1$ and $\partial_s^2 \widehat{\mathcal{F}}_0$ in \cite{msw08} (done for arbitrary two-cut spectral curves). The explicit comparison is rather lengthy, and we omit full details herein, but again shows perfect agreement with the generic next-to-leading order result presented in formula (3.70) of \cite{mss22}. Moving towards \textit{second order} in $\upmu$ the comparison\footnote{The direct computation of this matrix integral will appear in \cite{krsst26b}.} is now against the $(2|2)$ sector,
\begin{equation}
\frac{\CZ^{(2|2)} (t, g_{\text{s}})}{\CZ^{(0,0)} (t, g_{\text{s}})} = - \underbrace{\left( \frac{\upmu^2}{4\pi^2} \right)}_{\upmu \to 1} \left\{ \left( \frac{1}{g_{\text{s}}}\, \partial_s \widehat{\CF}_0 + g_{\text{s}}\, \partial_s \widehat{\CF}_1 + \cdots \right)^2 - \left( \partial_s^2 \widehat{\CF}_0 + g_{\text{s}}^2\, \partial_s^2 \widehat{\CF}_1 + \cdots \right) \right\}.
\end{equation}
\noindent
Moving on towards contributions just involving negative-instantons, these are easily obtainable from  \eqref{eq:pinchedcubic} by simply noticing that the sum is over $\ell \in \BZ$ hence it produces terms in $\sigma_2 \sim \frac{\upmu}{\uprho}$. For the $(0|1)$ sector, consider contributions with $\ell = -1$ but \textit{to first order} in $\upmu$. In fact, at lowest order in $\upmu$ we find zero in the expansion of the Gaussian partition-function appearing in \eqref{eq:pinchedcubic}, \textit{i.e.},
\begin{equation}
\label{eq:gaussian-zero}
\mathcal{Z}_{\text{G}} \left( \frac{\upmu}{2\pi\rmi}-1 \right) = 0 - \rmi \sqrt{\frac{g_{\text{s}}}{2\pi}}\, \upmu + \sqrt{g_{\text{s}}}\, \frac{1 - 2 \gamma_{\text{E}} + 2 \log g_{\text{s}}}{2 \left( 2\pi \right)^{3/2}}\, \upmu^2 + \mathcal{O} (\upmu^3).
\end{equation}
\noindent
This implies that the $\ell = -1$ contribution to \eqref{eq:pinchedcubic}, to leading order in $g_{\text{s}}$, is
\bea
\label{eq:neg-inst-test-matrix-integral}
\frac{1}{\uprho}\, \mathcal{Z}_{\text{G}} \left( \frac{\upmu}{2\pi\rmi}-1 \right) \exp \left\{ \widehat{\mathcal{F}} \left( t - g_{\text{s}} \left( \frac{\upmu}{2\pi\rmi}-1 \right), g_{\text{s}} \left( \frac{\upmu}{2\pi\rmi}-1 \right) \right) - \mathcal{F} (t) \right\} &=& \\
&&
\hspace{-275pt}
= \frac{\upmu}{\uprho} \left( \frac{\rmi}{\sqrt{2\pi}}\, \sqrt{g_{\text{s}}}\, \exp \left( \frac{1}{2} \partial_s^2 \widehat{\mathcal{F}}_0 \right) + \mathcal{O} (g_{\text{s}}^{3/2}) \right) + \frac{\upmu^2}{\uprho} \left( \cdots \right) + \frac{\upmu^3}{\uprho} \left( \cdots \right) + \cdots. \nonumber
\eea
\noindent
From the map to rectangular framing \eqref{eq:rectangular-map} we know that $\sigma_2$ is reproduced in the limit where $\uprho, \upmu \to + \infty$ whilst keeping their ratio fixed as $\sigma_2$. This further implies that the higher-order terms in $\upmu$ in \eqref{eq:neg-inst-test-matrix-integral} vanish in this limit, and we obtain for $(0|1)$ sector
\begin{equation}
\frac{\CZ^{(0|1)} (t, g_{\text{s}})}{\CZ^{(0|0)} (t, g_{\text{s}})} = \underbrace{\,\,\, \sigma_2 \,\,\,}_{\sigma_2 \to 1} \frac{\rmi}{\sqrt{2\pi}}\, \sqrt{g_{\text{s}}}\, \exp \left( \frac{1}{2} \partial_s^2 \widehat{\mathcal{F}}_0 \right) + \mathcal{O} (g_{\text{s}}^{3/2}),
\end{equation}
\noindent
which exactly agrees with the matrix integral prediction given in formula (3.54) in \cite{mss22} (and where, again, $\partial_s^2 \widehat{\mathcal{F}}_0$ is computed in \eqref{eq:msw08-ds2F0}). In the same way, we can also match higher negative-instanton contributions. For the $\ell$-th negative instanton, pick the $\uprho^{-\ell}$ contribution. In analogy with \eqref{eq:gaussian-zero} the corresponding Gaussian contribution will have a zero of order $\ell$---which then leads to transseries parameter $\sim \left( \frac{\upmu}{\uprho} \right)^{\ell} = \sigma_2^{\ell}$ as pursued. Setting $\sigma_2 \to 1$ yields at leading order in $g_{\text{s}}$,
\begin{equation}
\frac{\CZ^{(0|2)} (t, g_{\text{s}})}{\CZ^{(0|0)} (t, g_{\text{s}})} = \frac{g_{\text{s}}}{2\pi}\, \exp \left( 2 \partial_s^2 \widehat{\mathcal{F}}_0 \right), \qquad \frac{\CZ^{(0|3)} (t, g_{\text{s}})}{\CZ^{(0|0)} (t, g_{\text{s}})} = \frac{\rmi g_{\text{s}}^{\frac{9}{2}}}{\sqrt{2}\pi^{3/2}}\, \exp \left( \frac{9}{2} \partial_s^2 \widehat{\mathcal{F}}_0 \right), \qquad \cdots.
\end{equation}
\noindent
Comparing against formula (3.54) in \cite{mss22} yields perfect agreement. This concludes the tests of the ``boundaries'' and ``main diagonal'' of our resurgent lattice as in figure~\ref{fig:twopararectmtransseriesgrid}. The next step is hence addressing the off-diagonal inner-lattice. However, herein resonant sums start playing a major role, as thoroughly described in \cite{mss22}. On top, let us again emphasize how matrix integrals inherently contain \textit{fixed} Stokes data, which starts being highly intricate in the ``bulk'' of this resurgent lattice; whereas \eqref{eq:pinchedcubic} is written as a transseries with \textit{unfixed} transseries parameters. This leads to intricate comparisons/calculations, akin to the ones in \cite{mss22}, which we postpone to \cite{krsst26b} as that will be where generic Stokes data is addressed. Therein these Stokes data alongside \eqref{eq:pinchedcubic} will be seen to produce further and highly non-trivial matches.
\item \textbf{Cubic Matrix Model:} The cubic matrix model \eqref{eq:CubicMatrixModelPotential} has one-cut spectral-curve \eqref{eq:cubic-spectral-curve-one-cut}. We wish to construct its exact transseries partition-function following from our generic proposal \eqref{eq:pinchedcubic}, and compare the result to our earlier explicit derivation from its string equation---which culminated in \eqref{eq:CMMDiscreteFourierNC}. As the partition function as computed from the string equation is only defined up to a global pre-factor, the meaningful comparison to do concerns the ``discrete Fourier modes'', \textit{i.e.}, compare \eqref{eq:DFTKernelCNC} with the modes of \eqref{eq:pinchedcubic}. Let us start with our generic proposal \eqref{eq:pinchedcubic} and expand the modes inside the infinite sum in $g_{\text{s}}$ (see also formula \eqref{eq:zak-kernel}). We find
\bea
\mathcal{Z}_{\text{G}} (\nu)\, \exp \left\{ \widehat{\mathcal{F}} \left( t - \nu g_{\text{s}}, \nu g_{\text{s}} \right) \right\} &=& \exp \left\{ \frac{1}{g_{\text{s}}^2}\, \mathcal{F}_{-2} + \mathcal{F}_0 \right\}\, \exp \left( \frac{\nu^2}{2} \left. \partial^2_{s} \widehat{\mathcal{F}}_0 (t,s) \right|_{t_2=0} \right) \rme^{- \nu\, \frac{\widehat{A}(t)}{g_{\text{s}}}} \times \nonumber \\
&&
\hspace{-120pt}
\times \mathcal{Z}_{\text{G}} (\nu) \left\{ 1 - g_{\text{s}} \left( \nu \left. \partial_s \widehat{\mathcal{F}}_1 (t,s) \right|_{t_2=0} + \frac{\nu^3}{6} \left. \partial_s^{3} \widehat{\mathcal{F}}_0 (t,s) \right|_{t_2=0} \right) + \mathcal{O} (g_{\text{s}}^2) \right\},
\eea
\noindent
where the various free-energy derivatives at play may be computed from the aforementioned one-cut spectral-curve of the cubic-model, \eqref{eq:cubic-spectral-curve-one-cut}---see \eqref{eq:msw08-ds2F0}-\eqref{eq:msw08-d3sF0}-\eqref{eq:msw08-dsF1} and \cite{msw08, sst23}. Comparing the above with \eqref{eq:DFTKernelCNC} entails the following identities must hold:
\bea
\frac{1}{2} \left. \partial^2_{s} \widehat{\mathcal{F}}_0 (t,s) \right|_{t_2=0} &=& \log p_{\text{cubic}} (t), \\
1 &=& D_0 (\nu),\\
- \nu \left. \partial_s \widehat{\mathcal{F}}_1 (t,s) \right|_{t_2=0} - \frac{\nu^3}{6} \left. \partial_s^{3} \widehat{\mathcal{F}}_0 (t,s) \right|_{t_2=0} &=& D_1 (\nu),
\eea
\noindent
which indeed they do to yield perfect agreement (upon a lengthy computation and the final use of \eqref{eq:pcubicpP1}-\eqref{eq:CMM-Dk0}-\eqref{eq:CMM-Dk1}). Conversely, the above formulae yield matrix-integral expressions for the $D_{k} (\nu)$ polynomials. Note how by comparing Fourier modes we are indeed checking all instanton orders, up to and including the next-to-leading non-zero $g_{\text{s}}$-order of the partition-function transseries; a highly non-trivial matching.
\item \textbf{Painlev\'e~I:} The \PI~double-scaled model \eqref{eq:Painleve1Equation} has double-scaled spectral curve
\be
\label{eq:PI-sc-for-conjecture}
y (z) = -3 \sqrt{3} \left( z-1 \right) \sqrt{z}.
\ee
\noindent
To construct its exact transseries partition-function we follow the double-scaled discussion\footnote{The unpinching of double-scaled spectral-curves will be addressed in \cite{ss26}.} leading up to \eqref{eq:pinchedcubic-dsl} and again compute the Fourier modes as in \eqref{eq:dsl-fourier-kernel}, using our earlier double-scaled expressions \eqref{eq:msw08-dsF0-DSL}-\eqref{eq:msw08-d2sF0-DSL}-\eqref{eq:msw08-d3sF0-DSL}-\eqref{eq:msw08-dsF1-DSL}. These expressions for derivatives are evaluated directly from the spectral-curve input \eqref{eq:PI-sc-for-conjecture} and yield
\bea
\left. \partial_{s} \widehat{\mathcal{F}}^{\text{\PI}}_0 (s) \right|_{s=0} &=& \frac{8\sqrt{3}}{5}, \\
\exp \left( \left. \partial_{s}^2 \widehat{\mathcal{F}}^{\text{\PI}}_0 (s) \right|_{s=0} \right) &=& - \frac{1}{96 \sqrt{3}}, \\
\left. \partial_{s}^3 \widehat{\mathcal{F}}^{\text{\PI}}_0 (s) \right|_{s=0} &=& - \frac{47}{16 \sqrt{3}}, \\
\left. \partial_{s} \widehat{\mathcal{F}}^{\text{\PI}}_1 (s) \right|_{s=0} &=& - \frac{17}{192 \sqrt{3}},
\eea
\noindent
Comparison with \eqref{eq:DFTKernelPINC} and table~\ref{tab:PIDkPolynomials}, obtained constructively from the \PI~equation, yields indeed perfect\footnote{Recall that in table~\ref{tab:PIDkPolynomials} we are listing the $\widetilde{D}_{k}(\nu)$ polynomials, which relate via \eqref{eq:Dktilde-versus-Dknontilde}; and we rescaled $\upmu$.} agreement. One additional cross-check we can perform aligns this calculation with the previous cubic matrix model one: simply double-scale our earlier results back to the \PI~partition-function (this is basically done in equation (4.75) of \cite{msw08}). One finds the identifications
\begin{align}
\frac{1}{g_{\text{s}}}\, \partial_s \widehat{\mathcal{F}}_0 & \xrightarrow{\text{DSL}}A_{\text{\PI}} z^{\frac{5}{4}}, & \exp \left( \partial_s^2 \widehat{\mathcal{F}}_0 \right) &\xrightarrow{\text{DSL}} - \frac{1}{96 \sqrt{3}}, \\
g_{\text{s}}\, \partial_s \widehat{\mathcal{F}}_1 & \xrightarrow{\text{DSL}} -\frac{17}{192 \sqrt{3}} z^{-\frac{5}{4}}, & g_{\text{s}}\, \partial_s^3 \widehat{\mathcal{F}}_0 & \xrightarrow{\text{DSL}} - \frac{47}{16 \sqrt{3}} z^{-\frac{5}{4}}.
\end{align}
\noindent
We can compute the first two $D_{k} (\nu)$ polynomials in table~\ref{tab:PIDkPolynomials} and, upon comparison with \eqref{eq:DFTKernelPINC}, we find again perfect agreement. For higher-order checks one would need to run the two-cut topological recursion and compute higher-order derivatives of $\mathcal{F}$. We will perform further extensive checks which contain Stokes data in \cite{krsst26b}.
\end{itemize}

Our complete and exact transseries \eqref{eq:pinchedcubic} includes all resonant, nonperturbative corrections for problems with a \textit{single} saddle-point, $x^{\star}$. Of course we can be fully generic; and it is rather straightforward to extend \eqref{eq:pinchedcubic} so as to include any number of saddles. Consider a one-cut problem in the presence of $\kappa$ nonperturbative saddles (or pinched cycles; recall figure~\ref{fig:OneCutPotential}), denoted by $x_{1}^{\star}, \ldots, x_{\kappa}^{\star}$. Then \eqref{eq:pinchedcubic} simply becomes, with vector notation,
\bea
\label{eq:pinchedmulti}
\CZ \left( g_{\text{s}}; \boldsymbol{\uprho}, \boldsymbol{\upmu} \right) &=& \frac{1}{\prod\limits_{i=1}^{\kappa} \mathcal{Z}_{\text{G}} \left( \frac{\upmu_i}{2\pi\rmi} \right)}\, \sum_{\boldsymbol{\ell} \in \mathbb{Z}^{\kappa}} \boldsymbol{\uprho}^{\boldsymbol{\ell}}\, \prod_{i=1}^{\kappa} \mathcal{Z}_{\text{G}} \left( \frac{\upmu_i}{2\pi\rmi}+\ell_i \right) \times \\
&&
\hspace{75pt}
\times \exp \left\{ \widehat{\mathcal{F}} \left( t -g_{\text{s}} \sum_{i=1}^{\kappa} \left( \frac{\upmu_i}{2\pi\rmi}+\ell_i \right), g_{\text{s}} \left( \frac{\boldsymbol{\upmu}}{2\pi\rmi}+\boldsymbol{\ell} \right) \right) \right\}. \nonumber
\eea
\noindent
Herein, $\boldsymbol{\upmu} \in \BC^{\kappa}$ which is always coupled to $\boldsymbol{\ell} \in \BZ^{\kappa}$ is again interpreted as allowing for ``fractional'' eigenvalue tunneling in-between the many saddles. As in \eqref{eq:pinchedcubic}, there is an asymptotic-equality arrow establishing a ``bridge'' \cite{mss22} between matrix-integrals and string-equation transseries,
\be
\CZ \left( g_{\text{s}}; \boldsymbol{\uprho}, \boldsymbol{\upmu} \right) \quad \overset{\simeq}{\longleftrightarrow} \quad Z \left( g_{\text{s}}; \boldsymbol{\sigma} \right).
\ee
\noindent
The $(\kappa+1)$-cut free energy $\widehat{\CF} \left( t_1, t_2, \ldots, t_{\kappa+1} \right)$ has $\kappa$ pinched-cuts and is regularized at each of these pinches as in \eqref{eq:regularizedtwocut}. These are purely perturbative data, recursively computed via the topological recursion \cite{eo07a} starting from genus-zero which follows from special geometry as in \eqref{eq:specialgeometryrelations}. Here formula \eqref{eq:pinchedmulti} again implies a choice of cycles on the multi-cut curve. In addition we have normalized the exact partition-function with Gaussian pre-factors associated to each pinched-cycle, as this is the adequate normalization to exactly match against the matrix-integral computations which will be featured in \cite{krsst26b}. The diagonal transseries coefficients $(\boldsymbol{\uprho},\boldsymbol{\upmu})$ of course relate to the rectangular coefficients $\boldsymbol{\sigma}$ and undergo Stokes transitions. The ``moduli'' were introduced in \eqref{eq:PreDiagonalFramingTransseries} via $\upmu_i \equiv \sigma_i \widetilde{\sigma}_i$, and as explicit in the expression above control the size of the pinched cycles, from integer to complex values. The Fourier parameters $\boldsymbol{\uprho}$ are harder (as we will soon see in \eqref{eq:Fourier-parameters-MC-conjecture}), and the generic map to rectangular transseries parameters is spelled out in detail in subsection~\ref{subsec:KdV-solutions} (again via \eqref{eq:Fourier-parameters-MC-conjecture}).

Detailed (numerical) checks on our formulae will appear later in section~\ref{sec:checks-tests-numerics} as well as in \cite{krsst26b, krst26a}. For the moment we can do the same simple checks we did for \eqref{eq:pinchedcubic} now for our latest \eqref{eq:pinchedmulti}; \textit{i.e.}, by simple reversion of the analyses in the previous subsection~\ref{subsec:resurgent-Z-transasymptotics}, this time around on what concerns the quartic matrix model and \YL.
\begin{itemize}
\item \textbf{Quartic Matrix Model:} The quartic matrix model \eqref{eq:QuarticMatrixModelPotential} has one-cut spectral curve \eqref{eq:quartic-spectral-curve-one-cut}. As for the cubic, we wish to check our construction of its exact partition-function transseries, now following from \eqref{eq:pinchedmulti} due to its two nonperturbative saddles \eqref{eq:2-saddles-quartic}. This is done via comparison to our earlier explicit derivation from its string equation, resulting in \eqref{eq:QMMDiscreteFourierNC}. Just like we did in the example of the cubic model, we will only study the discrete Fourier modes \eqref{eq:quartic-dft-kernel} in the Zak transform \eqref{eq:QMMDiscreteFourierNC}, as global---possibly $g_{\text{s}}$-dependent---pre-factors as well as transseries parameters are not fixed by the string equation. In order to accommodate the two saddles \eqref{eq:2-saddles-quartic} in \eqref{eq:pinchedmulti} we set $\kappa = 2$, as well as $\uprho_1 = \uprho_2 \equiv \uprho$ and $\upmu_1 = \upmu_2 \equiv \upmu$; leading up to\footnote{Notice we have removed the $2\pi\rmi$-normalization as it is not needed in the following comparison.}
\bea
\label{eq:quartic-conjecture}
\CZ \left( g_{\text{s}}; \uprho, \upmu \right) &=& \frac{1}{\mathcal{Z}_{\text{G}} (\upmu)^2}\, \sum_{\boldsymbol{\ell} \in \mathbb{Z}^{2}} \uprho^{\ell_1+\ell_2}\, \mathcal{Z}_{\text{G}} (\upmu+\ell_1)\, \mathcal{Z}_{\text{G}} (\upmu+\ell_2) \times \\
&&
\times \exp \left\{ \widehat{\mathcal{F}} \big( t -g_{\text{s}} \left( 2\upmu+\ell_1+\ell_2 \right), g_{\text{s}} \left( \upmu+\ell_1 \right), g_{\text{s}} \left( \upmu+\ell_2 \right) \big) \right\}. \nonumber
\eea
\noindent
The relevant data to make this formula explicit are the derivatives of the pinched free-energy, and these may be computed from the quartic-model one-cut spectral-curve \eqref{eq:quartic-spectral-curve-one-cut}. Due to the $\BZ_2$-symmetry of the quartic potential the expansion in \eqref{eq:quartic-conjecture} may be simplified with the added identifications $\partial_{s_1}^{k} \widehat{F} (t,0,0) = \partial_{s_2}^{k}\widehat{F} (t,0,0)$ for any $k\in\mathbb{N}$. In comparing with the string-equation result, one needs to expand \eqref{eq:quartic-conjecture} in $g_{\text{s}}$ and study its lowest orders\footnote{The mixed $s_1, s_2$ derivatives of the quartic-model multi-cut free-energy quickly become quite involved. We refer the reader to \cite{krst26a} for explicit computations of these mixed derivatives of $\widehat{\mathcal{F}}_0$ for any spectral curve.}. This expansion is most conveniently presented by introducing the total instanton number $\ell=\ell_1+\ell_2$, so that
\be
\CZ \left( g_{\text{s}}; \uprho, \upmu \right) = \frac{1}{\mathcal{Z}_{\text{G}} (\upmu)^2}\, \sum_{\ell \in \mathbb{Z}} \uprho^{\ell}\, 
\begin{cases}
\mathrm{Z}_{\text{even}} \left( g_{\text{s}}; \ell, \upmu \right), \quad \ell \text{ even},\\
\mathrm{Z}_{\text{odd}} \left( g_{\text{s}}; \ell, \upmu \right), \quad \ell \text{ odd}.
\end{cases}
\ee
\noindent
The Fourier modes still encode the second instanton sum which is indexed by $\ell_1-\ell_2$, whose leading $g_{\text{s}}$ terms are explicitly written below, and it is here that the odd/even split from above appears. At leading (non-vanishing) order in $g_{\text{s}}$, one finds
\bea
\mathrm{Z}_{\text{even}} \left( g_{\text{s}}; \ell, \upmu \right) &=& \mathcal{Z}_{\text{G}}  \left( \upmu+\frac{\ell}{2} \right)^2 \exp \left\lbrace \frac{1}{g_{\text{s}}^2} \mathcal{F}_{-2} + \mathcal{F}_0 \right\rbrace \exp \left\lbrace - 2 \left( \upmu+\frac{\ell}{2} \right) \frac{\widehat{A}(t)}{g_{\text{s}}} \right\rbrace \times \\
&&
\hspace{-80pt}
\times \exp \left\lbrace \left( \upmu+\frac{\ell}{2} \right)^2 \left( \left. \partial^2_{s_1} \widehat{\mathcal{F}}_0 (t,s_1,s_2) \right|_{t_{2,3}=0} + \left. \partial_{s_1} \partial_{s_2} \widehat{\mathcal{F}}_0 (t,s_1,s_2) \right|_{t_{2,3}=0} \right) \right\rbrace\, \Big( 1 + \mathcal{O} (g_{\text{s}}) \Big), \nonumber \\
\mathrm{Z}_{\text{odd}} \left( g_{\text{s}}; \ell, \upmu \right) &=& 2 \cdot \mathcal{Z}_{\text{G}} \left( \upmu+\frac{\ell}{2}-\frac{1}{2} \right) \mathcal{Z}_{\text{G}} \left( \upmu+\frac{\ell}{2}+\frac{1}{2} \right) \exp \left\lbrace \frac{1}{g_{\text{s}}^2} \mathcal{F}_{-2} + \mathcal{F}_0 \right\rbrace \times \\
&&
\times \exp \left\lbrace - 2 \left( \upmu+\frac{\ell}{2} \right) \frac{\widehat{A}(t)}{g_{\text{s}}} \right\rbrace \exp \left\lbrace - \frac{1}{2} \left. \partial_{s_1}\partial_{s_2} \widehat{\mathcal{F}}_0 (t,s_1,s_2) \right|_{t_{2,3}=0} \right\rbrace \times \nonumber \\
&&
\hspace{-80pt}
\times \exp \left\lbrace \left( \upmu+\frac{\ell}{2}+\frac{1}{2} \right)^2 \left( \left. \partial^2_{s_1} \widehat{\mathcal{F}}_0 (t,s_1,s_2) \right|_{t_{2,3}=0} + \left. \partial_{s_1} \partial_{s_2} \widehat{\mathcal{F}}_0 (t,s_1,s_2) \right|_{t_{2,3}=0} \right) \right\rbrace\, \Big( 1 + \mathcal{O} (g_{\text{s}}) \Big). \nonumber
\eea
\noindent
We compare these formulae with the transeries construction \eqref{eq:QMMDiscreteFourierNC} by focusing on the Fourier modes \eqref{eq:quartic-dft-kernel}; which now entails the following identities must hold:
\bea
\left. \partial^2_{s_1} \widehat{\mathcal{F}}_0 (t,s_1,s_2) \right|_{t_{2,3}=0} + \left. \partial_{s_1} \partial_{s_2} \widehat{\mathcal{F}}_0 (t,s_1,s_2) \right|_{t_{2,3}=0} &=& 2 \log p_{\text{quartic}} (t), \\
- \frac{1}{2} \left. \partial_{s_1} \partial_{s_2} \widehat{\mathcal{F}}_0 (t,s_1,s_2) \right|_{t_{2,3}=0} &=& \log d,
\eea
\noindent
which indeed they do to yield perfect agreement at leading order in $g_{\text{s}}$ (upon a lengthy computation which requires the explicit expressions for the partial-derivatives of the multi-cut free-energy, computed in \cite{msw08} and also in \cite{krst26a} for the mixed derivative, and the final use of \eqref{eq:d2-app}-\eqref{eq:pquartic-app}). The same holds true for the instanton action. Notice how we perfectly reproduce the odd/even split found from our previous transasymptotic computation.
\item \textbf{Yang--Lee:} The \YL~double-scaled model \eqref{eq:YangLeeEquation} has double-scaled spectral-curve\footnote{Note that \eqref{eq:YL-sc-for-conjecture} is written in standard multicritical KdV conventions (see, \textit{e.g.}, \cite{gs21}) with the small change where instanton actions are computed from full cycles on the curve; \textit{e.g.}, \cite{ss03, emms22a, sst23}. This is a common extra normalization in double-scaled problems, which slightly differs from standard matrix-model conventions---where the action is a half-cycle associated with single eigenvalue tunneling; \textit{e.g.}, \cite{msw07, msw08, gs21, mss22}---by a factor of $1/2$.} (see, \textit{e.g.}, \cite{gs21})
\be
\label{eq:YL-sc-for-conjecture}
y (z) = \frac{1}{5} \left( 8 z^2 - 4z + 3 \right) \sqrt{z-1}.
\ee
\noindent
As we did in the earlier \PI~example, we would like to once again see how the exact partition-function transseries \eqref{eq:YLFullDFT} arises from a purely spectral-geometry discussion (now based upon \eqref{eq:pinchedmulti} due to the two saddles of this problem). Repeating the same\footnote{Albeit the two saddles would suggest a genus-two curve, the derivatives below are all with respect to \textit{only one} of the $s$-variables---hence we are free to pinch any other cycles and bring the curve back into genus-one; in which case the formulae to use are indeed the same as for \PI, \textit{i.e.}, equations \eqref{eq:msw08-dsF0-DSL}-\eqref{eq:msw08-d2sF0-DSL}-\eqref{eq:msw08-d3sF0-DSL}-\eqref{eq:msw08-dsF1-DSL}.} procedure as for \PI, we first compute\footnote{As mentioned in the discussion leading up to \eqref{eq:zak-kernel}, one is required to make a choice of cycles upon the spectral curve. In order to match with the string-equation solution we have obtained for \YL, it turns out that one must herein \textit{reverse} the canonical cycle choice (namely the one implied by \eqref{eq:InstantonActionmsw08}) for the first saddle, $x_1^{\star}$.} the derivatives
\begin{align}
\left. \partial_{s_1} \widehat{\mathcal{F}}^{\text{\YL}}_0 (s_1,s_2) \right|_{s_{1,2}=0} &= -A_1, & \left. \partial_{s_1} \widehat{\mathcal{F}}^{\text{\YL}}_1 (s_1,s_2) \right|_{s_{1,2}=0} &= - \frac{\sqrt{75750 - 46254 \rmi \sqrt{5}}}{4320}, \\
\exp \left( \left. \partial_{s_1}^2 \widehat{\mathcal{F}}^{\text{\YL}}_0 (s_1,s_2) \right|_{s_{1,2}=0} \right) &= \text{B}^{-1}, & \left. \partial_{s_1}^3 \widehat{\mathcal{F}}^{\text{\YL}}_0 (s_1,s_2) \right|_{s_{1,2}=0} &= - \frac{\sqrt{650310 - 433326 \rmi \sqrt{5}}}{360},
\end{align} 
\noindent
for the first saddle $x^{\star}_1 = \frac{1}{4} \left(1+\rmi \sqrt{5}\right)$; whilst for the second saddle $x^{\star}_2 = \frac{1}{4} \left(1-\rmi \sqrt{5}\right)$ we find
\begin{align}
\left. \partial_{s_2} \widehat{\mathcal{F}}^{\text{\YL}}_0 (s_1,s_2) \right|_{s_{1,2}=0} &= A_2, & \left. \partial_{s_2} \widehat{\mathcal{F}}^{\text{\YL}}_1 (s_1,s_2) \right|_{s_{1,2}=0} &= \frac{\sqrt{75750 + 46254 \rmi \sqrt{5}}}{4320}, \\
\exp \left( \left. \partial_{s_2}^2 \widehat{\mathcal{F}}^{\text{\YL}}_0 (s_1,s_2) \right|_{s_{1,2}=0} \right) &= -\bar{\text{B}}^{-1}, & \left. \partial_{s_2}^3 \widehat{\mathcal{F}}^{\text{\YL}}_0 (s_1,s_2) \right|_{s_{1,2}=0} &=  \frac{\sqrt{650310 + 433326 \rmi \sqrt{5}}}{360}.
\end{align} 
\noindent
Finally, we also require the mixed derivative term which is obtained in \cite{krst26a}, yielding 
\begin{equation}
\left. \partial_{s_2} \partial_{s_1} \widehat{\mathcal{F}}^{\text{\YL}}_0 (s_1,s_2) \right|_{s_{1,2}=0} = - \log \left( - \left( 11 + 2 \sqrt{30} \right) \right) \equiv - \log \left( - \text{C} \right). 
\end{equation}
\noindent
As always, we compare with the \YL~Fourier modes \eqref{eq:curlyZYangLeeFullExpression} rather than the discrete Fourier/Zak transform \eqref{eq:YLFullDFT}, as this is defined only up to normalization (and it is not relevant to compare arbitrary normalization with either string-equation or matrix-model origins). We find a complete structural match at leading order, as well as a match with four out of the six terms of the associated polynomial $D_1(\nu_1,\nu_2)$ dictating the next-to-leading order. In particular, we are able to match the coefficients of the terms which are linear and of third order in the $\nu_i$. The remaining two coefficients would require additional data from the matrix-model third-derivatives $\partial_{s_2 s_2 s_1}^3 \widehat{\mathcal{F}}^{\text{\YL}}_0$ and $\partial_{s_1 s_1 s_2}^3 \widehat{\mathcal{F}}^{\text{\YL}}_0$ .
\end{itemize}

\subsection{Comparing Approaches to Nonperturbative Partition Functions}
\label{subsec:from-theta-to-dual-to-transseries}

There were already a reasonable number of clues in the literature that, at least locally, \textit{i.e.}, at the level of the transseries itself, \eqref{eq:pinchedcubic} or \eqref{eq:pinchedmulti} had to be the final generic answer. Let us briefly review these results and compare them against our own. Partition functions written as discrete Fourier transforms have appeared in many diverse instances---from the Nekrasov--Okounkov dual partition-functions of gauge theory \cite{n02, no03} to the Eynard--Mari\~no background-independent partition-functions of random matrices \cite{e08, em08} to Liouville $c=1$ CFT conformal blocks \cite{gil12, gil13, ilt14, gl16, gl17, ddg20} to Painlev\'e tau-functions \cite{blmst16, gg18, i19, clt20, bgmt24, iilz25}---albeit with the exception of \cite{i19, clt20, im23} Stokes data was not fully addressed in these works; which we shall do from first principles in the matrix model \cite{krsst26b, krst26a, krst26b}. Further, as will be made clearer in \cite{ss26}, the aforementioned examples are mostly special cases of our general matrix-model framework.

Let us first compare with the background-independent partition-functions of \cite{em08}. Essentially building upon \eqref{eq:InstantonsOneParameter} \cite{bde00, e08, msw08} the discrete-Fourier-transform representation of the matrix-model partition-function in \cite{em08} was restricted to the non-resonant setting, \textit{i.e.}, all transseries sectors with anti-eigenvalues were missing (see figure~\ref{fig:twopararectmtransseriesgrid}). To be more explicit, this background-independent partition-function is expressed as, essentially, \eqref{eq:EM-2-cut-Z} with the theta-functions in \eqref{eq:theta-for-EM-2-cut-Z}. This expression was built starting from a sum over all eigenvalue distributions, as in \eqref{eq:BDE-sum}, and was shown to have the very nice property of being\footnote{Modular invariant in the sense that transseries coefficients have well-defined properties under symplectic transformations of the homology group basis of the spectral-curve underlying the matrix model.} modular invariant \cite{em08}. The matrix-model analysis in \cite{em08} restricted to the single\footnote{Whereas the modular invariance analysis remained fully general on what concerns theta-characteristics.} theta-function characteristic case, as in \eqref{eq:theta-for-EM-2-cut-Z}, where the resulting partition function matches the one-parameter instanton-expansion \eqref{eq:InstantonsOneParameter} from \cite{msw08}. Of course this misses a large part of the complete nonperturbative transseries, a part which will be shown in section~\ref{sec:checks-tests-numerics} and \cite{krsst26b} to be required if we wish to describe the matrix model in all its phases. On top of this, the partition-function in \cite{em08} was constructed directly as an expansion around a multi-cut background (no pinched cycles)---this would be a two-cut background when considering \eqref{eq:EM-2-cut-Z}---, in fact as already discussed in subsection~\ref{subsec:Z-phases}. Our novelties are two-fold: our proposal \eqref{eq:pinchedcubic}-\eqref{eq:pinchedmulti} includes all sectors, negative-instantons included; and, in contrast, our starting expansion point is around the simpler one-cut background---in fact, as our construction stems from string equations. A detailed exposition on the construction of the generic nonperturbative partition function \eqref{eq:EM-2-cut-Z} from all backgrounds will be addressed in \cite{ss26}.

As was already observed in \cite{e08, em08}, equation \eqref{eq:EM-2-cut-Z} (with both characteristics turned-on in \eqref{eq:theta-for-EM-2-cut-Z}) can be written\footnote{We are blending the notation in \cite{em08} with our own in \eqref{eq:pinchedcubic}, so that this expression is easily comparable to equation (3.57) in \cite{em08}.} in the form of a discrete Fourier transform as
\begin{equation}
\label{eq:compact-EM-theta-function}
\CZ \left( \upmu, \upnu \right) = \sum_{\ell\in\mathbb{Z}} \rme^{2\pi\rmi\ell\upnu}\,  \exp \Big\{ \mathcal{F} \left( t_1 - g_{\text{s}} \left( \upmu+\ell \right), t_2 + g_{\text{s}} \left( \upmu+\ell \right) \right) \Big\}.
\end{equation}
\noindent
The contact with \eqref{eq:pinchedcubic} is now immediate once we take this expression to the one-cut background where $t_1 \to t$ and $t_2 \to 0$ (of course when taking $t_2 \to 0$ the aforementioned regularization is required \cite{msw08, ms24}). In other words, \eqref{eq:compact-EM-theta-function} expanded around the one-cut background will match our nonperturbative proposal \eqref{eq:pinchedcubic}---as long as keeping \textit{both} characteristics turned on, and undergoing Stokes phenomena, unlike in \cite{em08}. Finally, it is interesting to note that as in \cite{em08} we can rewrite our partition functions in terms of theta-functions, \textit{e.g.}, as in \eqref{eq:PItheta} or \eqref{eq:cubictheta}, but they are of course \textit{not} the same theta-functions as in \cite{em08}, given by \eqref{eq:EM-2-cut-Z} and \eqref{eq:theta-for-EM-2-cut-Z}, due to the difference\footnote{In particular, we further need to expand the Gaussian (regularization) partition functions.} in reference backgrounds.

Let us next compare our \PI~formulation, in particular as it follows from our main expression \eqref{eq:pinchedcubic}, against the ``Painlev\'e/gauge theory correspondence'' of \cite{blmst16} which also featured a Zak-transform structure for the \PI~solution (albeit with greater emphasis on the relation to the linear isomonodromic system). Such construction was obtained within a larger context (too large to review herein), including, \textit{e.g.}, \cite{gil12, gil13, ilt14, gl16, gl17, ddg20}, which might open up parallels between our treatment and constructions based on isomonodromic deformations of linear systems. One instructive example is the connection between the Painlev\'e~III tau-function and the Nekrasov partition-function for pure $\CN=2$ $\text{SU}(2)$ gauge theory, which is also addressable by a Zak transform \cite{gl17}. In this example, the gauge-theory partition-function around weak-coupling is a convergent sum, whereas its analytic continuation into the strong-coupling regime yields an asymptotic series \cite{ggh23}. This latter divergent series\footnote{Notice that for the example of \PI, the weak-coupling analogue does not exist (see table~4 of \cite{blmst16}) and we only have the asymptotic series---which is exactly what was addressed in \cite{blmst16}.} is then the analogue of our multi-cut partition-function inside the discrete-Fourier/Zak transform in \eqref{eq:pinchedcubic}. Our general formulation \eqref{eq:pinchedmulti} greatly extends the \PI~problem; and it would hence be interesting to uncover what other gauge theories is it describing, in the spirit of \cite{blmst16}, which we shall discuss at length in \cite{ss26}.

An alternative approach to constructing solutions to this type of problems comes from the Riemann--Hilbert formulation (see, \textit{e.g.}, \cite{i03}). In this approach, solutions are constructed via the use of Fredholm determinants and very schematically written in the form (see, \textit{e.g.}, \cite{gl17} for the example of Painlev\'e~III)
\begin{equation}
\CZ \sim \det \left( \1 - \boldsymbol{\CK} \right),
\end{equation}
\noindent
where $\boldsymbol{\CK}$ is an operator that depends on the initial data and the Painlev\'e variable. Of course there are advantages and disadvantages associated to any formulation. In this particular type of formulation, while providing an inherently convergent rather than asymptotic, nonperturbative solution to the problem, it still requires the diagonalization of a complicated operator in order to produce an explicit result. Further, it somewhat obscures the semi-classical nonperturbative content of the theory at play, by naturally lacking the semi-classical decoding that the resurgent-transseries automatically provides and its direct relation to the string-theoretic expansion. It is in this context which we recall that our methods really come from minimal/topological string theory and their associated hermitian matrix models, which conversely means that a direct application to the Painlev\'e~III case is not immediately obvious nor may it necessarily provide any clean semi-classical decoding. Furthermore, although the Fredholm determinant formulation is very well established for a large number of examples, we are not aware of any explicit\footnote{For example, the construction which was used to compute the Fredholm determinant for Painlev\'e~III in \cite{gl17} does not directly work for \PI.} application to \PI, and much less to the large classes of extensions and matrix models which we consider herein. A generic construction of such tau-functions based on abelianization has recently been proposed in \cite{hn24}. If applied to our cases it would be complimentary to the methods presented in this paper and an explicit comparison would hence become very interesting.

\subsection{Partition-Function to Free-Energy to Specific-Heat: All KdV Solutions}
\label{subsec:KdV-solutions}

One very interesting consequence of \eqref{eq:pinchedcubic}-\eqref{eq:pinchedcubic-dsl}-\eqref{eq:pinchedmulti} is that these formulae for partition-functions necessarily yield formulae for free-energies hence necessarily yield formulae for string-equation solutions (or specific-heats)---in fact much akin to the relations that lead back from the Weierstrass $\upsigma$-function \eqref{eq:wp-sigma-function-def} to the usual Weierstrass $\wp$-function \eqref{eq:wp-function-def}, or else as in the explicit writing the Weierstrass $\wp$-function in terms of theta-functions as in \eqref{eq:generalizedweierstrass}; both described in appendix~\ref{app:elliptic-theta-modular}. In other words, our main conjecture should also yield \textit{general} and \textit{explicit} solutions\footnote{On what concerns \PI, a formal transseries solution was constructed in a similar fashion in \cite{i19}, starting from topological-recursion data. Extensions to other Painlev\'e type equations were addressed in \cite{mo19, eg19}.} for \text{all} non-linear string-equations along the KdV hierarchy\footnote{Of course it is well known from \cite{w91, k92} that the multicritical (in general, the two-dimensional topological gravity) partition-function is a tau-function of the KdV hierarchy, hence must give rise to all these solutions. What we do in here is to make these solutions fully explicit as transseries and encompassing Stokes data \cite{krsst26b, krst26a, krst26b}.}; a theme which will be further explored in \cite{krst26a, krst26b}. These solutions of course arise in the guise of resonant resurgent-transseries, built starting from the corresponding spectral curves (\textit{e.g.}, as listed in \cite{gs21} for the KdV multicritical hierarchy), and which still require Stokes data supplemental \cite{krsst26b}. In the following we will be more specific on what concerns the multicritical KdV hierarchy but, of course, all we say has natural extensions to minimal strings and finite-difference matrix-model string equations.

Focus on multicritical models \cite{gm90a, ds90, bk90, d90, gm90b}, as discussed in subsection~\ref{subsec:DSL-phases}, and follow \cite{gs21} for KdV conventions. As reviewed earlier, string equations for the order-$k$ multicritical model are non-linear ODEs of order $(2k-2)$ given by \eqref{eq:orderkstringequationfromgelfanddikii}. The associated spectral curves are written as
\be
\label{eq:k-multi-spectral-curve}
y_{k} (z) = 2k\, {}_2F_1 \left( 1-k, 1; \left. \frac{3}{2}\, \right| z+1 \right) \sqrt{z+1},
\ee
\noindent
where ${}_2F_1 (a,b; c|z)$ is the Gauss hypergeometric function \cite{olbc10} which, in this case, is actually denoting a hypergeometric \textit{polynomial} leading to $k-1$ nonperturbative saddles \cite{gs21}. Let us then apply our main partition-function result \eqref{eq:pinchedcubic}-\eqref{eq:pinchedcubic-dsl}-\eqref{eq:pinchedmulti} to the $k$-th multicritical theory and compare it with results obtained for these systems in \cite{krsst26b, krst26a}---soon to be reviewed below. These theories contain $k-1$ nonperturbative saddles from \eqref{eq:k-multi-spectral-curve}, hence $k-1$ (distinct) instanton actions which are generically complex albeit appearing in complex conjugate pairs \cite{gs21} (and on top, their symmetric pairs due to resonance \cite{mss22}). For subtle reasons relating to the Stokes data structure of multicritical theories, it turns out that there are many equivalent ways to write the discrete Fourier/Zak transform partition-function transseries for the multicritical hierarchy, associated to the use of different ``local coordinates'' for the transseries parameters based\footnote{Recall that, as was briefly alluded to in subsection~\ref{subsec:resurgent-Z}, also from the matrix-model perspective these different choices will correspond to distinct choice of cycles on the spectral curve.} on convenience. Leaving full details to our subsequent work \cite{krsst26b, krst26a, krst26b}, the main point is that there exist preferred ``local coordinates'' in which the action of Stokes automorphisms upon the partition-function (which, at the end-of-the-day, act upon transseries parameters) occurs in an universal manner (see as well \cite{im23}). Moreover, as explained in \cite{krst26a, krst26b, krsst26b}, Stokes automorphisms turn out to be \textit{asymmetrical}, in a way explicitly related to the arguments of the multicritical saddles. This has a direct consequence for the particular shape of these ``local coordinates'' and, as it turns out, it will be easiest to express our generic transseries solution if we label the actions $\{ A_i \}$ via the ordering of the corresponding \textit{saddle} arguments, such that $\arg x^{\star}_i > \arg x^{\star}_j$ whenever $i<j$. Having mentioned these subtleties, let us proceed with our main focus for this subsection.

Transseries solutions to the multicritical string-equations define the corresponding discrete Fourier/Zak transform partition-functions only up to normalizations. As such, we shall herein only make the comparison (between multi-cut free-energy derivatives and string-equation diagonal partition-function at leading $g_{\text{s}}$-order) up to these normalizations, while at the same time (for concreteness) present the solutions with unit normalization. We therefore write the $k$-th multicritical partition-function transseries solution as\footnote{The choice of transseries ``local coordinates'' for the string-equation partition-function solution made in this formula (and herein chosen for convenience of presentation) will be the one associated with the \textit{reverse} canonical cycle choice in comparison to \eqref{eq:InstantonActionmsw08}, for every instanton direction.}
\be
\label{eq:multicritical-DFT-conjecture}
\CZ_{\text{ds}} \left( z; \boldsymbol{\uprho}, \boldsymbol{\upmu} \right) = \sum_{\boldsymbol{\ell} \in \mathbb{Z}^{\kappa}} \boldsymbol{\uprho}^{\boldsymbol{\ell}}\, \mathcal{Z} \left( z; \ell_1- \frac{\alpha_1}{2}\, \upmu_1, \cdots, \ell_{\kappa}-\frac{\alpha_{\kappa}}{2}\, \upmu_{\kappa} \right),
\ee
\noindent
where in the above we have set $g_{\text{s}}=1$ (as is common to do for the multicritical hierarchy; see for example \cite{gs21} on why that is always possible to do). Further, we have chosen normalizations so as to match the multicritical normalization in \cite{gs21}, and we are using $\kappa=k-1$ as the genus of the pinched spectral-curve. This of course should first compare to \eqref{eq:pinchedcubic-dsl}, and then to the specific examples \eqref{eq:PIDiscreteFourierNC} and \eqref{eq:YLFullDFT}. Using analytic consistency requirements\footnote{In particular, by enforcing monodromy-closure for several additional elements along the KdV hierarchy \cite{krst26b}.} in \cite{krst26a}, the explicit form of the Fourier parameters is obtained; we simply state them here as:
\begin{equation}
\label{eq:Fourier-parameters-MC-conjecture}
\uprho_i = - \sigma_{2i}\, \sqrt{- \pi \alpha_i}\, \exp \left\{ -\rmi \pi \sum_{n=1}^{i-1} \alpha_n \upmu_n \right\}\, \prod_{\substack{j=1 \\ j \neq i}}^{\kappa} e_{ij}^{\alpha_j \upmu_j} \times  \frac{\widetilde{A}_i^{- \frac{\alpha_i}{2}\, \upmu_i}}{\Gamma \left( 1 - \frac{\alpha_i}{2}\, \upmu_i \right)}  
\end{equation}
\noindent
(this should compare with \eqref{eq:DFTrhoP1} and \eqref{eq:DFTrhosYL}). Let us briefly describe (albeit with full details postponed to \cite{krst26a}) how to obtain the various parameters appearing above. Firstly, we have more familiar\footnote{These parameters were already featured in \cite{bssv22}. In particular, do not confuse $\widetilde{A}$ with any instanton actions.} parameters, shown in \cite{krsst26b} to be given by
\bea
\alpha_i &=& 128\, \frac{\left( \kappa + 1 \right)^4}{\left( 2\kappa + 3 \right)^4} \left( x^{\star}_i + 1 \right)^2 \frac{V_{\text{h;eff}}'' (x^\star_i)}{V_{\text{h;eff}} (x^\star_i)^4}, \\
\widetilde{A}_i &=& - 16 \left( x^{\star}_i + 1 \right) V_{\text{h;eff}}'' (x^\star_i).
\eea
\noindent
Less familiar are the parameters $e_{ij}$, laid out in detail in \cite{krst26a}. These take a very simple form for the multicritical hierarchy, given in terms of the instanton actions, and explicitly given by
\begin{equation}
e_{ij} = \frac{A_j-A_i}{A_j+A_i}.
\end{equation}
\noindent
It is important to highlight that parameters $\tilde{A}_i, \alpha_i$, $e_{ij}$ are only defined up to reparametrization-invariance gauge-freedom \cite{asv11} of transseries parameters. As such, they can be chosen at will, simply amounting to appropriate rescalings of these transseries parameters. The above choices are those in which the multicritical KdV hierarchy has a particularly simple form of the transseries coefficients (see, \textit{e.g.}, \cite{asv11}, and the ensuing in-depth discussion of these parameters in \cite{krst26a}).

Finally, let us address the Fourier modes in \eqref{eq:multicritical-DFT-conjecture}, the main object of interest on what regards the structural match (up to normalization) with the double-scaled expression \eqref{eq:pinchedmulti}. They are given by
\bea
\mathcal{Z} \left( z; \nu_1, \cdots, \nu_{\kappa} \right) &=& z^{-\frac{\kappa}{24 \left(\kappa +1\right)}}\, \rme^{-\frac{\left(\kappa+1\right)^2}{2 \left(2\kappa+3\right) \left(\kappa+2\right)}\, z^{\frac{2\kappa+3}{\kappa+1}}}\, \prod_{i=1}^{\kappa} \prod_{j=i+1}^{\kappa} e_{ij}^{2\nu_i\nu_j} \times \\
&\times& \prod_{\ell=1}^{\kappa} \widetilde{A}_\ell^{-\frac{1}{2} \nu_\ell^2}\, \rme^{A_\ell z^{\frac{2\kappa+3}{2\kappa+2}} \nu_\ell} z^{-\frac{2\kappa+3}{2\kappa+2}\frac{\nu_\ell^2}{2}}\, \frac{G_2 \left(1+\nu_\ell\right)}{\left(2\pi\right)^{\frac{\nu_\ell}{2}}}\, \sum_{g=0}^{+\infty} D_g (\nu_1, \cdots, \nu_{\kappa})\, z^{-\frac{2\kappa+3}{2\kappa+2}\, g}, \nonumber
\eea
\noindent
which should compare with \eqref{eq:DFTKernelPINC} and \eqref{eq:curlyZYangLeeFullExpression}. In particular, herein the $-\frac{\left(\kappa+1\right)^2}{2 \left(\kappa+2\right) \left(2\kappa+3\right)}$ is the planar (perturbative) free energy and the $-\frac{\kappa}{24 \left(\kappa +1\right)}$ its genus-one; and $A_\ell = V_{\text{h;eff}} (x_\ell^\star)$ are the multicritical instanton actions. The $D_g (\nu_1, \cdots, \nu_{\kappa})$ are polynomials of degree $3g$ (with common variable), which are given for example by table~\ref{tab:PIDkPolynomials} for $k=2$ and equations \eqref{eq:DkPolynomialsYangLee-1}-\eqref{eq:DkPolynomialsYangLee-2} for $k=3$. To match with \eqref{eq:pinchedmulti}, we proceed by evaluating the adequate double-scaled expressions from subsection~\ref{subsec:resurgent-Z} as before. One finds a perfect match with these doubled-scaled quantities at leading order, given by\footnote{As we did for \YL, to obtain these formulae one must first rescale the multicritical spectral-curves \eqref{eq:k-multi-spectral-curve} by a factor of $\frac{1}{2}$. This accounts for the full- versus half-cycle convention difference discussed in subsection~\ref{subsec:resurgent-Z}.}
\bea
\exp \left( \left. \partial_{s_i}^2 \widehat{\mathcal{F}}_0 (\boldsymbol{s}) \right|_{\boldsymbol{s}=0} \right) &=& \widetilde{A}_i^{-1}, \\
\left. \partial_{s_i} \widehat{\mathcal{F}}_0 (\boldsymbol{s}) \right|_{\boldsymbol{s}=0} &=& A_i, \\
\left. \partial_{s_i} \partial_{s_j} \widehat{\mathcal{F}}_0 (\boldsymbol{s}) \right|_{\boldsymbol{s}=0} &=& \log e_{ij}^2.
\eea

\section{A Comment on the Interplay with the Topological Recursion}
\label{sec:topological-recursion}

Our main expressions, \eqref{eq:pinchedcubic} or \eqref{eq:pinchedmulti}, describe exact partition-function transseries constructed around the \textit{one}-cut spectral-geometry background, yet in their definitions they make direct use of the (regularized, perturbative) free-energy as computed around the (fully unpinched) \textit{multi}-cut background. In this way there are two, distinct, \textit{perturbative} free energies at play, the \textit{one}-cut and the \textit{multi}-cut, both computable by the topological recursion \cite{eo07a} albeit starting from \textit{different} spectral curves. Do these free energies talk to each other? Because the \textit{multi}-cut perturbative free energy defines the nonperturbative partition function, and, once back into rectangular framing, indeed the \textit{one}-cut perturbative free-energy sits inside this nonperturbative partition-function, the answer must be yes. We will see how this comes about in this section.

Another take on this very same point is to realize that we now have two independent analytic representations of the same object. On one side, we can construct the partition-function transseries starting from the one-cut problem and compute its diverse nonperturbative sectors iteratively from saddle-point expansion via (anti-)eigenvalue tunneling \cite{mss22} (no matter how unpractical and slow this might be). On the other side, we can simply grab the formulae in the present paper and expand in powers of the transseries parameters (which now relies on a single run of the perturbative topological-recursion, albeit on a more complicated multi-cut spectral curve). The very \textit{same} transseries sectors are hence obtained from \textit{different} origins. Equating the analytical results obtained in these two different paths then yields highly non-trivial relations between topological-recursion free-energies associated with \textit{distinct} spectral curves. In particular, such relations include novel analytic expressions for the multi-cut free-energy perturbative genus-expansion coefficients in terms of the one-cut free-energy. 

To set the stage, let us very briefly recall some basic formulae on the topological recursion \cite{e04, eo07a, eo08, eo09}. Let $\Sigma$ be a compact Riemann surface and consider two meromorphic functions $x,y : \Sigma \to \mathbb{C}$ such that
\be
\label{eq:toprec1}
\boldsymbol{\CE} ( x (\upzeta), y(\upzeta) ) = 0,
\ee
\noindent
where $\boldsymbol{\CE} (x,y) \in \BC \left[x,y\right]$. This algebraic relation defines an algebraic curve (see appendix~\ref{app:elliptic-theta-modular} as well), from where one defines a spectral curve as a tuple $\left\{ \Sigma, x, y, \CB \right\}$ where $\CB$ is a fundamental second-kind bilinear differential-form on $\Sigma$; see, \textit{e.g.}, \cite{ekr15}. The topological recursion is a framework which starts from such a spectral curve and iteratively computes a family of multi-differential forms $\left\{ \omega_{g,h}\right\}$ on $\Sigma$ with $g \in \BN_0$ and $h \in \BN$. This is a recursive procedure with respect to the Euler characteristic $\chi = 2 - 2g - h \leq 1$ which is basically implementing the loop equations that follow from the Virasoro constraints of the matrix model; and imply equivalent recursive relations between the perturbative coefficients of the multi-resolvent correlation functions \eqref{eq:genusgmultiresolvents}. These multi-differential forms are, among other things, symplectic invariants of the algebraic curve defined by \eqref{eq:toprec1}. They also are, essentially, the multi-resolvent correlation functions themselves, as (for $(g,h) \neq (0,1),(0,2)$)
\be
\omega_{g,h} \left( \upzeta_1, \ldots, \upzeta_h; t \right) = W_{g;h} \left( x_1, \ldots, x_h; t \right)\, \rmd x_1 \cdots \rmd x_h,
\ee
\noindent
where $x_i \equiv x (\upzeta_i)$. The initial data for this recursive procedure consists on the symplectic invariants
\be
\omega_{0,1} (\upzeta) = - \frac{1}{2}\, y (\upzeta)\, \rmd x (\upzeta), \qquad \omega_{0,2} (\upzeta_1, \upzeta_2) = \CB (\upzeta_1, \upzeta_2) = \frac{\rmd \upzeta_1\, \rmd \upzeta_2}{\left( \upzeta_1 - \upzeta_2 \right)^2},
\ee
\noindent
associated with the highest Euler characteristics. In the large $N$ limit hermitian one-matrix models give rise to spectral curves for which $\boldsymbol{\CE} (x,y)$ has degree two when considered as an element of $\BC[x][y]$. In this case, the map $x$ becomes a meromorphic double-cover of $\BC$ and the respective ramification points are given by the roots of the one-form $\rmd x$. Around each ramification point $\mathsf{r} \in \Sigma$, there exists a unique local involution map $\sigma_{\mathsf{r}}$, interchanging both sheets of $x$; see, \textit{e.g.}, \cite{ekr15}. Multi-differentials of lower Euler characteristic are then obtained through the recursive relations
\be
\label{eq:toprec2}
\omega_{g,h} \left( \upzeta_1,J\right) = \sum_{\mathsf{r} \in \mathsf{R}} \underset{\upzeta=\mathsf{r}}{\Res} \BK_{\mathsf{r}} ( \upzeta_1, \upzeta ) \Bigg\{ \omega_{g-1,h+1} (\upzeta,\sigma_{\mathsf{r}} (\upzeta),J) + \sum_{\substack{m+m'=g \\ I\sqcup I'=J}}' \omega_{m,|I|+1} (\upzeta,I)\, \omega_{m',|I'|+1} (\sigma_{\mathsf{r}} (\upzeta),I') \Bigg\}.
\ee
\noindent
Herein $J = \left\{ \upzeta_2, \ldots, \upzeta_h \right\}$, $\mathsf{R}$ is the set of all ramification points of $x$, and the primed-summation indicates that we should not include the index combinations $(I,m) = (J,g)$ and $(I',m') = (J,g)$. In the expression above, we further defined the recursion kernel
\be
\BK_{\mathsf{r}} (\upzeta_1,\upzeta_2) = \frac{1}{2}\, \frac{\int_{\sigma_{\mathsf{r}} (\upzeta_2)}^{\upzeta_2} \omega_{0,2} \left(\upzeta_{1},\bullet\right)}{\omega_{0,1} (\upzeta_2) - \omega_{0,1} (\sigma_{\mathsf{r}} (\upzeta_2))}.
\ee
\noindent
On top of the above $\omega_{g,n \ge 1}$ one needs one extra definition when $n=0$, for which one defines yet another family of symplectic invariants $\left\{ \omega_{g,0} \equiv F_g \right\}$ as
\be
\label{eq:toprec3}
F_g = \frac{1}{2-2g} \sum_{\mathsf{r} \in \mathsf{R}} \underset{\upzeta=\mathsf{r}}{\Res} \omega_{g,1} (\upzeta)\, \Phi(\upzeta),
\ee
\noindent
for all $g \ge 2$ and where $\rmd \Phi = \omega_{0,1}$ (the definitions of $F_0$ and $F_1$ are more complicated and we refer the reader to \cite{eo07a} for details). The coefficients $F_g$ are of course the free-energy genus-expansion. It is standard to schematically depict the topological recursion procedure \eqref{eq:toprec2} as in figure~\ref{fig:toprecfig1}.

\newcommand{\hole}[2]{
    \def\xcoordinate{#1};
\def\ycoordinate{#2};
   \draw[line width = 0pt,fill = white] (-0.5+\xcoordinate,0+\ycoordinate) to[out = -30, in = 180+30] (0.5+\xcoordinate,0+\ycoordinate)to[out = 180-50,in =50]cycle;
   \draw[line width = 2pt] (-0.5-0.1+\xcoordinate,0+0.05+\ycoordinate) to[out = -30, in = 180+30](0.5+0.1+\xcoordinate,0+0.05+\ycoordinate);
   \draw[line width = 2pt] (0.5+\xcoordinate,0+\ycoordinate)to[out = 180-50,in =50] (-0.5+\xcoordinate,0+\ycoordinate);
}
\begin{figure}
	\centering
	\begin{tikzpicture}[scale = 0.65]
	\def\boundaryangle{90};
	\def\hspace{-0.8};
\draw[line width=0pt,fill = purple, fill opacity = 0.3] (0+\hspace,1) .. controls (2+\hspace,1.5) .. (2+\hspace,4) to[out = 90, in = 180] (4+\hspace,6) to[out = -90-\boundaryangle, in = 90+\boundaryangle] (4+\hspace,5)to[out = 180, in = 90] (4-0.5+\hspace,5-0.5)to[out = -90, in = 180] (4+\hspace,4)to[out = -90-\boundaryangle, in = 90+\boundaryangle] (4+\hspace,3)to[out = 180, in =90] (4-0.5+\hspace,3-0.5)to[out =-90, in =180] (4+\hspace,2)to[out =-90-\boundaryangle, in =90+\boundaryangle] (4+\hspace,1)to[out =180, in =90] (4-0.5+\hspace,1-0.5)to[out =-90, in =180] (4+\hspace,0)to[out =-90-\boundaryangle, in =90+\boundaryangle] (4+\hspace,-1)to[out =180, in =90] (4-0.5+\hspace,-1-0.5)to[out =-90, in =180] (4+\hspace,-2)to[out =-90-\boundaryangle, in =90+\boundaryangle](4+\hspace,-3)to[out =180, in =90] (4-0.5+\hspace,-3-0.5)to[out =-90, in =180] (4+\hspace,-4)to[out =-90-\boundaryangle, in =90+\boundaryangle](4+\hspace,-5)to[out = 180, in = -90] (2+\hspace,-3) ..controls (2+\hspace,-0.5) .. (0+\hspace,0) to[out = 90-\boundaryangle, in = -90+\boundaryangle] cycle;
\draw[line width = 2pt,fill = purple, fill opacity = 0.5] (0+\hspace,0) to[out = 90-\boundaryangle, in = -90+\boundaryangle] (0+\hspace,1) to[out = 90+\boundaryangle, in = -90-\boundaryangle]cycle;
\draw[line width = 2pt,fill = purple, fill opacity = 0.5](4+\hspace,6) to[out = -90-\boundaryangle, in = 90+\boundaryangle] (4+\hspace,5) to[out = 90-\boundaryangle, in = -90+\boundaryangle]cycle;
\draw[line width = 2pt,fill = purple, fill opacity = 0.5](4+\hspace,4) to[out = -90-\boundaryangle, in = 90+\boundaryangle] (4+\hspace,3) to[out = 90-\boundaryangle, in = -90+\boundaryangle]cycle;
\draw[line width = 2pt,fill = purple, fill opacity = 0.5](4+\hspace,2) to[out = -90-\boundaryangle, in = 90+\boundaryangle] (4+\hspace,1) to[out = 90-\boundaryangle, in = -90+\boundaryangle]cycle;
\draw[line width = 2pt,fill = purple, fill opacity = 0.5](4+\hspace,0) to[out = -90-\boundaryangle, in = 90+\boundaryangle] (4+\hspace,-1) to[out = 90-\boundaryangle, in = -90+\boundaryangle]cycle;
\draw[line width = 2pt,fill = purple, fill opacity = 0.5](4+\hspace,-2) to[out = -90-\boundaryangle, in = 90+\boundaryangle] (4+\hspace,-3) to[out = 90-\boundaryangle, in = -90+\boundaryangle]cycle;
\draw[line width = 2pt,fill = purple, fill opacity = 0.5](4+\hspace,-4) to[out = -90-\boundaryangle, in = 90+\boundaryangle] (4+\hspace,-5) to[out = 90-\boundaryangle, in = -90+\boundaryangle]cycle;
\draw[line width = 2pt] (0+\hspace,1) .. controls (2+\hspace,1.5) .. (2+\hspace,4) to[out = 90, in = 180] (4+\hspace,6);
\draw[line width = 2pt](4+\hspace,5)to[out = 180, in = 90] (4-0.5+\hspace,5-0.5)to[out = -90, in = 180] (4+\hspace,4);
\draw[line width = 2pt](4+\hspace,3)to[out = 180, in = 90] (4-0.5+\hspace,3-0.5)to[out = -90, in = 180] (4+\hspace,2);
\draw[line width = 2pt](4+\hspace,1)to[out = 180, in = 90] (4-0.5+\hspace,1-0.5)to[out = -90, in = 180] (4+\hspace,0);
\draw[line width = 2pt](4+\hspace,-1)to[out = 180, in = 90] (4-0.5+\hspace,-1-0.5)to[out = -90, in = 180] (4+\hspace,-2);
\draw[line width = 2pt](4+\hspace,-3)to[out = 180, in = 90] (4-0.5+\hspace,-3-0.5)to[out = -90, in = 180] (4+\hspace,-4);
\draw[line width = 2pt](4+\hspace,-5)to[out = 180, in = -90] (2+\hspace,-3) ..controls (2+\hspace,-0.5) .. (0+\hspace,0);
\hole{1.6+\hspace}{0.9}
\hole{1.6+\hspace}{0.1}
\hole{2.7+\hspace}{0.5}
\hole{2.7+\hspace}{-0.5}
\hole{2.73+\hspace}{-1.5}
\hole{2.9+\hspace}{-2.5}
\hole{2.73+\hspace}{2.5}
\hole{2.95+\hspace}{3.5}
\def\hspace{0.5};
\draw[line width=0pt, fill = purple, fill opacity = 0.3] (0+5+\hspace,1) to[out = 0, in = 180] (6.7+\hspace,1.8) to [out = -\boundaryangle-90, in = 90+\boundaryangle](6.7+\hspace,0.8)to[out = 180, in = 90] (6.7-0.3+\hspace,0.8-0.3)to[out = -90, in = 180] (6.7+\hspace,0.2)to [out = -\boundaryangle-90, in = 90+\boundaryangle](6.7+\hspace,0.2-1)to[out = 180, in = 0](0+5+\hspace,0)to[out = 90-\boundaryangle, in = -90+\boundaryangle]cycle ;
\draw[line width=2pt,fill = purple, fill opacity = 0.5](6.7+\hspace,1.8) to [out = -\boundaryangle-90, in = 90+\boundaryangle](6.7+\hspace,0.8)to [out = -\boundaryangle+90, in = 90-\boundaryangle]cycle;
\draw[line width=2pt,fill = purple, fill opacity = 0.5](6.7+\hspace,0.2)to [out = -\boundaryangle-90, in = 90+\boundaryangle](6.7+\hspace,0.2-1)to [out = \boundaryangle-90, in = 90-\boundaryangle]cycle;
\draw[line width=2pt,fill = purple, fill opacity = 0.5](0+5+\hspace,0)to[out = 90-\boundaryangle, in = -90+\boundaryangle](0+5+\hspace,1)to[out = 90+\boundaryangle, in = -90-\boundaryangle]cycle;
\draw[line width=2pt] (0+5+\hspace,1) to[out = 0, in = 180] (6.7+\hspace,1.8);
\draw[line width=2pt](6.7+\hspace,0.8)to[out = 180, in = 90] (6.7-0.3+\hspace,0.8-0.3)to[out = -90, in = 180] (6.7+\hspace,0.2);
\draw[line width=2pt](6.7+\hspace,0.2-1)to[out = 180, in = 0](0+5+\hspace,0);
\draw [line width=0pt, fill = purple, fill opacity = 0.3] (7.8+\hspace,1.8) to [out = 0, in = 180] (9+\hspace,1.7)to [out = 0, in = -90] (9.5+\hspace,4)to [out = 90, in = 180] (11.5+\hspace,6)to [out = -\boundaryangle-90, in = 90+\boundaryangle](11.5+\hspace,5) to[out = 180, in = 90] (11+\hspace,4.5)to[out =-90, in = 180] (11.5+\hspace,4)to [out = -\boundaryangle-90, in = 90+\boundaryangle](11.5+\hspace,3) to[out = 180, in = 90] (11+\hspace,2.5)to[out =-90, in = 180] (11.5+\hspace,2)to [out = -\boundaryangle-90, in = 90+\boundaryangle](11.5+\hspace,1)to[out = 180, in = 90] (11+\hspace,0.5)to[out =-90, in = 180] (11.5+\hspace,0)to [out = -\boundaryangle-90, in = 90+\boundaryangle](11.5+\hspace,-1)to[out = 180, in = 90] (11+\hspace,-1.5)to[out =-90, in = 180] (11.5+\hspace,-2)to [out = -\boundaryangle-90, in = 90+\boundaryangle](11.5+\hspace,-3)to[out = 180, in = 90] (11+\hspace,-3.5)to[out =-90, in = 180] (11.5+\hspace,-4)to [out = -\boundaryangle-90, in = 90+\boundaryangle](11.5+\hspace,-5)to [out = 180, in = -90](9.5+\hspace,-3)to [out =90, in = 0](9+\hspace,-0.7)to [out =180, in = 0](7.8+\hspace,-0.8)to [out = -\boundaryangle+90, in = -90+\boundaryangle](7.8+\hspace,0.2)to [out =0, in = -90](7.8+0.3+\hspace,0.2+0.3)to [out =90, in = 0](7.8+\hspace,0.2+0.6)to [out = -\boundaryangle+90, in = -90+\boundaryangle]cycle;
\draw [line width=2pt] (7.8+\hspace,1.8) to [out = 0, in = 180] (9+\hspace,1.7)to [out = 0, in = -90] (9.5+\hspace,4)to [out = 90, in = 180] (11.5+\hspace,6);
\draw [line width=2pt] (11.5+\hspace,5) to[out = 180, in = 90] (11+\hspace,4.5)to[out =-90, in = 180] (11.5+\hspace,4);
\draw [line width=2pt] (11.5+\hspace,3) to[out = 180, in = 90] (11+\hspace,2.5)to[out =-90, in = 180] (11.5+\hspace,2);
\draw [line width=2pt] (11.5+\hspace,1) to[out = 180, in = 90] (11+\hspace,0.5)to[out =-90, in = 180] (11.5+\hspace,0);
\draw [line width=2pt] (11.5+\hspace,-1) to[out = 180, in = 90] (11+\hspace,-1.5)to[out =-90, in = 180] (11.5+\hspace,-2);
\draw [line width=2pt] (11.5+\hspace,-3) to[out = 180, in = 90] (11+\hspace,-3.5)to[out =-90, in = 180] (11.5+\hspace,-4);
\draw [line width=2pt](11.5+\hspace,-5)to [out = 180, in = -90](9.5+\hspace,-3)to [out =90, in = 0](9+\hspace,-0.7)to [out =180, in = 0](7.8+\hspace,-0.8);
\draw [line width=2pt](7.8+\hspace,0.2)to [out =0, in = -90](7.8+0.3+\hspace,0.2+0.3)to [out =90, in = 0](7.8+\hspace,0.2+0.6);
\draw [line width=2pt, fill = purple, fill opacity = 0.5](11.5+\hspace,6)to [out = -\boundaryangle-90, in = 90+\boundaryangle](11.5+\hspace,5)to [out = \boundaryangle-90, in = 90-\boundaryangle]cycle;
\draw [line width=2pt, fill = purple, fill opacity = 0.5](11.5+\hspace,4)to [out = -\boundaryangle-90, in = 90+\boundaryangle](11.5+\hspace,3)to [out = \boundaryangle-90, in = 90-\boundaryangle]cycle;
\draw [line width=2pt, fill = purple, fill opacity = 0.5](11.5+\hspace,2)to [out = -\boundaryangle-90, in = 90+\boundaryangle](11.5+\hspace,1)to [out = \boundaryangle-90, in = 90-\boundaryangle]cycle;
\draw [line width=2pt, fill = purple, fill opacity = 0.5](11.5+\hspace,0)to [out = -\boundaryangle-90, in = 90+\boundaryangle](11.5+\hspace,-1)to [out = \boundaryangle-90, in = 90-\boundaryangle]cycle;
\draw [line width=2pt, fill = purple, fill opacity = 0.5](11.5+\hspace,-2)to [out = -\boundaryangle-90, in = 90+\boundaryangle](11.5+\hspace,-3)to [out = \boundaryangle-90, in = 90-\boundaryangle]cycle;
\draw [line width=2pt, fill = purple, fill opacity = 0.5](11.5+\hspace,-4)to [out = -\boundaryangle-90, in = 90+\boundaryangle](11.5+\hspace,-5)to [out = \boundaryangle-90, in = 90-\boundaryangle]cycle;
\draw [line width=2pt,fill = purple, fill opacity = 0.5](7.8+\hspace,-0.8)to [out = -\boundaryangle+90, in = -90+\boundaryangle](7.8+\hspace,0.2)to [out = \boundaryangle+90, in = -90-\boundaryangle]cycle;
\draw [line width=2pt,fill = purple, fill opacity = 0.5](7.8+\hspace,-0.8+1.6)to [out = -\boundaryangle+90, in = -90+\boundaryangle](7.8+\hspace,0.2+1.6)to [out = \boundaryangle+90, in = -90-\boundaryangle]cycle;
\hole{9+\hspace}{1.1}
\hole{10+\hspace}{0.5}
\hole{9+\hspace}{-0.1}
\hole{10.35+\hspace}{-0.9}
\hole{10.35+\hspace}{-2.3}
\hole{10.35+\hspace}{1.9}
\hole{10.35+\hspace}{3.3}
\def\hspace{11};
\draw[line width=0pt, fill = purple, fill opacity = 0.3] (0+5+\hspace,1) to[out = 0, in = 180] (6.7+\hspace,1.8) to [out = -\boundaryangle-90, in = 90+\boundaryangle](6.7+\hspace,0.8)to[out = 180, in = 90] (6.7-0.3+\hspace,0.8-0.3)to[out = -90, in = 180] (6.7+\hspace,0.2)to [out = -\boundaryangle-90, in = 90+\boundaryangle](6.7+\hspace,0.2-1)to[out = 180, in = 0](0+5+\hspace,0)to[out = 90-\boundaryangle, in = -90+\boundaryangle]cycle ;
\draw[line width=2pt,fill = purple, fill opacity = 0.5](6.7+\hspace,1.8) to [out = -\boundaryangle-90, in = 90+\boundaryangle](6.7+\hspace,0.8)to [out = -\boundaryangle+90, in = 90-\boundaryangle]cycle;
\draw[line width=2pt,fill = purple, fill opacity = 0.5](6.7+\hspace,0.2)to [out = -\boundaryangle-90, in = 90+\boundaryangle](6.7+\hspace,0.2-1)to [out = \boundaryangle-90, in = 90-\boundaryangle]cycle;
\draw[line width=2pt,fill = purple, fill opacity = 0.5](0+5+\hspace,0)to[out = 90-\boundaryangle, in = -90+\boundaryangle](0+5+\hspace,1)to[out = 90+\boundaryangle, in = -90-\boundaryangle]cycle;
\draw[line width=2pt] (0+5+\hspace,1) to[out = 0, in = 180] (6.7+\hspace,1.8);
\draw[line width=2pt](6.7+\hspace,0.8)to[out = 180, in = 90] (6.7-0.3+\hspace,0.8-0.3)to[out = -90, in = 180] (6.7+\hspace,0.2);
\draw[line width=2pt](6.7+\hspace,0.2-1)to[out = 180, in = 0](0+5+\hspace,0);
\draw [line width=0pt, fill = purple, fill opacity = 0.3] (7.8+\hspace,1.8) to [out = 0, in = 180] (9+\hspace,1.7)to [out = 0, in = -90] (9.5+\hspace,4)to [out = 90, in = 180] (11.5+\hspace,6)to [out = -\boundaryangle-90, in = 90+\boundaryangle](11.5+\hspace,5) to[out = 180, in = 90] (11+\hspace,4.5)to[out =-90, in = 180] (11.5+\hspace,4)to [out = -\boundaryangle-90, in = 90+\boundaryangle](11.5+\hspace,3) to[out = 180, in = 90] (11+\hspace,2.5)to[out =-90, in = 180] (11.5+\hspace,2)to [out = -\boundaryangle-90, in = 90+\boundaryangle](11.5+\hspace,1)to[out = 180, in = 0](9+\hspace,0.5+0.1)to[out = 180, in = 0](7.8+\hspace,0.8)to [out = -\boundaryangle+90, in = -90+\boundaryangle]cycle;
\draw [line width=0pt, fill = purple, fill opacity = 0.3](7.8+\hspace,0.2)to[out = 0, in = 180](9+\hspace,0.5-0.1)to[out = 0, in = 180] (11.5+\hspace,0)to [out = -\boundaryangle-90, in = 90+\boundaryangle](11.5+\hspace,-1)to[out = 180, in = 90] (11+\hspace,-1.5)to[out =-90, in = 180] (11.5+\hspace,-2)to [out = -\boundaryangle-90, in = 90+\boundaryangle](11.5+\hspace,-3)to[out = 180, in = 90] (11+\hspace,-3.5)to[out =-90, in = 180] (11.5+\hspace,-4)to [out = -\boundaryangle-90, in = 90+\boundaryangle](11.5+\hspace,-5)to [out = 180, in = -90](9.5+\hspace,-3)to [out =90, in = 0](9+\hspace,-0.7)to [out =180, in = 0](7.8+\hspace,-0.8)to [out = -\boundaryangle+90, in = -90+\boundaryangle]cycle;
\draw [line width=2pt](7.8+\hspace,0.2)to[out = 0, in = 180](9+\hspace,0.5-0.1)to[out = 0, in = 180] (11.5+\hspace,0);
\draw [line width=2pt](11.5+\hspace,-1)to[out = 180, in = 90] (11+\hspace,-1.5)to[out =-90, in = 180] (11.5+\hspace,-2);
\draw [line width=2pt](11.5+\hspace,-3)to[out = 180, in = 90] (11+\hspace,-3.5)to[out =-90, in = 180] (11.5+\hspace,-4);
\draw [line width=2pt](11.5+\hspace,3)to[out = 180, in = 90] (11+\hspace,2.5)to[out =-90, in = 180] (11.5+\hspace,2);
\draw [line width=2pt](11.5+\hspace,5)to[out = 180, in = 90] (11+\hspace,4.5)to[out =-90, in = 180] (11.5+\hspace,4);
\draw [line width=2pt](11.5+\hspace,-5)to [out = 180, in = -90](9.5+\hspace,-3)to [out =90, in = 0](9+\hspace,-0.7)to [out =180, in = 0](7.8+\hspace,-0.8);
\draw [line width=2pt] (7.8+\hspace,1.8) to [out = 0, in = 180] (9+\hspace,1.7)to [out = 0, in = -90] (9.5+\hspace,4)to [out = 90, in = 180] (11.5+\hspace,6);
\draw [line width=2pt] (11.5+\hspace,1)to[out = 180, in = 0](9+\hspace,0.5+0.1)to[out = 180, in = 0](7.8+\hspace,0.8);
\draw [line width=2pt,fill = purple, fill opacity = 0.5](11.5+\hspace,6)to [out = -\boundaryangle-90, in = 90+\boundaryangle](11.5+\hspace,5)to[out = \boundaryangle-90, in = 90-\boundaryangle]cycle;
\draw [line width=2pt,fill = purple, fill opacity = 0.5](11.5+\hspace,4)to [out = -\boundaryangle-90, in = 90+\boundaryangle](11.5+\hspace,3)to[out = \boundaryangle-90, in = 90-\boundaryangle]cycle;
\draw [line width=2pt,fill = purple, fill opacity = 0.5](11.5+\hspace,2)to [out = -\boundaryangle-90, in = 90+\boundaryangle](11.5+\hspace,1)to[out = \boundaryangle-90, in = 90-\boundaryangle]cycle;
\draw [line width=2pt,fill = purple, fill opacity = 0.5](11.5+\hspace,0)to [out = -\boundaryangle-90, in = 90+\boundaryangle](11.5+\hspace,-1)to[out = \boundaryangle-90, in = 90-\boundaryangle]cycle;
\draw [line width=2pt,fill = purple, fill opacity = 0.5](11.5+\hspace,0-2)to [out = -\boundaryangle-90, in = 90+\boundaryangle](11.5+\hspace,-3)to[out = \boundaryangle-90, in = 90-\boundaryangle]cycle;
\draw [line width=2pt,fill = purple, fill opacity = 0.5](11.5+\hspace,0-4)to [out = -\boundaryangle-90, in = 90+\boundaryangle](11.5+\hspace,-5)to[out = \boundaryangle-90, in = 90-\boundaryangle]cycle;
\draw [line width=2pt,fill = purple, fill opacity = 0.5](7.8+\hspace,0.8)to [out = -\boundaryangle+90, in = -90+\boundaryangle](7.8+\hspace,1.8)to [out = \boundaryangle+90, in = -90-\boundaryangle]cycle;
\draw [line width=2pt,fill = purple, fill opacity = 0.5](7.8+\hspace,0.8-1.6)to [out = -\boundaryangle+90, in = -90+\boundaryangle](7.8+\hspace,1.8-1.6)to [out = \boundaryangle+90, in = -90-\boundaryangle]cycle;
\hole{10.3+\hspace}{1.5}
\hole{10.4+\hspace}{3}
\hole{10.4+\hspace}{3.9}
\hole{8.9+\hspace}{1.15}
\hole{8.9+\hspace}{1.15-1.3}
\hole{10.4+\hspace}{-0.7}
\hole{10.4+\hspace}{-2.1}
\hole{10.4+\hspace}{-2.9}
\node at (4.2,0.45){\scalebox{1.3}{$=$}};
\node at (14,0.2){\scalebox{1.3}{$\displaystyle+\sum_{\substack{m+m' = g \\ I \sqcup I' = J }}'$}};
\node at (6.4,0.5){$\BK_{\mathsf{r}}$};
\node at (16.9,0.5){$\BK_{\mathsf{r}}$};
\node at (0,3){$\omega_{g,h}$};
\node at (8.5,3){$\omega_{g-1,h+1}$};
\node at (19.2,3){$\omega_{m,|I|+1}$};
\node at (19.2,-3){$\omega_{m',|I'|+1}$};
	\end{tikzpicture}
    \caption{Standard schematic depiction of the topological recursion procedure in \eqref{eq:toprec2}.}
    \label{fig:toprecfig1}
\end{figure}
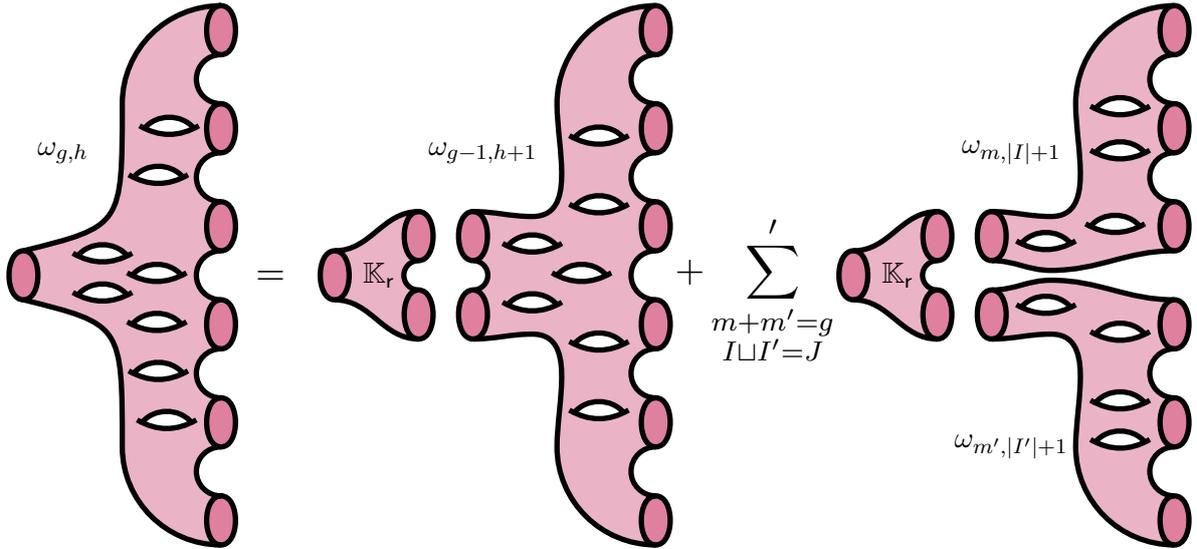

What we wish to show in this section is that there is a second expression for the coefficients $\left\{F_g\right\}$, obtained as a direct consequence of our generic transseries formulae \eqref{eq:pinchedcubic} or \eqref{eq:pinchedmulti}. For simplicity, in the following we first restrict our attention to the cubic matrix model \eqref{eq:CubicMatrixModelPotential}, as we compute explicit analytic expressions for all purely-forward instanton-sector coefficients of the partition-function transseries, using both aforementioned representations, in subsection~\ref{subsec:two-viewpoints-NP-sectors}. By equating the two results, we then obtain an alternative analytic expression for the two-cut free-energy perturbative genus-expansion coefficients, in subsection~\ref{subsec:secondformulaF}. By then it will be clear how the construction generalizes straightforwardly to free energies associated with multi-cut spectral curves, and we will briefly comment on this extension in the final subsection~\ref{subsec:multi-cuts-F}.

\subsection{Two Viewpoints Upon the Same Nonperturbative Sectors}
\label{subsec:two-viewpoints-NP-sectors}

As it shall soon be very clear, the calculations in this section are quite cumbersome. For simplicity (and without loss of generality) we hence focus on the purely-forward transseries sectors of the partition function which already fully illustrate our point. What this also implies is that using the formulation \cite{msw08} in \eqref{eq:InstantonsOneParameter} and \eqref{eq:regularizedtwocut} is enough herein. Let us then compute $\mathcal{Z}^{(\ell)}(t,g_{\text{s}})$ using the two alternative roads we alluded to earlier. In the first, we directly evaluate the matrix integral via an all-order saddle-point expansion. In the second, we evaluate the exact same sectors by suitably expanding the partition-function transseries \eqref{eq:InstantonsOneParameter}.

\paragraph{Matrix Integral Evaluation:}

The all-order matrix-integral saddle-point expansion is rather technical and we stash all details to appendix~\ref{appendix:Allordersaddlepoint}. Consider the cubic matrix model \eqref{eq:CubicMatrixModelPotential} with the single nonperturbative saddle $x^\star$ in \eqref{eq:CMMxstarsaddle} as discussed earlier in subsection~\ref{subsec:SG-phases}. The matrix-model $\ell$-instanton-sector is the integral where $\ell$ eigenvalues have tunneled to this saddle, and computable as \cite{msw07, mss22}
\be
\label{eq:toprec18}
\frac{\CZ^{(\ell)}}{\CZ^{(0)}} = \frac{1}{\ell!}\, \frac{\CZ^{(0)} (t-\ell g_{\text{s}})}{\CZ^{(0)} (t)} \int_{\CC^\star} \prod_{j=1}^{\ell} \frac{\rmd x_j}{2\pi}\, \rme^{-\frac{1}{g_{\text{s}}} V(x_j)}\, \Delta_{\ell}^2 \left( x_1, \ldots, x_\ell \right) \left\langle \prod_{k=1}^{\ell} \det \left( x_k - M \right)^2 \right\rangle_{N-\ell},
\ee
\noindent
where $\CC^\star$ denotes the steepest-descent contour through the saddle $x^\star$. Performing the change of variables \eqref{eq:step4}, followed by a series of straightforward power-series expansions and algebraic manipulations, one can analytically compute the saddle-point expansion-coefficients of the integrals above, to all orders in $g_{\text{s}}$. The interested reader can find these computations in appendix~\ref{appendix:Allordersaddlepoint}. Once everything is in place, we obtain
\be
\label{eq:toprec22}
\frac{\CZ^{(\ell)}}{\CZ^{(0)}} = \frac{1}{\ell!}\, \frac{g_{\text{s}}^{\frac{\ell^2}{2}}}{\left( 2\pi \right)^{\ell}}\, \exp \left( - \ell\, \frac{A}{g_{\text{s}}} + \ell^2\, \Big\{ \partial_{t} V_{\text{h;eff}} (x^\star) + 2 A_{0;2} (x^\star,x^\star) + \frac{1}{2}\, \partial^2_t \CF_0 (t) \Big\} \right) \sum_{g=0}^{+\infty} \mathcal{Z}^{\text{saddle}}_{\ell,g}\, g_{\text{s}}^{g},
\ee
\noindent
where $A_{0;2} \left(p_1,p_2\right)$ is the planar two-point \textit{integrated} multi-resolvent connected-correlator, and where the coefficients $\mathcal{Z}_{\ell,g}^{\text{saddle}}$ are defined in \eqref{eq:toprec15}. As advertised earlier, we shall now derive alternative analytical expressions for these very same transseries sectors by resorting to the generic partition-function transseries formula \eqref{eq:InstantonsOneParameter}.

\paragraph{Direct Transseries Evaluation:} 

Let us next look at the different road leading up to \eqref{eq:toprec22}; the road of the generic partition-function transseries \eqref{eq:InstantonsOneParameter}, which explicitly states\footnote{A word on notation (as the same symbol $\CF$ is used for both one- and two-cut free energies)---the arguments of the free-energies will tell them apart: $\CF (\bullet)$ denoting one-cut and $\CF (\bullet,\bullet)$ denoting two-cut free energies.}
\be
\frac{\CZ^{(\ell)}}{\CZ^{(0)}} = \CZ_{\text{G}}(\ell)\, \exp \left\{ \widehat{\CF} (t-\ell g_{\text{s}}, \ell g_{\text{s}}) - \mathcal{F} (t) \right\}.
\ee
\noindent
Following \cite{msw08} one may next perform a $g_{\text{s}}$-expansion---which in particular places all free-energy evaluations around the one-cut background $(t_1,t_2) = (t,0)$---that yields
\bea
\exp \left\{ \widehat{\CF} (t-\ell g_{\text{s}}, \ell g_{\text{s}}) - \mathcal{F} (t) \right\} &=& \exp \left( \sum_{\chi=-1}^{+\infty} g_{\text{s}}^\chi \sum_{\substack{(g,n) \in \BN_0 \times \BN \\ 2g+n-2 = \chi}} \frac{(-\ell)^n}{n!}\, \partial_s^n \widehat{\CF}_g (t,0) \right) = \\ 
&=& \exp \left( - \frac{\ell}{g_{\text{s}}}\, \partial_s \widehat{ \CF}_0 (t,0) + \frac{\ell^2}{2}\, \partial_s^2 \widehat{\CF}_0 (t,0) \right) \times \sum_{g=0}^{+\infty} \mathcal{Z}_{\ell,g}^{\text{transseries}}\, g_{\text{s}}^{g}, \nonumber
\eea
\noindent
where we recall $s = \frac{1}{2} \left( t_1 - t_2 \right)$ and where we have the explicit coefficients
\be
\label{eq:toprec7}
\mathcal{Z}_{\ell,0}^{\text{transseries}} = 1, \qquad
\mathcal{Z}_{\ell,g}^{\text{transseries}} = \sum_{\substack{(\boldsymbol{h},\boldsymbol{n}) \in \BN_0^p \times \BN^p \\ 2h+n-2p = g}}' \frac{1}{p!}\, \prod_{i=1}^p \frac{(-\ell)^{n_i}}{n_i!}\, \partial_s^{n_i} \widehat{\CF}_{h_i} (t,0), \quad g \in \mathbb{N}.
\ee
\noindent
We have used the notation \eqref{eq:toprec23} and the prime in the summation indicates we should exclude all $(\boldsymbol{h}, \boldsymbol{n}) \in \mathbb{N}_0^{p} \times \mathbb{N}^{p}$ for which $2h_i+n_i-2 \le 0$ for at least one $1 \le i \le s$. We can then write
\be
\label{eq:toprec6}
\frac{\CZ^{(\ell)}}{\CZ^{(0)}} = \frac{g_{\text{s}}^{\frac{\ell^2}{2}}}{\left( 2\pi \right)^{\frac{\ell}{2}}}\, G_{2} (\ell+1)\, \exp \left( - \frac{\ell}{g_{\text{s}}}\, \partial_s \widehat{\CF}_0 (t,0) + \frac{\ell^2}{2}\, \partial_s^2 \widehat{\CF}_0 (t,0) \right) \sum_{g=0}^{+\infty} \mathcal{Z}_{\ell,g}^{\text{transseries}}\, g_{\text{s}}^{g}.
\ee
\noindent
At first glance, this formula appears quite different from \eqref{eq:toprec22}. Nevertheless, they describe the very same object and must therefore be equivalent as we shall see next.
 
\subsection{Alternative Formulation of the Two-Cut Free-Energy Coefficients}
\label{subsec:secondformulaF}

As discussed earlier, from \eqref{eq:toprec3} the topological recursion provides a direct means to computing the free-energy genus-expansion coefficients for some given two-cut spectral problem. We shall now see how to derive a second, alternative, analytical formula for these same coefficients, but obtained starting from the considerations highlighted in the previous subsection.

The free-energy regularization in \eqref{eq:regularizedtwocut} ensures that $\widehat{\mathcal{F}}_g(t_1,t_2)$ is analytical around $(t_1,t_2) = (t,0)$ \cite{msw08}. As such, the power-series expansion holds:
\be
\label{eq:toprec11}
\widehat{\CF}_g (t_1,t_2) = \sum_{n=0}^{+\infty} \frac{s^n}{n!}\, \partial_s^n \widehat{\CF}_g (t,0).
\ee
\noindent
Deriving analytical expressions for all coefficients in this Taylor-series yields the aforementioned representation for the two-cut free-energy genus-expansion coefficients (via \eqref{eq:regularizedtwocut}). These expressions may be computed recursively by simply equating the right-hand-sides of equations \eqref{eq:toprec22} and \eqref{eq:toprec6}. Breaking down by $g_{\text{s}}$-order immediately yields, for all $g,\ell \in \BN$,
\bea
\label{eq:toprec8}
\partial_s \widehat{\CF}_0 (t,0) &=& A, \\
\label{eq:toprec9}
\frac{1}{\left( 2\pi \right)^{\frac{\ell}{2}}}\, G_2 (\ell+1)\, \exp \left( \frac{\ell^2}{2}\, \partial_s^2 \widehat{\CF}_0 (t,0) \right) &=& \frac{1}{\ell!}\, \frac{1}{\left( 2\pi \right)^\ell}\, \exp \left( \ell^2\, \Big\{ \partial_t V_{\text{h;eff}} (x^\star) + \right. \\
&&
\hspace{50pt}
\left. + 2 A_{0;2} (x^\star,x^\star) + \frac{1}{2}\, \partial_t^2 F_0 (t) \Big\} \right) \mathcal{Z}_{\ell,0}^{\text{saddle}}, \nonumber \\
\label{eq:toprec10}
\mathcal{Z}_{\ell,g}^{\text{transseries}} &=& \mathcal{Z}_{\ell,g}^{\text{saddle}},
\eea

In order to recursively solve this system, it is useful to assign an ``Euler characteristic'', $\chi$, to the above derivatives of the free-energy genus-expansion coefficients. More precisely, we define
\be
\label{eq:toprec17}
\chi \left( \partial_s^n \widehat{\CF}_g (t,0) \right) \equiv 2-2g-n,
\ee
\noindent
for all $g \in \BN_0$, $n \in \BN$. Note how \eqref{eq:toprec8} is solving for the free-energy derivative with Euler characteristic $\chi = 1$, while \eqref{eq:toprec9} can be used to solve for the free-energy derivative with Euler characteristic $\chi = 0$. These are the derivatives with the two highest Euler characteristic values. Finally, given the derivatives structure in \eqref{eq:toprec7}, for a fixed $g \in \BN$ \eqref{eq:toprec10} can be used to linearly solve for free-energy derivatives of Euler characteristic $\chi = -g$. There are 
\begin{eqnarray}
n_g = \left\lfloor \frac{g+2}{2} \right\rfloor+ g \mod 2
\end{eqnarray}
\noindent
such derivatives. Hence, the linear system
\be
\begin{lcases}
\, \mathcal{Z}_{1,g}^{\text{saddle}} &=  \mathcal{Z}_{1,g}^{\text{transseries}}, \\
\, \vdots& \\
\, \mathcal{Z}_{n_g,g}^{\text{saddle}} &= \mathcal{Z}_{n_g,g  }^{\text{transseries}},
\end{lcases}
\ee
\noindent
solves for the derivatives associated with Euler characteristic $-g$ in terms of derivatives with higher Euler characteristic. Resorting to this recursion, we find analytical expressions for all derivatives required in \eqref{eq:toprec11}. Let us see them explicitly. The first two read\footnote{In order to keep notation slightly lighter, as formulae will get rather intricate, all (multi) $x$-dependence in the remainder of this subsection is set to $x=x^{\star}$ and henceforth suppressed (see appendix~\ref{appendix:Allordersaddlepoint} as well).}
\bea
\partial_s \widehat{\CF}_0 (t,0) &=& A, \\
\partial_s^2 \widehat{\CF}_0 (t,0) &=& 2 \partial_t V_{\text{h;eff}} + 4 A_{0;2} + \partial_t^2 \CF_0 (t) - \log \partial_x^2 V_{\text{h;eff}}.
\eea
\noindent
In order to compute derivatives associated with Euler characteristic $\chi = -1$, we need to solve the system
\be
\label{eq:toprec14}
\begin{lcases}
\, \mathcal{Z}_{1,1}^{\text{saddle}} &= \mathcal{Z}_{1,1}^{\text{transseries}}, \\ 
\, \mathcal{Z}_{2,1}^{\text{saddle}} &= \mathcal{Z}_{2,1}^{\text{transseries}}.
\end{lcases}
\ee
\noindent
From their definition \eqref{eq:toprec7}, we have
\bea
\mathcal{Z}_{1,1}^{\text{transseries}} &=& -\frac{1}{6}\, \partial_s^3 \widehat{\CF}_0 (t,0) -\partial_s \widehat{\CF}_1 (t,0), \\
\mathcal{Z}_{2,1}^{\text{transseries}} &=& -\frac{8}{6}\, \partial_s^3 \widehat{\CF}_0 (t,0) - 2\, \partial_s \widehat{\CF}_1 (t,0).
\eea
\noindent
Solving the system \eqref{eq:toprec14}, then yields
\bea
\label{eq:toprec20}
\partial_s^3 \widehat{\CF}_0 (t,0) &=& 2\, \mathcal{Z}_{1,1}^{\text{saddle}} - \mathcal{Z}_{2,1}^{\text{saddle}}, \\
\label{eq:toprec21}
\partial_s \widehat{\CF}_1 (t,0) &=& \frac{1}{6} \left( \mathcal{Z}_{2,1}^{\text{saddle}} - 8\, \mathcal{Z}_{1,1}^{\text{saddle}} \right).
\eea
\noindent
All that remains are the coefficients $\mathcal{Z}^{\text{saddle}}_{\ell,g}$ which were computed in appendix~\ref{appendix:Allordersaddlepoint}, equation \eqref{eq:toprec15}. For $\mathcal{Z}_{1,1}^{\text{saddle}}$, we have
\be
\mathcal{Z}_{1,1}^{\text{saddle}} = a_{1,1} + b_{1,1}\, a_{1,0},
\ee
\noindent
where
\bea
a_{1,0} &=& \sqrt{2\pi}\left( \partial_x^2 V_{\text{h;eff}} \right)^{-\frac{1}{2}}, \\
b_{1,1} &=& - \frac{1}{6}\, \partial_t^3 \CF_0 (t) - \partial_t \CF_1 (t), \\
a_{1,1} &=& \sqrt{2\pi}\, \mathtt{p}_{1,2;0} \left( \partial_x^2 V_{\text{h;eff}} \right)^{-\frac{1}{2}} + \sqrt{2\pi}\, \mathtt{p}_{1,2;2} \left( \partial_x^2 V_{\text{h;eff}} \right)^{-\frac{3}{2}} + 3 \sqrt{2\pi}\, \mathtt{p}_{1,2;4} \left( \partial_x^2 V_{\text{h;eff}} \right)^{-\frac{5}{2}} + \nonumber \\
&&
+ 15 \sqrt{2\pi}\, \mathtt{p}_{1,2;6} \left( \partial_x^2 V_{\text{h;eff}} \right)^{-\frac{7}{2}}.
\eea
\noindent
One still needs the coefficients \eqref{eq:toprec16}, which for our particular evaluation are
\begin{align}
\mathtt{p}_{1,2;0} &= p_{1,2;0}+\mathsf{p}_{1,2;0}, & \mathtt{p}_{1,2;2} &= p_{1,2;2}+\mathsf{p}_{1,2;2}+p_{1,1;1}\mathsf{p}_{1,1;1}, \\
\mathtt{p}_{1,2;4} &= p_{1,2;4} +p_{1,1;3}\mathsf{p}_{1,1;1}, & \mathtt{p}_{1,2;6} &= p_{1,2;6},
\end{align}
\noindent
and where
\begin{align}
p_{1,1;1} &= \partial_{xt}^2 V_{\text{h;eff}}, \quad p_{1,1;3} = - \frac{1}{6}\, \partial_x^3 V_{\text{h;eff}}, & p_{1,2;0} &= - \frac{1}{2}\, \partial_t^2 V_{\text{h;eff}}, \\ 
p_{1,2;2} &= \frac{1}{2}\, \partial_{xxt}^3 V_{\text{h;eff}} + \frac{1}{2} \left( \partial_{xt}^2 V_{\text{h;eff}} \right)^2, & p_{1,2;4} &= - \frac{1}{24}\, \partial_x^4 V_{\text{h;eff}} - \frac{1}{6}\, \partial_x^3 V_{\text{h;eff}}\, \partial_{xt}^2 V_{\text{h;eff}}, \\ 
p_{1,2;6} &= \frac{1}{72} \left( \partial_x^3 V_{\text{h;eff}} \right)^2, & &
\end{align}
\noindent
as well as
\bea
\mathsf{p}_{1,1;1} &=& 4\, \partial_{x_1} A_{0;2}, \\
\mathsf{p}_{1,2;0} &=& \frac{8}{6}\, A_{0;3} + 2\, A_{1;1} - 2\, \partial_t A_{0;2}, \\
\mathsf{p}_{1,2;2} &=& 2\, \partial_{x_1}^2 A_{0;2} + 2\, \partial_{x_2 x_1}^2 A_{0;2} + 8 \left( \partial_{x_1} A_{0;2} \right)^2.
\eea
\noindent
Putting everything together yields rather long yet fully analytic formulae. One finds
\bea
\mathcal{Z}_{1;1}^{\text{saddle}} &=& \frac{1}{24} \left( \partial_x^2 V_{\text{h;eff}} \right)^{-3} \left\{ \partial_x^2 V_{\text{h;eff}} \left( 4\, \partial_x^2 V_{\text{h;eff}} \left\{ 3 \left[ \left( 4\, \partial_{x_1} A_{0;2} + \partial_{xt}^2 V_{\text{h;eff}} \right)^2 + 4\, \partial_{x_2 x_1}^2 A_{0;2} + \right. \right. \right. \right. \\
&&
\hspace{-25pt}
\left. \left.
+ 4\, \partial_{x_1}^2 A_{0;2} + \partial_{xxt}^3 V_{\text{h;eff}} \right] + \partial_x^2 V_{\text{h;eff}} \left( 8\, A_{0;3} - 12\, \partial_t A_{0;2} - 3\, \partial_t^2 V_{\text{h;eff}} + 12\, A_{1;1} \right) \right\} - \nonumber \\
&&
\hspace{-25pt}
\left. \left.
- 12\, \partial_x^3 V_{\text{h;eff}} \left( 4\, \partial_{x_1} A_{0;2} + \partial_{xt}^2 V_{\text{h;eff}} \right) - 3\, \partial_x^4 V_{\text{h;eff}} \right) + 5 \left( \partial_x^3 V_{\text{h;eff}} \right)^2 \right\} - \frac{1}{6}\, \partial_t^3 \CF_0 (t) - \partial_t \CF_1 (t). \nonumber
\eea
\noindent
Similarly, we also find
\bea
\mathcal{Z}_{2;1}^{\text{saddle}} &=& \frac{1}{12} \left( \partial_x^2 V_{\text{h;eff}} \right)^{-3} \left\{ \partial_x^2 V_{\text{h;eff}} \left( 16\, \partial_x^2 V_{\text{h;eff}} \left\{ 3 \left[ \left( 4\, \partial_{x_1} A_{0;2} + \partial_{xt}^2 V_{\text{h;eff}} \right)^2 + \partial_{x_2 x_1}^2 A_{0;2} + \right. \right. \right. \right. \\
&&
\hspace{-25pt}
\left. \left.
+ 2\, \partial_{x_1}^2 A_{0;2} + \partial_{xxt}^3 V_{\text{h;eff}} \right] + 6\, \partial_{x_1}^2 A_{0;2} + \partial_x^2 V_{\text{h;eff}} \left( 8\, A_{0;3} - 12\, \partial_t A_{0;2} - 3\, \partial_t^2 V_{\text{h;eff}} + 3\, A_{1;1} \right) \right\} - \nonumber \\
&&
\hspace{-25pt}
\left. \left.
- 48\, \partial_x^3 V_{\text{h;eff}} \left( 4\, \partial_{x_1} A_{0;2} + \partial_{xt}^2 V_{\text{h;eff}} \right) - 9\, \partial_x^4 V_{\text{h;eff}} \right) + 17 \left( \partial_x^3 V_{\text{h;eff}} \right)^2 \right\} - \frac{8}{6}\, \partial_t^3 \CF_0 (t) - 2\, \partial_t \CF_1 (t). \nonumber
\eea
\noindent
With these formulae in hand we can finally evaluate equations \eqref{eq:toprec20} and \eqref{eq:toprec21} as
\bea
\partial_s^3 \widehat{\CF}_0 (t,0) &=& \frac{1}{2} \left( \partial_x^2 V_{\text{h;eff}} \right)^{-3} \left\{ \partial_x^2 V_{\text{h;eff}} \left( - 2\, \partial_x^2 V_{\text{h;eff}}  \left\{ 3 \left[ \left( 4\, \partial_{x_1} A_{0;2} + \partial_{xt}^2 V_{\text{h;eff}} \right)^2 + \partial_{xxt}^3 V_{\text{h;eff}} \right] + \right. \right. \right. \nonumber \\
&&
\left.
+ 12\, \partial_{x_1}^2 A_{0;2} + \partial_x^2 V_{\text{h;eff}} \left( 8\, A_{0;3} - 12\, \partial_t A_{0;2} - 3\, \partial_t^2 V_{\text{h;eff}} - \partial_t^3 \CF_0 (t) \right) \right\} + \nonumber \\
&&
\left. \left.
+ 6\, \partial_x^3 V_{\text{h;eff}} \left( 4\, \partial_{x_1} A_{0;2} + \partial_{xt}^2 V_{\text{h;eff}} \right) + \partial_x^4 V_{\text{h;eff}} \right) - 2 \left( \partial_x^3 V_{\text{h;eff}} \right)^2 \right\}, \\
\partial_s \widehat{\CF}_1 (t,0) &=& - \frac{1}{24} \left( \partial_x^2 V_{\text{h;eff}} \right)^{-3} \left\{ 48\, \partial_{x_2 x_1}^2 A_{0;2} \left( \partial_x^2 V_{\text{h;eff}} \right)^2 - \partial_x^4 V_{\text{h;eff}}\, \partial_x^2 V_{\text{h;eff}} + \left( \partial_x^3 V_{\text{h;eff}} \right)^2 \right\} - \nonumber \\
&&
- 2\, A_{1;1} + \partial_t \CF_1 (t).
\eea
\noindent
Further derivatives, associated with lower Euler characteristics, can be computed recursively by following the tedious albeit straightforward procedure outlined above.

\subsection{On the Generalization to Multi-Cut Spectral-Curves} 
\label{subsec:multi-cuts-F}

As a direct consequence of our partition-function transseries-formula \eqref{eq:InstantonsOneParameter}, we have obtained novel analytical expressions for two-cut free-energy genus-expansion coefficients, by means of the power-series expansion \eqref{eq:toprec11}. Perhaps not surprisingly, these expansion coefficients follow a topological-recursion-like relation if we take the Euler characteristic definition \eqref{eq:toprec17} seriously. A good reason to do so follows immediately by recalling that \cite{eo07a}
\be
\partial_s^n \CF_h (t_1,t_2) = \oint_{B} \cdots \oint_{B} \omega_{h,n} \left( \bullet, \ldots, \bullet \right),
\ee
\noindent
where $B \subset \mathbb{C}$ is the $B$-cycle connecting both cuts along the clockwise direction. Indeed, the definition of Euler characteristic within the context of the topological recursion, and assigned to the correlation-functions on the right-hand-side of the above equation, is identical to the definition we provided for the derivatives featuring in its left-hand-side; at least if we disregard the Gaussian normalization. This suggests that the recursive procedure we derived to compute the multi-cut free-energy should be directly related to a suitable limit of the multi-cut topological recursion, where the underlying spectral curve is \textit{pinched}. If this is true, suitably reversing the argument might in fact yield a partial proof of our transseries construction, solely based upon the topological recursion. We leave this interesting exploration avenue for future research.

Finally, a generalization towards multi-cut spectral-curves follows immediately from the generic partition-function transseries formula \eqref{eq:pinchedmulti}, by computing saddle-point expansions associated with generic forward-instanton sectors, and then setting up linear-systems such as \eqref{eq:toprec14}. The resulting systems are larger but solvable in the same way as we did herein.

\section{The Local Solution: Analytics Versus Numerical Tests}
\label{sec:checks-tests-numerics}

We arrive at our final section having thoroughly discussed a large amount of nonperturbative features of matrix models and string theories in section~\ref{sec:strong-coupling-phases}, both at and off criticality; which should now be completely and exactly reproduced by the exact solutions we proposed and described at length in section~\ref{sec:resurgent-Z-transseries} (and further in section~\ref{sec:topological-recursion})---say, by either formulae \eqref{eq:pinchedcubic}, \eqref{eq:pinchedcubic-dsl}, or \eqref{eq:pinchedmulti}, depending on the specific example to consider. In short, we shall now attempt to use our \textit{analytical} partition-function transseries to reproduce all previously-obtained \textit{numerical} ``raw data''. For example, on what concerns matrix models, we ask to what extent our large-$N$ transseries can reproduce finite\footnote{Large-$N$ resurgent-transseries have been previously used to compute matrix-model finite-$N$ quantities in \cite{csv15}, albeit focusing exclusively on the quartic matrix model and solely in the non-resonant setting.} $N$ quantities---say, the oscillation patterns of orthogonal-polynomial recursion-coefficients $r_N (t)$ discussed throughout subsection~\ref{subsec:OP-phases}. On what concerns the double-scaling limit, we ask to what extent our large-$z$ transseries can reproduce non-trivial features of string-equation solutions---say, the precise location of their poles upon the complex plane, as discussed throughout subsection~\ref{subsec:DSL-phases}.

In order to achieve the aforementioned comparisons, one first needs to address two technicalities. The first concerns the resonant transseries: no matter how compactly we have managed to repackage them, at the end-of-the-day extracting actual numbers still requires resumming \textit{asymptotic} series, and, in particular, asymptotic series with \textit{both} exponential suppressed and exponentially enhanced contributions. How this is achieved in practice, and how that is in fact always a well-defined procedure, will be detailed in subsection~\ref{subsec:numerical-resummation-details}. The second concerns the numerical methods used to produce our ``raw data'', in particular the double-scaled pole-plots presented in subsection~\ref{subsec:DSL-phases}. We use the well-known Fornberg--Weideman numerical method \cite{fw11}, which we briefly review in subsection~\ref{subsec:numerical-dsl-details}. Having addressed the technical aspects we proceed to the comparisons: off-critical matrix models in subsection~\ref{subsec:matrix-model-numerics} and double-scaled theories in subsection~\ref{subsec:dsl-string-eq-numerics}. We almost succeed---but not quite. The punchline will appear in \cite{krsst26b}.

Resurgent-transseries descriptions of general functions are valid inside Stokes wedges, bounded by Stokes lines which induce discontinuous jumps of the transseries parameters $\boldsymbol{\sigma} \mapsto \underline{\pmb{\BS}}_{\theta} (\boldsymbol{\sigma}) \sim \boldsymbol{\sigma} + \boldsymbol{S}$ according to their Stokes data \cite{e81, e84, e93, bssv22}. We will see in subsections~\ref{subsec:matrix-model-numerics} and~\ref{subsec:dsl-string-eq-numerics} how in fact these Stokes transitions are an obstruction to achieving global analytic continuation of our transseries constructions---which implies we still need to present full Stokes automorphisms for all models at hand in order to have reached our \textit{globally} exact solutions; and which we shall do in \cite{krsst26b}. Regardless, it is still the case as we discuss in subsections~\ref{subsec:matrix-model-numerics} and~\ref{subsec:dsl-string-eq-numerics} that our resummed transseries are able to reproduce all of our ``raw data'' \textit{within individual Stokes wedges}. This should certainly make for a rather convincing argument that constructing Stokes automorphisms is in fact the \textit{only} remaining obstruction to constructing \textit{global solutions}. Finally, let us stress that even without Stokes data our exact solutions do correctly and completely describe our matrix models and string theories \textit{locally}. In particular, this implies that our complete discussions up to this point---from the inclusion of anti-eigenvalues \cite{mss22} or negative-branes \cite{sst23} up to the construction of diagonal-framing transasymptotics and their associated discrete Fourier/Zak transforms---all of these aspects are in fact \textit{necessary} for our local checks to go through. 

\subsection{On the Resummation and Evaluation of Resurgent Transseries}
\label{subsec:numerical-resummation-details}

Our main formulae \eqref{eq:pinchedcubic}-\eqref{eq:pinchedcubic-dsl}-\eqref{eq:pinchedmulti} yield exact albeit formal transseries representations for the partition-functions of wide classes of matrix models, minimal and topological strings (subsequently the same holds for a variety of closely related but still distinct observables: the free-energy, the orthogonal-polynomial recursion-coefficients or their corresponding double-scaled specific-heat). All these quantities are well-defined functions of their parameters (the string coupling, or $N$, and the 't~Hooft coupling for matrix models and topological strings; or else the variable $z$ in the double-scaled setting). Their transseries representations on the other hand are so far purely formal objects due to their underlying construction via asymptotic series. In other words, all the many transseries sectors which appeared throughout section~\ref{sec:resurgent-Z-transseries} are to this stage \textit{formal} genus-expansions which remain asymptotic, with zero radius of convergence. As such, they are not yet \textit{functions} to which we can plug numbers in and get finite numbers out. Additionally, one might worry of extra complications arising from resonant, exponentially \textit{enhanced} contributions, and what role do they play tackling the convergence of the associated instanton sums within Stokes wedges where these exponential weights are indeed turned on. Hence, for our exact transseries solutions to be proper representations of arbitrary finite-coupling quantities, we must now be precise on how to take the final (and simple) step from formal to finite.

Let us begin with the string equation solution discussed in subsection~\ref{subsec:OP-phases}, whose transseries was therein\footnote{Herein we have pulled the starting $g_{\text{s}}$ powers, $\sim g_{\text{s}}^{\beta_{\boldsymbol{n}}}$, out from the asymptotic sectors $R^{(\boldsymbol{n})} (t,g_{\text{s}})$, as these will play an important role in the convergence discussion which follows below.} of the form \eqref{eq:twoparameterresurgenttransseriesforR} (but see the specific cubic \eqref{eq:cubicMM-Rtransseries-explicit} and quartic \eqref{eq:quarticMM-Rtransseries-explicit} matrix-model transseries as well)
\be
R \left( t,g_{\text{s}}; \boldsymbol{\sigma} \right) = \sum_{\boldsymbol{n}=0}^{+\infty} \boldsymbol{\sigma}^{\boldsymbol{n}}\, \rme^{- \frac{\boldsymbol{n} \cdot \boldsymbol{A} (t)}{g_{\text{s}}}}\, R^{(\boldsymbol{n})} (t,g_{\text{s}}) = \sum_{\boldsymbol{n}=0}^{+\infty} \boldsymbol{\sigma}^{\boldsymbol{n}}\, \rme^{- N\, \frac{\boldsymbol{n} \cdot \boldsymbol{A} (t)}{t}}\, \frac{t^{\beta_{\boldsymbol{n}}}}{N^{\beta_{\boldsymbol{n}}}}\, R^{(\boldsymbol{n})} (t,N).
\ee
\noindent
Note how following \eqref{eq:gstoNtransmonomial} we have traded $g_{\text{s}}$ for $N$ and $t$, in order to carry through finite $N$ (and varying $t$) checks (see \cite{csv15} as well). As already discussed in section~\ref{sec:resurgent-Z-transseries}, there are two types of sums in this expression: the \textit{asymptotic} genus-expansion sums, and the \textit{non-asymptotic} instanton-expansion sums. Asymptotic sums are very simply dealt with via Borel resummation. Following, \textit{e.g.}, \cite{abs18}, picking any one asymptotic sector
\be
R^{(\boldsymbol{n})} (t,N) \simeq \sum_{g=0}^{+\infty} \frac{1}{N^g}\, t^g\, R^{(\boldsymbol{n})}_g (t),
\ee
\noindent
its Borel transform
\be
\mathcal{B} [R^{(\boldsymbol{n})}] (s) = \sum_{g=0}^{+\infty} \frac{s^{g}}{g!}\, t^g\, R^{(\boldsymbol{n})}_g (t)
\ee
\noindent
produces a new series with a finite non-zero radius of convergence---of which we are interested in its analytic continuation. In order to numerically evaluate the underlying function, one has to truncate this series at some order $g_{\text{max}}$, where for each instanton sector we set an individual cut-off based on our data availability. The numerical analytical continuation of the above power-series expansion, automatically incorporating numerical approximations to its Borel singularities, is achieved by approximating this truncated version of the Borel transform by the diagonal Pad\'{e}-approximant\footnote{The Pad\'{e}-approximant employed here is diagonal, which is to say that the polynomial degree in the numerator matches the denominator. Often it can also be numerically beneficial to employ slightly off-diagonal Pad\'{e}-approximants. The polynomial coefficients of the Pad\'{e} approximant are $a_g(t)$ and $b_g(t)$; see, \textit{e.g.}, \cite{abs18}.}
\be
\text{BP}_{g_{\text{max}}} [R^{(\boldsymbol{n})}] (s) = \frac{\sum_{g=0}^{\lfloor g_{\text{max}}/2\rfloor} a_g (t)\, s^g}{\sum_{g=0}^{\lfloor g_{\text{max}}/2\rfloor} b_g (t)\, s^g}.
\ee
\noindent
It is to the above rational function which we apply the Laplace transform in order to obtain the Borel--Pad\'e resummation of the initial asymptotic series $R^{(\boldsymbol{n})} (t,N)$, as
\be
\mathcal{S}_\theta \text{BP}_{g_{\text{max}}} [R^{(\boldsymbol{n})}] (t,N) = \int_{0}^{\rme^{\mathrm{i}\theta}\infty} \rmd s\, \text{BP}_{g_{\text{max}}} [R^{(\boldsymbol{n})}] (s)\, \rme^{-Ns}.
\ee
\noindent
Finite-$N$ tests will see $N$ as an integer, in which case it has vanishing argument and we set $\theta=0$. 

Moving-on towards the instanton sums, let us herein focus on the two-parameter case---with one type of instanton-action alongside its negative partner. The discussion with more resonant instanton actions is an immediate extrapolation, and all subtleties already occur in this simpler setting. Being explicit on the cubic and quartic matrix model cases, we now write the string-equation solution as
\bea
R \left( t,g_{\text{s}}; \sigma_1,\sigma_2 \right) &=& \sum_{n,m=0}^{+\infty} \sigma_1^n \sigma_2^m\, \rme^{- N \left(n-m\right) \frac{A (t)}{t}} \left[ f(t) \right]^{- \alpha \left(n-m\right) \upmu} g_{\text{s}}^{\beta_{nm}}\, \mathcal{S}_\theta \text{BP}_{g_{\text{max}}} [R^{(n,m)}] (t,N) = \nonumber \\
&=& \sum_{n,m=0}^{+\infty} \upxi_1^n \upxi_2^m\, g_{\text{s}}^{\beta_{nm}-\frac{n+m}{2}}\, \mathcal{S}_\theta \text{BP}_{g_{\text{max}}} [R^{(n,m)}] (t,N).
\label{eq:BorelResummedR}
\eea
\noindent
In the first line above $f(t)$ is a problem-dependent function and $\alpha$ is the logarithmic resummation constant. The reader may compare to the explicit cubic result in \eqref{eq:cubicMM-Rtransseries-explicit} (which is however prior to logarithmic resummation). In the second line the logarithmic dependence has been resummed and hence we introduced adequate rectangular-framing resummation-variables $\upxi_{1,2}$. The reader may compare to the corresponding explicit cubic variables in \eqref{eq:zetaxiCMM-1}-\eqref{eq:zetaxiCMM-2}, or the explicit quartic variables in \eqref{eq:upxi12forQuarticMAIN} or \eqref{eq:upxi12forQuartic} (which then yield the quartic transseries in \eqref{eq:quarticMM-Rtransseries-explicit}). Next, focus on the two instanton sums. As discussed early in subsection~\ref{subsec:resurgent-Z-transasymptotics}, summations along a chosen instanton direction whilst at fixed-genus are convergent within some (as yet unspecified) convergence radius (and this is what allowed for the transasymptotic resummations also discussed therein). Let us then make explicit how we estimate these convergence regions---regardless of dealing with exponentially-enhanced or exponentially-suppressed transmonomials. 

\begin{figure}
	\centering
	\includegraphics[width=.65\linewidth]{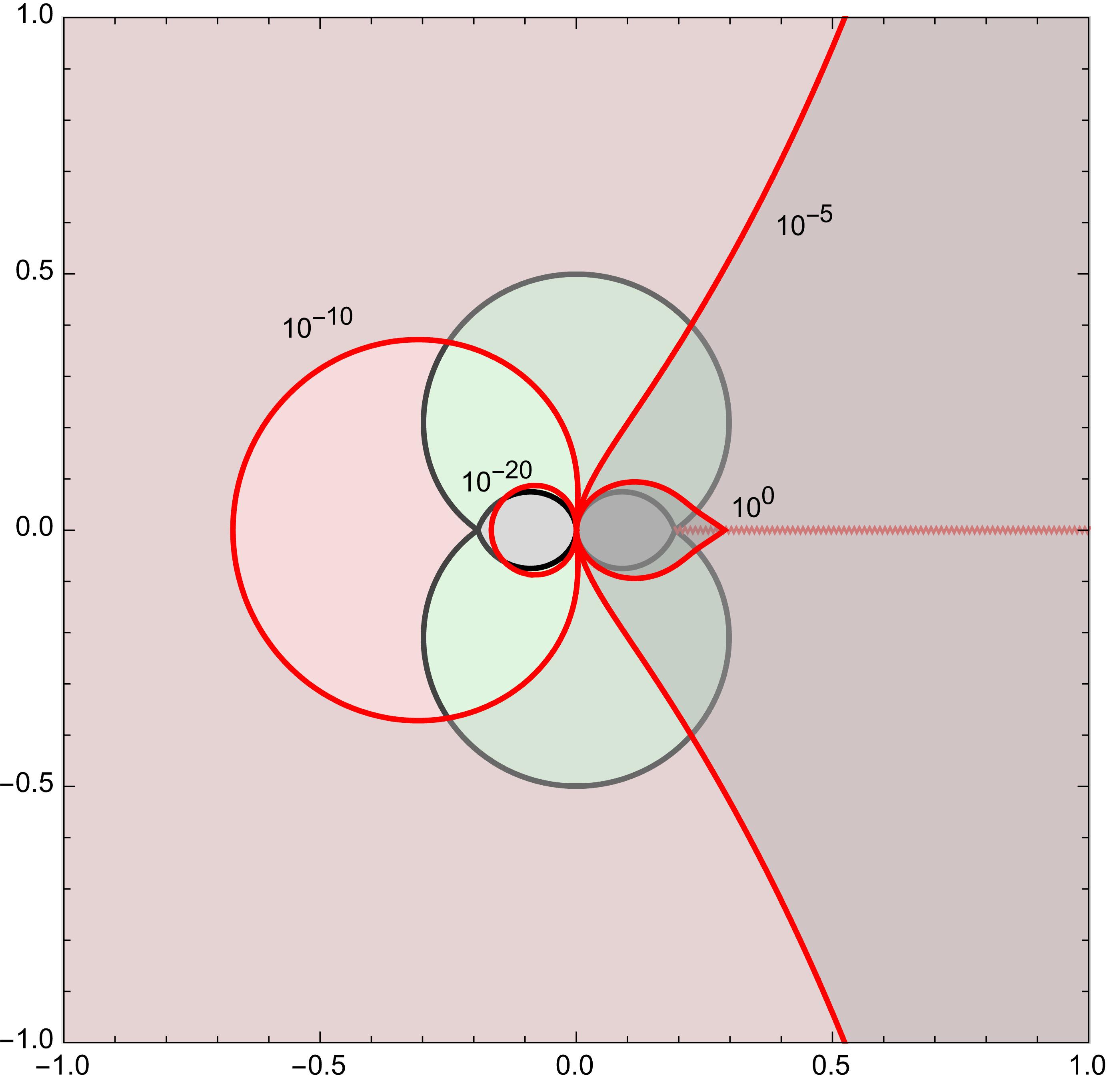}
	\caption{Regions of convergence, upon the 't~Hooft complex $t$-plane, of the planar positive-tension instanton sum of the cubic matrix model, for varying $\sigma_1$ as dictated by \eqref{eq:CMM Positive Tension Convergence Criterion} (where we have set $N=6$, $\lambda=1$, $\sigma_2=0$). Boundaries of regions of convergence (\textcolor{gray}{gray}-shaded increasingly darker) are given by the \textcolor{red}{red} lines, labeled with their corresponding values of $\sigma_1$. For the smallest value of $\sigma_1$ we used, the instanton sum converges everywhere except for a small region on the left-hand one-cut region, and the convergence regions shrink with increasing $\sigma_1$. Notice how due to the branch-cut the $\sigma_1=10^{0}$ boundary has a ``sharp edge'' on the positive real axis whereas the others ($\sigma_1=10^{-10},10^{-20}$) do not have one on the negative real axis.}
	\label{fig:HMM Convergence Regions}
\end{figure}

For the purpose of this analysis, first observe that the Borel resummation of any transseries sector is reasonably well approximated by the leading, planar contribution to its asymptotic expansion, as
\be
\mathcal{S}_\theta \text{BP}_{g_{\text{max}}} [R^{(n,m)}] (t,N) \approx R^{(n,m)}_0 (t) + \mathcal{O} (1/N).
\ee
\noindent
As such, one can estimate the convergence region of the \textit{full} Borel--Pad\'{e} resummed \textit{transseries} by simply considering its planar approximation. Now, in order to estimate this convergence region to a precision sufficient for our purposes, it is enough to enforce convergence of both the $n=0$ and $m=0$ sums. In this way, we consider the convergence properties of the two sums
\be
\sum_{n=0}^{+\infty} \upxi_1^{n}\, R^{(n,0)}_0 (t) \qquad \text{ and } \qquad \sum_{m=0}^{+\infty} \upxi_2^{m}\, R^{(0,m)}_0 (t).
\ee
\noindent
Their convergence properties may be studied via the linear transasymptotics methods explained in subsection~\ref{subsec:resurgent-Z-transasymptotics}. For example, for the cubic matrix model we found (compare with \eqref{eq:lineartransasymptoticssum} and \eqref{eq:linear-transasymptotics-example})
\be
\sum_{n=1}^{+\infty} \upxi_1^{n}\, R^{(n,0)}_0 (t) = - \frac{4}{\lambda^2} \left(1-3\lambda^2\, r\right) \sum_{n=1}^{+\infty} n \left( - \frac{1}{4}\, \frac{\lambda^2 \sqrt{r}}{\left( 1-3\lambda^2\, r \right)^{5/4}}\, \upxi_1 \right)^{n}.
\ee
\noindent
This geometric sum converges if and only if
\be
\label{eq:CMM Positive Tension Convergence Criterion}
\abs{\frac{1}{4}\, \frac{\lambda^2 \sqrt{r}}{\left( 1-3\lambda^2\, r\right)^{5/4}}\, \upxi_1} < 1,
\ee
\noindent
which is the convergence criterion we use. Carrying out the same exercise for negative-tension instantons, one finds the additional\footnote{This has essentially the same form as  $R^{(0,n)}_0=R^{(n,0)}_0$ due to their backward-forward symmetry \cite{asv11, bssv22}.} condition
\be
\label{eq:CMM Negative Tension Convergence Criterion}
\abs{\frac{1}{4}\, \frac{\lambda^2 \sqrt{r}}{\left( 1-3\lambda^2\, r\right)^{5/4}}\, \upxi_2} < 1.
\ee
\noindent
We illustrate the regions of convergence predicted by the first condition \eqref{eq:CMM Positive Tension Convergence Criterion} in figure~\ref{fig:HMM Convergence Regions}. In this figure, we vary $\sigma_1$ while fixing all other parameters, and illustrate the boundary of the convergence region upon the 't~Hooft complex $t$-plane of the cubic matrix model (recall its phase diagram in figure~\ref{fig:CMM Phase Diagram}). It is clear that the (increasingly darker) regions of convergence shrink as $\sigma_1$ increases. For small $\sigma_1$, convergence occurs almost everywhere on the $t$ plane save for a small region in the left-most one-cut phase, with convergence then getting worse for larger $\sigma_1$. 

\begin{figure}
     \centering
     \begin{subfigure}[b]{0.32\textwidth}
         \centering
         \includegraphics[width=\textwidth]{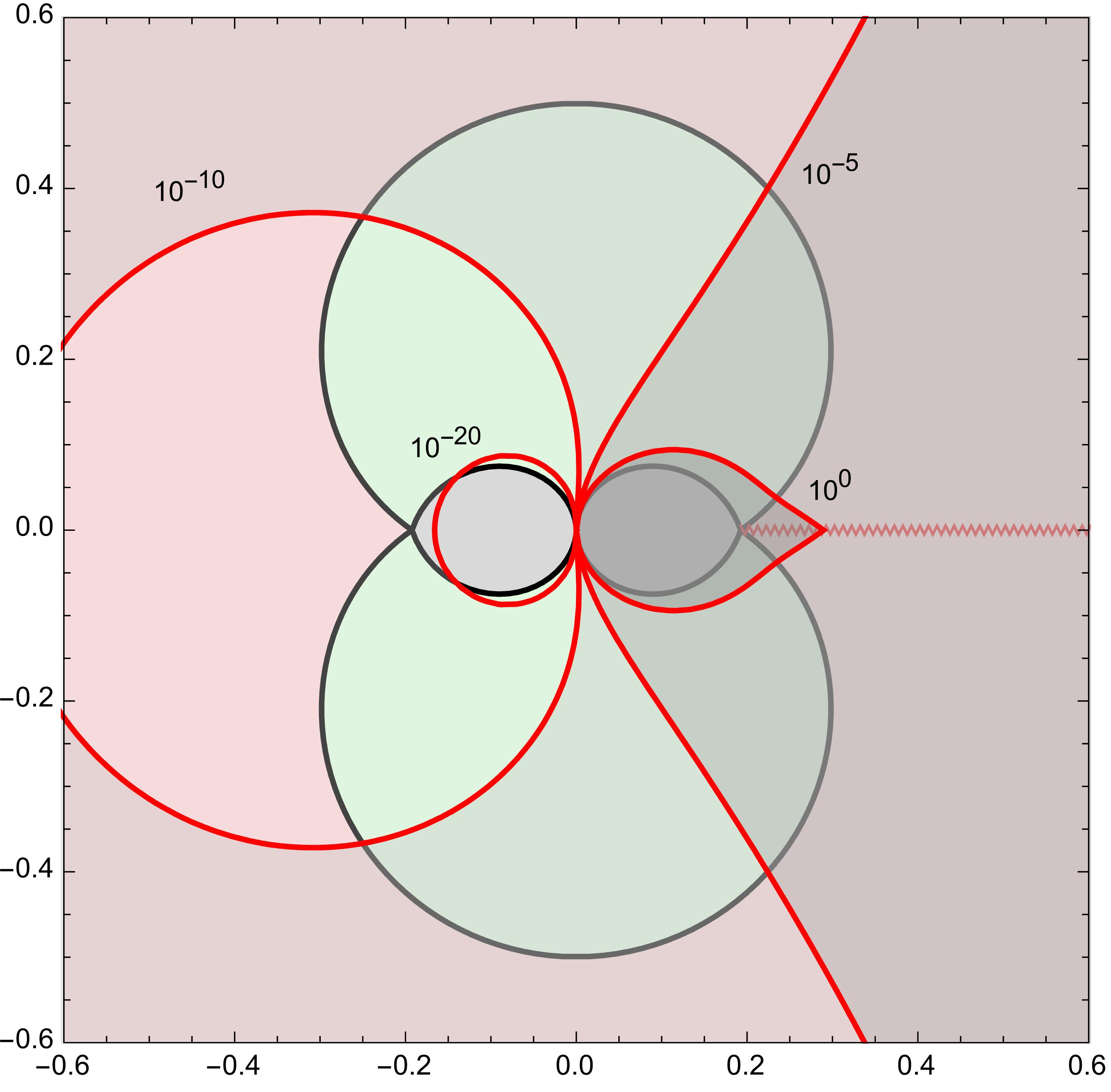}
         \caption{Positive tension only.}
         \label{fig:CMMPositiveTensionConvergenceRegions}
     \end{subfigure}
     \hfill
     \begin{subfigure}[b]{0.32\textwidth}
         \centering
         \includegraphics[width=\textwidth]{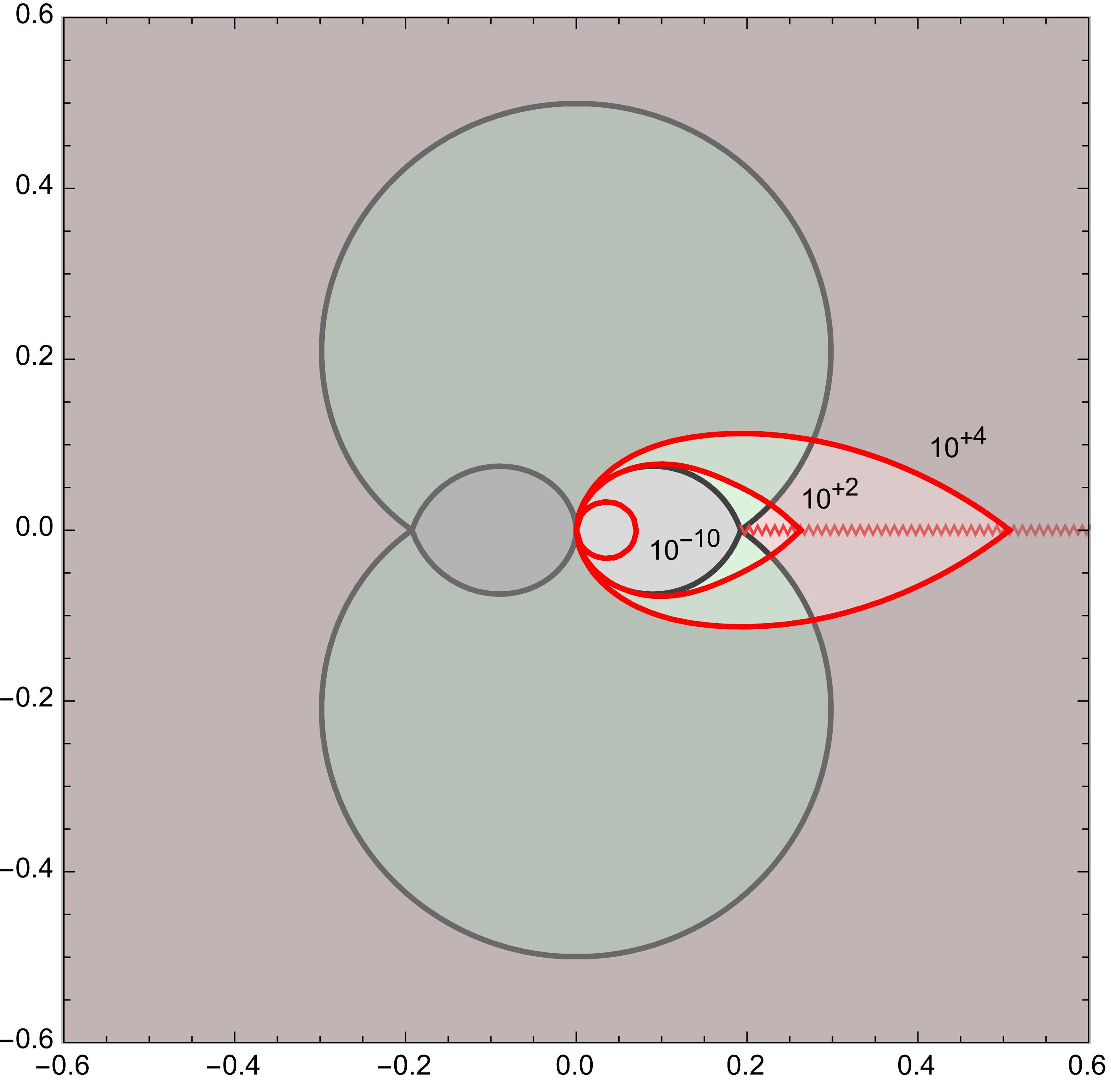}
         \caption{Negative tension only.}
         \label{fig:CMMNegativeTensionConvergenceRegions}
     \end{subfigure}
     \hfill
     \begin{subfigure}[b]{0.32\textwidth}
        \centering
         \includegraphics[width=\textwidth]{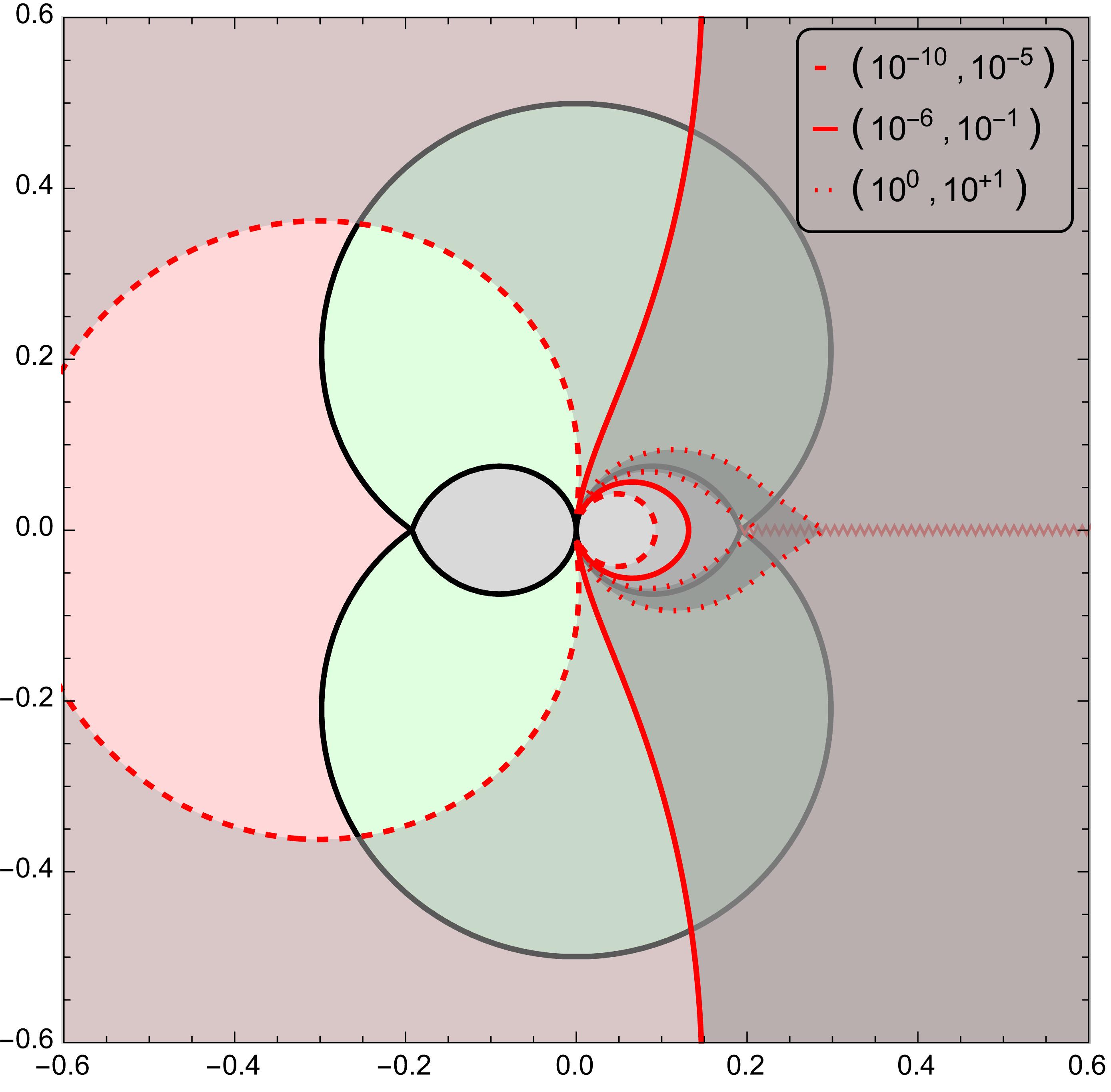}
        \caption{Both tensions.}
        \label{fig:CMMMixedTensionConvergenceRegions}
     \end{subfigure}
        \caption{Regions of convergence, upon the 't~Hooft complex $t$-plane, of the planar positive-, negative- and both-tension instanton sums of the cubic matrix model, for varying $\sigma_1$ and $\sigma_2$ as dictated by \eqref{eq:CMM Positive Tension Convergence Criterion} and \eqref{eq:CMM Negative Tension Convergence Criterion} (where we have set $N=6$, $\lambda=1$). Boundaries of regions of convergence are given by the \textcolor{red}{red} lines, and convergence occurs in the (increasingly darker) \textcolor{gray}{gray} shaded regions. The values of $\sigma_{1,2}$ we used for each line are indicated in the plot for figures~\ref{fig:CMMPositiveTensionConvergenceRegions} and~\ref{fig:CMMNegativeTensionConvergenceRegions}, and as an inset caption for figure~\ref{fig:CMMMixedTensionConvergenceRegions}.}
	\label{fig:CMMPosNegMixTensionConvergenceRegions}
\end{figure}

Albeit the consideration of transcendents with $\sigma_2=0$ is all we need for the purposes of the present paper, we may, nonetheless, make sure the discussion is complete (and further set-up machinery for our follow-up \cite{krsst26b}) by initiating the discussion of negative-tension convergence---essentially described by figure~\ref{fig:CMMPosNegMixTensionConvergenceRegions}. On top of figure~\ref{fig:HMM Convergence Regions}, which is reproduced as figure~\ref{fig:CMMPositiveTensionConvergenceRegions} for direct comparison purposes, we illustrate the regions of convergence predicted by the second condition above \eqref{eq:CMM Negative Tension Convergence Criterion} in figure~\ref{fig:CMMNegativeTensionConvergenceRegions}. In this figure, we vary $\sigma_2$ while fixing all other parameters, and using the same color-codes as in figure~\ref{fig:HMM Convergence Regions} or~\ref{fig:CMMPositiveTensionConvergenceRegions}. One immediately observes that negative-tension contributions are actually relatively mildly behaved: even for $\sigma_2$ as large as $100$, negative-tension instanton-sums converge everywhere except for a region that mostly consists of the one-cut region. The reason is that positive- and negative-tension exponential weights are symmetric, hence \textit{only one} of the two will be enhanced in any place upon the complex plane (the other suppressed). In particular, positive tension is enhanced \textit{outside} the one-cut region, whereas negative tension is enhanced \textit{inside}. This is simple to see, as the phase boundary separating the one-cut region from the other phases is exactly the condition $\Re(A/g_{\text{s}})=0$, indicating that within the one-cut region one has $\Re(A/g_{\text{s}})>0$ (positive-tension suppression and negative-tension enhancement) whereas outside it one has $\Re(A/g_{\text{s}})<0$ (positive-tension enhancement and negative-tension suppression). Finally, figure~\ref{fig:CMMMixedTensionConvergenceRegions} illustrates the regions of convergence of the instanton sums when both $\sigma_1$ and $\sigma_2$ are non-zero and we are hence enforcing both \eqref{eq:CMM Positive Tension Convergence Criterion} and \eqref{eq:CMM Negative Tension Convergence Criterion}. The qualitative behavior is the same: the positive-tension sum leads to divergences away from the standard one-cut region, whereas the negative-tension sum leads to divergences precisely inside that region. It should be clear by now how there really is no reason for concern regarding negative-tension transmonomial contributions: it is always easy to find regions where their instanton power-series converges (hence, where adequate truncations yield accurate results), but otherwise we can also simply proceed by straightforward analytic continuation given their geometric-series nature.

\begin{figure}
     \centering
     \begin{subfigure}[b]{0.32\textwidth}
         \centering
         \includegraphics[width=\textwidth]{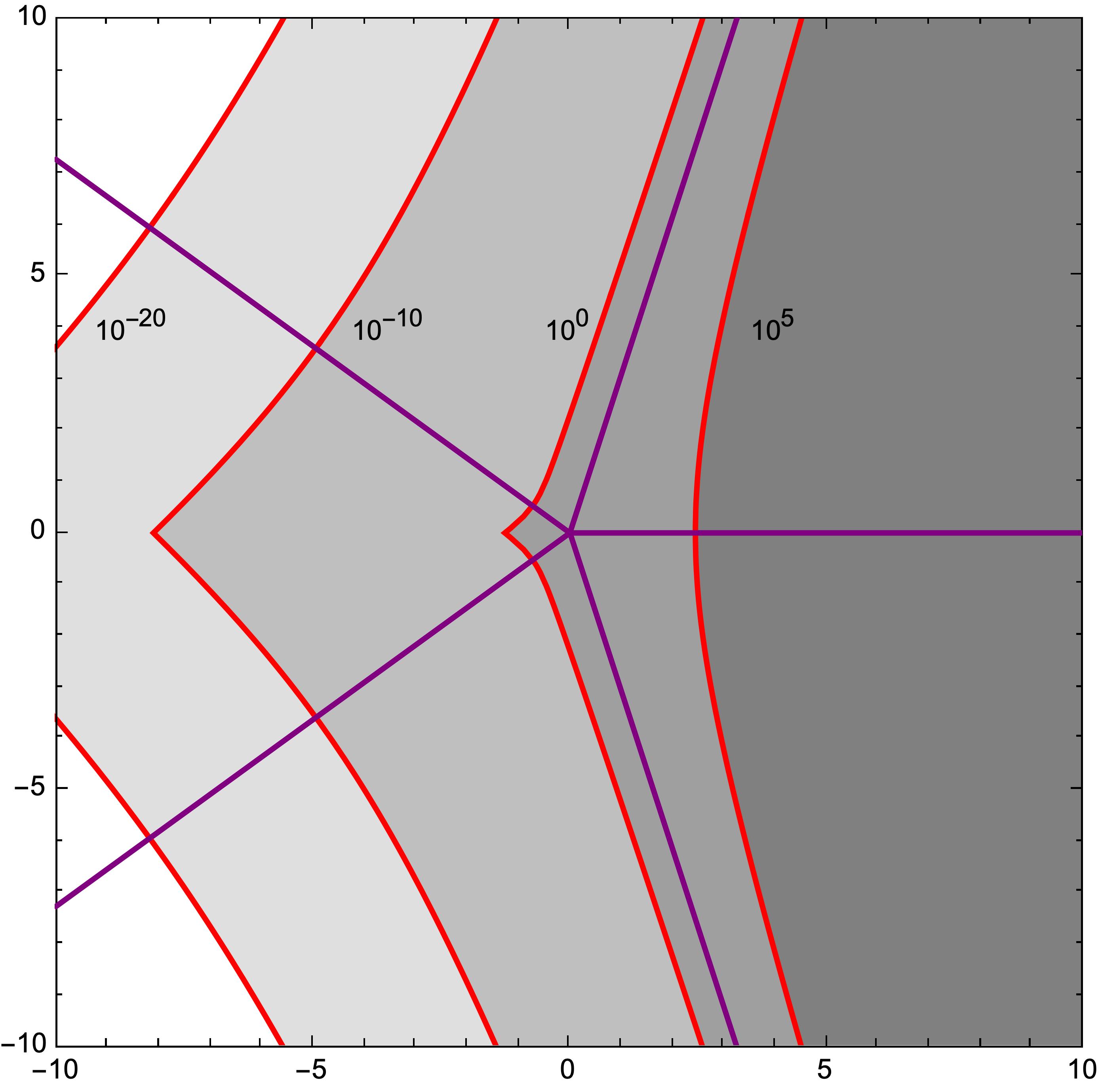}
         \caption{Positive tension only.}
         \label{fig:PIPositiveTensionConvergenceRegions}
     \end{subfigure}
     \hfill
     \begin{subfigure}[b]{0.32\textwidth}
         \centering
         \includegraphics[width=\textwidth]{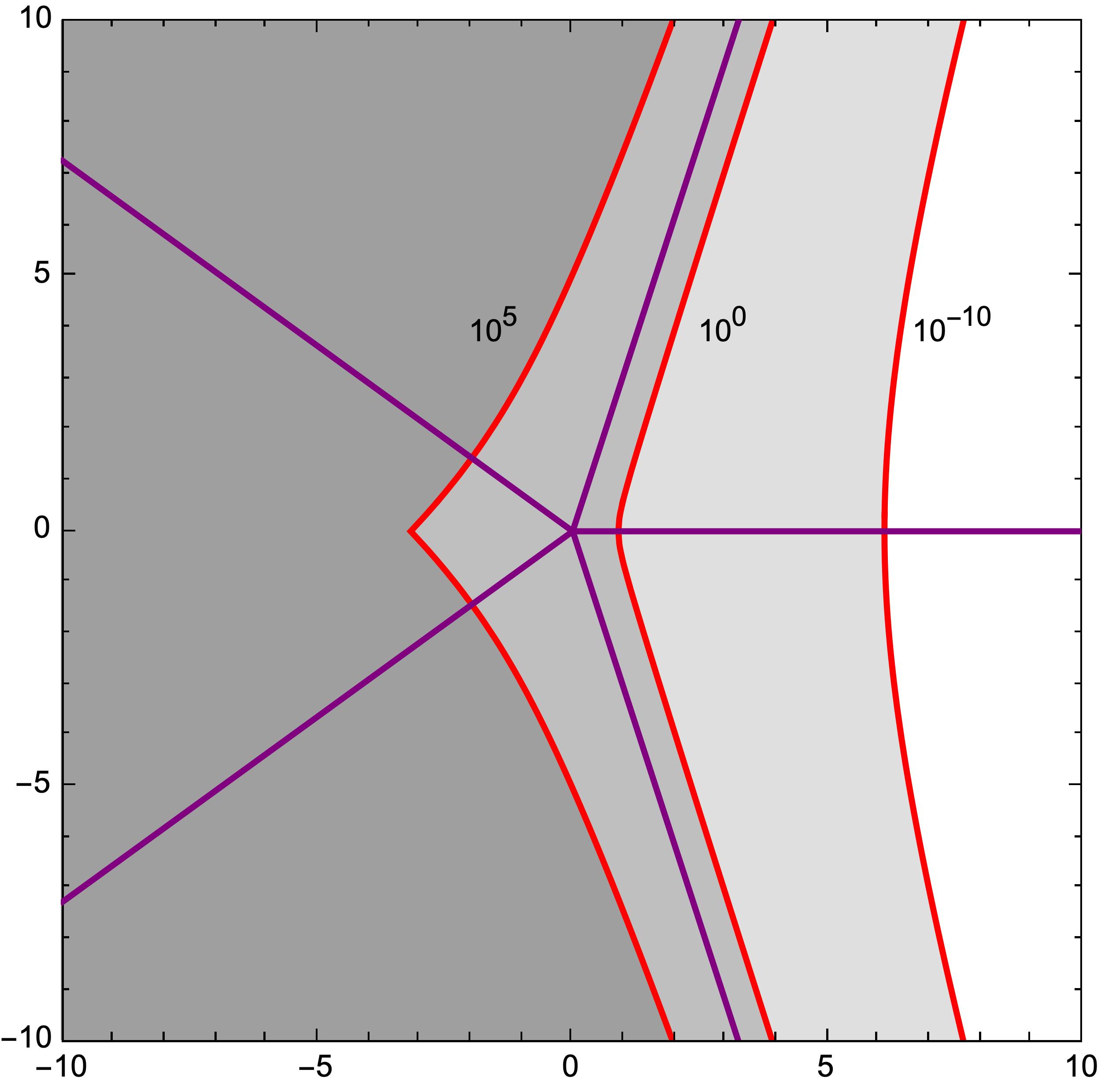}
         \caption{Negative tension only.}
         \label{fig:PINegativeTensionConvergenceRegions}
     \end{subfigure}
     \hfill
     \begin{subfigure}[b]{0.32\textwidth}
        \centering
         \includegraphics[width=\textwidth]{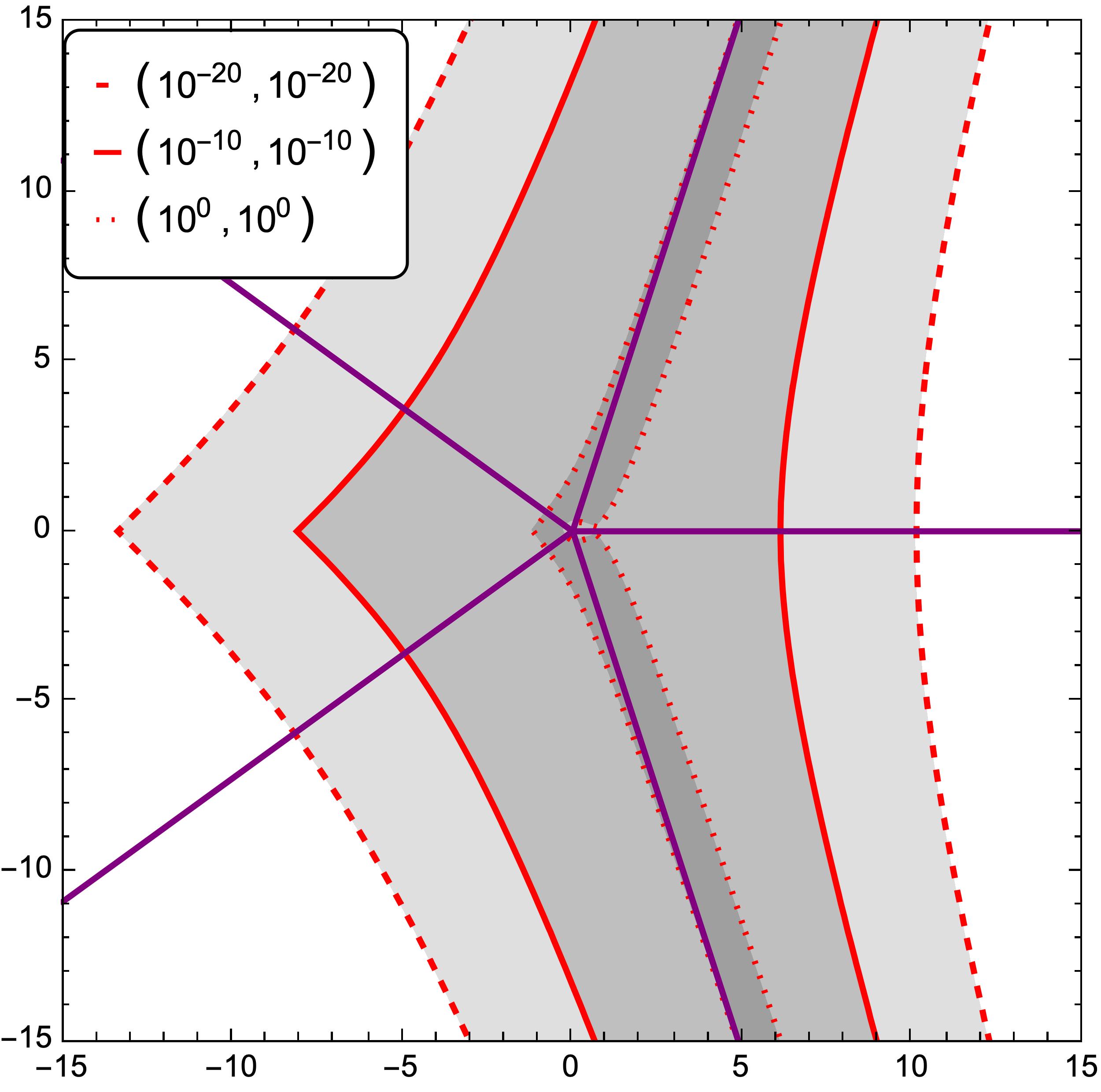}
        \caption{Both tensions.}
        \label{fig:PIMixedTensionConvergenceRegions}
     \end{subfigure}
        \caption{Regions of convergence on the complex $z$-plane, of the planar positive- and negative-tension instanton sums of the \PI~specific-heat transseries, for varying $\sigma_1$ and $\sigma_2$ as dictated by their convergence criteria. The \textcolor{violet}{purple} lines indicate Stokes/anti-Stokes lines (recall figure~\ref{fig:P1stokesAutomorphismsZplane}), and boundaries of regions of convergence are given by the \textcolor{red}{red} lines, with convergence occurring in the (increasingly darker) \textcolor{gray}{gray} shaded regions. The values of $\sigma_{1,2}$ we used for each line are indicated in the plot for figures~\ref{fig:PIPositiveTensionConvergenceRegions} and~\ref{fig:PINegativeTensionConvergenceRegions}, and as an inset caption for figure~\ref{fig:PIMixedTensionConvergenceRegions}. As in the previous matrix-model case, small $\sigma_{1,2}$ yields larger convergence regions.}
	\label{fig:PI Convergence Regions}
\end{figure}

The above discussion has a complete analogue in the double-scaled setting, where now the object to be resummed is the specific-heat $u(z)$, solution to the (differential) string equation. Just like discussed above, one can use Borel--Pad\'{e} resummation in order to make sense of all asymptotic sums. For the instanton sums, they are again not asymptotic, hence have non-trivial convergence regions. For example, in the \PI~case, the conditions to consider are just the double-scaling limits of \eqref{eq:CMM Positive Tension Convergence Criterion} and \eqref{eq:CMM Negative Tension Convergence Criterion}, which now read
\be
\abs{\frac{1}{12}\, z^{-5/8}\, \sigma_1\, \rme^{-A_{\text{\PI}} z^{5/4}}} < 1 \qquad \text { and } \qquad \abs{\frac{1}{12}\, z^{-5/8}\, \sigma_2\, \rme^{+A_{\text{\PI}} z^{5/4}}} < 1.
\ee
\noindent
The corresponding regions of convergence are illustrated in figure~\ref{fig:PI Convergence Regions}.

\begin{figure}
	\centering
	\includegraphics[width=.75\linewidth]{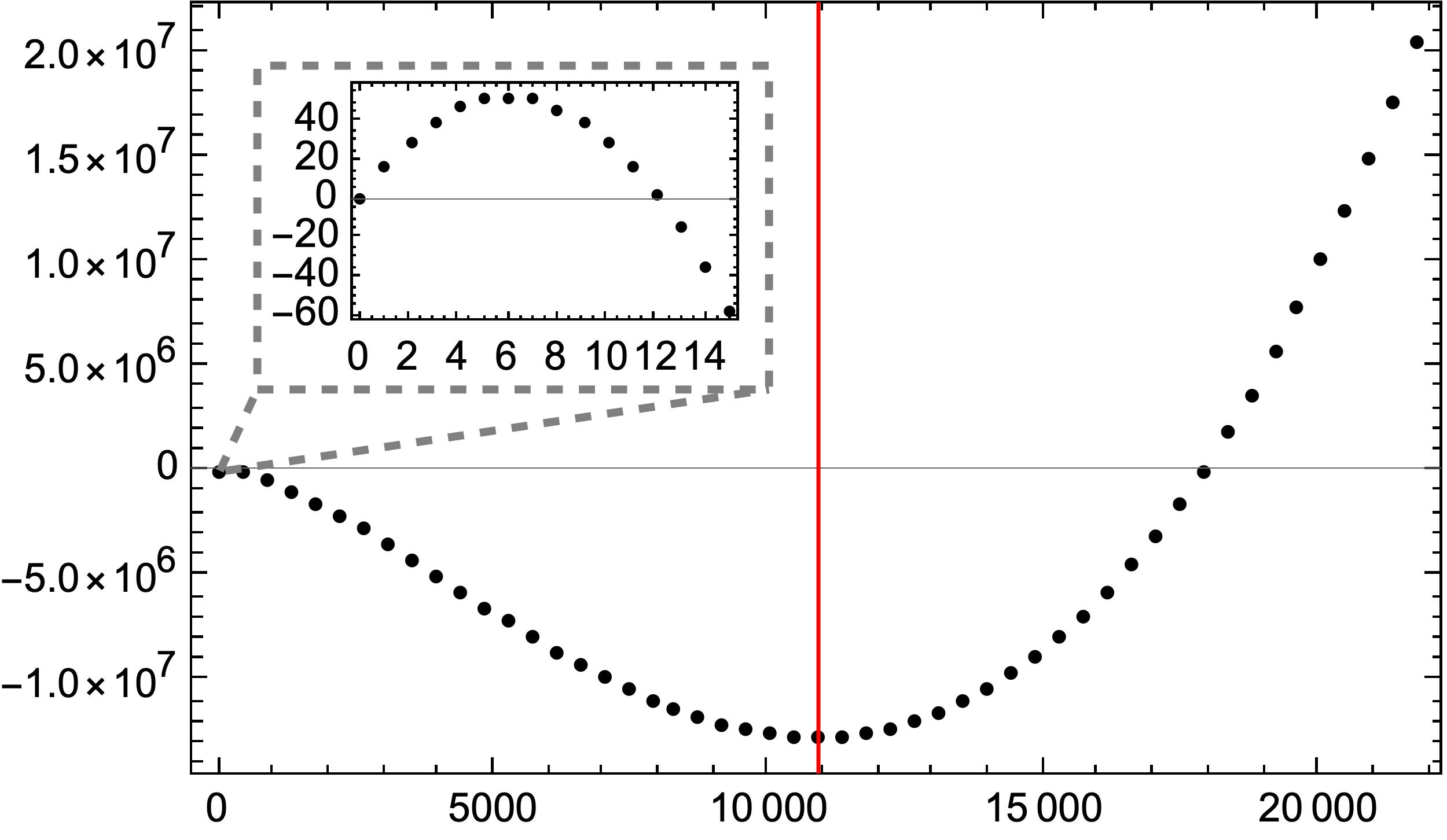}
	\caption{Relative magnitude of thousands of terms in the genus-zero instanton-sum \eqref{eq:CMM Instanton Sum Genus Zero} for the cubic matrix model (where we set $N=30$, $\lambda=1$, $\sigma_1=10^{-4}$, $\sigma_2=0$, $t=\rme^{-\rmi\pi/10}$). We plot $n$ on the horizontal axis, and the base-$10$ logarithm of the $n$-instanton term on the vertical axis. The main plot shows that the optimal truncation point, indicated by the \textcolor{red}{red} line, occurs at very high instanton number; and how its associated optimal-truncation error is extremely small, roughly a factor $\sim 10^{-1.4 \cdot 10^7}$ smaller than the largest term. The inset plot provides a zoomed-in view of the low-instanton range, showing how certain instanton contributions are dominant over the perturbative one. Furthermore, sub-leading terms become very small, very quickly, as even the smallest term in the inset plot is about $100$ orders of magnitude smaller than the dominant one. Truncation around the dominant term suffices to carry out high-precision numerical checks.}
	\label{fig:Z Instanton Sum Optimal Truncation}
\end{figure}

As should be evident from above, the value of $N$ and the transseries parameters $(\sigma_1,\sigma_2)$ will strongly influence the region on the complex $t$-plane where all instanton sums are converging, with convergence regions getting  smaller as $N$ gets larger whilst keeping $(\sigma_1,\sigma_2)$ fixed. Going forward one hence resorts to quadratic transasymptotics, as described in subsection~\ref{subsec:resurgent-Z-transasymptotics}. Note that in spite of being interested in reconstructing the recursion coefficients $r_N (t)$ from the large-$N$ point-of-view, say, in order to fully account for their oscillating behavior as described in subsection~\ref{subsec:OP-phases}, one can of course always do this starting from the partition function and directly using \eqref{eq:rsintermsofpartitionfunction}. Still focusing on the example of the cubic matrix model, we shall subsequently use the quadratic transasymptotic transseries \eqref{eq:cubicPartitionFunction}-\eqref{eq:cubicDFTKernel} or the discrete Fourier transform transseries \eqref{eq:CMMDiscreteFourierNC}-\eqref{eq:DFTKernelCNC} representation of the partition function. Start by setting $\sigma_2=0$ and let us be fully explicit on how to treat the cubic example (other examples and higher genera are treated identically). There is actually some computational efficiency to be gained here. First, of course we could always use Borel--Pad\'{e} resummation to make sense of the asymptotic sums. But in practice, though, since the $D_{k} (\nu)$ polynomials for the cubic model are known up to $k=2$---recall \eqref{eq:CMM-Dk0}-\eqref{eq:CMM-Dk1}-\eqref{eq:CMM-Dk2}---then using optimal truncation rather than Borel--Pad\'{e} makes no difference in terms of numerical precision and is in fact much faster. We can then consider the instanton sums organized genus-by-genus, as in \eqref{eq:cubicPartitionFunction}-\eqref{eq:cubicDFTKernel}. In the case of genus-zero we focus upon
\be
\label{eq:CMM Instanton Sum Genus Zero}
\sum_{n=0}^{+\infty} \sigma_1^{n}\, \rme^{- n \frac{A(t)}{g_{\text{s}}}\, }Z^{(n|0)}_{0} (t) = \sum_{n=0}^{+\infty} \sigma_1^{n}\, \rme^{-n \frac{A(t)}{g_{\text{s}}}}\, g_{\text{s}}^{\frac{n^2}{2}}\, p_{\text{cubic}}^{n^2} (t)\, G_2 \left(n+1\right).
\ee
\noindent
It is important to notice that due to the Barnes factor the behavior of these partition-function sums is \textit{distinct}\footnote{As already mentioned in subsection~\ref{subsec:resurgent-Z-transasymptotics}, the Barnes superfactorial growth combined with the $n^2$ exponent in the string-coupling conspire to now yield standard asymptotic growth for the partition-function instanton-sums. For completeness, the reader should check the small comment on this right after \eqref{eq:Z-trans-G2-asympt}.} from the one of the earlier $R$-transseries. To illustrate this point we plot the relative sizes of the different terms in this sum in figure~\ref{fig:Z Instanton Sum Optimal Truncation}. As a first observation, the perturbative $n=0$ term is not dominant. Instead, the terms initially grow in size, then reach a maximum, and go back down in size. After decreasing in size for a long time, they start growing again where some optimal-truncation point is reached. Observe that the sub-dominant terms become smaller very quickly, and that the instanton optimal truncation point occurs at very high instanton-number. This optimal truncation error is extremely small in comparison with the genus-sum truncation error; which for this problem is of order $N^{-3}$ as we have computed $D_{k} (\nu)$ polynomials up to $k=2$. The figure then suggests the best strategy for handling these instanton sums: simply make sure that the dominant contribution is factored in, along with enough sub-dominant terms to ensure sufficient precision. Adding \textit{five} terms on both sides of the optimal point was \textit{sufficient} for all the checks we have done. The optimal value $n_{\text{opt}}$ depends non-trivially on the model parameters, say $N$, $\lambda$, the 't~Hooft coupling $t$ and transseries parameter $\sigma_1$ (in fact, $n_{\text{opt}}=0$ in the one-cut phase), but the above strategy will always continue to work.

One might then wonder: if instanton sums are such a major practical issue, and quadratic transasymptotics is so effective in solving it; why should we ever bother to use either the $R$ or $u$ transseries at all? The answer to this question is purely pragmatic. While quadratic transasymptotics handles instanton issues very well, it is very difficult to carry out in practice to high genera. In fact, for \PI~we have $D_{k} (\nu)$ polynomials up to $k=45$, but for \YL, cubic, and quartic, we only have $k=2$, $k=2$, and $k=3$, respectively. In settings where having control over high-genus effects is important (\textit{e.g.}, when $g_{\text{s}}$ is large, or small $N$), quadratic transasymptotics is then not guaranteed to give enough precision. We will see some instances (\textit{e.g.}, the $N=3$ checks in figures~\ref{fig: CMM failure unfixed} and~\ref{fig: QMM failure unfixed}) where we in fact need dozens or even hundreds of genus orders, whilst just a few instanton orders suffices for our purposes. In this case, straightforward Borel--Pad\'{e} should be the strategy of choice. We conclude that the two methods complement each other very well, with one sometimes better suited to use in settings where the other is not.

One final comment is that, in the double-scaled context, the way we implemented our checks was to compare roots of the partition-function with poles of the numerical-solutions to the double-scaled string-equations (which in fact match, as already discussed in subsection~\ref{subsec:DSL-phases} and, \textit{e.g.}, \cite{bssv22}). For this, the above considerations in fact get us most of the way: we fix some instanton and genus sum cut-offs, fix a region of the complex $z$-plane of interest and probe for numerical root locations using the Newton method. As we shall see shortly, this leads to good local numerical agreement. We point out that we fix a constant genus cut-off for the entire region of interest, rather than using optimal truncation. On one hand, constant truncation is in fact sufficient for our purposes; and, on the other hand, one reason not to use optimal-truncation is that it destabilizes standard numerical root-finding algorithms. The reason for this is that the optimal-truncation order $g_{\text{opt}}(z)$ is piecewise continuous in $z$. When $g_{\text{opt}} (z)$ jumps, \textit{i.e.}, when optimal truncation instructs us to add/remove a term from the truncated sum, we change the number of terms we include in the sum in question; which then introduces discontinuous jumps in the sum approximation. But numerical root-finding algorithms are built on an assumption of continuity, hence they should not be expected to work well in this case.

\subsection{Numerical Solutions to Double-Scaled String Equations}
\label{subsec:numerical-dsl-details}

Let us say a few words on how we numerically obtained our ``raw data'' for double-scaled problems, as this method \cite{fw11} turns out to be very efficient and general. Since double-scaled string-equations are ODEs, one would first think that any ODE numerical solver would work in producing numerical approximations for $u(z)$. However, as discussed and illustrated in subsection~\ref{subsec:DSL-phases}, solutions to double-scaled string-equations have (fields of) singularities---which standard numerical-solving methods such as Taylor-metods do not handle well \cite{fw11}. This is particularly relevant to us because it is exactly the locations of these poles that we wish to study and in fact reproduce using our resurgent-transseries methods. The clean alternative is to adopt the numerical methods of \cite{fw11, fw14} (also shown in \cite{sv22} to yield good results in the \PI~setting), which moreover carry over directly to other double-scaled problems (such as \YL; or else the entire KdV hierarchy \cite{krst26a, krst26b}). This is the main tool producing the ``raw data'' numerical plots in subsection~\ref{subsec:DSL-phases}. In order to run this numerical algorithm one requires initial data, where we follow (and generalize) \cite{sv22} in producing such initial data via Borel--Pad\'{e} resummation of the specific-heat transseries (alongside however many transseries derivatives might be needed in order to specify enough boundary data, \textit{e.g.}, one for \PI~and three for \YL).

To illustrate the idea; say we know the value of the specific-heat $u(z)$ at some base point $z=z_0$ and we want to know $u(z_t)$ for some other target point $z=z_t$. Ordinary ODE solving strategies do this iteratively by constructing a Taylor series expansion of the solution around $z_0$ up to some finite order $K$, 
\be
u_K (z) = u (z_0) + u'(z_0) \left(z-z_0\right) + \cdots + \frac{1}{K!}\, u^{(K)}(z_0) \left(z-z_0\right)^K + \CO \left( \left(z-z_0\right)^{K+1} \right).
\ee
\noindent
One then evaluates this series and its first derivative at some point $z_1$, close\footnote{More precisely, when we say $z_1$ should be close to $z_0$, we mean that they should be close enough as to ensure that the number of available Taylor series terms yield sufficient numerical precision for our purposes.} to $z_0$, and uses $z_1$ as the next base point to repeat the whole process with. One repeats this procedure until the target point $z_t$ has been reached. However, this iteration becomes problematic if we stray too close to a pole of $u(z)$, as the Taylor series will not converge fast enough so as to maintain the required numerical precision. Perhaps this would not be too much of a problem if we were to know \textit{a priori} where the poles are, but the whole idea is precisely to use a numerical method that does not require such input. This issue was solved in \cite{fw11, fw14} by adding two additional steps to the iterative solving process. In the first step, convert the above Taylor series into a Pad\'{e}-approximant
\be
\text{P}_{K} [u_K] (z) = \frac{\sum_{k=0}^{K/2} a_k \left(z-z_0\right)^k}{\sum_{k=0}^{K/2} b_k \left(z-z_0\right)^k } + \CO \left( \left(z-z_0\right)^{K+1} \right),
\ee
\noindent
where its poles will approximate the poles in the vicinity of the base point $z_0$. In practice this has two benefits. First, one obtains high-precision approximations of pole locations (which are to be checked against partition-function transseries predictions). Second, the Pad\'{e}-approximant will be an accurate approximation to $u(z)$ in a much larger domain than the original Taylor series, even if the base point is close to a pole. We can then numerically solve the string equation in question on much larger domains than we could have using Taylor series, while also obtaining high-precision approximations to the pole locations. There is yet a second step, as \cite{fw11, fw14} observe that the accumulation of numerical errors can be further mediated by not getting too close to any poles in the iterative process. To implement this requirement, at every recursive step evaluate $\text{P}_{K} [u_K] (z)$ at a point near $z_0$ upon a line between $z_0$ and $z_t$, as well as at a couple of small angles, and then choose the next base point to be the one where $\abs{\text{P}_{K} [u_K] (z)}$ is the smallest. Keep repeating this process until one reaches a base point which is sufficiently close to $z_t$ (albeit it is not a good idea to force our algorithm to try to reach $z_t$ itself, as it could happen to be very close to a pole, which would imply we would accumulate a lot of numerical error in the last few steps). The output of this process is then a Pad\'{e}-approximant of the specific-heat $u(z)$ around the target-point $z_t$. We have solved string-equations on a domain in the $z$-plane by initializing a lattice of target points and carrying out the above process for each. This yields a Pad\'{e}-approximant for each target point, which we can then use to approximate the specific-heat anywhere on the solution domain. Having \cite{fw11, sv22} used this method to get high-precision approximations to \PI~solutions,  we will herein use this method on both \PI~and \YL~equations.

\subsection{Examples of Local Solutions: Hermitian Matrix Models}
\label{subsec:matrix-model-numerics}

Recall the cubic \eqref{eq:CubicMatrixModelPotential} and quartic \eqref{eq:QuarticMatrixModelPotential} matrix models, whose (nonperturbative) features were discussed at length in subsections~\ref{subsec:OP-phases} (\textit{e.g.}, their oscillations across different phases, as in figures~\ref{fig:CMMrNDataAndOPRoots} and~\ref{fig:QMMrNDataAndOPRoots}) and~\ref{subsec:SG-phases} (\textit{e.g.}, their corresponding phases, as in figures~\ref{fig:SpecGeofig:eigcmm}-\ref{fig:cubics-spectral-networks} and~\ref{fig:SpecGeofig:eigqmm}-\ref{fig:quartic-spectral-networks}, and phase diagrams, as in figures~\ref{fig:CMM Phase Diagram} and~\ref{fig:QMM Phase Diagram}). What we set out to solve back then was to construct an analytical large-$N$ resurgent-transseries which could reproduce all these features; basically amounting to reproducing the behavior of the orthogonal-polynomial coefficients in their different phases. As we shall unfold, our transseries construction matches the ``raw data'' locally, that is, on individual Stokes wedges, but we will need Stokes data to construct global solutions \cite{krsst26b}.

\begin{table}
\centering
\begin{tabular}{|c | c c c c c c c c c|} 
 \hline
 $n_1$ & 0 & 1 & 2 & 3 & 4 & 5 & 6 & 7 & 8 \\ [0.5ex] 
 \hline
 $g_{\text{max,cubic}}$ & 70 & 10 & 10 & 9 & 9 & 9 & 9 & 8 & 8 \\ 
 $g_{\text{max,quartic}}$ & 200 & 40 & 40 & 40 & 14 & 13 & 13 & 13 & 12 \\ 
 \hline
\end{tabular}
\caption{Instanton and genus orders up to which we have computed transseries data in the numerical checks for cubic and quartic models, presented in figures~\ref{fig: CMM failure unfixed} and~\ref{fig: QMM failure unfixed}. The quartic data up to $n_1=3$ was taken directly from \cite{csv15}, whereas the rest was produced in the present paper.}
\label{table: HMMs Available Data Table}
\end{table}

\begin{figure}[H]
	\centering
	\begin{subfigure}{.6\textwidth}
		\centering
		\includegraphics[width=0.9\textwidth]{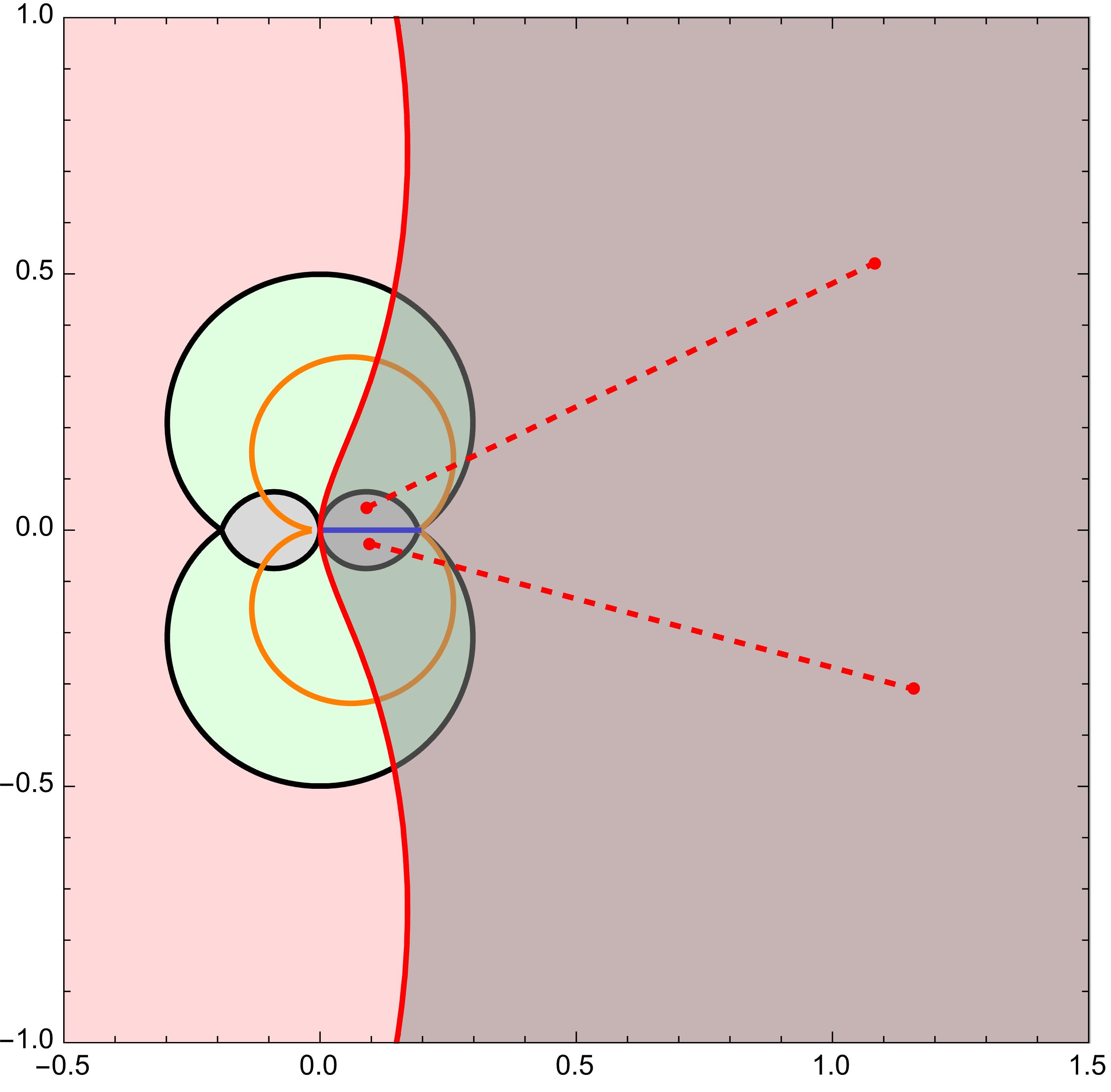}
		\caption{Choice of paths on the phase diagram.}
	\end{subfigure}
\break\\
    \begin{subfigure}{.49\textwidth}
		\centering    
		\includegraphics[width=0.9\textwidth]{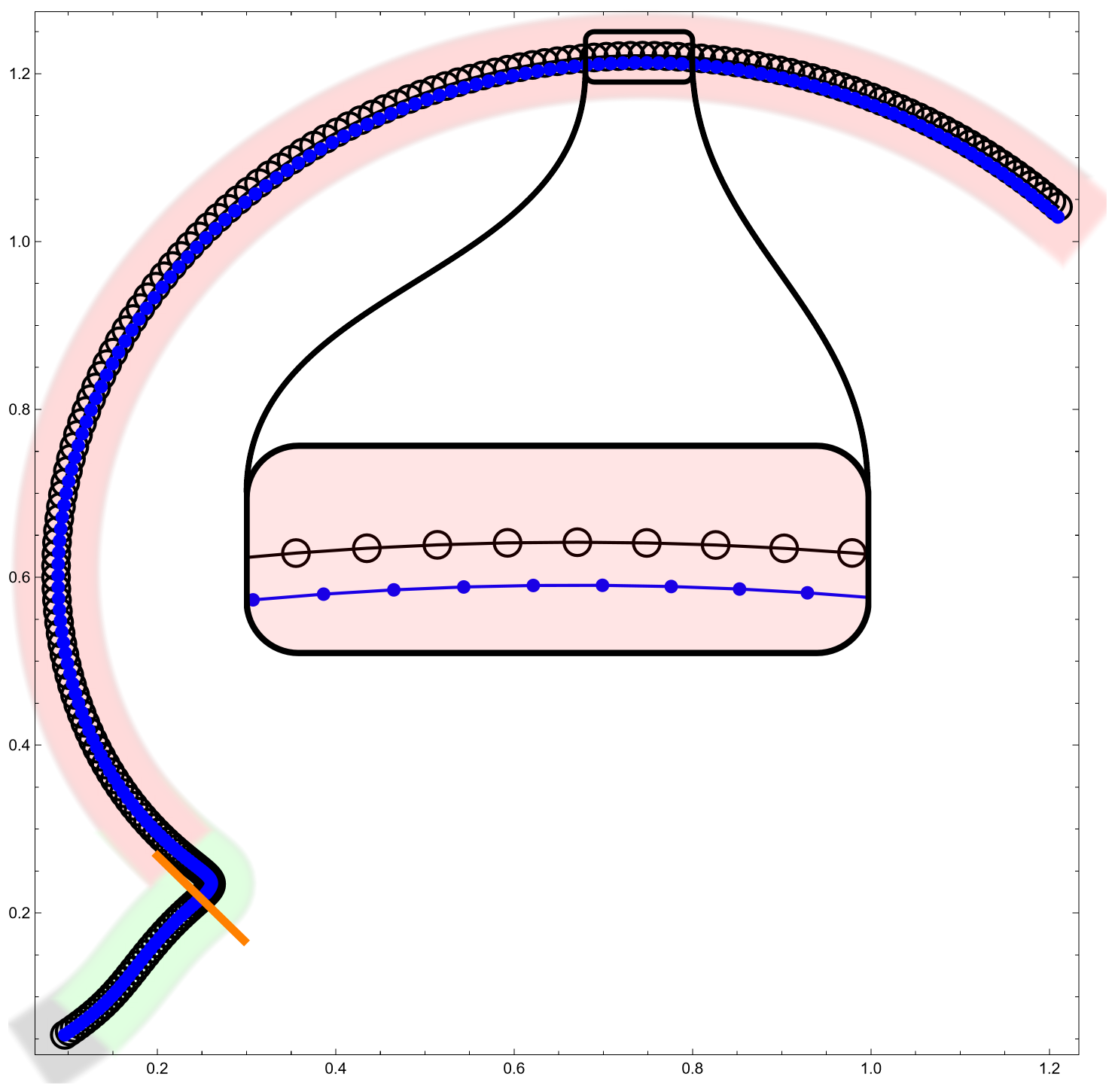}
		\caption{First-quadrant path.}
    \end{subfigure}       
    \begin{subfigure}{.49\textwidth}
		\centering
		\includegraphics[width=0.9\textwidth]{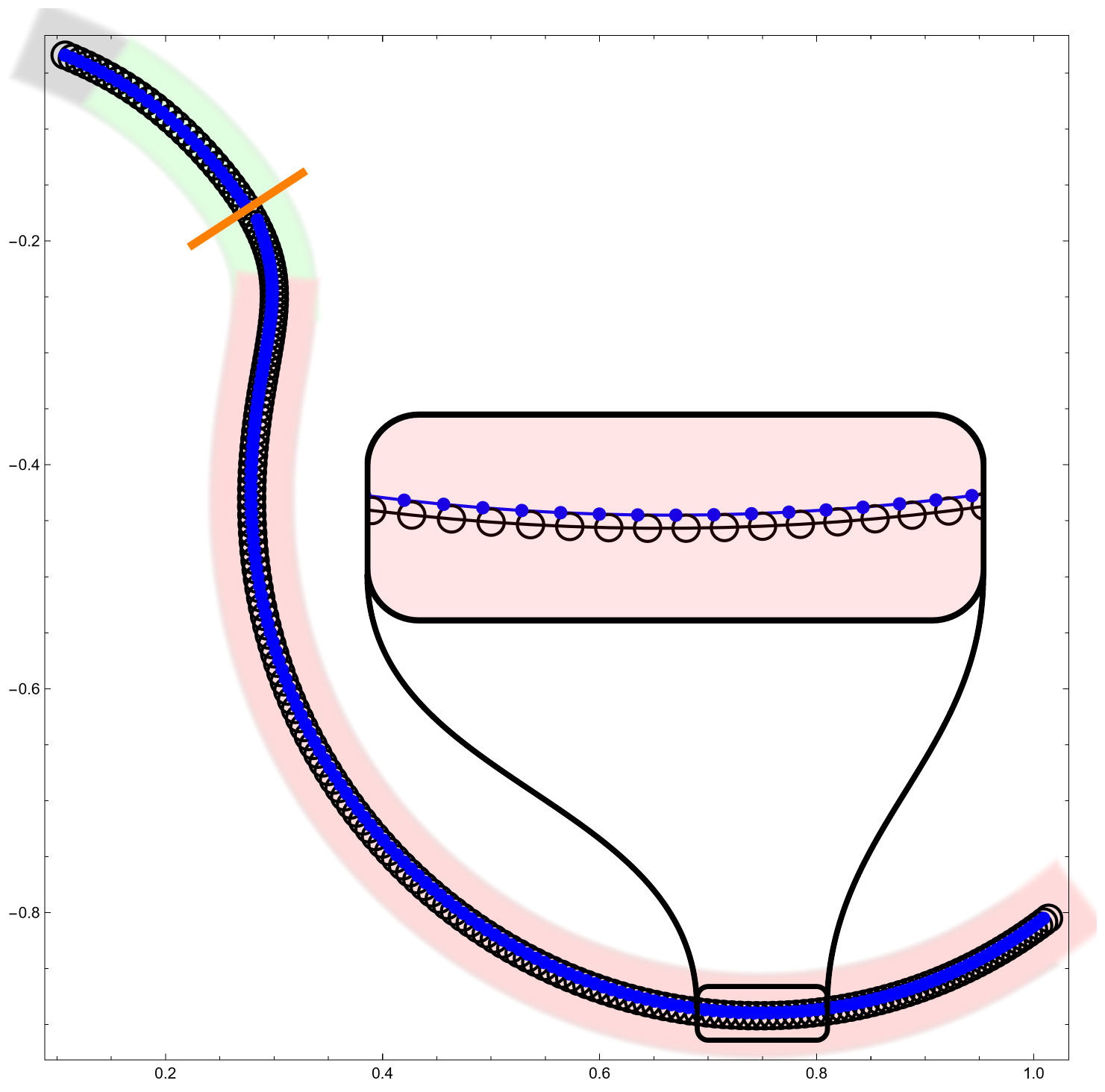}
		\caption{Fourth-quadrant path.}
	\end{subfigure}
	\caption{``Failure plots'' for the match between the cubic transseries and its $r_N$ data, for $N=3$, $\lambda=1$, and $(w_{\text{p}},w_{\text{np}})=(1,\frac{1}{80})$. The paths upon the complex $t$-plane along which we compute $r_N$ values are shown as \textcolor{red}{dashed red} lines on the phase diagram (top figure). Its \textcolor{gray}{darker gray} region (with \textcolor{red}{solid red} boundary) is the convergence region of the instanton sums, for our values of the parameters. On the two bottom figures we show $\Re$-$\Im$ plots of $r_N$ values (the black circles) versus the  transseries predictions (the \textcolor{blue}{blue dots}), for the two chosen paths. For ease of comparison, we add a ``phase shading'' nearby the $r_N$ values to indicate which phase the corresponding point is in. The match is excellent right up to the \textcolor{orange}{orange} Stokes line (illustrated in all plots), past which the transseries predictions visibly drift from the exact $r_N$ data.}
    \label{fig: CMM failure unfixed}
\end{figure}

\begin{figure}
    \centering
    \begin{subfigure}{.49\textwidth}
		\centering
		\includegraphics[width=0.9\textwidth]{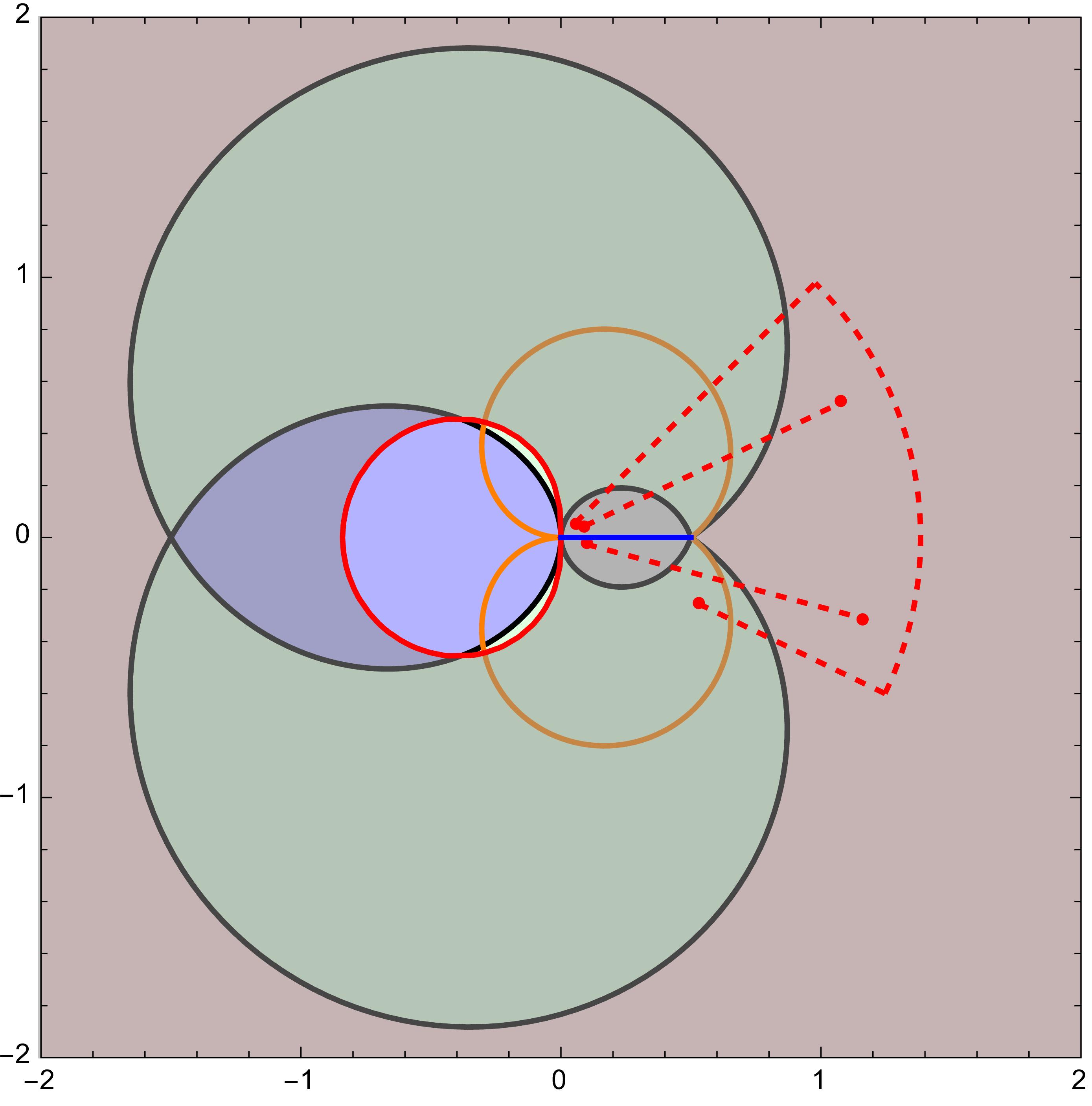}
        \caption{Choice of phase diagram paths.}
    \end{subfigure}
    \begin{subfigure}{.49\textwidth}
		\centering
		\includegraphics[width=0.9\textwidth]{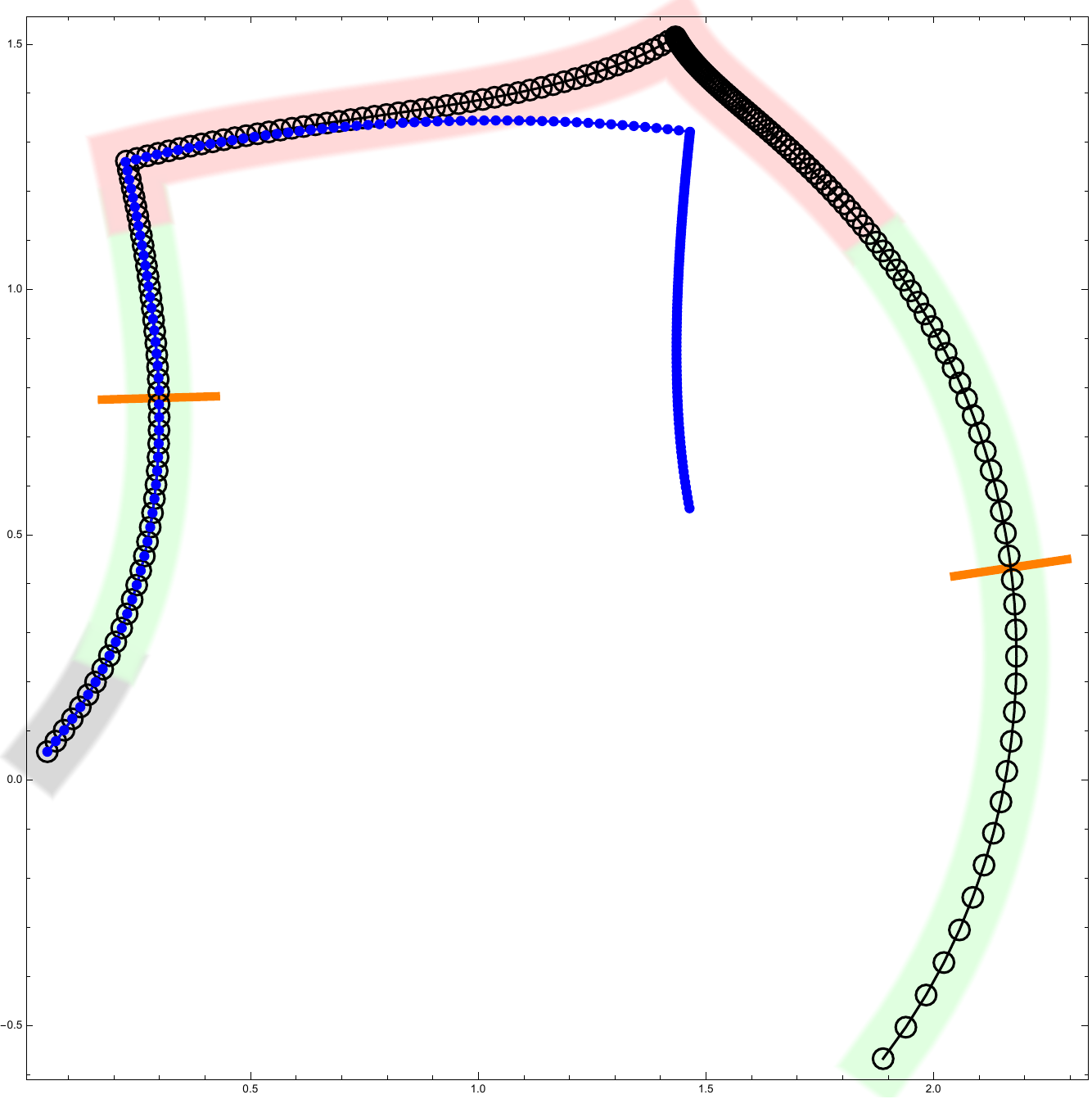}
        \caption{Mixed quadrants path.}
		\label{fig:QMMfailuremixedquadrants}
    \end{subfigure}
\break\\
    \begin{subfigure}{.49\textwidth}
		\centering
		\includegraphics[width=0.9\textwidth]{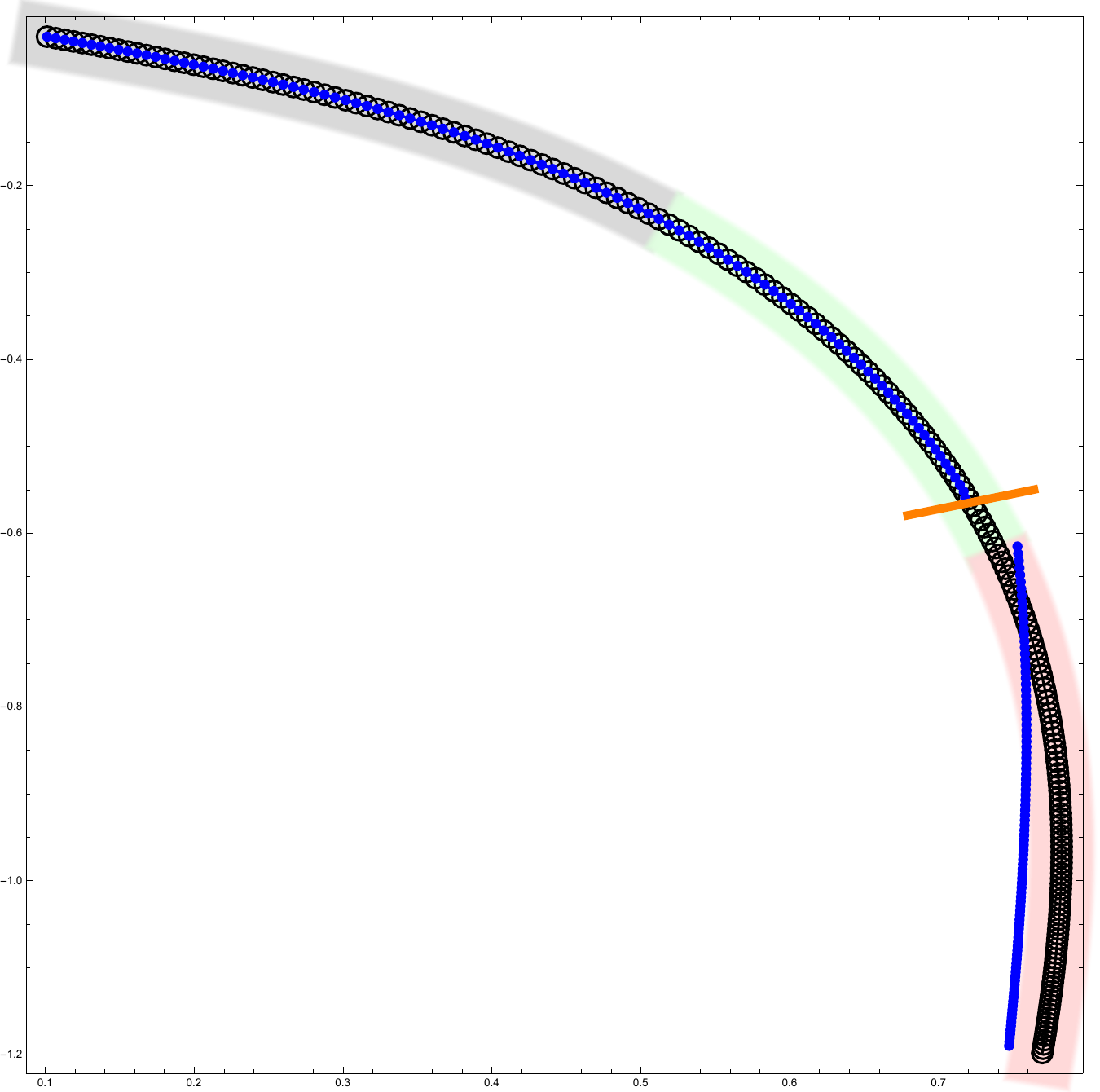}
        \caption{First-quadrant path.}
		\label{fig:QMM failure first quadrant}
    \end{subfigure}       
    \begin{subfigure}{.49\textwidth}
		\centering
		\includegraphics[width=0.9\textwidth]{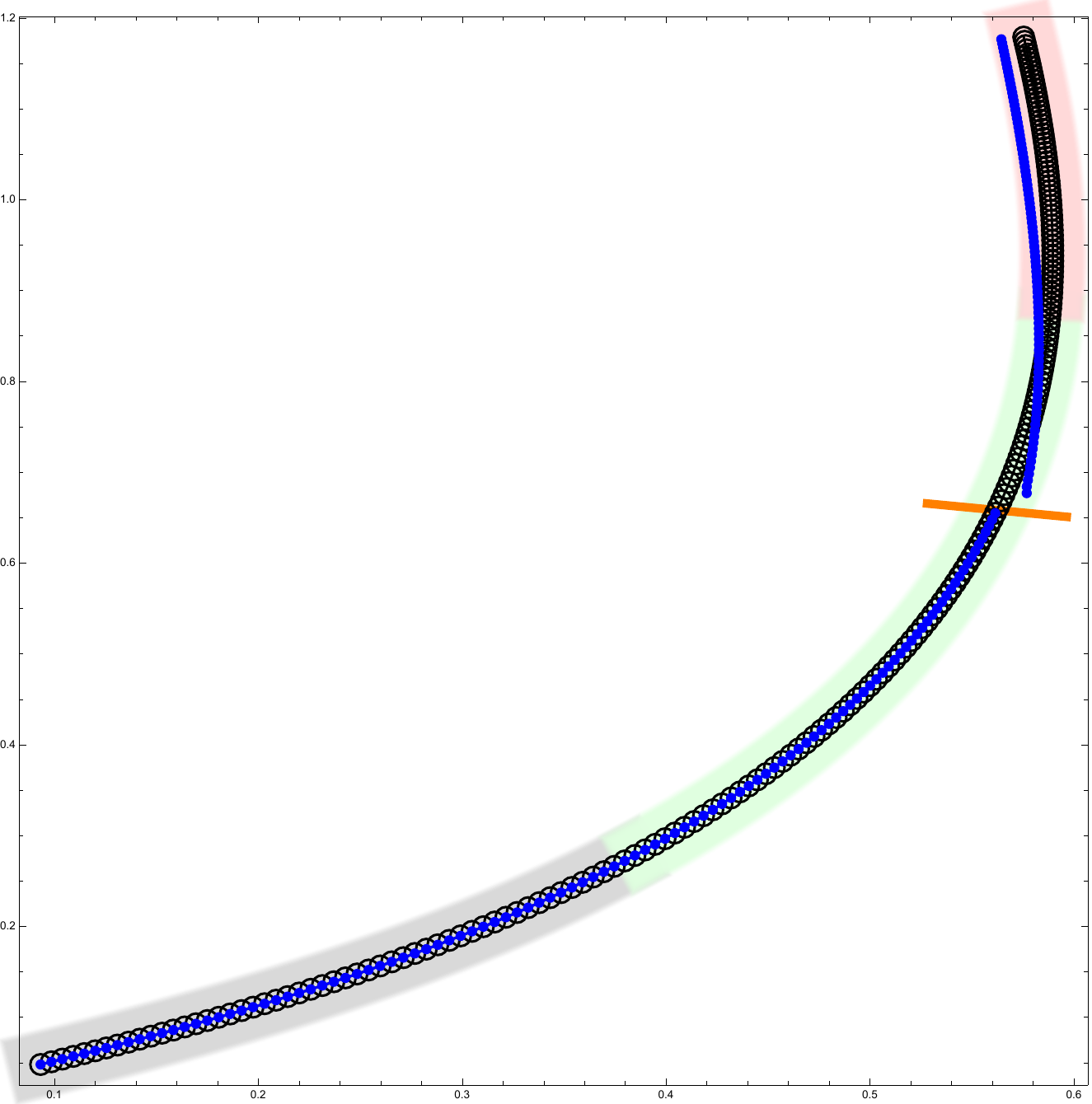}
        \caption{Fourth-quadrant path.}
		\label{fig:QMM failure fourth quadrant}
    \end{subfigure}
    \caption{``Failure plots'' for the match between the quartic transseries and its $r_N$ data, for $N=3$, $\lambda=1$, and $(w_{\text{p}},w_{\text{np}})=(1,10^{-5})$. The paths upon the complex $t$-plane along which we compute $r_N$ values are shown as \textcolor{red}{dashed red} lines on the phase diagram. Its \textcolor{gray}{darker gray} region (with \textcolor{red}{solid red} boundary) is the convergence region of the instanton sums, for our values of the parameters. On the remaining figures we show $\Re$-$\Im$ plots of $r_N$ values (the black circles) versus transseries predictions (the \textcolor{blue}{blue dots}). For ease of comparison, we add ``phase shading'' nearby the $r_N$ values to indicate which phase the corresponding point is in. The match is excellent right up to the \textcolor{orange}{orange} Stokes line (illustrated in all plots), past which the transseries predictions \textit{discontinuously jump} from the exact $r_N$ data, whose values in fact remain continuous.}
    \label{fig: QMM failure unfixed}
\end{figure}

Much like in \cite{csv15}, we begin with the one-parameter transseries (which one can trace back to \eqref{eq:twoparameterresurgenttransseriesforR} with $\sigma_2=0$). Our first checks involve computing $r_N(t)$ values, at fixed $N$ but varying $t$ across the several paths displayed in figure~\ref{fig: CMM failure unfixed} for the cubic matrix model, and figure~\ref{fig: QMM failure unfixed} for the quartic matrix model. As we are not addressing any Stokes data nor Stokes transitions, these checks must \textit{necessarily} eventually lead to mismatches (to be fully fixed in \cite{krsst26b}); in fact the whole point of the present section is to illustrate that our solutions cannot yet be complete precisely without these Stokes data. For the checks in the figures we have used the Borel--Pad\'e resummed $R$-transseries given in \eqref{eq:BorelResummedR}, with $N=3$, $\lambda=1$, and nonperturbative contour\footnote{Recall the cubic and quartic contours and their respective weights back from figure~\ref{fig:Cubic-AND-QuarticMatrixModelSteepestDescentContours}.} weights $w_{\text{np}}=1/80$ for the cubic and $w_{\text{np}}=10^{-5}$ for the quartic (with $w_{\text{p}}=1$ for both). Of course, we cannot match our transseries to finite $N$ data without specifying the transseries parameters $\boldsymbol{\sigma}$. We tested numerically that setting $(\sigma_1,\sigma_2)=(S_1 w_{\text{np}},0)$ in the one-cut region on the positive real line leads to a match for both cubic and quartic. This formula generalizes equation (3.10) of \cite{csv15} to arbitrary values of the nonperturbative contour weight. Fixing the transseries parameters in other wedges should then be done using Stokes automorphisms, which we reiterate will be given in the follow-up paper \cite{krsst26b}. The number of transseries-coefficients we used for each sector are shown in table~\ref{table: HMMs Available Data Table}. As stressed earlier, it is also very important to make sure we carry out these checks in the region of the complex 't~Hooft plane where the instanton sums converge\footnote{To give the reader some intuition on how this restricts where these checks may actually be carried out, recall how we illustrated the regions of convergence for varying values of $(N,\sigma_1)$ in figure~\ref{fig:HMM Convergence Regions}.}.

With all technical details specified, we now discuss our ``failure plots'' checks in figures~\ref{fig: CMM failure unfixed} and~\ref{fig: QMM failure unfixed}. We have selected several paths to follow within the complex 't~Hooft $t$-plane of the cubic and quartic matrix models, which clearly fall within the region of convergence of their respective instanton sums---and this is clearly illustrated in both figures~\ref{fig: CMM failure unfixed} and~\ref{fig: QMM failure unfixed}. In both examples, the match between finite $N$ and resurgent large $N$ is excellent within the one-cut (gray) region, where instantons are exponentially suppressed relative to the perturbative sector. As we cross into the anti-Stokes (green) regions (which is a two-cut phase for the cubic, and a symmetric three-cut phase for the quartic), standard positive-tension effects become exponentially enhanced and hence dominant. The match still remains excellent, albeit solely right up until the (backward) Stokes line is reached (illustrated by the orange lines throughout the several plots in figures~\ref{fig: CMM failure unfixed} and~\ref{fig: QMM failure unfixed}). The influence of this transition is most striking in the quartic-matrix-model checks shown in figures~\ref{fig:QMMfailuremixedquadrants}-\ref{fig:QMM failure first quadrant}-\ref{fig:QMM failure fourth quadrant}. In fact, we observe that the resurgent large $N$ predictions jump \textit{discontinuously} as the Stokes line is crossed, whilst on-the-other-hand the finite $N$ data remains continuous. This is completely in line with resurgence expectations: the effect of this Stokes line should be precisely to implement a Stokes transition which changes the transseries parameters $(\sigma_1,\sigma_2)$ describing our data from the resurgent large $N$ point-of-view. If we ignore such Stokes transitions, as we do herein, then the match \textit{must} fail. While this discontinuity is much less evident to the naked eye in the other figures, we can still observe that the discrepancy between finite $N$ and resurgent large $N$ gradually increases from the Stokes line onwards. We will show in our follow-up work \cite{krsst26b} that implementing Stokes transitions \textit{precisely fixes} all these mismatches. While we save the construction of the exact Stokes automorphisms for that future work, we can make one crucial comment herein, on the physical interpretation of these findings. The Stokes line beyond which our checks break down is a \textit{backward} Stokes transition, meaning that the instanton action that crosses the integration contour on the Borel plane is a negative-tension action. These checks then already indicate at this early stage that we need negative-tension effects to be included in order to make the precise match correct. Further observe that choosing a different path of analytic continuation towards the strong-coupling regime would not have changed any of this. After all, \textit{any} path from the one-cut regime (colored in gray) to the strongly-coupled regime (colored in pink) \textit{must} cross the orange Stokes line which turns-on negative-tension effects. 

Our numerical tests indicate that a one-parameter, positive-tension-only transseries is sufficient to \textit{fully} describe our finite $N$ ``raw data'' in \textit{certain} Stokes wedges. The next question is then to ask if it would be possible to generalize these tests, and check the full \textit{two}-parameter transseries \textit{without} using Stokes automorphisms. Indeed, from a purely transseries point-of-view, it is completely fine to go into the Stokes wedges we have looked at and switch-on negative-tension contributions by choosing a non-zero value for their weight, $\sigma_2$. However, it is not clear what finite $N$ ``raw data'' this transseries would describe. In the language of the Boutroux classification from subsection~\ref{subsec:DSL-phases}---adapted to the present off-critical context---the cubic and quartic (finite $N$) recursion coefficients $r_n$ appear to be ``tronqu\'{e}e solutions'' of their respective string equations, and it is not clear what (off-equilibrium) ``generic solutions'' are in this context.

\begin{figure}
    \centering
    \begin{subfigure}[b]{0.495\textwidth}
        \centering
        \includegraphics[width=\textwidth]{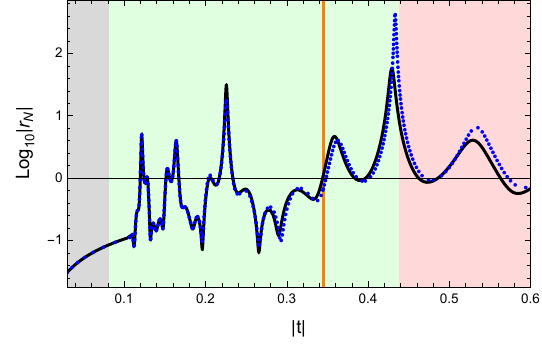}
        \caption{Cubic model: logarithmic norm.}
    \end{subfigure}
	\hfill
    \begin{subfigure}[b]{0.495\textwidth}
        \centering
        \includegraphics[width=\textwidth]{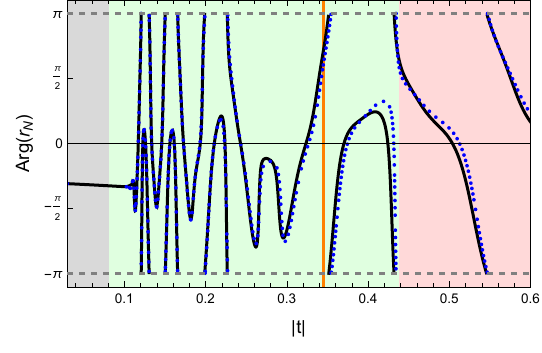}
        \caption{Cubic model: argument.}
    \end{subfigure}    
    \begin{subfigure}[b]{0.495\textwidth}
        \centering
        \includegraphics[width=\textwidth]{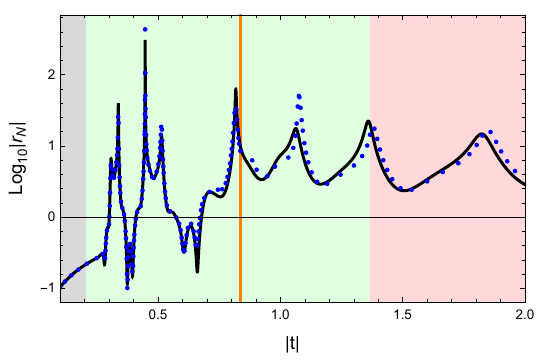}
        \caption{Quartic model: logarithmic norm.}
    \end{subfigure}
	\hfill
    \begin{subfigure}[b]{0.495\textwidth}
        \centering
        \includegraphics[width=\textwidth]{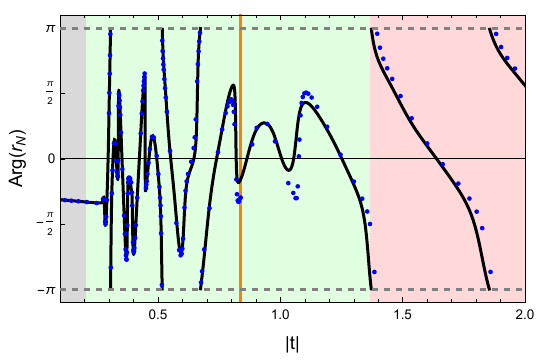}
        \caption{Quartic model: argument.}
    \end{subfigure}
    \caption{``Failure plots'' for the match between cubic and quartic transseries and their respective $r_{N} (t)$ data (compare with figures~\ref{fig:CMMrNDataAndOPRoots} and~\ref{fig:QMMrNDataAndOPRoots}). We used $N=20$, $\lambda=1$, $\arg(t)=-0.3\pi$ for both plots, and contour weights $w_{\text{np}}=10^{-8}$ for the cubic and $w_{\text{np}}=10^{-7}$ for the quartic. We plot $r_{20}(t)$ for varying $t$ as a black line, while the \textcolor{blue}{blue} dots represent large $N$ predictions as obtained via quadratic transasymptotics. We observe an excellent match on each plot, up to the \textcolor{orange}{orange} Stokes line, which then gradually becomes worse (due to both the lack of genus orders and Stokes data).}
    \label{fig:HMMFiniteNVsQuadraticTransasymptotics}
\end{figure}

Finally, let us go back to our opening discussion, illustrated in figures~\ref{fig:CMMrNDataAndOPRoots} and~\ref{fig:QMMrNDataAndOPRoots} from subsection~\ref{subsec:OP-phases} depicting the oscillations of the $r_n$ coefficients for the cubic and quartic matrix models. The capability that our local transseries has to reproduce those plots is shown in figure~\ref{fig:HMMFiniteNVsQuadraticTransasymptotics}, again for both cubic and quartic models; where in these plots we now compare our ``raw data'' with predictions from quadratic transasymptotics. We observe a very good match up to the Stokes line, which then gradually becomes worse as we go to stronger coupling. Such mismatch, of course, is due in part to not implementing Stokes automorphisms \cite{krsst26b}. These checks directly illustrate how our quadratic transasymptotics construction---and, by extension, our main proposal \eqref{eq:pinchedcubic}-\eqref{eq:pinchedcubic-dsl}-\eqref{eq:pinchedmulti}---not only produces exact, formal solutions to our models; but in fact it can very accurately predict the concrete numerical ``raw data'' we introduced back in section~\ref{sec:strong-coupling-phases}, in particular the oscillations of subsection~\ref{subsec:OP-phases}. Moreover, our quadratic-transasymptotics partition-function immediately\footnote{In spite of asymptotic, the $Z$-instanton-sums can be easily truncated and quickly applied to spit-out numerical outputs. In contrast, when the $R$-instanton-sums diverge, then they still require a resummation procedure.} produces meaningful predictions in parts of the 't~Hooft plane where the instanton sums of the $R$-transseries diverge---in fact, it can in principle be used \textit{everywhere} on the 't~Hooft complex plane, for any value of $N$ (albeit to varying degrees of precision, depending on availability of $D_k (\nu)$ polynomial data, as already commented in subsection~\ref{subsec:numerical-resummation-details}).

\subsection{Examples of Local Solutions: Double-Scaled Non-Critical Strings}
\label{subsec:dsl-string-eq-numerics}

Recall the \PI~\eqref{eq:Painleve1Equation} and \YL~\eqref{eq:YangLeeEquation} double-scaled non-critical string equations, whose (nonperturbative) features were discussed at length in subsection~\ref{subsec:DSL-phases} (\textit{e.g.}, their double-pole movable-singularities, as in figures~\ref{fig:PISolutionsNoZeros} and~\ref{fig:YLSolutionsNoZeros}). What we set out to solve back then was to construct an analytical large-$z$ resurgent-transseries which could reproduce all these features; basically amounting to reproducing the locations of these Painlev\'e-type singularities in different\footnote{Classified for \PI, as the tritronqu\'ee, tronqu\'ee, and generic solutions \cite{b13, b14}; still to be fully classified for \YL, with possible new solution-classes beyond the recently discovered tritronqu\'ee solutions \cite{gkk13}.} Boutroux classes. As we shall unfold, our transseries construction matches the ``raw data'' locally, that is, on individual Stokes wedges, but we will need Stokes data to construct global solutions \cite{krsst26b}.

\begin{table}
\centering
\begin{tabular}{|c | c |} 
 \hline
 $(n_1,n_2)$ & $g_{\text{max,\PI}}$ \\ [0.5ex] 
 \hline
 Up to $(5,0)$ and $(0,5)$ & 200 \\
 Up to $(5,5)$ & 100 \\
 \hline
\end{tabular}
\caption{Instanton and genus orders up to which we have computed transseries data in all our checks of the \PI~specific-heat transseries, presented in figure~\ref{fig:PIFailureSolutions}.}
\label{table: PI Available Data Table}
\end{table}

\begin{table}
\centering
\begin{tabular}{|c | c |} 
 \hline
 $(n_1,n_2,n_3,n_4)$ & $g_{\text{max,\YL}}$ \\ [0.5ex] 
 \hline
 Up to $(2,2,2,2)$, $(5,5,0,0)$ and $(0,0,5,5)$ & 100  \\
 Up to $(3,3,3,3)$ & 50  \\
 \hline
\end{tabular}
\caption{Instanton and genus orders up to which we have computed transseries data in all our checks of the \YL~specific-heat transseries, presented in figures~\ref{fig:YLFailureSolutionsNoStokesYesNeg} and~\ref{fig:YLFailureSolutionsNoStokesNoNeg}.}
\label{table: YL Available Data Table}
\end{table}

In the present double-scaled context, numerical contour-plots are obtained via the Fornberg--Weideman method \cite{fw11}, briefly described in subsection~\ref{subsec:numerical-dsl-details}, which accurately locates specific-heat poles (as illustrated in subsection~\ref{subsec:DSL-phases}). Analytic locations of these same specific-heat poles should be obtainable from the locations of the zeroes\footnote{Themselves related to zeroes of theta-functions as tabled in appendix~\ref{app:elliptic-theta-modular}.} of our exact partition-functions, for \PI~or \YL, \eqref{eq:PIDiscreteFourierNC} or \eqref{eq:YLFullDFT}. This is what we shall find in the following, with the sole caveat that the ensuing analysis will be purely \textit{local}, hence, necessarily limited (especially when we consider the effects of omitting negative-tension instantons). However, it will serve as a stepping stone toward the more refined \textit{global} reconstruction described in \cite{krsst26b}. To be fully explicit on what we have done, the transseries data\footnote{This very same data is also required to initialize the Fornberg--Weideman algorithm, translating between values of transseries parameters and the corresponding initial/boundary conditions of the multicritical string equations.} we computed for our examples is listed in tables~\ref{table: PI Available Data Table} for \PI, and~\ref{table: YL Available Data Table} for \YL.

\begin{figure}
\centering
\begin{subfigure}{0.325\textwidth}
    \centering
    \includegraphics[width=\textwidth]{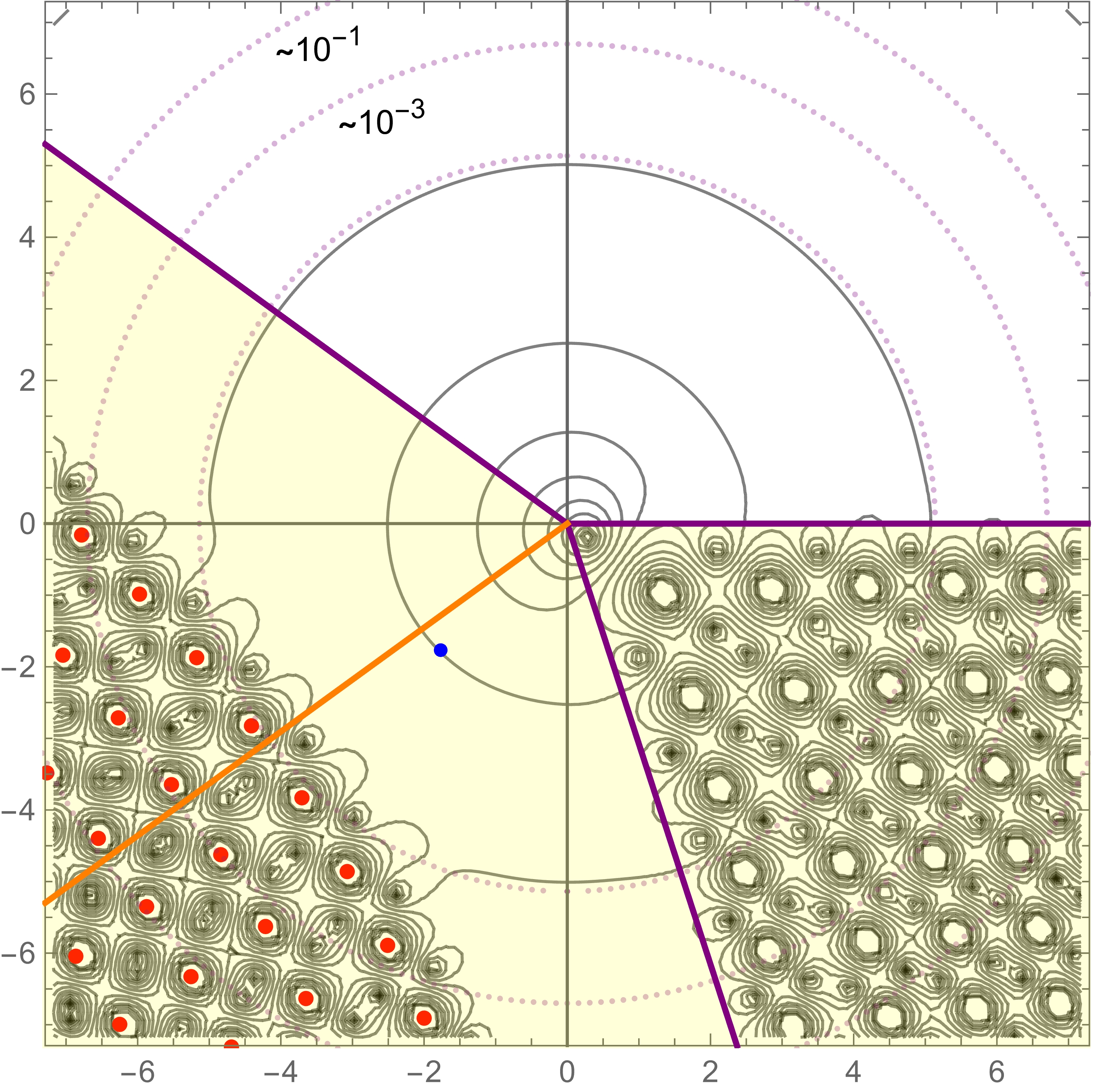}
    \caption{Tronqu\'ee: $z_0 = \frac{5}{2}\, \rme^{-\frac{3\pi\rmi}{4}}$, \\ $(\sigma_1,\sigma_2) \sim (10^{-8},-0.371)$.}
\end{subfigure}
\hfill
\begin{subfigure}{0.325\textwidth}
    \centering
    \includegraphics[width=\textwidth]{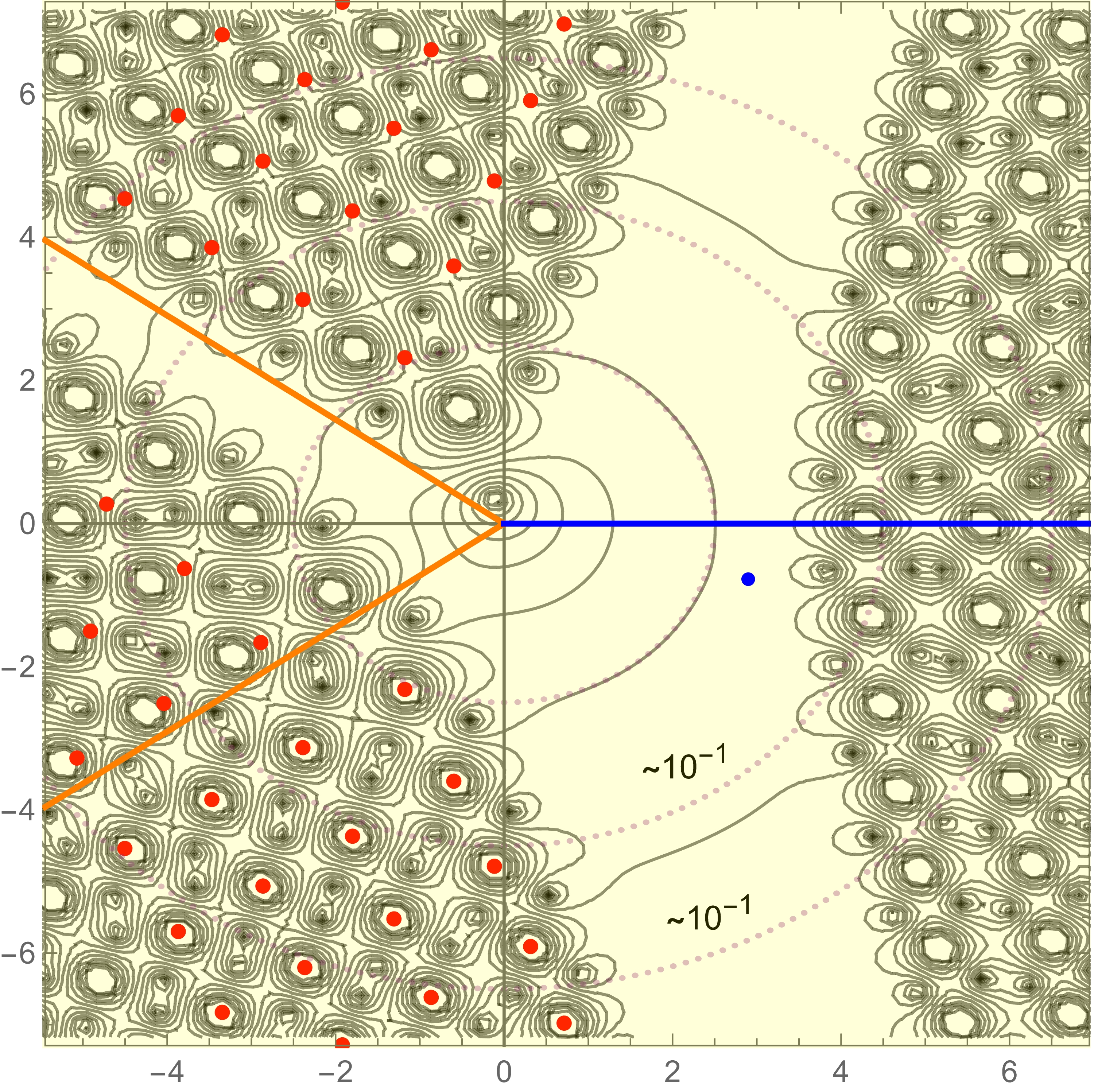}
    \caption{Generic: $z_0 = 3\, \rme^{-\frac{\rmi \pi}{12}}$, \\ $(\sigma_1,\sigma_2) = (10^{-2},10^{-6})$.}
\end{subfigure}
\hfill
\begin{subfigure}{0.325\textwidth}
    \centering
    \includegraphics[width=\textwidth]{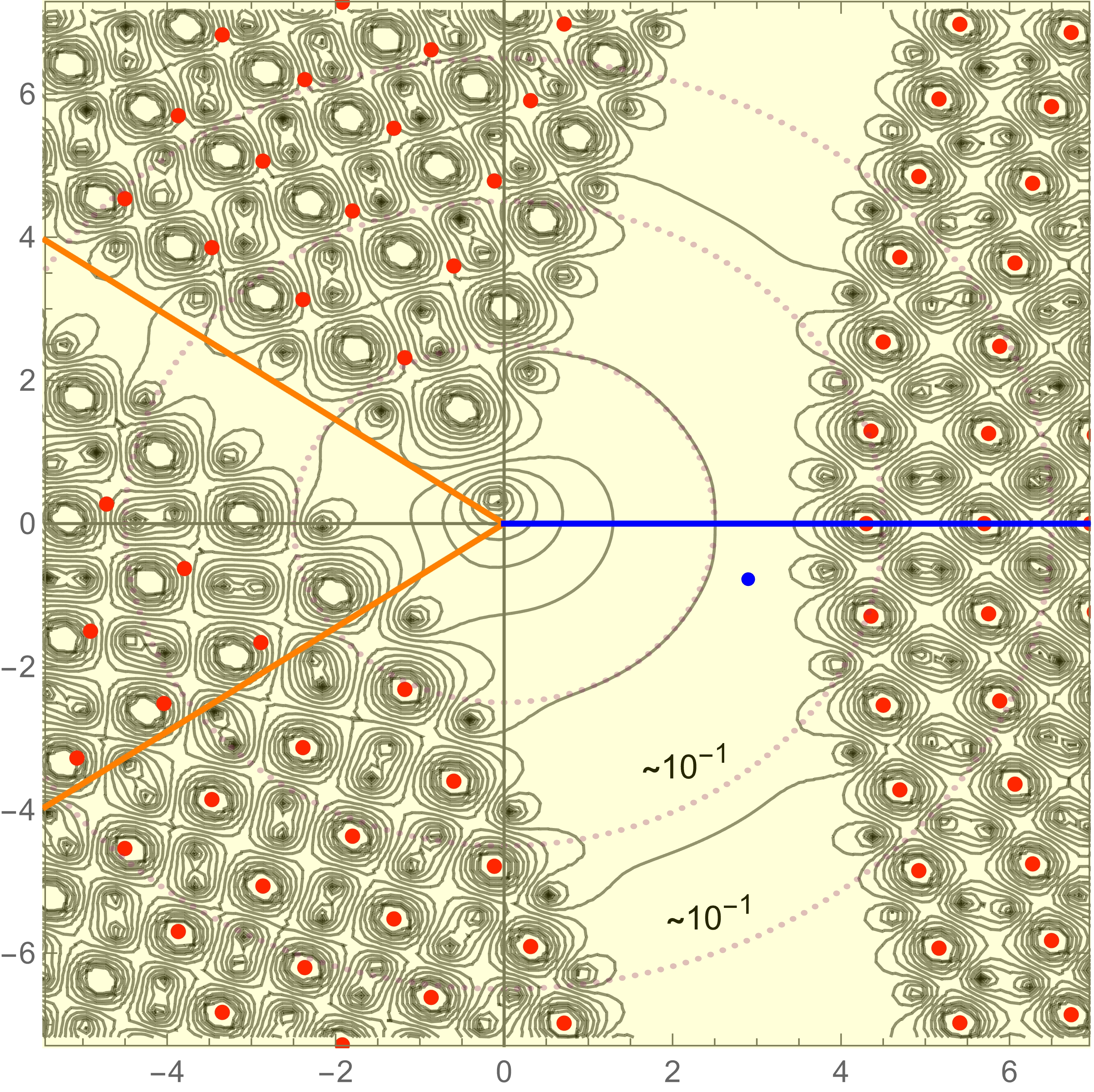}
    \caption{Generic: $z_0 = 3\, \rme^{-\frac{\rmi \pi}{12}}$, \\ $(\sigma_1,\sigma_2) = (10^{-2},10^{-6})$.}
\end{subfigure}
    \caption{Contour plots of numerical \PI~solutions, illustrating different Boutroux solution-classes accessible by our local transseries solutions (namely, without any implementation of Stokes automorphisms). As in figure~\ref{fig:PISolutionsNoZeros}, specific-heat poles appear as circles around which the contour-plot lines cluster. The initial point of the numerical evaluation is $z_0$ and is indicated by the \textcolor{blue}{blue} disk. Roots of the analytical partition-function are shown by the \textcolor{red}{red} dots. The anti-Stokes lines are highlighted in \textcolor{purple}{solid purple}, the \textcolor{blue}{blue} lines indicate forward Stokes transitions, and the \textcolor{orange}{orange} lines indicate backward Stokes transitions. The zeroes of the left-most plot are computed without considering negative-tension instantons, for the tronqu\'ee solution. The middle and right-most plots are without and with negative-tension instantons, respectively, resulting in a meaningful difference. These results indicate how there is only a partial match for tronqu\'ee and generic solutions, with the tritronqu\'ee solution not even being present (see the main text). We find that, due to the lack of any Stokes transitions, all solutions are at best only locally accurate. Interestingly enough, often, as shown in middle and left-most plots, the \textit{omission} of negative-tension instantons yields an even \textit{worse} match, with entire wedges of poles completely missing. However, one can still visually see the clear local validity of our transseries in describing these solutions upon inclusion of both positive and negative tension contributions. This is further \textit{quantitatively} clear by the matching-precision between analytical-zeroes and numerical-poles, given by the in-caption numbers (these are computed by averaging precision over the shown annuli, defined by the concentric circles displayed in \textcolor{purple}{dotted purple}).}
    \label{fig:PIFailureSolutions}
\end{figure}

We begin with \PI, for which the basic question is the same as always: how much (if at all) of the \textit{global} solution is our \textit{local} exact-transseries solution capable of reconstructing, if one does not implement Stokes automorphisms between one asymptotic wedge and the next, or if one simply ignored negative-tension instantons? Moreover, is there any evidence that this exact-transseries solution indeed gives us an exact \textit{local} solution to the problem at hand? Answers to both these questions may be obtained by carefully analyzing the contents of figure~\ref{fig:PIFailureSolutions} (which the reader may further compare with figure~\ref{fig:PISolutionsNoZeros}). The left-most plot shows a perfect agreement between the numerically-computed poles and the analytically-computed\footnote{Recall what we mean by this: computed by running the Newton method on a suitably truncated partition-function transseries, within the region(s) of interest on the complex plane.} zeroes from our transseries \textit{with} standard instantons but \textit{without} negative-tension instantons. This agreement only holds for the \textit{left wedge} of poles---indeed, there is not a single pole on the \textit{right wedge} which may be captured by our transseries. This shows how standard instantons alone\footnote{As already explained back in subsection~\ref{subsec:DSL-phases}, let us remark that although this transcendent is shown in the figure as a two-parameter transseries solution in the region near the initial-value point $z_0 = \frac{5}{2}\, \rme^{-\frac{3\pi\rmi}{4}}$, this is in fact still just a one-parameter transcendent, with transseries parameters $(\sigma_1,\sigma_2) = (10^{-8},0)$ (containing \textit{only} standard instantons) in the region past the backward Stokes line shown in the plot. This will be further discussed in \cite{krsst26b}.} are \textit{not enough} to reproduce this solution. In other words, this example shows how a physical tronqu\'ee solution, described in some region by a transseries with purely standard-instanton contributions, yields an \textit{incomplete} reconstruction of the full transcendent if one \textit{omits} negative-tension instanton contributions in a subsequent sector.

Moving-on to the remaining two plots in figure~\ref{fig:PIFailureSolutions}, it is evident how \textit{not} implementing Stokes automorphisms leads to a rather poor match for the poles in both left-middle and left-upper regions. On the other hand, the lower-left regions are still very-well described by our local transseries construction. Furthermore, the distinction between middle and right-most plots again illustrates how \textit{not} including negative-tension instantons leads to a complete lack of materialization of an entire wedge of poles. Finally, notice that the tritronqu\'ee plot is completely absent from these figures (recall figure~\ref{fig:PISolutionsNoZeros}). This is because \textit{none} of its poles are captured by our local transseries solutions. Even if we were to ignore the Stokes transitions and venture into neighboring Stokes wedges, we would fail to find any roots anywhere because, as we will see in \cite{krsst26b}, it requires \textit{negative} instantons to capture those poles.

One last point to address concerns the assessment we performed on the \textit{quantitative} agreement between the numerical-pole locations and the zeroes analytically-predicted by our transseries construction. This simply amounts to computing the absolute difference for each corresponding pole-zero pair; and then having these differences averaged over concentric annuli---illustrated by the purple dotted-circles in the plots of figure~\ref{fig:PIFailureSolutions}. Unsurprisingly, we found that although the left-most plot shows that the absolute errors lie in the range of $10^{-1}$ to $10^{-3}$, this is not the case for the remaining two plots which suffer from worse average-precision. This is clearly due to the large mismatch occurring in regions where zeroes are computed by transseries parameters which are incorrect in the sense that our constructions above are purely \textit{local}. However, in regions where zeroes are altogether absent (as in the left-most plot), due to no inclusion of negative-tension instantons, we get a much better precision. This reinforces the notion that the resurgent-transseries framework can overall capture the essential nonperturbative features of the solution, and hints at the dramatic improvements which are to be expected once the missing ingredients (the Stokes automorphisms and the negative-tension instantons) are included \cite{krsst26b}.

\begin{figure}[H]
\centering
\begin{subfigure}{0.45\textwidth}
    \centering
    \includegraphics[width=\textwidth]{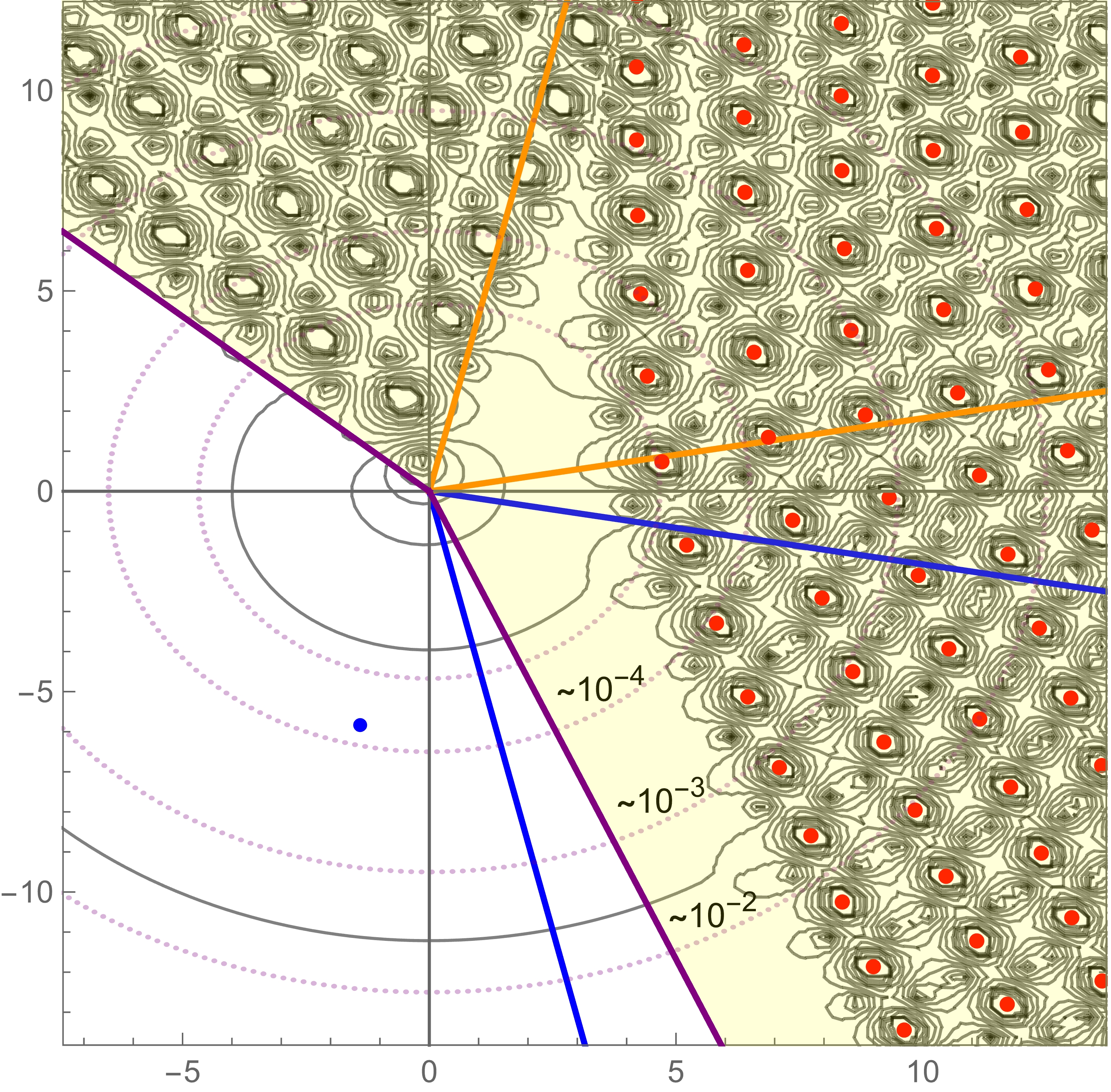}
    \caption{Tronqu\'ee: $z_0 = 6\, \rme^{- \frac{23\pi\rmi}{40}}$, \\ $\boldsymbol{\sigma} = \left( 0,0,0,\frac{1}{10^4} \right)$.}
\end{subfigure}
$\quad$
\begin{subfigure}{0.45\textwidth}
    \centering
    \includegraphics[width=\textwidth]{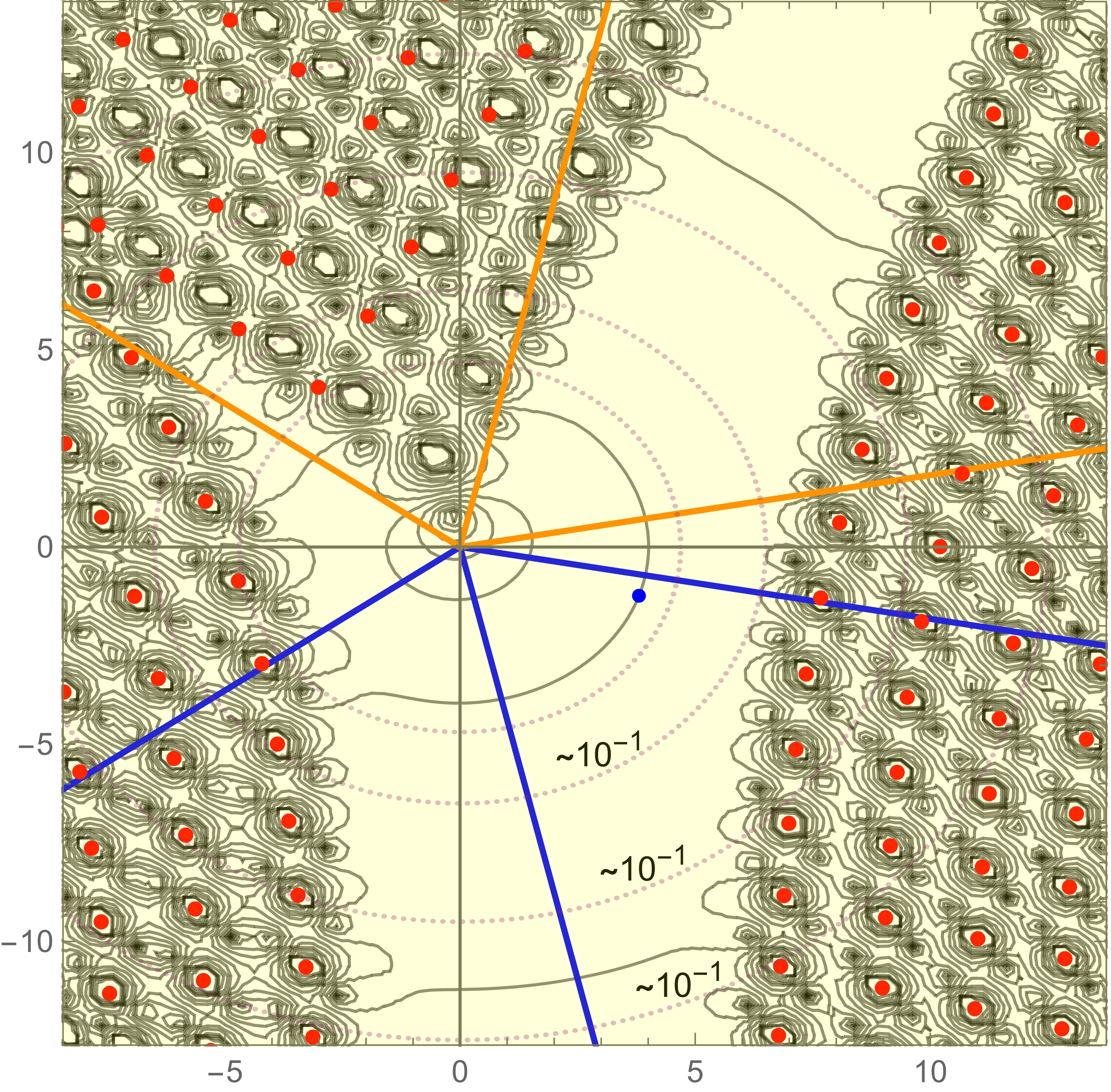}
    \caption{Two-parameter: $z_0 = 4\, \rme^{- \frac{\rmi\pi}{10}}$, \\ $\boldsymbol{\sigma} = \left( \frac{1}{10^4},\frac{1}{10^8},0,0 \right)$.}
\end{subfigure}
\break\\
\begin{subfigure}{0.45\textwidth}
    \centering
    \includegraphics[width=\textwidth]{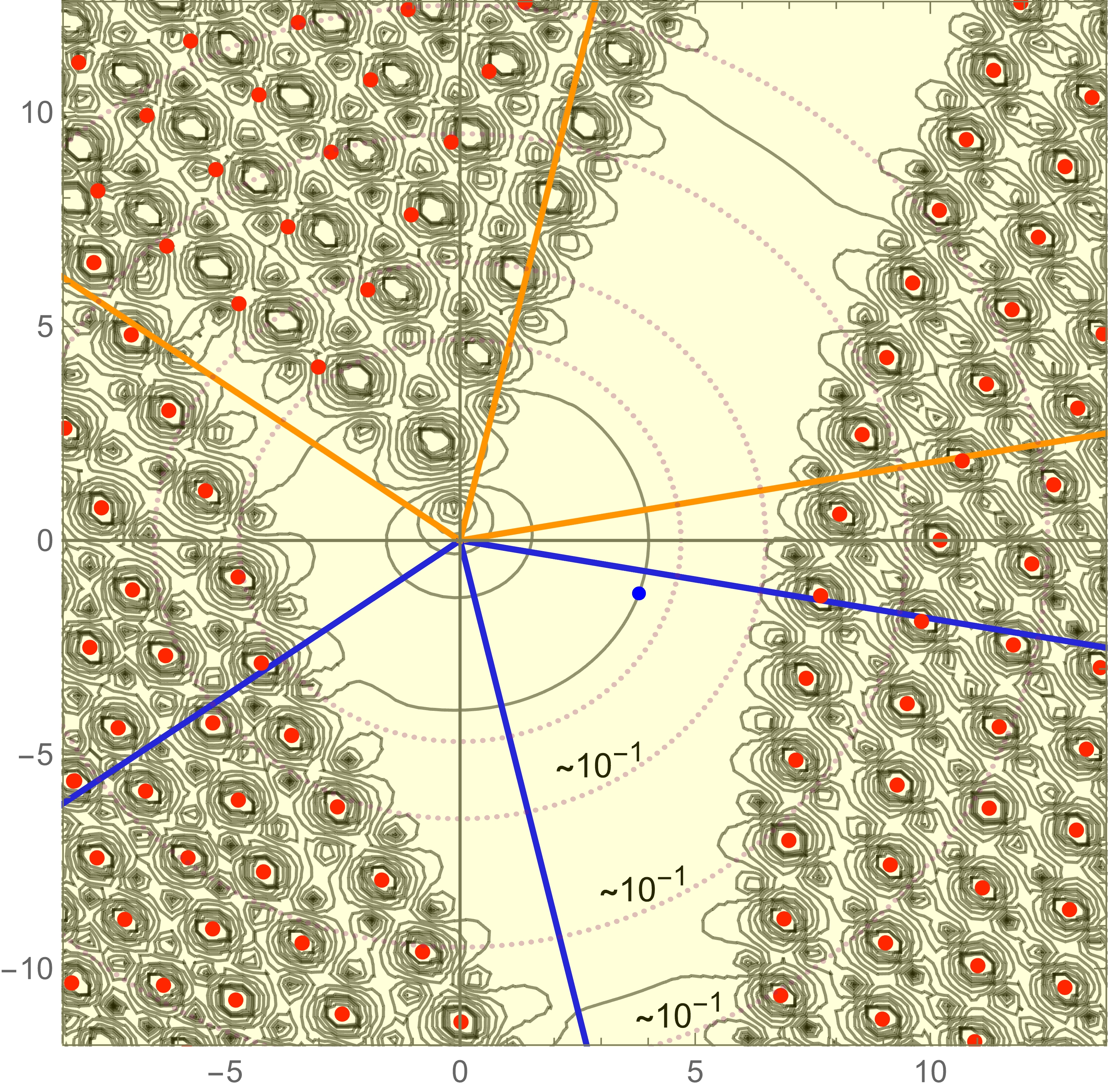}
    \caption{Three-parameter: $z_0 = 4\, \rme^{- \frac{\rmi\pi}{10}}$, \\ $\boldsymbol{\sigma} = \left( \frac{1}{10^4},\frac{1}{10^8},\frac{1}{10^5},0 \right)$.}
\end{subfigure}
$\quad$
\begin{subfigure}{0.45\textwidth}
    \centering
    \includegraphics[width=\textwidth]{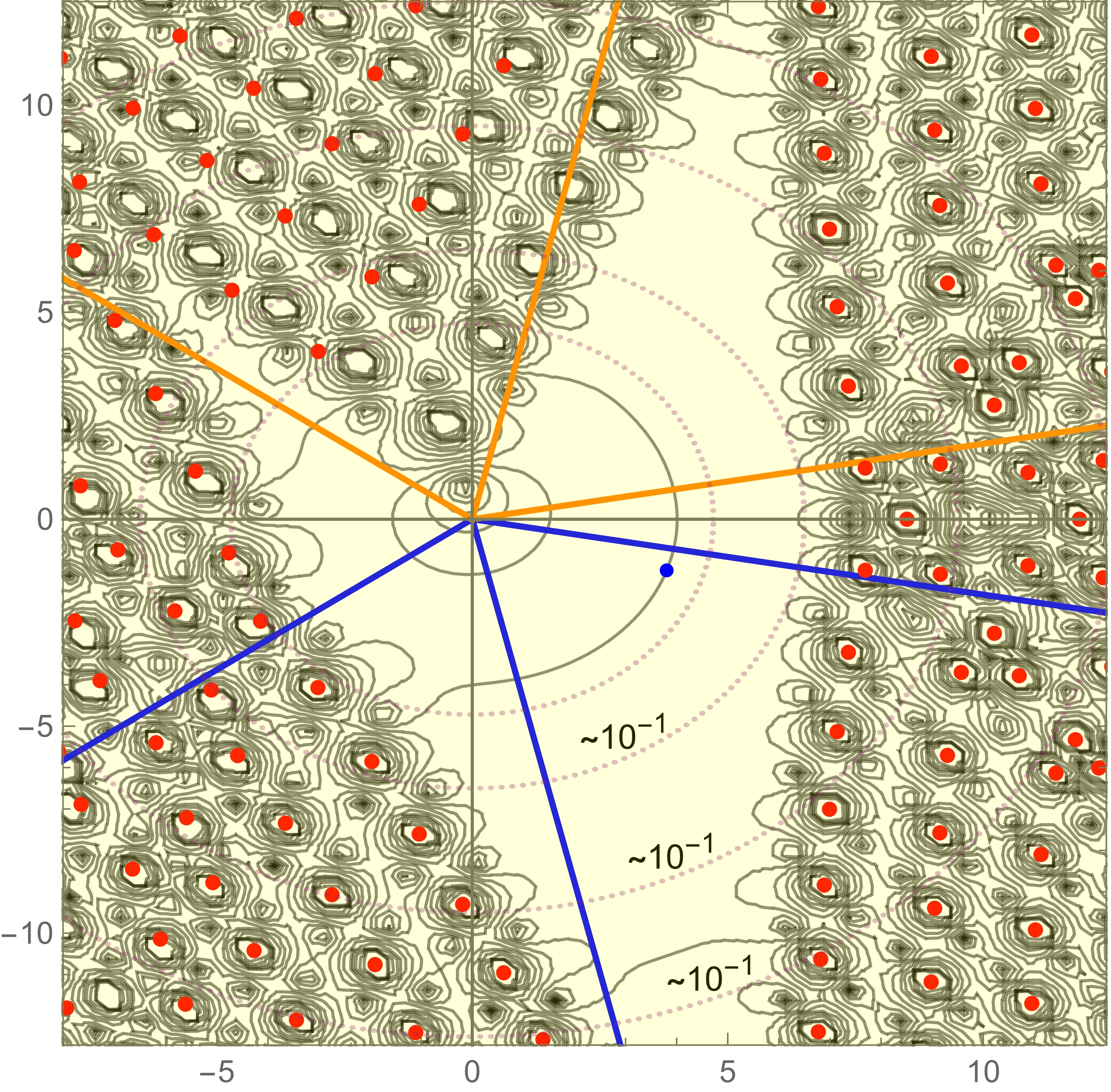}
    \caption{Four-parameter: $z_0 = 4\, \rme^{- \frac{\rmi\pi}{10}}$, \\ $\boldsymbol{\sigma} = \left( \frac{1}{10^4},\frac{1}{10^8},\frac{1}{10^4},\frac{1}{10^8} \right)$.}
\end{subfigure}
    \caption{Contour plots of numerical \YL~solutions. As before, specific-heat poles appear as circles around which the contour-plot lines cluster. The initial point $z_0$ for the algorithm is denoted by the \textcolor{blue}{blue} disk, and the analytical roots are shown as \textcolor{red}{red} dots. These zeroes were computed using a purely \textit{local} transseries; however, in this case, negative-tension instantons were included in the partition-function transseries-description. (In other words, we produced initial data with both positive- and negative-tension instantons, \textit{i.e.}, plots with $\sigma_2$ and/or $\sigma_4$ non-zero.) We indicate the anti-Stokes lines in \textcolor{purple}{purple}, the forward Stokes transitions in \textcolor{blue}{blue}, and the backward Stokes transitions in \textcolor{orange}{orange}. There is a clear match of pole-zero pairs within particular regions for each of the solutions. The in-caption numbers indicate the matching-precision between analytical-zeroes and numerical-poles (averaged over the dotted annuli regions shown in \textcolor{purple}{purple}).}
    \label{fig:YLFailureSolutionsNoStokesYesNeg}
\end{figure}

\begin{figure}
\centering
\begin{subfigure}{0.325\textwidth}
    \centering
    \includegraphics[width=\textwidth]{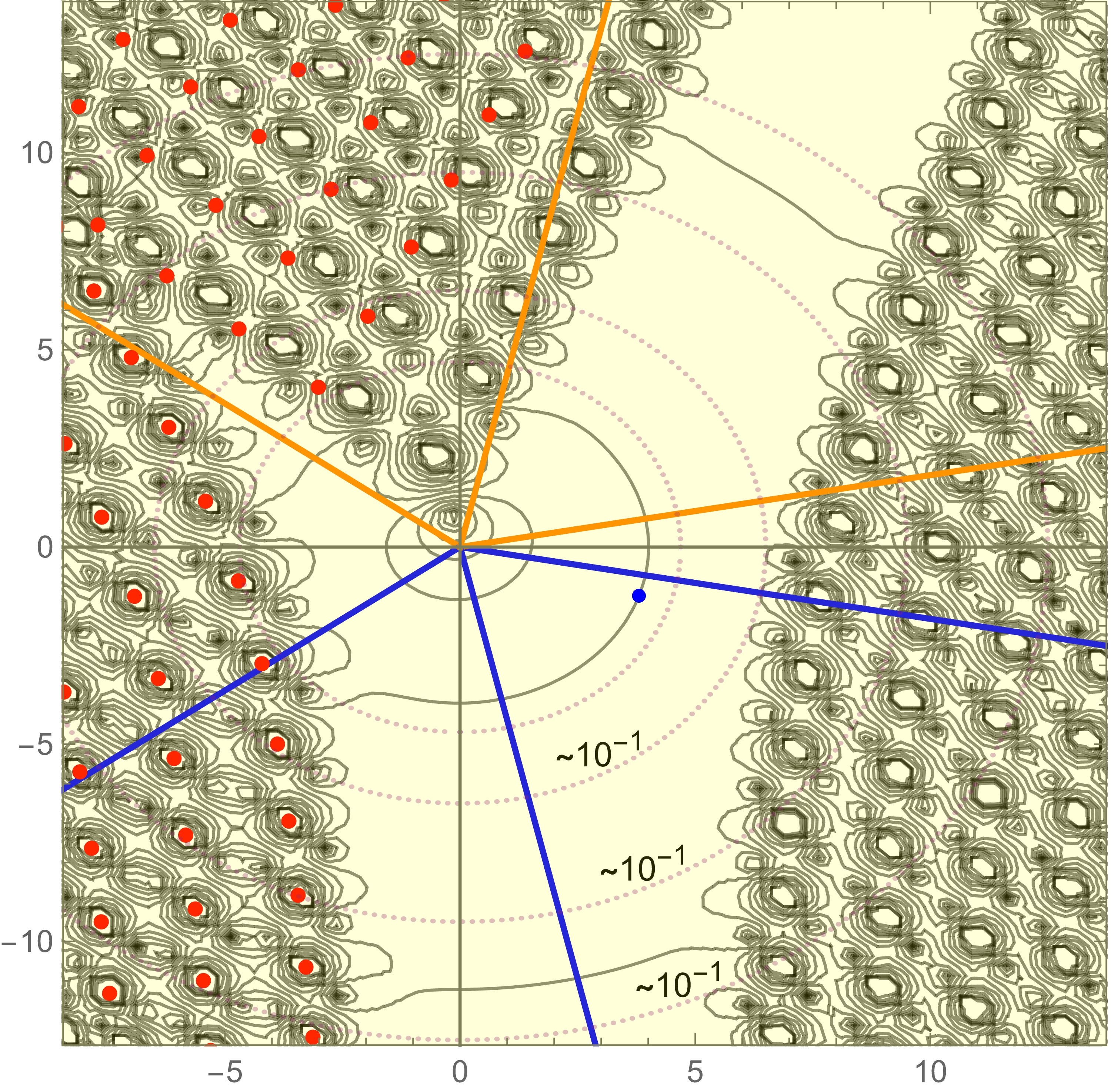}
    \caption{2-parameter: $z_0=4 \rme^{\frac{-\rmi \pi}{10}}$, \\  $\boldsymbol{\sigma}=\left(\frac{1}{10^4},\frac{1}{10^8},0,0\right)$}
\end{subfigure}
\hfill
\begin{subfigure}{0.325\textwidth}
    \centering
    \includegraphics[width=\textwidth]{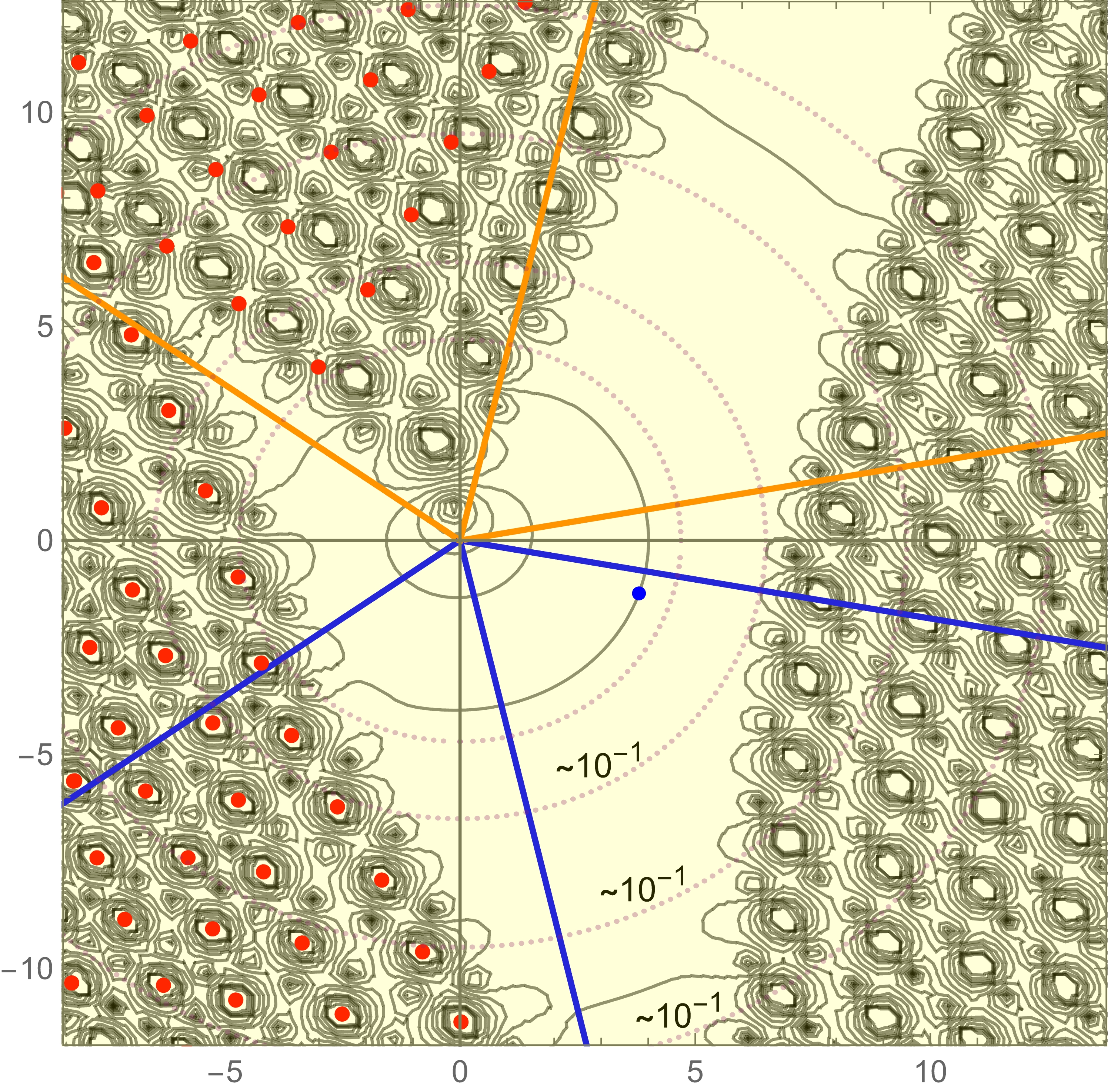}
    \caption{3-parameter: $z_0=4 \rme^{\frac{-\rmi \pi}{10}}$, \\  $\boldsymbol{\sigma}=\left(\frac{1}{10^4},\frac{1}{10^8},\frac{1}{10^5},0\right)$}
\end{subfigure}
\hfill
\begin{subfigure}{0.325\textwidth}
    \centering
    \includegraphics[width=\textwidth]{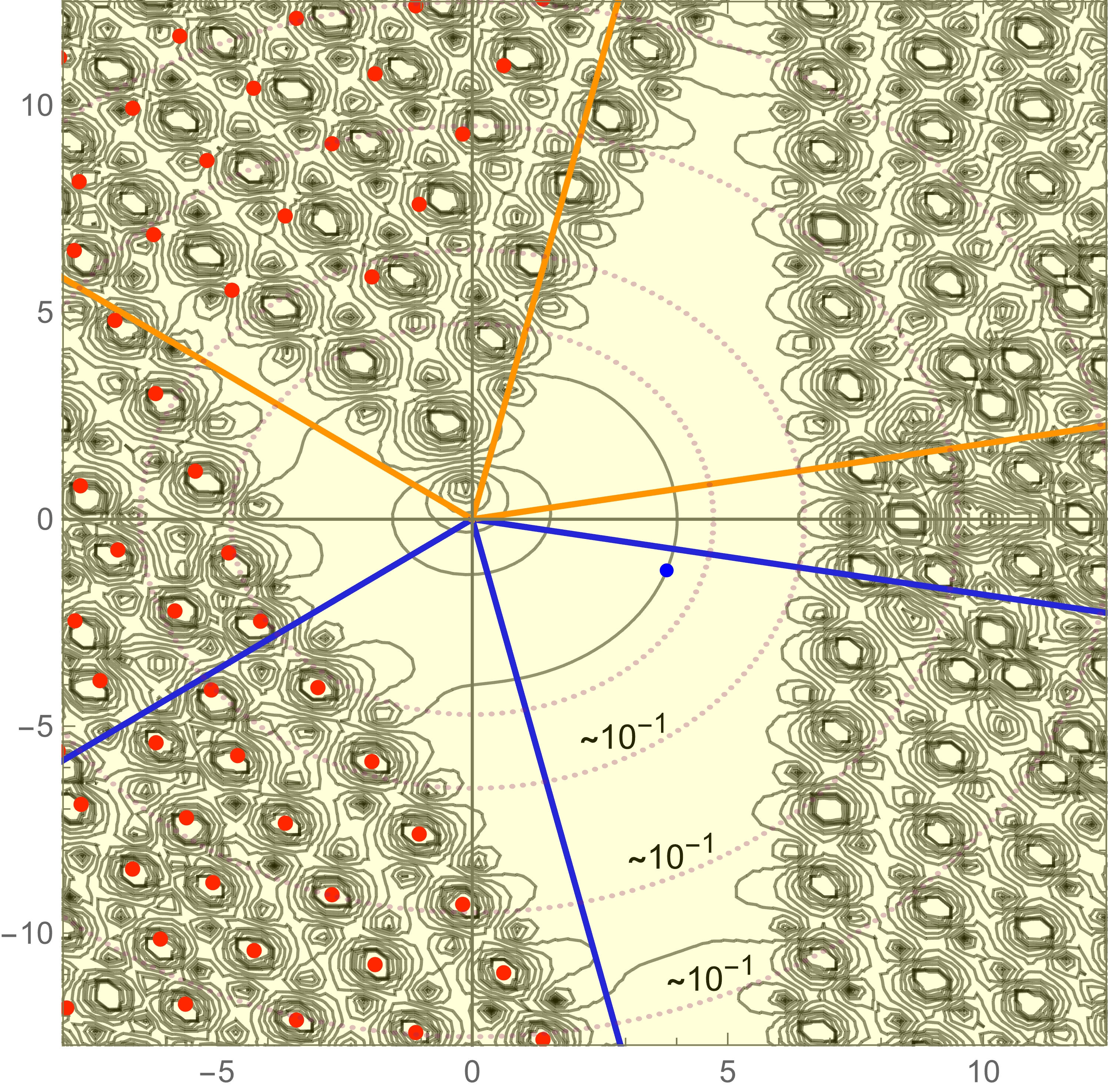}
    \caption{4-parameter: $z_0=4 \rme^{\frac{-\rmi \pi}{10}}$, \\  $\boldsymbol{\sigma}=\left(\frac{1}{10^4},\frac{1}{10^8},\frac{1}{10^4},\frac{1}{10^8}\right)$}
\end{subfigure}
    \caption{Contour plots of numerical \YL~solutions. As always, specific-heat poles are indicated by circles around which the contour-plot lines cluster, $z_0$ is denoted by the \textcolor{blue}{blue} disk, and the analytical roots are in \textcolor{red}{red}. The zeroes in these plots are still computed using a purely \textit{local} transseries, but now \textit{without} incorporating negative-tension instantons in the partition function. We indicate the anti-Stokes lines in \textcolor{purple}{purple}, the forward Stokes transitions in \textcolor{blue}{blue}, and the backward Stokes transitions in \textcolor{orange}{orange}. There is a clear match of pole-zero pairs within particular regions for each of the solutions. The in-caption numbers indicate the matching-precision between analytical-zeroes and numerical-poles (averaged over the dotted annuli regions in \textcolor{purple}{purple}).}
    \label{fig:YLFailureSolutionsNoStokesNoNeg}
\end{figure}

Let us next address \YL. Unlike \PI, which benefits from the well-established Boutroux classification \cite{b13, b14}, the \YL~equation is not as well understood. In particular, it exhibits richer and more intricate movable-pole structures, with many new types of transcendents; only a couple of which have already been classified \cite{gkk13}. Indeed, recall from subsection~\ref{subsec:DSL-phases}, in particular figure~\ref{fig:YLSolutionsNoZeros}, how our numerical contour-plots already partially demonstrated the existence of novel solution classes, as well as reproduced the recently understood type 1 and type 2 tritronqu\'ee solutions \cite{gkk13}. Our aim is now to present the first attempt at reconstructing all of these numerical results, via a local transseries approach, and in the process analyzing both successes and failures of this approach. As we saw for \PI, even a local solution with no negative-tension instantons is able to very accurately capture many (regions) of these transcendents. However, of course one will only obtain a complete, unequivocal, global match between the numerical evaluations and the analytical transseries, when one incorporates \textit{both} Stokes automorphisms \textit{and} negative-tension instantons into the transseries solution. Nonetheless, we can herein still demonstrate the partial match one finds when considering a local \YL~transseries, with and without negative-tension instantons. Again, our scenario is twofold: on the one hand, we obtain numerical contour-plots of the \YL~specific-heat solution via the Fornberg--Weideman \cite{fw11, fw14} method which accurately locate its poles; on the other hand, we employ the local partition-function transseries to analytically compute the corresponding zeroes. Illustrating all these partial matches is now more subtle for \YL~than it was for \PI. In fact, for \YL~there are more examples where there is a distinction between the plots in which the transseries only disregards Stokes automorphisms to those in which it also disregards negative-tension instantons. We illustrate the former in figure~\ref{fig:YLFailureSolutionsNoStokesYesNeg} while the latter is shown in figure~\ref{fig:YLFailureSolutionsNoStokesNoNeg}.

Since the structure of Stokes lines for \YL~is far more complicated than for \PI~(also discussed when comparing figure~\ref{fig:P1stokesAutomorphismsZplane} with figure~\ref{fig:YLstokesAutomorphismsZplane} back in subsection~\ref{subsec:DSL-phases}), we delay a more detailed analysis of \YL~plots to \cite{krsst26b}. Therein we shall see how a fully \textit{global} transseries, fully incorporating negative-tension instantons, will capture \textit{all} zeroes corresponding to numerical poles, for \textit{all} plots under consideration. In spite of that, there are however several things we can still understand herein: much like we saw for \PI, figures~\ref{fig:YLFailureSolutionsNoStokesYesNeg} and~\ref{fig:YLFailureSolutionsNoStokesNoNeg} now demonstrate that whilst the local transseries correctly captures certain features of the numerical-pole distribution, significant discrepancies remain in regions where Stokes automorphisms or negative-tension instantons or both are expected to play a decisive role. Both figures share this discrepancy---however, there is certainly a distinction between them: it is clear that not including negative-tension instantons in the local transseries leads to a huge drop in the number of numerical-poles which are correctly reproduced by the transseries analytical-zeroes. One can see entire wedges of accurately matching zero-pole pairs in figure~\ref{fig:YLFailureSolutionsNoStokesYesNeg} which are simply absent in figure~\ref{fig:YLFailureSolutionsNoStokesNoNeg}; demonstrating once again the clear necessity in including these nonperturbative contributions in our transseries. Finally, notice that there are some altogether absent plots as compared with what was initially tested back in figure~\ref{fig:YLSolutionsNoZeros}. This is simply because \textit{none} of their poles can be reproduce by a local transseries, which lacks their required Stokes automorphisms and/or negative-tension instanton contributions.

To complement the above visual comparisons, we have also performed a \textit{quantitative} error analysis. Like for \PI, we compute the absolute differences between the numerically determined pole locations and the analytical zeroes obtained from the local transseries resummation, averaged over concentric annuli. As before, we see that strong numerical agreements (ranging from $10^{-2}$ to $10^{-4}$) are to be found only in regions where misaligned transseries parameters are not present (whereas non-inclusion of negative-tension instantons simply yields a lack of zeroes, which thus does not spoil the average precision).

In summary, the \YL~analysis reinforce the conclusions drawn from the \PI~analysis: whereas the resurgent-transseries framework captures key nonperturbative features locally, a complete and globally consistent reconstruction of all solutions necessitates the inclusion of both Stokes automorphisms and negative-tension instanton contributions. We shall demonstrate in \cite{krsst26b} how once these ingredients are fully incorporated, any remaining discrepancies completely vanish, yielding a fully coherent description of all solutions for any values of the parameters. Nonetheless, the above partial matchings have still been highly instructive. On the one hand, the excellent local agreement in certain angular sectors reinforces our understanding that a resurgent transseries is indeed the correct tool to describe string-equation solutions (in fact, regardless of multicritical or off-criticality). On the other hand, the systematic misalignments observed in other regions clearly signal that a complete description requires the incorporation of both Stokes automorphisms (which allow us to continuously stitch together local transseries solutions across different asymptotic sectors) and negative-tension instanton sectors (which are essential in order to capture the full nonperturbative content). As we shall show in \cite{krsst26b}, it is only after the inclusion of \textit{both} of these ingredients that the remaining discrepancies vanish and a globally consistent match between numerical poles and analytic zeroes is achieved for these two cases (which would clearly generalize to the full KdV multicritical hierarchy and beyond \cite{krst26a, krst26b}). The tests in this section thus served a dual purpose: they provided further evidence that the transseries approach captures the correct asymptotic behavior for huge classes of solutions, while simultaneously highlighting the limitations of a naive purely-local reconstruction, or one that ignores the clear necessity of including \textit{negative-tension instantons} in our description of the solutions.

\begin{figure}
	\centering
	\includegraphics[width=.6\linewidth]{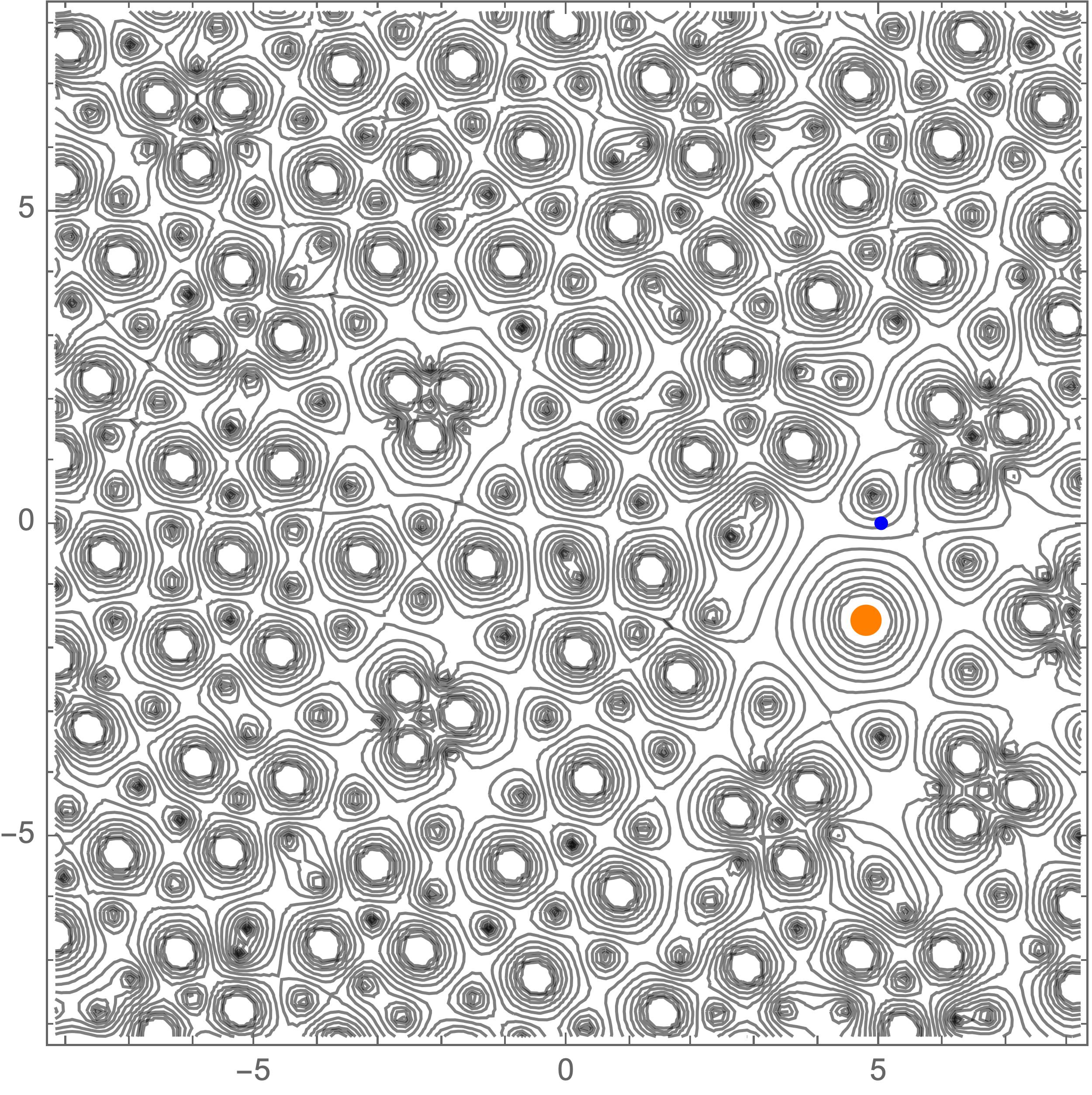}
	\caption{Numerical solution of \YL, with initial conditions set at $z_0 = 5$ through the Laurent expansion \eqref{eq:yangleeresidue6expansion}, with $z_p = 5\rme^{-\rmi\pi/10}$ and $\lambda_6=0=\lambda_8$. As always, this initial point is indicated in \textcolor{blue}{blue}. We can clearly see a distinctive singularity at $z=z_p$ as the \textit{large} circular contour-plot line which encircles it (shaded in \textcolor{orange}{orange}). This contour-plot line has a much larger radius than those showing the other pole locations, indicating its \textit{higher residue} value. This also makes clear that this type of singularity did not appear in other figures of this section.}
	\label{fig:yangleeresidue6}
\end{figure}

Let us end this section with a side-comment on the classes of solutions of the \YL~equation. Instead of constructing solutions via the use of resurgence and transseries, one may also construct them simply as Laurent power-series expansions around their movable double-pole singularities (this is reviewed for \PI~in \cite{bssv22}). For example, for \PI~one has
\be
u_{\text{\PI}} (z) = \frac{1}{\left(z-z_p\right)^2} + \frac{3z_p}{5} \left(z-z_p\right)^2 + \left(z-z_p\right)^3 + \lambda_4 \left(z-z_p\right)^4 + \frac{3z_p^2}{25} \left(z-z_p\right)^6 + \cdots.
\ee
\noindent
The two initial/boundary data for \PI~are now encoded in $(z_p,\lambda_4)$. This is the \textit{only} Laurent expansion that \PI~admits, \textit{i.e.}, all poles in the plots we have presented are double-poles with unit residue, abiding to the above expansion. Now, \YL~admits an analogous Laurent-series solution, which describes the local behavior of \textit{all} solutions we have presented so-far, around \textit{all} poles we have shown. It is
\bea
u_{\text{\YL}} (z) &=& \frac{2}{\left(z-z_p\right)^2} + \lambda_0 - \frac{3 \lambda_0^2}{2} \left(z-z_p\right)^{2} + \lambda_3 \left(z-z_p\right)^3 - \frac{5 \left(z_p+14\lambda_0^3\right)}{28} \left(z-z_p\right)^4  + \nonumber \\
&&
+ \frac{6\lambda_0\lambda_3-1}{8} \left(z-z_p\right)^5 + \lambda_6 \left(z-z_p\right)^6 - \frac{\lambda_0}{28} \left(z-z_p\right)^7 + \cdots,
\eea
\noindent
with initial/boundary data now encoded in the four parameters $(z_p,\lambda_0,\lambda_3,\lambda_6)$. Each pole of this type also gives rise to a \textit{simple zero}\footnote{Recall that simple partition-function roots come from residue $1$ poles for \PI~and residue $2$ poles in the \YL~setting. This factor of $2$ difference in residues is due to the factor of $2$ difference between their respective free-energy relations (recall, \textit{e.g.}, \eqref{eq:FdsEVENnormalization} and the discussion next to \eqref{eq:Painleve1Equation}).} of the partition-function. Now, as discussed in the footnote next to \eqref{eq:GDrecursionforpoles}, and very interestingly, the \YL~equation admits a \textit{second} expansion of this type \cite{gz90b}, but which is now given by
\be
\label{eq:yangleeresidue6expansion}
u_{\text{\YL}}(z) = \frac{6}{\left(z-z_p\right)^2} + \frac{5z_p}{252} \left(z-z_p\right)^4 + \frac{1}{24} \left(z-z_p\right)^5 + \lambda_6 \left(z-z_p\right)^6 + \lambda_8 \left(z-z_p\right)^8 + \cdots.
\ee
\noindent
As compared to the previous expression, we now only have \textit{three} initial/boundary data parameters $(z_p,\lambda_6,\lambda_8)$, and the double-pole residue is now $6$, hence giving rise to a partition-function with a \textit{triple root}. We stress that whilst one can readily check that \YL~admits this solution, \textit{none} of the plots we have presented actually feature a double-pole with this behavior. Of course, these solutions should be part of a Boutroux classification of string equation solutions. Further, one can always produce solutions of this type by simply developing the expansion \eqref{eq:yangleeresidue6expansion} to higher orders, numerically evaluating it as well as its first few derivatives at some base point, and then use those as initial conditions for our numerical solver. We present a solution obtained in this way in figure~\ref{fig:yangleeresidue6}. While this figure distinctively shows how these solutions do exist, it is not yet clear at this stage how to generate these solutions using the methods we have developed in this paper. As such, the problem of finding transseries-parameter values $(\sigma_1,...,\sigma_4)$ which give rise to a solution of this type is left for future work. Let us point out that the existence of several Laurent expansion solutions is not unique to \YL: instead, the $(2,2k-1)$ multicritical string string equation admits $k-1$ distinct Laurent expansions \cite{gz90b} that should all play a role in a classification of solutions of the type we are envisioning. The question of how to generate these solutions from the resurgent-transseries point-of-view is then a very relevant one.

\acknowledgments
We would like to thank David Berenstein, Marco Bertola, Bertrand Eynard, Alexander Goncharov, Alba Grassi, Qianyu Hao, Kohei Iwaki, Clifford Johnson, Ji Hoon Lee, Oleg Lisovyy, Raghu Mahajan, Marcos Mari\~no, Sebastiano Martinoli, Edward Mazenc, Ramon Miravitllas, Tomoki Nakanishi, Andy Neitzke, Jo\~ao Pimentel Nunes, Nicolas Orantin, Victor Rodriguez, Krishan Saraswat, David Sauzin, Marco Serone, Roberto Vega, Marcel Vonk, for useful discussions, comments and/or correspondence. RS would further like to thank In\^es Aniceto and Marcel Vonk, and to thank Ricardo Couso-Santamar\'\i a and Ricardo Vaz, for earlier unpublished collaborations related to the present topics. During those collaborations, RS further acknowledges Marco Bertola, Bertrand Eynard, Frank Ferrari, Marcos Mari\~no, Nicolas Orantin, Jo\~ao Penedones, Jo\~ao Pimentel Nunes, for useful discussions, comments and/or correspondence. JR would like to thank the University of California at Santa Barbara for extended hospitality, where parts of this work were conducted. JK was supported by the FCT-Portugal scholarship SFRH/BD/04742/2021 and by the CAMGSD scholarship BL178/2025-IST-ID. JR was supported by the FCT-Portugal scholarship UI/BD/151499/2021 and by the CAMGSD scholarship BL197/2025-IST-ID. MS was partially supported by the Swiss National Science Foundation under grant number 185723 and grant number 200021$\_$219267. NT was supported by a fellowship from ``la Caixa'' Foundation (ID 100010434) with code LCF/BQ/DI20/11780029 and by the CAMGSD scholarships BL297/2023-IST-ID and BL285/2024-IST-ID. This research was supported in part by CAMGSD/IST-ID via the FCT-Portugal grant UIDB/04459/2020 with DOI identifier 10-54499/UIDP/04459/2020. This research was partially funded by Funda\c c\~ao para a Ci\^encia e Tecnologia through grant UID/4459/2025. This paper is partly a result of the ERC-SyG project, Recursive and Exact New Quantum Theory (ReNewQuantum) funded by the European Research Council under the European Union's Horizon 2020 research and innovation programme, grant agreement 810573.

\newpage

\appendix

\section{Basics of (Hyper) Elliptic Curves and Theta Functions}
\label{app:elliptic-theta-modular}

Some elementary concepts of algebraic geometry and Riemann surfaces play an omnipresent role in our analyses, both at conceptual and computational levels. In order to keep this paper reasonably self-contained, we very briefly review these concepts in the present appendix, also setting up our conventions. For more details we refer the reader to some of the many introductions of different increasing depths available in the literature, \textit{e.g.}, \cite{olbc10, kz01, s04unpub, m.123.06, dk22, m.s07, bel12}.

Consider a \textit{lattice} $\BL = \left\{ m \upomega_A + n \upomega_B \left.\right| m,n \in \BZ \right\}$ over $\BC$, with basis $\upomega_A, \upomega_B \in \BC \setminus \{0\}$ and $\Im \frac{\upomega_B}{\upomega_A} \neq 0$ (which can be set $\Im \frac{\upomega_B}{\upomega_A} > 0$). This lattice naturally defines a torus, $\BT^2 \cong \BC/\BL$, where \textit{doubly-periodic} functions may be defined. If these doubly-periodic functions are meromorphic, \textit{i.e.}, if they are functions from the torus to the Riemann sphere $\BT^2 \to \BS^2 \cong \BC \cup \left\{ \infty \right\}$, then they are denoted by \textit{elliptic functions}. Changing the argument of an elliptic function by any element of $\BL$ leaves it unchanged, hence $\BL$ is also the set of \textit{periods} of the elliptic function. Generically, periods are complex numbers which may be written as integrals of rational functions, and which, as functions of their parameters, further satisfy \textit{Picard--Fuchs} linear differential equations. 

Riemann surfaces beyond the torus may be constructed as the image of multi-valued functions, and multi-valued functions may be constructed as solutions to algebraic equations. Consider an \textit{algebraic function} $y = y (x)$ defined via a degree-$n$ algebraic equation of the type $a_0 (x)\, y^{n} + a_1 (x)\, y^{n-1} + \cdots + a_n (x) = 0$ with $a_i (x)$ polynomials in $x$ with complex coefficients, and the analytic function $y(x)$ subsequently a multi-valued function of $x$. Our work solely requires degree $2$, denoted by \textit{hyperelliptic curves}, which we focus upon in the following. For such curves there is a change of variables which brings them to the form $y^2 = \left( x - e_1 \right) \left( x - e_2 \right) \cdots \left( x - e_k \right)$, with $\left\{ e_i \right\}$ the $k$ branch points. If the branch points are all distinct, the genus of the hyperelliptic curve is $g= \frac{1}{2} \left( k-1 \right)$ if $k$ is odd, and $g= \frac{1}{2} \left( k-2 \right)$ if $k$ is even (notice that if $k$ is odd then $\infty$ is also a branch point); and we may think of this algebraic curve as two copies of $\BS^2 \cong \BC \cup \left\{ \infty \right\}$ glued together along the $g+1$ branch cuts in-between the branch points.

Akin to the torus, hyperelliptic Riemann surfaces also have periods. At genus $g \ge 1$ there are $2g$ topologically distinct cycles on the surface, split into types $A$ and $B$ cycles just like in the torus, and hence leading to $2g$ ``fundamental'' periods. These cycles may be arranged into a canonical homology basis $\{ A_i, B_j \}$, with $i,j=1,\ldots,g$, satisfying
\be
A_i \cap A_j = 0 = B_i \cap B_k, \qquad A_i \cap B_j = \delta_{ij}.
\ee
\noindent
On top of this, the number of linearly independent holomorphic one-forms is $g$, given by
\be
\frac{x^j\, \rmd x}{y}, \qquad j = 0, \ldots, g-1,
\ee
\noindent
which may be normalized into a canonical basis $\left\{ \boldsymbol{\omega}_1, \ldots, \boldsymbol{\omega}_g \right\}$ satisfying
\be
\label{eq:omega-a-omega-b-period}
\oint_{A_i} \boldsymbol{\omega}_j = \delta_{ij}, \qquad \oint_{B_i} \boldsymbol{\omega}_j = \boldsymbol{\tau}_{ij}.
\ee
\noindent
Herein, the \textit{period matrix} $\boldsymbol{\tau}_{ij}$ of the hyperelliptic curve\footnote{In the elliptic case, the period ``matrix'' is simply the period \textit{modulus} $\tau \equiv \frac{\upomega_B}{\upomega_A}$ with $\Im \tau > 0$. In this case, two lattices $\BL$, $\BL'$ are similar iff
\be
\tau' = \frac{a \tau + b}{c \tau + d} \quad \text{ with } \quad \begin{bmatrix} a & b \\ c & d \end{bmatrix} \in \text{SL} (2,\BZ).
\ee
\noindent
In particular, the set of all complex structures which can be imposed on an elliptic curve is $\BH / \text{PSL} (2,\BZ)$, where $\BH = \left. \left\{ z \in \BC \,\right|\, \Im z > 0 \right\}$ is the upper-half plane and $\text{SL} (2,\BZ)$ the modular group.} is symmetric and $\Im \boldsymbol{\tau}_{ij}$ is positive definite. These integrals are in principle hard to evaluate directly, but, as already mentioned, there is an easier alternative approach to computing them. Depending on moduli\footnote{The moduli space of hyperelliptic curves of genus $g$ is $(2g-1)$-dimensional.}, iterated linear combinations of derivatives with respect to these moduli will eventually produce an exact form within the integrand, hence produce (Picard--Fuchs) linear differential equations (of at most order $2g$, and with coefficients solely moduli dependent) for the generic period integrals.

Going back to our lattice $\BL$, perhaps the simplest example\footnote{This is the \textit{doubly}-periodic analogue of the \textit{single}-periodic (trigonometric) function
\be
\pi \cot \pi x = \frac{1}{x} + \sum_{n \in \BZ \setminus \{0\}} \left( \frac{1}{x-n} + \frac{1}{n} \right).
\ee} of an elliptic function is the Weierstrass $\wp$-function: a doubly-periodic meromorphic even-function with a double-pole at the origin (in the fundamental domain of the lattice---due to double periodicity, this is all one needs to consider---albeit generically the double poles are at $z_{\text{dp}} = \BZ\, \upomega_A + \BZ\, \upomega_B$)
\be
\label{eq:wp-function-def}
\wp \left( z; \BL \right) = \frac{1}{z^2} + \sum_{\upomega \in \BL \setminus \{0\}} \left( \frac{1}{\left( z-\upomega \right)^2} - \frac{1}{\upomega^2} \right).
\ee
\noindent
This function satisfies the non-linear ordinary differential equation
\be
\label{eq:wp-function-ode}
\left( \wp' (z) \right)^2 = 4 \wp^3 (z) - g_2\, \wp (z) - g_3,
\ee
\noindent
where $g_2$ and $g_3$ are dictated\footnote{To be precise, $g_2 \equiv 60\, G_4 (\BL)$ and $g_3 \equiv 140\, G_6 (\BL)$, with $G_k (\BL)$ the Eisenstein series, but these are not directly relevant to our discussion hence we will leave them undefined.} by the lattice $\BL$, in which case it allows us to parametrize the cubic, or \textit{elliptic}, curve corresponding to the
torus. Indeed, a parametrization of the family of elliptic curves $y^2 = 4x^3 - g_2 x - g_3$ is immediately achievable via the Weierstrass function as $x \equiv \wp (z)$, $y \equiv \wp' (z)$, where $z$ is a \textit{uniformization parameter} for the Riemann surface specified by the pair $\left( x,y \right)$. The converse question is also of interest: if instead we are first given an elliptic curve, how does one recover its lattice of periods $\BL$? The holomorphic differential is simply given by $\rmd z = \frac{\rmd\wp}{\wp'} = \frac{\rmd x}{y}$ and the lattice is recovered from the curve via the periods $\upomega_{A,B} \equiv \oint_{A,B} \frac{\rmd x}{y}$.

Many interesting functions on an elliptic curve now follow. For instance, any elliptic function with respect to the lattice $\BL$ is a rational function of $\wp (z)$ and $\wp' (z)$. With direct relevance for our work, from the Weierstrass $\wp$-function one immediately defines the Weierstrass $\zeta$-function,
\be
\label{eq:wp-zeta-function-def}
\zeta' \left( z; \BL \right) = - \wp \left( z; \BL \right) \qquad \Rightarrow \qquad \zeta \left( z; \BL \right) = \frac{1}{z} + \sum_{\upomega \in \BL \setminus \{0\}} \left( \frac{1}{z-\upomega} + \frac{1}{\upomega} + \frac{z}{\upomega^2} \right).
\ee
\noindent
This function is no longer doubly-periodic but rather
\be
\zeta \left( z + m \upomega_A + n \upomega_B \right) = \zeta \left( z \right) + m\, \upeta_A + n\, \upeta_B,
\ee
\noindent
where $\upeta_{A,B} \equiv - \oint_{A,B} \frac{x\, \rmd x}{y}$. On top, one further defines the Weierstrass $\upsigma$-function,
\be
\label{eq:wp-sigma-function-def}
\upsigma \left( z; \BL \right) = \exp \int \rmd z\, \zeta \left( z; \BL \right) \qquad \Rightarrow \qquad \upsigma \left( z; \BL \right) = z \prod_{\upomega \in \BL \setminus \{0\}} \left( 1 - \frac{z}{\upomega} \right) \exp \left( \frac{z}{\upomega} + \frac{z^2}{2\upomega^2} \right),
\ee
\noindent
which is now quasi-periodic
\be
\upsigma \left( z + m \upomega_A + n \upomega_B \right) = (-1)^{m+n+mn}\, \rme^{\left( m\, \upeta_A + n\, \upeta_B \right) \left( z + \frac{1}{2} \left( m \upomega_A + n \upomega_B \right) \right)} \upsigma \left( z \right).
\ee
\noindent
Functions defined by \textit{elliptic integrals} also appear at several instances in this paper. The Legendre complete elliptic-integrals of the first, second, and third kind are, respectively:
\bea
\boldsymbol{K} (k) &=& \int_{0}^{1} \frac{\rmd t}{\sqrt{\left( 1 - t^2 \right) \left( 1 - k^2 t^2 \right)}}, \\
\boldsymbol{E} (k) &=& \int_{0}^{1} \rmd t \sqrt{\frac{1 - k^2 t^2}{1 - t^2}}, \\
\boldsymbol{\Pi} (\alpha,k) &=& \int_{0}^{1} \frac{\rmd t}{\left( 1 - \alpha^2 t^2 \right) \sqrt{\left( 1 - t^2 \right) \left( 1 - k^2 t^2 \right)}}.
\eea
\noindent
Herein, both $\boldsymbol{K} (k)$ and $\boldsymbol{E} (k)$ have branch-cuts on the complex $k^2$-plane from $1$ to $\infty$, $k$ is the \textit{elliptic modulus}, and $\alpha$ is the elliptic characteristic. For an elliptic curve explicitly written in terms of its branch points, as $y^2 = \left( x - e_1 \right) \left( x - e_2 \right) \left( x - e_3 \right) \left( x - e_4 \right)$, standard tables of integrals yield, along both $A$-cycles and the $B$-cycle, respectively,
\bea
\int_{e_1}^{e_2} \frac{\rmd x}{y} &=& - \frac{2 \rmi}{\sqrt{\left( e_1 - e_3 \right) \left( e_2 - e_4 \right)}}\, \boldsymbol{K} (k) = \int_{e_3}^{e_4} \frac{\rmd x}{y}, \\
\int_{e_2}^{e_3} \frac{\rmd x}{y} &=& \frac{2}{\sqrt{\left( e_1 - e_3 \right) \left( e_2 - e_4 \right)}}\, \boldsymbol{K} (k').
\eea
\noindent
Herein, the elliptic modulus is
\be
k^2 = \frac{\left( e_1 - e_2 \right) \left( e_3 - e_4 \right)}{\left( e_1 - e_3 \right) \left( e_2 - e_4 \right)},
\ee
\noindent
and the \textit{complementary modulus} $k'^2 = 1 - k^2$. As for the elliptic period modulus $\tau = \frac{\upomega_B}{\upomega_A}$, this is just the ratio of the above two integrals.

Tables of integrals are not always readily available; and most certainly not in the generic hyperelliptic problem where Picard--Fuchs equations need to grow to paramount importance. The simplest way to introduce and illustrate these equations is still within the elliptic realm, hence let us keep our focus on the Weierstrass family, now depending rationally on a single parameter $t \in \BC$. One expects that the two basic period integrals $\Uppi (t) = \oint \frac{\rmd x}{y}$ satisfy the second-order ordinary differential equation
\be
\frac{\rmd^2 \Uppi}{\rmd t^2} + a_1 (t)\, \frac{\rmd \Uppi}{\rmd t} + a_0 (t)\, \Uppi = 0,
\ee
\noindent
where $a_1 (t)$, $a_0 (t)$ are rational functions of $t$. These functions are determined by imposing that the one-form
\be
\boldsymbol{\eta} = \frac{\rmd^2 \boldsymbol{\omega}}{\rmd t^2} + a_1 (t)\, \frac{\rmd \boldsymbol{\omega}}{\rmd t} + a_0 (t)\, \boldsymbol{\omega}
\ee
\noindent
is closed (where $\boldsymbol{\omega} = \frac{\rmd x}{y}$), and, once fixed, the Stokes theorem immediately yields the Picard--Fuchs equations; \textit{e.g.}, in the particular example of \cite{s04unpub} this is
\be
\frac{\rmd^2 \Uppi}{\rmd t^2} + \frac{4t^3-1}{t \left( t^3-1 \right)}\, \frac{\rmd \Uppi}{\rmd t} + \frac{2t}{t^3-1}\, \Uppi = 0.
\ee
\noindent
In many examples the Picard--Fuchs equations have solutions of hypergeometric type. In the above example, a basis of solutions is given by a holomorphic function $\Uppi_1 (t)$ at $t=0$ (obtainable via power series) alongside $\Uppi_2 (t) = \Upomega (t) + \log t \cdot \Uppi_1 (t)$ (with $\Upomega (t)$ holomorphic at the origin).

One last player to introduce are \textit{theta-functions}. Generalizing some of our earlier discussion, consider now a higher-dimensional lattice, $\BL$, generated by $2g$ vectors in $\BC^g$ independent over $\BR$, and further consider the $g \times 2g$ matrix constructed from this basis of column vectors, $\boldsymbol{\Lambda}$. Generically, after some choices, one can write
\be
\label{eq:lattice}
\boldsymbol{\Lambda} = \left[ \boldsymbol{1} \; \boldsymbol{\tau} \right],
\ee
\noindent
where $\boldsymbol{1}$ is the $g \times g$ identity matrix and $\boldsymbol{\tau}$ is a $g \times g$ symmetric matrix with positive definite imaginary part. The set of all such matrices (denoted as Riemann matrices) is called the Siegel upper-half-space\footnote{When $g=1$ this is just the upper-half-plane $\BH \subset \BC$.} $\BH_g$,
\be
\label{eq:siegelupperhalfspace}
\BH_g = \left\{ \boldsymbol{\tau} \in \text{GL}(g,\BC) \, |\, \boldsymbol{\tau}^{\text{t}} = \boldsymbol{\tau} \, \wedge \, \Im \boldsymbol{\tau} > 0 \right\}.
\ee
\noindent
In this set-up, with $\BL = \BZ^g \oplus \boldsymbol{\tau}\, \BZ^g$, the quotient complex manifold
\be
\label{eq:torus}
\BT^g = \BC^g / \BL
\ee
\noindent
is a $g$-dimensional complex torus. Furthermore, classifying tori is now essentially the same as classifying lattices and these are fixed by a Riemann matrix in $\BH_g$ up to some redundancy: indeed, given the symplectic group\footnote{When $g=1$ this is just the modular group $\text{SL} (2, \BZ)$.}
\be
\label{eq:modulargroup}
\text{Sp} (2g, \BZ) = \left\{ \boldsymbol{\Gamma} \in \text{GL}(2g,\BZ) \, | \, \boldsymbol{\Gamma}^{\text{t}} \cdot \boldsymbol{J} \cdot \boldsymbol{\Gamma} = \boldsymbol{J} \right\},
\ee
\noindent
where $\boldsymbol{J}$ is the $2g \times 2g$ (antisymmetric) symplectic matrix, then $\boldsymbol{\tau}$ and $\boldsymbol{\tau}'$ will describe equivalent lattices if they are related by a fractional linear transformation of $\text{Sp} (2g, \BZ)$.

Our interest concerns (eventually non-trivial, multi-valued) functions defined on tori, $\BT^g$, where theta-functions are the canonical example. We define the $g$-dimensional Riemann theta-function as the holomorphic function
\be
\label{eq:riemanntheta}
\vartheta \left( \boldsymbol{z} \, | \, \boldsymbol{\tau} \right) = \sum_{\boldsymbol{n} \in \BZ^g} \exp \left\{ 2\pi\rmi \left( \frac{1}{2}\, \boldsymbol{n}^{\text{t}} \cdot \boldsymbol{\tau} \cdot \boldsymbol{n} + \boldsymbol{n}^{\text{t}} \cdot \boldsymbol{z} \right) \right\},
\ee
\noindent
with argument $\boldsymbol{z} \in \BC^g$ and Riemann matrix $\boldsymbol{\tau} \in \BH_g$. Quasi-periodicity (under translations of the argument) on the lattice turns $\vartheta \left( \boldsymbol{z} \, | \, \boldsymbol{\tau} \right)$ into a section of a line bundle over $\BT^g$,
\be
\vartheta \left( \boldsymbol{z} + \boldsymbol{m}_1 + \boldsymbol{\tau} \cdot \boldsymbol{m}_2 \, | \, \boldsymbol{\tau} \right) = \rme^{- 2 \pi \rmi \left( \frac{1}{2} \boldsymbol{m}_2^{\text{t}} \cdot \boldsymbol{\tau} \cdot \boldsymbol{m}_2 + \boldsymbol{m}_2^{\text{t}} \cdot \boldsymbol{z} \right)}\, \vartheta \left( \boldsymbol{z} \, | \, \boldsymbol{\tau} \right), \qquad \boldsymbol{m}_1, \boldsymbol{m}_2 \in \BZ^g. 
\ee
\noindent
In the simpler case where $g=1$, this definition matches the Jacobi theta-function $\theta_3 \left( z | \tau \right)$ defined on $\BT \times \BH$. In this case the modular group has two generators, under which the theta-function transforms under changes of the lattice parameters as
\bea
\vartheta \left( z | \tau+1 \right) &=& \vartheta \left. \left( z+\frac{1}{2} \, \right| \tau \right), \\
\vartheta \left( z \left| -\frac{1}{\tau} \right) \right. &=& \sqrt{- \rmi \tau}\, \rme^{\rmi \pi \tau z^2}\, \vartheta \left( - \tau z | \tau \right).
\eea
\noindent
In this way one may use theta-functions to construct modular forms of $\text{SL}(2,\BZ)$.

An alternative way to write the 1-dimensional Riemann theta-function is to use the \textit{elliptic nome}\footnote{Interestingly, the elliptic nome may be written in terms of the complete elliptic-integral of the first kind as
\be
q = \exp \left( - \pi\, \frac{\boldsymbol{K} (k')}{\boldsymbol{K} (k)} \right).
\ee
\noindent
Conversely, this complete elliptic-integral may be written in terms of a Jacobi theta-function, $\boldsymbol{K} (k) = \frac{\pi}{2}\, \theta_3^2 \left( 0 | \tau \right)$.} $q = \rme^{\rmi\pi\tau}$ and the argument $w = \rme^{\rmi\pi z}$ as
\be
\label{eq:jacobitheta}
\vartheta \left( w \, | \, q \right) = \sum_{n \in \BZ} q^{n^2} w^{2n}.
\ee
\noindent
If we consider $w \not = 0$ and $|q|<1$, then the Jacobi triple-product tells us that
\be
\label{eq:jacobitriple}
\sum_{n \in \BZ} q^{n^2} w^{2n} = \prod_{m=1}^{+\infty} \left( 1 - q^{2m} \right) \left( 1 + w^2\, q^{2m-1} \right) \left( 1 + w^{-2}\, q^{2m-1} \right).
\ee
\noindent
This is an extremely useful identity in order to compute the logarithm of the Riemann theta-function as
\be
\label{eq:logtheta}
\log \vartheta \left( w \, | \, q \right) = -\frac{1}{12} \log q + \log \eta (q)
 + \sum_{k=1}^{+\infty} \frac{(-1)^k}{k}\, \frac{w^{2k} + w^{-2k}}{q^k - q^{-k}},
\ee
\noindent
where $\eta (q)$ is Dedekin's eta function.

One of the uses of theta-functions is to uniformize complex tori by embedding them into complex projective space, $\BT^g \to \BC\BP^n$, in which case we actually need more than a single theta-function (\textit{e.g.}, when $g=1$ the four classical Jacobi theta-functions are required in order to provide the embedding $\BT \to \BC\BP^3$). One must hence introduce \textit{characteristics}.

It is simple to enlarge the definition \eqref{eq:riemanntheta} above to the inclusion of characteristics\footnote{Given $\boldsymbol{w} \in \BC^g$, one can always decompose it in the lattice basis as $\boldsymbol{w} = \boldsymbol{\tau} \cdot \boldsymbol{\alpha} + \boldsymbol{\beta}$. In this case, we denote the $g$-dimensional vectors $\boldsymbol{\alpha}, \boldsymbol{\beta} \in \BR^g$ as the \textit{characteristics} of $\boldsymbol{w} \in \BC^g$.} $\boldsymbol{\alpha}, \boldsymbol{\beta} \in \BR^g$, as
\be
\label{eq:riemannthetacharacteristics}
\vartheta { \begin{bmatrix} \boldsymbol{\alpha} \\ \boldsymbol{\beta} \end{bmatrix} } \left( \boldsymbol{z} \, | \, \boldsymbol{\tau} \right) = \sum_{\boldsymbol{n} \in \BZ^g} \exp \left\{ 2\pi\rmi \left( \frac{1}{2} \left( \boldsymbol{n} + \boldsymbol{\alpha} \right)^{\text{t}} \cdot \boldsymbol{\tau} \cdot \left( \boldsymbol{n} + \boldsymbol{\alpha} \right) + \left( \boldsymbol{n} + \boldsymbol{\alpha} \right)^{\text{t}} \cdot \left( \boldsymbol{z} + \boldsymbol{\beta} \right) \right) \right\}.
\ee
\noindent
Straightforward expansion of this expression yields
\be
\vartheta { \begin{bmatrix} \boldsymbol{\alpha} \\ \boldsymbol{\beta} \end{bmatrix} } \left( \boldsymbol{z} \, | \, \boldsymbol{\tau} \right) = \rme^{2\pi\rmi \left( \frac{1}{2} \boldsymbol{\alpha}^{\text{t}} \cdot \boldsymbol{\tau} \cdot \boldsymbol{\alpha} + \boldsymbol{\alpha}^{\text{t}} \cdot \left( \boldsymbol{z} + \boldsymbol{\beta} \right) \right)}\, \vartheta \left( \boldsymbol{z} + \boldsymbol{\tau} \cdot \boldsymbol{\alpha} + \boldsymbol{\beta} \, | \, \boldsymbol{\tau} \right),
\ee
\noindent
in which case it is also simple to check that one can relate the above Riemann theta-function \textit{with} characteristics to the theta-function \textit{without} characteristics. The quasi-periodicity in this case becomes, with $\boldsymbol{m}_1, \boldsymbol{m}_2 \in \BZ^g$,
\be
\label{eq:quasiperiodic}
\vartheta { \begin{bmatrix} \boldsymbol{\alpha} \\ \boldsymbol{\beta} \end{bmatrix} } \left( \boldsymbol{z} + \boldsymbol{m}_1 + \boldsymbol{\tau} \cdot \boldsymbol{m}_2 \, | \, \boldsymbol{\tau} \right) = \rme^{- 2 \pi \rmi \left( \frac{1}{2} \boldsymbol{m}_2^{\text{t}} \cdot \boldsymbol{\tau} \cdot \boldsymbol{m}_2 + \boldsymbol{m}_2^{\text{t}} \cdot \left( \boldsymbol{z} + \boldsymbol{\beta} \right) - \boldsymbol{m}_1^{\text{t}} \cdot \boldsymbol{\alpha} \right)}\, \vartheta { \begin{bmatrix} \boldsymbol{\alpha} \\ \boldsymbol{\beta} \end{bmatrix} } \left( \boldsymbol{z} \, | \, \boldsymbol{\tau} \right). 
\ee
\noindent
One may also show simple relations, such as
\be
\vartheta { \begin{bmatrix} \boldsymbol{\alpha} + \boldsymbol{m}_1 \\ \boldsymbol{\beta} + \boldsymbol{m}_2 \end{bmatrix} } \left( \boldsymbol{z} \, | \, \boldsymbol{\tau} \right) = \rme^{2\pi\rmi\, \boldsymbol{\alpha} \cdot \boldsymbol{m}_2}\, \vartheta { \begin{bmatrix} \boldsymbol{\alpha} \\ \boldsymbol{\beta} \end{bmatrix} } \left( \boldsymbol{z} \, | \, \boldsymbol{\tau} \right),
\ee
\noindent
illustrating how one may always restrict elements of $\boldsymbol{\alpha}$, $\boldsymbol{\beta}$ to $[0,1)$ without loss of generality. However, the modular properties of both the Riemann theta-function \eqref{eq:riemanntheta} and the Riemann theta-function with characteristics \eqref{eq:riemannthetacharacteristics}, for general $g$, are much more involved than in the aforementioned $g=1$ case (see, \textit{e.g.}, \cite{olbc10} for some explicit expressions).

Zeroes of theta-functions are rather simple to locate. For instance, for the Jacobi theta-function $\theta_3 \left( z | \tau \right)$ the vertices of the fundamental parallelogram upon the $z$-plane are at the usual $\left( 0, 1, 1 + \tau, \tau \right)$, in which case its $z$-zeroes are found to be at $\BZ + \frac{1}{2} + \left( \BZ + \frac{1}{2} \right) \tau$. If one considers the one-dimensional Riemann theta-function with characteristics, $\vartheta { \begin{bmatrix} \alpha \\ \beta \end{bmatrix} } \left( z  | \tau \right)$, then its $z$-zeroes are now characteristic-dependent and at $\left( \alpha + \frac{1}{2} \right) \tau + \left( \beta + \frac{1}{2} \right)$ modulo $\BL$. And for the $g$-dimensional Riemann theta-function with characteristics $\vartheta { \begin{bmatrix} \boldsymbol{\alpha} \\ \boldsymbol{\beta} \end{bmatrix} } \left( \boldsymbol{z} \, | \, \boldsymbol{\tau} \right)$, one finds its $z$-zeroes at
\be
\boldsymbol{\tau} \cdot \left( \boldsymbol{\alpha} + \frac{1}{2} \boldsymbol{e}_i \right) + \left( \boldsymbol{\beta} + \frac{1}{2} \boldsymbol{e}_i \right)
\ee
\noindent
(herein $\boldsymbol{e}_i$ with $i=1,\ldots,g$ is the usual canonical basis).

The $g$-dimensional Riemann theta-function with characteristics provides periodic solutions to the multidimensional heat (diffusion) equation. For example, it is simple to see that \eqref{eq:riemannthetacharacteristics} satisfies
\be
\label{eq:heatequation}
\frac{\partial^2}{\partial z_i \partial z_j} \vartheta { \begin{bmatrix} \boldsymbol{\alpha} \\ \boldsymbol{\beta} \end{bmatrix} } \left( \boldsymbol{z} \, | \, \boldsymbol{\tau} \right) = 4\pi\rmi\, \frac{\partial}{\partial \tau_{ij}} \vartheta { \begin{bmatrix} \boldsymbol{\alpha} \\ \boldsymbol{\beta} \end{bmatrix} } \left( \boldsymbol{z} \, | \, \boldsymbol{\tau} \right).
\ee
\noindent
Finally, one may use theta-functions to construct multiply periodic functions on the complex torus $\BT^g$. Indeed, via \eqref{eq:quasiperiodic}, it is simple to see that
\be
\label{eq:generalizedweierstrass}
\frac{\partial^2}{\partial z_i \partial z_j} \log \vartheta { \begin{bmatrix} \boldsymbol{\alpha} \\ \boldsymbol{\beta} \end{bmatrix} } \left( \boldsymbol{z} \, | \, \boldsymbol{\tau} \right)
\ee
\noindent
turns the quasi-periodic behavior into a purely periodic behavior with respect to the lattice. This function turns out to be a meromorphic function on the torus and, when $g=1$ and we consider the theta-function without characteristics, this is essentially the Weierstrass $\wp (z)$ function.

Although, as constructed above, theta-functions are defined for arbitrary tori, it is of special interest to consider those arising from a complex curve, \textit{i.e.}, to consider theta-functions associated to (hyperelliptic) Riemann surfaces $\Sigma_g$. For these surfaces, their period matrix $\boldsymbol{\tau}_{ij}$ in \eqref{eq:omega-a-omega-b-period} is a Riemann matrix, satisfying the \eqref{eq:siegelupperhalfspace} conditions, and can be used to define theta-functions associated to $\Sigma_g$. In fact, one can use the period matrix of $\Sigma_g$ to define the period lattice $\BL = \BZ^g \oplus \boldsymbol{\tau} \, \BZ^g$ and, in this way, define the $g$-dimensional complex torus
\be
\text{Jac} (\Sigma_g) = \BC^g / \BL
\ee
\noindent
as the Jacobian of the Riemann surface. One may also embed $\Sigma_g$ into its own Jacobian via the Jacobi map $\phi: \Sigma_g \to \text{Jac} (\Sigma_g)$. Thus, given the aforementioned embedding of the torus into complex projective space, provided by theta-functions with characteristics\footnote{To be fully precise, one needs to consider theta-functions with (high enough) level to have an embedding.} associated to $\Sigma_g$, this further allows for an embedding of the original complex curve into complex projective space.

Let us mention one final subtlety. One what concerns its $\BH_g$ dependence, a general theta-function depends on the number of independent components in its Riemann matrix $\boldsymbol{\tau}_{ij}$, corresponding to $\frac{1}{2} g (g+1)$ complex moduli for an arbitrary Riemann matrix. But the dimension of the moduli space of Riemann surfaces, as parameterized by period matrices, is actually smaller, $3g-3$ for $g>1$. This is known as the Schottky problem and can be phrased in equivalent ways: not all lattices are period lattices of some compact Riemann surface, \textit{i.e.}, not all Riemann matrices are period matrices of some complex curve, \textit{i.e.}, the number of ``general'' theta-functions is larger than the number of theta-functions which are associated to a given Riemann surface.

\section{Spectral Geometry of Cubic and Quartic Matrix Models}
\label{appendix:SpectralGeometryMatrixModels}

This appendix contains the details concerning the spectral geometry computations that went into the construction of the spectral curves, and their associated spectral networks, shown and discussed in section~\ref{sec:strong-coupling-phases}, mainly subsections~\ref{subsec:SG-phases} and~\ref{subsec:stokes-vs-phases}. These also include the computations of the locations of the end-points of the cuts, and of the corresponding order parameters.

\subsection{The Cubic Matrix Model}
\label{subappendix:cubicmatrixmodel}

Let us begin with the cubic matrix model \eqref{eq:CubicMatrixModelPotential} (see as well, \textit{e.g.}, \cite{biz80, msw08, kmr10, mss22}).

\begin{figure}
\centering
\begin{tikzpicture}
	\draw[line width=2pt] (2-0.4,-1+0.4) -- (4-0.4,1+0.4);
	\draw[line width=2pt,dashed] (2+0.4,-1-0.4) -- (4+0.4-0.105,1-0.4-0.105);
	\draw[line width=2pt](4+0.4-0.105,1-0.4-0.105) -- (4+0.4,1-0.4);
	\draw[line width=2pt]  (4+0.4,1-0.4) arc (-45:-45+180:0.5656);
	\draw[line width=2pt,dashed] (2+0.4,-1-0.4) arc (-45:-180:0.5656);
	\draw[line width=2pt] (2-0.4,-1+0.4) arc (45+90:180:0.5656);
	\node at (5+0.7,0+0.7){$A$};
	\draw[line width=2pt] (4+0.4,1+0.4) -- (6+0.4,-1+0.4);
	\draw[line width=2pt] (4-0.4,1-0.4) -- (6-0.4,-1-0.4);
	\draw[line width=2pt] (6-0.4,-1-0.4) arc (-90-45:45:0.5656);
	\draw[line width=2pt] (4-0.4,1-0.4) arc (-90-45:-90-180-45:0.5656);
	\draw[line width=2pt] (6-0.4,-1-0.4) arc (-90-45:45:0.5656);
	\node at (3-0.7,0+0.7){$B$};
	\draw[ decorate, decoration={snake, segment length=9, amplitude=4},line width=2pt, color = ForestGreen] (-1,-1)--(2,-1);
	\draw[ decorate, decoration={snake, segment length=9, amplitude=4},line width=2pt, color = ForestGreen] (4,1)--(6,-1);
	\node[ForestGreen] at (4,1.32){$x_3$};
	\node[ForestGreen] at (2+0.28,-1+0.28){$x_2$};
	\node[ForestGreen] at (-1-0.4,-1){$x_1$};
	\node[ForestGreen] at (6,-1-0.35){$x_4$};
	\filldraw[ForestGreen] (4,1) circle (1.1ex);
	\filldraw[ForestGreen] (6,-1) circle (1.1ex);
	\filldraw[ForestGreen] (-1,-1) circle (1.1ex);
	\filldraw[ForestGreen] (2,-1) circle (1.1ex);
\end{tikzpicture}
\caption{Schematic depiction of the two-cut spectral curve $y(z)$ for the cubic matrix model, including $A$ and $B$ cycles, as well as the end-points of the cuts $x_1,x_2,x_3,x_4$. The branch-cuts of $y(z)$ are illustrated by solid-\textcolor{ForestGreen}{green} wavy lines.}
\label{fig:SpecGeofig1}
\end{figure}
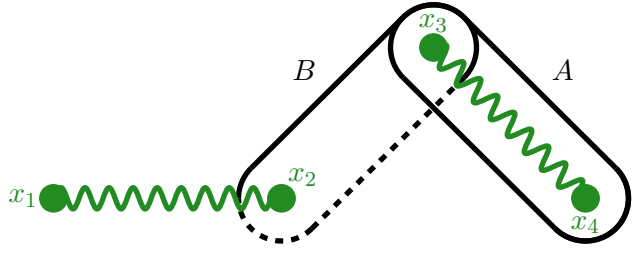

\paragraph{The Two-Cut Phase:} 

Let us address the cubic matrix model within a two-cut configuration. In this case, equations \eqref{eq:SpecGeo2a} and \eqref{eq:SpecGeo2b} explicitly read
\begin{eqnarray}
\label{eq:SpecGeo7a}
\oint_{\NCC} \frac{\rmd z}{2\pi\rmi}\, \frac{V'(z)}{\sqrt{(z-x_1)(z-x_2)(z-x_3)(z-x_4)}} &=& 0, \\
\label{eq:SpecGeo7b}
\oint_{\NCC} \frac{\rmd z}{2\pi\rmi}\, \frac{z\, V'(z)}{\sqrt{(z-x_1)(z-x_2)(z-x_3)(z-x_4)}} &=& 0, \\
\label{eq:SpecGeo7c}
\oint_{\NCC} \frac{\rmd z}{2\pi\rmi}\, \frac{z^2\, V'(z)}{\sqrt{(z-x_1)(z-x_2)(z-x_3)(z-x_4)}} &=& 2t,
\end{eqnarray}
\noindent
alongside
\begin{eqnarray}
\label{eq:SpecGeo7d}
\Re \frac{1}{t} \oint_{A} \rmd z\, \sqrt{(z-x_1)(z-x_2)(z-x_3)(z-x_4)} &=& 0, \\
\label{eq:SpecGeo7e}
\Re \frac{1}{t} \oint_{B} \rmd z\, \sqrt{(z-x_1)(z-x_2)(z-x_3)(z-x_4)} &=& 0.
\end{eqnarray}
\noindent
The cycles $A$ and $B$ in these last two integrals are schematically illustrated in figure~\ref{fig:SpecGeofig1}. Inverting the dummy variable in the integrands so as to circle the point at infinity and then using the residue theorem allows us to write the first three equations above algebraically as
\bea
&&
\lambda \left( x_1 + x_2 + x_3 + x_4 \right) - 4 = 0, \\
&&
3 \lambda \left( x_1^2 + x_2^2 + x_3^2 + x_4^2 \right) + 2 \lambda \left( x_1 \left( x_2 + x_3 + x_4 \right) + x_2 \left( x_3 + x_4 \right) + x_3 x_4 \right) - \nonumber \\
&&
\hspace{222pt}
- 8 \left( x_1 + x_2 + x_3 + x_4 \right) = 0, \\
&&
5 \lambda \left( x_1^3 + x_2^3 + x_3^3 + x_4^3 \right) + 3 \lambda \left( x_1^2 \left( x_2 + x_3 + x_4 \right) + x_2^2 \left( x_3 + x_4 \right) + x_3^2 x_4 \right) + \nonumber \\
&&
+ 3 \lambda \left( x_1 \left( x_2^2 + x_3^2 + x_4^2 \right) + x_2 \left( x_3^2 + x_4^2 \right) + x_3 x_4^2 \right) + 2 \lambda \left( x_1 x_2 \left( x_3 + x_4 \right) + \left( x_1 + x_2 \right) x_3 x_4 \right) - \nonumber \\
&&
- 12 \left( x_1^2 + x_2^2 + x_3^2 + x_4^2 \right) - 8 \left( x_1 \left( x_2 + x_3 + x_4 \right) + x_2 \left( x_3 + x_4 \right) + x_3 x_4 \right) = -64t.
\eea
\noindent
Moreover, resorting to several analytical expressions found in, \textit{e.g.}, \cite{bf13}, one can write the cycle integrals appearing in the Boutroux conditions as
\bea
&&
\frac{1}{t} \oint_{A} \rmd z\, \sqrt{\left(z-x_1\right) \left(z-x_2\right) \left(z-x_3\right) \left(z-x_4\right)} =  \frac{\rmi}{12 t \sqrt{\left(x_3-x_1\right) \left(x_4-x_2\right)}} \times \\
&&
\times \Bigg\{ \left( x_3-x_2 \right) \bigg[ \left( x_2-x_4 \right) \left( x_1^2 - 3 \left( x_2^2 - x_3^2 - x_4^2 \right) + 6 x_1 x_2 - 4 x_1 \left( x_3 + x_4 \right) - 2 x_3 x_4 \right) \boldsymbol{K} (k) + \nonumber \\
&&
+ 3 \left( x_1 - x_2 - x_3 + x_4 \right) \left( x_1 - x_2 + x_3 - x_4 \right) \left( x_1 + x_2 - x_3 - x_4 \right) \boldsymbol{\Pi} (\alpha_{A},k) \bigg] + \left( x_3-x_1 \right) \times \nonumber \\
&&
\times \left( x_4-x_2 \right) \bigg[ 3 \left( x_1^2 + x_2^2 + x_3^2 + x_4^2 \right) - 2 \left( x_1 \left( x_2 + x_3 + x_4 \right) + x_2 \left( x_3 + x_4 \right) + x_3 x_4 \right) \bigg] \boldsymbol{E} (k) \Bigg\}, \nonumber \\
&&
\label{eq:SpecGeo10}
\frac{1}{t} \oint_{B} \rmd z\, \sqrt{\left(z-x_1\right) \left(z-x_2\right) \left(z-x_3\right) \left(z-x_4\right)} = \frac{1}{12 t \sqrt{\left(x_3-x_1\right) \left(x_4-x_2\right)}} \times \\
&&
\times \Bigg\{ \left( x_2-x_1 \right) \bigg[ \left( x_3-x_1 \right) \left( x_4^2 - 3 \left( x_1^2 - x_2^2 - x_3^2 \right) + 6 x_1 x_4 - 4 x_4 \left( x_2 + x_3 \right) - 2 x_2 x_3 \right) \boldsymbol{K} (k') - \nonumber \\
&&
- 3 \left( x_1 - x_2 - x_3 + x_4 \right) \left( x_1 - x_2 + x_3 - x_4 \right) \left( x_1 + x_2 - x_3 - x_4 \right) \boldsymbol{\Pi} (\alpha_{B},k') \bigg] + \left( x_3-x_1 \right) \times \nonumber \\
&&
\times \left( x_4-x_2 \right) \bigg[ 3 \left( x_1^2 + x_2^2 + x_3^2 + x_4^2 \right) - 2 \left( x_1 \left( x_2 + x_3 + x_4 \right) + x_2 \left( x_3 + x_4 \right) + x_3 x_4 \right)
\bigg] \boldsymbol{E} (k') \Bigg\}. \nonumber
\eea
\noindent
Herein, $\boldsymbol{K} (k)$, $\boldsymbol{E} (k)$, and $\boldsymbol{\Pi} (\alpha,k)$ stand for the complete elliptic-integrals of the first, second, and third kinds, respectively (see appendix~\ref{app:elliptic-theta-modular} for definitions). The elliptic modulus $k$, complementary modulus $k'$, and elliptic characteristics $\alpha_{A,B}$, are
\be
k^2 = \frac{\left( x_1 - x_2 \right) \left( x_3 - x_4 \right)}{\left( x_1 - x_3 \right) \left( x_2 - x_4 \right)}, \quad k'^2 = \frac{\left( x_1 - x_4 \right) \left( x_2 - x_3 \right)}{\left( x_1 - x_3 \right) \left( x_2 - x_4 \right)}, \quad \alpha_{A} = \frac{x_4-x_3}{x_4-x_2}, \quad \alpha_{B} = \frac{x_3-x_2}{x_3-x_1}.
\ee
\noindent
Using all these results, algebraic and transcendental, we can then numerically solve the original system of equations in \eqref{eq:SpecGeo7a}-\eqref{eq:SpecGeo7e}, valid for any value of the 't~Hooft coupling $t$ within the two-cut phase of the cubic matrix model. In order to finally write down its spectral curve as in \eqref{eq:spectral-curve-moment-function}, all we need is the moment function which can be easily extracted from \eqref{eq:moment-function-multi-cut} and reads
\be
M(z) = - \frac{\lambda }{2}.
\ee

\paragraph{The Trivalent Phase:}

\begin{figure}
\centering
\begin{tikzpicture}
	\draw[line width=2pt] (2+0.4,-2+0.4) -- (0+0.4+0.1,0+0.4-0.1);
	\draw[line width=2pt] (2-0.4,-2-0.4) -- (0-0.4-0.05,0-0.4+0.05);
	\draw[line width=2pt,dashed] (0+0.4,0+0.4) arc (45:45+180:0.5656);
	\draw[line width=2pt] (2+0.4,-2+0.4) arc (45:45-180:0.5656);
	\node at (-0.7-0.5,0.7-0.5){``$B$''};
	\draw[line width=2pt] (-2-0.4,-2+0.4) -- (1-0.4,1+0.4);
	\draw[line width=2pt] (-2+0.4,-2-0.4) -- (0+0.4-0.05,0-0.4-0.05);
	\draw[line width=2pt,dashed] (0+0.4-0.05,0-0.4-0.05) -- (1+0.4,1-0.4);
	\draw[line width=2pt,dashed] (1+0.4,1-0.4) arc (-45:55:0.5656);
	\draw[line width=2pt] (1-0.4,1+0.4) arc (-45+180:45:0.5656);
	\draw[line width=2pt] (-2-0.4,-2+0.4) arc (90+45:90+45+180:0.5656);
	\node at (1+0.7,-1+0.7){``$A$''};
	\draw[ decorate, decoration={snake, segment length=9, amplitude=4},line width=2pt, color = ForestGreen] (-2,-2)--(0,0);
	\draw[ decorate, decoration={snake, segment length=9, amplitude=4},line width=2pt, color = ForestGreen] (2,-2)--(0,0);
	\draw[ decorate, decoration={snake, segment length=9, amplitude=4},line width=2pt, color = ForestGreen] (0,0)--(2,2);
	\node[ForestGreen] at(-0.05,0.35){$x_3$};
	\node[ForestGreen] at(2+0.26,2+0.26){$x_2$};
	\node[ForestGreen] at(-2-0.2,-2-0.26){$x_1$};
	\node[ForestGreen] at(2+0.1,-2-0.33){$x_4$};
	\filldraw[ForestGreen] (0,0) circle (1.1ex);
	\filldraw[ForestGreen] (2,-2) circle (1.1ex);
	\filldraw[ForestGreen] (-2,-2) circle (1.1ex);;
	\filldraw[ForestGreen] (2,2) circle (1.1ex);
\end{tikzpicture}
\caption{Schematic depiction of the trivalent-tree spectral curve $y(z)$ for the cubic matrix model, including a convenient choice of homologically independent ``$A$'' and ``$B$'' cycles, as well as the end-points (and node) of the cuts $x_1,x_2,x_3,x_4$. The cuts are the \textcolor{ForestGreen}{green} wavy lines.}
\label{fig:SpecGeofig2}
\end{figure}
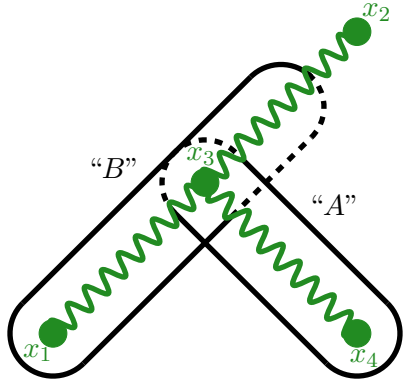

Next, consider the trivalent phase. This has the exact same number of branch points as the two-cut phase, hence it is described by essentially the same set of equations. In other words, we still need to solve the system of equations in \eqref{eq:SpecGeo7a}-\eqref{eq:SpecGeo7e}, but where, however, the cycles $A$ and $B$ are slightly more intricate\footnote{Recall the discussion in subsection~\ref{subsec:SG-phases}, where we mentioned how the splitting of contours into the canonical $A$ and $B$ cycles within the trivalent phase is slightly misleading and it is more convenient to use adequate homologically independent contours instead. We still keep these names just for compact nomenclature.}. A convenient choice of ``$A$'' and ``$B$'' cycles for the integrals associated to the Boutroux conditions is schematically illustrated in figure~\ref{fig:SpecGeofig2}.

\paragraph{The Order Parameter:}

The order parameter for two-cut and trivalent phases is given by \eqref{eq:SpecGeo9}, which itself is immediately computed by resorting to \eqref{eq:SpecGeo10}. This yields:
\bea
\NCO (t) &=& - \frac{\lambda}{24 t \sqrt{\left(x_3-x_1\right) \left(x_4-x_2\right)}} \Bigg\{ \left( x_2-x_1 \right) \bigg[ \left( x_3-x_1 \right) \left\{ x_4^2 - 3 \left( x_1^2 - x_2^2 - x_3^2 \right) + 6 x_1 x_4 - \right. \nonumber \\
&&
\left. - 4 x_4 \left( x_2 + x_3 \right) - 2 x_2 x_3 \right\} \boldsymbol{K} (k') - 3 \left( x_1 - x_2 - x_3 + x_4 \right) \left( x_1 - x_2 + x_3 - x_4 \right) \times \nonumber \\
&&
\times \left( x_1 + x_2 - x_3 - x_4 \right) \boldsymbol{\Pi} (\alpha_{B},k') \bigg] + \left( x_3-x_1 \right) \left( x_4-x_2 \right) \bigg[ 3 \left( x_1^2 + x_2^2 + x_3^2 + x_4^2 \right) - \nonumber \\
&&
- 2 \left( x_1 \left( x_2 + x_3 + x_4 \right) + x_2 \left( x_3 + x_4 \right) + x_3 x_4 \right) \bigg] \boldsymbol{E} (k') \Bigg\}.
\eea

\subsection{The Quartic Matrix Model}
\label{subappendix:quarticmatrixmodel}

Let us next address the quartic matrix model \eqref{eq:QuarticMatrixModelPotential} (see as well, \textit{e.g.}, \cite{biz80, msw07, m08, asv11, csv15, mss22}).

\begin{figure}
\centering
\begin{tikzpicture}
	\draw[line width=2pt,dashed] (4,-0.5656) -- (1.5,-0.5656);
	\draw[line width=2pt] (4,0.5656) -- (1.5,0.5656);
	\draw[line width=2pt] (4,0.5656) arc (90:0:0.5656);
    \draw[line width=2pt,dashed] (4+0.5656,0) arc (0:-90:0.5656);
    \draw[line width=2pt] (1.5,0.5656) arc (90:180:0.5656);
    \draw[line width=2pt,dashed] (1.5,-0.5656) arc (270:180:0.5656);
	\draw[line width=2pt] (4,0.5656) arc (90:270:0.5656);
	\draw[line width=2pt] (7,0.5656) arc (90:-90:0.5656);
	\draw[line width=2pt] (4,0.5656) -- (7,0.5656);
	\draw[line width=2pt] (4,-0.5656) -- (7,-0.5656);
	\node at (2.75,0.9){$B$};
	\node at (5.5,0.9){$A$};
	\draw[ decorate, decoration={snake, segment length=9, amplitude=4},line width=2pt, color = ForestGreen] (-1.5,0)--(1.5,0);
	\draw[ decorate, decoration={snake, segment length=9, amplitude=4},line width=2pt, color = ForestGreen] (4,0)--(7,0);
	\draw[ decorate, decoration={snake, segment length=9, amplitude=4},line width=2pt, color = ForestGreen] (-4,0)--(-7,0);
	\node[ForestGreen] at (1.5+0.4,0){$x_1$};
	\node[ForestGreen] at (-1.5-0.5,0){$-x_1$};
	\node[ForestGreen] at (4,0.33){$x_2$};
	\node[ForestGreen] at (-4,0.33){$-x_2$};
	\node[ForestGreen] at (7,0.33){$x_3$};
	\node[ForestGreen] at (-7,0.33){$-x_3$};
	\filldraw[ForestGreen] (-1.5,0) circle (1.1ex);
	\filldraw[ForestGreen] (1.5,0) circle (1.1ex);
	\filldraw[ForestGreen] (4,0) circle (1.1ex);
	\filldraw[ForestGreen] (7,0) circle (1.1ex);
	\filldraw[ForestGreen] (-4,0) circle (1.1ex);
	\filldraw[ForestGreen] (-7,0) circle (1.1ex);
\end{tikzpicture}
\caption{Schematic depiction of the $\BZ_2$-symmetric three-cut spectral curve $y(z)$ for the quartic matrix model, including $A$ and $B$ cycles, as well as the (symmetric) end-points of the cuts $\pm x_1,\pm x_2,\pm x_3$. The branch-cuts of $y(z)$ are illustrated by solid-\textcolor{ForestGreen}{green} wavy lines.}
\label{fig:SpecGeofig8}
\end{figure}
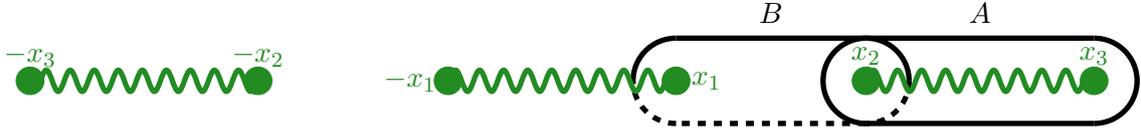

\paragraph{The Symmetric Three-Cut Phase:} 

Due to the $\BZ_2$-symmetry of the quartic potential \eqref{eq:QuarticMatrixModelPotential}, invariant under $x \leftrightarrow -x$, it turns out that some formulae are actually slightly simpler than in the cubic case. Let us address the quartic matrix model within a (symmetric) three-cut configuration. In this case, equations \eqref{eq:SpecGeo2a} and \eqref{eq:SpecGeo2b} explicitly read
\begin{eqnarray}
\label{eq:SpecGeo15a}
\oint_{\NCC} \frac{\rmd z}{2\pi\rmi}\, \frac{V'(z)}{\sqrt{(z^2-x_1^2)(z^2-x_2^2)(z^2-x_3^2)}} &=& 0, \\
\label{eq:SpecGeo15b}
\oint_{\NCC} \frac{\rmd z}{2\pi\rmi}\, \frac{z\, V'(z)}{\sqrt{(z^2-x_1^2)(z^2-x_2^2)(z^2-x_3^2)}} &=& 0, \\
\label{eq:SpecGeo15c}
\oint_{\NCC}\frac{\rmd z}{2\pi\rmi}\, \frac{z^2\, V'(z)}{\sqrt{(z^2-x_1^2)(z^2-x_2^2)(z^2-x_3^2)}} &=& 0, \\
\label{eq:SpecGeo15d}
\oint_{\NCC}\frac{\rmd z}{2\pi\rmi}\, \frac{z^3\, V'(z) }{\sqrt{(z^2-x_1^2)(z^2-x_2^2)(z^2-x_3^2)}} &=& 2t,
\end{eqnarray}
\noindent
alongside
\begin{eqnarray}
\label{eq:SpecGeo15e}
\Re \frac{1}{t} \oint_{A} \rmd z\, \sqrt{(z^2-x_1^2)(z^2-x_2^2)(z^2-x_3^2)} &=& 0, \\
\label{eq:SpecGeo15f}
\Re \frac{1}{t} \oint_{B} \rmd z\, \sqrt{(z^2-x_1^2)(z^2-x_2^2)(z^2-x_3^2)} &=& 0.
\end{eqnarray}
\noindent
The cycles $A$ and $B$ in these last two integrals are schematically illustrated in figure~\ref{fig:SpecGeofig8}. Inverting the dummy variable in the integrands so as to circle the point at infinity and then using the residue theorem allows us to write the second \eqref{eq:SpecGeo15b} and fourth \eqref{eq:SpecGeo15d} (the first \eqref{eq:SpecGeo15a} and third \eqref{eq:SpecGeo15c} yield trivially-satisfied relations due to the $\BZ_2$-symmetry of the quartic potential) equations above algebraically as
\bea
&&
\lambda \left( x_1^2 + x_2^2 + x_3^2 \right) - 12 = 0, \\
&&
3 \lambda \left( x_1^4 + x_2^4 + x_3^4 \right) - 24 \left( x_1^2 + x_2^2 + x_3^2 \right) + 2 \lambda \left( x_1^2 x_2^2 + x_1^2 x_3^2 + x_2^2 x_3^2 \right) = - 96 t.
\eea
\noindent
Moreover, resorting to several analytical expressions found in, \textit{e.g.}, \cite{bf13}, one can write the cycle integrals appearing in the Boutroux conditions as
\bea
&&
\frac{1}{t} \oint_A \rmd z\, \sqrt{\left( z^2-x_1^2 \right) \left( z^2-x_2^2 \right) \left( z^2-x_3^2 \right)} = \frac{\rmi}{8 t \sqrt{x_2^2 \left( x_3^2-x_1^2 \right)}} \times \\
&&
\times \Bigg\{ \left( x_2^2-x_1^2 \right) \Bigg[ \left( x_1^2-x_3^2 \right) \left(-x_1^2+x_2^2+x_3^2\right) \boldsymbol{K} (k) + \left( x_1 + x_2 - x_3 \right) \left( x_1 - x_2 + x_3 \right) \times \nonumber \\ 
&&
\times \left( x_1 - x_2 - x_3 \right) \left( x_1 + x_2 + x_3 \right) \boldsymbol{\Pi} (\alpha_{A},k) \Bigg] + x_2^2 \left( x_3^2-x_1^2 \right) \left( x_1^2 + x_2^2 + x_3^2 \right) \boldsymbol{E} (k) \Bigg\}, \nonumber \\
&&
\label{eq:SpecGeo16}
\frac{1}{t} \oint_B \rmd z\, \sqrt{\left( z^2-x_1^2 \right) \left( z^2-x_2^2 \right) \left( z^2-x_3^2 \right)} = \frac{1}{8 t \sqrt{x_2^2 \left( x_3^2-x_1^2 \right)}} \times \\
&&
\times \Bigg\{ x_1^2 \Bigg[ x_2^2 \left( x_1^2 + x_2^2 - 5 x_3^2 \right) \boldsymbol{K} (k') - \left( x_1 + x_2 - x_3 \right) \left( x_1 - x_2 + x_3 \right) \left( x_1 - x_2 - x_3 \right) \times \nonumber \\
&&
\times \left( x_1 + x_2 + x_3 \right) \boldsymbol{\Pi} (\alpha_{B},k') \Bigg] + x_2^2 \left( x_3^2-x_1^2 \right) \left( x_1^2 + x_2^2 + x_3^2 \right) \boldsymbol{E} (k') \Bigg\}. \nonumber
\eea
\noindent
Herein, $\boldsymbol{K} (k)$, $\boldsymbol{E} (k)$, and $\boldsymbol{\Pi} (\alpha,k)$ stand for the complete elliptic-integrals of the first, second, and third kinds, respectively (see appendix~\ref{app:elliptic-theta-modular} for definitions). The elliptic modulus $k$, complementary modulus $k'$, and elliptic characteristics $\alpha_{A,B}$, are
\be
k^2 = \frac{x_1^2 \left( x_2^2 - x_3^2 \right)}{x_2^2 \left( x_1^2 - x_3^2 \right)}, \quad k'^2 = \frac{x_3^2 \left( x_1^2 - x_2^2 \right)}{x_2^2 \left( x_1^2 - x_3^2 \right)}, \quad \alpha_{A} = \frac{x_3^2-x_2^2}{x_3^2-x_1^2}, \quad \alpha_{B} = \frac{x_2^2-x_1^2}{x_2^2}.
\ee
\noindent
Using all these results, algebraic and transcendental, we can then numerically solve the original system of equations in \eqref{eq:SpecGeo15a}-\eqref{eq:SpecGeo15f}, valid for any value of the 't~Hooft coupling $t$ within the symmetric three-cut phase of the quartic matrix model. In order to finally write down its spectral curve as in \eqref{eq:spectral-curve-moment-function}, all we need is the moment function which can be easily extracted from \eqref{eq:moment-function-multi-cut} and reads
\be
M(z) = -\frac{\lambda}{6}.
\ee

\paragraph{The Trivalent Phase:}

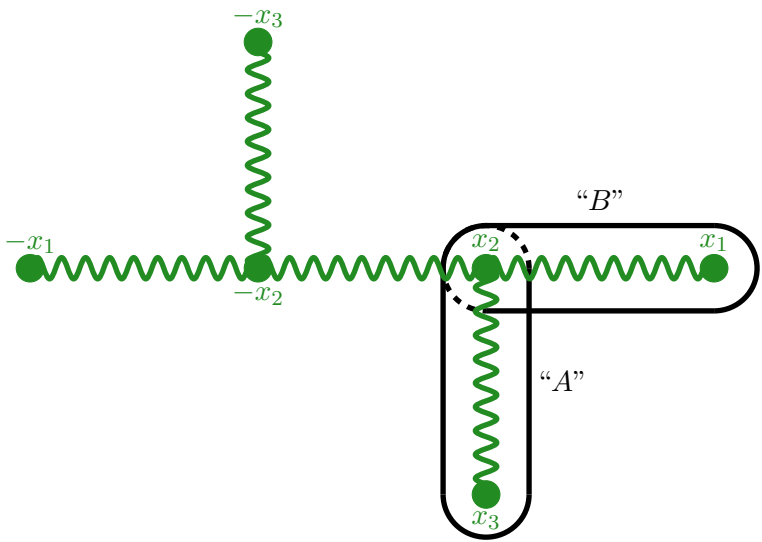
\begin{figure}
\centering
\begin{tikzpicture}
	\draw[line width=2pt] (4.5,0.5656) arc (90:-90:0.5656);
	\draw[line width=2pt] (1.5,0.5656) arc (90:180:0.5656);
	\draw[line width=2pt,dashed] (1.5-0.5656,0) arc (180:270:0.5656);
	\draw[line width=2pt] (1.5,-0.5656) --  (4.5,-0.5656);
	\draw[line width=2pt] (1.5,0.5656) --  (4.5,0.5656);
	\draw[line width=2pt] (1.5+0.5656,-3) arc (0:-180:0.5656);
	\draw[line width=2pt,dashed] (1.5+0.5656,0) arc (0:180:0.5656);
	\draw[line width=2pt] (1.5+0.5656,0) -- (1.5+0.5656,-3);
	\draw[line width=2pt] (1.5-0.5656,0) -- (1.5-0.5656,-3);
	\node at (3,0.9){``$B$''};
	\node at (1.6+0.9,-1.5){``$A$''};
	\draw[ decorate, decoration={snake, segment length=9, amplitude=4},line width=2pt, color = ForestGreen] (-4.5,0)--(-1.5,0);
	\draw[ decorate, decoration={snake, segment length=9, amplitude=4},line width=2pt, color = ForestGreen] (-1.5,0)--(1.5,0);
	\draw[ decorate, decoration={snake, segment length=9, amplitude=4},line width=2pt, color = ForestGreen] (1.5,0)--(4.5,0);
	\draw[ decorate, decoration={snake, segment length=9, amplitude=4},line width=2pt, color = ForestGreen] (-1.5,0)--(-1.5,3);
	\draw[ decorate, decoration={snake, segment length=9, amplitude=4},line width=2pt, color = ForestGreen] (1.5,0)--(1.5,-3);
	\node[ForestGreen] at (1.5,0.34){$x_2$};
	\node[ForestGreen] at (-1.5,-0.34){$-x_2$};
	\node[ForestGreen] at (4.5,0.34){$x_1$};
	\node[ForestGreen] at (-4.5,0.34){$-x_1$};
	\node[ForestGreen] at (1.5,-3-0.34){$x_3$};
	\node[ForestGreen] at (-1.5,3+0.34){$-x_3$};
	\filldraw[ForestGreen] (-4.5,0) circle (1.1ex);
	\filldraw[ForestGreen] (-1.5,0) circle (1.1ex);
	\filldraw[ForestGreen] (1.5,0) circle (1.1ex);
	\filldraw[ForestGreen] (4.5,0) circle (1.1ex);
	\filldraw[ForestGreen] (1.5,-3) circle (1.1ex);
	\filldraw[ForestGreen] (-1.5,3) circle (1.1ex);
\end{tikzpicture}
\caption{Schematic depiction of the trivalent-tree spectral curve $y(z)$ for the quartic matrix model, including a convenient choice of homologically independent ``$A$'' and ``$B$'' cycles, as well as the end-points (and nodes) of the cuts $\pm x_1,\pm x_2,\pm x_3$. The cuts are the \textcolor{ForestGreen}{green} wavy lines.}
\label{fig:SpecGeofig9}
\end{figure}

Next, consider the trivalent phase. This has the exact same number of branch points as the three-cut phase, hence it is described by essentially the same set of equations. In other words, we still need to solve the system of equations in \eqref{eq:SpecGeo15a}-\eqref{eq:SpecGeo15f}, but where, however, the cycles $A$ and $B$ are slightly more intricate (recall the comments concerning homologically independent contours in subsections~\ref{subsec:SG-phases} and~\ref{subappendix:cubicmatrixmodel}). A convenient choice of ``$A$'' and ``$B$'' cycles for the integrals associated to the Boutroux conditions is illustrated in figure~\ref{fig:SpecGeofig9}.

\paragraph{The Symmetric Two-Cut Phase:}

One phase of the quartic matrix model we do consider but do not discuss at such great lengths as the other phases is the symmetric \textit{two}-cut phase. This phase was thoroughly discussed in \cite{sv13}, and is a standard ``Stokes phase''. In fact, we need not resort to the Boutroux conditions in order to determine the location of the associated branch points. As in the familiar one-cut case, also now it is enough to use the equations arising from the asymptotics of the planar resolvent:
\begin{eqnarray}
\oint_{\NCC} \frac{\rmd z}{2\pi\rmi}\, \frac{V'(z)}{\sqrt{(z^2-x_1^2)(z^2-x_2^2)}} &=& 0, \\ 
\oint_{\NCC} \frac{\rmd z}{2\pi\rmi}\, \frac{z\, V'(z)}{\sqrt{(z^2-x_1^2)(z^2-x_2^2)}} &=& 0, \\ 
\oint_{\NCC} \frac{\rmd z}{2\pi\rmi}\, \frac{z^2\, V'(z)}{\sqrt{(z^2-x_1^2)(z^2-x_2^2)}} &=& 2t.
\end{eqnarray}
\noindent
In the usual fashion, one can write the first and third equations (the second one yields a trivially-satisfied relation) above algebraically as
\bea
&&
\lambda \left( x_1^2 + x_2^2 \right) - 12 = 0, \\
&&
3 \lambda \left( x_1^4 + x_2^4 \right) + 2 \lambda x_1^2 x_2^2 - 24 \left( x_1^2 + x_2^2 \right) = - 96 t.
\eea
\noindent
This system has a very simple explicit solution,
\bea
x_1 (t) &=& - \sqrt{\frac{2}{\lambda}}\, \sqrt{3 - \sqrt{-6\lambda t}}, \\ 
x_2 (t) &=& - \sqrt{\frac{2}{\lambda}}\, \sqrt{3 + \sqrt{-6\lambda t}}.
\eea
\noindent
Finally, the moment function reads
\be
M (z) = -\frac{\lambda}{6}\, z.
\ee
\noindent
This has a simple root at the origin, corresponding to the location of the pinched saddle \cite{sv13}.

\paragraph{The Order Parameter:} 

The order parameter for symmetric three-cut and trivalent phases is given by \eqref{eq:SpecGeo9}, which itself is immediately computed by resorting to \eqref{eq:SpecGeo16}. This yields:
\bea
\NCO (t) &=& - \frac{\lambda}{48 t \sqrt{x_2^2 \left( x_3^2-x_1^2 \right)}} \Bigg\{ x_1^2 \Bigg[ x_2^2 \left( x_1^2 + x_2^2 - 5 x_3^2 \right) \boldsymbol{K} (k') - \left( x_1 + x_2 - x_3 \right) \left( x_1 - x_2 + x_3 \right) \times \nonumber \\
&&
\times \left( x_1 - x_2 - x_3 \right) \left( x_1 + x_2 + x_3 \right) \boldsymbol{\Pi} (\alpha_{B},k') \Bigg] + x_2^2 \left( x_3^2-x_1^2 \right) \left( x_1^2 + x_2^2 + x_3^2 \right) \boldsymbol{E} (k') \Bigg\}.
\eea

\section{On the Construction of Transasymptotic Resummations}
\label{app:transasymptotic-transseries}

This appendix contains most calculations which were relevant for the results discussed in section~\ref{sec:resurgent-Z-transseries} concerning rectangular versus diagonal transseries framings, as well as quadratic transasymptotic resummations, discrete Fourier transforms and theta functions, for our several examples. We will begin by addressing the \PI~specific-heat, loaded with computational details, and likewise for its free energy and partition function. That will be enough information to then do the explicit two-parameter transasymptotic resummation. As we move to the examples of \YL, as well as cubic and quartic matrix models, we will sometimes present slightly less computational details.

\subsection{The Painlev\'e~I Equation}
\label{subapp:transasymptotic-transseries-PI}

Let us explicitly address the rewriting of the \PI~transseries into its diagonal framing formulation, for specific heat, free energy, and partition function. In this sub-appendix we will largely use the conventions\footnote{Recall the caveat on the \PI~normalization pointed out in the main text, next to \eqref{eq:Painleve1Equation} in subsection~\ref{subsec:DSL-phases}.} in \cite{asv11, bssv22}, which we recall as $u^2 (z) - \frac{1}{6} u^{\prime\prime} (z) = z$ and $x=z^{-5/4}$.

\paragraph{Specific-Heat Diagonal-Framing:}

In the standard rectangular-framing, the \PI~transseries \eqref{eq:Painleve1SOlution} has the form\footnote{The coefficients we use for the specific-heat herein are shifted with respect to the coefficients used in \cite{asv11}, as
\begin{equation}
\left. u^{(n|m)[k]}_{2g} \right|_{\text{\tiny{here}}} = \left. u^{(n|m)[k]}_{2g+2\beta^{[k]}_{nm}} \right|_{\text{\tiny{there}}}.
\end{equation}
\noindent
This basically implies that our coefficients always start at zero for each transseries sector.}
\begin{equation}
\label{eq:SpecificHeatRectangularFraming}
u \left( x; \sigma_1,\sigma_2 \right) = x^{-2/5} \sum_{n,m=0}^{+\infty} \sigma_{1}^{n} \sigma_{2}^{m}\, \rme^{- \left(n-m\right) \frac{A}{x}} \sum_{k=0}^{k_{nm}} \left( \frac{\log x}{2} \right)^{k}\, \underbrace{\sum_{g=0}^{+\infty} u^{(n|m)[k]}_{2g}\, x^{g+\beta^{[k]}_{nm}}}_{\equiv \Phi_{(n|m)}^{[k]}},
\end{equation}
\noindent
where in comparison with formula \eqref{eq:Painleve1SOlution} from the main text we have expanded the $x^{(\dots)}$ terms in order to make the logarithmic sectors explicit. This form is the original, resurgent version of the transseries, \textit{i.e.}, the one where its sectors $\Phi_{(n|m)}^{[k]}$ are the ones appearing in the corresponding bridge equations \cite{asv11}. The linear $\beta$-structure is \cite{asv11}
\begin{equation}
\label{eq:specificheatP1-beta}
\beta^{[k]}_{nm} = \frac{n+m}{2} - \left\lfloor \frac{k_{nm}+k}{2} \right\rfloor, \qquad k_{nm} = \text{min} \left(n,m\right) - n\, \delta_{nm}.
\end{equation}
\noindent
Furthermore, we have the following relation between logarithmic sectors \cite{asv11}
\begin{equation}
\label{eq:ASV-log-resonance}
\Phi_{(n|m)}^{[k]} = \frac{1}{k!}\, \Big( 2 \alpha\left(m-n\right) \Big)^k\, \Phi_{(n-k|m-k)}^{[0]}, \qquad \alpha=\frac{2}{\sqrt{3}}.
\end{equation}
\noindent
It was already shown in \cite{asv11} that it is possible to reformulate the logarithmic dependence of the above series. Herein, we want to use this fact to introduce new variables that remove it altogether. These are the diagonal framing variables, $\upmu = \sigma_{1}\sigma_{2}$, alongside
\begin{equation}
\label{eq:defzeta}
\upzeta_{1} = \sigma_{1}\, \rme^{-\frac{A}{x}}\, x^{-\frac{2}{\sqrt{3}} \upmu}, \qquad \upzeta_{2} = \sigma_{2}\, \rme^{\frac{A}{x}}\, x^{\frac{2}{\sqrt{3}} \upmu}.
\end{equation}
\noindent
It is useful to further define
\begin{equation}
\label{eq:XiMuRelation}
\upxi_{1} = \sqrt{x}\, \upzeta_{1}, \qquad \upxi_{2} = \sqrt{x}\, \upzeta_{2}, \qquad \upxi_{1} \upxi_{2} = x\, \upmu.
\end{equation}
\noindent
To reformulate the transseries \eqref{eq:SpecificHeatRectangularFraming}, first split the $n$ and $m$ sums into the following three cases:
\begin{itemize}
\item $n=m$, which is the ``main diagonal''. There are no logarithmic sectors, $\beta^{[0]}_{nm} = n$, and we can spell out:
\begin{equation}
\sum_{n=0}^{+\infty} \left( \sigma_{1} \sigma_{2} \right)^{n} \sum_{g=0}^{+\infty} u^{(n|n)[0]}_{2g}\, x^{g+n} = \sum_{g=0}^{+\infty} x^{g}\, \underbrace{\sum_{n=0}^{g} \upmu^{n}\, u^{(n|n)[0]}_{2g-2n}}_{\equiv P_{\text{diag}, g}^{(0)}(\upmu)}.
\end{equation}
\item $n>m$, which are the ``forward sectors''. Let us perform this calculation in full detail, as it stands as the backbone of diagonal framing. Using $\alpha = n-m$, $\beta^{[k]}_{\alpha+m,m} = \frac{\alpha+2m}{2} - \left\lfloor \frac{m+k}{2} \right\rfloor$, and \eqref{eq:ASV-log-resonance}, we may successively start unwinding the Cauchy product and remove logarithms as:
\begin{align}
&
\sum_{n>m}^{+\infty} \sigma_{1}^{n} \sigma_{2}^{m}\, \rme^{- (n-m)\, \frac{A}{x}}\, \sum_{k=0}^{k_{nm}} \left( \frac{\log x}{2} \right)^{k} \Phi_{(n|m)}^{[k]} = \nonumber \\
&
= \sum_{\alpha=1}^{+\infty} \sum_{m=0}^{+\infty} \sum_{k=0}^{m} \sigma_{1}^{\alpha} \left( \sigma_{1} \sigma_{2} \right)^{m-k}\, \rme^{-\alpha\, \frac{A}{x}}\, \left( \frac{\log x}{2} \right)^{k} \frac{1}{k!} \left( - \frac{4\alpha}{\sqrt{3}} \right)^k \left( \sigma_{1} \sigma_{2} \right)^{k} \Phi_{(\alpha+m-k|m-k)}^{[0]} = \nonumber \\
&
= \sum_{\alpha=1}^{+\infty} \sum_{m=0}^{+\infty} \sigma_{1}^{\alpha} \left( \sigma_{1} \sigma_{2} \right)^{m} \rme^{-\alpha\, \frac{A}{x}}\, \Phi_{(\alpha+m|m)}^{[0]} \sum_{k=0}^{+\infty} \frac{1}{k!} \left( - \frac{2\alpha}{\sqrt{3}}\, \log (x)\, \sigma_{1} \sigma_{2} \right)^{k} = \nonumber \\
&
= \sum_{\alpha=1}^{+\infty} \sigma_{1}^{\alpha}\, \rme^{-\alpha\, \frac{A}{x}}\, x^{-\frac{2}{\sqrt{3}}\, \alpha\, \sigma_{1}\sigma_{2}} \sum_{m=0}^{+\infty} \left( \sigma_{1} \sigma_{2} \right)^{m} \Phi_{(\alpha+m|m)}^{[0]} = \nonumber \\
&
\equiv \sum_{\alpha=1}^{+\infty} \upzeta_{1}^{\alpha} \sum_{m=0}^{+\infty} \upmu^{m}\, \Phi_{(\alpha+m|m)}^{[0]}.
\end{align}
\noindent
These manipulations have completely reformulated the logarithmic dependence in \eqref{eq:SpecificHeatRectangularFraming}, and in the process naturally introduced the diagonal variables. To do proper transasymptotics, we still want to organize the above primarily by powers of $x$. As such, expand the asymptotic sectors and again rewrite the Cauchy product as:
\begin{align}
\label{eq:SpecificHeatRectangularFramingNoLogarithms}
&
\sum_{\alpha=1}^{+\infty} \upzeta_{1}^{\alpha} \sum_{m=0}^{+\infty} \upmu^{m}\, \Phi_{(\alpha+m|m)}^{[0]} = \nonumber \\
&
= \sum_{\alpha=1}^{+\infty} \upzeta_{1}^{\alpha} \sum_{m=0}^{+\infty} \upmu^{m} \sum_{g=0}^{+\infty} u^{(\alpha+m|m)[0]}_{2g}\, x^{g + \beta^{[0]}_{\alpha+m, m}} = \nonumber \\
&
= \sum_{\alpha=1}^{+\infty} \upxi_{1}^{\alpha} \sum_{m=0}^{+\infty} \upmu^{m} \sum_{g=0}^{+\infty} u^{(\alpha+m|m)[0]}_{2g}\, x^{g + m - \left\lfloor\frac{m}{2}\right\rfloor} = \nonumber \\
&
= \sum_{\alpha=1}^{+\infty} \upxi_{1}^{\alpha} \sum_{g=0}^{+\infty} \left\{ \sum_{m=0}^{+\infty} \upmu^{2m}\, u^{(\alpha+2m|2m)[0]}_{2g}\, x^{g+m} + \sum_{m=0}^{+\infty} \upmu^{2m+1}\, u^{(\alpha+2m+1|2m+1)[0]}_{2g}\, x^{g+m+1} \right\} = \nonumber \\
&
= \sum_{\alpha=1}^{+\infty} \upxi_{1}^{\alpha} \sum_{g=0}^{+\infty} \left\{ \sum_{m=0}^{g} \upmu^{2m}\, u^{(\alpha+2m|2m)[0]}_{2g-2m}\, x^{g} + \sum_{m=0}^{g-1} \upmu^{2m+1}\, u^{(\alpha+2m+1|2m+1)[0]}_{2g-2m-2}\, x^{g} \right\} = \nonumber \\
&
= \sum_{g=0}^{+\infty} x^{g} \sum_{\alpha=1}^{+\infty} \upxi_{1}^{\alpha} \underbrace{\left\{ \sum_{m=0}^{g} \upmu^{2m}\, u^{(\alpha+2m|2m)[0]}_{2g-2m} + \sum_{m=0}^{g-1} \upmu^{2m+1}\, u^{(\alpha+2m+1|2m+1)[0]}_{2g-2m-2} \right\}}_{\equiv P^{(\alpha)}_{\text{forw}, g}(\upmu)}.
\end{align}
\noindent
Note that the sum over powers of $\upmu$ is now \textit{finite}---hence \textit{polynomial}.
\item $n<m$, which are the ``backward sectors''. This is a completely analogous calculation:
\begin{align}
&
\sum_{n<m}^{+\infty} \sigma_{1}^{n} \sigma_{2}^{m}\, \rme^{- (n-m)\,\frac{A}{x}}\, \sum_{k=0}^{k_{nm}} \left( \frac{\log x}{2} \right)^{k} \Phi_{(n|m)}^{[k]} = \nonumber \\
&
= \sum_{g=0}^{+\infty} x^{g} \sum_{\alpha=1}^{+\infty} \upxi_{2}^{\alpha} \underbrace{\left\{ \sum_{m=0}^{g} \upmu^{2m}\, u^{(2m|\alpha+2m)[0]}_{2g-2m} + \sum_{m=0}^{g-1} \upmu^{2m+1}\, u^{(2m+1|\alpha+2m+1)[0]}_{2g-2m-2} \right\}}_{\equiv P^{(\alpha)}_{\text{back}, g}(\upmu)}.
\end{align}
\end{itemize}

After all this, we may put the transseries \eqref{eq:SpecificHeatRectangularFraming} back together. It now has the form
\begin{equation}
\label{eq:PreDiagonalSpecificHeatThreeVariables}
u \left( x; \upxi_{1},\upxi_{2},\upmu \right) = x^{-2/5} \sum_{g=0}^{+\infty} x^{g} \left\{ P^{(0)}_{\text{diag}, g}(\upmu) + \sum_{\alpha=1}^{+\infty} \upxi_{1}^{\alpha}\, P^{(\alpha)}_{\text{forw}, g}(\upmu) + \sum_{\alpha=1}^{+\infty} \upxi_{2}^{\alpha}\, P^{(\alpha)}_{\text{back}, g}(\upmu) \right\}.
\end{equation}
\noindent
This result can still be further simplified. The specific-heat rectangular-framing asymptotic-coefficients enjoy a reflection symmetry \cite{asv11} (which is the backward-forward symmetry discussed in \cite{bssv22}),
\begin{equation}
u^{(n|m)[0]}_{2g} = (-1)^{g+\left\lfloor\frac{m}{2}\right\rfloor}\, u^{(m|n)[0]}_{2g}, \qquad n>m,
\end{equation}
\noindent
in which case this translates to diagonal framing as
\begin{equation}
P_{\text{forw}, g}^{(\alpha)}(\upmu) = (-1)^{g}\, P_{\text{back},g}^{(\alpha)}(-\upmu).
\end{equation} 
\noindent
For convenience, further define
\begin{equation}
P^{(\alpha)}_{g}(\upmu) \equiv 
\begin{cases}
P_{\text{forw},g}^{(\alpha)} (\upmu), \quad & \alpha>0, \\[3pt]
P_{\text{diag},g}^{(0)} (\upmu), \quad & \alpha=0, \\[3pt]
P_{\text{back},g}^{(-\alpha)} (\upmu), \quad & \alpha<0.
\end{cases}
\end{equation}
\noindent
The backward-forward symmetry is now rewritten simply as:
\begin{equation}
P_{g}^{(\alpha)}(\upmu) = (-1)^{g}\, P_{g}^{(-\alpha)}(-\upmu), \qquad \alpha > 0.
\end{equation}
\noindent
It is also useful to know how to invert this change of variables and get back the rectangular coefficients from the diagonal ones. This simply reads\footnote{The notation $\left. \bullet \right|_{m}$ selects the degree-$m$ term in $\upmu$, out from the $\bullet$-expression.}
\bea
u_{2g}^{(n|m)[0]} &=& \left. P^{(n-m)}_{g+\left\lfloor\frac{m+1}{2}\right\rfloor} (\upmu) \right|_{m}, \qquad n-m>0, \\
u_{2g}^{(n|n)[0]} &=& \left. P^{(0)}_{g+n} (\upmu) \right|_{n}.
\eea
\noindent
In this way, the explicit writing of the specific-heat in \eqref{eq:PreDiagonalSpecificHeatThreeVariables} has simplified, to
\begin{equation}
\label{eq:DiagonalSpecificHeatThreeVariables}
u \left( x; \upxi_{1},\upxi_{2},\upmu \right) = x^{-2/5} \sum_{g=0}^{+\infty} x^{g} \left\{ \sum_{\alpha=0}^{+\infty} \upxi_{1}^{\alpha}\, P^{(\alpha)}_{g} (\upmu) + \sum_{\alpha=1}^{+\infty} \upxi_{2}^{\alpha}\, P^{(-\alpha)}_{g} (\upmu) \right\}.
\end{equation}

Note, however, how the above transseries is explicitly depending upon three boundary/initial data, or diagonal transseries parameters. Of course, this set can obviously be reduced as we started off with only two, $\sigma_1$ and $\sigma_2$. This is simple to do using \eqref{eq:XiMuRelation} and performing yet another Cauchy product on the last term of \eqref{eq:DiagonalSpecificHeatThreeVariables} as
\bea
u \left( x; \upxi_{1},\upxi_{2},\upmu \right) &=& x^{-2/5} \left\{\, \cdots\, \right\} + x^{-2/5} \sum_{g=0}^{+\infty} \sum_{\alpha=1}^{+\infty} \frac{\upmu^{\alpha}\, x^{g+\alpha}}{\upxi_1^{\alpha}}\, P^{(-\alpha)}_{g} (\upmu) = \nonumber \\
&=& x^{-2/5} \left\{\, \cdots\, \right\} + x^{-2/5} \sum_{g=0}^{+\infty} \sum_{\alpha=-g}^{-1}x^{g}\, \upxi_{1}^{\alpha}\, \upmu^{-\alpha}\, P^{(\alpha)}_{g+\alpha} (\upmu).
\eea
\noindent
With this result in hand we may finally define what we shall call the diagonal-framing  coefficients for the specific-heat, as
\begin{equation}
\mathsf{u}^{(\alpha)}_{g} (\upmu) \equiv 
\begin{cases}
P^{(\alpha)}_{g} (\upmu), \qquad & \alpha \geq 0, \\
\upmu^{-\alpha} P^{(\alpha)}_{g+\alpha} (\upmu), \qquad & -g \leq \alpha \leq -1, \\
0, \qquad & \text{otherwise}.
\end{cases}
\end{equation}
\noindent
We list a few explicit polynomial-coefficients for completeness; both along the main diagonal where $\deg \mathsf{u}^{(0)}_{g} = g$,
\begin{align}
\mathsf{u}^{(0)}_0 &= 1, \qquad \mathsf{u}^{(0)}_1 = -\upmu, \qquad \mathsf{u}^{(0)}_2 = -\frac{1}{48}-\frac{5}{6}\upmu^2, \qquad \mathsf{u}^{(0)}_3 = -\frac{75}{512}\upmu-\frac{235}{192}\upmu^3,\\ \mathsf{u}^{(0)}_4 &= -\frac{49}{4608}-\frac{54425}{82944}\upmu^2-\frac{270095}{124416}\upmu^4,
\end{align}
\noindent
as well as some off-diagonal entries where now $\deg \mathsf{u}^{(\alpha)}_{g} = 2g$,
\begin{align}
\mathsf{u}^{(1)}_0 &= 1, \qquad \mathsf{u}^{(1)}_1 = - \frac{5}{64\sqrt{3}} + \frac{11}{72}\, \upmu - \frac{47}{24\sqrt{3}}\, \upmu^2, \\
\mathsf{u}^{(1)}_2 &= \frac{75}{8192} - \frac{985}{4608\sqrt{3}}\, \upmu + \frac{4213}{20736}\, \upmu^2 - \frac{2317}{1296\sqrt{3}}\, \upmu^3 + \frac{2209}{3456}\, \upmu^4, \\
\mathsf{u}^{(2)}_0 &= \frac{1}{6}, \qquad \mathsf{u}^{(2)}_1 = - \frac{55}{576\sqrt{3}} + \frac{3}{16}\, \upmu - \frac{47}{72\sqrt{3}}\, \upmu^2, \\
\mathsf{u}^{(2)}_2 &= \frac{1325}{36864} - \frac{4355}{10368\sqrt{3}}\, \upmu + \frac{54415}{124416}\, \upmu^2 - \frac{9569}{7776\sqrt{3}}\, \upmu^3 + \frac{2209}{5184}\, \upmu^4, \\
\mathsf{u}^{(3)}_0 &= \frac{1}{48}, \qquad \mathsf{u}^{(3)}_1 = - \frac{215}{9216\sqrt{3}} + \frac{101}{2304}\, \upmu - \frac{47}{384\sqrt{3}}\, \upmu^2, \\
\mathsf{u}^{(3)}_2 &= \frac{40775}{3538944} - \frac{15361}{147456\sqrt{3}}\, \upmu + \frac{7795}{55296}\, \upmu^2 - \frac{58157}{165888\sqrt{3}}\, \upmu^3 + \frac{2209}{18432}\, \upmu^4.
\end{align}
\noindent
For these coefficients, the backward-forward symmetry manifests itself as
\begin{equation}
\mathsf{u}^{(-\alpha)}_{g} (\upmu) = (-1)^{g-\alpha}\, \upmu^{\alpha}\, \mathsf{u}^{(\alpha)}_{g-\alpha} (-\upmu), \qquad \alpha > 0.
\end{equation}
\noindent
The specific-heat transseries in diagonal framing is finally written as
\begin{equation}
\label{eq:PainleveISolutionDiagonalFraming}
u \left( x; \upxi_{1}, \upmu \right) = x^{-2/5} \sum_{g=0}^{+\infty} \sum_{\alpha=-g}^{+\infty} \upxi_{1}^{\alpha}\, \mathsf{u}^{(\alpha)}_{g} (\upmu)\, x^{g}.
\end{equation}
\noindent
It is worth stressing once again that the coefficients $\mathsf{u}^{(\alpha)}_{g} (\upmu)$ are polynomials in $\upmu$. It is also useful to know how to invert this formula to get back the rectangular coefficients from the above diagonal ones; which reads
\bea
u_{2g}^{(n|m)[0]} &=& \left. \mathsf{u}^{(n-m)}_{g+\left\lfloor\frac{m+1}{2}\right\rfloor} (\upmu) \right|_{m}, \qquad n-m > 0, \\
u_{2g}^{(n|n)[0]} &=& \left. \mathsf{u}^{(0)}_{g+n} (\upmu) \right|_{n}.
\eea

\paragraph{A Diagonal Recursion:}

As a quick side note, let us remark that one could also have directly solved \PI~in diagonal framing. To see how this goes, rewrite the \PI~equation as given\footnote{Where $w = x^{1/2} = z^{-5/8}$ and $u(w) \equiv \left. \frac{u(z)}{\sqrt{z}} \right|_{z=w^{-8/5}}$.} in \cite{asv11},
\begin{equation}
u(w)^2 + \frac{1}{24} w^4\, u(w) - \frac{25}{384} w^5\, u^{\prime}(w) - \frac{25}{384} w^6\, u^{\prime\prime}(w) = 1,
\end{equation}
\noindent
and make the \textit{ansatz}
\begin{equation}
u(w) = \sum_{g=0}^{+\infty} w^{2g} \left\{ \sum_{s=-g}^{+\infty} \upzeta_1^{s}\, \mathsf{u}_{g}^{(s)} (\upmu) \right\},
\end{equation}
\noindent
where $\upzeta_1$ herein keeps the definition in \eqref{eq:defzeta}. This leads to the recursive relation
\begin{equation}
\label{eq:DiagonalRecursionRelation}
\sum_{\tilde{g}=0}^{g} \sum_{\tilde{s} = - \tilde{g}}^{s+g-\tilde{g}} \mathsf{u}^{(s-\tilde{s})}_{g-\tilde{g}} (\upmu)\, \mathsf{u}^{(\tilde{s})}_{\tilde{g}}(\upmu) + p_{0,g}^{(s)}\, \mathsf{u}_{g}^{(s)}(\upmu) + p_{1,g}^{(s)}\, \mathsf{u}_{g-1}^{(s)}(\upmu) + p_{2,g}^{(s)}\, \mathsf{u}_{g-2}^{(s)}(\upmu) = \delta_{s0}\,\delta_{g0},
\end{equation}
\noindent
where we have defined
\bea
p_{0,g}^{(s)} &=& - \frac{25}{384}\, c_{6, g}^{(s)}(\upmu), \\
p_{1,g}^{(s)} &=& - \frac{25}{384} \left( 2 s A + c_{4,g-1}^{(s)}(\upmu) \right), \\
p_{2,g}^{(s)} &=& \frac{1}{24} - \frac{25}{384} \left( 2 \left( g-2 \right) + s \left( 1-\frac{4}{\sqrt{3}}\, \upmu \right) + c_{2,g-2}^{(s)}(\upmu) \right),
\eea
\noindent
alongside the coefficients
\begin{align}
c_{2,g}^{(s)} &= 2 g \left( 2g-1 \right) + 4 g s \left( 1-\frac{4}{\sqrt{3}}\, \upmu \right) + s \left( s-1 \right) \left( 1-\frac{4}{\sqrt{3}}\, \upmu \right)^2 + s \left( \frac{16}{3}\, \upmu^2-\frac{4}{\sqrt{3}}\, \upmu \right), \\
c_{4,g}^{(s)} &= 8 g s A + 4 s^2 A \left( 1-\frac{4}{\sqrt{3}}\, \upmu \right) - 6 s A, \\
c_{6,g}^{(s)} &= \left( 2 s A \right)^2.
\end{align}
\noindent
Indeed this recursion produces exactly the diagonal coefficients above.

\paragraph{Free-Energy Diagonal-Framing:}

The \PI~free energy is simply given by double-integration of the specific-heat as in\footnote{Again, recall the caveat on the \PI~normalization pointed out in the main text, next to \eqref{eq:Painleve1Equation} in subsection~\ref{subsec:DSL-phases}.} \eqref{eq:FdsEVENnormalization},
\begin{equation}
F_{\text{\PI}} \left( z; \sigma_1,\sigma_2 \right) = - \iint \rmd^2 z\, u \left( z; \sigma_1,\sigma_2 \right).
\end{equation}
\noindent
The structure of this free energy was already addressed in \cite{asv11}. Herein we again work with the $x$-variable (as for the specific heat). In this case, the free energy has a finite number of terms in inverse powers of $x$ (which we remove from the general sum) alongside the usual logarithmic dependence in $x$ (which we keep in the general sum). Let us make this explicit in equations, writing the free-energy rectangular-framing transseries\footnote{Note that, and analogously to what happened for the specific-heat, the coefficients we use herein for the free energy are not exactly the same as those in \cite{asv11}, but relate as
\begin{equation}
\left. F^{(n|m)[0]}_{2g} \right|_{\text{\tiny{here}}} = \left. F^{(n|m)}_{g} \right|_{\text{\tiny{there}}}.
\end{equation}} as\footnote{Herein ``A'' stands for ``atypical'' as these terms usually require distinct analysis from all (infinite) others.}
\bea
\label{eq:freeEnergyPainleveI}
F \left( x; \sigma_1,\sigma_2 \right) &=& F_{\text{A}} \left( x; \sigma_{1}\sigma_{2} \right) + \\
&&
+ \sum_{n,m=0}^{+\infty} \sigma_{1}^{n} \sigma_{2}^{m}\, \rme^{- \left(n-m\right) \frac{A}{x}} \sum_{k=0}^{k_{nm}} \left( \frac{\log x}{2} \right)^{k}\, \underbrace{\sum_{g=0}^{+\infty} F^{(n|m)[k]}_{2g}\, x^{g+\beta^{[k]}_{{\tiny{F}},nm}}}_{\equiv F_{(n|m)}^{[k]}}, \nonumber
\eea
\noindent
where the free-energy $\beta$-structure is basically dictated by the one from the specific-heat in \eqref{eq:specificheatP1-beta}, as
\begin{equation}
\beta^{[k]}_{F,nm} = 
\begin{cases}
\beta^{[k]}_{nn} - 2, \qquad & n=m,\\[3pt]
\beta^{[k]}_{nm}, \qquad & n\neq m.
\end{cases}
\end{equation}
\noindent
Further, we have gathered all singular terms in $x$ in the pre-factor $F_{\text{A}} \left( x; \sigma_{1}\sigma_{2} \right)$. This includes terms arising from both $(0|0)$ and the $(1|1)$ sectors. Explicitly, it is given by
\begin{equation}
\label{eq:FAxmu-sum}
F_{\text{A}} \left( x; \sigma_{1}\sigma_{2} \right) = - \frac{4}{15 x^2} + \frac{16}{5x}\, \upmu + \frac{4}{5} \left( \frac{1}{48} + \frac{5}{6}\, \upmu^2 \right) \log x.
\end{equation}
\noindent
In addition we have the backward-forward relation for the free energy,
\begin{equation}
\label{eq:back-forw-PI-rectangularF}
F^{(n|m)[0]}_{2g} = (-1)^{g+\left\lfloor\frac{m}{2}\right\rfloor}\, F^{(m|n)[0]}_{2g}, \qquad n>m.
\end{equation}
\noindent
In complete analogy with what we did for the specific-heat we can straightforwardly introduce diagonal-framing for the free-energy as:
\begin{align}
P^{(\alpha)}_{F,g}(\upmu) \equiv
\begin{cases}
\sum\limits_{m=0}^{g} \upmu^{2m}\, F^{(\alpha+2m|2m)[0]}_{2g-2m} + \sum\limits_{m=0}^{g-1} \upmu^{2m+1}\, F^{(\alpha+2m+1|2m+1)[0]}_{2g-2m-2}, \qquad & \alpha>0,\\
\sum\limits_{n=0}^{g+2} \upmu^n\, F^{(n|n)[0]}_{2g-2n+4}, \qquad & \alpha=0,\\
\sum\limits_{m=0}^{g} \upmu^{2m}\, F^{(2m|-\alpha+2m)[0]}_{2g-2m} + \sum\limits_{m=0}^{g-1} \upmu^{2m+1}\, F^{(2m+1|-\alpha+2m+1)[0]}_{2g-2m-2}, \qquad & \alpha<0,
\end{cases}
\end{align}
\noindent
and where the backward-forward relation reads
\begin{equation}
P_{F,g}^{(\alpha)}(\upmu) = (-1)^{g}\, P_{F,g}^{(-\alpha)}(-\upmu), \qquad \alpha>0.
\end{equation}
\noindent
With this, it is now immediate to rewrite the free energy \eqref{eq:freeEnergyPainleveI} in diagonal-framing,
\be
F \left( x; \sigma_1,\sigma_2 \right) = F_{\text{A}} \left( x; \upmu \right) + \sum_{g=0}^{+\infty} x^{g} \left\{ \sum_{\alpha=0}^{+\infty} \upxi_{1}^{\alpha}\, P^{(\alpha)}_{F,g}(\upmu) + \sum_{\alpha=1}^{+\infty} \upxi_{2}^{\alpha}\, P^{(-\alpha)}_{F,g}(\upmu) \right\}.
\ee
\noindent
We can now repeat \textit{ipsis verbis} the diagonal-framing discussion of the specific-heat, expressing $\upxi_2$ via $\upxi_1$. Introduce
\begin{equation}
\mathsf{F}^{(\alpha)}_{g} (\upmu) \equiv \begin{cases}
P^{(\alpha)}_{F,g} (\upmu), \qquad & \alpha \geq 0, \\
\upmu^{-\alpha}\, P^{(\alpha)}_{F,g+\alpha} (\upmu), \qquad & -g \leq \alpha \leq -1, \\
0, \qquad & \text{otherwise},
\end{cases}
\end{equation}
\noindent
with backward-forward symmetry manifesting itself as
\begin{equation}
\label{eq:back-forw-PI-diagonalF}
\mathsf{F}^{(-\alpha)}_{g} (\upmu) = (-1)^{g-\alpha}\, \upmu^{\alpha}\, \mathsf{F}^{(\alpha)}_{g-\alpha} (-\upmu), \qquad \alpha > 0.
\end{equation}
\noindent
The free-energy transseries in diagonal framing is finally written as
\begin{equation}
\label{eq:PainleveIFreeEnergyDiagonalFraming}
F \left( x; \upxi_{1}, \upmu \right) = F_{\text{A}} \left( x; \upmu \right) + \sum_{g=0}^{+\infty} \sum_{\alpha=-g}^{+\infty} \upxi_{1}^{\alpha}\, \mathsf{F}^{(\alpha)}_{g} (\upmu)\, x^{g}.
\end{equation}
\noindent
All we lack is an inversion of these formulae back to rectangular framing. This is done via
\bea
F_{2g}^{(n|m)[0]} &=& \left. \mathsf{F}^{(n-m)}_{g+\left\lfloor\frac{m+1}{2}\right\rfloor} (\upmu) \right|_{m}, \qquad n-m > 0, \\
F_{2g}^{(n|n)[0]} &=& \left. \mathsf{F}^{(0)}_{g+n-2} (\upmu) \right|_{n}.
\eea
\noindent
Note in addition that any $\mathsf{F}^{(\alpha)}_g$ with a negative $g$-entry vanishes.

\paragraph{Partition-Function Diagonal-Framing:}

The calculations are, again, essentially the same. The partition-function in rectangular-framing reads
\begin{equation}
\label{eq:PainleveIPartitionFunctionRectangularApp}
Z \left( x; \sigma_1,\sigma_2 \right) = \rme^{F_{\text{A}} \left(x;\upmu\right)}\, \sum_{n,m=0}^{+\infty} \sigma_{1}^{n} \sigma_{2}^{m}\, \rme^{- \left(n-m\right) \frac{A}{x}} \sum_{k=0}^{k_{nm}} \left( \frac{\log x}{2} \right)^{k} \sum_{g=0}^{+\infty} Z^{(n|m)[k]}_{2g}\, x^{g+\beta^{[k]}_{nm}},
\end{equation}
\noindent
where for the moment we will naively assume that the partition function keeps the same $\beta$-structure as for the specific heat or free energy. We will soon refine upon this structure. Then, in exactly the same way as for the two previous diagonal-framing calculations, we can spell out the diagonal-framing version of the partition function via
\begin{align}
P^{(\alpha)}_{Z,g}(\upmu) \equiv
\begin{cases}
\sum\limits_{m=0}^{g} \upmu^{2m}\, Z^{(\alpha+2m|2m)[0]}_{2g-2m} + \sum\limits_{m=0}^{g-1} \upmu^{2m+1}\, Z^{(\alpha+2m+1|2m+1)[0]}_{2g-2m-2}, \qquad & \alpha>0, \\
\sum\limits_{n=0}^{g} \upmu^n\, Z^{(n|n)[0]}_{2g-2n}, \qquad & \alpha=0, \\
\sum\limits_{m=0}^{g} \upmu^{2m}\, Z^{(2m|-\alpha+2m)[0]}_{2g-2m} + \sum\limits_{m=0}^{g-1} \upmu^{2m+1}\, Z^{(2m+1|-\alpha+2m+1)[0]}_{2g-2m-2}, \qquad & \alpha<0,
\end{cases}
\end{align}
\noindent
and where the backward-forward relation reads
\begin{equation}
P_{Z,g}^{(\alpha)}(\upmu) = (-1)^{g}\, P_{Z,g}^{(-\alpha)}(-\upmu), \qquad \alpha>0.
\end{equation}
\noindent
Having done this, it is straightforward to rewrite the partition function \eqref{eq:PainleveIPartitionFunctionRectangularApp} in diagonal framing,
\be
Z \left( x; \sigma_1,\sigma_2 \right) = \rme^{F_{\text{A}} \left( x; \upmu \right)}\, \sum_{g=0}^{+\infty} x^{g} \left\{ \sum_{\alpha=0}^{+\infty} \upxi_{1}^{\alpha}\, P^{(\alpha)}_{Z,g}(\upmu) + \sum_{\alpha=1}^{+\infty} \upxi_{2}^{\alpha}\, P^{(-\alpha)}_{Z,g}(\upmu) \right\}.
\ee
\noindent
It is also straightforward to express $\upxi_2$ via $\upxi_1$, and hence introduce
\begin{equation}
\mathsf{Z}^{(\alpha)}_{g} (\upmu) \equiv 
\begin{cases}
P^{(\alpha)}_{Z,g} (\upmu), \qquad & \alpha \geq 0, \\
\upmu^{-\alpha}\, P^{(\alpha)}_{Z,g+\alpha} (\upmu), \qquad & -g \leq \alpha \leq -1, \\
0, \qquad & \text{otherwise},
\end{cases}
\end{equation}
\noindent
for which the backward-forward symmetry manifests as
\begin{equation}
\mathsf{Z}^{(-\alpha)}_{g} (\upmu) = (-1)^{g+\alpha}\, \upmu^{\alpha}\, \mathsf{Z}^{(\alpha)}_{g-\alpha} (-\upmu), \qquad \alpha>0.
\end{equation}
\noindent
The partition-function transseries in diagonal framing is finally written as
\begin{equation}
\label{eq:PainleveIPartitionFunctionDiagonalFramingApp}
Z \left( x; \upxi_{1}, \upmu \right) = \rme^{F_{\text{A}} \left( x; \upmu \right)}\, \sum_{g=0}^{+\infty} \sum_{\alpha=-g}^{+\infty} \upxi_{1}^{\alpha}\, \mathsf{Z}^{(\alpha)}_{g} (\upmu)\, x^{g}.
\end{equation}
\noindent
As always, we end with formulae for the inverse procedure, which are simply
\bea
Z_{2g}^{(n|m)[0]} &=& \left. \mathsf{Z}^{(n-m)}_{g+\left\lfloor\frac{m+1}{2}\right\rfloor} (\upmu) \right|_{m}, \qquad n-m > 0, \\
Z_{2g}^{(n|n)[0]} &=& \left. \mathsf{Z}^{(0)}_{g+n} (\upmu) \right|_{n}.
\eea

Having discussed the general structure, we now turn to the explicit (recursive) evaluation of the partition function. Because it arises from the free energy by exponentiation, we can explicitly split it as
\begin{equation}
Z \left( x; \upxi_{1}, \upmu \right) = \rme^{F_{\text{A}} \left( x; \upmu \right)}\, \rme^{\mathsf{F}_{0} \left( \upxi_{1}, \upmu \right)}\, \exp \left( \sum_{g=1}^{+\infty} \sum_{\alpha=-g}^{+\infty} \upxi_{1}^{\alpha}\, \mathsf{F}^{(\alpha)}_{g} (\upmu)\, x^{g} \right),
\end{equation}
\noindent
where we use
\begin{equation}
\mathsf{F}_{0} \left( \upxi_{1}, \upmu \right) = \sum_{\alpha=0}^{+\infty} \upxi_{1}^{\alpha}\, \mathsf{F}^{(\alpha)}_{0} (\upmu).
\end{equation}
\noindent
We can address the above three terms separately, as:
\begin{itemize}
\item $\rme^{F_{\text{A}} \left( x; \upmu \right)}$: This is just a finite sum in $x$ and $\upmu$, as already explicitly evaluated in \eqref{eq:FAxmu-sum}. We leave it as a general pre-factor, which will turn out to be very convenient in the following.
\item $\rme^{\mathsf{F}_{0} \left( \upxi_{1}, \upmu \right)}$: At genus $g=0$ this contribution of the free energy can be simply resummed
\begin{equation}
\mathsf{F}_{0} \left( \upxi_{1}, \upmu \right) = \log \left( 1 + \frac{\upxi_1}{12} \right),
\end{equation}
\noindent
with no $\upmu$-dependence at this lowest order, and from where immediately follows
\be
\rme^{\mathsf{F}_{0} \left( \upxi_{1}, \upmu \right)} = 1 + \frac{\upxi_1}{12}.
\ee
\item $\rme^{\big( \,\cdots\, \big)}$: For this term, we have to explicitly perform the exponentiation as
\begin{equation}
\exp \left( \sum_{g=1}^{+\infty} \sum_{\alpha=-g}^{+\infty} \upxi_{1}^{\alpha}\, \mathsf{F}^{(\alpha)}_{g} (\upmu)\, x^{g} \right) = \sum_{g=0}^{+\infty} \sum_{\alpha=-g}^{+\infty} \upxi_1^\alpha\, \mathsf{b}^{(\alpha)}_{g} (\upmu)\, x^{g}.
\end{equation}
\noindent
Herein, the coefficients $\mathsf{b}^{(\alpha)}_{g}$ are calculated via a Bell-polynomial-like recursion,
\begin{equation}
\mathsf{b}^{(\alpha)}_{n} (\upmu) = \sum_{i=0}^{n-1} \frac{i+1}{n}\, \sum_{m=0}^{\alpha+n} \mathsf{b}^{(m+i+1-n)}_{n-1-i} (\upmu)\, \mathsf{F}^{(\alpha+n-m-i-1)}_{i+1} (\upmu),
\end{equation}
\noindent
with the initial recursion coefficients
\begin{equation}
\mathsf{b}^{(0)}_{0} = 1, \qquad \mathsf{b}^{(\alpha)}_{0} = 0, \quad \alpha \geq 1.
\end{equation}
\end{itemize}
\noindent
The last two expressions may actually be combined as
\begin{equation}
\rme^{\mathsf{F}_{0} \left( \upxi_{1}, \upmu \right)}\, \exp \left( \sum_{g=1}^{+\infty} \sum_{\alpha=-g}^{+\infty} \upxi_{1}^{\alpha}\, \mathsf{F}^{(\alpha)}_{g} (\upmu)\, x^{g} \right) \equiv \sum_{g=0}^{+\infty} \sum_{\alpha=-g}^{+\infty} \upxi_{1}^{\alpha}\, \mathsf{Z}^{(\alpha)}_{g} (\upmu)\, x^{g},
\end{equation}
\noindent
where, at the component level, we can relate coefficients as
\begin{equation}
\mathsf{Z}^{(\alpha)}_{g} (\upmu) = \mathsf{b}^{(\alpha)}_{g} (\upmu) - \frac{1}{12}\, \mathsf{b}^{(\alpha-1)}_{g} (\upmu),
\end{equation}
\noindent
and where $\mathsf{Z}^{(\alpha)}_{g} (\upmu)$ are the coefficients from \eqref{eq:PainleveIPartitionFunctionDiagonalFramingApp}. Analogously to the $\mathsf{b}^{(\alpha)}_{g}$ coefficients, the $\mathsf{Z}^{(\alpha)}_{g}$ coefficients can be calculated via a recursion
\begin{equation}
\mathsf{Z}^{(\alpha)}_{n} (\upmu) = \sum_{i=0}^{n-1} \frac{i+1}{n}\, \sum_{k=0}^{\alpha+n} \mathsf{Z}^{(\alpha-k+i+1)}_{n-1-i} (\upmu)\, \mathsf{F}^{(k-i-1)}_{i+1} (\upmu),
\end{equation}
\noindent
with the fixed initial data
\begin{equation}
\mathsf{Z}^{(\alpha)}_{0} = \delta_{\alpha 0} - \frac{1}{12}\, \delta_{\alpha 1}.
\end{equation} 
\noindent
For the convenience of the reader we present some explicit coefficients herein, where\footnote{The main information arises from $\alpha=0$ as this alone dictates all the $\widetilde{D}_{k} (\nu)$ polynomials in \eqref{eq:diagonalframingresultapp}.}, \textit{e.g.}, $\deg \mathsf{Z}^{(0)}_{g} = 3g$,
\begin{align}
\label{eq:Zmu-poly-examples-1}
\mathsf{Z}^{(0)}_{0} (\upmu) &= 1, \qquad \mathsf{Z}^{(0)}_{1} (\upmu) = \frac{17}{288}\, \upmu + \frac{47}{108}\, \upmu^3, \\
\label{eq:Zmu-poly-examples-2}
\mathsf{Z}^{(0)}_{2} (\upmu) &= \frac{7}{5760} + \frac{4999}{55296}\, \upmu^2 + \frac{2129}{7776}\, \upmu^4 + \frac{2209}{23328}\, \upmu^6, \\
\label{eq:Zmu-poly-examples-3}
\mathsf{Z}^{(0)}_{3} (\upmu) &= \frac{36275}{2654208}\, \upmu + \frac{124622833}{716636160}\, \upmu^3 + \frac{8713739}{29859840}\, \upmu^5 + \frac{254317}{2239488}\, \upmu^7 + \frac{103823}{7558272}\, \upmu^9, \\
\label{eq:Zmu-poly-examples-4}
\mathsf{Z}^{(1)}_{0} (\upmu) &= - \frac{1}{12}, \qquad \mathsf{Z}^{(1)}_{1} (\upmu) = \frac{37}{768\sqrt{3}} - \frac{299}{3456}\, \upmu - \frac{47}{288\sqrt{3}}\, \upmu^2 - \frac{47}{1296}\, \upmu^3, \\
\label{eq:Zmu-poly-examples-5}
\mathsf{Z}^{(1)}_{2} (\upmu) &= - \frac{96943}{4423680} + \frac{109633}{663552\sqrt{3}}\, \upmu - \frac{117307}{663552}\, \upmu^2 + \frac{19561}{62208\sqrt{3}}\, \upmu^3 - \frac{41651}{373248}\, \upmu^4 + \\
\label{eq:Zmu-poly-examples-6}
&
+ \frac{2209}{31104\sqrt{3}}\, \upmu^5 - \frac{2209}{279936}\, \upmu^6, \qquad
\mathsf{Z}^{(2)}_{1} (\upmu) = - \frac{1}{13824\sqrt{3}} - \frac{1}{20736}\, \upmu, \\
\label{eq:Zmu-poly-examples-7}
\mathsf{Z}^{(2)}_{2} (\upmu) &= - \frac{131}{1327104} - \frac{1931}{3981312\sqrt{3}}\, \upmu + \frac{1709}{5971968}\, \upmu^2 - \frac{329}{1492992\sqrt{3}}\, \upmu^3 + \frac{47}{2239488}\, \upmu^4, \\
\label{eq:Zmu-poly-examples-8}
\mathsf{Z}^{(3)}_{3} (\upmu) &= \frac{1}{2293235712\sqrt{3}} - \frac{5}{6879707136}\, \upmu + \frac{1}{859963392\sqrt{3}}\, \upmu^2 - \frac{1}{5159780352}\, \upmu^3, \\
\label{eq:Zmu-poly-examples-9}
\mathsf{Z}^{(-1)}_{1} (\upmu) &= - \frac{1}{12}\, \upmu, \qquad \mathsf{Z}^{(-1)}_{2} (\upmu) = -\frac{37}{768\sqrt{3}}\, \upmu - \frac{299}{3456}\, \upmu^2 - \frac{47}{288\sqrt{3}}\, \upmu^3 - \frac{47}{1296}\, \upmu^4.
\end{align}
\noindent
Lower undisplayed coefficients above vanish, \textit{e.g.}, $\mathsf{Z}^{(2)}_{0} = 0 = \mathsf{Z}^{(3)}_{0} = \mathsf{Z}^{(3)}_{1} = \mathsf{Z}^{(3)}_{2} = \mathsf{Z}^{(-1)}_{0}$. We have computed coefficients for $-8 \leq \alpha \leq 8$ with $0 \leq g < 39$. In addition we have computed the main diagonal to higher order, with $g \leq 45$. Notice that this equally translates to data for free-energy and specific-heat for which we have computed the corresponding coefficients.

As previously advertised, it is now that we can check that the diagonal coefficients of the partition function have a starting genus which is more restricted than the one we observed for the free energy. In fact, it is now \textit{quadratic} and given by
\begin{equation}
\beta_{\alpha} = \frac{1}{2}\, \alpha \left(\alpha-1\right).
\end{equation}
\noindent
In this way, it makes sense to restrict the above $\alpha$-sum to simply
\begin{align}
Z \left( x; \upxi_{1}, \upmu \right) = \rme^{F_{\text{A}} \left( x; \upmu \right)}\, \sum_{g=0}^{+\infty} x^{g} \sum_{\alpha = \lceil\frac{1}{2}-\sqrt{\frac{1}{4}+2g}\rceil}^{\lfloor\frac{1}{2}+\sqrt{\frac{1}{4}+2g}\rfloor} \upxi_{1}^{\alpha}\, \mathsf{Z}_g^{(\alpha)} (\upmu).
\end{align}

\paragraph{Quadratic Transasymptotics:}

Now that we have realized the quadratic $\beta$-structure of the partition function, we want to make full use of this structure and also sum quadratically. This may be achieved by reformulating the partition function in a way which makes this structure obvious, by writing
\be
\label{eq:PreparedFormForSummingQuadratically}
Z \left( x; \upxi_{1}, \upmu \right) = \rme^{F_{\text{A}} \left( x; \upmu \right)}\, \sum_{g=0}^{+\infty}\, \sum_{\alpha \in \BZ} \upxi_{1}^{\alpha}\, \mathsf{Z}_{g + \frac{1}{2} \alpha \left(\alpha-1\right)}^{(\alpha)} (\upmu)\, x^{g + \frac{1}{2} \alpha \left(\alpha-1\right)}.
\ee
\noindent
Alternatively, there also is a natural, more theta-function-like (see appendix~\ref{app:elliptic-theta-modular}), reformulation,
\be
\label{eq:PreparedFormForSummingQuadratically}
Z \left( x; \upxi_{1}, \upmu \right) = \rme^{F_{\text{A}} \left( x; \upmu \right)}\, \sum_{g=0}^{+\infty}\, \sum_{\alpha \in \BZ} \upzeta_{1}^{\alpha}\, \mathsf{Z}_{g + \frac{1}{2} \alpha \left(\alpha-1\right)}^{(\alpha)} (\upmu)\, x^{g + \frac{1}{2} \alpha^2}.
\ee
\noindent
Summing over sections of constant genus in these expressions precisely implies summing in a quadratic way. On top, due to the features of diagonal framing, instead of having to perform the (original) transasymptotic resummations for a two-parameter transseries, one now simply needs to evaluate a one-parameter transseries resummation (albeit depending on $\upmu$). The reason this ends up being possible lies in the fact that the $\mathsf{Z}_g^{(\alpha)} (\upmu)$ are known functions---\textit{polynomials} in $\upmu$ in our examples. We have calculated large amounts\footnote{As listed above: coefficients for $-8 \leq \alpha \leq 8$ with $0 \leq g < 39$ generically and $g \leq 45$ along the main diagonal.} of coefficients for the partition function, recursively, and analyzed their patterns in light of the structures we just discussed. Our conclusion is the following closed-form expression for the partition function:
\bea
\label{eq:diagonalframingresultapp}
Z \left( x; \upxi_{1}, \upmu \right) &=& \rme^{- \frac{4}{15 x^{2}} + \frac{16}{5 x}\, \upmu}\, x^{\frac{1}{60} + \frac{2}{3} \upmu^2}\, \sum_{k=0}^{+\infty}\, \sum_{n \in\mathbb{Z}} \upxi_{1}^{n}\, x^{ \frac{1}{2} n \left(n-1\right) + k}\, \times \\
&&
\hspace{-60pt}
\times \underbrace{\widetilde{D}_{k} \left(n-\frac{2}{\sqrt{3}}\upmu\right) \left\{ \left(-1\right)^{k} \left(-\frac{1}{12}\right)^{n+\frac{k}{2}} \left(-96\sqrt{3}\right)^{-\frac{1}{2} n \left(n-1\right)} \frac{G_2 \left(1+n-\frac{2}{\sqrt{3}}\upmu\right)}{G_2 \left(1-\frac{2}{\sqrt{3}}\upmu\right) \Gamma \left(1-\frac{2}{\sqrt{3}}\upmu\right)^n} \right\}}_{\equiv \mathsf{Z}^{(n)}_{k+\frac{1}{2} n \left(n-1\right)} (\upmu)}, \nonumber
\eea
\noindent
The role of the Barnes and Gamma functions is discussed in the main text. As to the $\widetilde{D}_k (\nu)$, these are degree $3k$ polynomials which follow from the partition-function in diagonal-framing. To the extent that they have been computed, they exactly match the (same notation) polynomials in \cite{blmst16} (more on this relation in the main text). It is clearly convenient to change their normalization, which we often do, as
\begin{equation}
\widetilde{D}_{k} (\nu) \equiv \rmi^{k}\, 2^{k}\, 3^{\frac{k}{2}}\, D_{k} (\nu) \equiv \rmi^{k}\, 2^{k}\, 3^{\frac{k}{2}}\, \mathsf{Z}_{k}^{(0)} \left(-\sqrt{\frac{3}{4}}\, \nu\right).
\end{equation}
\noindent
Note how these polynomials may be computed just from partition-function diagonal-coefficients.

\subsection{The Yang--Lee Equation}
\label{subapp:transasymptotic-transseries-YL}

Let us explicitly address the \YL~partition-function transseries in rectangular framing, which is used to compare against its discrete-Fourier-transform formulation presented in the main text, subsection~\ref{subsec:resurgent-Z-transasymptotics}. Start with recalling the \YL~equation in \eqref{eq:YangLeeEquation} alongside its specific-heat transseries solution in \eqref{eq:YangLeeSolution}. Being a fourth-order ODE there are four transseries parameters encoding boundary/initial data. Being a resonant transseries, it has logarithmic sectors which are however not-independent as was the case in the previous \PI~sub-appendix, say, \eqref{eq:ASV-log-resonance}. For \YL~there are two logarithmic resonance conditions, to be discussed in \cite{krst26a} as part of the \YL~large-order analysis therein, and which may be resumed so as to yield the aforementioned rectangular-framing specific-heat transseries-solution
\be
u \left( x; \boldsymbol{\sigma} \right) = x^{-2/7} \sum_{\boldsymbol{n}=0}^{+\infty} \boldsymbol{\sigma}^{\boldsymbol{n}}\, \rme^{- \left( n_1-n_2 \right) \frac{A_1}{x}}\, \rme^{- \left( n_3-n_4 \right) \frac{A_2}{x}}\, x^{\frac{\alpha}{2} \left( n_1-n_2 \right) \sigma_1 \sigma_2}\, x^{\frac{\bar{\alpha}}{2} \left( n_3-n_4 \right) \sigma_3 \sigma_4} \sum_{g=0}^{+\infty} u_{g}^{(\boldsymbol{n})}\, x^{g+\beta_{\boldsymbol{n}}^{[0]}},
\ee
\noindent 
with instanton actions $A_1= \frac{6}{7} \sqrt{5+\rmi \sqrt{5}}$ and $A_2= \frac{6}{7} \sqrt{5-\rmi \sqrt{5}}$ \cite{gs21}, and the resummation parameter of the resonant logarithmic sectors $\alpha=2\rmi \sqrt{1+\frac{\rmi}{\sqrt{5}}}$. Moreover, there are two symmetries of the coefficients (which can be directly proven at the level of their non-linear recursion) that correspond to backward-forward and complex-conjugacy symmetries; respectively given by
\be
\label{eq:YLuSymmetries}
u_{g}^{\left(n_1|n_2|n_3|n_4\right)} = (-1)^{g+\beta_{\boldsymbol{n}}-\frac{1}{2} \left(n_1+n_2+n_3+n_4\right)}\, u_{g}^{\left(n_2|n_1|n_4|n_3\right)} = \overline{u_{g}^{\left(n_3|n_4|n_1|n_2\right)}}.
\ee

\renewcommand{\arraystretch}{1.2}
\begin{table}
\centering
\begin{minipage}[t]{0.31\textwidth}
\vspace{0pt}
\begin{tabular}{c c c c | c}
 $n_1$ & $n_2$ & $n_3$ & $n_4$ & $\beta^{[0]}_{Z,\boldsymbol{n}}$ \\
\hline
 0 & 0 & 0 & 0 & 0 \\
 1 & 0 & 0 & 0 & $\frac{1}{2}$ \\
 1 & 0 & 1 & 0 & 1 \\
 1 & 1 & 0 & 0 & 1 \\
 1 & 1 & 1 & 0 & $\frac{3}{2}$ \\
 1 & 1 & 1 & 1 & 2 \\
 2 & 0 & 0 & 0 & 2 \\
 2 & 0 & 1 & 0 & $\frac{5}{2}$ \\
 2 & 0 & 2 & 0 & 4 \\
 2 & 1 & 0 & 0 & $\frac{3}{2}$ \\
 2 & 1 & 1 & 0 & 2 \\
 2 & 1 & 1 & 1 & $\frac{3}{2}$ \\
 2 & 1 & 2 & 0 & $\frac{7}{2}$ \\
 2 & 1 & 2 & 1 & 2 \\
 2 & 2 & 0 & 0 & 2 \\
 2 & 2 & 1 & 0 & $\frac{3}{2}$ \\
 2 & 2 & 1 & 1 & 1 \\
\end{tabular}
\end{minipage}
\hfill
\begin{minipage}[t]{0.31\textwidth}
\vspace{0pt}
\begin{tabular}{c c c c|c}
 $n_1$ & $n_2$ & $n_3$ & $n_4$ & $\beta^{[0]}_{Z,\boldsymbol{n}}$ \\
\hline
 2 & 2 & 2 & 0 & 3 \\
 2 & 2 & 2 & 1 & $\frac{3}{2}$ \\
 2 & 2 & 2 & 2 & 2 \\
 3 & 0 & 0 & 0 & $\frac{9}{2}$ \\
 3 & 0 & 1 & 0 & 5 \\
 3 & 0 & 2 & 0 & $\frac{13}{2}$ \\
 3 & 1 & 0 & 0 & 2 \\
 3 & 1 & 1 & 0 & $\frac{5}{2}$ \\
 3 & 1 & 1 & 1 & 3 \\
 3 & 1 & 2 & 0 & 4 \\
 3 & 1 & 2 & 1 & $\frac{7}{2}$ \\
 3 & 1 & 3 & 0 & $\frac{13}{2}$ \\
 3 & 1 & 3 & 1 & 4 \\
 3 & 2 & 0 & 0 & $\frac{3}{2}$ \\
 3 & 2 & 1 & 0 & 2 \\
 3 & 2 & 1 & 1 & $\frac{3}{2}$ \\
\end{tabular}
\end{minipage}
\hfill
\begin{minipage}[t]{0.31\textwidth}
\vspace{0pt}
\begin{tabular}{c c c c|c}
 $n_1$ & $n_2$ & $n_3$ & $n_4$ & $\beta^{[0]}_{Z,\boldsymbol{n}}$ \\
\hline
 3 & 2 & 2 & 0 & $\frac{7}{2}$ \\
 3 & 2 & 2 & 1 & 2 \\
 3 & 2 & 2 & 2 & $\frac{5}{2}$ \\
 3 & 2 & 3 & 0 & 6 \\
 3 & 2 & 3 & 1 & $\frac{7}{2}$ \\
 3 & 2 & 3 & 2 & 3 \\
 3 & 3 & 0 & 0 & 1 \\
 3 & 3 & 1 & 0 & $\frac{3}{2}$ \\
 3 & 3 & 1 & 1 & 2 \\
 3 & 3 & 2 & 0 & 3 \\
 3 & 3 & 2 & 1 & $\frac{5}{2}$ \\
 3 & 3 & 2 & 2 & 3 \\
 3 & 3 & 3 & 0 & $\frac{11}{2}$ \\
 3 & 3 & 3 & 1 & 3 \\
 3 & 3 & 3 & 2 & $\frac{5}{2}$ \\
 3 & 3 & 3 & 3 & 2 \\
\end{tabular}
\end{minipage}
\caption{Starting powers $\beta^{[0]}_{Z,\boldsymbol{n}}$ of \YL~partition-function transseries-sectors.}
\label{tab:betaZYangLee}
\end{table}

Our main focus herein is the \YL~partition-function, constructed in the usual way along the multicritical KdV hierarchy from the specific-heat by double-integration for the free-energy and then exponentiation; see, \textit{e.g.}, \cite{gs21}. This partition function was written in the main text, in the form of a discrete Fourier transform \eqref{eq:YLFullDFT}. That may be verified by direct comparison against the rectangular-framing partition-function following from the string-equation construct. This comparison is simplest when performed against the logarithmic-resummed partition-function, which is
\be
\label{eq:YangLeePartitionFunctionRectangularLogResummed}
Z \left( x; \boldsymbol{\sigma} \right) = \rme^{F_{\text{A}} \left(x;\upmu_1,\upmu_2\right)} \sum_{\boldsymbol{n}=0}^{+\infty} \boldsymbol{\sigma}^{\boldsymbol{n}}\, \rme^{- \left( n_1-n_2 \right) \frac{A_1}{x}}\, \rme^{- \left( n_3-n_4 \right) \frac{A_2}{x}}\, x^{\frac{\alpha}{2} \left( n_1-n_2 \right) \sigma_1 \sigma_2}\, x^{\frac{\bar{\alpha}}{2} \left( n_3-n_4 \right) \sigma_3 \sigma_4}\, Z^{(\boldsymbol{n})} (x),
\ee
\noindent
and where
\be
Z^{(\boldsymbol{n})} (x) \simeq \sum_{g=0}^{+\infty} Z_{g}^{(\boldsymbol{n}) [0]}\, x^{g+\beta_{Z,\boldsymbol{n}}^{[0]}}.
\ee
\noindent
While we have no closed-form expression for the above $\beta$-structure $\beta^{[0]}_{Z,\boldsymbol{n}}$, we list values for sectors we have computed in table~\ref{tab:betaZYangLee}. Analogously to the specific heat, the partition-function coefficients satisfy the symmetries
\be
\label{eq:YLZSymmetries}
Z_{g}^{\left(n_1|n_2|n_3|n_4\right)} = (-1)^{g+\beta_{Z,\boldsymbol{n}}^{[0]}-\frac{1}{2} \left(n_1+n_2+n_3+n_4\right)}\, Z_{g}^{\left(n_2|n_1|n_4|n_3\right)} = \overline{Z_{g}^{\left(n_3|n_4|n_1|n_2\right)}}.
\ee
\noindent
We have computed partition-function coefficients up to the $\left(3|3|3|3\right)$ sector for genera $g \leq 2$, in order to explicitly compare between rectangular-framing and discrete-Fourier-transform rewritings. To illustrate, the first few sectors of the form $Z^{(n|0|0|0)} (x)$ are given by
\bea
Z^{(0|0|0|0)} (x) &\simeq& 1 + \frac{1}{240}\, x^2 +\frac{98989}{21772800}\, x^4 + \frac{655764967}{47029248000}\, x^6 + \frac{52667231867107}{948109639680000}\, x^8 + \cdots, \\
Z^{(1|0|0|0)} (x) &\simeq& \frac{\rmi}{60} \left( \sqrt{5} + 5\, \rmi \right) \sqrt{x} + \frac{\sqrt{\frac{1}{3} \left( - 17035 - 30583\, \rmi \sqrt{5} \right)}}{2880}\, x^{3/2} - \frac{2170 - 11397\, \rmi \sqrt{5}}{460800}\, x^{5/2} + \cdots, \\
Z^{(2|0|0|0)} (x) &\simeq& - \frac{\sqrt{\frac{1}{6} \left( 355 - 119\, \rmi \sqrt{5} \right)}}{129600}\, x^2 + \frac{1975 - 929\, \rmi \sqrt{5}}{10368000}\, x^3 + \cdots.
\eea
\noindent
The first few sectors of the form $Z^{(n|1|0|0)} (x)$ are given by
\bea
Z^{(1|1|0|0)} (x) &\simeq& - \left( \frac{1}{45} + \frac{3\, \rmi}{16 \sqrt{5}} \right) x + \frac{378393 - 170879\, \rmi \sqrt{5}}{12441600}\, x^3 - \frac{61174359557 + 10880193204\, \rmi \sqrt{5}}{125411328000}\, x^5 + \cdots, \\
Z^{(2|1|0|0)} (x) &\simeq& \frac{4205 + 2237\, \rmi \sqrt{5}}{43200}\, x^{3/2} - \frac{\sqrt{\frac{1}{3} \left( 5384055881591 + \frac{8322417243583\, \rmi}{\sqrt{5}} \right)}}{6220800}\, x^{5/2} + \cdots, \\
Z^{(3|1|0|0)} (x) &\simeq& - \frac{1 + 4\, \rmi \sqrt{5}}{129600}\, x^2 + \frac{\sqrt{\frac{1}{6} \left( - 9559136735 + 11041034179\, \rmi \sqrt{5}\right)}}{93312000}\, x^3 + \cdots.
\eea
\noindent
For a few more non-trivial sectors, consider the first few sectors of the form $Z^{(n|1|1|0)} (x)$, which are given by
\bea
Z^{(1|1|1|0)} (x) &\simeq& \frac{613 + 299\, \rmi \sqrt{5}}{8640}\, x^{3/2} + \frac{167  \sqrt{ \frac{1}{3} \left( - 37755505 - 3155243\, \rmi \sqrt{5} \right)}}{2073600}\, x^{5/2} + \cdots, \\
Z^{(2|1|1|0)} (x) &\simeq& \frac{29072 - 5312 \sqrt{30} + \left( 7539 \sqrt{5} - 6858 \sqrt{6}
\right) \rmi}{86400}\, x^2 + \cdots, \\
Z^{(3|1|1|0)} (x) &\simeq& \frac{5 \left( - 241 + 44 \sqrt{30} \right) + 7 \left( 241 \sqrt{5} - 220 \sqrt{6} \right) \rmi}{2592000}\, x^{5/2} + \cdots.
\eea
\noindent
Finally, let us illustrate a few sectors on the main diagonal; namely of the form $Z^{(n|n|n|n)} (x)$. One finds
\bea
Z^{(1|1|1|1)} (x) &\simeq& \frac{301789}{518400}\, x^4 - \frac{1007072371}{497664000}\, x^6 + \cdots, \\
Z^{(2|2|2|2)} (x) &\simeq& - \frac{1879}{600}\, x^2 + \frac{2426135961583}{71663616000}\, x^4 - \frac{87089705423425540757}{97519848652800000}\, x^6 + \cdots, \\
Z^{(3|3|3|3)} (x) &\simeq& \frac{191893}{108000}\, x^2 -\frac{31735120439869}{403107840000}\, x^4 + \frac{556331699755797126612551}{175535727575040000000}\, x^{6} + \cdots.
\eea
\noindent
Using the above coefficients back in \eqref{eq:YangLeePartitionFunctionRectangularLogResummed}, alongside an expansion of its pre-factor, one can compare with the discrete Fourier transform presented in subsection~\ref{subsec:resurgent-Z-transasymptotics}, in particular in \eqref{eq:YLFullDFT}. One indeed finds a precise match, for all our $g \leq 2$ coefficients of all sectors up to $(3|3|3|3)$.

\subsection{The Cubic Matrix Model}
\label{subapp:transasymptotic-transseries-CMM}

Moving towards full fledged matrix models, let us begin by presenting data concerning the cubic matrix model with potential \eqref{eq:CubicMatrixModelPotential} (the data herein should be supplemented with the one presented in \cite{mss22}). The string equation for the cubic model is given by \eqref{eq:cubicstringequationthooftlimit}, and in standard rectangular-framing its transseries solution is of the form \eqref{eq:twoparameterresurgenttransseriesforR} (with the free energy subsequently constructed via \eqref{eq:freeenergyfromstringequation}). Being more detailed than \eqref{eq:twoparameterresurgenttransseriesforR} on what concerns logarithmic sectors and $\beta$-structure (which is the same as for \PI; see \eqref{eq:specificheatP1-beta}), the transseries solutions to \eqref{eq:cubicstringequationthooftlimit} and \eqref{eq:freeenergyfromstringequation} are of the form
\begin{equation}
\label{eq:cubicMM-Rtransseries-explicit}
R \left( t,g_{\text{s}}; \sigma_1,\sigma_2 \right) = \sum_{n,m=0}^{+\infty} \sum_{k=0}^{\min(n, m)} \sum_{g=\beta_{nm}^{[k]}}^{+\infty} \sigma_1^n \sigma_2^m\, \rme^{- \left(n-m\right) \frac{A (t)}{g_{\text{s}}}} \left( \log \frac{f(t)}{g_{\text{s}}^2} \right)^k R^{(n|m)[k]}_g (t)\, g_{\text{s}}^{g},
\end{equation}
\noindent
for the solution to the string equation; and
\begin{equation}
F \left( t,g_{\text{s}}; \sigma_1,\sigma_2 \right) = \sum_{n,m=0}^{+\infty} \sum_{k=0}^{\min(n, m)} \sum_{g=\beta_{nm}^{[k]}}^{+\infty} \sigma_1^n \sigma_2^m\, \rme^{- \left(n-m\right) \frac{A (t)}{g_{\text{s}}}} \left( \log \frac{f(t)}{g_{\text{s}}^2} \right)^k F^{(n|m)[k]}_g (t)\, g_{\text{s}}^{g},
\end{equation}
\noindent
for its associated free energy. Herein, the instanton action $A(t)$ is encoded in \eqref{eq:cubicaction} (see as well sub-appendix~\ref{subappendix:cubicmatrixmodel}) and follows from \cite{msw08, mss22}
\begin{equation}
2 - 5 \lambda^2\, r = \lambda^2\, r\, \cosh \left( A^{\prime}(t) \right).
\end{equation}
\noindent
The function $f(t)$ inside the logarithm is given by 
\begin{equation}
f(t) = \frac{\left( 1 - 3 \lambda^2\, r \right)^5}{108 \lambda^8\, r^2}.
\end{equation}
\noindent
For convenience, above we used the classical string solution from \eqref{eq:cubic-MM-classical-string-eq} with $r \equiv r(t)$.

\begin{table}
\centering
\begin{tabular}{c|c c c c c c c c c  }
$g_{\text{max}}$ & $n=0$ & $1$ & $2$ & $3$ & $4$ & $5$ & $6$ & $7$ & $8$ \\\hline 
$m=0$ & 70 & 10 & 10 & 9 & 9 & 9 & 9 & 8 & 8 \\
$1$ & 10 & 11 & 9 & 9 & 8 & 8 & 8 & 7 & 7 \\
$2$ & 10 & 9 & 9 & 8 & 8 & 7 & 7 & 6 & 6 \\
$3$ & 9 & 9 & 8 & 9 & 7 & 7 & 6 & 5 & 5 \\
$4$ & 9 & 8 & 8 & 7 &  7 & 6 & 5 &4 & 4 \\
$5$ & 9 & 8 & 7 & 7 & 6 &  7 & 4 & 3 & 3 \\
$6$ & 9 & 8 & 7 & 6 & 5 & 4 & 5 & 2 & 2 \\
$7$ & 8 & 7 & 6 & 5 & 4 &3 & 2 & 3 & 1 \\
$8$ & 8 & 7 & 6 & 5 & 4 & 3 & 2 & 1 & 1  \\
\end{tabular}
\caption{Explicit values of the maximum genus $g_{\text{max}}$ up to which we have computed associated transseries sectors labelled by $(n|m)$ for both $R(t,g_{\text{s}};\sigma_1,\sigma_2)$ and $F(t,g_{\text{s}};\sigma_1,\sigma_2)$ transseries.}
\label{tab:gForRandF}
\end{table}

Plugging-in the above transseries \textit{ans\"atze} into their corresponding finite-difference equations, and extracting as much data as possible is a straightforward albeit very laborious and computationally intense procedure. We have carried through such procedure up to the data schematically indicated in table~\ref{tab:gForRandF}, and list a few illustrative coefficients of both transseries solutions in the following. Starting with $R(t)$, with $R^{(0|0)}_0 = r$, for example the first few $(n|0)$ sectors are given by
\bea
R^{(0|0)} &\simeq& r - \frac{\lambda^4 r \left( 9 \lambda^2 r - 5 \right)}{8 \left( 3 \lambda^2 r - 1 \right)^4}\, g_{\text{s}}^2 - \frac{3 \lambda^8 r \left( 162 \lambda^6 r^3 + 1017 \lambda^4 r^2 - 1316 \lambda^2 r + 385 \right)}{128 \left( 3 \lambda^2 r - 1 \right)^9}\, g_{\text{s}}^4 + \cdots, \\
R^{(1|0)} &\simeq& \frac{\sqrt{r}}{\sqrt[4]{1 - 3 \lambda^2 r}} - \frac{9 \lambda^4 r^2 - 12 \lambda^2 r + 8}{96 \sqrt{r} \left( 1 - 3 \lambda^2 r \right)^{11/4}}\, g_{\text{s}} + \\
&&
+ \frac{37017 \lambda^8 r^4 - 41688 \lambda^6 r^3 + 11808 \lambda^4 r^2 - 192 \lambda^2 r + 64}{18432 r^{3/2} \left( 1 - 3 \lambda^2 r \right)^{21/4}}\, g^2_{\text{s}} + \cdots, \nonumber \\
R^{(2|0)} &\simeq& - \frac{\lambda^2 r}{2 \left( 1 - 3 \lambda^2 r \right)^{3/2}} - \frac{\lambda^2 \left( 195 \lambda^4 r^2 - 96 \lambda^2 r - 8 \right)}{96 \left( 3 \lambda^2 r - 1 \right)^4}\, g_{\text{s}} + \\
&&
- \frac{\lambda^2 \left( 101277 \lambda^8 r^4 - 140256 \lambda^6 r^3 + 42240 \lambda^4 r^2 + 1536 \lambda^2 r + 64 \right)}{9216 r \left( 1 - 3 \lambda^2 r \right)^{13/2}} + \cdots. \nonumber
\eea
\noindent
Moving on, list a couple of ``bulk'' sectors,
\bea
R^{(1|1)} &\simeq& - \frac{3 \lambda^2 r - 2}{\left( 1 - 3 \lambda^2 r \right)^{3/2}} + \frac{\lambda^4 \left( 6561 \lambda^6 r^3 - 8370 \lambda^4 r^2 + 2256 \lambda^2 r + 160 \right)}{128 \left( 1 - 3 \lambda^2 r \right)^{13/2}}\, g_{\text{s}}^2 - \\
&&
- \frac{3 \lambda^8 }{32768 \left( 1 - 3 \lambda^2 r\right)^{23/2}}\, \Big( 28005507 \lambda^{10} r^5 - 13016538 \lambda^8 r^4 - 40077216 \lambda^6 r^3 + \nonumber \\
&&
+ 36043968 \lambda^4 r^2 - 7736064 \lambda^2 r - 197120 \Big)\, g_{\text{s}}^4 + \cdots, \nonumber \\
R^{(2|1)} &\simeq& \frac{\sqrt{r}}{2 \sqrt[4]{1 - 3 \lambda^2 r}}\, \log \left( \frac{\left( 1 - 3 \lambda^2 r \right)^5}{108 g_\text{s}^2\, r^2} \right) \frac{1}{\sqrt{g_{\text{s}}}} + \frac{1}{192 \sqrt{r} \left( 1 - 3 \lambda^2 r \right)^{11/4}}\, \Big(\, \big\{ - 9 \lambda^4 r^2 + \nonumber \\
&&
+ 12 \lambda^2 r - 8\, \big\}\, \log \frac{\left( 1 - 3 \lambda^2 r\right)^5}{108 g_\text{s}^2\,  r^2} - 24 \left\{ 7 \lambda^4 r^2 + 6 \lambda^2 r - 4 \right\} \Big)\, \sqrt{g_{\text{s}}} + \cdots, \\
R^{(2|2)} &\simeq& - \frac{3 \lambda^2 \left( 9 \lambda^4 r^2 + 11 \lambda^2 r - 8 \right)}{4 \left( 3 \lambda^2 r - 1 \right)^4} + \\
&&
+ \frac{3 \lambda^6 \left( 90906 \lambda^8 r^4 - 232363 \lambda^6 r^3 + 160104 \lambda^4 r^2 - 26212 \lambda^2 r - 3584 \right)}{256 \left( 3 \lambda^2 r - 1 \right)^9}\, g_{\text{s}}^2 + \cdots. \nonumber
\eea
\noindent
Turning to the free energy, for example the first few $(n|0)$ sectors are given by\footnote{Each diagonal sector includes unfixed integration constants. These may be fixed by setting the double-scaling limit to precisely match against \PI~data (see as well \cite{asv11, mss22}).}
\bea
F^{(0|0)} &\simeq& \left( \frac{1}{4} r^2 \left( 2 \lambda^2 r - 1 \right) \log \left( 1 - 2 \lambda^2 r \right) - \frac{3 \lambda^6 r^3 + 2 \lambda^4 r^2 - 2 \lambda^2 r + 1}{6 \lambda^4} \right) \frac{1}{g_{\text{s}}^2} + \nonumber \\
&&
+ \frac{1}{24} \log \frac{1 - 2 \lambda^2 r}{6 \lambda^2 r - 2} + \cdots, \\
F^{(1|0)} &\simeq& \frac{\lambda^2 \sqrt{r}}{4 \left( 1 - 3 \lambda^2 r \right)^{5/4}}\, \sqrt{g_{\text{s}}} - \frac{8 \lambda^2 - 423 \lambda^6 r^2 + 228 \lambda^4 r}{384 \sqrt{r} \left( 1 - 3 \lambda^2 r \right)^{15/4}}\, g_{\text{s}}^{3/2} + \frac{\lambda^2}{73728 r^{3/2} \left( 1 - 3 \lambda^2 r \right)^{25/4}} \times \nonumber \\
&& \times \Big( 371385 \lambda^8 r^4 - 604152 \lambda^6 r^3 + 206496 \lambda^4 r^2 + 3648 \lambda^2 r + 64 \Big)\, g_{\text{s}}^{5/2} + \cdots, \\
F^{(2|0)} &\simeq& - \frac{\lambda^4 r}{32 \left( 1 - 3 \lambda^2 r \right)^{5/2}}\, g_{\text{s}} +\frac{\lambda^4 \left( 429 \lambda^4 r^2 - 228 \lambda^2 r - 8 \right)}{1536 \left( 3 \lambda^2 r - 1 \right)^5}\, g_{\text{s}}^2 - \frac{\lambda ^4}{147456 r \left( 1 - 3 \lambda^2 r \right)^{15/2}} \times \nonumber \\
&&
\times \Big( 293409 \lambda^8 r^4 - 407304 \lambda^6 r^3 + 125184 \lambda^4 r^2 + 3648 \lambda^2 r + 64 \Big)\, g_{\text{s}}^3 + \cdots. 
\eea
\noindent
Moving on, list a couple of ``bulk'' sectors,
\bea
F^{(1|1)} &\simeq& \left( \frac{2 \sqrt{1 - 3 \lambda^2 r}}{3 \lambda^2} - 2 r \sqrt{1 - 2 \lambda^2 r}\, \log \frac{\sqrt{1 - 2 \lambda^2 r} + \sqrt{1 - 3 \lambda^2 r}}{\lambda^2 \sqrt{r}} \right) \frac{1}{g_{\text{s}}} - \\
&&
- \frac{63 \lambda^4 r^2 - 60 \lambda^2 r + 8}{96 r \left( 1 - 3 \lambda^2 r \right)^{5/2}}\, g_{\text{s}} + \frac{1}{122880 r^3 \left( 1 - 3 \lambda^2 r\right)^{15/2}}\, \Big( 2350215 \lambda^{12} r^6 - 3911220 \lambda^{10} r^5 + \nonumber \\
&&
+ 2254680 \lambda^8 r^4 - 694080 \lambda^6 r^3 + 163200 \lambda^4 r^2 - 19968 \lambda^2 r + 1024 \Big)\, g_{\text{s}}^3 + \cdots, \nonumber \\
F^{(2|1)} &\simeq& \frac{\lambda^2 \sqrt{r}}{8 \left( 1 - 3 \lambda^2 r \right)^{5/4}}\, \log \left( \frac{\left( 1 - 3 \lambda^2 r \right)^5}{108 g_{\text{s}}^2 r^2} \right) \sqrt{g_{\text{s}}} + \frac{\lambda^2}{768 \sqrt{r} \left( 1 - 3 \lambda^2 r\right)^{15/4}} \times \\
&&
\times \left( \left\{ 3 \lambda^2 r \left( 141 \lambda^2 r - 76 \right) - 8 \right\} \log \frac{\left( 1 - 3 \lambda^2 r\right)^5}{108 g_{\text{s}}^2\, r^2} + 24 \left\{ 7 \lambda^2 r \left( 6 - 13 \lambda^2 r \right) + 4 \right\} \right) g_{\text{s}}^{3/2} + \nonumber \\
&&
+ \frac{\lambda ^2}{147456 r^{3/2} \left( 1 - 3 \lambda^2 r \right)^{25/4}}\, \Bigg( \Big\{ 3 \lambda^2 r \left( 9 \lambda^2 r \left( \lambda^2 r \left( 13755 \lambda^2 r - 22376 \right) + 7648 \right) + \right. \nonumber \\
&&
\left.
+ 1216 \right) + 64 \Big\} \log \frac{\left( 1 - 3 \lambda^2 r\right)^5}{108 g_{\text{s}}^2\, r^2} - 48 \Big\{ \lambda^2 r \left( 3 \lambda^2 r \left( 3 \lambda^2 r \left( 8271 \lambda^2 r - 11242 \right) + 9500 \right) + \right. \nonumber \\
&&
\left.
+ 2080 \right) - 32 \Big\} \Bigg)\, g_{\text{s}}^{5/2} + \cdots \nonumber \\
F^{(2|2)} &\simeq& - \frac{1}{4} \log \frac{\left( 1 - 3 \lambda^2 r \right)^5}{108 g_{\text{s}}^2\, r^2} +\frac{843 \lambda^8 r^4 + 3444 \lambda^6 r^3 - 2720 \lambda^4 r^2 + 416 \lambda^2 r - 32}{768 r^2 \left( 3 \lambda^2 r - 1 \right)^5}\, g_{\text{s}}^2 + \cdots .
\eea

\subsection{The Quartic Matrix Model}
\label{subapp:transasymptotic-transseries-QMM}

\begin{table}
\centering
\begin{tabular}{c|c c c c c c c c c c c}
$\beta_{nm}^{[0]}$ & $n=0$ & $1$ & $2$ & $3$ & $4$ & $5$ & $6$ & $7$ & $8$ & $9$ & $10$\\\hline 
$m=0$ & 0 & 0 & 0 & 0 & 0 & 0 & 0 & 0 & 0 & 0 & 0\\
$1$ & 0 & 0 & 0 & 0 & 0 & 0 & 0 & 0 & 0 & 0 & 0\\
$2$ & 0 & 0 & 0 & -1 & -1 & -1 & -1 & -1 & -1 & -1 & -1\\
$3$ & 0 & 0 & -1 & 0 & -1 & -1 & -1 & -1 & -1 & -1 & -1\\
$4$ & 0 & 0 & -1 & -1 &   & -2 & -2 & -2 & -2 & -2 & -2\\
$5$ & 0 & 0 & -1 & -1 & -2 &   & -2 & -2 & -2 & -2 & -2\\
$6$ & 0 & 0 & -1 & -1 & -2 & -2 &  & -3 & -3 & -3 & -3\\
$7$ & 0 & 0 & -1 & -1 & -2 & -2 & -3 &  &  &  & \\
$8$ & 0 & 0 & -1 & -1 & -2 & -2 & -3 &  &  &  & \\
$9$ & 0 & 0 & -1 & -1 & -2 & -2 & -3 &  &  &  & \\
$10$ & 0 & 0 & -1 & -1 & -2 & -2 & -3 &  &  &  & \\
\end{tabular}
\caption{Explicit values for $\beta_{nm}^{[0]}$ as observed from our calculated data.}
\label{tab:BetasForR}
\end{table}

Finally, let us address the quartic matrix model with potential \eqref{eq:QuarticMatrixModelPotential} (the data herein should be supplemented with the one presented in \cite{asv11, csv15, mss22}, whose conventions we follow below). The string equation for the quartic model is \eqref{eq:quarticstringequationthooftlimit}, with its free energy following via \eqref{eq:freeenergyfromstringequation}, and in rectangular-framing the corresponding transseries solution is of the form \eqref{eq:twoparameterresurgenttransseriesforR}. Note that the quartic model is slightly more intricate than the cubic due to the existence of a symmetric configuration of saddles. Immediately following the discussion of \PI, in our earlier sub-appendix~\ref{subapp:transasymptotic-transseries-PI}, the transseries sectors have a logarithmic dependence in $g_{\text{s}}$ which may be conveniently resummed. As such, the transseries may be written as (this expression matches equation (6.81) in \cite{asv11})
\begin{equation}
\label{eq:quarticMM-Rtransseries-explicit}
R \left( t,g_{\text{s}}; \upxi_{1},\upxi_{2} \right) = \sum_{n,m=0}^{+\infty} \sum_{g=0}^{+\infty} \upxi_{1}^{n} \upxi_{2}^{m}\, R_{g+\beta_{nm}}^{(n|m)} (t)\, g_{\text{s}}^{g+\beta_{nm}}.
\end{equation}
\noindent
Herein, the logarithmic dependence has already been resummed via (compare with \eqref{eq:defxi}, or \eqref{eq:defzeta} and \eqref{eq:XiMuRelation})
\begin{equation}
\label{eq:upxi12forQuartic}
\upxi_{1} = \sqrt{g_{\text{s}}}\, \sigma_{1}\, \rme^{-\frac{A(t)}{g_{\text{s}}}} \left( \frac{f(t)}{5184 \lambda^2\, g_{\text{s}}^2} \right)^{\frac{\lambda}{12} \upmu}, \qquad \upxi_{2} = \sqrt{g_{\text{s}}}\, \sigma_{2}\, \rme^{\frac{A(t)}{g_{\text{s}}}} \left( \frac{f(t)}{5184 \lambda^2\, g_{\text{s}}^2}\right)^{-\frac{\lambda}{12} \upmu},
\end{equation}
\noindent
where
\begin{equation}
f(t) = \frac{\left( 3 -\lambda\, r \right)^3 \left( 3 - 3 \lambda\, r \right)^5}{15552 \lambda^6\,  r^4}
\end{equation}
\noindent
and we use the classical string solution from \eqref{eq:quartic-MM-classical-string-eq} with $r \equiv r(t)$ as usual. Interestingly, we find a refinement on the $\beta$-structure in comparison to \cite{asv11}, in the sense that this starting genus can in fact be restricted even more than therein. Some explicit values for the $\beta$-structure of the $R_{g}^{(n|m)}$ sectors are given in table~\ref{tab:BetasForR}, from which one quickly concludes that
\begin{equation}
\beta_{nm} = - \left\lfloor \frac{\min \left(n,m\right)}{2} \right\rfloor.
\end{equation}

Moving towards diagonal-framing, with the usual $\upmu = \sigma_1 \sigma_2$ and $\upxi_1 \upxi_2 = g_{\text{s}}\, \upmu$, we follow the \PI~strategy in sub-appendix~\ref{subapp:transasymptotic-transseries-PI} and hence begin by addressing sectors $n\geq m$. With the usual $\alpha=n-m$,
\begin{align}
R_{n \geq m} &= \sum_{n \geq m} \upxi_{1}^{n-m}\, \upmu^{m} \sum_{g=0}^{+\infty} R_{g+\beta_{nm}}^{(n|m)} (t)\, g_{\text{s}}^{g+\beta_{nm}+m} = \\
&
= \sum_{\alpha=0}^{+\infty} \upxi_{1}^{\alpha} \sum_{m=0}^{+\infty} \upmu^{m} \sum_{g=0}^{+\infty} R_{g-\lfloor\frac{m}{2}\rfloor}^{(\alpha+m|m)} (t)\, g_{\text{s}}^{g+m-\lfloor\frac{m}{2}\rfloor} = \nonumber \\
&
= \sum_{\alpha=0}^{+\infty} \upxi_{1}^{\alpha} \sum_{g=0}^{+\infty} \left\{ \sum_{m=0}^{+\infty} \upmu^{2m}\, R_{g-m}^{(\alpha+2m|2m)} (t)\, g_{\text{s}}^{g+m} + \sum_{m=0}^{+\infty} \mu^{2m+1}\, R_{g-m}^{(\alpha+2m+1|2m+1)} (t)\, g_{\text{s}}^{g+m+1} \right\} = \nonumber \\ 
&
= \sum_{g=0}^{+\infty} g_{\text{s}}^{g} \sum_{\alpha=0}^{+\infty} \upxi_{1}^{\alpha} \underbrace{ \left\{ \sum_{m=0}^{g} \upmu^{2m}\, R_{g-2m}^{(\alpha+2m|2m)} (t) + \sum_{m=0}^{g-1} \upmu^{2m+1}\, R_{g-2m-1}^{(\alpha+2m+1|2m+1)} (t) \right\}}_{\equiv P_{\text{diag}, g}^{(0)} (t,\upmu) + P_{\text{forw}, g}^{(\alpha)} (t,\upmu)}. \nonumber
\end{align}
\noindent
A completely analogous calculation holds for the $n<m$ case,
\begin{align}
R_{n < m} &= \sum_{n < m} \upxi_{2}^{m-n}\, \upmu^{n} \sum_{g=0}^{+\infty} R_{g+\beta_{nm}}^{(n|m)} (t)\, g_{\text{s}}^{g+\beta_{nm}+n} = \\
&
= \sum_{g=0}^{+\infty} g_{\text{s}}^{g} \sum_{\alpha=0}^{+\infty} \upxi_{2}^{\alpha} \underbrace{\left\{ \sum_{n=0}^{g} \upmu^{2n}\, R_{g-2n}^{(2n|\alpha+2n)} (t) + \sum_{n=0}^{g-1} \upmu^{2n+1}\, R_{g-2n-1}^{(2n+1|\alpha+2n+1)} (t) \right\}}_{\equiv P_{\text{back}, g}^{(\alpha)} (t, \upmu)}. \nonumber
\end{align}
\noindent
Reassembling the transseries, we again have (compare with \eqref{eq:PreDiagonalSpecificHeatThreeVariables} for \PI)
\begin{equation}
R \left( t,g_{\text{s}}; \upxi_{1},\upxi_{2},\upmu \right) = \sum_{g=0}^{+\infty} g_{\text{s}}^{g} \left\{ P_{\text{diag}, g}^{(0)} (t,\upmu) + \sum_{\alpha=1}^{+\infty} \upxi_{1}^{\alpha}\, P_{\text{forw}, g}^{(\alpha)} (t,\upmu) + \sum_{\alpha=1}^{+\infty} \upxi_{2}^{\alpha}\, P_{\text{back}, g}^{(\alpha)} (t,\upmu) \right\}.
\end{equation}
\noindent
As for \PI, this result can still be further simplified. The backward-forward symmetry of \cite{bssv22} has essentially the same structure,
\begin{equation}
P_{\text{forw}, g}^{(\alpha)} (t,\upmu) = (-1)^{g}\, P_{\text{back}, g}^{(\alpha)} (t,-\upmu),
\end{equation}
\noindent
in which case with the same definition as for \PI,
\begin{equation}
P^{(\alpha)}_{g} (t,\upmu) \equiv
\begin{cases}
P_{\text{forw},g}^{(\alpha)} (t,\upmu), \quad & \alpha>0, \\[3pt]
P_{\text{diag},g}^{(0)} (t,\upmu), \quad & \alpha= 0, \\[3pt]
P_{\text{back},g}^{(-\alpha)} (t,\upmu), \quad & \alpha<0,
\end{cases}
\end{equation}
\noindent
for which the backward-forward relation becomes (matching with corresponding results in \cite{asv11})
\begin{equation}
P^{(\alpha)}_{g} (t,\upmu) = (-1)^g\, P^{(-\alpha)}_{g} (t,-\upmu),
\end{equation}
\noindent
one immediately arrives at the transseries rewriting of (compare with \eqref{eq:DiagonalSpecificHeatThreeVariables})
\begin{equation}
\label{eq:DiagonalQuarticRThreeVariables}
R \left( t,g_{\text{s}}; \upxi_{1},\upxi_{2},\upmu \right) = \sum_{g=0}^{+\infty} g_{\text{s}}^{g}\left\{ \sum_{\alpha=0}^{+\infty} \upxi_{1}^{\alpha}\, P_{g}^{(\alpha)} (t,\upmu) + \sum_{\alpha=1}^{+\infty} \upxi_{2}^{\alpha}\, P_{g}^{(-\alpha)} (t,\upmu) \right\}.
\end{equation}
\noindent
In order to go back to the original number of two rather than three transseries parameters, we again follow the \PI~calculations and reformulate the second sum above as
\bea
R \left( t,g_{\text{s}}; \upxi_{1},\upxi_{2},\upmu \right) &=& \left\{\, \cdots\, \right\} + \sum_{g=0}^{+\infty} \sum_{\alpha=1}^{+\infty} \frac{\upmu^{\alpha}\, g_{\text{s}}^{g+\alpha}}{\upxi_{1}^{\alpha}}\, P_{g}^{(-\alpha)} (t,\upmu) = \nonumber \\
&=& \left\{\, \cdots\, \right\} + \sum_{g=0}^{+\infty} \sum_{\alpha=-g}^{-1} g_{\text{s}}^{g}\, \upxi_{1}^{\alpha}\, \upmu^{-\alpha}\, P_{g+\alpha}^{(\alpha)} (t,\upmu).
\eea
\noindent
Keep following our \PI~example whom-knows-it-all, and define the quartic-model diagonal-framing coefficients as
\begin{equation}
\mathsf{R}^{(\alpha)}_{g} (t,\upmu) \equiv
\begin{cases}
P^{(\alpha)}_{g} (t,\upmu), \qquad & \alpha \geq 0, \\
\upmu^{-\alpha} P^{(\alpha)}_{g+\alpha} (t,\upmu), \qquad & - g \leq \alpha \leq -1,\\
0, \qquad & \text{otherwise},
\end{cases}
\end{equation}
\noindent
which allow us to finally write the quartic-model transseries in diagonal framing as (compare with \eqref{eq:PainleveISolutionDiagonalFraming})
\begin{equation}
\label{eq:QMMSolutionDiagonalFraming}
R \left( t,g_{\text{s}}; \upxi_{1},\upmu \right) = \sum_{g=0}^{+\infty} \sum_{\alpha=-g}^{+\infty} \upxi_{1}^{\alpha}\, \mathsf{R}^{(\alpha)}_{g}  (t,\upmu)\, g_{\text{s}}^{g}.
\end{equation}
\noindent
We list a few\footnote{We have computed up to $g=50$ when $\alpha=0$, up to $g=10$ for $1 \leq \alpha \leq 5$, and up to $g=6$ for $6 \leq \alpha \leq 14$.} explicit polynomial-coefficients for completeness,
\bea
\mathsf{R}^{(0)}_{0} (t,\upmu) &=& r, \\
\mathsf{R}^{(0)}_{1} (t,\upmu) &=& - \frac{\lambda \left( \lambda\, r - 2 \right)}{\sqrt{9 - 3 \lambda\, r} \left( 1 - \lambda\, r \right)^{3/2}}\, \upmu, \\
\mathsf{R}^{(0)}_{2} (t,\upmu) &=& \frac{\lambda^2\, r}{6 \left( \lambda\, r - 1 \right)^4} - \frac{\left( \lambda^3 \left( \lambda\, r - 2 \right) \left(\lambda\, r \left( 5 \lambda\, r - 21 \right) + 36 \right) \right)}{36 \left( \lambda\, r - 3 \right)^2 \left( \lambda\, r - 1 \right)^4}\, \upmu^2, \\
\mathsf{R}^{(0)}_{3} (t,\upmu) &=& \frac{\lambda^3 \left( \lambda\, r \left( \lambda\, r \left( \lambda\, r \left( 4 \lambda\, r \left( \lambda\, r - 12 \right) + 207 \right) - 380 \right) + 220 \right) + 72 \right)}{8 \sqrt{3 - 3 \lambda\, r} \left( 3 - \lambda\, r \right)^{7/2} \left( \lambda\, r - 1 \right)^6}\, \upmu - \nonumber \\
&&
- \frac{5 \left(\lambda^5 \left( \lambda\, r - 2 \right) \left( \lambda\, r \left( \lambda\, r \left( \lambda\, r \left( 11 \lambda\, r - 74 \right) + 276 \right) - 576 \right) + 504 \right) \right)}{432 \sqrt{3 - 3 \lambda\, r} \left( 3 - \lambda\, r \right)^{7/2} \left( \lambda\, r - 1 \right)^6}\, \upmu^3, \\
\mathsf{R}^{(1)}_{0} (t,\upmu) &=& \frac{\sqrt{\lambda\, r}}{\sqrt[4]{3 - 3 \lambda\, r}\, \sqrt[4]{3 - \lambda\, r}}, \\
\mathsf{R}^{(1)}_{1} (t,\upmu) &=& - \frac{\lambda \left( \lambda\, r \left( 2 \lambda\, r \left( \lambda\, r - 3 \right) + 3 \right) + 6 \right)}{8 \sqrt{\lambda\, r} \left( 3 - 3  \lambda\, r \right)^{3/4} \left( 3 - \lambda\, r \right)^{7/4} \left( \lambda\, r - 1 \right)^2} + \\
&&
\hspace{-35pt}
+ \frac{\left( \lambda\, r \right)^{3/2} \left( 2 \lambda\, r - 3 \right) \left( \lambda\, r \left( 4 \lambda\, r + 3 \right) - 18 \right)}{4 r^2 \left( 3 - 3 \lambda\, r \right)^{11/4} \left( 3 - \lambda\, r \right)^{7/4}}\, \upmu + \frac{\left( \lambda\, r \right)^{5/2} \left( \lambda\, r \left( \lambda\, r \left( 90 - 29 \lambda\, r \right) - 36 \right) - 72 \right)}{16 r^3 \left( 3 - 3 \lambda\, r \right)^{11/4} \left( 3 - \lambda\, r \right)^{7/4}}\, \upmu^2, \nonumber \\
\mathsf{R}^{(2)}_{0} (t,\upmu) &=& - \frac{\lambda^2\, r}{2 \left( 3 - 3 \lambda\, r \right)^{3/2} \sqrt{3 - \lambda\, r}}, \\
\mathsf{R}^{(2)}_{1} (t,\upmu) &=& \frac{\lambda^2 \left( \lambda\, r \left( 2 \lambda\, r \left( 11 \lambda\, r - 51 \right) + 117 \right) + 18 \right)}{216 \left( \lambda\, r - 3 \right)^2 \left( \lambda\, r - 1 \right)^4} - \\
&&
\hspace{-35pt}
- \frac{\left( \lambda^3 \left( \lambda\, r \left( \lambda\, r \left( 11 \lambda\, r - 38 \right) + 18 \right) + 36 \right) \right)}{216 \left( \lambda\, r - 3 \right)^2 \left( \lambda\, r - 1 \right)^4}\, \upmu + \frac{\lambda^4 \left( \lambda\, r \left( \lambda\, r \left( 29 \lambda\, r - 90 \right) + 36 \right) + 72 \right)}{1296 \left( \lambda\, r - 3 \right)^2 \left( \lambda\, r - 1 \right)^4}\, \upmu^2. \nonumber
\eea

Direct formulae for translating between rectangular and diagonal coefficients are, as always, straightforward,
\begin{itemize}
\item if $n \geq m$, we have
\begin{equation}
\mathsf{R}^{(\alpha)}_{g} (t,\upmu) = \sum_{m=0}^{g} \upmu^{2m}\, R_{g-2m}^{(\alpha+2m|2m)} (t) + \sum_{m=0}^{g-1} \upmu^{2m+1}\, R_{g-2m-1}^{(\alpha+2m+1|2m+1)} (t),
\end{equation}
\noindent
with inverse map\footnote{The notation $\left. \bullet \right|_{m}$ selects the degree-$m$ term in $\upmu$, out from the $\bullet$-expression.}
\begin{equation}
R^{(n|m)}_{g} (t) = \left. \mathsf{R}^{(n-m)}_{g+m} (t,\upmu) \right|_{m};
\end{equation}
\item if $n < m$, we have
\begin{equation}
\mathsf{R}^{(\alpha)}_{g} (t,\upmu) = \sum_{m=0}^{g+\alpha} \upmu^{2m-\alpha}\, R_{g+\alpha-2m}^{(2m|2m-\alpha)} (t) + \sum_{m=0}^{g+\alpha-1} \upmu^{2m+1-\alpha}\, R_{g+\alpha-2m-1}^{(2m+1|2m+1-\alpha)} (t),
\end{equation}
\noindent
with inverse map
\begin{equation}
R^{(n|m)}_{g} (t) = \left. \mathsf{R}^{(n-m)}_{g+m} (t,\upmu) \right|_{m}
\end{equation}
\end{itemize}

\paragraph{Diagonal-Framing Double-Scaling Limit:}

It should be clear by now how similar \PI, cubic, and quartic matrix models all are at the structural transseries level---with the matrix models being a ``deformation'' of \PI, and with the quartic being more intricate due to the $\BZ_2$ symmetry. As such, a very good check of the (diagonal framing) matrix model results is their natural double-scaling back to \PI, which we will now check for the quartic (the same was done for the cubic; discussed in the main text, subsection~\ref{subsec:resurgent-Z-transasymptotics}).

A very direct way to understand this relation is to compare the $P^{(\alpha)}_{g}$ functions defined above, for \PI~in \eqref{eq:DiagonalSpecificHeatThreeVariables} and for the quartic in \eqref{eq:DiagonalQuarticRThreeVariables}. This suffices as the operations where we go back to the original number of two, $\upxi_1$ and $\upmu$, rather than three, $\upxi_1$, $\upxi_2$ and $\upmu$, transseries parameters are identical for either matrix model or \PI. For the quartic matrix model one has
\begin{equation}
P_{\text{quartic},\text{d}+\text{f}, g}^{(\alpha)} (t,\upmu) = \sum_{m=0}^{g} \upmu^{2m}_{\text{quartic}}\, R_{g-2m}^{(\alpha+2m|2m)} (t) + \sum_{m=0}^{g-1} \upmu^{2m+1}_{\text{quartic}}\, R_{g-2m-1}^{(\alpha+2m+1|2m+1)} (t),
\end{equation}
\noindent
which needs to double-scale to the \PI~result
\begin{equation}
P_{\text{\PI},\text{d}+\text{f}, g}^{(\alpha)} (\upmu) = \sum_{m=0}^{g} \upmu^{2m}_{\text{\PI}}\, u_{2g-2m}^{(\alpha+2m|2m)} + \sum_{m=0}^{g-1} \upmu^{2m+1}_{\text{\PI}}\, u_{2g-2m-2}^{(\alpha+2m+1|2m+1)}.
\end{equation}
\noindent
From \cite{asv11} one recalls how matrix-model transseries-coefficients double-scale to \PI, via
\begin{equation}
\frac{\sigma_{\text{quartic},i}}{C} \longrightarrow \sigma_{\text{\PI},i},
\end{equation}
\noindent
where $C = - \frac{2 \cdot 3^{1/4}}{\sqrt{\lambda}}$. Our earlier $\upxi_{1,2}$ variables in \eqref{eq:upxi12forQuartic} are not very double-scaling friendly due to their $\sqrt{g_{\text{s}}}$ factor, in which case and in analogy with the \PI~analysis in \eqref{eq:defzeta}-\eqref{eq:XiMuRelation} we introduce
\be
\upzeta_{1} = \frac{\upxi_{1}}{\sqrt{g_{\text{s}}}} = \sigma_{1}\, \rme^{- \frac{A(t)}{g_{\text{s}}}} \left( \frac{f(t)}{5184 \lambda^2\, g_{\text{s}}^2} \right)^{\frac{\lambda}{12} \sigma_{1} \sigma_2}, \qquad \upzeta_{2} = \frac{\upxi_{2}}{\sqrt{g_{\text{s}}}} = \sigma_{2}\, \rme^{\frac{A(t)}{g_{\text{s}}}} \left( \frac{f(t)}{5184 \lambda^2\, g_{\text{s}}^2} \right)^{- \frac{\lambda}{12} \sigma_{1} \sigma_{2}}.
\ee
\noindent
With these variables the double-scaling limit is almost obvious,
\bea
\label{eq:DSL-quartic-to-P1-mu}
\frac{\upmu_{\text{quartic}}}{C^2} &\longrightarrow& \upmu_{\text{\PI}}, \\
\label{eq:DSL-quartic-to-P1-zeta}
\frac{\upzeta_{\text{quartic}, 1,2}}{C} &\longrightarrow& \sigma_{\text{\PI}, 1,2}\, \rme^{\mp A_{\text{\PI}}\, z^{\frac{5}{4}}} \left( z^{\frac{5}{4}} \right)^{\pm \frac{2}{\sqrt{3}} \upmu_{\text{\PI}}} = \upzeta_{\text{PI}, 1,2}.
\eea
\noindent
Our focus herein is on double-scaling the relevant combination above,
\begin{equation}
g_{\text{s}}^{g}\, \upxi_{1}^{\alpha}\, P_{\text{quartic},\text{d}+\text{f}, g}^{(\alpha)} (t,\upmu).
\end{equation}
\noindent
This amounts to double-scaling the following two terms,
\bea
g_{\text{s}}^{g}\, \upzeta_{\text{quartic}, 1}^{\alpha}\, \sqrt{g_{\text{s}}}^{\alpha}\, \upmu_{\text{quartic}}^{2m}\, R_{g-2m}^{(\alpha+2m|2m)} (t) &=& \\
&&
\hspace{-150pt}
= \left( \frac{\upzeta_{\text{quartic}, 1}}{C} \right)^{\alpha} \left( \frac{\upmu_{\text{quartic}}}{C^{2}} \right)^{2m} \left( C^{\alpha+4m}\, g_{\text{s}}^{g+\frac{\alpha}{2}}\, R_{g-2m}^{(\alpha+2m|2m)} (t) \right) \longrightarrow \nonumber \\
&&
\hspace{-150pt}
\longrightarrow  \upzeta_{\text{\PI}, 1}^{\alpha}\, \upmu_{\text{\PI}}^{2m}\, z^{-\frac{10g+5\alpha-4}{8}}\, u_{2g-2m}^{(\alpha+2m|2m)[0]} = \sqrt{z}\, z^{-\frac{5}{4}g}\, \upxi_{\text{\PI}, 1}^{\alpha}\, \upmu_{\text{\PI}}^{2m}\, u_{2g-2m}^{(\alpha+2m|2m)[0]}, \nonumber
\eea
\noindent
and
\bea
g_{\text{s}}^{g}\, \upzeta_{\text{quartic}, 1}^{a\alpha}\, \sqrt{g_{\text{s}}}^{\alpha}\, \upmu_{\text{quartic}}^{2m+1}\, R_{g-2m-1}^{(\alpha+2m+1|2m+1)} (t) &=& \\ 
&&
\hspace{-200pt}
= \left( \frac{\upzeta_{\text{quartic}, 1}}{C} \right)^{\alpha} \left( \frac{\upmu_{\text{quartic}}}{C^{2}} \right)^{2m+1} \left( C^{\alpha+4m+2}\, g_{\text{s}}^{g+\frac{\alpha}{2}}\, R_{g-2m-1}^{(\alpha+2m+1|2m+1)} (t) \right) \longrightarrow \nonumber \\
&&
\hspace{-200pt}
\longrightarrow \upzeta_{\text{\PI}, 1}^{\alpha}\, \upmu_{\text{\PI}}^{2m+1}\, z^{-\frac{10g+5\alpha-4}{8}}\, u_{2g-2m-2}^{(\alpha+2m+1|2m+1)[0]} = \sqrt{z}\, z^{-\frac{5}{4}g}\, \upxi_{\text{\PI}, 1}^{\alpha}\, \upmu_{\text{\PI}}^{2m+1}\, u_{2g-2m-2}^{(\alpha+2m+1|2m+1)[0]}. \nonumber
\eea
\noindent
The above double-scalings immediately yield the sought-for double-scaling of the diagonal coefficients, as
\begin{equation}
g_{\text{s}}^{g}\, \upxi_{\text{quartic}, 1}^{\alpha}\, P_{\text{quartic},\text{d}+\text{f}, g}^{(\alpha)} (t,\upmu_{\text{quartic}}) \longrightarrow \sqrt{z}\, z^{-\frac{5}{4}g}\, \upxi_{\text{\PI}, 1}^{\alpha}\, P_{\text{\PI},\text{d}+\text{f}, g}^{(\alpha)} (\upmu_{\text{\PI}})
\end{equation}
\noindent
The $P_{\text{back}, g}^{(\alpha)}$ contributions can be treated in a completely analogous fashion. As to the square-root in the above expressions, it corresponds to the global factor that was taken out in the \PI~discussion in \cite{asv11} (the overall $z$-square-root in \eqref{eq:Painleve1SOlution} or \eqref{eq:SpecificHeatRectangularFraming}). We conclude that diagonal framing is consistent with the double-scaling limit term-by-term.

\paragraph{Free-Energy Diagonal-Framing:}

The structure of the quartic free-energy in rectangular-framing was addressed in \cite{asv11}. It can be written as\footnote{Our perturbative free energy starts at $g=-2$, and our notation hence slightly differs from the one in \cite{asv11}.}
\begin{equation}
F \left( t, g_{\text{s}}; \sigma_1, \sigma_2 \right) = \sum_{n,m=0}^{+\infty} \sigma_{1}^{n} \sigma_{2}^{m}\, \rme^{- \left(n-m\right) \frac{A(t)}{g_{\text{s}}}}\, \sqrt{g_{\text{s}}}^{\,n+m} \left( \frac{f(t)}{5184 \lambda^2\, g_{\text{s}}^2} \right)^{\frac{\lambda}{12} \left(n-m\right) \sigma_{1} \sigma_{2}} \sum_{g=\beta_{nm}^{F}}^{+\infty} F_{g}^{(n|m)[0]} (t)\, g_{\text{s}}^{g}.
\end{equation}
\noindent
The above transseries coefficients are generically rational. The following ones are exceptional (and we shall single them out)
\begin{equation}
F^{(0|0)}_{-2}, \quad F^{(0|0)}_{0}, \quad F^{(1|1)}_{-1}, \quad F^{(2|2)}_{0},
\end{equation}
\noindent
but for all others it is possible to understand their generic structure. As functions of the classical string solution from \eqref{eq:quartic-MM-classical-string-eq}, they factorize into a rational $\CQ^{(n|m)}_{g}$ and a polynomial $\CP^{(n|m)}_{g}$ term, as \cite{asv11}
\begin{equation}
\label{eq:FQMMQPsplit}
F^{(n|m)}_{g} (t) = \CQ^{(n|m)}_{g} (r)\, \CP^{(n|m)}_{g} (\lambda\, r),
\end{equation}
\noindent
where\footnote{Interestingly the structure of the present coefficients of the free energy is very similar to the structure of the coefficients of the string equation in formula (6.51) of \cite{asv11}. This can be made explicit by rewriting the above expression in the precise same notation as was used for $R$ in equation (6.51) of \cite{asv11},
\begin{equation}
\CQ^{(n|m)}_{g} (r) = \frac{\left( \lambda\, r \right)^{p_1+1}}{r^{p_2+1} \left( 3 - 3 \lambda\, r \right)^{p_3+1} \left( 3 - \lambda\, r \right)^{p_4+1-\delta}}.
\end{equation}}
\begin{equation}
\CQ^{(n|m)}_{g} (r) = 
\begin{cases}
\frac{\left( \lambda\, r \right)^{g+1}}{r^{g} \left( 2 - \lambda\, r \right)^{g} \left( 3 - 3 \lambda\, r \right)^{\frac{5}{2}g}}, \quad & n = m = 0, \\
\frac{\left( \lambda\, r \right)^n}{r^{g+2n} \left( 3 - 3 \lambda\, r \right)^{\frac{5}{2} \left(g+n\right)} \left( 3 - \lambda\, r \right)^{\frac{3}{2} \left(g+n\right)}}, \quad & n = m \neq 0, \\
\frac{\left( \lambda\, r \right)^{\frac{3n-m}{2}}}{r^{g+n+m} \left( 3 - 3 \lambda\, r \right)^{\frac{5}{2} \left(g+\frac{n+m}{2}\right)} \left( 3 - \lambda\, r \right)^{\frac{3}{2} \left(g+\frac{n+m}{2}\right)-\frac{\delta}{2}}}, \quad & \text{otherwise},
\end{cases},
\end{equation}
\noindent
and where the $n=m$ cases are clearly just a special case of the last formula. As to the polynomials $\CP^{(n|m)}_{g}$, they have to be computed recursively \cite{asv11}, with their degree given by
\begin{equation}
\deg P^{(n|m)}_{g} = 
\begin{cases}
\frac{1}{2} \left( 3g-2 \right), \quad & n = m = 0, \\
3 \left( g+n \right), \quad & n = m \neq 0, \\
\frac{1}{2} \left( 6g+n+5m-\delta \right), \quad & \text{otherwise},
\end{cases}
\end{equation}
\noindent
and where we have $\delta = \left(n+m\right) \mod 2$. The off diagonal $\beta$-structure for the free energy corresponds to the one of the string equation, with
\begin{equation}
\beta_{nm}^{F} = \beta_{nm}, \qquad n \neq m.
\end{equation} 
\noindent
What this implies is that the free-energy diagonal-framing structure is completely analogous to what we did above (and earlier for \PI). We have (compare with \eqref{eq:PainleveIFreeEnergyDiagonalFraming} for \PI, and with \eqref{eq:QMMSolutionDiagonalFraming} for the quartic string equation)
\begin{equation}
F \left( t,g_{\text{s}}; \upxi_{1},\upmu \right) = F_{\text{A}} \left( g_{\text{s}}; \upmu \right) + \sum_{g=0}^{+\infty} \sum_{\alpha=-g}^{+\infty} \upxi_{1}^{\alpha}\, \mathsf{F}^{(\alpha)}_{g} (t,\upmu)\, g_{\text{s}}^{g},
\end{equation}
\noindent
with
\be
\label{eq:FAQMM}
F_{\text{A}} \left( g_{\text{s}}; \sigma_1 \sigma_2 \right) = \frac{1}{g_{\text{s}}^2}\, F^{(0|0)}_{-2} (t) + F^{(0|0)}_{0} (t) + \frac{1}{g_{\text{s}}}\, F^{(1|1)}_{-1} (t)\, \upmu + F^{(2|2)}_{0} (t)\, \upmu^2,
\ee
\noindent
and where the relation between diagonal and rectangular framing is given by:
\begin{itemize}
\item $n \geq m$, where we have
\begin{equation}
\mathsf{F}^{(\alpha)}_{g} (t,\upmu) = \sum_{m=0}^{g} \upmu^{2m}\, F_{g-2m}^{(\alpha+2m|2m)} (t) + \sum_{m=0}^{g-1} \upmu^{2m+1}\, F_{g-2m-1}^{(\alpha+2m+1|2m+1)} (t),
\end{equation}
\noindent
with the inversion given by 
\begin{equation}
F^{(n|m)}_{g} (t) = \left. \mathsf{F}^{(n-m)}_{g+m} (t,\upmu) \right|_{m};
\end{equation}
\item $n < m$, where we have
\begin{equation}
\mathsf{F}^{(\alpha)}_{g} (t,\upmu) = \sum_{m=0}^{g+\alpha} \upmu^{2m-\alpha}\, F_{g+\alpha-2m}^{(2m|2m-\alpha)} (t) + \sum_{m=0}^{g+\alpha-1} \upmu^{2m+1-\alpha}\, F_{g+\alpha-2m-1}^{(2m+1|2m+1-\alpha)} (t),
\end{equation}
\noindent
with the inversion given by
\begin{equation}
F^{(n|m)}_{g} (t) = \left. \mathsf{F}^{(n-m)}_{g+m} (t,\upmu) \right|_{m}.
\end{equation}
\end{itemize}

All coefficients above fulfill some form of the backward-forward relation. For the rectangular coefficients we have (compare with \eqref{eq:back-forw-PI-rectangularF} for \PI)
\begin{equation}
F^{(n|m)}_{g} (t) = (-1)^{g}\, F^{(m|n)}_{g} (t).
\end{equation}
\noindent
This translates to the polynomials in \eqref{eq:FQMMQPsplit} as
\begin{equation}
\CP^{(n|m)}_{g} (\lambda\, r) = (-1)^{g} \left( \lambda\, r \right)^{2(m-n)} \CP^{(m|n)}_{g} (\lambda\, r).
\end{equation}
\noindent
Finally, for the diagonal coefficients (with $n>m$) we have (compare with \eqref{eq:back-forw-PI-diagonalF} for \PI)
\begin{equation}
\mathsf{F}^{(\alpha)}_{g} (t,\upmu) = (-1)^{g}\, \mathsf{F}^{(-\alpha)}_{g - \alpha} (t,-\upmu).
\end{equation}

\paragraph{A Diagonal Recursion:}

Recall from \eqref{eq:freeenergyfromstringequation} that the free energy follows from the string equation via the Euler--MacLaurin formula (see, \textit{e.g.}, \cite{m04}),
\begin{equation}
F (t+g_{\text{s}}) - 2 F (t) + F (t-g_{\text{s}}) = \log \frac{R (t)}{t}
\end{equation}
\noindent
Interestingly, it is possible to tackle this expression directly in diagonal framing. Start by noting that the logarithm on the right-hand-side has an expansion of the form (itself defining the $\mathsf{L}_{g}^{(\alpha)} (t)$ coefficients)
\begin{equation}
\log \frac{R (t)}{t} = \sum_{g=0}^{+\infty} \sum_{\alpha=-g}^{+\infty} \upxi^{\alpha} \mathsf{L}_{g}^{(\alpha)} (t)\, g_{\text{s}}^{g}.
\end{equation}
\noindent
Further, it is possible to directly resum the genus-zero contribution for $R$ as
\begin{equation}
R_{0} \left(t; \upxi, \upmu \right) = \sum_{\alpha=0}^{+\infty} \upxi^{\alpha}\, \mathsf{R}^{(\alpha)}_{0} (t,\upmu),
\end{equation}
\noindent
which does not explicitly depend on $\upmu$, and from where one may define the following quantities:
\bea
I^{(k)} (t) &=& \frac{1}{k!} \left. \frac{\rmd^{k}}{\rmd\upxi^k} \frac{1}{R_{0} (t; \upxi, \upmu)} \right|_{\upxi=0}, \\
\ell^{(\alpha)} (t) &=& \frac{1}{\alpha!} \left. \frac{\rmd^{\alpha}}{\rmd\upxi^{\alpha}} \log \frac{R_{0} (t; \upxi, \upmu)}{t} \right|_{\upxi=0}, \\
a_{g}^{(\alpha)} &=& g!\, \sum_{k=0}^{\alpha+g} \mathsf{R}^{(\alpha-k)}_{g} (t,\upmu)\, I^{(k)} (t).
\eea
\noindent
Next, recursively define the Bell polynomial-like quantities
\begin{equation}
B_{n,k}^{(\alpha)} = \sum_{i=1}^{n-k+1}{{n-1}\choose{i-1}} \sum_{\beta=-i}^{\alpha+n-i} a_{i}^{(\beta)}\, B_{n-i,k-1}^{(\alpha-\beta)},
\end{equation}
\noindent
with initial conditions
\begin{equation}
B^{(0)}_{0,0} = 1; \qquad B_{0,0}^{(\alpha)} = 0,\,\, \alpha \neq 0; \qquad B_{n,0}^{(\alpha)} = 0,\,\, n \geq 1,\, \forall \alpha; \qquad B_{0,k}^{(\alpha)} = 0,\,\, k \geq 1,\, \forall \alpha.
\end{equation}
\noindent
Using these recursive coefficients, the above logarithm coefficients $\mathsf{L}_{g}^{(\alpha)} (t)$ may be calculated as
\be
\mathsf{L}_{g}^{(\alpha)} (t) = 
\begin{cases}
\log \frac{\mathsf{R}^{(0)}_{0} (t,\upmu)}{t}, \quad & g = 0 = \alpha, \\
\ell^{(\alpha)} (t), \quad & g = 0,\, \alpha \neq 0, \\
\frac{1}{g!} \sum\limits_{k=1}^{g} (-1)^{k+1} \left(k-1\right)!\, B^{(\alpha)}_{g,k}, \quad & \text{otherwise}.
\end{cases}
\ee

Having determined the logarithm coefficients $\mathsf{L}_{g}^{(\alpha)} (t)$ we may next compute the free energy coefficients directly in diagonal framing. Let us begin with sectors without transmonomial contribution, and then address the ones with transmonomial contributions.

\subparagraph{$\alpha=0$ contribution:}

If there are no transmonomial contributions, then, and for each $g \geq 0$, one can compute the recursion relation
\begin{equation}
\label{eq:main-diag-FE-recursion}
\sum_{k=1}^{\lfloor\frac{g+2}{2}\rfloor} \frac{1}{\left(2k\right)!}\, \frac{\rmd^{2k}}{\rmd t^{2k}} \mathsf{F}_{g-2k}^{(0)} (t) = \frac{1}{2}\, \mathsf{L}_{g}^{(0)} (t).
\end{equation} 
\noindent
This recursion can be integrated, and for $g \geq -2$ we find for all values of $g$
\begin{equation}
\frac{\rmd^{2}}{\rmd t^{2}} \mathsf{F}^{(0)}_{g} (t) = \sum_{k=0}^{\lfloor\frac{g+2}{2}\rfloor} \alpha_{k}\, \frac{\rmd^{2k}}{\rmd t^{2k}} \mathsf{L}^{(0)}_{g-2k+2} (t),
\end{equation}
\noindent
where the first few $\alpha_k$ are given by
\begin{equation}
\alpha_{0} = 1, \qquad \alpha_{1} = - \frac{1}{12}, \qquad \alpha_{2} = \frac{1}{240}, \qquad \alpha_{3} = \frac{1}{6048}, \qquad \cdots.
\end{equation}
\noindent
We have computed these coefficients up to $k=50$ from inverting \eqref{eq:main-diag-FE-recursion}. This is the diagonal-framing version of equation (6.123) for the perturbative sector of the free energy given in \cite{asv11}.

\subparagraph{$\alpha\neq 0$ contribution:}

If there are transmonomial contributions, we need a bit more work. In order to write down our result we first define, for convenience, the following quantities\footnote{Note the overall minus sign for $a^{(\alpha)}_{k}$, which is needed to get the correct double-scaling limit \cite{asv11}. In fact this overall sign-ambiguity is the choice of sheet that makes the off-diagonal sectors match \PI, hence makes the double-scaling limit work. It also appears in the instanton action, as
\begin{equation}
A^{\prime}(x) = - \text{arcosh} \left(\frac{3}{\lambda\, r}-2\right).
\end{equation}}
\bea
a_{k}^{(\alpha)} (t,\pm) &=& - \left(\pm 1\right)^{k+1} \frac{\alpha}{k+1}\, \frac{\rmd^{k+1}}{\rmd t^{k+1}} A(t), \qquad b_{k}^{(\alpha)} (t,\pm) = \frac{\lambda}{12}\, \alpha\, \upmu \left. \frac{\rmd^{k}}{\rmd g_{\text{s}}^{k}} \log \left( \frac{f \left( t \pm g_{\text{s}} \right)}{5184 \lambda^2\,  g_{\text{s}}^2} \right) \right|_{g_{\text{s}}=0}, \\
\varphi_{k}^{(\alpha)} (t,\pm) &=& \rme^{\mp \alpha\, A^{\prime} (t)}\, \sum_{n=0}^{k} \frac{1}{\left( k-n \right)!\, n!} \times \nonumber \\
&& \times B_{k-n} \left( a_1^{(\alpha)} (t,\pm), \ldots, a_{k-n}^{(\alpha)} (t,\pm) \right) B_{n} \left( b_1^{(\alpha)} (t,\pm), \ldots, b_{n}^{(\alpha)} (t,\pm) \right),
\eea
\noindent
where the $B_n$ denote the $n$th complete exponential Bell polynomial. Using these definitions, the recursion relation for the free-energy $\alpha \neq 0$ diagonal-coefficients reads
\begin{equation}
\sum_{k=0}^{\min(g,g+\alpha)} \sum_{n=0}^{k} \left\{ \varphi_{k-n}^{(\alpha)} (t,+) + (-1)^n\, \varphi_{k-n}^{(\alpha)} (t,-) \right\} \frac{1}{n!}\, \frac{\rmd^{n}}{\rmd t^{n}} \mathsf{F}^{(\alpha)}_{g-k} (t) - 2 \mathsf{F}^{(\alpha)}_{g} (t) = \mathsf{L}^{(\alpha)}_{g} (t).
\end{equation}
\noindent
For the sole purpose of helping the reader reproduce all of our calculations and efficiently compute free-energy data, it is instructive to spell out the two possible cases explicitly,
\begin{itemize}
\item $\alpha>0$, in which case we find for $g \geq 0$
\bea
\left\{ \varphi^{(\alpha)}_{0} (t,+) + \varphi^{(\alpha)}_{0} (t,-) - 2 \right\} \mathsf{F}^{(\alpha)}_{g} (t) &+& \\
&&
\hspace{-100pt}
+ \sum_{k=1}^{g} \sum_{n=0}^{k} \left\{ \varphi_{k-n}^{(\alpha)} (t,+) + (-1)^n\, \varphi_{k-n}^{(\alpha)} (t,-) \right\} \frac{1}{n!}\, \frac{\rmd^{n}}{\rmd t^{n}} \mathsf{F}^{(\alpha)}_{g-k} (t) = \mathsf{L}^{(\alpha)}_{g} (t); \nonumber
\eea
\item $\alpha<0$, in which case we find for $g \geq 0$ and $-g \leq \alpha < 0$
\bea
\left\{ \varphi^{(\alpha)}_{0} (t,+) + \varphi^{(\alpha)}_{0} (t,-) - 2 \right\} \mathsf{F}^{(\alpha)}_{g} (t) &+& \\
&&
\hspace{-100pt}
+ \sum_{k=1}^{g+\alpha} \sum_{n=0}^{k} \left\{ \varphi_{k-n}^{(\alpha)} (t,+) + (-1)^n\, \varphi_{k-n}^{(\alpha)} (t,-) \right\} \frac{1}{n!}\, \frac{\rmd^{n}}{\rmd t^{n}} \mathsf{F}^{(\alpha)}_{g-k} (t) = \mathsf{L}^{(\alpha)}_{g} (t). \nonumber
\eea
\end{itemize}

\paragraph{Partition-Function Diagonal-Framing:}

To wrap things up, all we have to do is calculate the diagonal-coefficients of the partition-function, which may be obtained by exponentiation of the free-energy in diagonal-framing. Explicitly, and in analogy with \PI,
\begin{equation}
Z \left( g_{\text{s}}; \upxi_{1}, \upmu \right) = \rme^{F_{\text{A}} \left( g_{\text{s}}; \upmu \right)}\, \rme^{\mathsf{F}_{0} \left( \upxi_{1}, \upmu \right)}\, \exp \left( \sum_{g=1}^{+\infty} \sum_{\alpha=-g}^{+\infty} \upxi_{1}^{\alpha}\, \mathsf{F}^{(\alpha)}_{g} (t,\upmu)\, g_{\text{s}}^{g} \right),
\end{equation}
\noindent
where we used \eqref{eq:FAQMM} for $F_{\text{A}} \left( g_{\text{s}}; \upmu \right)$ and denoted
\begin{equation}
\mathsf{F}_{0} \left( \upxi_{1}, \upmu \right) = \sum_{\alpha=0}^{+\infty} \upxi_{1}^{\alpha}\, \mathsf{F}^{(\alpha)}_{0} (t,\upmu).
\end{equation}
\noindent
We can next address the above three terms separately, as (compare with the \PI~discussion in sub-appendix~\ref{subapp:transasymptotic-transseries-PI})
\begin{itemize}
\item $\rme^{F_{\text{A}} \left( g_{\text{s}}; \upmu \right)}$: This term is just a finite sum in $g_{\text{s}}$ and $\upmu$, as already explicitly evaluated in \eqref{eq:FAQMM}. We leave it as a general pre-factor, which will be very convenient in the following.
\item $\rme^{\mathsf{F}_{0} \left( \upxi_{1}, \upmu \right)}$: In complete analogy\footnote{And, naturally, it also double scales to the corresponding $g=0$ part in the \PI~free-energy.}
 with \PI, at genus $g=0$ this contribution of the free energy can be simply resummed (and is $\upmu$-independent)
\begin{equation}
\mathsf{F}_{0} \left( \upxi_{1}, \upmu \right) = \log \left( 1 + \frac{\left( \lambda\, r \right)^{\frac{3}{2}}}{2 r \left( 3 - 3 \lambda\, r \right)^{\frac{5}{4}} \left( 3 - \lambda\,  r \right)^{\frac{1}{4}}}\, \upxi_{1} + \frac{\left( \lambda\, r \right)^{4}}{8 r^{2} \left( 3 - 3 \lambda\, r \right)^{\frac{5}{2}} \left( 3 - \lambda\, r \right)^{\frac{3}{2}}}\, \upxi_{1}^2 \right),
\end{equation}
\noindent
from where immediately follows
\be
\rme^{\mathsf{F}_{0} \left( \upxi_{1}, \upmu \right)} = 1 + \underbrace{\frac{\left( \lambda\, r \right)^{\frac{3}{2}}}{2 r \left( 3 - 3 \lambda\, r \right)^{\frac{5}{4}} \left( 3 - \lambda\,  r \right)^{\frac{1}{4}}}}_{\equiv c_1 (r)} \upxi_{1} +  \underbrace{\frac{\left( \lambda\, r \right)^{4}}{8 r^{2} \left( 3 - 3 \lambda\, r \right)^{\frac{5}{2}} \left( 3 - \lambda\, r \right)^{\frac{3}{2}}}}_{\equiv c_2 (r)} \upxi_{1}^2.
\ee
\item $\rme^{\big( \,\cdots\, \big)}$: Again in complete analogy with \PI, for this term we have to explicitly perform the exponentiation as
\begin{equation}
\exp \left( \sum_{g=1}^{+\infty} \sum_{\alpha=-g}^{+\infty} \upxi_{1}^{\alpha}\, \mathsf{F}^{(\alpha)}_{g} (t,\upmu)\, g_{\text{s}}^{g} \right) = \sum_{g=0}^{+\infty} \sum_{\alpha=-g}^{+\infty} \upxi_{1}^\alpha\, \mathsf{b}^{(\alpha)}_{g} (t,\upmu)\, g_{\text{s}}^{g}.
\end{equation}
\noindent
Herein, the coefficients $\mathsf{b}^{(\alpha)}_{g}$ are again calculated via a Bell-polynomial-like recursion,
\begin{equation}
\mathsf{b}^{(\alpha)}_{n} (t,\upmu) = \sum_{i=0}^{n-1} \frac{i+1}{n}\, \sum_{m=0}^{\alpha+n} \mathsf{b}^{(m+i+1-n)}_{n-1-i} (t,\upmu)\, \mathsf{F}^{(\alpha+n-m-i-1)}_{i+1} (t,\upmu),
\end{equation}
\noindent
with the initial recursion coefficients
\begin{equation}
\mathsf{b}^{(0)}_{0} = 1, \qquad \mathsf{b}^{(\alpha)}_{0} = 0, \quad \alpha \geq 1.
\end{equation}
\end{itemize}
\noindent
The last two expressions may actually be combined as
\begin{equation}
\rme^{\mathsf{F}_{0} \left( \upxi_{1}, \upmu \right)}\, \exp \left( \sum_{g=1}^{+\infty} \sum_{\alpha=-g}^{+\infty} \upxi_{1}^{\alpha}\, \mathsf{F}^{(\alpha)}_{g} (t,\upmu)\, g_{\text{s}}^{g} \right) \equiv \sum_{g=0}^{+\infty} \sum_{\alpha=-g}^{+\infty} \upxi_{1}^{\alpha}\, \mathsf{Z}^{(\alpha)}_{g} (t,\upmu)\, g_{\text{s}}^{g},
\end{equation}
\noindent
where, at the component level, we can relate coefficients as
\begin{equation}
\mathsf{Z}^{(\alpha)}_{g} (t,\upmu) = \mathsf{b}^{(\alpha)}_{g} (t,\upmu) + c_1 (r)\, \mathsf{b}^{(\alpha-1)}_{g} (t,\upmu) + c_2 (r)\, \mathsf{b}^{(\alpha-2)}_{g} (t,\upmu).
\end{equation}
\noindent
Just like for \PI, this allows us to spell-out the partition-function in the compact form
\begin{equation}
Z \left( g_{\text{s}}; \upxi_{1}, \upmu \right) = \rme^{F_{\text{A}} \left( g_{\text{s}}; \upmu \right)}\, \sum_{g=0}^{+\infty} \sum_{\alpha=-g}^{+\infty} \upxi_{1}^{\alpha}\, \mathsf{Z}^{(\alpha)}_{g} (t,\upmu)\, g_{\text{s}}^{g}.
\end{equation}
\noindent
Analogously to the $\mathsf{b}^{(\alpha)}_{g}$ coefficients, the $\mathsf{Z}^{(\alpha)}_{g}$ coefficients can be calculated via a recursion
\begin{equation}
\mathsf{Z}^{(\alpha)}_{n} (t,\upmu) = \sum_{i=0}^{n-1} \frac{i+1}{n}\, \sum_{k=0}^{\alpha+n} \mathsf{Z}^{(\alpha-k+i+1)}_{n-1-i} (t,\upmu)\, \mathsf{F}^{(k-i-1)}_{i+1} (t,\upmu),
\end{equation}
\noindent
with the fixed initial data
\begin{equation}
\mathsf{Z}^{(\alpha)}_{0} = \delta_{\alpha 0} + c_1 (r)\, \delta_{\alpha 1} + c_2 (r)\, \delta_{\alpha 2}.
\end{equation} 

Finally---and again in complete analogy with \PI---the diagonal-framing coefficients of the partition-function have a starting genus, or $\beta$-structure, which is more restricted than the one we observed for the free energy. It is quadratic and given by\footnote{Note that the starting genus for the quartic-model differs from the one for \PI. This is due to the $\BZ_2$-symmetry of the quartic which induces a factor of $2$ in its free energy in comparison to the case of \PI; hence making the quartic partition-function double-scale to the \textit{square} of the \PI~partition-function. Thus, the different starting genus.}
\begin{equation}
\beta_{(\alpha)} = \left\lceil \frac{\alpha \left( \alpha-2 \right)}{4} \right\rceil.
\end{equation}
\noindent
Then we can finally write
\begin{equation}
Z \left( g_{\text{s}}; \upxi_{1}, \upmu \right) = \rme^{F_{\text{A}} \left( g_{\text{s}}; \upmu \right)}\, \sum_{\alpha \in \BZ}\, \sum_{g=\beta_{(\alpha)}}^{+\infty} \upxi_{1}^{\alpha}\, \mathsf{Z}^{(\alpha)}_{g} (t,\upmu)\, g_{\text{s}}^{g}.
\end{equation}

Let us list a few explicit examples for these diagonal partition-function polynomial-coefficients (where we set $\lambda=1$ for brevity---this $\lambda$ dependence may always be restored by dimensional analysis). One finds\footnote{We have computed diagonal partition-function coefficients (albeit truncated at $\upmu$-order $5$) for $-6<\alpha<6$ up to the first $5$ non-vanishing $g_{\text{s}}$ orders, and for $-10<\alpha<10$ up to the first $2$ non-vanishing $g_{\text{s}}$ orders. In addition we have computed diagonal partition-function coefficients (now truncated at $\upmu$-order $9$) for $-3< \alpha < 3$ up to the first $3$ non-vanishing $g_{\text{s}}$ contributions.}, along the main diagonal, \textit{e.g.},
\bea
\label{eq:Zmu-poly-examples-QMM-1}
\mathsf{Z}^{(0)}_{0} (t,\upmu) &=& 1, \\
\label{eq:Zmu-poly-examples-QMM-2}
\mathsf{Z}^{(0)}_{1} (t,\upmu) &=& \frac{\left( r-3 \right) \left( r \left( - 4 r^2 + 30 r - 63 \right) + 18 \right)}{72 \sqrt{3} \left( \left( r-3 \right) \left( r-1 \right) \right)^{5/2} r}\, \upmu + \frac{\left( r \left( r \left( 29r-90 \right) + 36 \right) + 72 \right)}{144 \left( 3-3r \right)^{5/2} \left( 3-r \right)^{3/2} r}\, \upmu^3, \\
\label{eq:Zmu-poly-examples-QMM-3}
\mathsf{Z}^{(0)}_{2} (t,\upmu) &=& - \frac{r \left( 41 r^2 - 185 r + 200 \right)}{2880 \left( r-2 \right)^2 \left( r-1 \right)^5} + \nonumber \\
&&
+ \frac{22 r^6 - 1920 r^5 + 17856 r^4 - 55908 r^3 + 66285 r^2 - 21708 r + 4212}{31104 \left( r-3 \right)^3 \left( r-1 \right)^5 r^2}\, \upmu^2 + \nonumber \\
&&
+ \frac{2261 r^6 - 16320 r^5 + 46314 r^4 - 52128 r^3 - 6912 r^2 + 46656 r - 10368}{559872 \left( r-3 \right)^3 \left( r-1 \right)^5 r^2}\, \upmu^4 + \nonumber \\
&&
+ \frac{\left( 29 r^3 - 90 r^2 + 36 r + 72 \right)^2}{10077696 \left( r-3 \right)^3 \left( r-1 \right)^5 r^2}\, \upmu^6,
\eea
\noindent
as well as off the main diagonal, \textit{e.g.},
\bea
\label{eq:Zmu-poly-examples-QMM-4}
\mathsf{Z}^{(1)}_{0} (t,\upmu) &=& \frac{\sqrt{r}}{2 \left( 3-3 r \right)^{5/4} \sqrt[4]{3-r}}, \\
\label{eq:Zmu-poly-examples-QMM-5}
\mathsf{Z}^{(1)}_{1} (t,\upmu) &=& \frac{r \left(2 \left( 27-5 r \right) r-75 \right)-6}{48\, 3^{3/4} \left( r-1 \right)^2 \left( \left( r-3 \right) \left( r-1 \right) \right)^{7/4} \sqrt{r}} + \nonumber \\
&&
+ \frac{r \left( r \left( 59 r-252 \right) + 261 \right) + 90}{432\, 3^{3/4} \left( r-1 \right)^2 \left( \left( r-3 \right) \left( r-1 \right) \right)^{7/4} \sqrt{r}}\, \upmu + \nonumber \\
&&
+ \frac{r \left( \left( 90-29 r \right) r-36 \right) - 72}{32 \left( 3-3 r \right)^{15/4} \left( 3-r \right)^{7/4} \sqrt{r}}\, \upmu^2 + \frac{r \left( r \left( 29 r-90 \right) + 36 \right) + 72}{288 \left( 3-3 r \right)^{15/4} \left( 3-r \right)^{7/4} \sqrt{r}}\, \upmu^3, \\
\label{eq:Zmu-poly-examples-QMM-6}
\mathsf{Z}^{(2)}_{0} (t,\upmu) &=& \frac{r^2}{8 \left( 3-3 r \right)^{5/2} \left( 3-r \right)^{3/2}}, \\
\label{eq:Zmu-poly-examples-QMM-7}
\mathsf{Z}^{(2)}_{1} (t,\upmu) &=& - \frac{r \left( r \left( 62 r^2 - 210 r + 135 \right) + 126 \right)}{2592 \left( r-3 \right)^3 \left( r-1 \right)^5} + \frac{r \left( r \left( 2 r \left( 89 r-285 \right) + 279 \right) + 414 \right)}{15552 \left( r-3 \right)^3 \left( r-1 \right)^5}\, \upmu - \nonumber \\
&&
- \frac{r \left( r \left( r \left( 29 r-90 \right) + 36 \right) + 72 \right)}{15552 \left( r-3 \right)^3 \left( r-1 \right)^5}\, \upmu^2 + \frac{r \left( r \left( r \left( 29 r-90 \right) + 36 \right) + 72 \right)}{279936 \left( r-3 \right)^3 \left( r-1 \right)^5}\, \upmu^3, \\
\label{eq:Zmu-poly-examples-QMM-8}
\mathsf{Z}^{(2)}_{1} (t,\upmu) &=& - \frac{r^{9/2}}{15552 \sqrt[4]{3-3 r} \left( 3-r \right)^{13/4} \left( r-1 \right)^6} + \frac{r^{9/2}}{93312 \sqrt[4]{3-3 r} \left( 3-r \right)^{13/4} \left( r-1 \right)^6}\, \upmu, \\
\label{eq:Zmu-poly-examples-QMM-9}
\mathsf{Z}^{(-1)}_{1} (t,\upmu) &=& \frac{\sqrt{r}}{2 \left( 3-3 r \right)^{5/4} \sqrt[4]{3-r}}\, \upmu.
\eea
\noindent
Lower undisplayed coefficients in this list vanish, \textit{i.e.}, $\mathsf{Z}^{(2)}_{0} (t,\upmu) = 0 = \mathsf{Z}^{(-1)}_{0} (t,\upmu)$.

Given all our data, some illustrated just above, we may proceed to write the quadratic-transasymptotics partition-function in closed-form. There are two extra facts to take into account herein:
\begin{itemize}
\item Due to the $\BZ_2$-symmetry of the quartic matrix model, its partition function should have a structure which is as in ``two copies'' of the cubic partition function.
\item Likewise, given the known double-scaling limit of the cubic partition-function, the quartic partition-function should now double-scale to the \textit{square} of the \PI~partition function.
\end{itemize}
\noindent
In light of this, and with the experience we have acquired so far, it is immediate to conjecture the closed-form
\be
\label{eq:QMM-quadratic-transasymptotics-APP}
Z \left( t, g_{\text{s}}; \upzeta_{1}, \upmu \right) = \rme^{F_{\text{A}} \left( g_{\text{s}}; \upmu \right)}\, \sum_{g=0}^{+\infty}\, \sum_{\alpha=-g}^{+\infty} \upzeta_1^\alpha\, \mathsf{Z}_{g}^{(\alpha)} (t,\upmu),
\ee
\noindent
with an explicit even/odd split
\be
\mathsf{Z}_{g}^{(\alpha)} (t,\upmu) \equiv 
\begin{cases}
\sum\limits_{\tilde{g}=0}^{g} \sum\limits_{s=-\tilde{g}}^{\alpha+g-\tilde{g}} \mathsf{Y}^{(\alpha-s)}_{\text{even}, g-\tilde{g}} (\upmu)\, \mathsf{Y}^{(s)}_{\text{even}, \tilde{g}} (\upmu), \quad & \alpha \text{ even}, \\
d \cdot \sum\limits_{\tilde{g}=0}^{g} \sum\limits_{s=-\tilde{g}}^{\alpha+g-\tilde{g}} \mathsf{Y}^{(\alpha-s)}_{\text{odd}, g-\tilde{g}} (\upmu)\, \mathsf{Y}^{(s)}_{\text{odd}, \tilde{g}} (\upmu), \quad  & \alpha \text{ odd},
\end{cases}
\ee
\noindent
and where we have defined
\begin{equation}
\label{eq:d2-app}
d^2 = \frac{3-\lambda r}{2 \lambda r}
\end{equation}
\noindent
and (compare with \eqref{eq:diagonalframingresult} and, specially, with \eqref{eq:cubicDFTKernel})
\bea
\mathsf{Y}_{\text{even}/\text{odd},\, g + \frac{1}{2} \alpha \left(\alpha-1\right)}^{(\alpha)} (t,\upmu) &=& \left( \rmi\, \frac{2^{1/2}}{3^{1/4} C} \right)^\alpha \left( p_{\text{quartic}} (t)\, g_{\text{s}}^{\frac{1}{2}} \right)^{\alpha^2} \left(- \frac{\rmi}{2 \sqrt{3}}\, g_{\text{s}} \right)^{g} \times \\
&&
\hspace{-60pt}
\times D_{\text{even}/\text{odd},\, g} \left( \alpha-\frac{2}{\sqrt{3}}\, \frac{\upmu_{\text{quartic}}}{C^2} \right) \frac{G_2 \left( 1 + \alpha - \frac{2}{\sqrt{3}} \frac{\upmu_{\text{quartic}}}{C^2} \right)}{G_2 \left( 1 - \frac{2}{\sqrt{3}} \frac{\mu_{\text{quartic}}}{C^2} \right) \Gamma \left( 1 - \frac{2}{\sqrt{3}} \frac{\upmu_{\text{quartic}}}{C^2} \right)^\alpha}. \nonumber
\eea
\noindent
This expression is ready to be double-scaled to \PI, much like we did in the case of the cubic. We are now denoting $C = - \frac{2 \cdot 3^{1/4}}{\sqrt{\lambda}}$, alongside the (quartic/\PI) quantities
\be
\label{eq:pquartic-app}
p_{\text{quartic}} (t) = \frac{\rmi \sqrt{3}\, \lambda^{3/2} r}{2 \left( 3 - 3 \lambda r \right)^{5/4} \left( 3 - \lambda r \right)^{3/4}}, \qquad  p_{\text{\PI}} = \frac{\rmi}{2^{5/2}\, 3^{3/4}},
\ee
\noindent
so that the relevant limits are \eqref{eq:DSL-quartic-to-P1-mu}-\eqref{eq:DSL-quartic-to-P1-zeta} supplemented by
\bea
g_{\text{s}}^{\frac{1}{2}}\, p_{\text{quartic}} &\longrightarrow& z^{-\frac{5}{8}}\, p_{\text{\PI}} = z^{-\frac{5}{8}}\, \frac{\rmi}{2^{5/2}\, 3^{3/4}},\\
d &\longrightarrow& 1.
\eea
\noindent
Finally, the quartic-model $D_g (\nu)$ polynomials such as \eqref{eq:quartic-dk-e} and \eqref{eq:quartic-dk-o} double-scale to the corresponding ones in \PI, as
\begin{equation}
g_{\text{s}}^{g}\, D_{\text{quartic}, \text{even}/\text{odd}, g} (\nu) \,\longrightarrow\, z^{-\frac{5}{4}g}\, D_{\text{\PI}, g} (\nu),
\end{equation}
\noindent
where the distinction between even and odd polynomials disappears in the limit, exactly as expected for \PI. For some explicit $D_{\text{even}/\text{odd}, k}$ see \eqref{eq:quartic-dk-e}-\eqref{eq:quartic-dk-o}.

\section{All-Order Saddle-Point Expansion of the Cubic Matrix Model}
\label{appendix:Allordersaddlepoint}

This appendix contains the relevant calculations for the matrix-integral evaluation of the $\ell$-instanton sector $\mathcal{Z}^{(\ell)}(t,g_{\text{s}})$ of the cubic matrix model in subsection~\ref{subsec:two-viewpoints-NP-sectors}; \textit{i.e.}, all intermediate steps from equation \eqref{eq:toprec18} to \eqref{eq:toprec22}. In particular, this yields a fully-analytical, closed-form expression for all expansion-coefficients in \eqref{eq:toprec22}. Starting-off with \eqref{eq:toprec18}, we begin by evaluating the multi-determinant correlation-function as
\bea
\left\langle \prod_{i=1}^\ell \det \left( x_i - M \right)^2 \right\rangle_{N} &=& \exp \left( \sum_{\substack{\boldsymbol{n} \in \BN_0^\ell}}' \left\{ \prod_{j=1}^\ell \frac{2^{n_j}}{n_j!} \right\} A_{n} ( \underbrace{x_1,\cdots,x_1}_{n_1}, \ldots, \underbrace{x_\ell,\cdots,x_\ell}_{n_\ell} ) \right) = \\
&=& \exp \left( \sum_{\substack{\boldsymbol{n} \in \BN_0^\ell}}' \left\{ \prod_{j=1}^\ell \frac{2^{n_j}}{n_j!} \right\} \sum_{g=0}^{+\infty} A_{g;n} ( \underbrace{x_1,\cdots,x_1}_{n_1}, \ldots, \underbrace{x_\ell,\cdots,x_\ell}_{n_\ell} )\, g_{\text{s}}^{2g+n-2} \right). \nonumber
\eea
\noindent
This follows via the usual cumulant expansion, where the $A$-functions in the first line above are connected correlation-functions obtained by integration of the multi-resolvent correlation-functions \eqref{eq:multiresolventcorrelationfunctionshermitianmatrix} (and, in the second line above, of their perturbative coefficients \eqref{eq:genusgmultiresolvents}, respectively; see the discussions and formulae in \cite{msw07, mss22}), the prime in the summation indicates we should exclude the index $\boldsymbol{n} = \boldsymbol{0}$, and we used the compact notation
\begin{equation}
\label{eq:toprec23}
n = \sum_{i=1}^\ell n_i
\end{equation}
\noindent
for any $\boldsymbol{n} \in \mathbb{Z}^\ell$. Changing the matrix-integral by removing $\ell$ eigenvalues shifts its 't~Hooft coupling by $t \to t - \ell g_{\text{s}}$; hence upon Taylor $g_{\text{s}}$-expansion this yields
\be
\left\langle \prod_{i=1}^\ell \det \left( x_i - M \right)^2 \right\rangle_{N-\ell} = \exp \left( \sum_{\substack{\boldsymbol{n} \in \BN_0^\ell}}' \left\{ \prod_{j=1}^\ell \frac{2^{n_j}}{n_j!} \right\} \sum_{g,m \in \mathbb{N}_0} \frac{(-\ell)^m}{m!}\, \partial^m_t A_{g;n}\, g_{\text{s}}^{2g+n-2+m} \right),
\ee
\noindent
where (the derivatives of) the $A_{g;n}$ are still evaluated at $\underbrace{x_1,\cdots,x_1}_{n_1}, \ldots, \underbrace{x_\ell,\cdots,x_\ell}_{n_\ell}$ just like above, but we omit notation to simplify formulae. Equation \eqref{eq:toprec18} hence becomes
\bea
\frac{\CZ^{(\ell)}}{\CZ^{(0)}} &=& \frac{1}{\ell!}\, \frac{\CZ^{(0)} (t-\ell g_{\text{s}})}{\CZ^{(0)} (t)} \int_{\CC^\star} \prod_{j=1}^\ell \frac{\rmd x_j}{2\pi}\, \exp \left( -\frac{1}{g_{\text{s}}} \sum_{m=0}^{+\infty} \frac{(-\ell)^m}{m!}\, \partial^m_t V_{\text{h;eff}} (x_j)\, g_{\text{s}}^m \right) \times \\
&&
\times \prod_{\substack{1\le k<i \le \ell}} \left( x_i - x_k \right)^2 \times \exp \left( \sum_{\substack{\boldsymbol{n} \in \BN_0^\ell}}' \left\{ \prod_{j=1}^\ell \frac{2^{n_j}}{n_j!} \right\} \sum_{\substack{g,m \in \mathbb{N}_0 \\ (g,n)\neq(0,1)}} \frac{(-\ell)^m}{m!}\, \partial^m_t A_{g;n}\, g_{\text{s}}^{2g+n-2+m} \right). \nonumber
\eea
\noindent
Next, consider the change of variables
\be
\label{eq:step4}
x_j = x^\star + \sqrt{g_{\text{s}}}\, t_j,
\ee
\noindent
suitable for saddle-point expansion around our nonperturbative saddle. Performing this change of variables followed by another Taylor $g_{\text{s}}$-expansion yields, in the first exponential term above,
\bea
\exp \left( -\frac{1}{g_{\text{s}}} \sum_{m=0}^{+\infty} \frac{(-\ell)^m}{m!}\, \partial^m_t V_{\text{h;eff}} (x_j)\, g_{\text{s}}^m \right) &=& \exp \left( - \frac{1}{g_{\text{s}}}\, V_{\text{h;eff}}(x^\star) \right) \times \\
&&
\hspace{-50pt}
\times \exp \left( -\frac{1}{2}\, \frac{\partial^2 V_{\text{h;eff}}}{\partial x^2} (x^\star)\, t_j^2 + \ell\, \frac{\partial V_{\text{h;eff}}}{\partial t} (x^\star) \right) \sum_{g=0}^{+\infty} p_{\ell,g} (t_j)\, g_{\text{s}}^{\frac{g}{2}}; \nonumber
\eea
\noindent
it further yields, in the Vandermonde term,
\be
\prod_{\substack{1\le k<i \le \ell}} \left( x_i - x_k \right)^2 = g_{\text{s}}^{\binom{\ell}{2}}\prod_{\substack{1\le k<i \le \ell}} \left( t_i - t_k \right)^2;
\ee
\noindent
and finally yields in the second exponential term above,
\bea
\exp \left( \sum_{\substack{\boldsymbol{n} \in \BN_0^\ell}}' \left\{ \prod_{j=1}^\ell \frac{2^{n_j}}{n_j!} \right\} \sum_{\substack{g,m \in \mathbb{N}_0 \\ (g,n)\neq(0,1)}} \frac{(-\ell)^m}{m!}\, \partial^m_t A_{g;n}\, g_{\text{s}}^{2g+n-2+m} \right) &=& \\
&&
\hspace{-75pt}
= \rme^{2 \ell^2\, A_{0;2} (x^\star,x^\star)}\, \sum_{g=0}^{+\infty} \mathsf{p}_{\ell,g} (t_1,\ldots,t_\ell)\, g_{\text{s}}^{\frac{g}{2}}. \nonumber
\eea
\noindent
In these expansions we introduced several polynomials: a set of single-variable polynomials
\be
\label{eq:acoef}
p_{\ell,0} (t_j) = 1, \qquad p_{\ell,g} (t_j) = \sum_{\substack{(\boldsymbol{m},\boldsymbol{k}) \in \BN_0 ^p \times \BN_0 ^p \\ k+2m-2p=g}}' \frac{1}{p!}\, \prod_{i=1}^p \left\{ -\frac{(-\ell)^{m_i}}{k_i!m_i!}\, \frac{\partial^{k_i+m_i} V_{\text{h;eff}}}{\partial x^{k_i}\, \partial t^{m_i}} (x^\star)\, t_j^{k_i} \right\}, \quad g \in \mathbb{N},
\ee
\noindent
where the prime in the summation indicates we should exclude all indices $(\boldsymbol{m},\boldsymbol{k}) \in \BN_0^p \times \BN_0^p$ for which $k_i+2m_i-2<1$ for some $1 \le i \le p$; and a set of multi-variable polynomials
\bea
\mathsf{p}_{\ell,0} (t_1,\ldots,t_\ell) &=& 1, \\
\mathsf{p}_{\ell,g} (t_1,\ldots,t_\ell) &=& \sum_{\substack{(\boldsymbol{k}_1,\boldsymbol{n}_1,m_1,h_1), \ldots, (\boldsymbol{k}_p,\boldsymbol{n}_p,m_p,h_p) \in \BN^\ell_0 \times \BN^\ell_0 \times \BN_0 \times \BN_0 \\ (k_1+2m_1+4h_1+2n_1-4) + \cdots + (k_p+2m_p+4h_p+2n_p-4) = g}}' \frac{1}{p!}\, \prod_{i=1}^p \left( \left\{ \prod_{j=1}^\ell \frac{2^{(n_i)_j}}{(n_i)_j!}\, \frac{t_j^{(k_i)_j}}{(k_i)_j!} \right\} \times \right. \nonumber \\
&&
\hspace{-80pt}
\left. \times \frac{(-\ell)^{m_i}}{m_i!}\, \frac{\partial^{k_i+m_i}}{\partial x_1^{(k_i)_1} \cdots \partial x_\ell^{(k_i)_\ell}\, \partial t^{m_i}} A_{h_i;n_i} (\underbrace{x_1,\cdots,x_1}_{(n_i)_1}, \ldots, \underbrace{x_\ell,\cdots,x_\ell}_{(n_i)_\ell}) \Bigg\vert_{x_1 = \cdots = x_\ell = x^\star} \right),\, g \in \mathbb{N},
\label{eq:bcoef}
\eea
\noindent
where the prime in the summation indicates we should exclude indices such that $k_i+2m_i+4h_i+2n_i-4<2$ or $(n_i,h_i)=(1,0)$ or $n_i=0$ for some $1 \le i \le p$. Having gone through all these expansions, we can finally write for \eqref{eq:toprec18}
\bea
\frac{\CZ^{(\ell)}}{\CZ^{(0)}} &=& \frac{1}{\ell!}\, \frac{\CZ^{(0)} (t-\ell g_{\text{s}})}{\CZ^{(0)} (t)}\, g_{\text{s}}^{\frac{\ell^2}{2}}\, \exp \left( -\frac{\ell}{g_{\text{s}}}\, V_{\text{h;eff}}(x^\star) +\ell^2\, \Big\{ \partial_t V_{\text{h;eff}} (x^\star) + 2 A_{0;2} (x^\star,x^\star) \Big\} \right) \times \nonumber \\
&&
\times \prod_{j=1}^{\ell} \int_{-\infty}^{+\infty} \frac{\rmd t_j}{2\pi}\, \exp \left( - \frac{1}{2}\, \frac{\partial^2 V_{\text{h;eff}}}{\partial x^2} (x^\star)\, t_j^2 \right) \times \sum_{g=0}^{+\infty} \mathtt{p}_{\ell,g} (t_1,\ldots,t_{\ell})\, g_{\text{s}}^{\frac{g}{2}},
\label{eq:step1}
\eea
\noindent
where we further defined the ``new'' multi-variable polynomials essentially given by the ``older'' ones:
\be
\mathtt{p}_{\ell,g} (t_1,\ldots,t_\ell) = \prod_{\substack{1 \le j < i \le \ell}} \left( t_i - t_j \right)^2 \times \sum_{\substack{ (\boldsymbol{n},i) \in \BN_0^\ell \times \BN_0 \\ n+i = g}} \left\{ \prod_{j=1}^\ell p_{\ell,n_j} (t_j)\right\} \mathsf{p}_{\ell,i} (t_1,\ldots,t_\ell).
\ee
\noindent
Next, in order to perform all the Gaussian integrations in \eqref{eq:step1}, we now have to make the coefficients of these polynomials explicit. To this end, we begin by writing them explicitly as
\bea
\label{eq:step2}
p_{\ell,g} (t_j) &=& \sum_{k=0}^{g+2} p_{\ell,g;k}\, t_j^k, \\ 
\mathsf{p}_{\ell,g} (t_1,\ldots,t_\ell) &=& \sum_{\boldsymbol{i} \in \mathbb{Z}^\ell_{g+1}} \mathsf{p}_{\ell,g;\boldsymbol{i}}\, \boldsymbol{t}^{\boldsymbol{i}}, \label{eq:step3}
\eea
\noindent
where $\mathbb{Z}_g = \left\{0,\cdots,g-1\right\}$ stands for the Abelian group of integers modulo $g$ and where the polynomial coefficients $ p_{\ell,g;k}$ in \eqref{eq:step2} and $\mathsf{p}_{\ell,g;\boldsymbol{i}}$ in \eqref{eq:step3} follow directly from \eqref{eq:acoef} and \eqref{eq:bcoef}, respectively. We also need the coefficients from
\be
\prod_{\substack{1 \le j < i \le \ell}} \left( t_i - t_j \right)^2 = \sum_{\substack{\boldsymbol{i} \in \mathbb{Z}^\ell_{\ell}}} \Delta_{\ell;\boldsymbol{i}}\, \boldsymbol{t}^{\boldsymbol{i}},
\ee
\noindent
where $\Delta_{\ell;\boldsymbol{i}} = \operatorname{sgn} (\sigma)$, $\boldsymbol{i} = \sigma \left[ (0,1,\ldots,\ell-1) \right]$ for any permutation $\sigma \in \mathsf{S}_\ell$ and $\Delta_{\ell;\boldsymbol{i}} = 0$ otherwise. We can then finally write
\be
\mathtt{p}_{\ell,g} (t_1,\ldots,t_\ell) = \sum_{\boldsymbol{i} \in \mathbb{Z}_{2g+\ell+2}^\ell}\mathtt{p}_{\ell,g;\boldsymbol{i}}\, \boldsymbol{t}^{\boldsymbol{i}},
\ee
\noindent
with the explicit coefficients
\be
\label{eq:toprec16}
\mathtt{p}_{\ell,g;\boldsymbol{i}} = \sum_{\substack{ (k,\boldsymbol{v},\boldsymbol{n},\boldsymbol{r},\boldsymbol{m}) \in \BN_{0} \times\BN_{0}^\ell \times \BZ_{k+1}^\ell\times \BZ_{g+3}^\ell \times \BZ_{\ell}^\ell \\ v +k = g,\, \boldsymbol{n}+\boldsymbol{m}+\boldsymbol{r} = \boldsymbol{i}}} \left\{ \prod_{j=1}^\ell p_{\ell,v_j;r_j} \right\} \mathsf{p}_{\ell,k;\boldsymbol{n}}\, \Delta_{\ell;\boldsymbol{m}}, \qquad g \in \mathbb{N}_0.
\ee
\noindent
All Gaussian integrations may now be explicitly evaluated in \eqref{eq:step1}, in which case one obtains
\be
\frac{\CZ^{(\ell)}}{\CZ^{(0)}} = \frac{1}{\ell!}\, \frac{g_{\text{s}}^{\frac{\ell^2}{2}}}{\left( 2\pi\right)^{\ell}}\, \frac{\CZ^{(0)} (t-\ell g_{\text{s}})}{\CZ^{(0)} (t)}\, \exp \left( -\frac{\ell}{g_{\text{s}}}\, V_{\text{h;eff}}(x^\star) +\ell^2\, \Big\{ \partial_t V_{\text{h;eff}} (x^\star) + 2 A_{0;2} (x^\star,x^\star) \Big\} \right)\sum_{g=0}^{+\infty} a_{\ell,g}\, g_{\text{s}}^{g},
\ee
\noindent
where the $a_{\ell,g}$ coefficients are, explicitly,
\be
\label{eq:toprec19}
a_{\ell,g} = \sum_{\boldsymbol{i}\in \mathbb{Z}^\ell_{2g+\ell}}' \mathtt{p}_{\ell,g;\boldsymbol{i}} \left\{ \prod_{j=1}^\ell \frac{\Gamma \left( \frac{1+2i_j}{2} \right)}{\left( \frac{1}{2}\, \frac{\partial^2 V_{\text{h;eff}}}{\partial x^2} (x^\star) \right)^{\frac{1+2i_j}{2}}} \right\}, \qquad g \in \mathbb{N}_0.
\ee
\noindent
Herein the prime in the summation indicates we should not include indices $\boldsymbol{i} \in \mathbb{Z}^\ell_{2g+\ell}$ featuring odd components, as such terms integrate to zero. Finally, and directly following \cite{msw07}, all we have left to do is to evaluate
\be
\frac{\CZ^{(0)} (t-\ell g_{\text{s}})}{\CZ^{(0)} (t)} = \exp \left( \sum_{n=0}^{+\infty} \CG_{\ell,n}\, g_{\text{s}}^{n-1} \right), \qquad \CG_{\ell,n} = \sum_{k=0}^{\left[\frac{n}{2}\right]} \frac{(-\ell)^{n-2k+1}}{\left( n-2k+1 \right)!}\, \partial_t^{n-2k+1} \CF_k (t).
\ee
\noindent
Keep following \cite{msw07} to rewrite the above as
\be
\frac{\CZ^{(0)} (t-\ell g_{\text{s}})}{\CZ^{(0)} (t)} = \exp\left( \frac{\ell}{g_{\text{s}}}\, V_{\text{h;eff}}(x_2) + \frac{\ell^2}{2}\, \partial_t^2 \CF_0 (t) \right)\, \sum_{g=0}^{+\infty} b_{\ell,g}\, g_{\text{s}}^{g},
\ee
\noindent
where $x_2$ is the end-point of the cubic-matrix-model single-cut $\NCC = (x_1,x_2)$ in \eqref{eq:cubic-spectral-curve-one-cut}, and where we used the equation (see \cite{msw07})
\be
\partial_t \CF_0 (t) = - V_{\text{h;eff}} (x_2).
\ee
\noindent
As to the $b_{\ell,g}$ coefficients, they are given by
\be
\label{eq:toprec19b}
b_{\ell,0} = 1, \qquad b_{\ell,g} = \sum_{\substack{\boldsymbol{n} \in \BN_{0}^p \\ n-p = g}}' \frac{1}{p!} \prod_{i=1}^p \CG_{\ell,n_i}, \quad g \in \mathbb{N},
\ee
\noindent
where the prime in the summation indicates we should not include indexes $\boldsymbol{n} \in \BN_{0}^p$ such that $n_i < 2$ for any $1 \le i \le g$. This result has hence led us to our final expression
\bea
\frac{\CZ^{(\ell)}}{\CZ^{(0)}} &=& \frac{1}{\ell!}\, \frac{g_{\text{s}}^{\frac{\ell^2}{2}}}{\left( 2\pi\right)^{\ell}}\, \exp \left( - \frac{\ell}{g_{\text{s}}} \Big( V_{\text{h;eff}} (x^\star) - V_{\text{h;eff}} (x_2) \Big) + \ell^2\, \Big\{ \partial_t V_{\text{h;eff}} (x^\star) + 2 A_{0;2} (x^\star,x^\star) + \right. \nonumber \\
&&
\left.
+ \frac{1}{2}\, \partial_t^2 \CF_0 (t) \Big\} \right) \times \sum_{g=0}^{+\infty} \mathcal{Z}^{\text{saddle}}_{\ell,g}\, g_{\text{s}}^{g},
\label{eq:toprec5}
\eea
\noindent
where the coefficients $\mathcal{Z}_{\ell,g}^{\text{saddle}}$ which appear in \eqref{eq:toprec22} in the main text have now been explicitly computed and are given by
\be
\label{eq:toprec15}
\mathcal{Z}^{\text{saddle}}_{\ell,g} = \sum_{\substack{j,h \in \BZ_{g+1} \\ j+h = g}} a_{\ell,h}\, b_{\ell,j}, \qquad g \in \mathbb{N}_0,
\ee
\noindent
with the $a_{\ell,g}$ from \eqref{eq:toprec19} and the $b_{\ell,g}$ from \eqref{eq:toprec19b}. In the main text, subsection~\ref{subsec:secondformulaF}, we will explicitly evaluate some such coefficients. Overall, this also illustrates how the formulae obtained in this appendix may be used to derive exact analytical expressions for the saddle-point expansion coefficients of the matrix integral.

\newpage

\bibliographystyle{plain}

\begin{thebibliography}{10}

\bibitem{s16}
K.~Schwarzschild,
\textit{\"Uber das Gravitationsfeld eines Massenpunktes nach der Einsteinschen Theorie}, Sitzber.\ Deut.\ Akad.\ Wiss.\ \textbf{7} (1916) 189.

\bibitem{s26}
E.~Schr\"odinger, 
\textit{Quantisierung als Eigenwertproblem},
Annalen\ Phys.\ \textbf{384} (1926) 361.

\bibitem{m97}
J.M.~Maldacena,
\textit{The Large N Limit of Superconformal Field Theories and Supergravity},
Int.\ J.\ Theor.\ Phys.\ \textbf{38} (1999) 1113,
Adv.\ Theor.\ Math.\ Phys.\ \textbf{2} (1998) 231,
\texttt{arXiv:\arxivlink{hep-th/9711200}}.

\bibitem{w55}
E.P.~Wigner,
\textit{Characteristic Vectors of Bordered Matrices with Infinite Dimensions},
Ann.\ Math.\ \textbf{62} (1955) 548,
\texttt{DOI:\doilink{10.2307/1970079}}.

\bibitem{bgs84}
O.~Bohigas, M.J.~Giannoni, C.~Schmit,
\textit{Characterization of Chaotic Quantum Spectra and Universality of Level Fluctuation Laws},
Phys.\ Rev.\ Lett.\ \textbf{52} (1984) 1,
\texttt{DOI:\doilink{10.1103/PhysRevLett.52.1}}.

\bibitem{th74}
G.~'t~Hooft,
\textit{A Planar Diagram Theory for Strong Interactions},
Nucl.\ Phys.\ \textbf{B72} (1974) 461,
\texttt{DOI:\doilink{10.1016/0550-3213(74)90154-0}}.

\bibitem{gm90a}
D.J.~Gross, A.A.~Migdal,
\textit{Nonperturbative Two-Dimensional Quantum Gravity},
Phys.\ Rev.\ Lett.\ \textbf{64} (1990) 127,
\texttt{DOI:\doilink{10.1103/PhysRevLett.64.127}}.

\bibitem{ds90}
M.R.~Douglas, S.H.~Shenker,
\textit{Strings in Less Than One-Dimension},
Nucl.\ Phys.\ \textbf{B335} (1990) 635,
\texttt{DOI:\doilink{10.1016/0550-3213(90)90522-F}}.

\bibitem{bk90}
E.~Br\'ezin, V.A.~Kazakov,
\textit{Exactly Solvable Field Theories of Closed Strings},
Phys.\ Lett.\ \textbf{B236} (1990) 144,
\texttt{DOI:\doilink{10.1016/0370-2693(90)90818-Q}}.

\bibitem{d90}
M.R.~Douglas,
\textit{Strings in Less Than One Dimension and the Generalized KdV Hierarchies},
Phys.\ Lett.\ \textbf{B238} (1990) 176,
\texttt{DOI:\doilink{10.1016/0370-2693(90)91716-O}}.

\bibitem{gm90b}
D.J.~Gross, A.A.~Migdal,
\textit{A Nonperturbative Treatment of Two-Dimensional Quantum Gravity},
Nucl.\ Phys.\ \textbf{B340} (1990) 333,
\texttt{DOI:\doilink{10.1016/0550-3213(90)90450-R}}.

\bibitem{dv02a}
R.~Dijkgraaf, C.~Vafa,
\textit{Matrix Models, Topological Strings, and Supersymmetric Gauge Theories},
Nucl.\ Phys.\ \textbf{B644} (2002) 3,
\texttt{arXiv:\arxivlink{hep-th/0206255}}.

\bibitem{dv02b}
R.~Dijkgraaf, C.~Vafa,
\textit{On Geometry and Matrix Models},
Nucl.\ Phys.\ \textbf{B644} (2002) 21,
\texttt{arXiv:\arxivlink{hep-th/0207106}}.

\bibitem{emo07}
B.~Eynard, M.~Mari\~no, N.~Orantin,
\textit{Holomorphic Anomaly and Matrix Models},
JHEP\ \textbf{0706} (2007) 058,
\texttt{arXiv:\arxivlink{hep-th/0702110}}.

\bibitem{ss03}
N.~Seiberg, D.~Shih,
\textit{Branes, Rings and Matrix Models in Minimal (Super)String Theory},
JHEP\ \textbf{02} (2004) 021,
\texttt{arXiv:\arxivlink{hep-th/0312170}}.

\bibitem{kopss04}
D.~Kutasov, K.~Okuyama, J.-W.~Park, N.~Seiberg, D.~Shih,
\textit{Annulus Amplitudes and ZZ Branes in Minimal String Theory},
JHEP\ \textbf{08} (2004) 026,
\texttt{arXiv:\arxivlink{hep-th/0406030}}.

\bibitem{mmss04}
J.M.~Maldacena, G.W.~Moore, N.~Seiberg, D.~Shih,
\textit{Exact vs. Semiclassical Target Space of the Minimal String},
JHEP\ \textbf{10} (2004) 020,
\texttt{arXiv:\arxivlink{hep-th/0408039}}.

\bibitem{ss04b}
N.~Seiberg, D.~Shih,
\textit{Flux Vacua and Branes of the Minimal Superstring},
JHEP\ \textbf{0501} (2005) 055,
\texttt{arXiv:\arxivlink{hep-th/0412315}}.

\bibitem{mss15}
J.~Maldacena, S.H.~Shenker, D.~Stanford,
\textit{A Bound on Chaos},
JHEP\ \textbf{1608} (2016) 106,
\texttt{arXiv:\arxivlink{1503.01409}[hep-th]}.

\bibitem{t83}
C.~Teitelboim,
\textit{Gravitation and Hamiltonian Structure in Two Space-Time Dimensions},
Phys.\ Lett.\ \textbf{126B} (1983) 41,
\texttt{DOI:\doilink{10.1016/0370-2693(83)90012-6}}.

\bibitem{j85}
R.~Jackiw,
\textit{Lower Dimensional Gravity},
Nucl.\ Phys.\ \textbf{B252} (1985) 343,
\texttt{DOI:\doilink{10.1016/0550-3213(85)90448-1}}.

\bibitem{ap14}
A.~Almheiri, J.~Polchinski,
\textit{Models of $\text{AdS}_{2}$ Backreaction and Holography},
JHEP\ \textbf{1511} (2015) 014,
\texttt{arXiv:\arxivlink{1402.6334}[hep-th]}.

\bibitem{sss19}
P.~Saad, S.H.~Shenker, D.~Stanford,
\textit{JT Gravity as a Matrix Integral},
\texttt{arXiv:\arxivlink{1903.11115}[hep-th]}.

\bibitem{sw19}
D.~Stanford, E.~Witten,
\textit{JT Gravity and the Ensembles of Random Matrix Theory},
Adv.\ Theor.\ Math.\ Phys.\ \textbf{24} (2020) 1475,
\texttt{arXiv:\arxivlink{1907.03363}[hep-th]}.

\bibitem{tw23}
G.J.~Turiaci, E.~Witten,
\textit{$\CN=2$ JT Supergravity and Matrix Models},
JHEP\ \textbf{12} (2023) 003,
\texttt{arXiv:\arxivlink{2305.19438}[hep-th]}.

\bibitem{d91}
F.~David,
\textit{Phases of the Large $N$ Matrix Model and Nonperturbative Effects in 2D Gravity},
Nucl.\ Phys.\ \textbf{B348} (1991) 507,
\texttt{DOI:\doilink{10.1016/0550-3213(91)90202-9}}.

\bibitem{d92}
F.~David,
\textit{Nonperturbative Effects in Matrix Models and Vacua of Two-Dimensional Gravity},
Phys.\ Lett.\ \textbf{B302} (1993) 403,
\texttt{arXiv:\arxivlink{hep-th/9212106}}.

\bibitem{gp88}
D.J.~Gross, V.~Periwal,
\textit{String Perturbation Theory Diverges}
Phys.\ Rev.\ Lett.\ \textbf{60} (1988) 2105,
\texttt{DOI:\doilink{10.1103/PhysRevLett.60.2105}}.

\bibitem{s90}
S.H.~Shenker,
\textit{The Strength of Nonperturbative Effects in String Theory},
in ``The Large N Expansion in Quantum Field Theory and Statistical Physics'' (1990) 809,
\texttt{DOI:\doilink{10.1142/9789814365802$\_$0057}}.

\bibitem{p94}
J.~Polchinski,
\textit{Combinatorics of Boundaries in String Theory},
Phys.\ Rev.\ \textbf{D50} (1994) R6041,
\texttt{arXiv:\arxivlink{hep-th/9407031}}.

\bibitem{p95}
J.~Polchinski,
\textit{Dirichlet Branes and Ramond-Ramond charges},
Phys.\ Rev.\ Lett.\ \textbf{75} (1995) 4724,
\texttt{arXiv:\arxivlink{hep-th/9510017}}.

\bibitem{zz01}
A.B.~Zamolodchikov, Al.B.~Zamolodchikov,
\textit{Liouville Field Theory on a Pseudosphere},
in ``6th Workshop on Supersymmetries and Quantum Symmetries'' (2001) 280,
\texttt{arXiv:\arxivlink{hep-th/0101152}}.

\bibitem{fzz00}
V.~Fateev, A.B.~Zamolodchikov, Al.B.~Zamolodchikov,
\textit{Boundary Liouville Field Theory I: Boundary State and Boundary Two Point Function},
\texttt{arXiv:\arxivlink{hep-th/0001012}}.

\bibitem{t00}
J.~Teschner,
\textit{Remarks on Liouville Theory with Boundary},
in ``4th Annual European TMR Conference on Integrability Nonperturbative Effects and Symmetry in Quantum Field Theory'' (2000) 041,
\texttt{arXiv:\arxivlink{hep-th/0009138}}.

\bibitem{gz90b}
P.H.~Ginsparg, J.~Zinn-Justin,
\textit{Action Principle and Large Order Behavior of Non-Perturbative Gravity},
in ``Random Surfaces and Quantum Gravity'' NATO\ ASI\ Series\ B:\ Phys.\ \textbf{262} (1990) 85,
\texttt{DOI:\doilink{10.1007/978-1-4615-3772-4$\_$7}}.

\bibitem{gz91}
P.H.~Ginsparg, J.~Zinn-Justin,
\textit{Large Order Behaviour of Nonperturbative Gravity},
Phys.\ Lett.\ \textbf{B255} (1991) 189,
\texttt{DOI:\doilink{10.1016/0370-2693(91)90234-H}}.

\bibitem{ez93}
B.~Eynard, J.~Zinn-Justin,
\textit{Large Order Behavior of 2D Gravity Coupled to $d<1$ Matter},
Phys.\ Lett.\ \textbf{B302} (1993) 396,
\texttt{arXiv:\arxivlink{hep-th/9301004}}.

\bibitem{e81}
J.~\'Ecalle,
\textit{Les Fonctions R\'esurgentes},
Publications Math\'ematiques d'Orsay\ \textbf{81-05} (1981), \textbf{81-06} (1981), \textbf{85-05} (1985).

\bibitem{e84}
J.~\'Ecalle,
\textit{Cinq Applications des Fonctions R\'esurgentes: Singularit\'es Irr\'egulieres et R\'esurgence Multiple},
Publications Math\'ematiques d'Orsay\ \textbf{84-62} (1984).

\bibitem{e93}
J.~\'Ecalle,
\textit{Six Lectures on Transseries, Analysable Functions and the Constructive Proof of Dulac's Conjecture},
in ``Bifurcations and Periodic Orbits of Vector Fields'' NATO ASI Series \textbf{408} (1993) 75,
\texttt{DOI:\doilink{10.1007/978-94-015-8238-4$\_$3}}.

\bibitem{abs18}
I.~Aniceto, G.~Ba\c sar, R.~Schiappa,
\textit{A Primer on Resurgent Transseries and Their Asymptotics},
Phys.\ Rept.\ \textbf{809} (2019) 1,
\texttt{arXiv:\arxivlink{1802.10441}[hep-th]}.

\bibitem{m06}
M.~Mari\~no,
\textit{Open String Amplitudes and Large-Order Behavior in Topological String Theory},
JHEP\ \textbf{0803} (2008) 060,
\texttt{arXiv:\arxivlink{hep-th/0612127}}.

\bibitem{msw07}
M.~Mari\~no, R.~Schiappa, M.~Weiss,
\textit{Nonperturbative Effects and the Large-Order Behavior of Matrix Models and Topological Strings},
Commun.\ Number\ Theor.\ Phys.\ \textbf{2} (2008) 349,
\texttt{arXiv:\arxivlink{0711.1954}[hep-th]}.

\bibitem{msw08}
M.~Mari\~no, R.~Schiappa, M.~Weiss,
\textit{Multi-Instantons and Multi-Cuts},
J.\ Math.\ Phys.\ \textbf{50} (2009) 052301,
\texttt{arXiv:\arxivlink{0809.2619}[hep-th]}.

\bibitem{ps09}
S.~Pasquetti, R.~Schiappa,
\textit{Borel and Stokes Nonperturbative Phenomena in Topological String Theory and $c=1$ Matrix Models},
Ann.\ Henri\ Poincar\'e \textbf{11} (2010) 351,
\texttt{arXiv:\arxivlink{0907.4082}[hep-th]}.

\bibitem{m08}
M.~Mari\~no,
\textit{Nonperturbative Effects and Nonperturbative Definitions in Matrix Models and Topological Strings},
JHEP\ \textbf{0812} (2008) 114,
\texttt{arXiv:\arxivlink{0805.3033}[hep-th]}.

\bibitem{gikm10}
S.~Garoufalidis, A.~Its, A.~Kapaev, M.~Mari\~no,
\textit{Asymptotics of the Instantons of Painlev\'e~I},
Int.\ Math.\ Res.\ Notices\ \textbf{2012} (2012) 561,
\texttt{arXiv:\arxivlink{1002.3634}[math.CA]}.

\bibitem{kmr10}
A.~Klemm, M.~Mari\~no, M.~Rauch,
\textit{Direct Integration and Non-Perturbative Effects in Matrix Models},
JHEP\ \textbf{1010} (2010) 004,
\texttt{arXiv:\arxivlink{1002.3846}[hep-th]}.

\bibitem{dmp11}
N.~Drukker, M.~Mari\~no, P.~Putrov,
\textit{Nonperturbative Aspects of ABJM Theory},
JHEP\ \textbf{1111} (2011) 141,
\texttt{arXiv:\arxivlink{1103.4844}[hep-th]}.

\bibitem{cesv13}
R.~Couso-Santamar\'\i a, J.D.~Edelstein, R.~Schiappa, M.~Vonk,
\textit{Resurgent Transseries and the Holomorphic Anomaly},
Ann.\ Henri\ Poincar\'e\ \textbf{17} (2016) 331,
\texttt{arXiv:\arxivlink{1308.1695}[hep-th]}.

\bibitem{gmz14}
A.~Grassi, M.~Mari\~no, S.~Zakany,
\textit{Resumming the String Perturbation Series},
JHEP\ \textbf{1505} (2015) 038,
\texttt{arXiv:\arxivlink{1405.4214}[hep-th]}.

\bibitem{cesv14}
R.~Couso-Santamar\'\i a, J.D.~Edelstein, R.~Schiappa, M.~Vonk,
\textit{Resurgent Transseries and the Holomorphic Anomaly: Nonperturbative Closed Strings in Local $\BC\BP^2$},
Commun.\ Math.\ Phys.\ \textbf{338} (2015) 285,
\texttt{arXiv:\arxivlink{1407.4821}[hep-th]}.

\bibitem{c15}
R.~Couso-Santamar\'\i a,
\textit{Universality of the Topological String at Large Radius and NS-Brane Resurgence},
Lett.\ Math.\ Phys.\ \textbf{107} (2017) 343,
\texttt{arXiv:\arxivlink{1507.04013}[hep-th]}.

\bibitem{csv16}
R.~Couso-Santamar\'\i a, R.~Schiappa, R.~Vaz,
\textit{On Asymptotics and Resurgent Structures of Enumerative Gromov--Witten Invariants},
Commun.\ Number\ Theor.\ Phys.\ \textbf{11} (2017) 707,
\texttt{arXiv:\arxivlink{1605.07473}[math.AG]}.

\bibitem{cms16}
R.~Couso-Santamar\'\i a, M.~Mari\~no, R.~Schiappa,
\textit{Resurgence Matches Quantization},
J.\ Phys.\ \textbf{A50} (2017) 145402,
\texttt{arXiv:\arxivlink{1610.06782}[hep-th]}.

\bibitem{cms17}
S.~Codesido, M.~Mari\~no, R.~Schiappa,
\textit{Nonperturbative Quantum Mechanics from Nonperturbative Strings},
Ann.\ Henri\ Poincar\'e\ \textbf{20} (2019) 543,
\texttt{arXiv:\arxivlink{1712.02603}[hep-th]}.

\bibitem{gm21}
J.~Gu, M.~Mari\~no,
\textit{Peacock Patterns and New Integer Invariants in Topological String Theory},
SciPost\ Phys.\ \textbf{12} (2022) 058,
\texttt{arXiv:\arxivlink{2104.07437}[hep-th]}.

\bibitem{gm22a}
J.~Gu, M.~Mari\~no,
\textit{Exact Multi-Instantons in Topological String Theory},
SciPost\ Phys.\ \textbf{15} (2023) 179,
\texttt{arXiv:\arxivlink{2211.01403}[hep-th]}.

\bibitem{gm22b}
J.~Gu, M.~Mari\~no,
\textit{On the Resurgent Structure of Quantum Periods},
SciPost\ Phys.\ \textbf{15} (2023) 035,
\texttt{arXiv:\arxivlink{2211.03871}[hep-th]}.

\bibitem{gkkm23}
J.~Gu, A.-K.~Kashani-Poor, M.~Mari\~no, A.~Klemm,
\textit{Nonperturbative Topological String Theory on Compact Calabi--Yau 3-Folds},
SciPost\ Phys.\ \textbf{16} (2024) 079,
\texttt{arXiv:\arxivlink{2305.19916}[hep-th]}.

\bibitem{im23}
K.~Iwaki, M.~Mari\~no,
\textit{Resurgent Structure of the Topological String and the First Painlev\'e Equation},
SIGMA\ \textbf{20} (2024) 028,
\texttt{arXiv:\arxivlink{2307.02080}[hep-th]}.

\bibitem{ms24}
M.~Mari\~no, M.~Schwick,
\textit{Large $N$ Instantons, BPS States, and the Replica Limit},
\texttt{arXiv:\arxivlink{2403.14462}[hep-th]}.

\bibitem{asv11}
I.~Aniceto, R.~Schiappa, M.~Vonk,
\textit{The Resurgence of Instantons in String Theory},
Commun.\ Number\ Theor.\ Phys.\ \textbf{6} (2012) 339,
\texttt{arXiv:\arxivlink{1106.5922}[hep-th]}.

\bibitem{sv13}
R.~Schiappa, R.~Vaz,
\textit{The Resurgence of Instantons: Multi-Cut Stokes Phases and the Painlev\'e~II Equation},
Commun.\ Math.\ Phys.\ \textbf{330} (2014) 655,
\texttt{arXiv:\arxivlink{1302.5138}[hep-th]}.

\bibitem{as13}
I.~Aniceto, R.~Schiappa,
\textit{Nonperturbative Ambiguities and the Reality of Resurgent Transseries},
Commun.\ Math.\ Phys.\ \textbf{335} (2015) 183,
\texttt{arXiv:\arxivlink{1308.1115}[hep-th]}.

\bibitem{gs21}
P.~Gregori, R.~Schiappa,
\textit{From Minimal Strings towards Jackiw--Teitelboim Gravity: On their Resurgence, Resonance, and Black Holes},
Class.\ Quant.\ Grav.\ \textbf{41} (2024) 115001,
\texttt{arXiv:\arxivlink{2108.11409}[hep-th]}.

\bibitem{bssv22}
S.~Baldino, R.~Schiappa, M.~Schwick, R.~Vega,
\textit{Resurgent Stokes Data for Painlev\'e Equations and Two-Dimensional Quantum (Super) Gravity},
Commun.\ Number\ Theor.\ Phys.\ \textbf{17} (2023) 385,
\texttt{arXiv:\arxivlink{2203.13726}[hep-th]}.

\bibitem{mss22}
M.~Mari\~no, R.~Schiappa, M.~Schwick,
\textit{New Instantons for Matrix Models},
\texttt{arXiv:\arxivlink{2210.13479}[hep-th]}.

\bibitem{sst23}
R.~Schiappa, M.~Schwick, N.~Tamarin,
\textit{All the D-Branes of Resurgence},
\texttt{arXiv:\arxivlink{2301.05214}[hep-th]}.

\bibitem{eggls23}
B.~Eynard, E.~Garcia-Failde, P.~Gregori, D.~Lewa\'nski, R.~Schiappa,
\textit{Resurgent Asymptotics of Jackiw--Teitelboim Gravity and the Nonperturbative Topological Recursion},
Ann.\ Henri\ Poincar\'e\ \textbf{25} (2024) 4121,
\texttt{arXiv:\arxivlink{2305.16940}[hep-th]}.

\bibitem{ss26}
R.~Schiappa, M.~Schwick,
\textit{Exact Solutions to Matrix Models and String Theories: Background Independence of the Large $N$ Expansions},
(2026).

\bibitem{krsst26b}
J.~Kager, J.~Rodrigues, R.~Schiappa, M.~Schwick, N.~Tamarin,
\textit{Exact Solutions to Matrix Models and String Theories: The Global Construction},
(2026).

\bibitem{ad95}
P.C.~Argyres, M.R.~Douglas,
\textit{New Phenomena in $\text{SU}(3)$ Supersymmetric Gauge Theory},
Nucl.\ Phys.\ \textbf{B448} (1995) 93,
\texttt{arXiv:\arxivlink{hep-th/9505062}}.

\bibitem{apsw95}
P.C.~Argyres, M.R.~Plesser, N.~Seiberg, E.~Witten,
\textit{New $\CN=2$ Superconformal Field Theories in Four-Dimensions},
Nucl.\ Phys.\ \textbf{B461} (1996) 71,
\texttt{arXiv:\arxivlink{hep-th/9511154}}.

\bibitem{krst26a}
J.~Kager, J.~Rodrigues, R.~Schiappa, N.~Tamarin,
\textit{Resurgence on the KdV Hierarchy: Multicritical Models, Minimal Strings, and Argyres--Douglas Theories},
(2026).

\bibitem{krst26b}
J.~Kager, J.~Rodrigues, R.~Schiappa, N.~Tamarin,
\textit{From Resurgence to Monodromy Along the KdV Hierarchy},
(2026).

\bibitem{krs26}
J.~Kager, J.~Rodrigues, R.~Schiappa,
\textit{Exact Nonperturbative Amplitudes for FZZT D-Branes},
(2026).

\bibitem{ccfks26}
F.~Cominelli, L.~Cu\'ellar, D.~Fragoso, J.~Kager, R.~Schiappa,
\textit{Nonperturbative Correlation Functions of Matrix Models and String Theories},
(2026).

\bibitem{gd75}
I.M.~Gel'fand, L.A.~Dikii,
\textit{Asymptotic Behavior of the Resolvent of Sturm--Liouville Equations and the Algebra of the Kortweg--de~Vries Equations},
Russ.\ Math.\ Surv.\ \textbf{30} (1975) 77,
\texttt{DOI:\doilink{10.1070/RM1975v030n05ABEH001522}}.

\bibitem{mss91}
G.W.~Moore, N.~Seiberg, M.~Staudacher,
\textit{From Loops to States in 2D Quantum Gravity},
Nucl.\ Phys.\ \textbf{B362} (1991) 665,
\texttt{DOI:\doilink{10.1016/0550-3213(91)90548-C}}.

\bibitem{dw18}
R.~Dijkgraaf, E.~Witten,
\textit{Developments in Topological Gravity},
Int.\ J.\ Mod.\ Phys.\ \textbf{A33} (2018) 1830029,
\texttt{arXiv:\arxivlink{1804.03275}[hep-th]}.

\bibitem{hmo25}
Y.~Hatsuda, T.~Matsumoto, K.~Okuyama,
\textit{Non-Perturbative Effects in JT Gravity from KdV Equations},
\texttt{arXiv:\arxivlink{2505.16433}[hep-th]}.

\bibitem{kms03b}
I.R.~Klebanov, J.M.~Maldacena, N.~Seiberg,
\textit{Unitary and Complex Matrix Models as 1d Type-0 Strings},
Commun.\ Math.\ Phys.\ \textbf{252} (2004) 275,
\texttt{arXiv:\arxivlink{hep-th/0309168}}.

\bibitem{v23}
R.~Vega,
\textit{Parametric Resurgences of the Second Painlev\'e Equation and Minimal Superstrings},
\texttt{arXiv:\arxivlink{2311.03103}[hep-th]}.

\bibitem{emms22a}
D.S.~Eniceicu, R.~Mahajan, C.~Murdia, A.~Sen,
\textit{Normalization of ZZ Instanton Amplitudes in Minimal String Theory},
JHEP\ \textbf{07} (2022) 139,
\texttt{arXiv:\arxivlink{2202.03448}[hep-th]}.

\bibitem{emms22b}
D.S.~Eniceicu, R.~Mahajan, C.~Murdia, A.~Sen,
\textit{Multi-Instantons in Minimal String Theory and in Matrix Integrals},
JHEP\ \textbf{10} (2022) 065,
\texttt{arXiv:\arxivlink{2206.13531}[hep-th]}.

\bibitem{e23}
D.S.~Eniceicu,
\textit{Comments on the Giant-Graviton Expansion of the Superconformal Index},
\texttt{arXiv:\arxivlink{2302.04887}[hep-th]}.

\bibitem{emm23}
D.S.~Eniceicu, R.~Mahajan, C.~Murdia,
\textit{Complex Eigenvalue Instantons and the Fredholm Determinant Expansion in the Gross--Witten--Wadia Model},
JHEP\ \textbf{01} (2024) 129,
\texttt{arXiv:\arxivlink{2308.06320}[hep-th]}.

\bibitem{cemm24}
V.~Chakrabhavi, D.S.~Eniceicu, R.~Mahajan, C.~Murdia,
\textit{Normalization of ZZ Instanton Amplitudes in Type 0B Minimal Superstring Theory},
JHEP\ \textbf{09} (2024) 114,
\texttt{arXiv:\arxivlink{2406.16867}[hep-th]}.

\bibitem{cmt24}
Y.~Chen, R.~Mahajan, H.~Tang,
\textit{Giant Graviton Expansion from Eigenvalue Instantons},
\texttt{arXiv:\arxivlink{2407.08155}[hep-th]}.

\bibitem{emt24}
D.S.~Eniceicu, C.~Murdia, A.~Torchylo,
\textit{The Complete Non-Perturbative Partition Function of Minimal Superstring Theory and JT Supergravity},
JHEP\ \textbf{06} (2025) 178,
\texttt{arXiv:\arxivlink{2412.08698}[hep-th]}.

\bibitem{dt00}
E.~Delabaere, D.T.~Trinh,
\textit{Spectral Analysis of the Complex Cubic Oscillator},
J.\ Phys.\ Math.\ Gen.\ \textbf{A33} (2000) 8771,
\texttt{DOI:\doilink{10.1088/0305-4470/33/48/314}}.

\bibitem{m09}
D.~Masoero,
\textit{Poles of Int\'egrale Tritronqu\'ee and Anharmonic Oscillators: A WKB Approach},
J.\ Phys.\ Math.\ Theor.\ \textbf{A43} (2010) 095201,
\texttt{arXiv:\arxivlink{0909.5537}[math.CA]}.

\bibitem{m10}
D.~Masoero,
\textit{Poles of Int\'egrale Tritronqu\'ee and Anharmonic Oscillators: Asymptotic Localization from WKB Analysis},
Nonlinearity\ \textbf{23} (2010) 2501,
\texttt{arXiv:\arxivlink{1002.1042}[math.CA]}.

\bibitem{b13}
P.~Boutroux,
\textit{Recherches sur les Transcendantes de M.~Painlev\'e et l'\'Etude Asymptotique des \'Equations Diff\'erentielles du Second Ordre},
Ann.\ Sci.\ \'Ecole\ Norm.\ Sup.\ \textbf{30} (1913) 255,
\texttt{Numdam:\href{http://www.numdam.org/item/?id=ASENS_1913_3_30__255_0}{\texttt{ASENS$\_$1913$\_$3$\_$30$\_\_$255$\_$0}}}.

\bibitem{b14}
P.~Boutroux,
\textit{Recherches sur les Transcendantes de M.~Painlev\'e et l'\'Etude Asymptotique des \'Equations Diff\'erentielles du Second Ordre (Suite)},
Ann.\ Sci.\ \'Ecole\ Norm.\ Sup.\ \textbf{31} (1914) 99,
\texttt{Numdam:\href{http://www.numdam.org/item/?id=ASENS_1914_3_31__99_0}{\texttt{ASENS$\_$1914$\_$3$\_$31$\_\_$99$\_$0}}}.

\bibitem{bde00}
G.~Bonnet, F.~David, B.~Eynard,
\textit{Breakdown of Universality in Multicut Matrix Models},
J.\ Phys.\ \textbf{A33} (2000) 6739,
\texttt{arXiv:\arxivlink{cond-mat/0003324}}.

\bibitem{e08}
B.~Eynard,
\textit{Large $N$ Expansion of Convergent Matrix Integrals, Holomorphic Anomalies, and Background Independence},
JHEP\ \textbf{03} (2009) 003,
\texttt{arXiv:\arxivlink{0802.1788}[hep-th]}.

\bibitem{em08}
B.~Eynard, M.~Mari\~no,
\textit{A Holomorphic and Background Independent Partition Function for Matrix Models and Topological Strings},
J.\ Geom.\ Phys.\ \textbf{61} (2011) 1181,
\texttt{arXiv:\arxivlink{0810.4273}[hep-th]}.

\bibitem{eo07a}
B.~Eynard, N.~Orantin,
\textit{Invariants of Algebraic Curves and Topological Expansion},
Commun.\ Num.\ Theor.\ Phys.\ \textbf{1} (2007) 347,
\texttt{arXiv:\arxivlink{math-ph/0702045}}.

\bibitem{fw11}
B.~Fornberg, J.A.C.~Weideman,
\textit{A Numerical Methodology for the Painlev\'e Equations},
J.\ Comput.\ Phys.\ \textbf{230} (2011) 5957,
\texttt{DOI:\doilink{10.1016/j.jcp.2011.04.007}}.

\bibitem{fw14}
B.~Fornberg, J.A.C.~Weideman,
\textit{A Computational Exploration of the Second Painlev\'e Equation},
Found.\ Comput.\ Math.\ \textbf{14} (2014) 985,
\texttt{DOI:\doilink{10.1007/s10208-013-9156-x}}.

\bibitem{sv22}
A.~van~Spaendonck, M.~Vonk,
\textit{Painlev\'e~I and Exact WKB: Stokes Phenomenon for Two-Parameter Transseries},
J.\ Phys.\ \textbf{A55} (2022) 454003,
\texttt{arXiv:\arxivlink{2204.09062}[hep-th]}.

\bibitem{mpp09}
M.~Mari\~no, S.~Pasquetti, P.~Putrov,
\textit{Large $N$ Duality Beyond the Genus Expansion},
JHEP\ \textbf{07} (2010) 074,
\texttt{arXiv:\arxivlink{0911.4692}[hep-th]}.

\bibitem{dgz93}
P.~Di~Francesco, P.H.~Ginsparg, J.~Zinn-Justin,
\textit{2D Gravity and Random Matrices},
Phys.\ Rept.\ \textbf{254} (1995) 1,
\texttt{arXiv:\arxivlink{hep-th/9306153}}.

\bibitem{m04}
M.~Mari\~no,
\textit{Les Houches Lectures on Matrix Models and Topological Strings},
\texttt{arXiv:\arxivlink{hep-th/0410165}}.

\bibitem{ekr15}
B.~Eynard, T.~Kimura, S.~Ribault,
\textit{Random Matrices},
\texttt{arXiv:\arxivlink{1510.04430}[math-ph]}.

\bibitem{biz80}
D.~Bessis, C.~Itzykson and J.B.~Zuber,
\textit{Quantum Field Theory Techniques in Graphical Enumeration},
Adv.\ Appl.\ Math.\ \textbf{1} (1980) 109,
\texttt{DOI:\doilink{10.1016/0196-8858(80)90008-1}}.

\bibitem{iz92}
C.~Itzykson, J.B.~Zuber,
\textit{Combinatorics of the Modular Group II: The Kontsevich Integrals},
Int.\ J.\ Mod.\ Phys.\ \textbf{A7} (1992) 5661,
\texttt{arXiv:\arxivlink{hep-th/9201001}}.

\bibitem{bipz78}
E.~Br\'ezin, C.~Itzykson, G.~Parisi, J.B.~Zuber,
\textit{Planar Diagrams},
Commun.\ Math.\ Phys.\ \textbf{59} (1978) 35,
\texttt{ProjectEuclid:{\href{https://projecteuclid.org/euclid.cmp/1103901558}{euclid.cmp/1103901558}}}.

\bibitem{ackm93}
J.~Ambj\o rn, L.~Chekhov, C.F.~Kristjansen, Yu.~Makeenko,
\textit{Matrix Model Calculations Beyond the Spherical Limit},
Nucl.\ Phys.\ \textbf{B404} (1993) 127 [\textit{Erratum:} Nucl.\ Phys.\ \textbf{B449} (1995) 681],
\texttt{arXiv:\arxivlink{hep-th/9302014}}.

\bibitem{csv15}
R.~Couso-Santamar\'\i a, R.~Schiappa, R.~Vaz,
\textit{Finite N from Resurgent Large $N$},
Annals\ Phys.\ \textbf{356} (2015) 1,
\texttt{arXiv:\arxivlink{1501.01007}[hep-th]}.

\bibitem{olbc10}
F.W.J.~Olver, D.W.~Lozier, R.F.~Boisvert, C.W.~Clark,
\textit{NIST Handbook of Mathematical Functions},
Cambridge University Press (2010),
\href{http://dlmf.nist.gov}{\texttt{http://dlmf.nist.gov}}.

\bibitem{ddjt90}
K.~Demeterfi, N.~Deo, S.~Jain, C.I.~Tan,
\textit{Multiband Structure and Critical Behavior of Matrix Models},
Phys.\ Rev.\ \textbf{D42} (1990) 4105,
\texttt{DOI:\doilink{10.1103/PhysRevD.42.4105}}.

\bibitem{l92}
O.~Lechtenfeld,
\textit{On Eigenvalue Tunneling in Matrix Models},
Int.\ J.\ Mod.\ Phys.\ \textbf{A7} (1992) 2335,
\texttt{DOI:\doilink{10.1142/S0217751X92001046}}.

\bibitem{j91}
J.~Jurkiewicz,
\textit{Chaotic Behavior in One Matrix Models},
Phys.\ Lett.\ \textbf{B261} (1991) 260,
\texttt{DOI:\doilink{10.1016/0370-2693(91)90325-K}}.

\bibitem{s92}
D.~S\'en\'echal,
\textit{Chaos in the Hermitian One Matrix Model},
Int.\ J.\ Mod.\ Phys.\ \textbf{A7} (1992) 1491,
\texttt{DOI:\doilink{10.1142/S0217751X9200065X}}.

\bibitem{bdjt93}
R.C.~Brower, N.~Deo, S.~Jain, C.I.~Tan,
\textit{Symmetry Breaking in the Double--Well hermitian 1-matrix Models},
Nucl.\ Phys.\ \textbf{B405} (1993) 166,
\texttt{arXiv:\arxivlink{hep-th/9212127}}.

\bibitem{m03}
E.J.~Martinec,
\textit{The Annular Report on Noncritical String Theory},
\texttt{arXiv:\arxivlink{hep-th/0305148}}.

\bibitem{dw90}
R.~Dijkgraaf, E.~Witten,
\textit{Mean Field Theory, Topological Field Theory, and Multimatrix Models},
Nucl.\ Phys.\ \textbf{B342} (1990) 486,
\texttt{DOI:\doilink{10.1016/0550-3213(90)90324-7}}.

\bibitem{w91}
E.~Witten,
\textit{Two-Dimensional Gravity and Intersection Theory on Moduli Space},
Surveys\ Diff.\ Geom.\ \textbf{1} (1991) 243,
\texttt{DOI:\doilink{10.4310/SDG.1990.v1.n1.a5}}.

\bibitem{k92}
M.~Kontsevich,
\textit{Intersection Theory on the Moduli Space of Curves and the Matrix Airy Function},
Commun.\ Math.\ Phys.\ \textbf{147} (1992) 1,
\texttt{DOI:\doilink{10.1007/BF02099526}}.

\bibitem{bpz84}
A.A.~Belavin, A.M.~Polyakov, A.B.~Zamolodchikov,
\textit{Infinite Conformal Symmetry in Two-Dimensional Quantum Field Theory},
Nucl.\ Phys.\ \textbf{B241} (1984) 333,
\texttt{DOI:\doilink{10.1016/0550-3213(84)90052-X}}.

\bibitem{cgmps06}
N.~Caporaso, L.~Griguolo, M.~Mari\~no, S.~Pasquetti, D.~Seminara,
\textit{Phase Transitions, Double-Scaling Limit, and Topological Strings},
Phys.\ Rev.\ \textbf{D75} (2007) 046004,
\texttt{arXiv:\arxivlink{hep-th/0606120}}.

\bibitem{cogp91}
P.~Candelas, X.C.~De~La~Ossa, P.S.~Green, L.~Parkes,
\textit{A Pair of Calabi--Yau Manifolds as an Exactly Soluble Superconformal Theory},
Nucl.\ Phys.\ \textbf{B359} (1991) 21,
\texttt{DOI:\doilink{10.1016/0550-3213(91)90292-6}}.

\bibitem{bcov93b}
M.~Bershadsky, S.~Cecotti, H.~Ooguri, C.~Vafa,
\textit{Kodaira--Spencer Theory of Gravity and Exact Results for Quantum String Amplitudes},
Commun.\ Math.\ Phys.\ \textbf{165} (1994) 311,
\texttt{arXiv:\arxivlink{hep-th/9309140}}.

\bibitem{gv95}
D.~Ghoshal, C.~Vafa,
\textit{$c=1$ String as the Topological Theory of the Conifold},
Nucl.\ Phys.\ \textbf{B453} (1995) 121,
\texttt{arXiv:\arxivlink{hep-th/9506122}}.

\bibitem{w93}
E.~Witten,
\textit{Phases of $\CN=2$ Theories in Two-Dimensions},
Nucl.\ Phys.\ \textbf{B403} (1993) 159,
\texttt{arXiv:\arxivlink{hep-th/9301042}}.

\bibitem{agm93a}
P.S.~Aspinwall, B.R.~Greene, D.R.~Morrison,
\textit{Calabi--Yau Moduli Space, Mirror Manifolds and Space-Time Topology Change in String Theory},
Nucl.\ Phys.\ \textbf{B416} (1994) 414,
\texttt{arXiv:\arxivlink{hep-th/9309097}}.

\bibitem{agm93b}
P.S.~Aspinwall, B.R.~Greene, D.R.~Morrison,
\textit{Measuring Small Distances in $\CN=2$ Sigma Models},
Nucl.\ Phys.\ \textbf{B420} (1994) 184,
\texttt{arXiv:\arxivlink{hep-th/9311042}}.

\bibitem{t07}
K.~Takasaki,
\textit{Hamiltonian Structure of \PI~Hierarchy},
SIGMA\ \textbf{3} (2007) 042,
\texttt{arXiv:\arxivlink{nlin/0610073}}.

\bibitem{p02}
P.~Painlev\'e,
\textit{Sur les \'Equations Diff\'erentielles du Second Ordre et d'Ordre Sup\'erieur dont l'Int\'egrale G\'en\'erale est Uniforme},
Acta\ Math.\ \textbf{25} (1902) 1,
\texttt{DOI:\doilink{10.1007/BF02419020}}.

\bibitem{p06}
P.~Painlev\'e,
\textit{Sur les \'Equations Diff\'erentielles du Second Ordre \`a Points Critiques Fix\'es},
Comptes\ Rendus\ Acad.\ Sci.\ Paris\ \textbf{143} (1906) 1111.

\bibitem{d16}
E.~Delabaere,
\textit{Divergent Series, Summability and Resurgence III: Resurgent Methods and the First Painlev\'e Equation},
Lec.\ Notes\ Math.\ \textbf{2155} (2016),
\texttt{DOI:\doilink{10.1007/978-3-319-29000-3}}.

\bibitem{bmp90}
E.~Br\'ezin, E.~Marinari, G.~Parisi,
\textit{A Nonperturbative Ambiguity Free Solution of a String Model},
Phys.\ Lett.\ \textbf{B242} (1990) 35,
\texttt{DOI:\doilink{10.1016/0370-2693(90)91590-8}}.

\bibitem{c17}
S.~Codesido,
\textit{On the Resummation of the Lee--Yang Edge Singularity Coupled to Gravity},
\texttt{arXiv:\arxivlink{1712.02752}[hep-th]}.

\bibitem{s04}
S.~Shimomura,
\textit{A Certain Expression of the First Painlev\'e Hierarchy},
Proc.\ Japan\ Acad.\ \textbf{80A} (2004) 105.
\texttt{DOI:\doilink{10.3792/pjaa.80.105}}.

\bibitem{s64}
G.G.~Stokes,
\textit{On the Discontinuity of Arbitrary Constants which Appear in Divergent Developments},
Trans.\ Camb.\ Phil.\ Soc.\ \textbf{10} (1864) 106.

\bibitem{k94}
A.V.~Kitaev,
\textit{Elliptic Asymptotics of the First and the Second Painlev\'e Transcendents},
Russ.\ Math.\ Surv.\ \textbf{49} (1994) 81,
\texttt{DOI:\doilink{10.1070/RM1994v049n01ABEH002133}}.

\bibitem{c97}
O.~Costin,
\textit{Correlation between Pole Location and Asymptotic Behavior for Painlev\'e~I Solutions},
Commun.\ Pure\ App.\ Math.\ \textbf{52} (1999) 461,
\texttt{arXiv:\arxivlink{math/9709223}}.

\bibitem{jk01}
N.~Joshi, A.V.~Kitaev,
\textit{On Boutroux's Tritronqu\'ee Solutions of the First Painlev\'e Equation},
Stud.\ Appl.\ Math.\ \textbf{107} (2001) 253,
\texttt{DOI:\doilink{10.1111/1467-9590.00187}}.

\bibitem{n13}
V.Yu.~Novokshenov,
\textit{Special Solutions of the First and Second Painlev\'e Equations and Singularities of the Monodromy Data Manifold},
Proc.\ Steklov\ Inst.\ Math.\ \textbf{281} (2013) 105,
\texttt{DOI:\doilink{10.1134/S0081543813050106}}.

\bibitem{cch13}
O.~Costin, R.D.~Costin, M.~Huang,
\textit{Tronqu\'ee Solutions of the Painlev\'e Equation \PI},
Constr.\ Approx.\ \textbf{41} (2015) 467,
\texttt{arXiv:\arxivlink{1310.5330}[math.CA]}.

\bibitem{ly52a}
C.N.~Yang, T.D.~Lee,
\textit{Statistical Theory of Equations of State and Phase Transitions 1: Theory of Condensation},
Phys.\ Rev.\ \textbf{87} (1952) 404,
\texttt{DOI:\doilink{10.1103/PhysRev.87.404}}.

\bibitem{ly52b}
T.D.~Lee, C.N.~Yang,
\textit{Statistical Theory of Equations of State and Phase Transitions 2: Lattice Gas and Ising Model},
Phys.\ Rev.\ \textbf{87} (1952) 410,
\texttt{DOI:\doilink{10.1103/PhysRev.87.410}}.

\bibitem{f65}
M.E.~Fisher,
\textit{The Nature of Critical Points},
in ``Lectures in Theoretical Physics VII-C'' (1965) 1.

\bibitem{gkk13}
T.~Grava, A.~Kapaev, C.~Klein,
\textit{On the Tritronqu\'ee Solutions of \PI$^2$},
Constr.\ Approx.\ \textbf{41} (2015) 425,
\texttt{arXiv:\arxivlink{1306.6161}[math-ph]}.

\bibitem{hm80}
S.P.~Hastings, J.B.~McLeod,
\textit{A Boundary Value Problem Associated with the Second Painlev\'e Transcendent and the Korteweg--de~Vries Equation},
Arch.\ Ration.\ Mech.\ Anal.\ \textbf{73} (1980) 31,
\texttt{DOI:\doilink{10.1007/BF00283254}}.

\bibitem{a96}
G.~Akemann,
\textit{Higher Genus Correlators for the Hermitian Matrix Model with Multiple Cuts},
Nucl.\ Phys.\ \textbf{B482} (1996) 403,
\texttt{arXiv:\arxivlink{hep-th/9606004}}.

\bibitem{e04}
B.~Eynard,
\textit{Topological Expansion for the Hermitian 1-Matrix Model Correlation Functions},
JHEP\ \textbf{11} (2004) 031,
\texttt{arXiv:\arxivlink{hep-th/0407261}}.

\bibitem{bm06}
M.~Bertola, M.Y.~Mo
\textit{Commuting Difference Operators, Spinor Bundles and the Asymptotics of Orthogonal Polynomials with Respect to Varying Complex Weights},
Adv.\ Math.\ \textbf{220} (2009) 154,
\texttt{arXiv:\arxivlink{math-ph/0605043}}.

\bibitem{bt16}
M.~Bertola, A.~Tovbis,
\textit{On Asymptotic Regimes of Orthogonal Polynomials with Complex Varying Quartic Exponential Weight},
SIGMA\ \textbf{12} (2016) 118,
\texttt{arXiv:\arxivlink{1612.08732}[nlin.SI]}.

\bibitem{b07}
M.~Bertola,
\textit{Boutroux Curves with External Field: Equilibrium Measures without a Minimization Problem},
Anal.\ Math.\ Phys.\ \textbf{1} (2011) 167,
\texttt{arXiv:\arxivlink{0705.3062}[nlin.SI]}.

\bibitem{b.s07}
B.~Simon,
\textit{Equilibrium Measures and Capacities in Spectral Theory},
Inv.\ Prob.\ Imag.\ \textbf{1} (2007) 713,
\texttt{arXiv:\arxivlink{0711.2700}[math.SP]}.

\bibitem{bt11}
M.~Bertola, A.~Tovbis,
\textit{Asymptotics of Orthogonal Polynomials with Complex Varying Quartic Weight: Global Structure, Critical Point Behaviour and the First Painlev\'e Equation},
Constr.\ Approx.\ \textbf{41} (2015) 529,
\texttt{arXiv:\arxivlink{1108.0321}[nlin.SI]}.

\bibitem{aam13b}
G.~\'Alvarez, L.M.~Alonso, E.~Medina,
\textit{Determination of $S$-Curves with Applications to the Theory of Nonhermitian Orthogonal Polynomials},
J.\ Stat.\ Mech.\ (2013) P06006,
\texttt{arXiv:\arxivlink{1305.3028}[math-ph]}.

\bibitem{hkl13}
D.~Huybrechs, A.~Kuijlaars, N.~Lejon,
\textit{Zero Distribution of Complex Orthogonal Polynomials with Respect to Exponential Weights},
J.\ Approx.\ Theo.\ \textbf{184} (2014) 28,
\texttt{arXiv:\arxivlink{1312.4376}[math.CA]}.

\bibitem{bs16}
P.M.~Bleher, G.L.F.~Silva,
\textit{The Mother Body Phase Transition in the Normal Matrix Model}
Memoirs\ AMS\ \textbf{265} (2020) 1289,
American Mathematical Society (2020),
\texttt{arXiv:\arxivlink{1601.05124}[math-ph]}.

\bibitem{bgm21}
P.~Bleher, R.~Gharakhloo, K.T.-R.~McLaughlin,
\textit{Phase Diagram and Topological Expansion in the Complex Quartic Random Matrix Model}
Comm.\ Pure\ Appl.\ Math.\ \textbf{77} (2024) 1405,
\texttt{arXiv:\arxivlink{2112.09412}[math-ph]}.

\bibitem{gjk21}
P.~Gao, D.L.~Jafferis, D.K.~Kolchmeyer,
\textit{An Effective Matrix Model for Dynamical End of the World Branes in Jackiw--Teitelboim Gravity},
JHEP\ \textbf{01} (2022) 038,
\texttt{arXiv:\arxivlink{2104.01184}[hep-th]}.

\bibitem{bdl05}
I.~Bena, M.~Droz, A.~Lipowski,
\textit{Statistical Mechanics of Equilibrium and Nonequilibrium Phase Transitions: The Yang--Lee Formalism},
Int.\ J.\ Mod.\ Phys.\ \textbf{B19} (2005) 4269,
\texttt{arXiv:\arxivlink{cond-mat/0510278}}.

\bibitem{sk84}
W.~van~Saarloos, D.A.~Kurtze,
\textit{Location of Zeros in the Complex Temperature Plane: Absence of Lee--Yang Theorem},
J.\ Phys.\ \textbf{A17} (1984) 1301,
\texttt{DOI:\doilink{10.1088/0305-4470/17/6/026}}.

\bibitem{sc84}
J.~Stephenson, R.~Couzens,
\textit{Partition Function Zeros for the Two-Dimensional Ising Model},
Physica\ \textbf{129A} (1984) 201,
\texttt{DOI:\doilink{10.1016/0378-4371(84)90028-1}}.

\bibitem{ajjmm17}
M.~Assis, J.L.~Jacobsen, I.~Jensen, J.-M.~Maillard, B.M.~McCoy,
\textit{Analyticity of the Ising Susceptibility: An Interpretation},
J.\ Phys.\ \textbf{A50} (2017) 365203,
\texttt{arXiv:\arxivlink{1705.02541}[math-ph]}.

\bibitem{admt17}
N.G.~Antoniou, F.K.~Diakonos, X.N.~Maintas, C.E.~Tsagkarakis,
\textit{Condensation of Lee--Yang Zeros in Scalar Field Theory},
Phys.\ Rev.\ \textbf{E95} (2017) 052145,
\texttt{arXiv:\arxivlink{1705.00262}[hep-th]}.

\bibitem{bj17}
N.R.~Beaton, E.J.~Janse~van~Rensburg,
\textit{Partition Function Zeros of Adsorbing Dyck Paths},
J.\ Phys.\ \textbf{A51} (2018) 114002,
\texttt{arXiv:\arxivlink{1711.00945}[math-ph]}.

\bibitem{ps93}
C.~Pisani, E.R.~Smith,
\textit{Lee--Yang Zeros and Stokes Phenomenon in a Model with a Wetting Transition},
J.\ Stat.\ Phys.\ \textbf{72} (1993) 51,
\texttt{DOI:\doilink{10.1007/BF01048040}}.

\bibitem{h81}
O.~Haan,
\textit{Large $N$ as a Thermodynamic Limit},
Phys.\ Lett.\ \textbf{B106} (1981) 207.

\bibitem{i19}
K.~Iwaki,
\textit{2-Parameter $\tau$-Function for the First Painlev\'e Equation: Topological Recursion and Direct Monodromy Problem via Exact WKB Analysis},
Commun.\ Math.\ Phys.\ \textbf{377} (2020) 1047,
\texttt{arXiv:\arxivlink{1902.06439}[math-ph]}.

\bibitem{jr26}
C.V.~Johnson, J.~Rodrigues,
\textit{Non-Perturbative Data for Weil--Petersson Volumes and Intersection Numbers using Ordinary Differential Equations},
\texttt{arXiv:\arxivlink{2601.03351}[hep-th]}.

\bibitem{cc01}
O.~Costin, R.~Costin,
\textit{On the Formation of Singularities of Solutions of Nonlinear Differential Systems in Anti-Stokes Directions}, 
Invent.\ Math.\ \textbf{145} (2001) 425,
\texttt{arXiv:\arxivlink{math/0202234}}.

\bibitem{asv17a}
I.~Aniceto, R.~Schiappa, M.~Vonk,
\textit{Physics and Mathematics of 2d Gravity: Movability and Modularity in Painlev\'e~I},
at ``KITP Conference: Resurgence in Gauge and String Theory'' (2017),
(\href{http://online.kitp.ucsb.edu/online/resurgent_c17/vonk}{\texttt{kitp.ucsb.edu/online/resurgent$\_$c17/vonk}}).

\bibitem{asv17b}
I.~Aniceto, R.~Schiappa, M.~Vonk,
\textit{Physics and Mathematics of 2d Gravity: Stokes and Large $N$ Anti-Stokes},
at ``KITP Conference: Resurgence in Gauge and String Theory'' (2017),
(\href{http://online.kitp.ucsb.edu/online/resurgent_c17/aniceto}{\texttt{kitp.ucsb.edu/online/resurgent$\_$c17/aniceto}}).

\bibitem{blmst16}
G.~Bonelli, O.~Lisovyy, K.~Maruyoshi, A.~Sciarappa, A.~Tanzini,
\textit{On Painlev\'e/Gauge Theory Correspondence},
Lett.\ Math.\ Phys.\ \textbf{107} (2017) 2359,
\texttt{arXiv:\arxivlink{1612.06235}[hep-th]}.

\bibitem{eo08}
B.~Eynard, N.~Orantin,
\textit{Algebraic Methods in Random Matrices and Enumerative Geometry},
\texttt{arXiv:\arxivlink{0811.3531}[math-ph]}.

\bibitem{eo09}
B.~Eynard, N.~Orantin,
\textit{Topological Recursion in Enumerative Geometry and Random Matrices},
J.\ Phys.\ \textbf{A42} (2009) 293001,
\texttt{DOI:\doilink{10.1088/1751-8113/42/29/293001}}.

\bibitem{n02}
N.A.~Nekrasov,
\textit{Seiberg--Witten Prepotential from Instanton Counting},
Adv.\ Theor.\ Math.\ Phys.\ \textbf{7} (2003) 831,
\texttt{arXiv:\arxivlink{hep-th/0206161}}.

\bibitem{no03}
N.A.~Nekrasov, A.~Okounkov,
\textit{Seiberg--Witten Theory and Random Partitions},
Prog.\ Math.\ \textbf{244} (2006) 525,
\texttt{arXiv:\arxivlink{hep-th/0306238}}.

\bibitem{gil12}
O.~Gamayun, N.~Iorgov, O.~Lisovyy,
\textit{Conformal Field Theory of Painlev\'e~VI},
JHEP\ \textbf{10} (2012) 038,
\texttt{arXiv:\arxivlink{1207.0787}[hep-th]}.

\bibitem{gil13}
O.~Gamayun, N.~Iorgov, O.~Lisovyy,
\textit{How Instanton Combinatorics Solves Painlev\'e~VI,~V and~IIIs},
J.\ Phys.\ \textbf{A46} (2013) 335203,
\texttt{arXiv:\arxivlink{1302.1832}[hep-th]}.

\bibitem{ilt14}
N.~Iorgov, O.~Lisovyy, J.~Teschner,
\textit{Isomonodromic Tau-Functions from Liouville Conformal Blocks},
Commun.\ Math.\ Phys.\ \textbf{336} (2015) 671,
\texttt{arXiv:\arxivlink{1401.6104}[hep-th]}.

\bibitem{gl16}
P.~Gavrylenko, O.~Lisovyy,
\textit{Fredholm Determinant and Nekrasov Sum Representations of Isomonodromic Tau Functions},
Commun.\ Math.\ Phys.\ \textbf{363} (2018) 1,
\texttt{arXiv:\arxivlink{1608.00958}[math-ph]}.

\bibitem{gl17}
P.~Gavrylenko, O.~Lisovyy,
\textit{Pure $\text{SU}(2)$ Gauge Theory Partition Function and Generalized Bessel Kernel},
Proc.\ Symp.\ Pure\ Math.\ \textbf{18} (2018) 181,
\texttt{arXiv:\arxivlink{1705.01869}[math-ph]}.

\bibitem{ddg20}
F.~Del~Monte, H.~Desiraju, P.~Gavrylenko,
\textit{Isomonodromic Tau Functions on a Torus as Fredholm Determinants, and Charged Partitions},
Commun.\ Math.\ Phys.\ \textbf{398} (2023) 1029,
\texttt{arXiv:\arxivlink{2011.06292}[math-ph]}.

\bibitem{gg18}
A.~Grassi, J.~Gu,
\textit{Argyres--Douglas Theories, Painlev\'e~II and Quantum Mechanics},
JHEP\ \textbf{02} (2019) 060,
\texttt{arXiv:\arxivlink{1803.02320}[hep-th]}.

\bibitem{clt20}
I.~Coman, P.~Longhi, J.~Teschner,
\textit{From Quantum Curves to Topological String Partition Functions II},
Ann.\ Henri\ Poincar\'e\ \textbf{26} (2025) 4271,
\texttt{arXiv:\arxivlink{2004.04585}[hep-th]}.

\bibitem{bgmt24}
G.~Bonelli, P.~Gavrylenko, I.~Majtara, A.~Tanzini,
\textit{Surface Observables in Gauge Theories, Modular Painlev\'e Tau Functions and Non-Perturbative Topological Strings},
\texttt{arXiv:\arxivlink{2410.17868}[hep-th]}.

\bibitem{iilz25}
N.~Iorgov, K.~Iwaki, O.~Lisovyy, Y.~Zhuravlov,
\textit{Many-Faced Painlev\'e~I: Irregular Conformal Blocks, Topological Recursion, and Holomorphic Anomaly Approaches},
\texttt{arXiv:\arxivlink{2505.16803}[math-ph]}.

\bibitem{ggh23}
P.~Gavrylenko, A.~Grassi, Q.~Hao,
\textit{Connecting Topological Strings and Spectral Theory via Non-Autonomous Toda Equations},
\texttt{arXiv:\arxivlink{2304.11027}[hep-th]}.

\bibitem{i03}
A.R.~Its,
\textit{The Riemann--Hilbert Problem and Integrable Systems},
Notices\ Amer.\ Math.\ Soc.\ \textbf{50} (2003) 1389,
\texttt{Notices:\href{http://www.ams.org/notices/200311/fea-its.pdf}{200311}}.

\bibitem{hn24}
Q.~Hao, A.~Neizke,
\textit{A New Construction of $c=1$ Virasoro Blocks},
\texttt{arXiv:\arxivlink{2407.04483}[hep-th]}.

\bibitem{mo19}
O.~Marchal, N.~Orantin,
\textit{Isomonodromic Deformations of a Rational Differential System and Reconstruction with the Topological Recursion: The $\mathfrak{sl}_2$ Case},
J.\ Math.\ Phys.\ \textbf{61} (2020) 061506,
\texttt{arXiv:\arxivlink{1901.04344}[math-ph]}.

\bibitem{eg19}
B.~Eynard, E.~Garcia-Failde,
\textit{From Topological Recursion to Wave Functions and PDEs Quantizing Hyperelliptic Curves},
Forum\ Math.\ Sigma\ \textbf{11} (2023) 1,
\texttt{arXiv:\arxivlink{1911.07795}[math-ph]}.

\bibitem{kz01}
M.~Kontsevich, D.~Zagier,
\textit{Periods},
in ``Mathematics Unlimited --- 2001 and Beyond'' (2001) 771,
\texttt{DOI:\doilink{10.1007/978-3-642-56478-9$\_$39}}.

\bibitem{s04unpub}
C.~Schnell,
\textit{On Computing Picard--Fuchs Equations},
(unpublished, 2004),
\href{https://www.math.stonybrook.edu/~cschnell/pdf/notes/picardfuchs.pdf}{\texttt{//www.math.stonybrook.edu/$\sim$cschnell/picardfuchs}}.

\bibitem{m.123.06}
D.~Mumford,
\textit{Tata Lectures on Theta I, II \& III},
Modern Birkh\"auser Classics (2006).

\bibitem{dk22}
E.~D'Hoker, J.~Kaidi,
\textit{Lectures on Modular Forms and Strings},
\texttt{arXiv:\arxivlink{2208.07242}[hep-th]}.

\bibitem{m.s07}
M.~Schlichenmaier,
\textit{An Introduction to Riemann Surfaces, Algebraic Curves and Moduli Spaces},
Springer (2007),
\texttt{DOI:\doilink{10.1007/978-3-540-71175-9}}.

\bibitem{bel12}
V.M.~Buchstaber, V.Z.~Enolski, D.V.~Leykin,
\textit{Multi-Dimensional Sigma-Functions},
\texttt{arXiv:\arxivlink{1208.0990}[math-ph]}.

\bibitem{bf13}
P.F.~Byrd, M.D.~Friedman,
\textit{Handbook of Elliptic Integrals for Engineers and Scientists},
Grundlehren der Mathematischen Wissenschaften \textbf{67} (1971),
Springer-Verlag,
\texttt{DOI:\doilink{10.1007/978-3-642-65138-0}}.

\end{thebibliography}

\end{document}